 \documentclass[final,3p,times,sort&compress]{elsarticle}

\usepackage{amssymb}

\journal{Physics Reports}

\pdfoutput=1

\usepackage{amsmath}
\usepackage{bbold}
\usepackage{mathtools}
\usepackage{slashed}
\usepackage{hyperref}
\usepackage{graphicx}
\usepackage{pdfpages}
\usepackage{comment}
\usepackage{booktabs}
\usepackage{subfig}
\usepackage{nccmath} 
\usepackage{enumitem}   
\usepackage{etoolbox}

\makeatletter
\patchcmd{\upbracefill}{\m@th}{\scriptstyle\m@th}{}{}
\patchcmd{\upbracefill}{$\braceld$}{$\scriptstyle\braceld$}{}{}
\patchcmd{\upbracefill}{\bracelu}{\bracelu\mkern-1mu}{}{}
\patchcmd{\upbracefill}{\hfill\braceru}{\hfill\mkern-1mu\braceru}{}{}
\makeatother

\newenvironment{subfigure}[2][c]{\begin{minipage}[#1]{#2}}{\end{minipage}}
\newcommand{\subcaption}[2]{\\[1ex]{\small (#1) #2}}
\newcommand{\Vcenter}[1]{\ensuremath{\vcenter{\hbox{#1}}}\xspace}

\input{newcommands}

\def\draftdate{\relax}
\newcommand{\mpar}[1]{\relax}
\def\mda{\relax}
\def\mua{\relax}
\def\mla{\relax}
\def\mdad{\relax}
\def\muad{\relax}
\def\mlad{\relax}
\def\draft{
\def\thtystars{******************************}
\def\sixtystars{\thtystars\thtystars}
\typeout{}
\typeout{\sixtystars**}
\typeout{* Draft mode!
         For final version remove \protect\draft\space in source file *}
\typeout{\sixtystars**}
\typeout{}
\def\draftdate{\today}
\def\musd{\marginpar[\boldmath\hfil$\Uparrow$SD]%
                   {\boldmath$\Uparrow$SD\hfil}%
                    \typeout{marginpar: $\uparrow$}\ignorespaces}
\def\mdsd{\marginpar[\boldmath\hfil$\Downarrow$SD]%
                   {\boldmath$\Downarrow$SD\hfil}%
                    \typeout{marginpar: $\downarrow$}\ignorespaces}
\def\mlsd{\marginpar[\boldmath\hfil$\Rightarrow$SD]%
                   {\boldmath$\Leftarrow $SD\hfil}%
                    \typeout{marginpar: $\leftrightarrow$}\ignorespaces}
\def\muad{\marginpar[\boldmath\hfil$\Uparrow$AD]%
                   {\boldmath$\Uparrow$AD\hfil}%
                    \typeout{marginpar: $\uparrow$}\ignorespaces}
\def\mdad{\marginpar[\boldmath\hfil$\Downarrow$AD]%
                   {\boldmath$\Downarrow$AD\hfil}%
                    \typeout{marginpar: $\downarrow$}\ignorespaces}
\def\mlad{\marginpar[\boldmath\hfil$\Rightarrow$AD]%
                   {\boldmath$\Leftarrow $AD\hfil}%
                    \typeout{marginpar: $\leftrightarrow$}\ignorespaces}
\def\mua{\marginpar[\boldmath\hfil$\uparrow$]%
                   {\boldmath$\uparrow$\hfil}%
                    \typeout{marginpar: $\uparrow$}\ignorespaces}
\def\mda{\marginpar[\boldmath\hfil$\downarrow$]%
                   {\boldmath$\downarrow$\hfil}%
                    \typeout{marginpar: $\downarrow$}\ignorespaces}
\def\mla{\marginpar[\boldmath\hfil$\rightarrow$]%
                   {\boldmath$\leftarrow $\hfil}%
                    \typeout{marginpar: $\leftrightarrow$}\ignorespaces}
\def\Mua{\marginpar[\boldmath\hfil$\Uparrow$]%
                   {\boldmath$\Uparrow$\hfil}%
                    \typeout{marginpar: $\uparrow$}\ignorespaces}
\def\Mda{\marginpar[\boldmath\hfil$\Downarrow$]%
                   {\boldmath$\Downarrow$\hfil}%
                    \typeout{marginpar: $\downarrow$}\ignorespaces}
\def\Mla{\marginpar[\boldmath\hfil\textcolor{red}{$\Rightarrow$}]%
                   {\boldmath\textcolor{red}{$\Leftarrow $}\hfil}%
                    \typeout{marginpar: $\leftrightarrow$}\ignorespaces}
\def\mpar##1{\marginpar{\hbadness10000%
                      \sloppy\hfuzz10pt\boldmath\bf{##1}}%
                      \typeout{marginpar: ##1}\ignorespaces}
\overfullrule 5pt
\oddsidemargin 15mm
\oddsidemargin 0mm
\oddsidemargin -12mm
\marginparwidth 25mm
}

\begin{document}


\begin{frontmatter}



\title{Electroweak Radiative Corrections for Collider Physics}
\tnotetext[]{We dedicate the review to the late Manfred B\"ohm
  who introduced us to the field of electroweak radiative corrections.}


\author[wu]{Ansgar Denner}
\ead{Ansgar.Denner@physik.uni-wuerzburg.de}

\author[fr]{Stefan Dittmaier\corref{mycorrespondingauthor}}
\cortext[mycorrespondingauthor]{Corresponding author}
\ead{Stefan.Dittmaier@physik.uni-freiburg.de}

\address[wu]{%
        Universit\"at W\"urzburg, %
        Institut f\"ur Theoretische Physik und Astrophysik,  %
         Emil-Hilb-Weg 22, 
        97074 W\"urzburg, %
        Germany%
}

\address[fr]{
        Albert-Ludwigs-Universit\"at Freiburg, %
        Physikalisches Institut,  %
        Hermann-Herder-Str. 3, %
        79104 Freiburg, %
        Germany
}

\address{}

\begin{abstract}
  Current particle phenomenology is characterized by the spectacular
  agreement of the predictions of the Standard Model of particle
  physics (SM) with all results from collider experiments
  and by the absence of significant signals of non-standard physics,
  despite the fact that we know that the SM cannot be the ultimate
  theory of nature.  In this situation, confronting theory and
  experiment with high precision is a promising direction to look for
  potential traces of physics beyond the SM.  On the theory side, the
  calculation of radiative corrections of the strong and electroweak
  interactions is at the heart of this task, a field that has seen
  tremendous conceptual and technical progress in the last decades.
  This review aims at a coherent introduction to the field of {\it
    electroweak corrections} and tries to fill gaps in the literature
  between standard textbook know\-ledge and the current state of the art.
  The SM and the machinery for its perturbative evaluation are
  reviewed in detail, putting particular emphasis on renormalization,
  on one-loop techniques, on modern amplitude methods and tools, on
  the separation of infrared singularities in real-emission
  corrections, on electroweak issues connected with hadronic initial
  or final states in collisions, and on the issue of unstable
  particles in quantum field theory together with corresponding
  practical solutions.
\end{abstract}

\begin{keyword}
Electroweak theory
\sep
electroweak corrections
\sep
NLO calculations
\sep
Standard Model
\sep
renormalization
\sep
unstable particles

\PACS
12.15.-y      
\sep
12.15.Lk      
\sep
12.38.Bx      
\sep
13.40.Ks      

\end{keyword}

\end{frontmatter}


\tableofcontents


\clearpage

\section{Introduction}

\noindent
{\it The Standard Model of particle physics}
\\[.3em]
The {\it Standard Model (SM) of particle physics} provides a
successful theory for three out of the four known interactions of
fundamental particles: The strong interaction is described by {\it
  Quantum Chromodynamics
  (QCD)}~\cite{Fritzsch:1973pi,Gross:1973ju,Politzer:1973fx,Gross:1974cs},
and the description of the electromagnetic and weak interactions is
unified in the {\it Glashow--Salam--Weinberg (GSW) model} of the
electroweak (EW) interaction, also called the {\em Electroweak
  Standard Model (EWSM).}  
In this article we review the salient
features of the EWSM and the basic concepts that are required to
prepare precise predictions for EW phenomena that can be tested at
particle colliders.  Since perturbation theory is the method of choice
in such precision calculations, {\it electroweak radiative
  corrections} are at the heart of this task and represent the main
theme of this review.  Effects of the strong interaction only play a
minor role, and gravitational effects no role in the following.

The modern era of the theory of EW interaction began in the 1960s when
the previously suggested phenomenological model of intermediate
massive vector bosons was turned into a Yang--Mills gauge theory by
Glashow~\cite{Glashow:1961tr}, Weinberg~\cite{Weinberg:1967tq} and
Salam~\cite{Salam:1968rm}.  The obstacle that pure Yang--Mills gauge
fields predict massless gauge bosons, but experimental facts require
the force carriers of the weak interaction to be massive, was overcome
by spontaneously breaking the EW $\SU(2)_{\rw} \times \U(1)_{Y}$
gauge symmetry down to electromagnetic $\U(1)_{\text{em}}$ invariance.
This spontaneous symmetry breaking is driven by the gauge interaction
with an elementary scalar field that develops a non-vanishing
vacuum expectation value (vev)---a mechanism nowadays known as
{\em Brout--Englert--Higgs} or simply {\em Higgs mechanism} that
was actually developed in a series of papers by several 
groups~\cite{Englert:1964et,Higgs:1964pj,Higgs:1964ia,Guralnik:1964eu,%
Higgs:1966ev}.
Specifically, the GSW model employs a complex scalar doublet to
break EW symmetry, so that three out of the four scalar degrees of freedom
deliver the longitudinal polarizations of the massive weak gauge
bosons $\PW^\pm$ and $\PZ$, while the photon remains massless.
The fourth scalar degree of freedom corresponds to a neutral,
massive boson, the SM {\em Higgs boson}.

The dynamical generation of masses for the elementary particles%
\footnote{In this context {\em elementary} means {\em point-like},
in contrast to composite particles like nucleons. 
For the latter, the binding energy even delivers the
major part of the mass.}
by interaction with the Higgs field also offers a solution to
another theoretical problem 
in the fermionic sector of the theory. 
Owing to observation of parity (P) violation, 
fermions are {\em chiral}, \ie
left- and right-handed fermions interact differently with the 
weak gauge bosons, a fact that forbids the introduction of plain
fermion mass terms in the underlying Lagrangian due to the
gauge-invariance requirement. 
The so-called {\em Yukawa couplings} between left- and right-handed 
fermions and the Higgs doublet introduce both fermion masses
and Higgs-boson--fermion interactions in a consistent manner.
For leptons, this structure was already part of the original GSW
model, where neutrinos were taken as massless.%
\footnote{The observation of neutrino
  oscillations~\cite{Cleveland:1998nv,Fukuda:1998mi,Ahmad:2002jz}
  requires the introduction of neutrino masses, which are usually not
  considered part of the EWSM. These are, however, irrelevant for
  physics at high-energy colliders and not considered in this review.}
The description of hadronic degrees of freedom in EW interactions was
suggested by Glashow, Iliopoulos and Maiani~\cite{Glashow:1970gm} in
the early 1970s.  Up to currently achievable energies (TeV range),
both leptons and quarks appear in three generations.  In particular,
from the experiments at LEP1 we know that there are exactly three
with light neutrinos \cite{Decamp:1989tu,ALEPH:2005ab}.  While leptons
of different generations do not mix as long as neutrinos are mass
degenerate (e.g.\ if taken massless), mixing of the three quark
generations occurs and is extremely well described by the
{\it Cabibbo--Kobayashi--Maskawa (CKM)
matrix}~\cite{Cabibbo:1963yz,Kobayashi:1973fv}, which contains the only
source of CP~violation in the SM.  Note that the introduction of
fermion masses via Yukawa couplings to the single Higgs doublet field
leads to a distinctive phenomenological imprint in the coupling
structure, \ie all fermionic couplings to the Higgs boson are
proportional to the mass of the corresponding fermion.

The breakthrough of the SM came in the early 1970s with the
proof of the renormalizability of gauge theories with and
without spontaneous symmetry breaking given by 
't~Hooft and Veltman~\cite{tHooft:1971akt,tHooft:1971qjg,tHooft:1972tcz,tHooft:1972qbu}
and Lee and Zinn-Justin~\cite{Lee:1972fj,Lee:1974zg,Lee:1973fn,Lee:1973rb}.
Moreover, the fermionic matter content of the SM renders the
model anomaly free.
In summary, the SM is a
mathematically consistent quantum field theory,
a fact that serves as our basis to work out predictions
for collider experiments at a level of precision that is
essentially only limited by our technical capabilities to evaluate
higher orders in perturbation theory and by the level to which
we can extract necessary non-perturbative information 
(parton distribution functions, hadronic vacuum polarization, etc.)
from experiment.

\vspace{.5em}
\noindent
{\it Electroweak precision tests and theoretical concepts}
\\*[.3em]
After their discovery at the UA1 and UA2 experiments in 1983
\cite{Arnison:1983rp,Banner:1983jy,Arnison:1983mk,Bagnaia:1983zx}, the
precise experimental investigation of the EW gauge bosons started at
the CERN Large Electron--Positron Collider (LEP) and at the Stanford
Linear Collider (SLC) at SLAC in the late 1980s.  In the first LEP
phase (LEP1, 1989--1995) and at the SLC, 
millions of Z~bosons were
produced as $s$-channel resonances, \ie\ in the reaction
$\Pep\Pem\to\PZ/\gamma^*\to f\bar f$, providing extremely precise
measurements of the Z-boson mass, decay widths, cross sections, and
asymmetries.  From these observables, pseudo-observables such as
effective couplings of the Z~boson to various fermions were deduced,
in particular the {\it effective weak mixing angle} which determines
the EW coupling strength at the Z-boson resonance.  On the theory
side, the concepts for perturbative precision calculations had to be
worked out, a process that started in the late 1970s with technical
articles from 't~Hooft and Veltman~\cite{'tHooft:1978xw} and Passarino
and Veltman~\cite{Passarino:1978jh}, and that produced the concept of
renormalization in the EWSM in different
formulations~\cite{Ross:1973fp,Sirlin:1980nh,Aoki:1982ed,Bohm:1986rj,%
  Jegerlehner:1991dq,Denner:1991kt} in the work-up and early phase of
LEP1 and the SLC.  Moreover, a concept was worked out to describe the
Z-boson resonance and its (pseudo-)observables to high precision (see,
for instance, the
reviews~\cite{Altarelli:1989hv,Bardin:1997xq,Bardin:1999ak,Bardin:1999gt}
and references therein).

The second phase of LEP (LEP2, 1996--2000), provided first direct
experimental access to the non-abelian gauge-boson interactions
between photons, Z, and W~bosons as well as precision measurements of
the W-boson mass via the reaction $\Pep\Pem\to\PWp\PWm\to4f$ above the
W-boson pair production threshold.  On the theory side, this implied
more challenges, too.  Precision calculations for four (and more)
particles in the final state triggered great advances in the
development of Monte Carlo integrators and generators (see, e.g.,
\citeres{Bardin:1997gc,Grunewald:2000ju}).  In particular, the
technique of adaptive {\em multi-channel Monte Carlo
  integration}~\cite{Berends:1984gf,Hilgart:1992xu,Berends:1994pv,Kleiss:1994qy}
was developed, the method that 
has served as basis for almost all major
Monte Carlo generators for collider experiments in the previous two
decades.  Moreover, the inclusion of real-emission effects in the
description of multi-particle processes lead to more involved patterns
in the structure of infrared (IR) singularities caused by phase-space
regions of soft and/or collinear photon (or QCD parton) emission.  To
resolve this issue, techniques to isolate IR singularities via {\em
  phase-space slicing} \cite{Harris:2001sx,Giele:1991vf} or {\em
  subtraction} \cite{Frixione:1995ms,Catani:1996vz,Dittmaier:1999mb}
were further developed and formulated in an algorithmic way that
allows for automated evaluation.  Finally, a more general concept for
describing resonances in combination with radiative corrections but
without violating basic principles like gauge invariance was required.
Systematic expansions of matrix elements about the resonance poles
\cite{Stuart:1991xk,Aeppli:1993rs}
and approximations based on the leading terms
provided a successful framework for describing W-pair production
at LEP2 (see, e.g., \citere{Grunewald:2000ju}),
but also triggered further developments towards a {\em theory of
unstable particles}. 
The most flexible method that emerged from this learning process,
in our view, is the 
{\em complex-mass scheme}~\cite{Denner:1999gp,Denner:2005fg},
which provides a gauge-invariant, uniform description of cross sections in
resonant and non-resonant regions at next-to-leading order (NLO) in
perturbation theory.

After the millennium, the $\Pp\bar\Pp$~collider Tevatron at Fermilab
continued and further tightened the EW precision tests, in particular,
by providing precision measurements of the W-boson and top-quark masses.
The SM passed all those tests with very little tension between
observations and predictions (see, e.g., \citeres{Schael:2013ita,Erler:2019hds}).
It should be emphasized that this agreement between experiment and theory
was (and still is) only seen if higher-order radiative corrections
of QCD and EW origin are properly taken into account in predictions,
\ie the SM is truly tested as quantum field theory.
The major outcome of the SM fit to all precision data was 
a prediction of a preferred range for the mass of the Higgs boson
well before its discovery in 2012 by the ATLAS~\cite{Aad:2012tfa}
and CMS~\cite{Chatrchyan:2012xdj} collaborations at the
CERN Large Hadron Collider (LHC).

\vspace{.5em}
\noindent
{\it Electroweak phenomenology and theory challenges in the LHC era}
\\[.3em]
Among the global search for new phenomena and the detailed
confrontation of all observations with SM predictions, one of the
central questions of present-day particle physics is whether or to
which extent the discovered Higgs boson represents the SM Higgs boson.
Or more generally, is EW symmetry breaking realized in nature as
described by the Higgs mechanism; if so, is the Higgs sector realized
in the minimal form as in the SM, or are there more Higgs bosons; if
not, what is the right formulation of EW symmetry breaking, and what
is the role of the discovered Higgs boson in it; is the observed Higgs
boson a composite state, etc.?  Since practically all SM extensions
modify the sector of EW symmetry breaking of the SM, Higgs-boson
physics is certainly the right place to look for potential deviations
from the SM.  Up to now, no significant discrepancies between observed
properties of the Higgs boson (production cross sections, decay
branching ratios, Higgs couplings) and SM predictions have been found,
showing that any potential differences are small and subtle.  Given 
the fact that no spectacular signals of physics beyond the SM
are in sight, the ultimate answer to this question can only be given
via {\it precision} in theory and
experiment~\cite{Dittmaier:2011ti,Dittmaier:2012vm,%
Heinemeyer:2013tqa,deFlorian:2016spz}.

Again, switching from EW physics at LEP, the SLC, and Tevatron to the
LHC, brought further theoretical challenges.  While only a few
processes at Tevatron, such as the Drell--Yan-like production of Z or
W~bosons, required the inclusion of EW radiative corrections in
predictions, the higher luminosity and the deeper energy reach of the
LHC render EW corrections to all major processes important or at least
relevant in the upcoming high-luminosity phase.  For a proper account
of EW corrections, their consistent inclusion in the perturbative
evaluation of cross sections in the QCD-improved parton model for
hadronic collisions is necessary. In particular, the photon appears as
additional parton in the colliding protons with an own photon
distribution function.  Moreover, multi-particle final states play a
much more important role at the LHC. On the one hand, the higher
scattering energy at the LHC globally leads to higher jet
multiplicities in association with any signature of colour-neutral
particles, and the number of partonic channels grows enormously.  On
the other hand, new interesting classes of EW processes with
multi-particle final states become accessible, including EW
vector-boson scattering, $\Pp\Pp\to4\Pl+2\,\mathrm{jets}+X$, or
triple-EW-gauge-boson production such as
$\Pp\Pp\to\PW\PW\PW\to3\Pl\,3\nu+X$.  To make one-loop calculations of
such multi-particle amplitudes possible and numerically efficient, it
was necessary to develop techniques and automated amplitude generators
that are not based on individual Feynman diagrams anymore, but proceed
recursively on the basis of appropriate substructures of amplitudes
such as off-shell currents or subamplitudes based on generalized
unitarity.  For the automated calculation of NLO EW corrections, the
programs {\sc MadLoop}~\cite{Alwall:2014hca},
{\Openloops}~\cite{Cascioli:2011va,Kallweit:2014xda,Buccioni:2019sur},
and \Recola~\cite{Actis:2012qn,Actis:2016mpe} are the most powerful
amplitude generators, which can deal with up to $\sim8{-}9$ external
particles in one-loop amplitudes.

\vspace{.5em}
\noindent
{\it Structure of the review}
\\[.3em]
The above statements have illustrated how the conceptual and technical
progress towards EW precision for present-day collider physics step by
step came along with new collider experiments with increasing
precision and energy reach.  We are not aware of any comprehensive
document or textbook that collects the corresponding theoretical
progress of the previous 30 years in the field of EW radiative
corrections in a coherent form. This situation renders it particularly
difficult for newcomers to grow into the field; typically they have to
read series of papers with some redundancy or with gradual
improvements or generalizations from paper to paper. 
It is the aim of this review to
improve on this situation by presenting the major concepts for the
calculation of EW corrections in a unified manner.  Of course, some
disclaimer is in order at this point.  No review can be fully
comprehensive, and the selection of topics and their emphasis is
necessarily subjective to some extent.  We have tried to be
as complete as possible, and if we did not spell out all techniques,
methods, or concepts in detail, we at least intend to give the most
important references for alternatives.  Generally, we put more
emphasis on topics, both theoretical and phenomenological, where we
see the more urgent need in the literature.  For further links to the
current research frontier in higher-order calculations and
phenomenological applications, we recommend to consult dedicated
working-group reports like those of the Les Houches workshops on TeV
colliders~\cite{Andersen:2014efa,Badger:2016bpw,Bendavid:2018nar}, of
the LHC Higgs Cross Section Working
Group~\cite{Dittmaier:2011ti,Dittmaier:2012vm,%
  Heinemeyer:2013tqa,deFlorian:2016spz}, or reviews and books on
phenomenological applications like
\citeres{Djouadi:2005gi,Dittmaier:2012nh,Schorner-Sadenius:2015cga,%
Spira:2016ztx,Campbell:2017hsr}.

In detail, the structure of the review is as follows:
\begin{myitemize}
\item
In \refse{se:sm} we introduce the SM, defining all sectors in detail.
Its quantization is performed with the standard Faddeev--Popov method
and, alternatively, employing back\-ground-field quantization, which
is beneficial for some subsequent sections where gauge-invariance
properties play a role. Moreover, some Slavnov--Taylor identities
are recalled that are used in the renormalization and in the formulation
of the Goldstone-boson equivalence theorem.
Finally, \refse{se:sm} concludes with a brief introduction
to the {\em Standard Model Effective Theory} which is used in current
analyses of LHC data to look for potential deviations from SM predictions in
a widely model-independent way.
\item Section~\ref{se:virt} provides an introduction to various
  concepts and techniques for the calculation of virtual one-loop EW
  corrections.  We provide a detailed description of one-loop
  renormalization of the EWSM in the on-shell scheme 
both in the conventional and in the background-field formalism and
discuss techniques for calculating one-loop integrals.  Moreover,
methods for evaluating one-loop amplitudes and different one-loop
amplitude generators are briefly introduced.
\item
Section~\ref{se:real} turns to real-emission corrections,
starting from a general discussion of the IR (soft and collinear) limits
of one-particle emission amplitudes relevant for NLO EW calculations
and followed by a brief introduction into the techniques of
{\em two-cutoff slicing} and {\em dipole subtraction}  for the isolation of
IR singularities in real EW corrections.  
Subsequently,
we turn to specific issues connected with hadrons in the initial or
final states of collisions: 
electromagnetic corrections to parton distribution functions and
the separation of photons and jets in the
final state.  Finally, we end the section 
with a discussion of enhanced photonic corrections induced by
collinear photon--lepton splittings and the evolution of their leading
logarithmic behaviour beyond NLO.
\item Section~\ref{se:general} is devoted to some general aspects of
  EW NLO corrections. We discuss the EW input-parameter schemes and
  the possibilities to classify EW corrections into gauge-independent
  subcontributions.  We sketch the Goldstone-boson
  equivalence theorem including the correction factors appearing in
  higher orders.  Section~\ref{se:general} closes with a discussion
  of the structure of EW corrections at high scattering energies,
  which is particularly important for new-particle searches at the
  LHC.

\item Section~\ref{se:unstable} picks up the issue of unstable
  particles in quantum field theory and methods to describe resonances
  in perturbative calculations. The discussion starts with a
  description of the general problem, the naive {\em narrow-width
    approximation} and possible improvements, the issue of gauge
  invariance, and the precise definition of mass and width of a
  resonance in perturbation theory. The rest of the section is devoted
  to the most frequently applied schemes to treat resonances in EW
  higher-order calculations: the {\em pole scheme} and {\em pole
    approximation}, 
 the {\em complex-mass  scheme},  
and brief comments on further schemes.
\item In \refse{se:conclusions} we give our conclusions and a brief
  outlook.
\item The appendices provide further useful details, such as a
  complete set of Feynman rules for the SM, our conventions for Green
  functions, as well as a derivation of
  the Ward identity used to simplify the renormalization of the
  electric charge.
\end{myitemize}

\section{The Standard Model of particle physics}
\label{se:sm}

\subsection{Lagrangian of the Standard Model}
\label{se:lagrangian}

We formulate the classical Lagrangian of the EWSM in a form in
which the gauge symmetry is manifest and then pass to the
representation in terms of fields corresponding to charge and mass 
eigenstates and experimentally accessible parameters so that its
particle content and the physical meaning of its free parameters
become visible. In the following we refer to this form as
``physical basis'' of fields and parameters.

\subsubsection{Lagrangian of the Electroweak Standard Model in the
  symmetric basis}
\label{se:symmetric_lagrangian}

As the gauge group is not simple but rather a product of $\SU(2)_\rw$ and
$\U(1)_Y$, EW interactions are described by two
{\em gauge coupling constants}: $g_{2}$ for the {\em weak isospin}
group $\SU(2)_{\rw}$ and $g_{1}$ for the {\em weak hypercharge} group
$\U(1)_{Y}$.  We denote the three generators of weak isospin by
$I^{a}_{\rw}$ $(a=1,2,3)$
and the generator of weak hypercharge by $Y_{\rw}$. The
generator~$Q$ of the electric charge is defined via the
{Gell-Mann--Nishijima relation}
\beq
Q=I_{\rw}^{3}+\frac{Y_{\rw}}{2}.
\label{Gell-Mann-Nishijima}
\eeq 

According to the dimension of the gauge group $\SU(2)_{\rw}\times
\U(1)_{Y}$ 
there are four EW gauge fields, which transform according to
the adjoint representation.  The gauge fields belonging to the
weak-isospin group $\SU(2)_{\rw}$ are denoted $\FW^{a}_{\mu}(x)$
and the one belonging to the weak-hypercharge group
$\U(1)_{Y}$ is called $\FB_{\mu}(x)$.
The corresponding field strengths read
\begin{equation}\label{eq:EW_field_strengths}
\FB_{\mu\nu} ={} \partial_{\mu }\FB_{\nu }-\partial_{\nu}\FB_\mu, \qquad
\FW^{a}_{\mu\nu} ={} \partial_{\mu }\FW^{a}_{\nu }-\partial_{\nu 
}\FW^{a}_\mu + g_{2}\eps^{abc}\FW^{b}_{\mu }\FW^{c}_{\nu },
\end{equation}
where $\eps^{abc}$ are the totally antisymmetric structure constants
of $\SU(2)$.
The gauge interaction of the matter fields is determined by the covariant
derivative%
\footnote{\label{fn:conventions} We adopt the conventions of
  \citeres{Bohm:1986rj,Hollik:1988ii,Denner:1991kt,Bohm:2001yx}. Different sign
  conventions are used in the literature. Often the form $D_{\mu} =
  \partial_{\mu} +\ri g_{2} I^{a}_{\rw} \FW^{a}_\mu +\ri g_{1}
    \frac{Y_{\rw}}{2}\FB_\mu$ is used
  \cite{Gunion:1989we,Grzadkowski:2010es}, which differs in the sign
  of the $\SU(2)_\rw$ gauge fields $\FW^{a}_\mu$. In the field basis 
  corresponding to charge and mass eigenstates this
  corresponds to a sign change in the $\FWpm$ and $\FZ$ fields. Peskin
  and Schroeder~\cite{Peskin:1995ev} and Schwartz~\cite{Schwartz:2013pla},
  on the other hand, use the
  convention $D_{\mu} = \partial_{\mu} -\ri g_{2} I^{a}_{\rw}
    \FW^{a}_\mu -\ri g_{1} \frac{Y_{\rw}}{2}\FB_\mu$, which is
  related to our convention by a sign change in the $\U(1)_Y$ gauge
  field $\FB_\mu$ and the photon field $\FA_\mu$. Parameter relations are
  not affected by these transformations. However, the signs of Feynman
  rules and Green functions change.}%
\beq\label{eq:covdergsw}
D_{\mu} = \partial_{\mu} -\ri
    g_{2} I^{a}_{\rw} \FW^{a}_\mu +\ri g_{1}
    \frac{Y_{\rw}}{2}\FB_\mu.
\eeq

In order to generate mass terms for the weak gauge bosons, the Higgs
mechanism is employed to break the $\SU(2)_{\rw}\times\U(1)_Y$
symmetry in such a way that the electromagnetic symmetry
$\U(1)_{\mathrm{em}}$ remains exact.  To this end, a weak-isospin
doublet $\Phi(x)$ of two complex scalar fields, the Higgs doublet, is
introduced which couples in a gauge-invariant way to the vector bosons
($I^{a}_{\rw}=\tau^{a}/2$ with Pauli matrices $\tau^{a}$, $a=1,2,3$).
To allow for an electrically neutral component with non-vanishing {\it
  vacuum expectation value} (vev), 
its hypercharge must be
$Y_{\rw,\Phi}=\pm1$.  Using the convention $Y_{\rw,\Phi}=+1$, we can
write the Higgs doublet as $\Phi(x)=\left(\phi^{+}(x),
  \phi^{0}(x)\right){}^\rT$, where the upper indices indicate the
electric charges of the components fixed by the Gell-Mann--Nishijima
relation \refeqf{Gell-Mann-Nishijima}.  The choice of a weak-isospin
doublet with hypercharge $Y_{\rw,\Phi}=\pm1$ also allows for {\it
  Yukawa couplings}, \ie gauge-invariant couplings between scalar and
fermion fields, which are necessary for the generation of fermion
masses via spontaneous symmetry breaking.

The left-handed fermions of each lepton ($L$) and quark ($Q$)
generation are grouped into $\SU(2)_{\rw}$ doublets (we suppress the
colour index of the quark fields)
\beq
\FL_{j}^{\prime\rL}=\omega _{-}  \FL'_{j} =
\left( \barr{l} {\nu}_{j}^{\prime\rL} \\ \Fl_{j}^{\prime\rL} \earr \right) , \qquad
\FQ_{j}^{\prime\rL}=\omega _{-}  \FQ'_{j} =
\left( \barr{l} \Fu_{j}^{\prime\rL} \\ \Fd_{j}^{\prime\rL} \earr \right) 
\eeq
 with representation
matrices $I^{a}_{\rw} = \tau^{a}/2$, 
and the right-handed fermions into singlets ($I^{a}_{\rw} = 0$)
\beq
\Fl_{j}^{\prime\rR}=\omega _{+}  \Fl'_{j}, \qquad
\Fu_{j}^{\prime\rR}=\omega _{+}  \Fu'_{j}, \qquad
\Fd_{j}^{\prime\rR}=\omega _{+}  \Fd'_{j},
\eeq
where 
\beq
\omega _{\pm}=\frac{1}{2}(1\pm\gamma _{5})
\eeq 
is the projector on right- and left-handed fields, respectively,
$j=1,2,3$ is the generation index, and $\nu$, $\Fl$, $\Fu$, and $\Fd$
stand for neutrinos, charged leptons, up-type quarks, and down-type
quarks, respectively.  The weak hypercharges of the right- and
left-handed multiplets are chosen in such a way that the known
electric charges of the fermions are reproduced by the
Gell-Mann--Nishijima relation \refeqf{Gell-Mann-Nishijima}, leading to
\begin{align}
\setlength{\arraycolsep}{1pt}
\begin{array}[b]{rccclrcccl}
              & & Y_{\rw,L,i} + 1 &=& 2 Q_{\nu,i} = 0 ,& \qquad
Y_{\rw,l,i}   &=& Y_{\rw,L,i} - 1 &=& 2 Q_{l,i} =  -2 ,\\[.3em]
Y_{\rw,u,i}   &=& Y_{\rw,Q,i} + 1 &=& 2 Q_{u,i} =  \frac{4}{3}, & \qquad
Y_{\rw,d,i}   &=& Y_{\rw,Q,i} - 1 &=& 2 Q_{d,i} = -\frac{2}{3}.
\end{array}
\end{align}
We do not include right-handed neutrinos, since these are irrelevant
for collider physics.%
\footnote{Right-handed Dirac neutrinos can be easily included in the
  SM without affecting its basic structure. This is in fact necessary
  in order to allow for the description of the experimental results on
  neutrino oscillations. However, if the right-handed neutrinos are
  Majorana particles, additional Majorana mass terms appear (see,
  for instance,
  \citeres{Bilenky:2014ema,Hernandez:2017txl,Tanabashi:2018oca}).}  
The
  primes at the fermion fields indicate eigenstates of the EW
  interaction, \ie the covariant derivatives are diagonal with respect
  to the generation indices in this basis. These states are not
  necessarily mass eigenstates. Because left- and right-handed fields
  transform according to different representations of the symmetry
  group $\SU(2)_{\rw}\times\U(1)_Y$, the theory is chiral, and
  explicit mass terms for the fermions are forbidden.

The classical Lagrangian ${ \cal L}_{\mathrm{class}}$ of the GSW theory 
consists of all renormalizable terms that respect the  $\SU(2)_{\rw}\times
\U(1)_{Y}$ symmetry:
\begin{align}
\cL_{\mathrm{class}} ={}& {-\frac{1}{4}}\FB_{\mu\nu}\FB^{\mu\nu} 
               -\frac{1}{4}\FW^{a}_{\mu\nu}\FW^{a,\mu\nu} \nl
& + (D_{\mu}\Phi)^{\dag}(D^{\mu}\Phi) 
+ \mu^2 \left(\Phi^{\dag}\Phi\right) 
-\frac{\lambda}{4}\left(\Phi^{\dag}\Phi\right)^2 
\nl
&+ \sum_{i}\,\left(
 \longbar{\FL}_{i}^{\prime\rL} \ri \slashed{D} \FL_{i}^{\prime\rL}
+\longbar{\FQ}_{i}^{\prime\rL} \ri \slashed{D} \FQ_{i}^{\prime\rL}
\right) 
 + \sum_{i}\,\left(
 \bar{\Fl}_{i}^{\prime\rR} \ri \slashed{D} \Fl_{i}^{\prime\rR}
+\bar{\Fu}_{i}^{\prime\rR} \ri \slashed{D} \Fu_{i}^{\prime\rR}
+\bar{\Fd}_{i}^{\prime\rR} \ri \slashed{D} \Fd_{i}^{\prime\rR}
\right)\nl
& -\sum_{i,j} \left(
 \longbar{\FL}_{i}^{\prime\rL}G^{\Fl}_{ij}\Fl_{j}^{\prime\rR}\Phi
+\longbar{\FQ}_{i}^{\prime\rL}G^{\Fu}_{ij}\Fu_{j}^{\prime\rR}\Phi^{\mathrm{c}}
+\longbar{\FQ}_{i}^{\prime\rL}G^{\Fd}_{ij}\Fd_{j}^{\prime\rR}\Phi
+ \hc \right),
\label{eq:LSM_sym}
\end{align}
where $\hc$ means hermitean conjugate.
The first line of \refeq{eq:LSM_sym} represents the pure gauge-field
Lagrangian. The second line describes the kinetic terms of the scalar
doublet, its interaction with the gauge fields, and the {\em Higgs
  potential}, involving the quartic coupling $\la$, which is positive
as required by vacuum stability, and the mass parameter $\mu^2$. The
third line summarizes the kinetic terms of the fermions and their
gauge interaction. Finally, the last line represents the Yukawa
interactions of the scalar doublet with the fermions, where $G^{f}$
denote Yukawa coupling matrices. Since the two-dimensional
representation of $\SU(2)$ is equivalent to its complex conjugate
representation, the scalar doublet allows for Yukawa couplings to
 down-type right-handed fermion fields via $\Phi$ and to
up-type right-handed fermion fields via the
charge-conjugate field
\beq\label{eq:phi_cc}
\Phi^{\mathrm{c}}(x) = \ri\tau^2\Phi^*=
(\phi^{0*}(x),-\phi^{-}(x))^{\rT}. 
\eeq
For $\mu^2>0$, the Higgs field acquires a non-vanishing vev that leads
to mass terms for the gauge bosons and---arising from the Yukawa
couplings---also for the fermions.

The Lagrangian \refeqf{eq:LSM_sym} is invariant under the infinitesimal
{\it gauge transformations},
\begin{align}\label{eq:gtsym}
\FW_\mu^{a}(x) \to{}& \FW_\mu^a(x) + \partial_\mu \de\theta^a(x) +
\eps^{abc} g_2 \FW_\mu^b \de\theta^c(x), \nl
\FB_\mu^{}(x) \to{}& \FB_\mu(x) + \partial_\mu \de\theta^Y(x), \nl
\Phi(x) \to{}& \left[1-\ri \frac{1}{2}g_1\de\theta^Y(x) + 
\ri \frac{\tau^{a}}{2}g_2\de\theta^{a}(x)\right] \Phi(x),\nl
F^{\prime\rL}_i(x) \to{}& 
\left[1-\ri \frac{Y_{\rw,F,i}}{2}g_1\de\theta^Y(x) + 
\ri \frac{\tau^{a}}{2}g_2\de\theta^{a}(x)\right] F^{\prime\rL}_i(x),
\quad F=\FL,\FQ, \nl
\Ff^{\prime\rR}_i(x) \to{}& 
\left[1-\ri \frac{Y_{\rw,f,i}}{2}g_1\de\theta^Y(x)\right]
\Ff^{\prime\rR}_i(x), \quad \Ff=\Fl,\Fu,\Fd,
\end{align} 
where $\de\theta^{a}$ and $\de\theta^{Y}$ are the parameters of the gauge
transformations corresponding to the groups $\SU(2)_{\rw}$ and
$\U(1)_Y$, respectively.%
\footnote{The signs of the parameters $\de\theta$ are linked to the
  corresponding fields and should be adapted to different conventions
  analogously as specified in footnote \ref{fn:conventions}.}

The Lagrangian ${\cal L}_{\mathrm{class}}$  depends on the
two gauge couplings $g_{1}$ and $g_{2}$, the Yukawa couplings $G^{f}_{ij}$,
and the parameters $\mu^{2}$ and $\lambda$ of the scalar potential.
 

\subsubsection{Lagrangian of the Electroweak Standard Model in the physical basis}
\label{se:physical_lagrangian}

The scalar-field self-interaction is chosen in such a way that the
classical ground state appears for non-vanishing scalar field, \ie the
scalar potential has a minimum for
\begin{equation} \label{eq:vdef}
|\langle\Phi\rangle|^{2} = \frac{2\mu ^{2}}{\lambda } =
\frac{\varv^{2}}{2} \ne 0  . 
\end{equation}
We choose a ground state $\Phi_0$ 
that is annihilated by the electric charge operator $Q$,
\beq \label{eq:Qphi0}
Q \Phi_0 = \left(\frac{\tau^3}{2} + \frac{1}{2}Y_{\rw,\Phi}\right)\Phi_0 =
\left(\barr{l@{\ }r} 1 \quad & 0 \\ 0 & 0 \earr\right) \Phi_0 = 0,
\eeq
so that the remaining (unbroken) symmetry is the one of
electromagnetic gauge transformations $\U(1)_{\mathrm{em}}$.  Using
the solution of \refeqs{eq:vdef} and \refeqf{eq:Qphi0}, 
\begin{equation} \label{eq:phi0_ug}
\Phi_{0} = \left( \begin{array}{c}
                    0\\
                   \frac{\varv}{\sqrt{2}}
                  \end{array} \right),
\end{equation}
which is unique up to a phase,
we can parametrize the scalar doublet as 
\begin{equation} \label{eq:phidec}
\Phi (x) = \left( \begin{array}{c}
\phi ^{+}(x) \\ \frac{1}{\sqrt2}\bigl[\varv +\FH(x) +\ri\chi (x) \bigr]
\end{array} \right)  , \qquad \phi^-(x) = [\phi^+(x)]^{\dagger},
\end{equation}
where $\varv>0$ and
$\FH(x)$, $\chi(x)$, and $\phi^\pm(x)$ have vanishing vev.
The fields $\phi^+(x)$, $\phi^-(x)$, and
$\chi(x)$, the {\it would-be Goldstone fields}, turn out to be
unphysical degrees of freedom and can be eliminated by a transition to
the {\it unitary gauge}, where $\phi^{\pm}=\chi=0$ and
the physical content of the EWSM can be extracted most easily.

The physical degrees of freedom can be classified as eigenstates of
electric charge and mass.  Inserting \refeq{eq:phidec} into the
Lagrangian \refeq{eq:LSM_sym} and diagonalizing the resulting mass
matrices one obtains the following fields corresponding to mass
eigenstates
\begin{alignat}{3}
\FW_{\mu }^{\pm}  ={} & \frac{1}{\sqrt{2}}
\left( \FW_{\mu }^{1} \mp \ri \FW_{\mu }^{2} \right) , &\qquad
\left(\barr{l} \FZ_{\mu } \\ \FA_{\mu } \earr \right) =& {}
\left(\barr{rr} \cw & \sw \\ -\sw & \cw  \earr \right)
\left(\barr{l} \FW_{\mu }^{3} \\ \FB_{\mu } \earr \right) , \nl
\Ff^{\rL}_{i} ={} & \sum_k U_{ik}^{f,\rL}f_{k}^{\prime\rL} , &\qquad
\Ff^{\rR}_{i} = &{} \sum_k U_{ik}^{f,\rR}f_{k}^{\prime\rR} , 
\label{eq:physfields}
\end{alignat}
where 
\beq
\cw=\cos\theta_{\rw}=\frac{g_{2}}{\sqrt{g_{2}^{2}+g_{1}^{2}}},
\qquad \sw=\sin\theta_{\rw} ,
\label{eq:ewangle1}
\eeq
with the {\em weak mixing angle} $\theta _{\rw}$, and $f$ stands for
$\nu $, $l$, $u$, or $d$.  The resulting masses read
\begin{alignat}{5}
\MW & {}= \frac{1}{2}g_{2}\varv,\qquad  &
\MZ & {}= \frac{1}{2}\sqrt{g_{1}^{2}+g_{2}^{2}} \; \varv, 
& m_{\gamma } & = 0,
\nl
\MH & = \sqrt{2\mu^2} = \sqrt{\frac{{\la}}{{2}}}\,\varv,\qquad &
m_{\Pf,i}  &=  \frac{\varv}{\sqrt{2}}
\sum_{k,m} U_{ik}^{f,\rL} G_{km}^{f} U_{mi}^{f,\rR\dagger},\qquad&& 
\label{eq:SMmass}
\end{alignat}
where $\MH$ is the mass of the {\em Higgs boson}.  
The mass terms for the fermions result from the Yukawa
interactions and can be diagonalized by a bi-unitary transformation
with $U_{ik}^{f,\rL}$ and $U_{ik}^{f,\rR}$ for left-handed and
right-handed fermion fields, respectively [see \refeq{eq:physfields}].
The neutrinos stay massless in accordance with  the absence of right-handed
neutrinos.  With \refeqs{eq:ewangle1} and \refeqf{eq:SMmass} we find
\beq
\cw = \frac{\MW}{\MZ}
\eeq
for the weak mixing angle.  The photon remains massless as a
consequence of the unbroken electromagnetic gauge invariance.

After diagonalization of the fermion mass matrices, the
interaction of the Higgs field $\FH(x)$ with fermions is
diagonal in flavour space,  and the associated couplings $\Mf{}_{,i}/\varv$
are proportional to the fermion masses.

The W-boson fields transform according to the adjoint representation 
of $\SU(2)_{\rw}$ (\ie $I_{\rw}=1$, $Y_{\rw}=0$), and
the fields defined in \refeq{eq:physfields} correspond to
eigenstates of the electric charge operator, since
\begin{equation}
Q \FW^\pm_\mu(x) ={} I^{3}_{\rw}  \FW^\pm_\mu(x) = \pm \FW^\pm_\mu(x), \qquad
Q \FZ_\mu(x) = Q \FA_\mu(x) = 0. 
\end{equation}
Thus, we have one positively charged gauge boson, $\PW^+$, one
negatively charged one, $\PW^-$, and two neutral ones, $\PZ$ and~$\PA$.

Identifying the coupling of the photon field $\FA_{\mu}$ to the fermions
with their electrical charge $\Qf e$ 
we can relate the coupling constants $g_1$ and $g_2$ to
the {elementary charge} $e$,
\beq\label{eq:charge}
\sqrt{4\pi\al} = e  = \frac{g_1g_2}{\sqrt{g_1^2+g_2^2}}= g_2 \sw= g_1 \cw,
\eeq
with $\alpha=\alpha(0)=1/137.0\dots$ denoting the fine-structure constant.

Owing to their unitarity, the matrices $U_{ik}^{f,\rL}$ and
$U_{ik}^{f,\rR}$ drop out in the interaction terms of fermions and
neutral gauge bosons, \ie there are no flavour-changing neutral
currents at tree level. In the quark--\PW-boson interaction a
non-trivial matrix remains, the unitary {\em quark-mixing matrix} or
{\em Cabibbo--Kobayashi--Maskawa matrix} \cite{Kobayashi:1973fv}
\beq\label{42VKMdef}
\VKM=  U^{u,\rL} U^{d,\rL\dagger}  .
\eeq
Since we do not include right-handed neutrinos, all neutrinos remain
massless and thus mass degenerate, and we can choose
$U^{\nu,\rL}= U^{l,\rL}$ to eliminate the mixing matrix in the
lepton--\PW-boson interaction 
rendering  their charged-current interaction flavour diagonal.%
\footnote{If one includes right-handed neutrinos, a mixing matrix in
  the lepton sector appears, the {\em Pontecorvo--Maki--Nakagawa--Sakata
  (PMNS) matrix} \cite{Pontecorvo:1957qd,Maki:1962mu}.}

Replacing the original set of parameters and fields by the 
``physical
parameters'' $e$, $\MW$, $\MZ$, $\MH$, $\Mf{}_{,i}$, and $\VKM_{ij}$ and
by the fields corresponding to charge and mass eigenstates,
$\FA_\mu$, $\FZ_\mu$, $\FW^\pm_\mu$, $\FH$, $\Fl$, $\nu$, $\Fu$, $\Fd$, 
and the would-be Goldstone fields $\phi^\pm$ and $\chi$, 
we can write the complete classical Lagrangian $\cL_{\mathrm{class}}$
of the EWSM as follows:
\begin{align} \label{eq:Lphys}
\cL_{\mathrm{class}} ={}&
\sum_{f=\Fl,\nu,\Fu,\Fd}\sum_i \left( \bar f_i (\ri\slashed{\partial} - \Mf{}_{,i})f_i
                          - e \Qf \bar f_i \ga^\mu f_i  \FA_\mu \right) 
  + \sum_{f=l,\nu,\Fu,\Fd}\sum_i \frac{e}{\sw\cw} 
                         \left( \If \bar f_i^{\rL} \ga^\mu f_i^{\rL} 
                          - \sw^2 \Qf \bar f_i \ga^\mu f_i\right)\, \FZ_\mu \nl
&{} + \sum_{i,j} \frac{e}{\sqrt2\sw} 
                      \left( \bar\Fu_i^{\rL} \ga^\mu \VKM_{ij} \Fd_j^{\rL} \FW_\mu^+
                      + \bar\Fd_i^{\rL} \ga^\mu \VKM_{ij}^{\dagger} \Fu_j^{\rL} 
                         \FW_\mu^-\right) 
 + \sum_{i} \frac{e}{\sqrt2\sw} 
                      \left( \bar \nu_i^{\rL} \ga^\mu \Fl_i^{\rL} \FW^+_\mu
                      + \bar\Fl_i^{\rL} \ga^\mu \nu_i^{\rL} \FW^-_\mu\right) \nl
&{} - \frac{1}{4} \left( \partial_\mu \FA_\nu - \partial_\nu \FA_\mu
            - \ri e(\FW^-_\mu \FW^+_\nu - \FW^-_\nu \FW^+_\mu)
          \right)^2 
- \frac{1}{4} \left( \partial_\mu \FZ_\nu - \partial_\nu \FZ_\mu
    + \ri e\frac{\cw}{\sw}(\FW^-_\mu \FW^+_\nu - \FW^-_\nu \FW^+_\mu) \right)^2 \nl
&{} - \frac{1}{2} \left| \partial_\mu \FW^+_\nu - \partial_\nu \FW^+_\mu
    - \ri e(\FW^+_\mu \FA_\nu - \FW^+_\nu \FA_\mu) 
    + \ri e\frac{\cw}{\sw}(\FW^+_\mu \FZ_\nu - \FW^+_\nu \FZ_\mu) \right|^2 \nl
&{} + \frac{1}{2} \left| \partial_\mu (\FH + \ri\chi) 
                  - \ri  \frac{e}{\sw}\FW^-_\mu \phi^+ + \ri \MZ\FZ_\mu 
        + \ri \frac{e}{2\cw\sw} \FZ_\mu(\FH + \ri\chi) \right|^2 \nl
&{} + \left| \partial_\mu \phi^+ + \ri e \FA_\mu \phi^+ 
        - \ri e \frac{\cw^2 - \sw^2}{2\cw\sw} \FZ_\mu \phi^+ - \ri \MW\FW^+_\mu
        - \ri \frac{e}{2\sw} \FW^+_\mu(\FH + \ri\chi) \right|^2 \nl
&{} - \frac{1}{2} \MH^2 \FH^2 - e\frac{\MH^2}{2\sw\MW} \FH 
         \left(\phi^-\phi^+ + \frac{1}{2} |\FH + \ri \chi|^2 \right)
    - e^2\frac{\MH^2}{8\sw^2\MW^2} 
         \left(\phi^-\phi^+ + \frac{1}{2} |\FH + \ri \chi|^2 \right)^2 \nl
&{} - \sum_{f=\Fl,\Fu,\Fd}\sum_i e\frac{m_{\Pf,i}}{2\sw\MW} 
            \left( \bar f_i f_i \FH -2\If  \ri\bar f_i \ga_5 f_i\chi\right) \nl
&{} + \sum_{i,j} \frac{e}{\sqrt2\sw\MW} 
       \left[ 
       m_{u,i}   \bar\Fu_i^{\rR} \VKM_{ij} \Fd_j^{\rL} \phi^+
       + m_{u,j} \bar\Fd_i^{{\rL}} \VKM_{ij}^{\dagger} \Fu_j^{\rR} \phi^- 
       - m_{d,j} \bar\Fu_i^{\rL} \VKM_{ij} \Fd_j^{\rR} \phi^+
       - m_{d,i} \bar\Fd_i^{\rR} \VKM_{ij}^{\dagger} \Fu_j^{\rL} \phi^-
       \right] \nl
&{} - \sum_{i} 
           \frac{e}{\sqrt2\sw\MW} m_{l,i}\left(\bar \nu_i^{\rL} \Fl_i^{\rR} \phi^+
           +\bar\Fl_i^{\rR} \nu_i^{\rL}\phi^- \right) ,
\end{align}
where we have omitted an irrelevant constant term.
Here $\If$ and $\Qf$ are the third component of the weak isospin and
the relative charge of the fermion $\Pf$, respectively.




The Lagrangian \refeqf{eq:Lphys} is invariant under 
gauge transformations that follow from \refeq{eq:gtsym}
by inserting the fields and parameters of the physical basis,
\begin{align} \label{eq:gtphf}
\delta \FA_{\mu } ={}& 
\partial_{\mu } \de\theta^{\FA}
+ \ri e (\FW_{\mu }^{+} \de\theta^{-} - \FW_{\mu }^{-}\de\theta^{+}),
\notag\\
\delta \FZ_{\mu } ={}& 
\partial_{\mu }\de\theta^{\FZ}
- \ri e \frac{\cw}{\sw} 
(\FW_{\mu }^{+} \de\theta^{-} - \FW_{\mu }^{-} \de\theta^{+}),
\notag \\
\delta \FW_{\mu }^{\pm}={}& 
\partial_{\mu } \de\theta^{\pm } 
\mp\ri \frac{e}{\sw} \left[
\FW_{\mu }^{\pm}\left(\sw\de\theta^{\FA}-\cw\de\theta^{\FZ}\right)
-(\sw \FA_{\mu } - \cw \FZ_{\mu }) \de\theta^{\pm }
\right],
\notag \\
\delta \FH ={}& 
\frac{e}{2\sw\cw} \chi \de\theta^{\FZ} 
+ \frac{\ri e}{2\sw} (\phi^{+} \de\theta^{-} - \phi^{-} \de\theta^{+}),
\notag \\
\delta \chi ={}& 
{- \frac{e}{2\sw\cw}} (\varv + \FH) \de\theta^{\FZ}
+ \frac{e}{2\sw} (\phi^{+} \de\theta^{-} + \phi^{-} \de\theta^{+}), \nl
\delta \phi ^{\pm} ={}& 
{\mp \ri} e \phi^{\pm}\biggl(\de\theta^{\FA} - 
\frac{\cw^{2} - \sw^{2}}{2\cw\sw} \de\theta^{\FZ} \biggr)
\pm \frac{\ri e}{2\sw} (\varv + \FH {\pm } \ri {\chi } ) \de\theta^{\pm},
\notag \\
\delta \FF_{\pm,i}^{\rL} = {}&
{-\ri} e \left[ Q_{\pm}\delta\theta^{A} +
\frac{\sw}{\cw}\left( Q_{\pm} \mp \frac{1}{2\sw^2}\right)
\delta\theta^Z \right] \FF_{\pm,i}^{\rL} 
+\ri \frac{e}{\sqrt{2}\sw} \delta\theta^\pm  \sum_{j} \varv^{\pm}_{ij} \FF_{\mp,j}^{\rL} ,
\notag \\
\delta \Ff_i^{\rR} ={}& 
{-\ri} e  Q_{\Pf} \left(\delta\theta^{A} +
\frac{\sw}{\cw}\delta\theta^Z \right) \Ff_i^{\rR},
%
\end{align}
with the parameters
\begin{equation}\label{eq:gtp}
\de\theta^\pm ={}  \frac{1}{\sqrt{2}}\left(\de\theta^1 \mp\ri\de\theta^2\right), \qquad
\de\theta^\FA   ={} \cw\de\theta^Y - \sw\de\theta^3, 
\qquad
\de\theta^\FZ   = \cw\de\theta^3 + \sw\de\theta^Y
\end{equation}
of the gauge transformation, the generic fermion fields
\beq\label{eq:defpsipmi}
(\FF_{+,i},\FF_{-,i}) = (\Fu_i,\Fd_i) \text{\quad or\quad } (\nu_i,\Fl_i),
\eeq
and 
\beq\label{eq:defvij}
\varv^+_{ij}= \VKM_{ij}, \quad \varv^-_{ij}= \VKM^\dagger_{ij}
\quad \text{for quarks}, \qquad  \varv^{\pm}_{ij}= \de_{ij}\quad  \text{for leptons}. 
\eeq

\subsubsection{Quantization}

Quantization of ${\cL}_{\mathrm{class}}$ requires the specification of
a gauge. A general linear renormalizable gauge is given by the 
gauge-fixing functionals
\begin{equation}\label{eq:gf}
C^{\pm} ={} \partial^{\mu } W^{\pm}_{\mu } \mp \ri \MW \xi'_\FW \phi ^{\pm}, \qquad
C^{\FZ} ={} \partial ^{\mu} Z_{\mu } - \MZ \xi'_\FZ \chi, \qquad
C^{\FA} ={} \partial^{\mu} A_{\mu }
\end{equation}
and the gauge-fixing Lagrangian
\begin{equation} \label{eq:Lfix}
  {\cal L}_{\mathrm{fix}} = - \frac{1}{2\xi_\FA} \left(C^{\FA}\right)^{2}
  - \frac{1}{2\xi_\FZ} \left(C^{\FZ}\right)^{2} 
- \frac{1}{\xi_\FW} C^{+} C^{-}.
\end{equation}
In the {\em 't~Hooft gauge}, where $\xi'_\FW=\xi_\FW$ and $\xi'_\FZ=\xi_\FZ$
and which is used in the following, the terms involving the would-be
Goldstone-boson fields in the gauge-fixing functions are chosen in
such a way
that they cancel the mixing terms $V_\mu\,\partial^{\mu} \phi$ in the
classical Lagrangian \refeqf{eq:Lphys} up to irrelevant total
derivatives.  On the other hand, these gauge-fixing terms give rise to
masses for the would-be Goldstone-boson fields,
\beq
M_\phi = \sqrt{\xi_\FW}\MW, \qquad M_\chi = \sqrt{\xi_\FZ}\MZ,
\eeq
and equivalent masses for the scalar components [polarization
vector $\veps(k)\propto k$] of the vector-boson fields.

The corresponding Faddeev--Popov ghost Lagrangian is obtained as
\begin{equation}\label{eq:Lghost}
{\cal L}_{\mathrm{FP}} = 
-\int\rd^4y\, \bar{u}^{a }(x) \frac{\delta C^{a}(x)}{\delta \theta^{b}(y)} u^{b}(y),
\end{equation}
where ${\delta C^{a}(x)}/{\delta \theta^{b}(y)}$ is the variation
of the gauge-fixing function $C^{a }$ under infinitesimal gauge
transformations characterized by $\theta ^{b}(y)$, and $u^{a}(x)$,
$\bar{u}^{a}(x)$, $a=\FA ,\FZ,\pm$, are Faddeev--Popov ghost fields.
The masses for the Faddeev--Popov fields coincide with those of the
corresponding   would-be Goldstone-boson fields.

The 't~Hooft gauge leads to propagators that behave as $1/k^2$ for
large momenta and thus to a renormalizable Lagrangian for finite
values of the gauge parameters $\xi_a$. For $\xi_a\to\infty$ the
masses of the unphysical degrees of freedom (except for $u^A$) tend to
infinity, the unphysical degrees of freedom decouple, and we are left
with the theory in the unitary gauge, where renormalizability is not
obvious.

For $\xi_a=1$, $a=\FA, \FZ, \FW$, the {\it 't~Hooft--Feynman gauge},
pole positions and residues of the longitudinal parts of the
gauge-boson propagators coincide with those of the transverse parts,
so that $k_\mu k_\nu$ parts of the gauge-boson propagators are absent.
Moreover, the masses of the unphysical Faddeev--Popov ghost and
would-be Goldstone-boson fields equal those of the corresponding gauge
fields in the physical basis.  Therefore, the 't~Hooft--Feynman gauge
is most convenient for practical higher-order calculations.

Adding up all terms of \refeqs{eq:Lphys}, \refeqf{eq:Lfix}, and
\refeqf{eq:Lghost}, we obtain the complete {\it quantized Lagrangian}
of the EWSM suitable for higher-order calculations,
\begin{equation}\label{eq:LGSW}
{\cL}_{\EW} = {\cL}_{\mathrm{class}} + {\cL}_{\mathrm{fix}} + {\cL}_{\mathrm{FP}}  .
\end{equation} 
The corresponding Feynman rules are given in \refapp{se:feynman_rules}.

\subsubsection{Inclusion of QCD}
\label{se:QCD}

The strong interaction described by the colour gauge group
$\SU(3)_{\rc}$ can be easily incorporated in the SM.  We denote the
generators of $\SU(3)_{\rc}$ by $T^A$, $A=1,\ldots,8$, and the
corresponding gauge fields, the {\it gluon fields,} by $G^A$. The
gluon gauge field strength reads
\beq\label{eq:QCD_field_strength}
G^{A}_{\mu\nu} = \partial_{\mu }G^{A}_{\nu }-\partial_{\nu 
}G^{A}_\mu + \gs f^{ABC}G^{B}_{\mu }G^{C}_{\nu },
\eeq
where $\gs$ is the strong gauge coupling and $f^{ABC}$ are the structure
constants of $\SU(3)$.
The covariant derivative \refeqf{eq:covdergsw} is extended to
\beq\label{eq:covdersm}
D_{\mu} = 
  \partial_{\mu} 
  -\ri \gs T^{A} G^{A}_\mu 
  -\ri g_{2} I^{a}_{\rw} W^{a}_\mu 
  +\ri g_{1} \frac{Y_{\rw}}{2}\FB_\mu
  .
\eeq
Quarks transform according to the fundamental representation of
$\SU(3)_{\rc}$ with the generators $T^A=\la^A/2$, where $\la^A$ are
the Gell-Mann matrices. The leptons, the Higgs-doublet, and the EW
gauge fields are singlets with respect to $\SU(3)_{\rc}$ ($T^A=0$).
Conversely, the gluons do not carry weak isospin and hypercharge.

Apart from the extension of the covariant derivative, the Lagrangian
\refeqf{eq:LSM_sym} gets an additional kinetic term for the $\SU(3)_{\rc}$
gauge fields,
\begin{equation}
\cL_{\mathrm{class}} \to \cL_{\mathrm{class}} 
 -\frac{1}{4}G^{A}_{\mu\nu}G^{A,\mu\nu}.
\end{equation}
Corresponding changes apply to the Lagrangian \refeqf{eq:Lphys} in the
physical basis.

Quantization of the $\SU(3)_{\rc}$ part requires to include the usual
gauge-fixing and Faddeev--Popov ghost
terms~\cite{Sterman:1994ce,Peskin:1995ev,Bohm:2001yx,Schwartz:2013pla}.

\subsubsection{\texorpdfstring{$\theta$-terms}{theta-terms} and strong CP violation}
\label{se:Ltheta}


The requirements of $\SU(3)_{\rc}\times \SU(2)_{\rw}\times \U(1)_Y$
gauge invariance and renormalizability admit the presence of the
following terms in the SM Lagrangian (see, \eg,
\citere{Schwartz:2013pla}, which is compatible with our conventions in
this subsection),
\beq\label{eq:thetadef}
\cL_\theta = -\theta_\QCD \frac{\gs^2}{16\pi^2} \tilde G^a_{\mu\nu}G^{a,\mu\nu}
-\theta_2 \frac{g_2^2}{16\pi^2} \tilde W^a_{\mu\nu}W^{a,\mu\nu}
-\theta_1 \frac{g_1^2}{16\pi^2} \tilde B_{\mu\nu}B^{\mu\nu}
\eeq
with the dual field-strength tensors
\beq
\label{eq:dual-field-strengths}
\widetilde X_{\mu\nu}=
-\frac{1}{2}\eps_{\mu\nu\rho\sigma} X^{\rho\sigma}, \qquad
X=\FG^A,\FW^a,\FB, 
\eeq
which we define with $\eps^{0123}=+1$.  Since $\cL_\theta$ can be
written as total derivative $\partial^\mu \Theta_\mu(x)$ of some
quantity $\Theta_\mu(x)$ (known as {\it Chern--Simons current}),
effects of $\cL_\theta$ can never show up in perturbative
calculations.  For non-abelian gauge fields, however, topologically
non-trivial, non-perturbative field configurations {\it (instantons)}
can potentially lead to observable effects caused by $\cL_\theta$,
while abelian gauge fields do not admit such configurations (see
\citeres{Weinberg:1996kr,Dine:2000cj} and references therein).  While
the parameter $\theta_1$ could, thus, be set to 0, we nevertheless
keep it in the following and show that it is unphysical, as it can
be eliminated via field redefinitions.
To answer the
question of observability for 
the $\theta$ parameters in \refeq{eq:thetadef}
it is necessary to
inspect the behaviour of the complete Lagrangian under phase
transformations of the fermion fields, because the functional integral
of the fermion fields shows a non-trivial behaviour under chiral phase
transformations.  In detail, the global phase transformations
\beq
\label{eq:fphasetrafo}
f_\rR(x) \;\to\; \re^{\ri\theta_{f_\rR}} f_\rR(x), \qquad
f_\rL(x) \;\to\; \re^{\ri\theta_{f_\rL}} f_\rL(x)
\eeq
effectively modify the Lagrangian of the SM by the 
term~\cite{Fujikawa:1979ay}
\beq
\cL_{\mathrm{phase}} = 
-\sum_q \frac{1}{2}\left( \theta_{q_\rR}-\theta_{q_\rL} \right)
\frac{\gs^2}{16\pi^2} \tilde G^a_{\mu\nu}G^{a,\mu\nu}
+\sum_f \frac{1}{2}\theta_{f_\rL}
\frac{g_2^2}{16\pi^2} \tilde W^a_{\mu\nu}W^{a,\mu\nu}
-\sum_f \frac{1}{4}
\left( Y_{\rw,f_\rR}^2\theta_{f_\rR}-Y_{\rw,f_\rL}^2\theta_{f_\rL} \right)
\frac{g_1^2}{16\pi^2} \tilde B^a_{\mu\nu}B^{a,\mu\nu},
\eeq
where the sum over $q$ runs over all six quark flavours and the one
over $f$ over all fermion flavours of the SM.  Note that the phases
$\theta_{f_{\rR/\rL}}$ are not all independent. We should rather
ensure that the SM Lagrangian $\cL_{\mathrm{class}}$
\refeqf{eq:LSM_sym} is left unchanged by the phase transformations
\refeqf{eq:fphasetrafo}.  The transformations in the quark and lepton
sectors can be considered independently.

Focusing on the quark sector with three generations, we have 12
independent phase transformations. To keep the CKM matrix unchanged in
the hadronic charged-current interaction, only one global phase change
by some angle $\theta_{Q_{\rL}}$ is still allowed, \ie 5 out of the
6~phases of the $q_\rL$ fields are fixed relative to the phase of one
of those.  These constraints can be formulated as
\beq
\label{eq:quarkphasetrafo1}
\theta_{u_{i,\rL}}=\theta_{d_{i,\rL}}\equiv\theta_{Q_{\rL}}, 
\eeq
where $i=1,2,3$ is the generation index.  This is exactly the
reduction in the number~9 of real degrees of freedom of a general
unitary matrix to the number~4 of physically relevant ones in the CKM
matrix.  To keep the hadronic Yukawa couplings in
$\cL_{\mathrm{class}}$ invariant as well, and allowing for a phase
transformation of the Higgs doublet by an angle $\theta_\Phi$, we get
the further constraints
\beq
\label{eq:quarkphasetrafo2}
0=-\theta_{Q_\rL}+\theta_{d_{i,\rR}}+\theta_\Phi, \qquad
0=-\theta_{Q_\rL}+\theta_{u_{i,\rR}}-\theta_\Phi,
\eeq
which fix all the phases of the right-handed quark fields $q_\rR$ in
terms of $\theta_{Q_\rL}$ and $\theta_\Phi$.  Respecting the
constraints \refeqf{eq:quarkphasetrafo1} and
\refeqf{eq:quarkphasetrafo2}, the parameters $\theta_\QCD$ and
$\theta_2$ change as follows,
\begin{align}
\theta_\QCD &{}\to \theta_\QCD 
+\sum_q \frac{1}{2}\left( \theta_{q_\rR}-\theta_{q_\rL} \right)
=
\theta_\QCD +\sum_{i=1}^3 \frac{1}{2}\left( \theta_{u_{i,\rR}}-\theta_{u_{i,\rL}} 
+ \theta_{d_{i,\rR}}-\theta_{d_{i,\rL}} \right) = \theta_\QCD,
\\
\theta_2    &{}\to \theta_2 - \sum_q \frac{1}{2}\theta_{q_\rL} 
= \theta_2-3\theta_{Q_\rL},
\\
\theta_1    &{}\to \theta_1 
+\sum_q \frac{1}{4}\left( Y_{\rw,q_\rR}^2\theta_{q_\rR}
-Y_{\rw,q_\rL}^2\theta_{q_\rL} \right)
= \theta_1 
+\sum_i \frac{1}{9}\left( 4\theta_{u_{i,\rR}} +\theta_{d_{i,\rR}} 
\right)
-\frac{1}{6}\theta_{Q_\rL}.
\end{align}
Thus, $\theta_\QCD$ is uniquely fixed by the phase choice of the CKM
matrix and the fermion masses, which are taken real and non-negative,
while $\theta_2$ can be transformed to zero by a phase
transformation that leaves the SM Lagrangian invariant. In other
words, $\theta_\QCD$ is a physically measurable parameter, whereas
$\theta_2$ is not.  If at least one of the quarks, say $q$, was
massless, the phase of $q_\rR$ would not be constrained, so that
$\theta_\QCD$ could be transformed to zero as well; this possibility
is, however, ruled out by experiment.  
Since the leptons do not
contribute to the $\theta_\QCD$ term in $\cL_\theta$, taking into
account phase transformations of lepton fields does not change the
above conclusions on $\theta_\QCD$, and $\theta_2$ can still be transformed
to zero by adjusting $\theta_{Q_\rL}$.
Since phase changes of the right-handed lepton fields neither influence
$\theta_\QCD$ nor $\theta_2$, the phase of those can be adjusted
to render $\theta_1$ zero.%
\footnote{For massless neutrinos, where no generation mixing exists in
  the lepton sector, the only constraints on the phase changes for
  leptons are
  $\theta_{l_{i,\rL}}=\theta_{\nu_{i,\rL}}\equiv\theta_{L_{i,\rL}}$
  and $0=-\theta_{L_{i,\rL}}+\theta_{l_{i,\rR}}+\theta_\Phi$, so that
  the phases $\theta_{l_{i,\rR}}$ can be adjusted to transform
  $\theta_1$ to zero.  For massive neutrinos with mixing, we have to
  include right-handed neutrino fields which bring in new phases
  $\theta_{\nu_{i,\rR}}$, but also new constraints:
  $\theta_{l_{i,\rL}}=\theta_{\nu_{i,\rL}}\equiv\theta_{L_{\rL}}$,
  $0=-\theta_{L_{\rL}}+\theta_{l_{i,\rR}}+\theta_\Phi$, and
  $0=-\theta_{L_\rL}+\theta_{\nu_{i,\rR}}-\theta_\Phi$.  Here the
  freedom in choosing $\theta_{\nu_{i,\rR}}$ is sufficient to
  transform $\theta_1$ to zero.}

The Lagrangian $\cL_\theta$ is particularly interesting, since it
violates CP~symmetry {\it (strong CP violation)}.  Experimentally the
upper bound on $|\theta_\QCD|$ is of the order $10^{-10}$
\cite{Dragos:2019oxn}, which is deduced from experimentally
constraining the neutron electric dipole
moment~\cite{Afach:2015sja,Graner:2016ses}. Within the SM, there is no
explanation of this fine-tuning of $\theta_\QCD$, which asks for a
solution by physics beyond the SM, as, e.g., suggested by axion models
(see, e.g., \citeres{Weinberg:1996kr,Schwartz:2013pla} and references
therein).  Since $\cL_\theta$ does not play any role in collider
physics, we will not consider it any further in this review.

\subsection{\texorpdfstring{Slavnov--Taylor}{Slavnov-Taylor} 
and Ward identities}
\label{se:wi}\label{se:sti}

The quantized Lagrangian \refeqf{eq:LGSW} of the EWSM, ${\cal L}_{\EW}$,
is invariant under {\it Becchi--Rouet--Stora (BRS)
  transformations} \cite{Becchi:1974md}, defined as
\beq
\delta_{\mathrm{BRS}} \Psi_I = \de\lambda\, s\Psi_I,
\eeq
where $\de\la$ is an infinitesimal anticommuting constant and $\Psi_I$
represents any of the fields $V^a_\mu$, $u^a$, $\ubar^a$
($a=A,Z,\pm$), $\phi^\pm$, $\chi$, $\FH$, $\FF_{\pm,i}^{\rL}$, or
$\Ff_i^{\rR}$.  The BRS transformations $s\Psi_I$ in the physical basis read
\begin{align} \label{eq:BRSphf}
sA_{\mu } ={}& 
\partial_{\mu } u^{\FA} + \ri e (W_{\mu }^{+} u^{-} - W_{\mu }^{-}u^+),
\nl
sZ_{\mu } ={}& \partial_{\mu }u^{\FZ}
- \ri e \frac{\cw}{\sw} (W_{\mu }^{+} u^{-} - W_{\mu }^{-} u^{+}),
\nl
sW_{\mu }^{\pm}={}& 
\partial_{\mu } u^{\pm } 
\mp \ri e \left[
W_{\mu }^{\pm}\left(u^{\FA}-\frac{\cw}{\sw} u^{\FZ}\right)
-\left(A_{\mu } - \frac{\cw}{\sw} Z_{\mu }\right) u^{\pm }
\right] ,\nl
s\FH ={}& 
\frac{e}{2\sw\cw} \chi u^{\FZ} 
+ \frac{\ri e}{2\sw} (\phi^{+} u^{-} - \phi^{-} u^{+})
,\nl
s\chi ={}&
{- \frac{e}{2\sw\cw}} (\varv + \FH) u^{\FZ}
+ \frac{e}{2\sw} (\phi^{+} u^{-} + \phi^{-} u^{+}) ,\nl
s\phi^{\pm} ={}&
{\mp \ri} e \phi^{\pm}\biggl(u^{\FA} - 
\frac{\cw^{2} - \sw^{2}}{2\cw\sw} u^{\FZ} \biggr)
\pm \frac{\ri e}{2\sw} (\varv + \FH {\pm } \ri {\chi } ) u^{\pm}
,\nl
s\FF_{\pm,i}^{\rL} ={}& 
{-\ri} e \left[ Q_{\pm}u^{\FA} +
\frac{\sw}{\cw}\left( Q_{\pm} \mp \frac{1}{2\sw^2}\right)
u^Z \right] \FF_{\pm,i}^{\rL} 
+\ri \frac{e}{\sqrt{2}\sw} u^\pm  \sum_j \varv^{\pm}_{ij} \FF_{\mp,j}^{\rL} ,\nl
s\Ff_i^{\rR} ={}& 
{-\ri} e  Q_{\Pf} \left(u^{\FA} +
\frac{\sw}{\cw}u^\FZ \right) \Ff_i^{\rR},
\notag \\
s u^\pm ={}& {\pm\ri} \frac{e}{\sw} u^\pm(\sw u^\FA - \cw u^\FZ),\qquad
s u^\FZ ={} -\ri e\frac{\cw}{\sw} u^- u^+ ,\qquad
s u^\FA ={}  \,\ri e  u^- u^+,\nl
s\ubar^\pm ={}&  {-\frac{1}{\xi_\FW}} C^{\mp},\qquad
s\ubar^\FZ =  - \frac{1}{\xi_\FZ} C^{\FZ},\qquad
s\ubar^\FA =  -\frac{1}{\xi_\FA} C^{\FA}, 
\end{align}
with $\FF_{\pm,i}$ defined in \refeq{eq:defpsipmi} and
$\varv^{\pm}_{ij}$ in \refeq{eq:defvij}.

The BRS symmetry gives rise to relations between different Green
functions, called {\em Slavnov--Taylor identities}. The identities for
the full (reducible) Green functions
$G_{\rc}^{\Psi_{I_1}\dots}=\langle T\prod_l\Psi_{I_l}\rangle$
(symbolically written as vevs of time-ordered fields) are obtained
from the invariance of Green functions under BRS transformations (see,
for instance, \citeres{Collins:1984xc,Bohm:2001yx}):
\beq\label{eq:WIGF}
0= \frac{\de_{\mathrm{BRS}}}{\de\la} \Bigl\langle T \prod_l \Psi_{I_l} \Bigr\rangle.
\eeq

Applying this to Green functions involving an antighost field
$\ubar^a$ and arbitrary physical, on-shell (OS) fields
$\Psi_{I_l}^{\text{phys}}$, one obtains
\beq\label{eq:sti_c+phys}
0 = \Bigl\langle T C^a \prod_l \Psi_{I_l}^{\text{phys}}
\Bigr\rangle,
\eeq
up to terms that vanish after truncation, since BRS variations of the
physical components of asymptotic OS fields vanish.
Using the equations of motion of the antighost fields, this can be
generalized to Green functions with more gauge-fixing terms:
\begin{equation}\label{eq:sti_nc+phys}
 \Bigl\langle T 
C^a(x) \, \Bigl(\prod_m C^{a_m}(x_m)\Bigr) 
\prod_l \Psi_{I_l}^{\text{phys}} \Bigr\rangle 
{}= -\ri \xi_a
\sum_k \de^{aa_k} \de^{(4)}(x-x_k)\Bigl\langle T \Bigl(\prod_{m\ne k} C^{a_m}(x_m)\Bigr)
\prod_l \Psi_{I_l}^{\text{phys}} \Bigr\rangle.
\end{equation}
An important special case of this relation reads
\beq
\left\langle T C^a(x) C^b(y)\right\rangle =
-\ri \xi_a\de^{ab} \de^{(4)}(x-y),
\eeq
which is known as {\it Slavnov identity} in the case of unbroken gauge
symmetries.  Using the gauge-fixing functions \refeqf{eq:gf} 
in the 't~Hooft gauge ($\xi'_a=\xi_a$)
and
transforming to momentum space, we find the following relations for
the propagators of the gauge bosons,
\begin{align}\label{eq:WIVV}
{-\ri}\xi_\FW ={}&
k^{\mu} k^{\nu} G^{\FW^+\FW^-}_{\mu \nu} (k,-k) 
- \xi_\FW \MW k^{\mu} G_{\mu}^{\FW^+\phi^-} (k,-k)
- \xi_\FW \MW k^{\mu} G_{\mu}^{\phi^+\FW^-} (k,-k) 
+ \xi_\FW^2 \MW^{2} G^{\phi^+\phi^-}(k,-k) ,
\nl
{-\ri}\xi_\FZ ={}&
k^{\mu} k^{\nu} G^{\FZ\FZ}_{\mu \nu} (k,-k) - 2 \ri \xi_\FZ \MZ k^{\mu}
G_{\mu}^{\FZ\chi} (k,-k) + \xi_\FZ^2\MZ^{2} G^{\chi\chi}(k,-k),
\nl
0 ={}&
k^{\mu} k^{\nu} G^{\FA\FZ}_{\mu \nu} (k,-k) - \ri\xi_\FZ\MZ k^{\mu}
G_{\mu}^{\FA\chi} (k,-k),
\nl
{-\ri}\xi_\FA ={}&
k^{\mu} k^{\nu} G^{\FA\FA}_{\mu \nu} (k,-k) .
\end{align}
According to our conventions 
(see \refapp{se:conventions})
the field labels in non-truncated Green functions indicate the
outgoing fields, but those in truncated Green functions or vertex
functions the incoming fields.  The momentum arguments denote incoming
momenta corresponding to the fields in the superscripts.  The
relations \refeqf{eq:WIVV} between the longitudinal parts of the
gauge-boson propagators, the gauge-field--would-be Goldstone-field
mixing propagators, and the propagators of the would-be Goldstone
bosons are exact. They show, in particular, that the poles of the
unphysical parts of the propagators coincide. The relation for the
photon propagator ensures that its longitudinal part gets no
higher-order corrections.

\subsection{Background-field quantization}
\label{se:bfm}

\begin{sloppypar}
  Gauge-invariance relations between Green functions can be greatly
  simplified within the background-field method (BFM), which is a
  modified version of the conventional formalism for the quantization
  of gauge fields (see, e.g.,
  \citeres{Weinberg:1996kr,Bohm:2001yx,Schwartz:2013pla}).  The BFM
  was introduced originally for quantum gravity in
  \citere{DeWitt:1967ub} and subsequently generalized to QCD in
  \citeres{DeWitt:1980jv,tHooft:1975uxh,Boulware:1980av,Abbott:1980hw}.
  Its application to the EWSM was formulated in
  \citere{Denner:1994xt}.
As in the latter reference, we neglect quark mixing in this section.%
\footnote{As shown in \citere{Denner:1994nn}, the vertex functions of
  the BFM in the Feynman gauge coincide with the results of the {\it
    pinch technique}
  \cite{Cornwall:1981zr,Cornwall:1989gv,Papavassiliou:1989zd}
  sometimes used in the literature.}
\end{sloppypar}

\subsubsection{Quantized Lagrangian in the background-field method}

In the conventional formalism the fields appearing in the classical
Lagrangian are quantized.  Instead, in the BFM, the fields of the
classical Lagrangian are additively split into classical background fields
(denoted with a caret), $\Psihat$, and quantum fields, $\Psi$,
\beq
\label{eq:quantLc}
{\cL}_{\mathrm{class}}(\Psi) \rightarrow {\cL}_{\mathrm{class}}(\Psihat + \Psi) .
\eeq
While the quantum fields are the variables of integration in the
functional integral, \ie quantized, the classical fields serve as
external sources.  Invariance of the classical Lagrangian with respect
to the transformations \refeqf{eq:gtsym} then implies its invariance
under two kinds of transformations: {\it quantum gauge
  transformations}
\begin{alignat}{5}\label{eq:gtsym_qu}
\de \FW_\mu^{a}={}& \partial_\mu \de\theta^a +
g_2 \eps^{abc} (\FW_\mu^b + \FWhat_\mu^b )\,\de\theta^c, \qquad &
\de \FB_\mu^{} ={}& \partial_\mu \de\theta^Y, \qquad &
\de\Phi ={}& \left(-\frac{\ri}{2}g_1\de\theta^Y + 
\ri \frac{\tau^{a}}{2}g_2\de\theta^{a}\right) (\Phi+\Phihat),
\nl
\de \FWhat_\mu^{a}={}& 0, &
\de \FBhat_\mu^{} ={}& 0, &
\de\Phihat ={}& 0,
\end{alignat} 
and {\it background gauge transformations}
\begin{alignat}{5}\label{eq:gtsym_bg}
\de \FWhat_\mu^{a}={}& \partial_\mu \de\thetahat^a +
g_2 \eps^{abc} \FWhat_\mu^b\,\de\thetahat^c, \qquad &
\de \FBhat_\mu^{} ={}& \partial_\mu \de\thetahat^Y, \qquad &
\de\Phihat ={}& \left(- \frac{\ri}{2}g_1\de\thetahat^Y + 
\ri \frac{\tau^{a}}{2}g_2\de\thetahat^{a}\right) \Phihat,\nl
\de \FW_\mu^{a}={}& 
g_2 \eps^{abc} \FW_\mu^b\,\de\thetahat^c, &
\de \FB_\mu^{} ={}& 0, &
\de\Phi ={}& \left(- \frac{\ri}{2}g_1\de\thetahat^Y + 
\ri \frac{\tau^{a}}{2}g_2\de\thetahat^{a}\right) \Phi,
\end{alignat}
where we mark the gauge-transformation parameters of the latter by a caret.
Since we do not split the fermion fields, they transform 
under both types of transformations as in \refeq{eq:gtsym}.%
\footnote{For fields that do not enter the gauge-fixing term,
  quantization in the BFM is equivalent to the conventional formalism,
  and the Feynman rules for background and quantum fields are
  identical.}  The splitting of the Higgs-doublet field is done in
such a way that the background Higgs field $\Phihat$ has the usual
non-vanishing vev $\varv$, while the one of the
quantum Higgs field $\Phi$ is zero,
\begin{equation}
\label{eq:Phi}
\Phihat(x) = \left( \begin{array}{c}
\phihat ^{+}(x) \\ \frac{1}{\sqrt{2}}\bigl[\varv + {\hat \FH}(x) +\ri
\chihat (x) \bigr]
\end{array} \right) , \qquad
\Phi (x) ={} \left( \begin{array}{c}
\phi ^{+}(x) \\ \frac{1}{\sqrt{2}}\bigl[\FH(x) +\ri \chi (x) \bigr]
\end{array} \right) .
\end{equation}

A gauge-fixing term is added that breaks the invariance under quantum
gauge transformations, but retains the invariance with respect to
background gauge transformations:
\begin{align}\label{eq:tHgfbfm}
\cL_{\mathrm{fix,BFM}} ={}& {-\frac{1}{2\xi_\FW}}
\biggl[\left(\de^{ac}\partial_\mu + g_2 \eps^{abc}\FWhat^b_\mu\right)W^{c,\mu}
-\ri g_2\xi_\FW'\frac{1}{2}\left(\Phihat^\dagger\tau^a\Phi
                  - \Phi^\dagger\tau^a\Phihat\right)\biggr]^2 \nl
                 & - \frac{1}{2\xi_\FB}\biggl[\partial_\mu \FB^{\mu}
       +\ri g_1\xi_\FB'\frac{1}{2}\left(\Phihat^\dagger \Phi
                  - \Phi^\dagger\Phihat\right)\biggr]^2.
\end{align}
Invariance under background gauge transformations requires that the
background gauge fields appear only within a covariant derivative in
the gauge-fixing term, so that the terms in square brackets transform
according to the adjoint representation of the gauge group. Only four
independent gauge parameters are allowed, $\xi_\FW$ and $\xi_\FW'$ for
$\SU(2)_{\rw}$ and $\xi_\FB$ and $\xi_\FB'$ for $\U(1)_{Y}$.  Mixing
between gauge bosons and the corresponding would-be Goldstone bosons
is eliminated by choosing $\xi_\FW'=\xi_\FW$ and $\xi_\FB'=\xi_\FB$,
resulting in the {\it 't~Hooft gauge} for the background-field
formalism. In order to avoid tree-level mixing between the quantum
photon and $\PZ$-boson fields, we set $\xi=\xi_\FW=\xi_\FB$.  Note
that \refeq{eq:tHgfbfm} translates to the conventional gauge-fixing
term \refeqf{eq:Lfix} for $\xi_\FW=\xi_\FZ=\xi_\FA$ upon replacing the
background Higgs field by its vev and omitting the background
$\SU(2)_{\rw}$ triplet field $\hat W^a_\mu$.

Expressing the gauge-fixing Lagrangian in terms of fields in the
physical basis yields
\beq
\cL_{\mathrm{fix,BFM}} = - \frac{1}{2 \xi}
\left[ \left(C^{A}\right)^{2} + \left(C^{\FZ}\right)^{2} + 2 C^{+} C^{-} \right] ,
\label{eq:LGFph}
\eeq
where
\begin{align}\label{eq:gfbfm}
C^{\FA} ={}& \partial^{\mu } \FA_{\mu } 
 + \ri e (\FWhat^+ _{\mu} \FW^{- \mu} - W^+ _{\mu} \FWhat^{- \mu})
 + \ri e \xi (\phihat^- \phi^+ - \phihat^+ \phi^-) , \nl
C^{\FZ} ={}& \partial^{\mu } \FZ_{\mu } 
 - \ri e \frac{\cw}{\sw} (\FWhat^+ _{\mu} \FW^{- \mu} - \FW^+ _{\mu}
  \FWhat^{- \mu})
 - \ri e \xi \frac{\cw^2 - \sw^2}{2 \cw \sw} 
  (\phihat^- \phi^+ - \phihat^+ \phi^-) 
 - \xi \MZ \chi+ e \xi \frac{1}{2 \cw \sw} 
   (\chihat \FH - {\FHhat} \chi) , \nl
C^{\pm} ={}& \partial^{\mu } \FW^{\pm}_{\mu }
 \pm \ri e \left(\FAhat^{\mu} - \frac{\cw}{\sw} \FZhat^{\mu}\right) \FW^{\pm}_{\mu}
 \mp \ri e \left(\FA^{\mu} - \frac{\cw}{\sw} \FZ^{\mu}\right) \FWhat^{\pm}_{\mu} 
\mp \ri \xi \MW \phi^\pm
\mp \ri e \xi \frac{1}{2 \sw} 
 \left[ (\FHhat \mp \ri \chihat) \phi^{\pm} 
  - (\FH \mp \ri \chi) \phihat^{\pm} \right] .
\end{align}

The Faddeev--Popov Lagrangian $\cL_{\mathrm{FP,BFM}}$ can be derived from \refeq{eq:Lghost}
using the variation of the gauge-fixing functions $C^{a }$ 
\refeqf{eq:gfbfm} under the infinitesimal quantum gauge transformations
\refeqf{eq:gtsym_qu}~\cite{Denner:1994xt}.

Starting from the Lagrangian
\beq
\cL_{\mathrm{BFM}}(\Psi,\Psihat)= \cL_{\mathrm{class}}(\Psi+\Psihat)
+\cL_{\mathrm{fix,BFM}}(\Psi,\Psihat) + \cL_{\mathrm{FP,BFM}}(\Psi,\Psihat,u,\bar{u}) ,
\eeq
the generating functionals for Green functions and connected Green
functions can be defined as usual with background fields as additional
external sources. A Legendre transformation with respect to the
sources of the quantum fields leads to an effective action depending
on quantum fields and background fields. From this the
background-field effective action
$\Gammahat[\Psihat]=\Gamma[\tilde\Psi=0,\Psihat]$ is obtained upon
setting the Legendre transforms $\tilde\Psi$ of the sources of the
quantum fields to zero (apart from those for the fermion fields). Note
that $\Gammahat[\Psihat]$ is invariant with respect to background
gauge transformations \refeqf{eq:gtsym_bg}. This procedure is
described in \citeres{Abbott:1980hw,Abbott:1981ke,Bohm:2001yx}, where
it is also shown that the gauge-invariant effective action is
equivalent to a conventional effective action in a particular gauge.

Upon differentiating the gauge-invariant effective action
$\Gammahat[\Psihat]$ with respect to its arguments, the
background-field vertex functions are generated.  These can be
calculated from Feynman rules that distinguish between quantum and
background fields. While the quantum fields appear only inside loops,
the background fields are associated with the external lines of vertex
functions.  Apart from doubling the gauge and Higgs fields, the BFM
Feynman rules differ from the conventional ones only owing to the
gauge-fixing and ghost terms, which affect only vertices involving
both background and quantum fields.  Since the gauge-fixing functions
are non-linear in the fields, the gauge parameters enter also the
gauge-boson vertices.  As the gauge-fixing Lagrangian is quadratic in
the quantum fields, only vertices that involve exactly two quantum
fields or Faddeev--Popov fields are different from the conventional
ones. The fermions, which are not split into background and quantum
fields, appear both inside loops and on external lines, and the
Feynman rules involving fermion fields are the same as in the
conventional formalism.

The $S$~matrix is constructed in the usual way by forming trees with
vertices from $\Gammahat[\Psihat]$ which are connected by lowest-order
background-field propagators. The definition of the background-field
propagators requires the introduction of a gauge fixing for the
background fields, which is independent of the gauge fixing for the
quantum fields. It has been shown that the resulting $S$~matrix is
identical to the one of the conventional formalism
\cite{Abbott:1983zw,Rebhan:1984bg}. After introducing the gauge fixing
for the background fields a generating functional 
$\hat{T}_{\mathrm{c}}$ 
of connected background-field Green functions can be
obtained via a Legendre transformation \cite{Denner:1996gb}.

The BFM can be used to simplify calculations in
the EWSM. This is partly due to the fact that the gauge fixing of the
background fields is totally unrelated to the gauge fixing of the
quantum fields, which allows to choose a particularly suitable
background gauge, \eg the unitary gauge.  Moreover, in the
't~Hooft--Feynman gauge ($\xi=1$) for the quantum fields, many
vertices simplify with respect to the conventional formalism
\cite{Denner:1994nn,Denner:1994xt}.

\subsubsection{Background-field Ward identities} 
\label{se:bfWI}

An important property of the background-field formalism is the
invariance of the background-field effective action under
background-field gauge transformations with associated group
parameters $\thetahat^a$ ,
\beq\label{eq:WIGamma}
\frac{\delta\Gammahat[\Psihat]}{\delta\hat\theta^a} \;=\; 0, \quad
a=A,Z,\pm.
\eeq
This invariance implies ghost-free identities for the vertex functions
that are precisely the {\it Ward identities} related to the classical
Lagrangian.  For the EWSM, these identities were derived originally
for one-particle irreducible (1PI) vertex functions
$\GaBFM^{\ldots}_{\mathrm{1PI}}$ \cite{Denner:1994xt}, which are
obtained by taking functional derivatives of $\Gammahat[\Psihat]$
w.r.t.\ background fields and setting all background fields to zero
afterwards.  In this formulation the Ward identities for the bosonic
2-point functions read
\begin{align}
& k^\mu \GaBFM^{\FAhat\FAhat}_{\mathrm{1PI},\mu\nu}(k,-k) ={} 0, \qquad
\label{eq:gaAZ1}
k^\mu \GaBFM^{\FAhat\FZhat}_{\mathrm{1PI},\mu\nu}(k,-k) = 0, \qquad
k^\mu \GaBFM^{\FAhat\chihat}_{\mathrm{1PI},\mu}(k,-k) ={} 0, \qquad
k^\mu \GaBFM^{\FAhat\FHhat}_{\mathrm{1PI},\mu}(k,-k) ={} 0, \qquad
\\
& k^\mu \GaBFM^{\FZhat\FZhat}_{\mathrm{1PI},\mu\nu}(k,-k) -\ri\MZ \GaBFM^{\chihat\FZhat}_{\mathrm{1PI},\nu}(k,-k) 
= 0, \qquad
 k^\mu \GaBFM^{\FZhat\chihat}_{\mathrm{1PI},\mu}(k,-k) -{}\ri\MZ \GaBFM^{\chihat\chihat}_{\mathrm{1PI}}(k,-k)
+\frac{\ri e}{2\sw\cw} 
T^{\FHhat}
= 0,  
\\
\qquad
& k^\mu \GaBFM^{\FWhat^\pm\FWhat^\mp}_{\mathrm{1PI},\mu\nu}(k,-k)
\mp{}\MW \GaBFM^{\phihat^\pm\FWhat^\mp}_{\mathrm{1PI},\nu}(k,-k) = 0, \qquad
k^\mu \GaBFM^{\FWhat^\pm\phihat^\mp}_{\mathrm{1PI},\mu}(k,-k)
\mp{}\MW \GaBFM^{\phihat^\pm\phihat^\mp}_{\mathrm{1PI}}(k,-k)
\pm\frac{e}{2\sw} 
T^{\FHhat}
= 0,
\label{eq:gaW}
\end{align}
where fields and momenta are defined as incoming and
$T^{\FHhat}= \GaBFM^{\FHhat}_{\mathrm{1PI}}(0)$ (see \refse{se:tadpoles}).

Note that the vertex functions derived from the background-field
effective action $\Gammahat[\Psihat]$ do not involve contributions of
gauge-fixing terms for the background fields.  For ``full'' vertex
functions $\GaBFM^{\ldots}$ (involving also tadpole loop diagrams)
some of the corresponding identities change; in one-loop approximation
those are given by
\begin{alignat}{3}
& k^\mu \GaBFM^{\FZhat\FZhat}_{\mu\nu}(k,-k) 
-{}\ri\MZ 
r_{\FHhat}
\GaBFM^{\chihat\FZhat}_\nu(k,-k) 
= 0, &&
k^\mu \GaBFM^{\FZhat\chihat}_{\mu}(k,-k) -{}\ri\MZ
r_{\FHhat}
\GaBFM^{\chihat\chihat}(k,-k)
=0,   
 \nn\\  
& 
k^\mu \GaBFM^{\FWhat^\pm\FWhat^\mp}_{\mu\nu}(k,-k)
\mp{}\MW
r_{\FHhat}
\GaBFM^{\phihat^\pm\FWhat^\mp}_\nu(k,-k)
= 0, &\qquad&
 k^\mu \GaBFM^{\FWhat^\pm\phihat^\mp}_{\mu}(k,-k)
\mp{}\MW  
r_{\FHhat}
\GaBFM^{\phihat^\pm\phihat^\mp}(k,-k)
= 0,
\label{eq:BFMWI2ptfctfull}
\end{alignat}
where the background Higgs field $\Phihat$ is set to its vev
(including higher-order corrections) after taking functional
derivatives.  We give our precise definition of the vertex functions
in \refapp{se:conventions} and refer to Appendix~E of
\citere{Denner:2018opp} and to \refse{se:tadpoles} for a discussion of
the different tadpole schemes.  The extra factors
\beq
r_{\FHhat} = 1+ \frac{T^{\FHhat}}{\MH^2\varv}
\label{eq:rHhatFJTS}
\eeq
can be interpreted as describing the shift from the bare vev to the
correct vev.%
\footnote{When we take into account tadpole renormalization constants
in vertex functions in \refse{se:tadpoles}, \refeqs{eq:BFMWI2ptfctfull} 
and the subsequent equations of this section
remain valid with \refeq{eq:rHhatFJTS} in the FJTS variant described there.
In the PRTS variant described in \refse{se:tadpoles}, 
those identities hold with $r_{\FHhat}=1$.}

For the photon--fermion and the photon--\PW-boson vertices,
QED-like Ward identities are found,
\begin{align}
\label{WIAff}
& k^\mu \GaBFM^{\FAhat\Ffbar\Ff}_{\mu}(k,\bar p, p) ={} -e\Qf
\left[\GaBFM^{\Ffbar\Ff}(\bar p,-\bar{p}) - \GaBFM^{\Ffbar\Ff}(-p,p)\right],\\ 
& k^\mu \GaBFM^{\FAhat\FWhat^+\FWhat^-}_{\mu\rho\si}(k,k_+,k_-) ={}
e \left[\GaBFM^{\FWhat^+\FWhat^-}_{\rho\si}(k_+,-k_+) -
\GaBFM^{\FWhat^+\FWhat^-}_{\rho\si }(-k_-,k_-)\right], \\ 
& k_+^\rho \GaBFM^{\FAhat\FWhat^+\FWhat^-}_{\mu\rho\si}(k,k_+,k_-) -{} {}
\MW 
r_{\FHhat}
\GaBFM^{\FAhat\hat\phi^+\FWhat^-}_{\mu\si}(k,k_+,k_-) 
={} 
e\left[\GaBFM^{\FWhat^+\FWhat^-}_{\mu\si}(-k_-,k_-) -
\GaBFM^{\FAhat\FAhat}_{\mu\si}(k,-k)
+\frac{\cw}{\sw} \GaBFM^{\FAhat\FZhat}_{\mu\si}(k,-k)\right] ,
\label{eq:BFMWIAWW2full}
\end{align}
which hold for full ($\GaBFM^{\dots}$) and as well for 1PI vertex
functions ($\GaBFM^{\dots}_{\mathrm{1PI}}$) after setting
$r_{\FHhat}=1$.
From \refeq{WIAff} the universality of the electric charge in the
EWSM can be derived in the same way as for {\it Quantum Electrodynamics (QED)}
\cite{Aoki:1980ix,Bohm:2001yx}.

The Ward identities \refeqf{eq:WIGamma} for the effective action
$\Gammahat[\Psihat]$ translate into identities for the generating
functional $\hat{T}_\mathrm{c}$
of connected Green functions, which
were explicitly derived in \citere{Denner:1996gb}.  The connected
2-point functions involving neutral gauge bosons, for instance, obey
\begin{align}\label{eq:WIVV_BFM}
&k^\mu G^{\FAhat\FAhat}_{\rc,\mu\nu}(k,-k) =  \frac{-\ri\hat\xi_\FA k_\nu}{k^2}, \qquad
 k^\mu G^{\FAhat\FZhat}_{\rc,\mu\nu}(k,-k) = 0, \nl
&k^\mu G^{\FZhat\FZhat}_{\rc,\mu\nu}(k,-k) 
+\ri\hat\xi_\FZ\MZ G^{\chihat\FZhat}_{\rc,\nu}(k,-k)
= \frac{-\ri\hat\xi_\FZ k_\nu}{k^2-\hat\xi_\FZ\MZ^2
r_{\FHhat}
},\nl
&k^\mu G^{\FZhat\chihat}_{\rc,\mu}(k,-k) 
+\ri\hat\xi_\FZ\MZ G^{\chihat\chihat}_{\rc}(k,-k)
=\frac{-\hat\xi_\FZ\MZ
r_{\FHhat}
}
{k^2-\hat\xi_\FZ\MZ^2
r_{\FHhat}
},\nl
&k^\mu G^{\FWpmhat\FWmphat}_{\rc,\mu\nu}(k,-k) 
\pm\hat\xi_\FW\MW G^{\phipmhat\FWmphat}_{\rc,\nu}(k,-k)
= \frac{-\ri\hat\xi_\FW k_\nu}{k^2-\hat\xi_\FW\MW^2
r_{\FHhat}
},\nl
&k^\mu G^{\FWpmhat\phimphat}_{\rc,\mu}(k,-k) 
\pm\hat\xi_\FW\MW G^{\phipmhat\phimphat}_{\rc}(k,-k)
=\frac{\pm\ri\hat\xi_\FW\MW
r_{\FHhat}
}
{k^2-\hat\xi_\FW\MW^2
r_{\FHhat}
},
\end{align}
where $\hat\xi_a$ are the gauge parameters of the background-field
't~Hooft gauge-fixing terms, which are assumed to have the form of 
\refeqs{eq:gf} and \refeqf{eq:Lfix} for background fields.
The identities in \refeq{eq:WIVV_BFM} can be combined to derive
\refeq{eq:WIVV}, where all explicit tadpole contributions drop out. 
The photon--fermion vertex fulfils the same Ward
identity as in QED,
\begin{equation}
-\frac{\ri}{\hat\xi_\FA}k^2 k^\mu G^{\FAhat\Ff\Ffbar}_{\rc,\mu}(k,\bar{p},p) 
={} -e\Qf
\left[G^{\Ff\Ffbar}_{\rc}(-p,p)-G^{\Ff\Ffbar}_{\rc}(\bar{p},-\bar{p})\right].
\end{equation}

For truncated OS Green functions (defined in \refapp{se:conventions})
the Ward identities simplify, for instance, to~\cite{Denner:1996gb}
\begin{equation}\label{eq:WIET}
k^\nu G^{\FAhat\ldots}_{\mathrm{trunc},\nu} = 0, \qquad
k^\nu G^{\FZhat\ldots}_{\mathrm{trunc},\nu} = 
\ri\MZ
r_{\FHhat}
G^{\chihat\ldots}_{\mathrm{trunc}}, \qquad
k^\nu G^{\FWhat^\pm\ldots}_{\mathrm{trunc},\nu} 
= \pm\MW
r_{\FHhat}
G^{\phihat^\pm\ldots}_{\mathrm{trunc}},
\end{equation}
where the ellipses stand for any OS fields, i.e.\ the
corresponding legs are truncated, put on shell, and contracted with
wave functions.%
\footnote{Recall that the field indices denote outgoing fields for the
  connected Green functions $G^{\dots}_{\rc}$,
  but incoming fields for the truncated Green
  functions~$G^{\dots}_{\mathrm{trunc}}$.}
  While the first of these identities expresses
electromagnetic current conservation, the others imply the
Goldstone-boson equivalence theorem, as spelled out in \refse{se:gbet}.

We finally mention that the validity of
the background-field Ward identities for connected Green functions
can even be sustained in finite orders of perturbation theory if the
resummed propagators are used to connect the vertex functions and the
inverse propagators are calculated in the same order of perturbation
theory as all the other vertex functions \cite{Denner:1996gb}.  
The necessary prerequisite for this property is 
the linearity of the background-field Ward
identities for vertex functions, as featured by the
identities \refeqf{eq:gaAZ1}--\refeqf{eq:gaW} for the 1PI vertex functions
and by the identities \refeqf{eq:BFMWI2ptfctfull} for the
vertex functions including reducible tadpole contributions.%
\footnote{Note that the identities \refeqf{eq:BFMWI2ptfctfull} are only
linear in the BFM vertex functions if the constant $r_{\FHhat}$ 
is interpreted as part of the parameters and not as vertex function.}
As a consequence, these Ward identities and all identities derived
therefrom hold exactly in finite orders of perturbation theory.
In contrast, Slavnov--Taylor identities of the conventional Faddeev--Popov
quantization in general hold for
connected Green functions in a given order of perturbation theory only
if all contributions, including the propagators, are expanded up to
this order.

\subsection{Standard-Model effective theory}
\label{se:smeft}

Presently, there is no significant evidence for physics beyond the SM
in the TeV region, 
so that new-physics effects at the LHC are expected to be small.  
In this situation, deviations from the SM Lagrangian can
conveniently and consistently be described within an effective
Lagrangian framework.  The effective Lagrangian results from the
Lagrangian of a more comprehensive theory by integrating out the heavy
degrees of freedom, so that the different terms in the effective
Lagrangian are obtained from a systematic expansion in inverse powers
of the heavy scale $\Lambda$
of new physics.  In this way, the effective
Lagrangian provides a parametrization of possible deviations from SM
predictions.  It should be noted, however, that such an approach
assumes that no light degrees of freedom beyond those of the SM are
present.

The description of BSM physics based on an effective Lagrangian in
terms of the SM fields was pioneered by Buchm\"uller and Wyler
\cite{Buchmuller:1985jz}, who provided a list of operators of
dimensions~5 and 6 in the linear parametrization of the Higgs sector
with a Higgs doublet. In the sequel various authors have considered
subsets of this operator basis or introduced different sets of
operators adapted to specific goals. A complete minimal basis of
dimension-6 operators was presented in \Bref{Grzadkowski:2010es},
where {\it minimal} means that equations of motion are used as far as
possible to reduce the number of operators. Note that operators that
can be expressed in terms of others via equations of motion do not
additionally affect physical $S$-matrix elements.  Other bases were
proposed, and different authors prefer to use different sets of
operator bases, motivated by different expectations for the form of
new
physics~\cite{Giudice:2007fh,Bonnet:2011yx,Corbett:2012dm,%
Bonnet:2012nm,Passarino:2012cb,Corbett:2012ja,Contino:2013kra,Heinemeyer:2013tqa,deFlorian:2016spz}.
In this review, we restrict ourselves to the {\it Warsaw basis} of
\Bref{Grzadkowski:2010es} which is based on a linear representation of
the Higgs sector involving a single Higgs doublet, like the usual
formulation of the SM. For a more comprehensive discussion of the SM
effective field theory (SMEFT) we refer to \citere{Brivio:2017vri}.

We define an effective Lagrangian based on a linear representation of
the EW gauge symmetry with a Higgs-doublet field $\Phi$ following
closely the framework introduced in \Bref{Buchmuller:1985jz} and
further developed in \Bref{Grzadkowski:2010es}.%
\footnote{We slightly adapt the conventions in order to conform with our
  conventions for the SM Lagrangian \refeqf{eq:LSM_sym}. To this end, we have to change
  the signs of the $\SU(2)_\rw$ $\FW^a$ fields and of the $\SU(3)_{\rc}$ fields
  $\FG^A$. Moreover, we use the convention
  $\eps^{0123}=-\eps_{0123}=+1$, while \Bref{Grzadkowski:2010es}
  uses $\eps_{0123}=+1$.}
The effective Lagrangian has the general form
\beq\label{eq:LEFT}
\cL_{\mathrm{eff}}= \cL^{(4)}_{\mathrm{SM}}
+ \frac{1}{\Lambda}\sum_k \Lcoeff^{(5)}_k \Lop^{(5)}_k
+ \frac{1}{\Lambda^2}\sum_k \Lcoeff^{(6)}_k \Lop^{(6)}_k
+ \cO\left( \frac{1}{\Lambda^3}\right),
\eeq
where $\cL^{(4)}_{\mathrm{SM}}$ is the symmetric SM Lagrangian
\refeqf{eq:LSM_sym}, including also QCD as described in
\refse{se:QCD}, $\Lop^{(5)}_k$ denotes dimension-5 operators and
$\Lcoeff^{(5)}_k$ the corresponding Wilson coefficients, and
$\Lop^{(6)}_k$ denotes dimension-6 operators with corresponding Wilson
coefficients $\Lcoeff^{(6)}_k$. The operators $\Lop^{(D)}_k$ are
invariant under $\SU(3)_{\rc}\times \SU(2)_{\rw}\times \U(1)_Y$
transformations.  Since the effective Lagrangian must be hermitean, in
\refeq{eq:LEFT} for each non-hermitean operator $\Lop_k$ the
hermitean-conjugate operator $\Lop_k^\dagger$ appears with the
complex-conjugate Wilson coefficient $\Lcoeff_k^*$.

\subsubsection{Conventions and definition of the effective operator basis}

We employ the conventions for the SM used above as far as possible.
Colour and weak-isospin indices in the fundamental representations are
denoted as $\alpha=1,2,3$ and $\si=1,2$, respectively, those in the
adjoint representations as $A=1,\ldots,8$ and $a=1,2,3$, while the
generation indices are written as $i=1,2,3$. The matter fields of the
SM comprise the left-handed lepton doublets $L_{\si,i}$, the
right-handed charged leptons $l_i$, the left-handed quark doublets
$Q_{\al,\si,i}$, the right-handed quarks $\Fu_{\al,i}$, $\Fd_{\al,i}$,
and the Higgs doublet $\Phi_\si$ with hypercharges
$Y_{\rw}=-1,-2,1/3,4/3,-2/3,1$, respectively.  Right-handed neutrinos are
not included.  The fermion fields appearing in this section are the
interaction eigenstates and should therefore be primed, but we
suppress the primes in this section.  The charge-conjugate Higgs field
$\Phi^{\mathrm{c}}$ is given in \refeq{eq:phi_cc}.  The field-strength
tensors and the covariant derivative are defined in
\refeqs{eq:EW_field_strengths}, \refeqf{eq:QCD_field_strength}, and
\refeqf{eq:covdersm}.

Using the dual field-strength tensors defined in
\refeq{eq:dual-field-strengths}, we introduce hermitean derivatives
\beq
\Phi^\dagger\ri \overset\leftrightarrow{D}_\mu\Phi
=\ri\left[\Phi^\dagger D_\mu\Phi - (D_\mu\Phi)^\dagger\Phi\right],
\qquad
\Phi^\dagger\ri {\overset\leftrightarrow{D}}{}^a_\mu\Phi
=\ri\left[\Phi^\dagger\tau^a D_\mu\Phi - (D_\mu\Phi)^\dagger \tau^a\Phi\right],
\eeq
and $\si^{\mu\nu}=\ri(\gamma^\mu\gamma^\nu-\gamma^\nu\gamma^\mu)/2$.
Finally, $\eps_{\si\tau}$ is totally antisymmetric with $\eps_{12}=+1$.

The SM gauge symmetry allows for only one dimension-5 operator up to
hermitean conjugation and flavour assignments \cite{Weinberg:1979sa},
\beq
\Lop_{\nu\nu} = (\Phi^{\mathrm{c},\dagger}\FL_i)^{\rT}\, C\, (\Phi^{\mathrm{c},\dagger}\FL_j),
\eeq
where $C$ is the charge-conjugation matrix in Dirac space. 
Here and in the following, generation indices $i,j,$ etc.\ are
suppressed in the operator names.
The operator $\Lop_{\nu\nu}$ violates lepton number and generates neutrino masses and
mixing after EW symmetry breaking.

The independent dimension-6 operators allowed by the SM gauge
symmetries are listed in \reftas{ta:Lops1} and \ref{ta:Lops2}. They
are obtained by transforming the results of \Bref{Grzadkowski:2010es}
to our conventions.  The generation indices are suppressed on the
left-hand sides of the equations. Dirac indices are contracted within
the parentheses and suppressed. This is also done for colour and
isospin indices as far as possible. In addition to the operators in
\reftas{ta:Lops1} and \ref{ta:Lops2}, for each non-hermitean operator
its hermitean conjugate must be included. The Wilson coefficients of
these operators are in general complex, while those of the hermitean
operators are real.
\begin{table}
\centering{$\displaystyle
\renewcommand{\arraystretch}{1.4}\begin{array}{r@{}l@{\qquad}r@{}l@{\qquad}r@{}l}
\hline
\multicolumn{2}{l}{\Phi^6\quad \text{and}\quad \Phi^4D^2} & 
\multicolumn{2}{l}{\psi^2\Phi^3} & 
\multicolumn{2}{l}{X^3}
\\
\hline
\Lop_\Phi &{}= (\Phi^\dagger\Phi)^3   &
\Lop_{\Fl\Phi}&{}=(\Phi^\dagger\Phi)(\longbar{\FL}_i\, \Fl_j \Phi) 
&\Lop_{\FG} &{}= -f^{ABC}\FG^{A\nu}_\mu \FG^{B\rho}_\nu \FG^{C\mu}_\rho
\\
\Lop_{\Phi\Box} &{}= (\Phi^\dagger\Phi)\Box (\Phi^\dagger\Phi)   &
\Lop_{\Fu\Phi}&{}=(\Phi^\dagger\Phi)(\longbar{\FQ}_i\, \Fu_j \Phi^{\mathrm{c}}) 
&\Lop_{\widetilde{G}} &{}= -f^{ABC}\widetilde{G}^{A\nu}_\mu \FG^{B\rho}_\nu \FG^{C\mu}_\rho
\\
\Lop_{\Phi D} &{}= (\Phi^\dagger D^\mu\Phi)^*(\Phi^\dagger D_\mu\Phi) &
\Lop_{\Fd\Phi}&{}=(\Phi^\dagger\Phi)(\longbar{\FQ}_i\, \Fd_j \Phi) 
&\Lop_{\FW} &{}= -\eps^{abc}\FW^{a\nu}_\mu \FW^{b\rho}_\nu \FW^{c\mu}_\rho
\\
&&&&\Lop_{\widetilde{\FW}} &{}= -\eps^{abc}\widetilde{\FW}^{a\nu}_\mu \FW^{b\rho}_\nu \FW^{c\mu}_\rho
\rule[-1.5ex]{0pt}{3ex}\\
\hline
\multicolumn{2}{l}{X^2\Phi^2} &\multicolumn{2}{l}{\psi^2 X\Phi} & \multicolumn{2}{l}{\psi^2\Phi^2D} \\
\hline\rule{0pt}{3.5ex}
\Lop_{\Phi\FG} &{}= (\Phi^\dagger\Phi)\FG^A_{\mu\nu}\FG^{A\mu\nu} &
\Lop_{\Fu\FG} &{}= -(\longbar{\FQ}_i\sigma^{\mu\nu}\frac{\lambda^A}{2} \Fu_j)
\,\Phi^{\mathrm{c}}\, \FG^A_{\mu\nu} &
\Lop_{\Phi \FL}^{(1)}&{}=(\Phi^\dagger\ri\overset\leftrightarrow{D}_\mu\Phi)
(\longbar{\FL}_i\gamma^\mu \FL_j)
\\
\Lop_{\Phi \widetilde{\FG}} &{}= (\Phi^\dagger\Phi)\widetilde{\FG}^A_{\mu\nu}\FG^{A\mu\nu} &
\Lop_{\Fd\FG} &{}= -(\longbar{\FQ}_i\sigma^{\mu\nu}\frac{\lambda^A}{2} \Fd_j)
\,\Phi\, \FG^A_{\mu\nu} &
\Lop_{\Phi \FL}^{(3)}&{}=(\Phi^\dagger\ri\overset\leftrightarrow{D}{}_\mu^a\Phi)
(\longbar{\FL}_i\gamma^\mu \tau^a \FL_j)
\\
\Lop_{\Phi\FW} &{}= (\Phi^\dagger\Phi)\FW^a_{\mu\nu}\FW^{a\mu\nu} &
\Lop_{\Fl\FW} &{}= -(\longbar{\FL}_i\sigma^{\mu\nu} \Fl_j) {\tau^a}
\Phi\FW^a_{\mu\nu} &
\Lop_{\Phi\Fl}&{}=(\Phi^\dagger\ri\overset\leftrightarrow{D}_\mu\Phi)
(\bar{\Fl}_i\gamma^\mu \Fl_j)
\\
\Lop_{\Phi \widetilde{\FW}} &{}= (\Phi^\dagger\Phi)\widetilde{\FW}^a_{\mu\nu}\FW^{a\mu\nu} &
\Lop_{\Fu\FW} &{}= -(\longbar{\FQ}_i\sigma^{\mu\nu}\Fu_j){\tau^a}\Phi^{\mathrm{c}}\FW^a_{\mu\nu} &
\Lop_{\Phi \FQ}^{(1)}&{}=(\Phi^\dagger\ri\overset\leftrightarrow{D}_\mu\Phi)
(\longbar{\FQ}_i\gamma^\mu \FQ_j)
\\
\Lop_{\Phi \FB} &{}= (\Phi^\dagger\Phi)\FB_{\mu\nu}\FB^{\mu\nu} &
\Lop_{\Fd\FW} &{}= -(\longbar{\FQ}_i\sigma^{\mu\nu} \Fd_j) {\tau^a}
\Phi \FW^a_{\mu\nu} &
\Lop_{\Phi \FQ}^{(3)}&{}=(\Phi^\dagger\ri\overset\leftrightarrow{D}{}_\mu^a\Phi)
(\longbar{\FQ}_i\gamma^\mu \tau^a \FQ_j)
\\
\Lop_{\Phi \widetilde{\FB}} &{}= (\Phi^\dagger\Phi)\widetilde{\FB}_{\mu\nu}\FB^{\mu\nu} &
\Lop_{\Fl \FB} &{}= (\longbar{\FL}_i\sigma^{\mu\nu} \Fl_j)\Phi\FB_{\mu\nu} &
\Lop_{\Phi \Fu}&{}=(\Phi^\dagger\ri\overset\leftrightarrow{D}_\mu\Phi)
(\bar{\Fu}_i\gamma^\mu \Fu_j)
\\
\Lop_{\Phi{\FW \FB}} &{}= -(\Phi^\dagger\tau^a\Phi){\FW}^a_{\mu\nu}\FB^{\mu\nu} &
\Lop_{\Fu \FB} &{}= (\longbar{\FQ}_i\sigma^{\mu\nu} \Fu_j\Phi^{\mathrm{c}})\FB_{\mu\nu} &
\Lop_{\Phi \Fd}&{}=(\Phi^\dagger\ri\overset\leftrightarrow{D}_\mu\Phi)
(\bar{\Fd}_i\gamma^\mu \Fd_j)
\\
\Lop_{\Phi \widetilde{\FW}\FB} &{}= -(\Phi^\dagger\tau^a\Phi)\widetilde{\FW}^a_{\mu\nu}\FB^{\mu\nu} &
\Lop_{\Fd \FB} &{}= (\longbar{\FQ}_i\sigma^{\mu\nu} \Fd_j)\Phi\FB_{\mu\nu} &
\Lop_{\Phi \Fu\Fd}&{}=(\Phi^{\mathrm{c}\dagger}\ri{D}_\mu\Phi)
(\bar{\Fu}_i\gamma^\mu \Fd_j)
\rule[-1.5ex]{0pt}{3ex}\\
\hline
\end{array}$}
\caption{Dimension-6 operators involving the Higgs-doublet field or
  gauge-boson fields. For all
$\psi^2\Phi^3$, $\psi^2 X\Phi$ operators and for $\Lop_{\Phi \Pu\Fd}$ the
hermitean conjugates must be included as well (taken from \citere{Grzadkowski:2010es}).}
\label{ta:Lops1}
\end{table}


%
\begin{table}
\centering{$\displaystyle
\renewcommand{\arraystretch}{1.4}\begin{array}{r@{}l@{\qquad}r@{}l@{\qquad}r@{}l}
\hline
\multicolumn{2}{l}{(\bar LL)(\bar LL)} & 
\multicolumn{2}{l}{(\bar RR)(\bar RR)} &
\multicolumn{2}{l}{(\bar LL)(\bar RR)}\\
\hline
\Lop_{\FL\FL}        ={}& (\bar \FL_i \gamma_\mu \FL_j)(\bar \FL_k \gamma^\mu \FL_l) &
\Lop_{\Fl\Fl}        ={}& (\bar \Fl_i \gamma_\mu \Fl_j)(\bar \Fl_k \gamma^\mu \Fl_l) &
\Lop_{\FL\Fl}        ={}& (\bar \FL_i \gamma_\mu \FL_j)(\bar \Fl_k \gamma^\mu \Fl_l) \\
\Lop_{\FQ\FQ}^{(1)}  ={}& (\bar \FQ_i \gamma_\mu \FQ_j)(\bar \FQ_k \gamma^\mu \FQ_l) &
\Lop_{\Fu\Fu}        ={}& (\bar \Fu_i \gamma_\mu \Fu_j)(\bar \Fu_k \gamma^\mu \Fu_l) &
\Lop_{\FL\Fu}        ={}& (\bar \FL_i \gamma_\mu \FL_j)(\bar \Fu_k \gamma^\mu \Fu_l) \\
\Lop_{\FQ\FQ}^{(3)}  ={}& (\bar \FQ_i \gamma_\mu \tau^a \FQ_j)(\bar \FQ_k \gamma^\mu \tau^a \FQ_l) &
\Lop_{\Fd\Fd}        ={}& (\bar \Fd_i \gamma_\mu \Fd_j)(\bar \Fd_k \gamma^\mu \Fd_l) &
\Lop_{\FL\Fd}        ={}& (\bar \FL_i \gamma_\mu \FL_j)(\bar \Fd_k \gamma^\mu \Fd_l) \\
\Lop_{\FL\FQ}^{(1)}  ={}& (\bar \FL_i \gamma_\mu \FL_j)(\bar \FQ_k \gamma^\mu \FQ_l) &
\Lop_{\Fl\Fu}        ={}& (\bar \Fl_i \gamma_\mu \Fl_j)(\bar \Fu_k \gamma^\mu \Fu_l) &
\Lop_{\FQ\Fl}        ={}& (\bar \FQ_i \gamma_\mu \FQ_j)(\bar \Fl_k \gamma^\mu \Fl_l) \\
\Lop_{\FL\FQ}^{(3)}  ={}& (\bar \FL_i \gamma_\mu \tau^a \FL_j)(\bar \FQ_k \gamma^\mu \tau^a \FQ_l) &
\Lop_{\Fl\Fd}        ={}& (\bar \Fl_i \gamma_\mu \Fl_j)(\bar \Fd_k\gamma^\mu \Fd_l) &
\Lop_{\FQ\Fu}^{(1)}  ={}& (\bar \FQ_i \gamma_\mu \FQ_j)(\bar \Fu_k \gamma^\mu \Fu_l) \\ 
&&
\Lop_{\Fu\Fd}^{(1)}  ={}& (\bar \Fu_i \gamma_\mu \Fu_j)(\bar \Fd_k \gamma^\mu \Fd_l) &
\Lop_{\FQ\Fu}^{(8)}  ={}& \left(\bar \FQ_i \gamma_\mu \frac{\la^A}{2} \FQ_j\right)\left(\bar \Fu_k \gamma^\mu \frac{\la^A}{2} \Fu_l\right) \\ 
&& 
\Lop_{\Fu\Fd}^{(8)}  ={}& \left(\bar \Fu_i \gamma_\mu \frac{\la^A}{2} \Fu_j\right)\left(\bar \Fd_k \gamma^\mu \frac{\la^A}{2} \Fd_l\right) &
\Lop_{\FQ\Fd}^{(1)}  ={}& (\bar \FQ_i \gamma_\mu \FQ_j)(\bar \Fd_k \gamma^\mu \Fd_l) \\
&&&&
\Lop_{\FQ\Fd}^{(8)}  ={}& \left(\bar \FQ_i \gamma_\mu \frac{\la^A}{2} \FQ_j\right)\left(\bar \Fd_k \gamma^\mu \frac{\la^A}{2} \Fd_l\right)\\
\hline
\multicolumn{2}{l}{(\bar LR)(\bar RL)\text{ and }(\bar LR)(\bar LR)} &
\multicolumn{2}{l}{\text{B-violating}}\\\hline
\Lop_{\FL\Fl\Fd\FQ} &{}= (\bar \FL_i^\si \Fl_j)(\bar \Fd_k \FQ_l^\si) &
\Lop_{\Fd\Fu\FQ} & \multicolumn{3}{@{}l}{{}= \eps^{\alpha\beta\gamma} \eps_{\si\tau} 
 \left[ (d^\alpha_i)^{\rT} C u^\beta_j \right]\left[(\FQ^{\gamma\si}_k)^{\rT} C \FL^\tau_l\right]}\\
\Lop_{\FQ\Fu\FQ\Fd}^{(1)} &{}= (\bar \FQ_i^\si \Fu_j) \eps_{\si\tau} (\bar \FQ_k^\tau \Fd_l) &
\Lop_{\FQ\FQ\Fu}& \multicolumn{3}{@{}l}{ {}= \eps^{\alpha\beta\gamma} \eps_{\si\tau} 
  \left[ (\FQ^{\alpha\si}_i)^{\rT} C \FQ^{\beta\tau}_j \right]\left[(u^\gamma_k)^{\rT} C \Fl_l\right]}\\
\Lop_{\FQ\Fu\FQ\Fd}^{(8)} &{}= \left(\bar \FQ_i^\si \frac{\la^A}{2} \Fu_j\right) \eps_{\si\tau} \left(\bar \FQ_k^\tau \frac{\la^A}{2} \Fd_l\right) &
\Lop_{\FQ\FQ\FQ}& \multicolumn{3}{@{}l}{ {}=\eps^{\alpha\beta\gamma} \eps_{\si\tau} \eps_{\si'\tau'} 
  \left[ (\FQ^{\alpha\si}_i)^{\rT} C \FQ^{\beta\tau}_j \right]\left[(\FQ^{\gamma\si'}_k)^{\rT} C \FL^{\tau'}_l\right]}\\
\Lop_{\FL\Fl\FQ\Fu}^{(1)} &{}= (\bar \FL_i^\si \Fl_j) \eps_{\si\tau}
(\bar \FQ_k^\tau \Fu_l) &
\Lop_{\Fd\Fu\Fu} & \multicolumn{3}{@{}l}{ {}=\eps^{\alpha\beta\gamma} 
  \left[ (\Fd^\alpha_i)^{\rT} C u^\beta_j \right]\left[(u^\gamma_k)^{\rT} C \Fl_l\right]}\\
\Lop_{\FL\Fl\FQ\Fu}^{(3)} &{}= (\bar \FL_i^\si \sigma_{\mu\nu} \Fl_j)
\eps_{\si\tau} \rlap{$(\bar \FQ_k^\tau \sigma^{\mu\nu} \Fu_l)$}\\
\hline
\end{array}$}
\caption{Dimension-6 operators involving four fermion fields (taken from \citere{Grzadkowski:2010es}).
}
\label{ta:Lops2}
\end{table}

Counting the number of dimension-6 operators in \reftas{ta:Lops1} and
\ref{ta:Lops2} that conserve baryon and lepton number, we find 15
bosonic operators, 19 operators with a single fermionic current, and
25 B-conserving four-fermion operators. For three generations, the
dimension-6 Lagrangian involves 1350 CP-even and 1149 CP-odd B- and
L-conserving operators.

\subsubsection{Translation to the physical basis}

The higher-dimensional operators do not only introduce new coupling
structures, they also contribute to the linear, to the bilinear, and
to the interaction terms of the Lagrangian.  This leads to
modifications of the relations between properly normalized fields and
parameters of the physical basis and their counterparts of the
symmetric Lagrangian with respect to the original SM relations.  We
outline these modifications in the following, but restrict ourselves
to the B-conserving dimension-6 operators and to terms linear in the
Wilson coefficients.  The discussion is largely taken from
\citeres{Alonso:2013hga,Passarino:2016pzb}, but translated to our
conventions.

The operator $\Lop_\Phi$ changes the scalar potential at order
$\varv^2/\Lambda^2$ to
\beq
V(\Phi)= -\mu^2(\Phi^\dagger\Phi) + \frac{1}{4}\la(\Phi^\dagger\Phi)^2
-\frac{\Lcoeff_{\Phi}}{\Lambda^2}(\Phi^\dagger\Phi)^3,
\eeq
leading to the new minimum
\begin{equation} \label{eq:vdef_SMEFT}
|\langle\Phi\rangle|^{2} = \frac{2\mu ^{2}}{\lambda } 
\left(1+\frac{12\Lcoeff_{\Phi}\mu^2}{\Lambda^2\la^2}\right)
=
\frac{\varv^{2}}{2}
\end{equation}
by expanding the exact solution to first order in $\Lcoeff_{\Phi}$.

The Higgs-doublet kinetic terms are modified by the operators
$\Lop_{\Phi\Box}$ and $\Lop_{\Phi D}$. By parametrizing the scalar doublet field as
\begin{equation} \label{eq:phidec_smeft}
\Phi (x) = \left( \begin{array}{c}
\phi ^{+}(x) \\ \frac{1}{\sqrt2}\bigl[\varv +(1+c_{\FH,\mathrm{kin}})\FH(x) +\ri(1+c_{\chi,\mathrm{kin}})\chi (x) \bigr]
\end{array} \right),
\end{equation}
with the coefficients 
\begin{equation}
c_{\FH,\mathrm{kin}} ={}
\left(\Lcoeff_{\Phi\Box}-\frac{1}{4}\Lcoeff_{\Phi
    D}\right)\frac{\varv^2}{\La^2},\qquad
c_{\chi,\mathrm{kin}} ={}
-\frac{1}{4}\Lcoeff_{\Phi D}\frac{\varv^2}{\La^2},
\end{equation}
the scalar kinetic terms are properly normalized when dimension-6
terms are included.  The Higgs-boson mass is obtained as
\beq
\MH^2=\frac{1}{2}\la \varv^2\left(1-\frac{6\Lcoeff_{\Phi}}{\la}\frac{\varv^2}{\La^2} +
2c_{\FH,\mathrm{kin}}\right).
\eeq

The operators $\Lop_{\Phi G}$, $\Lop_{\Phi W}$, $\Lop_{\Phi B}$, and
$\Lop_{\Phi WB}$ affect the kinetic terms of the gauge fields,
resulting in the broken theory in the following contribution to the Lagrangian
\begin{align}
&
{-\frac{1}{2}} \FW^+_{\mu\nu}\FW^{-\mu\nu}
-\frac{1}{4} \FW^3_{\mu\nu}\FW^{3,\mu\nu}
-\frac{1}{4} \FB_{\mu\nu}\FB^{\mu\nu}
-\frac{1}{4} \FG^A_{\mu\nu}\FG^{A,\mu\nu} 
\nl&
+\frac{1}{2} \frac{\varv^2}{\La^2} \left({
  {\Lcoeff_{\Phi W}} \FW^a_{\mu\nu}\FW^{a,\mu\nu} 
+ {\Lcoeff_{\Phi B}} \FB_{\mu\nu}\FB^{\mu\nu} 
+ {\Lcoeff_{\Phi WB}} \FW^3_{\mu\nu}\FB^{\mu\nu}
+ \Lcoeff_{\Phi G}} \FG^A_{\mu\nu}\FG^{A,\mu\nu} 
\right),
\end{align} 
\ie the gauge fields are not canonically normalized.
The gauge-boson mass terms are obtained as
\begin{equation}
\frac{1}{4} g_2^2\varv^2\FW^+_{\mu}\FW^{-\mu}
+\frac{1}{8} \varv^2\left(1+\frac{1}{2}\frac{\varv^2}{\La^2}\Lcoeff_{\Phi D}\right)
(g_2\FW^{3}_{\mu}+g_1\FB_\mu)^2.
\end{equation} 

Properly normalized kinetic terms are obtained upon redefining the
gauge fields as
\beq\label{eq:vector_fields_smeft}
\FG^A_\mu=\FGn^A_\mu\left(1+\frac{\Lcoeff_{\Phi\FG}}{\La^2}\varv^2\right),\qquad
\FW^a_\mu=\FWn^a_\mu\left(1+\frac{\Lcoeff_{\Phi\FW}}{\La^2}\varv^2\right),\qquad
\FB_\mu=\FBn_\mu\left(1+\frac{\Lcoeff_{\Phi\FB}}{\La^2}\varv^2\right).
\eeq
It is customary to introduce the modified coupling constants
\beq
\gn_\rs=\gs\left(1+\frac{\Lcoeff_{\Phi\FG}}{\La^2}\varv^2\right),\qquad
\gn_2=g_2\left(1+\frac{\Lcoeff_{\Phi\FW}}{\La^2}\varv^2\right),\qquad
\gn_1=g_1\left(1+\frac{\Lcoeff_{\Phi\FB}}{\La^2}\varv^2\right),
\eeq
so that the products $g_\rs\FG_\mu^A=\gn_\rs\FGn_\mu^A$, etc. are
unchanged and the form of the covariant derivative is not modified.

The transformation from the $\FWn^3,\FBn$ fields to the $\FZ,\FA$
fields can be written as
\begin{equation}
\left(\barr{l} \FWn^3_{\mu } \\ \FBn_{\mu } \earr \right) = {}
\left(\barr{cc} 1 & \frac{1}{2}\frac{\varv^2}{\La^2}\Lcoeff_{\Phi\FW\FB}\\ 
 \frac{1}{2}\frac{\varv^2}{\La^2}\Lcoeff_{\Phi\FW\FB} & 1 \earr \right)
\left(\barr{rr} \cwn & -\swn \\ \swn & \cwn  \earr \right)
\left(\barr{l} \FZn_{\mu } \\ \FAn_{\mu } \earr \right) ,
\end{equation}
with the rotation angle $\thetawn$ determined from 
\beq
\cwn=\cos\thetawn=\frac{\gn_{2}}{\sqrt{\gn_{2}^{2}+\gn_{1}^{2}}}
\left(1-\frac{1}{2}\frac{\varv^2}{\La^2}\frac{\gn_1}{\gn_2}
\frac{\gn_2^2-\gn_1^2}{\gn_2^2+\gn_1^2} \Lcoeff_{\Phi\FW\FB}\right)
,
\qquad \swn=\sin\thetawn .
\label{eq:ewanglen}
\eeq

The masses of the W and Z~bosons result in 
\begin{equation}
\MW^2 = \frac{1}{4}\gn_{2}^2 \varv^2,  
\qquad \MZ^2  = \frac{1}{4}(\gn_{1}^{2}+\gn_{2}^{2}) \varv^2
+\frac{1}{8}(\gn_1^2+\gn_2^2) \frac{\varv^4}{\La^2} \Lcoeff_{\Phi D}
+\frac{1}{2}\gn_1\gn_2 \frac{\varv^4}{\La^2} \Lcoeff_{\Phi\FW\FB}.
\end{equation}
The EW part of the covariant derivative becomes
\beq
D_\mu=\partial_\mu-\ri\frac{\gn_2}{\sqrt{2}}
\left( I_{\rw}^+ \FWn^+_\mu + I_{\rw}^-\FWn^-_\mu\right)
-\ri{\gn_\FZ}\left(I_{\rw}^3-\swn^2 Q\right)\FZn_\mu+\ri\gen Q\FAn_\mu,
\eeq
with the effective couplings
\begin{equation}
\gen ={} \frac{\gn_1\gn_2}{\sqrt{\gn_1^2+\gn_2^2}}
\left(1-\frac{\gn_1\gn_2}{\gn_1^2+\gn_2^2} \frac{\varv^2}{\La^2}
    \Lcoeff_{\Phi\FW\FB} \right),
\qquad 
\gn_\FZ ={} \frac{\gen}{\swn\cwn}
\left(1+\frac{\gn_1^2+\gn_2^2} {2\gn_1\gn_2}\frac{\varv^2}{\La^2}
    \Lcoeff_{\Phi\FW\FB} \right) ,
\end{equation}
and $I_{\rw}^\pm=I_{\rw}^1\pm\ri I_{\rw}^2.$
The $\rho$ parameter, defined as the ratio of charged and neutral
currents at low energies \cite{Ross:1975fq}, results in
\beq
\rho=\frac{\gn_2^2\MZ^2}{\gn_\FZ^2\MW^2} = 1+ \frac{1}{2}\frac{\varv^2}{\La^2} \Lcoeff_{\Phi D}.
\eeq

The Yukawa coupling matrices and the fermion mass matrices are
modified by the presence of $\psi^2\Phi^3$ operators. The fermion mass
matrices are obtained as 
\beq
M^f_{ij} =
\frac{1}{\sqrt{2}}\varv\left(G^\Ff_{ij}-\frac{1}{2}\frac{\varv^2}{\La^2}\Lcoeff^{\Ff\Phi}_{ij}\right),
\quad \Ff=\Fu,\Fd,\Fl
\eeq
after EW symmetry breaking. The coupling matrices of the Higgs boson to
the fermions appearing in $-\FH\Ffbar \Gn^\Ff \Ff/\sqrt{2}$, 
on the other hand, read
\beq
\Gn^f_{ij} = 
G^f_{ij}\left(1+c_{\FH,\mathrm{kin}}\right)
-\frac{3}{2}\frac{\varv^2}{\La^2}\Lcoeff^{\Ff\Phi}_{ij}
 = \frac{\sqrt{2}}{\varv}
M^f_{ij}\left(1+c_{\FH,\mathrm{kin}}\right)
-\frac{\varv^2}{\La^2}\Lcoeff^{\Ff\Phi}_{ij}.
\eeq
The Higgs--fermion couplings are thus not necessarily proportional to
the fermion mass matrices as in the SM and will in general not be
simultaneously diagonalizable. Consequently, the Yukawa couplings do
not necessarily have to be suppressed by the corresponding fermion masses.

Expressing the Lagrangian in terms of the redefined physical basis of 
fields and parameters, the quadratic part of the Lagrangian looks
exactly as in the SM. After adding the usual 't~Hooft--Feynman
gauge-fixing term (note that $\FA$ stands for the photon field and $D$
for the colour index of the gluon field)
\beq
\cL_{\mathrm{fix}}=-C^+C^- -\frac{1}{2}(C^{\FZ})^2-\frac{1}{2}(C^{\FA})^2
-\frac{1}{2}C^{\FG,D}C^{\FG,D}
\eeq
with
\beq
C^{\FG,D}=\partial_\mu\FGn^{D\mu},\qquad
C^{\FA}=\partial_\mu\FAn^\mu,\qquad
C^{\FZ}=\partial_\mu\FZn^\mu - \MZ\chi,\qquad
C^\pm=\partial_\mu\FWn^{\pm\mu} \mp \ri\MW\phi^{\pm}
\eeq
in terms of the fields and parameters of the new physical basis, we
obtain the quadratic part of the Lagrangian
\begin{align}
\label{eq:Lquad}
\cL^{(4)}_{\mathrm{SM},0} ={}& 
-\frac{1}{2}(\partial_\mu\FGn^D_{\nu})
(\partial^\mu\FGn^{D\nu})
-(\partial_\mu\FWn^+_{\nu})
(\partial^\mu\FWn^{-\nu})
-\frac{1}{2}(\partial_\mu\FZn_{\nu})
(\partial^\mu\FZn^{\nu})
-\frac{1}{2}(\partial_\mu\FAn_{\nu})
(\partial^\mu\FAn^{\nu})
\notag\\&{}
+\MW^2\FWn^+_\mu \FWn^{-\mu} +\frac{1}{2}\MZ^2\FZn_\mu \FZn^{\mu}
+\frac{1}{2}(\partial_\mu \FHn) (\partial^\mu\FHn)
-\frac{1}{2}\MH^2 \FHn^2
\notag\\&{}
+(\partial_\mu \phicn^+) (\partial^\mu\phicn^-)
-\MW^2 \phicn^+\phicn^-
+\frac{1}{2}(\partial_\mu \phinn) (\partial^\mu\phinn)
-\frac{1}{2}\MZ^2 \phinn^2
\notag\\&{}
+ \sum_{\Ff=\Fl,\Fu,\Fd}\sum_i \bar\Ff_i(\ri\slashed{\partial}-\Mf{}_{,i})\Ff_i
+\sum_i \bar\nu_i\ri\slashed{\partial}\nu_i.
\end{align}

The value of the vev can be obtained in the SM from the measurement of
the Fermi coupling in $\mu$ decay. In SMEFT (neglecting lepton masses)
this relation is modified to
\beq 
\frac{4\GF}{\sqrt{2}} = \frac{2}{\varv^2} - \frac{1}{\La^2}(
\Lcoeff_{\FL\FL,\mu \Fe\Fe\mu} + \Lcoeff_{\FL\FL,\Fe\mu\mu \Fe} 
-2  \Lcoeff^{(3)}_{\Phi\FL,\Fe\Fe} -2 \Lcoeff^{(3)}_{\Phi\FL,\mu\mu}).
\eeq
Similarly the determination of the modified gauge coupling $\gn_{1,2}$
from $\MZ$ and $\MW$ or from $\MZ$ and $\alpha(0)$ is affected by
contributions of dimension-6 operators.

After applying the field transformations to canonically normalized
fields, defined in \refeqs{eq:phidec_smeft} and
\refeqf{eq:vector_fields_smeft}, to the full Lagrangian, the Feynman
rules can be derived as usual, and observables can be calculated.
At this point, we emphasize that SMEFT predictions for observables
should be parametrized by a minimal set of input parameters
like, e.g., $\alpha=\bar e^2/(4\pi)$,
$\alphas=\bar\gs^2/(4\pi)$, $\MW$, $\MZ$, $\MH$, $m_{f,i}$,
$V_{ij}$, and all SMEFT Wilson coefficients $C_{\dots}/\Lambda^n$
and consistently linearized in the latter.
Otherwise the appearance of quadratic or higher terms in $C_{\dots}/\Lambda^n$
could spoil the consistency of the predictions.

As an example for couplings modified by dimension-6 operators, we consider
the CP-even couplings of the Higgs boson to the weak gauge
bosons, which receive contributions from the Lagrangian terms
\begin{equation}
(D_{\mu}\Phi)^{\dag}(D^{\mu}\Phi) 
+\frac{\Lcoeff_{\Phi\FW}}{\Lambda^2}\Lop_{\Phi\FW}
+\frac{\Lcoeff_{\Phi\FB}}{\Lambda^2}\Lop_{\Phi\FB}
+\frac{\Lcoeff_{\Phi\FW\FB}}{\Lambda^2}\Lop_{\Phi\FW\FB}
+\frac{\Lcoeff_{\Phi D}}{\Lambda^2}\Lop_{\Phi D}.
\end{equation}
After translation to the physical basis these lead to the interactions
\begin{align}
\cL_{HVV,\mathrm{CP}} ={}&
\frac{1}{2}\gn_2^2\varv\FH\FWn^+_\mu\FWn^{-\mu}(1+c_{\FH,\mathrm{kin}})
+ \frac{1}{4}(\gn_2^2+\gn_1^2){\varv} \left(
1+c_{\FH,\mathrm{kin}}+\frac{\varv^2}{\Lambda^2}\Lcoeff_{\Phi D}
+\frac{2\gn_1\gn_2}{\gn_2^2+\gn_1^2} \frac{\varv^2}{\Lambda^2} 
\Lcoeff_{\Phi\FW\FB}
\right) \FH\FZn^\mu\FZn_\mu
\nl
& {}
+2\frac{\Lcoeff_{\Phi\FW}}{\Lambda^2}\varv\FH\FWn^+_{\mu\nu}\FWn^{-\mu\nu}
+ \frac{\alpha_{ZZ}}{\Lambda^2} \varv\FH\FZn^{\mu\nu}\FZn_{\mu\nu}
- \frac{\alpha_{AZ}}{\Lambda^2} \varv\FH\FAn^{\mu\nu}\FZn_{\mu\nu}
+ \frac{\alpha_{AA}}{\Lambda^2} \varv\FH\FAn^{\mu\nu}\FAn_{\mu\nu},
\end{align}
with the coefficients
\begin{align}
\alpha_{ZZ} &{}= \cwn^2 \Lcoeff_{\Phi\FW} + \swn^2 \Lcoeff_{\Phi\FB}
                 +\cwn\swn \Lcoeff_{\Phi\FW\FB},
\nl
\alpha_{AZ} &{}= 2\cwn\swn(\Lcoeff_{\Phi\FW}-\Lcoeff_{\Phi\FB})
                 -(\cwn^2-\swn^2)\Lcoeff_{\Phi\FW\FB},
\nl
\alpha_{AA} &{}= \swn^2 \Lcoeff_{\Phi\FW} + \cwn^2 \Lcoeff_{\Phi\FB}
                 -\cwn\swn \Lcoeff_{\Phi\FW\FB}.
\end{align}
While the operators $\Lop_{\Phi\Box}$ and
$\Lop_{\Phi D}$ change only the normalization of the couplings, the
operators $\Lop_{\Phi\FW}$, $\Lop_{\Phi \FB}$, and $\Lop_{\Phi\FW\FB}$
yield extra momentum-dependent contributions.  The former drop out in
the ratio of the Higgs-boson couplings to W and Z~bosons, but the
latter do not.

The dimension-6 operators contribute to practically all interaction
terms. Their contributions to anomalous triple gauge couplings are,
for instance,  given in \citere{Passarino:2016pzb}.

\subsubsection{Applications}

The theoretical preparation of phenomenological predictions towards a
full fit of all SMEFT Wilson coefficients to experimental data from
the LHC and previous colliders is a very active line of research.  In
this context, the evaluation of SMEFT predictions with NLO QCD and EW
corrections is quite a challenge.

The renormalization-group evolution of the complete set of SM
dimension-6 operators was presented in
\citeres{Jenkins:2013zja,Jenkins:2013wua,Alonso:2013hga}.
Results for operator renormalization in the Warsaw basis were
elaborated in \citeres{Passarino:2012cb,Ghezzi:2015vva}.

The calculation of NLO corrections in SMEFT was discussed in
\citeres{deFlorian:2016spz,Passarino:2016pzb}.
The first pioneering calculations in this framework were for the
processes $\mu\to\Pe\ga$ \cite{Pruna:2014asa}, $\PH\to\ga\ga$
\cite{Ghezzi:2015vva,Hartmann:2015aia,Hartmann:2015oia},
and $\PH\to\ga\PZ,\PZ\PZ,\PW\PW$ \cite{Ghezzi:2015vva}. More
recently, NLO QCD corrections were calculated in SMEFT for
instance for W-pair production at the LHC \cite{Baglio:2018bkm} (see
also references therein) and the Higgs decay in bottom quarks
\cite{Cullen:2019nnr}.  NLO EW corrections to Z-boson decays were
investigated in \citeres{Hartmann:2016pil,Dawson:2018jlg}, while those
to Higgs-boson decays were calculated in \citere{Dawson:2018liq} and
references therein. EW and QCD corrections to $\PZ$ and $\PW$ pole
observables within SMEFT were computed in \citere{Dawson:2019clf}.
NLO EW corrections to  $t\bar{t}$ Production at the LHC were recently
published \cite{Martini:2019lsi}.
An overview of SMEFT calculations and tools can be found in 
\citere{Brivio:2019irc}.

\section{Electroweak radiative \texorpdfstring{corrections---virtual}
{corrections - virtual} effects}
\label{se:virt}

\subsection{Renormalization of the Electroweak Standard Model in the on-shell scheme}
\label{se:ren_sm}

\subsubsection{Historical development and variants} 
\label{se:ren_conv}

As a non-abelian gauge theory with spontaneous symmetry breaking the
SM of particle physics including the EWSM and QCD is
renormalizable
\cite{tHooft:1971akt,tHooft:1971qjg,tHooft:1972tcz,tHooft:1972qbu,%
Lee:1972fj,Lee:1973fn,Lee:1973rb,Lee:1974zg,Becchi:1974md}.
All ultraviolet (UV) singularities can be absorbed into renormalization
constants
that are generated by a renormalization of the independent input
parameters and the fields and/or the external wave functions.

In order to define a {\it renormalization scheme} one has to choose a
set of independent parameters. One possibility is to start from the
Lagrangian in its symmetric form \refeqf{eq:LSM_sym}. To absorb all
UV~singularities of the effective action, one needs to introduce
renormalization
constants for all parameters of the theory in the symmetric basis, a
renormalization of the vacuum expectation value (vev) of the Higgs
field, and a rescaling of the field multiplets \cite{Lee:1973rb}.  In
this framework, the calculation of $S$-matrix elements requires the
inclusion of finite wave-function renormalization constants
for the external states.%
\footnote{By wave-function renormalization constants we denote the
  renormalization factors of the external states, which are also known
  as {\em LSZ factors}. The wave-function renormalization has to be applied
  to obtain correctly normalized $S$-matrix elements. It is not needed,
  if the field renormalization is performed appropriately.}


For practical calculations it is more convenient to fix the
renormalization constants by renormalization conditions 
imposed directly on
the parameters in the physical basis.  It has become customary to
require the renormalized masses of the gauge bosons, of the Higgs
boson, and of the fermions to equal the physical masses, ``physical''
in the sense of being derived from the locations of the poles of the
propagators.  Moreover, the renormalized electric charge can be fixed
in such a way that it coincides with the one measured in the {\it
  Thomson limit}, \ie in the limit of low-energy Compton scattering of
on-shell (OS) particles.  This is the so-called {\it on-shell
  renormalization scheme} which was originally proposed by Ross and
Taylor \cite{Ross:1973fp} and later worked out and used by many
authors
\cite{Sirlin:1980nh,Marciano:1980pb,Bardin:1980fe,Fleischer:1980ub,%
Sakakibara:1980hw,Aoki:1980ix,Sirlin:1981yz,Bardin:1981sv,Aoki:1982ed,%
Bohm:1986rj,Hollik:1988ii,Denner:1991kt}.
The OS scheme is distinguished by the fact that all renormalized
parameters have a direct physical meaning and can be measured most directly
in suitable
experiments.  
For most of the particle masses this is evident.%
\footnote{This does not hold for the masses of the light quarks owing
  to the presence of the strong interaction and confinement.  However,
  their contributions can often be absorbed into directly measurable
  and thus more appropriate input quantities.}  It is also true for
the definition of the electric charge, because in the Thomson limit of
Compton scattering all higher-order corrections vanish, and the
lowest-order QED cross section known as {\it Thomson cross section} becomes
exact in all orders of perturbation theory
\cite{Thirring:1950cy,Dittmaier:1997dx}.  The renormalization of the
parameters together with the renormalization of the external wave
functions, which is dictated by the correct normalization of
$S$-matrix elements, is sufficient to obtain UV-finite predictions
without any additional field renormalization.  This fact has been
exploited for instance by Passarino and
Veltman~\cite{Passarino:1978jh} and Sirlin~\cite{Sirlin:1980nh}.

In order to obtain finite Green functions, field renormalization is
required in addition. When renormalizing the Lagrangian in
its symmetric form, it is sufficient to introduce only one field
renormalization constant for each multiplet.  This minimal field
renormalization \cite{Bohm:1986rj} of the EW sector of the SM requires
one renormalization constant for the triplet of $\SU(2)_{\rw}$ gauge
fields, one for the $\U(1)_{Y}$ gauge field, one for the complex
Higgs-doublet field, one for each left-handed fermion doublet, and
one for each right-handed fermion singlet.  In this scheme,
additional UV-finite wave-function renormalizations have to be applied
when calculating $S$-matrix elements.

Another possibility is to renormalize the fields in the physical 
basis~\cite{Aoki:1982ed,Denner:1991kt}.  Since fields that carry identical
conserved quantum numbers in general mix, the field renormalization
constants (and possibly the wave-function renormalization constants)
take the form of matrices. This allows us to fix the field
renormalization in such a way that the renormalized fields do not mix
on shell and that no wave-function renormalization is required.
Alternatively, one can combine non-diagonal mass renormalization
constants and field-renormalization matrices involving fewer non-zero,
non-diagonal entries, as has been done for the renormalization of the
photon--Z-boson system in \citeres{Bardin:1981sv,Jegerlehner:1991dq}.

In this review, we restrict our account of renormalization to the one-loop
level.  Two-loop renormalization of the EWSM is, for instance, discussed in
\citeres{Freitas:2002ja,Awramik:2002vu}, and a complete renormalization
framework for the EWSM at the two-loop level is laid out in
\citeres{Actis:2006ra,Actis:2006rb,Actis:2006rc}.  In this section, we put
our emphasis on OS renormalization in the EW sector, which is the most
frequent choice in the calculation of EW corrections; we briefly sketch EW
\MSbar\ renormalization in \refse{se:input_msbar}.

In this section we sometimes assume CP conservation. This is
correct when neglecting the CP-violating phase of the quark-mixing
matrix, an approximation that is justified for the majority of
processes accessible at high-energy colliders. In this respect, we
note that at least four quark-mixing-matrix elements in an appropriate
combination are required to construct CP-violating observables
\cite{Jarlskog:1985ht,Jarlskog:1985cw}.

\subsubsection{Renormalization transformation in the on-shell scheme}
\label{se:rtsm}
We perform the renormalization directly for the parameters and fields of
the physical basis introduced in \refse{se:physical_lagrangian}.  The {\it
  bare} parameters are split into {\it renormalized parameters} and {\it
  counterterms} as follows (bare quantities are denoted by a subscript~0):
\begin{align} \label{eq:Pren}
\notag \MWb^{2} ={}&  \MW^{2} + \delta \MW^{2} , \qquad
\MZb^{2} =  \MZ^{2} + \delta \MZ^{2}    ,\qquad
\MHb^{2} ={}  \MH^{2} + \delta \MH^{2}     ,\nl
\notag m_{0,f,i}\;  ={}&  m_{f,i}   + \delta m_{f,i}       ,\qquad
\notag V_{0,ij}\;   ={}  
V_{ij} + \delta V_{ij} ,\nl
e_{0} ={}&  Z_{e} e \;=\; (1+\delta Z_{e})e \;=\; e+\delta e.
\end{align}
Here $\MW$, $\MZ$, $\MH$, and $m_{f,i}$ are the renormalized masses of
the corresponding particles. In OS renormalization schemes these
correspond to the physical masses defined by the locations of the
poles of the propagators ({\em OS} or {\it pole masses}).%
\footnote{The subtle difference between OS and pole masses will be discussed
in the context of unstable-particle effects in \refse{se:mass_width_unstableparticles}.}
The renormalization of tadpoles is discussed in \refse{se:tadpoles}.

We introduce field renormalization for fields with equal quantum numbers
via renormalization matrices.  We assume that the physical Higgs field does
not mix with other fields, which is valid in the one-loop approximation.
More generally, this holds if CP conservation is required, since then the
physical Higgs field has CP parity~$+$ while the other neutral bosons have
CP parity $-$.  The bare fields in the physical basis are split according
to
\begin{align} \label{eq:field_ren}
W_{0}^{\pm}  ={}& Z_{W}^{1/2} W^{\pm}=
\left(1+\frac{1}{2}\delta Z_{W}\right) W^{\pm}  ,\nl
\left(\barr{l} Z_{0} \\ A_{0} \earr \right)  ={}&
\left(\barr{ll} Z_{ZZ}^{1/2} & Z_{ZA}^{1/2}  \nl[1ex]
                Z_{AZ}^{1/2} & Z_{AA}^{1/2}  \earr \right)
\left(\barr{l} Z \\ A \earr \right)   =
\left(\barr{cc} 1 + \frac{1}{2}\delta Z_{ZZ} & \frac{1}{2}
\delta Z_{ZA}
\nl
\frac{1}{2}\delta Z_{AZ}  & 1 + \frac{1}{2}\delta Z_{AA} \earr \right)
\left(\barr{l} Z \\[1ex] A \earr \right)   ,\nl
\FH_{0} ={}& Z_{\FH}^{1/2} \FH = 
\left(1+\frac{1}{2}\delta Z_{\FH}\right) \FH ,\nl
f_{0,i}^{\rL} ={}& \sum_j \left(Z^{f,\rL}_{ij}\right)^{1/2} \! f_{j}^{\rL} 
=\sum_j\left(\delta _{ij}+\frac{1}{2}\delta Z_{ij}^{f,\rL}\right) f_{j}^{\rL} ,\qquad
f_{0,i}^{\rR} ={} \sum_j\left(Z^{f,\rR}_{ij}\right)^{1/2} \!f_{j}^{\rR} 
=\sum_j\left(\delta _{ij}+\frac{1}{2}\delta Z_{ij}^{f,\rR}\right) f_{j}^{\rR},
\end{align}
where the last expressions in each line are valid in first-order
approximation.  This general transformation allows us to define the fields
in such a way that their quanta are mass eigenstates also in the presence
of higher-order corrections.

The renormalization constants introduced so far are sufficient to render
$S$-matrix elements and Green functions involving only physical external
states UV~finite.  A complete renormalization of the EWSM in addition
requires the renormalization of the unphysical sector. This amounts to the
field renormalization of the would-be Goldstone fields,
\begin{align}
\chi_{0} ={}& Z_{\chi}^{1/2} \chi = 
\left(1+\frac{1}{2}\delta Z_{\chi}\right) \chi ,\qquad
\phi^\pm_{0} ={} Z_{\phi}^{1/2} \phi^\pm = 
\left(1+\frac{1}{2}\delta Z_{\phi}\right) \phi^\pm ,
\end{align}
of the Faddeev--Popov ghost fields,
\begin{align}
u_{0}^{\pm}  ={}& \tilde Z_{\pm} u^{\pm}=
\left(1+\delta \tilde Z_{\pm}\right) u^{\pm},  \nl 
\left(\barr{l} u^Z_{0} \\ u^A_{0} \earr \right)  ={}&
\left(\barr{ll} \tilde Z_{ZZ} & \tilde Z_{ZA}  \\
                \tilde Z_{AZ} & \tilde Z_{AA}  \earr \right)
\left(\barr{l} u^Z \\ u^A \earr \right)   =
\left(\barr{cc} 1 + \delta \tilde Z_{ZZ} & \delta \tilde Z_{ZA} \\
\delta \tilde Z_{AZ}  & 1 + \delta \tilde Z_{AA}
\earr \right)
\left(\barr{l} u^Z \\[1ex] u^A \earr \right),  
\end{align} 
and of the renormalization of the gauge parameters,
\beq\label{eq:rcxi}
\xi_{0,W}^{(\prime)} = Z_{\xi_{W}^{(\prime)}}\xi_{W}^{(\prime)}, \qquad
\xi_{0,Z}^{(\prime)} = Z_{\xi_{Z}^{(\prime)}}\xi_{Z}^{(\prime)}, \qquad
\xi_{0,A}          = Z_{\xi_{A}}         \xi_{A}.
\eeq
We use the convention not to renormalize the Faddeev--Popov antighost
fields $\ubar$. This is possible, since the antighost fields always
appear together with the ghost fields, so that all related
UV~divergences can be absorbed via a renormalization of the
Faddeev--Popov ghost fields $\Fu$.

By choosing $\xi_\FW'=\xi_\FW$ and $\xi_\FZ'=\xi_\FZ$, the would-be
Goldstone bosons decouple from the scalar gauge bosons, and the poles of
their propagators are located at $p^2=\xi_\FW\MW^2$ and $\xi_\FZ\MZ^2$ in
lowest order, respectively.  In order to conserve these features at higher
orders, one has to renormalize $\xi'_a$ and $\xi_a$ independently.  The
renormalization of $\xi'_a$ and $\xi_a$ can be performed in such a way that the
gauge-fixing functionals remain unchanged, so that they do not generate
counterterms at all.

Upon inserting the renormalization transformations into the bare
Lagrangian \refeqf{eq:LGSW} and writing $Z=1+\delta Z$ for the
multiplicative renormalization constants (matrices) we can split the
bare Lagrangian as
\beq 
{\cal L}(\Psi_0,p_0) = {\cL}(\Psi,p) + \cL_{\mathrm{ct}}(\Psi,p,\de Z),  
\eeq 
where $\Psi_0$ and $p_0$ represent the bare fields and parameters and
$\Psi$ and $p$ their renormalized counterparts.  The renormalized
Lagrangian ${\cL}(\Psi,p)$ has the same functional form as ${\cal
  L}(\Psi_0,p_0) $, but with unrenormalized parameters and fields replaced
by renormalized ones.  The {\em counterterm Lagrangian}
${\cL}_{\mathrm{ct}}(\Psi,p,\de Z)$ gives rise to the counterterms
containing the renormalization constants $\de Z$.
If not stated otherwise, we restrict ourselves to one-loop corrections and
consistently neglect terms of order $(\delta Z)^{2}$ in the following.

\subsubsection {Renormalization conditions}
\label{se:rcsm}

The renormalization constants are fixed by imposing {\it renormalization
conditions}. These consist of three sets: the conditions that
define the renormalized physical parameters, those that define the
renormalized fields corresponding to physical particle states, 
and those that fix the renormalization in
the unphysical sector.  While only the first set is relevant for
the calculation of observables ($S$-matrix elements), a clever choice of the
second set allows us not only to eliminate the explicit wave-function
renormalization of the external particles, but also to simplify the
explicit form of the renormalization conditions for the physical
parameters. The choice of the third set determines the
form of the renormalized Slavnov--Taylor identities.

The renormalized mass parameters of the particles are fixed by the
requirement that they are equal to the physical masses, which are
defined via the locations of the poles of the corresponding
propagators. These poles are equivalent to the zeros of the inverse
connected 2-point functions projected onto physical states, \ie on
polarization vectors $\veps^\mu(k)$ for gauge bosons and on spinors
$u(p)$ and $\varv(p)$ for fermions.%
\footnote{For issues concerning the renormalization in the presence of
  unstable particles we refer to \refse{se:unstable}.}  In case of
mass matrices these conditions have to be fulfilled by the
corresponding eigenvalues, in general resulting in complicated
expressions. These can be considerably simplified by requiring the
OS conditions for the field renormalization matrices in addition
\cite{Aoki:1982ed,Denner:1991kt}.  The latter conditions require that
close to its pole each propagator is given by its lowest-order
expression with the bare mass replaced by the renormalized mass.  As a
consequence, an OS particle does not mix with others, and its
propagator has residue one, \ie the quanta of the renormalized fields
are canonically normalized mass eigenstates.  Thus, we arrive at the
following renormalization conditions for the renormalized 2-point
functions $\Gamma_{\ren}$ for OS 
fields for physical external states:
\begin{align}\label{eq:RCM}
\left. \left[\rRe\Gamma^{V^{\prime\dagger}V}_{\ren,\mu\nu} (-k,k)\right]
\veps ^{\nu }(k)
\right\vert_{k^{2}= M_{V}^{2}} ={}& 0, 
\quad V,V' = W,Z,A
,\nl
\lim_{k^{2}\to M_{V}^{2}} \frac{1}{k^{2}-M_{V}^{2}}
\left[\rRe\Gamma^{V^{\dagger}V}_{\ren,\mu\nu}(-k,k)\right] \veps^{\nu}(k)
={}& -\veps_{\mu}(k)
,\quad V = W,Z,A
  ,\nl
\left. \rRe\Gamma_{\ren}^{\FH\FH } (-k,k)
\right\vert_{k^{2}= \MH^{2}} ={}& 0  ,\nl
\lim_{k^{2}\to \MH^{2}} \frac{1}{k^{2}-\MH^{2}}\,
\rRe \Gamma_{\ren}^{\FH\FH} (-k,k)    ={}& 1,\nl
\left. \left[\rRe\Gamma_{\ren,ij}^{\Ffbar\Ff} (-p,p)\right] u_{j}(p)
\right\vert_{p^{2}= m_{f,j}^{2}} ={}& 0 
,\nl
\lim_{p^{2} \to m_{f,i}^{2}} \frac{\slashed{p}+m_{f,i}}{p^{2}-m_{f,i}^{2}}
\left[\rRe\Gamma^{\Ffbar\Ff}_{\ren,ii} (-p,p)\right] u_{i}(p) ={}& u_i(p) .
\end{align}
The polarization vectors and spinors of the external fields are
denoted by $\veps^{\mu}(k)$ and $u_i(p)$, respectively.  The definition of
$\rRe$ depends on the scheme. In the traditional OS scheme, it takes the
real part of the vertex functions.  In the presence of genuinely complex
couplings, \eg resulting from a phase in the quark-mixing matrix, it takes
the real part of the loop integrals, \ie it eliminates the absorptive
parts, but does not affect the complex couplings. Thus in the CP-conserving
SM, $\rRe$ can be replaced by $\Re$ everywhere. Finally, in the
complex-mass scheme discussed in \refse{se:unstable} the real part is
not taken at all.

The renormalization conditions \refeqf{eq:RCM} can be simplified upon
inserting the Lorentz decompositions of the 2-point functions,
\begin{align}\label{eq:Lorentz_SE}
\Gamma^{V^{\prime\dagger}V}_{\mu \nu }(-k,k) ={}&
\left(g_{\mu \nu} -\frac{k_{\mu} k_{\nu }}{k^{2}}\right)
\Gamma^{V^{\prime\dagger}V}_{\rT}(k^{2})
+\frac{k_{\mu} k_{\nu }}{k^{2}} 
\Gamma^{V^{\prime\dagger}V}_{\rL}(k^{2}),  \notag\\
\Gamma_{ij}^{\Ffbar\Ff}(-p,p) ={}& 
    \slashed{p} \frac{1-\ga_5}{2} \Ga_{ij}^{f,\rL}(p^{2})
    +\slashed{p} \frac{1+\ga_5}{2} \Ga_{ij}^{f,\rR}(p^{2}) 
    + \frac{1-\ga_5}{2} \Ga_{ij}^{f,\rl}(p^{2}) 
    + \frac{1+\ga_5}{2} \Ga_{ij}^{f,\rr}(p^{2}),
\end{align}
which hold for renormalized vertex functions $\Gamma_\rR$ analogously.  If
we neglect absorptive parts, which are irrelevant for the (one-loop)
renormalization, the hermiticity of the Lagrangian implies the
hermiticity of the effective action. Therefore, the fermionic 2-point
functions must have the symmetry
\beq
 \rRe\Gamma_{ij}^{\Ffbar\Ff}(-p,p) = \rRe \ga^0  \Bigl(\Gamma_{ji}^{\Ffbar\Ff}(-p,p)\Bigr)^\dagger
 \ga^0,
\eeq
which implies 
\begin{align}\label{eq:Gaherm}
  \rRe\Ga_{ij}^{f,\rL}(p^{2}) ={}& \rRe \left(\Ga_{ji}^{f,\rL}(p^{2})\right)^*, \qquad
  \rRe\Ga_{ij}^{f,\rR}(p^{2}) =  \rRe\Bigl(\Ga_{ji}^{f,\rR}(p^{2})\Bigr)^*,\qquad
  \rRe\Ga_{ij}^{f,\rl}(p^{2}) =  \rRe\Bigr(\Ga_{ji}^{f,\rr}(p^{2})\Bigl)^*,
\end{align}
where $\rRe$ eliminates the absorptive parts
in the loop integrals, but has no effect on complex couplings and
Dirac matrices.
CP symmetry,
which holds in the SM for the one-loop 2-point functions or for real
$V_{ij}$ in general, implies
\begin{align}\label{eq:GaCP}
  \Ga_{ij}^{f,\rL}(p^{2}) ={}& \Ga_{ji}^{f,\rL}(p^{2}), \qquad
  \Ga_{ij}^{f,\rR}(p^{2}) =  \Ga_{ji}^{f,\rR}(p^{2}),\qquad
  \Ga_{ij}^{f,\rl}(p^{2}) =  \Ga_{ji}^{f,\rr}(p^{2}).
\end{align}
Moreover, at the one-loop level in the SM the scalar coefficients of
the fermion self-energy take the form
\beq\label{eq:GaSrel}
\Ga_{ij}^{f,\rl}(p^{2}) = m_{f,i} \Ga_{ij}^{f,\rS}(p^{2}), \qquad
\Ga_{ij}^{f,\rr}(p^{2}) = m_{f,j} \Ga_{ij}^{f,\rS}(p^{2}), 
\eeq
\ie $\Ga_{ij}^{f,\rl}$ and $\Ga_{ij}^{f,\rr}$ are proportional to the
single function $\Ga_{ij}^{f,\rS}$.  This will be used in later sections.
The relations \refeqs{eq:Gaherm}--\refeqf{eq:GaSrel} can be maintained
for the renormalized self-energies as well.

Inserting the Lorentz decompositions \refeqf{eq:Lorentz_SE}, we obtain%
\footnote{The subscript $\ren$ for renormalization should be be
  confused with the superscript $\rR$ for right handed fermions.}
from \refeq{eq:RCM}
\begin{align} \label{eq:RCG}
&\rRe\Ga^{V^{\prime\dagger}V}_{\ren,\rT}(M_{V}^{2}) = 0 ,\qquad
\left. \rRe
\frac{\partial \Ga^{V^{\dagger}V}_{\ren,\rT}(k^{2})}{\partial k^{2}}
\right\vert_{k^{2}=M_{V}^{2}}  ={} -1 ,\\
&\rRe\Ga^{\FH\FH}_{\ren} (\MH^{2}) = 0, \qquad 
\left. \rRe \frac{\partial \Ga^{\FH\FH }_{\ren}(k^{2})}{\partial k^{2}}
\right\vert_{k^{2} = \MH^{2}}  ={}1 ,\\
& m_{f,j}\rRe\Ga_{\ren,ij}^{\Ff,\rL}(m_{f,j}^{2}) +
\rRe\Ga_{\ren,ij}^{f,\rr}(m_{f,j}^{2}) ={}0 ,\qquad
 m_{f,j}\rRe\Ga_{\ren,ij}^{f,\rR}(m_{f,j}^{2})
+\rRe\Ga_{\ren,ij}^{f,\rl}(m_{f,j}^{2}) ={}0 ,
\label{eq:fmassrencond} 
\\
&\rRe\biggl\{\Ga_{\ren,ii}^{\Ff,\rR}(m_{f,i}^{2})
+ \Ga_{\ren,ii}^{f,\rL}(m_{f,i}^{2}) 
{}+2\frac{\partial}{\partial p^{2}}\left[
m_{f,i}^{2}\left(\Ga_{\ren,ii}^{\Ff,\rR}(p^{2})
+ \Ga_{\ren,ii}^{\Ff,\rL}(p^{2})\right)
\left. 
+ m_{f,i}\left(\Ga_{\ren,ii}^{\Ff,\rr}(p^{2})
+ \Ga_{\ren,ii}^{f,\rl}(p^{2})\right)
\right]\right\vert_{p^{2}=m_{f,i}^{2}} \biggr\}={} 2.
\end{align}
We mention that \refeq{eq:fmassrencond} together with the relations
\refeqf{eq:Gaherm} or alternatively \refeqf{eq:GaCP} for the
renormalized self-energies implies
\begin{align}
m_{f,i}\rRe\Ga_{\ren,ii}^{f,\rR}(m_{f,i}^{2})
=m_{f,i}\rRe\Ga_{\ren,ii}^{f,\rL}(m_{f,i}^{2}) 
=-\rRe\Ga_{\ren,ii}^{f,\rr}(m_{f,i}^{2})
=-\rRe\Ga_{\ren,ii}^{f,\rl}(m_{f,i}^{2}).
\end{align}
Finally, note that the condition $\Ga^{AA}_{\ren,\rT}(0) = 0$ is automatically
fulfilled as a consequence
of the Slavnov--Taylor identities and the analyticity properties%
\footnote{Since $\Gamma^{AA}_{\mu \nu }(k,-k)$ cannot develop a pole for
  $k^2\to0$, we have $\Ga^{AA}_{\ren,\rT}(0)=\Ga^{AA}_{\ren,\rL}(0)$.  On
  the other hand, the last identity of \refeq{eq:WIVV} implies
  $\Ga^{AA}_{\ren,\rL}(k^2)\equiv0$ and thus $\Ga^{AA}_{\ren,\rT}(0) = 0$.}
of the 2-point functions and does not imply a constraint on the
counterterms.

\subsubsection{Charge renormalization}
\label{se:charge_ren}

The electric charge is defined as the full $\Pe\Pe\gamma$ coupling for
OS electrons in the {\it Thomson limit}, \ie for
vanishing photon momentum. The full $\Pe\Pe\gamma$ coupling consists of
the corresponding vertex function including possible wave-function
renormalization constants. Owing to the photon--\PZ-boson mixing, also
the $\Pe\Pe\PZ$ coupling enters in general. In the complete OS
renormalization scheme the field renormalization is chosen in such a
way that the OS photon does not mix with the \PZ~boson and that
no wave-function renormalization is needed. Then, the
{\it charge renormalization condition} takes the simple form
\beq
\left.\bar{u}(p)\Gamma^{\FA\Febar\Fe}_{\ren,\mu }(0,-p,p)u(p)\right\vert_{
p^{2}=\Me^{2}}
=e \bar{u}(p)\gamma_{\mu }u(p)
\label{eq:RCE}
\eeq
for the (truncated) vertex function
$\Gamma^{\FA\Febar\Fe}_{\ren,\mu}(k,\bar{p},p)$.  
Owing to {\it charge universality} we could impose the above
renormalization condition on any charged particle to obtain the same
renormalized charge~$e$.

\subsubsection{Renormalization of the quark-mixing matrix}
\label{se:renCKM}

For a non-trivial {\it quark-mixing matrix} $V$, also the
corresponding parameters need to be renormalized. There is a vast
literature on the renormalization of $V_{ij}$, and various
prescriptions for its renormalization have been advocated
\cite{Denner:1990yz,Gambino:1998ec,Balzereit:1998id,Diener:2001qt,Yamada:2001px,Pilaftsis:2002nc,Denner:2004bm,Kniehl:2006rc,Kniehl:2009kk}.
For scattering processes at high-energy colliders, the renormalization
of the quark-mixing matrix in the SM is practically irrelevant, since
in higher-order corrections neglecting the quark mixing is a good
approximation.

Therefore, we here sketch only the original, simple renormalization
prescription~\cite{Denner:1990yz} for $V_{ij}$. This prescription can be
motivated as follows. The bare quark-mixing matrix $V_{0}$ is given by
\refeq{42VKMdef},
\beq
V_{0,ij}= \sum_k U^{u,\rL}_{0,ik} U^{d,\rL\dagger}_{0,kj},
\eeq
where the matrices $U^{f,\rL}_0$ transform the bare fields $\Ff'_{0}$
corresponding to weak-interaction eigenstates to the fields $\Ff_{0}$
of the bare mass eigenstates,
\beq
\sum_j U^{\Ff,\rL\dagger}_{0,ij}\Ff^{\rL}_{0,j}= \Ff^{\prime\rL}_{0,i}.
\eeq
In the OS renormalization scheme, the fields of the fermion mass
eigenstates are related in higher orders to their bare counterparts
through the field renormalization constants of the fermions,
\beq
f^{\rL}_{0,i}=\sum_j Z^{1/2,f,\rL}_{ij}f^{\rL}_{j} .
\eeq
The renormalized quark-mixing matrix is defined in analogy to the
unrenormalized one through the rotation from the weak-interaction
basis to the renormalized mass basis.
Since both the bare and renormalized quark-mixing matrices have to be
unitary, only the unitary part of the field renormalization matrices
can enter.

\newcommand{\AH}{\mathrm{AH}}
In one-loop approximation the rotation contained in the fermion
wave-function renormalization $\mathbb{1}+\frac{1}{2}\delta Z^{\rL}$
is simply given by its antihermitean part
\beq
\delta Z^{f,\rL,\AH}_{ij}=\frac{1}{2} \left(\delta Z^{f,\rL}_{ij}-\delta
Z^{f,\rL\dagger}_{ij}\right) .
\eeq
This leads us to define the renormalized quark-mixing matrix as
\begin{align}\label{eq:RCV}
\notag V_{ij} ={}& 
\sum_{k,n} \left(\delta_{ik} + {\textstyle\frac{1}{2}}\delta Z^{u,\rL,\AH\dagger}_{ik}\right)
V_{0,kn} \left(\delta_{nj} +  {\textstyle\frac{1}{2}}\delta Z^{d,\rL,\AH}_{nj}\right) \nl
={}& V_{0,ij} +  {\textstyle\frac{1}{2}}\sum_k \left(\delta
  Z^{u,\rL,\AH\dagger}_{ik}V_{0,kj}
+V_{0,ik}Z^{d,\rL,\AH}_{kj}\right)
= V_{0,ij} - \de  V_{ij} .
\end{align}
This condition absorbs all one-loop UV~divergences, yields $V_{ij} =
V_{0,ij}$ in the limit of degenerate up- or down-type quark masses
\cite{Denner:1990yz}, and treats all quarks on the same footing.

It has been criticised that the resulting renormalization constant depends
on the gauge, if the fermion wave-function renormalization constants fixed
by the OS conditions \refeqf{eq:RCG} are used to fix the renormalization
constant of the quark-mixing matrix \cite{Gambino:1998ec}. However, $\de
V_{ij}$ can be fixed by definition in a specific gauge
\cite{Yamada:2001px,Pilaftsis:2002nc}. If this value of $\de V_{ij}$ is
kept in other gauges, the $S$-matrix elements in fact depend on
renormalized input parameters in a gauge-independent way.  In the light of
this fact, the original renormalization condition for the
quark-mixing matrix \cite{Denner:1990yz} emerges as a consistent and simple
recipe \cite{Denner:2018opp}.

\subsubsection{Tadpole renormalization}
\label{se:tadpoles}

When calculating higher-order corrections in spontaneously broken
gauge theories like the SM, so-called tadpole diagrams
arise, \ie Feynman diagrams containing one or more subdiagrams of the
form
\beq
T^\FH \,=\,
\vcenter{\hbox{\includegraphics[page=1,scale=1]{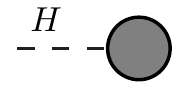}}} .
\label{eq:tadpolediags}
\eeq
The vertex functions, defined via a Legendre transformation from the
connected Green functions, involve such tadpole contributions if the
splitting $\bar{\varv}+H(x)$ of the physical Higgs field into a
constant shift $\bar{\varv}$ and a field excitation $H(x)$ does not
provide an expansion of the effective Higgs potential about its true
minimum (see for instance App.~C of \citere{Denner:2018opp}).  The
location of the minimum is quantified by the vev $\varv$, which itself
is not a free parameter, but determined by the free parameters of the
theory.  Minimizing the Higgs potential to define $\varv$ in
perturbation theory necessarily leads to tadpole-like contributions
beyond tree level, just by the perturbative ordering and truncation of
contributions.  Technically, it is desirable to organize the
perturbative bookkeeping by appropriate parameter and field definition and
renormalization in such a way that the occurrence of tadpole
contributions is widely suppressed.

At tree level, tadpole contributions can be easily eliminated upon
demanding that the classical ground state $\langle\Phi\rangle$
of the Higgs field minimizes the tree-level
potential of the Lagrangian \refeqf{eq:LSM_sym}. In the SM, this leads to the
condition
\beq\label{eq:bare_vev}
|\langle\Phi\rangle|^{2} = \frac{2\mu_0^{2}}{\lambda_0} =
\frac{\varv_0^{2}}{2}, \qquad
\varv_0= 2\sqrt{\frac{\mu_0^2}{\la_0}}.
\eeq
Keeping the definition of $\bar{\varv}=\varv_0$ at least to leading
order, implies that the tadpole subdiagrams~\refeqf{eq:tadpolediags}
involve at least one loop.  The freedom of defining the shift
$\bar{\varv}$ can be exploited to generate a counterterm contribution
$\delta t\, H$ in the counterterm Lagrangian $\delta {\cal L}$ in
order to compensate all explicit loop subdiagrams of the
form~\refeqf{eq:tadpolediags} which define the unrenormalized tadpole
contribution $T^\FH$ (in momentum space) to the Higgs-field 1-point
vertex function $\Ga^\FH$.  Demanding that $T^\FH$ is fully cancelled
by the tadpole counterterm $\delta t$ can, thus, be expressed by the
following renormalization condition for the renormalized 1-point
function $\Ga_{\ren}^\FH$,
\beq\label{eq:RCT} 
\Ga_{\ren}^\FH = \Ga^\FH_{\mathrm{1PI}} + \delta t = T^\FH + \delta t = 0 
\quad\Rightarrow\quad 
\de t = - T^\FH. 
\eeq
If this condition is fulfilled the corresponding $\bar{\varv}$ is
equal to the true minimum of the effective Higgs potential in the
considered order, \ie $\bar{\varv}=\varv$.  As a consequence of the
condition \refeqf{eq:RCT} no Feynman diagrams involving tadpoles as
subdiagrams need to be considered in actual calculations.  Instead,
$\delta t$ enters several counterterm structures of the theory, i.e.\ 
$\delta t$ enters $\delta {\cal L}$ not just as the single term
$\delta t\, H$. Effectively the introduction of $\delta t$ just
redistributes tadpole contributions. Since the definition of the
higher-order contributions to $\bar{\varv}$ is a matter of convention,
the technical details of how $\bar{\varv}$ is defined and how $\delta
t$ is introduced should not affect any predictions of observables.

In the following we describe several different {\it tadpole schemes} for the
definition of $\bar{\varv}$ and $\delta t$.
All of them make use of \refeq{eq:RCT} to calculate $\delta t$
from the explicit diagrammatic tadpole contribution $T^\FH$, but the
additional tadpole contributions to counterterms are different.
Since $T^\FH$ and, thus, also $\delta t$ are gauge-dependent quantities,
the discussion of gauge dependences in the various tadpole schemes 
is interesting.
\myparagraph{(a) Fleischer--Jegerlehner tadpole scheme (FJTS)}

In the FJTS~\cite{Fleischer:1980ub,Krause:2016oke,Denner:2016etu}, the bare
vev $\varv_0$ from \refeq{eq:bare_vev} is used to define the bare masses in
terms of the original bare parameters of the Lagrangian,
\begin{alignat}{3}
  \MWb &= \frac{1}{2}g_{0,2}\varv_0,\qquad & \MZb & =
  \frac{1}{2}\sqrt{g_{0,1}^{2}+g_{0,2}^{2}} \; \varv_0,
\nl
\MHb^2 & = 2\mu_0^2 = {\frac{\la_0}{2}}\varv_0^2,\qquad &
m_{0,f,i}  &=  \frac{\varv_0}{\sqrt{2}}
\sum_{k,m} U_{0,ik}^{f,\rL} G_{0,km}^{f} U_{0,mi}^{f,\rR\dagger}.\qquad&& 
\label{eq:SMmass_FJ}
\end{alignat}
The tadpole counterterm is introduced by the field transformation
\beq
\FH_0 \,\to\, \FH_0 + \Delta\varv,
\label{eq:Hshift}
\eeq
with a constant $\Delta\varv$. This transformation can be interpreted
as a change of the integration variable $\FH_0$ in the functional integral
(with unit Jacobian), which has no physical effect, so that $\Delta\varv$
can be chosen arbitrarily. 
The field shift~\refeqf{eq:Hshift} introduces $\Delta\varv$ terms 
in the Lagrangian wherever the field $H_0$ occurs. 
At one-loop order, only the terms linear in $\Delta\varv$ are
relevant, so that
\begin{align}
\delta \cL_{\Delta\varv} ={}& -\frac{1}{4}\Delta\varv \la_0 \biggl[
2\FH_0 \varv_0^2
+\varv_0\Bigl(2\phi_0^-\phi_0^+ + 3\FH_0^2+\chi_0^2\Bigr) 
+\FH_0\Bigl(2\phi_0^-\phi_0^+ + \FH_0^2+\chi_0^2\Bigr) 
\biggr]
\nl &
+ \text{terms from kinetic and Yukawa parts of the Higgs Lagrangian}.
\label{eq:LDev}
\end{align}
Diagrammatically the Feynman rules for the $\Delta\varv$ terms result
from replacing external $\FH$~lines by $\Delta\varv$ in the original
Feynman rules, so that in particular quartic couplings with
$\Delta\varv$ do not exist. On the other hand, tadpole contributions
arise in all bosonic vertices with less than four fields and in all
2-point functions.

The term in \refeq{eq:LDev} that is linear in $\FH_0$ can be directly
identified with the desired $\de t\,\FH$ counterterm for the explicit
tadpole diagrams, so that
\beq
\de t = (\varv_0+\Delta\varv)\left(\mu_0^2 
- \frac{1}{4}\la_0 (\varv_0+\Delta \varv)^2\right)
=\Delta \varv\left(\mu_0^2 - \frac{3}{4}\la_0 \varv_0^2\right) 
+ \order{\Delta \varv^2}
=-\frac{1}{2}\Delta \varv\la_0 \varv_0^2 + \order{\Delta \varv^2}.
\label{eq:deltat}
\eeq
For later convenience we did not yet make use of $\varv_0=
2\sqrt{\mu_0^2/\la_0}$ and the linearization in $\Delta\varv$ in the first
equation.  Solving this relation for $\Delta \varv$ in one-loop
approximation and using \refeq{eq:SMmass_FJ}, fixes $\Delta \varv$ to
\beq\label{eq:def_Delta_v}
\Delta \varv = -\frac{\de t}{\MH^2}.
\eeq
Using this and \refeq{eq:SMmass_FJ} to eliminate $\varv_0$ in favour of
$\MWb$ in \refeq{eq:LDev}, the tadpole contributions to the Lagrangian can
be written as
\begin{align}
\delta \cL_{\delta t}^{\mathrm{FJTS}} ={}& 
\de t\,\FH
+ \frac{1}{4}\frac{\de t\, e}{\MW\sw} \Bigl[2\phi^-\phi^+ +
 3\FH^2+\chi^2\Bigr] 
+ \frac{1}{8}\frac{\de t\, e^2}{\MW^2\sw^2} 
 \FH\Bigl[2\phi^-\phi^+ + \FH^2+\chi^2\Bigr]
\nl &
+ \text{terms from kinetic and Yukawa parts of the Higgs Lagrangian}.
\end{align}
Since $\de t$ is already a one-loop correction, all bare quantities in
$\delta \cL_{\delta t}$ can be replaced by their renormalized
counterparts.  In the FJTS, all parameter counterterms are gauge
independent. To this end, it is crucial that we use self-energies that
are defined from complete inverse connected 2-point Green functions
including tadpole contributions
to determine the counterterms.%
\footnote{At variance with
  \citeres{Denner:1991kt,Denner:1994xt,Bohm:2001yx}, where all
  self-energies $\Sigma$ are 1PI, we use self-energies $\Sigma$ based on
  complete inverse connected 2-point functions and denote their 1PI parts
  by $\Sigma_{\mathrm{1PI}}$, see \refeq{eq:definition_se}.}

As an alternative to the field shift~\refeqf{eq:Hshift}, 
the $\Delta\varv$ contributions to $\delta {\cal L}$ can be
obtained via the parameter shift~\cite{Denner:2016etu}
\beq\label{eq:tadpoleshift_FJ}
\varv_0\to\varv_0 +\Delta\varv = \varv_0 - \frac{\de t}{\MH^2}
\eeq
before the vev $\varv_0$ is fixed by minimizing the scalar potential
and thus related to $\la_0$ and $\mu_0^2$ via \refeq{eq:bare_vev}.

Recall that without the field shift~\refeqf{eq:Hshift},
there is no tadpole counterterm and all tadpole diagrams would have to be
included explicitly. Thus, the FJTS
is equivalent to the inclusion of all tadpole diagrams.
Finally, we also mention that the FJTS scheme is equivalent to the
$\beta_t$ scheme of \citere{Actis:2006ra}.

\myparagraph{(b) Parameter-renormalized tadpole scheme (PRTS)}

The tadpole scheme of \citere{Denner:1991kt}
  generates the tadpole counterterm as part of the parameter
  renormalization transformation without touching the fields.
  Following \citere{Denner:2018opp}, we call this scheme {\em
    parameter renormalized tadpole scheme (PRTS)}.  Introducing a
  shifted neutral scalar field $\bar{\varv}+H_0(x)$ into the bare
  Lagrangian~\refeqf{eq:LSM_sym} without any constraint on
  $\bar{\varv}$, leads to a linear term $t_0\FH_0$ in the Lagrangian
  with
\beq
t_0= \bar{\varv}\left(\mu_0^2-\frac{1}{4}\lambda_0 \bar{\varv}^2\right).
\label{eq:t0_PRTS}
\eeq
Instead of setting the bare tadpole parameter $t_0$ to zero, 
as done in the FJTS in \refeq{eq:bare_vev},
this parameter is directly taken as tadpole counterterm, i.e.\ $\de
t=t_0$, leading to the identification $\bar\varv=\varv$.
This is equivalent to the statement of setting the renormalized 
tadpole $t$ to zero after the tadpole renormalization transformation
$t_0=t+\de t$, and
\beq
\delta t= {\varv}\left(\mu_0^2-\frac{1}{4}\lambda_0 {\varv}^2\right).
\label{eq:dt_PRTS}
\eeq
Isolating the terms in the Lagrangian that are quadratic in $\FH_0$,
the bare Higgs-boson mass can be related to the original input parameters,
\beq
\MHb^2 = -\mu_0^2+\frac{3}{4}\la_0\varv^2.
\label{eq:MHb_PRTS}
\eeq
Similarly, the bare masses of the other particles are defined as the
coefficients of the terms in the Lagrangian quadratic in the fields, \ie
\beq
  \MWb = \frac{1}{2}g_{0,2}\varv,\qquad  \MZb  =
  \frac{1}{2}\sqrt{g_{0,1}^{2}+g_{0,2}^{2}} \; \varv, \qquad
m_{0,f,i}  =  \frac{\varv}{\sqrt{2}}
\sum_{k,m} U_{0,ik}^{f,\rL} G_{0,km}^{f} U_{0,mi}^{f,\rR\dagger}. 
\label{eq:SMmass_PRTS}
\eeq
Using the relations \refeqf{eq:dt_PRTS}, \refeqf{eq:MHb_PRTS}, and
\refeqf{eq:SMmass_PRTS} together with \refeqs{eq:charge} and
\refeqf{42VKMdef} for bare parameters, the bare
Lagrangian~\refeqf{eq:LSM_sym} can be expressed in terms of the bare
masses, the bare coupling $e_0$, the quark-mixing matrix $V_0$, and the
tadpole counterterm $\delta t$. Then, renormalized parameters and
fields are introduced using \refeqs{eq:Pren} and \refeqf{eq:field_ren}.

An alternative way to generate the $\delta t$ terms can be found by
inspecting the two relations~\refeqf{eq:dt_PRTS} and
\refeqf{eq:MHb_PRTS}, which are linear in $\mu_0^2$ and $\lambda_0$
and the only source for $\de t$ terms. Thus, the $\de t$ terms of the
PRTS can be generated by making the substitutions
\cite{Denner:2016etu},
\beq
\la_0\to\la_0+2\frac{\de t}{\varv^3}, \qquad
\mu^2_0\to\mu^2_0+\frac{3}{2}\frac{\de t}{\varv}
\eeq
in the bare Lagrangian with all relations between bare and
renormalized quantities in \refse{se:rtsm} without any tadpole
contributions.  This results in
\begin{align}
\delta \cL_{\delta t}^{\mbox{\scriptsize PRTS}} 
={}& 
  \de t\,\FH
+ \frac{1}{4}\frac{\de t\, e}{\MW\sw} \Bigl[2\phi^-\phi^+ + \chi^2\Bigr] 
-\frac{1}{8}\frac{\de t\, e^2}{\MW^2\sw^2} 
 \FH\Bigl[2\phi^-\phi^+ + \FH^2+\chi^2\Bigr]
\nl &
-\frac{1}{64}\frac{\de t\, e^3}{\MW^3\sw^3} 
 \Bigl[4(\phi^-\phi^+)^2  +
 4\phi^-\phi^+\Bigl(\FH^2+\chi^2\Bigr)
 + \Bigl(\FH^2+\chi^2\Bigr)^2
\Bigr].
\end{align}
Note that tadpole contributions appear only in vertex functions with
external scalar fields, and by definition no tadpole contribution
appears in the 2-point function of the physical Higgs field.

Concerning the issue of gauge dependences, the PRTS shows the
unpleasant feature that the mass counterterms of all particles
become gauge dependent. This is due to the fact that 
\refeq{eq:dt_PRTS} relates the parameter $\varv$ not only to the
gauge-independent bare parameters $\mu_0^2$ and $\lambda_0$, but also
to the gauge-dependent term $t_0=\de t$, and this gauge dependence
is transferred to all mass renormalization constants via $\varv$.

\myparagraph{(c) \texorpdfstring{$\beta_h$}{beta_h} tadpole scheme}

This tadpole scheme was introduced in \citere{Actis:2006ra}. The tadpole
counterterm and the masses of the vector bosons and fermions are introduced
as in the PRTS in \refeqs{eq:dt_PRTS} and \refeqf{eq:SMmass_PRTS}, however,
the bare Higgs-boson mass is defined as
\beq
\MHb^2 = \frac{1}{2}\la_0\varv^2.
\label{eq:MHb_betah}
\eeq
The bare Lagrangian~\refeqf{eq:LSM_sym} can be expressed in terms of
the bare masses, the bare coupling $e_0$, the bare quark-mixing matrix $V_0$,
and the tadpole counterterm $\delta t$ in a similar way as for the
PRTS, and renormalized quantities can be introduced thereafter via
\refeqs{eq:Pren} and \refeqf{eq:field_ren}.  Alternatively, the
tadpole contributions are obtained from the bare Lagrangian via the
shift~\cite{Denner:2016etu}
\beq
\mu^2_0\to\mu^2_0+\frac{\de t}{\varv_0}.
\eeq
The resulting tadpole contributions to the counterterm
Lagrangian are given by
\begin{align}
\delta \cL_{\delta t}^{\beta_h} ={}& 
\de t\,\FH
+\frac{1}{4}\frac{\de t\, e}{\MW\sw} \Bigl[2\phi^-\phi^+ +
\chi^2 + \FH^2 \Bigr].
\end{align}
In this scheme, tadpole counterterms appear only in 2-point functions
involving scalar fields.  Concerning gauge dependences, the same
comments as for the PRTS apply, i.e.\ mass renormalization constants
become gauge dependent.

\myparagraph{(d) Tadpole scheme of \citere{Aoki:1980ix}  }

Finally, we mention that \citere{Aoki:1980ix} uses yet another tadpole
scheme.  It is constructed in a similar way as the PRTS and the
$\beta_h$ scheme, but introduces the bare Higgs-boson mass via
\beq
\MHb^2 = 2\mu^2_0.
\label{eq:MHb_Aoki}
\eeq
\vspace{1em}

In the PRTS~\cite{Denner:1991kt} and the $\beta_h$ scheme of
\citere{Actis:2006ra} all bare masses are gauge dependent, since the
bare masses are related to the gauge-independent bare parameters of
the Lagrangian via the gauge-dependent tadpole terms.%
\footnote{This holds also for the tadpole scheme of
  \citere{Aoki:1980ix} with the exception of the bare Higgs-boson mass.}
This leads to a
gauge-dependent parametrization of $S$-matrix elements in terms of
bare input parameters. However, this gauge dependence cancels in
physical quantities if all parameters of the theory are defined by
OS renormalization conditions as is
the case in the renormalization of the EWSM described above.%
\footnote{The renormalization conditions for the quark-mixing matrix
do not involve tadpole contributions, so that the argument is not 
spoiled if the CKM matrix is not renormalized by OS conditions.}  
The gauge dependence of the
counterterms results from tadpole contributions that are momentum
independent.  Thus, these contributions cancel in OS or
momentum-subtraction schemes, where the corresponding quantity is
subtracted at some point in momentum space. On the other hand, when
some parameters are renormalized in the $\MSbar$ scheme, the extra
tadpole contributions do not cancel and lead to a possible gauge
dependence in the $S$-matrix. This issue becomes relevant in
extensions of the SM like the Two-Higgs-Doublet Model or the
Higgs-Singlet Extension of the SM, where usually some parameters are
renormalized within the $\MSbar$ scheme, as discussed in
\citere{Denner:2018opp}
(see also references therein).

\subsubsection{Explicit form of the 
renormalization constants for parameters and fields 
in the physical basis}
\label{se453eforc}

The renormalized vertex functions consist of unrenormalized loop
contributions and counterterms.  The renormalization conditions allow
us to express the counterterms in terms of unrenormalized vertex
functions at specific external momenta. While the charge
renormalization constant is fixed from a condition on the
photon--fermion--fermion vertex function, a Ward identity allows us to
express it in terms of self-energies as well.  Following the
convention of \citere{Denner:2018opp}, we define self-energies
$\Sigma(k^2)$ to comprise 1PI contributions
$\Sigma_{\mathrm{1PI}}(k^2)$, 2-point tadpole counterterms
$\Sigma_{\de t,2}$, (reducible) tadpole loop contributions
$\Sigma_{\mathrm{tad}}$, and 1-point tadpole counterterms $\Sigma_{\de
  t,1}$.  As indicated, only the 1PI part depends on the virtuality
$k^2$ of the transferred momentum~$k$ of the 2-point function.  At the
one-loop level, the various contributions are illustrated as follows,
\begin{alignat}{8}\label{eq:definition_se}
\Sigma(k^2) &\;{}=\; 
\Sigma_{\mathrm{1PI}}(k^2) &&\;{}+\;
\;\;\Sigma_{\de t,2} &&\;{}+\;
\;\;\Sigma_{\mathrm{tad}} &&\;{}+\;
\;\;\Sigma_{\de t,1}\;,
\\*[.2em]
\raisebox{-.35em}{\includegraphics[scale=.8]{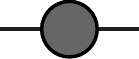}}  &\;{}=\;
\raisebox{-.35em}{\includegraphics[scale=.8]{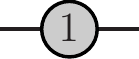}} &&\;{}+\;
\raisebox{-.35em}{\includegraphics[scale=.8]{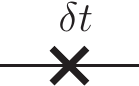}}   &&\;{}+\;
\raisebox{-.35em}{\includegraphics[scale=.8]{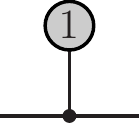}}  &&\;{}+\;
\raisebox{-.35em}{\includegraphics[scale=.8]{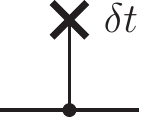}}
\nn
\end{alignat}
where the ``1'' in the blobs stands for one loop.
In the FJTS, the tadpole contributions $\Sigma_{\de t,1}$ and
$\Sigma_{\de t,2}$ cancel exactly so that \refeq{eq:definition_se} is
the usual definition of the self-energy. The introduction of these
terms allows us to eliminate the explicit tadpole contributions
$\Sigma_{\mathrm{tad}}$ in favour of the implicit tadpoles
$\Sigma_{\de t,2}$. In the PRTS, the terms $\Sigma_{\de t,2}$ are
absent by definition, with the exception of the self-energies of the
would-be Goldstone bosons, while $\Sigma_{\de t,1}$ cancels
$\Sigma_{\mathrm{tad}}$. As a consequence, all but the self-energies 
of the would-be Goldstone-boson fields only
consist of 1PI parts.


The renormalized 2-point functions entering the renormalization
conditions for the parameters and fields in the physical basis
are decomposed into lowest-order
contributions, unrenormalized self-energies, and counterterms
(in addition to the implicitly contained $\de t$ terms)
according to
\begin{align}\label{eq:defsigma}
\Gamma^{\FWp\FWm}_{\ren,\rT}(k^2) ={}& 
-(k^{2}-\MW^{2}) -\Sigma^{\FW}_{\rT}(k^{2}) 
-(k^{2}-\MW^{2}) \delta Z_{\FW} +\delta
   \MW^{2},\nl
\Gamma^{VV^{\prime}}_{\ren,\rT}(k^2) ={}& 
-\delta_{VV'} (k^{2}-M^{2}_{V}) -\Sigma^{VV^{\prime}}_{\rT}(k^{2}) 
-\left[ \frac{1}{2} (k^{2}-M_{V}^{2}) \delta Z_{VV'}
+ \frac{1}{2}(k^{2}-M_{V'}^{2}) \delta Z_{V'V} -\delta _{VV'} \delta
   M_{V}^{2}\right],\quad V,V' = \FA,\FZ, \nl
\Gamma^{\FH\FH}_{\ren}(k^2) 
={}&k^{2}-\MH^{2} + \Sigma^\FH(k^{2})
+(k^{2}-\MH^{2}) \delta Z_{\FH} -\delta \MH^{2},\nl
\Gamma_{\ren,ij}^{f,\rL}(p^2)={}& \delta_{ij}
+\Sigma_{ij}^{f,\rL}(p^{2})
+ \frac{1}{2}\left(\delta Z_{ij}^{f,\rL}+\delta
  Z_{ij}^{f,\rL\dagger}\right), \nl
\Gamma_{\ren,ij}^{f,\rR}(p^2)={}& \delta_{ij}
+\Sigma_{ij}^{f,\rR}(p^{2})
+ \frac{1}{2}\left(\delta Z_{ij}^{f,\rR}+\delta Z_{ij}^{f,\rR\dagger}\right),\nl
\Gamma _{\ren,ij}^{f,\rl}(p^2)={}& -m_{f,i}\delta_{ij}
+\Sigma_{ij}^{f,\rl}(p^{2}) 
- \frac{1}{2}\left(m_{f,i}\delta Z_{ij}^{f,\rL}
+ m_{f,j}\delta Z_{ij}^{f,\rR\dagger}\right) 
-\delta_{ij}\de m_{f,i}, \nl
\Gamma _{\ren,ij}^{f,\rr}(p^2)={}& -m_{f,i}\delta_{ij}
+\Sigma_{ij}^{f,\rr}(p^{2}) 
- \frac{1}{2}\left(m_{f,i}\delta Z_{ij}^{f,\rR}
+ m_{f,j}\delta Z_{ij}^{f,\rL\dagger}\right) 
-\delta_{ij}\de m_{f,i}. 
\end{align}
The renormalization of the longitudinal parts of the gauge-boson
self-energies belongs to the unphysical sector, discussed briefly in
\refse{se:ren_unphys}.

Inserting these equations into the renormalization conditions, we find
for the mass renormalization constants
\begin{align}\label{eq:CTM}
\delta \MW^2 ={}& \rRe\SIW(\MW^2),\qquad
\delta \MZ^2  =  \rRe\SIZZ(\MZ^2) ,\qquad
\delta \MH^2 ={} \rRe\SIH(\MH^2) ,\nl
 \delta m_{f,i} ={}& \frac{1}{2}\rRe\left[
m_{f,i}\left(\Sigma^{f,\rL}_{ii}(m_{f,i}^{2})
+\Sigma^{f,\rR}_{ii}(m_{f,i}^{2})\right)
{}+\Sigma^{f,\rl}_{ii}(m_{f,i}^{2})+\Sigma^{f,\rr}_{ii}(m_{f,i}^{2})\right].
\end{align}
The mass counterterms are real even for a complex quark mixing matrix,
which is not affected by $\rRe$.
The field renormalization constants of the boson fields are obtained as
\begin{alignat}{3}\label{eq:FR_bos}
\delta Z_{\FW} ={}& \left.
-\rRe\frac{\partial\Sigma_{\rT}^{\FW}(k^{2})}{\partial k^{2}}
\right\vert_{k^{2}=\MW^{2}} , & \qquad 
\delta Z_{\FV\FV} ={}& \left.
-\rRe\frac{\partial\Sigma_{\rT}^{\FV\FV}(k^{2})}{\partial k^{2}}
\right\vert_{k^{2}=M^{2}_{\FV}} ,\quad \FV = \FA,\FZ ,\nl
\delta Z_{\FA\FZ} ={}&
-2\rRe\frac{\Sigma_{\rT}^{\FA\FZ}(\MZ^2)}{\MZ^2}, &\qquad 
\delta Z_{\FZ\FA} ={}& 2\frac{\Sigma_{\rT}^{\FA\FZ}(0)}{\MZ^2} ,\nl
\delta Z_{\FH} ={}& \left. -\rRe\frac{\partial\Sigma^{\FH}(k^{2})}
{\partial k^{2}} \right\vert_{k^{2} = \MH^{2}} ,
\end{alignat}
and those of the fermion fields read
\begin{align}\label{eq:CTF}
\delta Z_{ii}^{f,\rL} ={}&
-\rRe\Sigma _{ii}^{f ,\rL}(m_{f,i}^{2}) 
- m_{f,i}\frac{\partial}{\partial p^{2}} \rRe\left[
 m_{f,i}\left(\Sigma^{f,\rL}_{ii}(p^{2})
   +\Sigma^{f,\rR}_{ii}(p^{2})\right)
\left.
{} + \Sigma^{f,\rl}_{ii}(p^{2})+ \Sigma^{f,\rr}_{ii}(p^{2})\right]
\right\vert_{p^{2}=m_{f,i}^{2}} , \nl
\delta Z_{ii}^{f,\rR} ={}&
-\rRe\Sigma _{ii}^{f,\rR}(m_{f,i}^{2}) 
- m_{f,i}\frac{\partial}{\partial p^{2}} \rRe\left[
 m_{f,i}\left(\Sigma^{f,\rL}_{ii}(p^{2})
   +\Sigma^{f,\rR}_{ii}(p^{2})\right)
\left.{} 
+ \Sigma^{f,\rl}_{ii}(p^{2})+  \Sigma^{f,\rr}_{ii}(p^{2})
\right]\right\vert_{p^{2}=m_{f,i}^{2}} 
,\nl 
\delta Z_{ij}^{f,\rL} ={}& \frac{2}{m_{f,i}^{2}-m_{f,j}^{2}}\,
\rRe\left[
 m_{f,j}^{2}\Sigma^{f,\rL}_{ij}(m_{f,j}^{2})
+m_{f,i}m_{f,j}\Sigma^{f,\rR}_{ij}(m_{f,j}^{2})
{}+m_{f,i}\Sigma^{f,\rl}_{ij}(m_{f,j}^{2})
 +m_{f,j}\Sigma^{f,\rr}_{ij}(m_{f,j}^{2})
\right], \qquad i\ne j,\nl
\delta Z_{ij}^{f,\rR} ={}& \frac{2}{m_{f,i}^{2}-m_{f,j}^{2}}\,
\rRe\left[
 m_{f,j}^{2}\Sigma^{f,\rR}_{ij}(m_{f,j}^{2})
+m_{f,i}m_{f,j}\Sigma ^{f,\rL}_{ij}(m_{f,j}^{2})
{}+m_{f,j}\Sigma^{f,\rl}_{ij}(m_{f,j}^{2})
 +m_{f,i}\Sigma^{f,\rr}_{ij}(m_{f,j}^{2})\right],
\qquad i\ne j 
.
\end{align}
Owing to \refeq{eq:Gaherm},
\beq
\delta Z^{\Ff,\si\dagger}_{ij}=\delta Z^{\Ff,\si}_{ij} 
\Big|_{m_i \leftrightarrow m_j, \,
\Sigma^{f,\rr}_{ij} \leftrightarrow \Sigma^{f,\rl}_{ij}} \,, 
\quad \si=\rL,\rR,
\eeq
where the interchange $m_i \leftrightarrow m_j$ applies to all 
explicit masses in \refeq{eq:CTF}.
For $i=j$ the renormalization conditions \refeq{eq:RCG} fix only the
hermitean part $\delta Z^{\Ff,\si}_{ii}+\delta Z^{\Ff,\si
  \dagger}_{ii}$ of the fermion field renormalization constant, while
its antihermitean part corresponds to the free global phase of the
fermion fields \cite{Aoki:1980ix,Denner:2004bm}. It is customary to
require that the antihermitean part vanishes, so that
\beq
\delta Z^{\Ff,\si\dagger}_{ii}=\delta Z^{\Ff,\si}_{ii}. 
\eeq
To be precise, defining this relation for $\sigma=\rR$, it is automatically
fulfilled for $\sigma=\rL$, or vice versa.

Since there is no generation mixing in the lepton sector, all (one-loop)
lepton self-energies are diagonal in generation space, and the off-diagonal
lepton wave-function renormalization constants are zero. The same holds for
the quark sector if one replaces the quark-mixing matrix~$V$ by the unit
matrix, as often done in calculations of radiative corrections for
high-energy processes.

The renormalization constant for the quark-mixing matrix $V$ can be
directly read off from \refeq{eq:RCV},
\beq  \label{eq:DV}
\delta V_{ij}=\frac{1}{4} \Bigl[(\delta Z^{u,\rL}_{ik}-\delta
Z^{u,\rL\dagger}_{ik}) V_{kj} -
V_{ik}(\delta Z^{d,\rL}_{kj}-\delta Z^{d,\rL\dagger}_{kj})\Bigr] .
\eeq
Inserting the fermion-field renormalization constants \refeqf{eq:CTF} yields
\begin{align}
\delta V_{ij} ={}&
\frac{1}{2}\rRe \sum_k \Biggl\{\frac{1}{m_{u,i}^{2}-
m_{u,k}^{2}} 
\biggl[
m_{u,i}^{2}\Sigma ^{\Fu,\rL}_{ik}(m_{u,i}^{2})
+m_{u,k}^{2}\Sigma ^{u,\rL}_{ik}(m_{u,k}^{2})
+m_{u,i}m_{u,k}(\Sigma ^{\Fu,\rR}_{ik}(m_{u,i}^{2})+\Sigma ^{\Fu,\rR}_{ik}(m_{u,k}^{2}))
\nl&\qquad{}
+m_{u,i}\Sigma^{\Fu,\rl}_{ik}(m_{u,k}^{2})
+m_{u,i}\Sigma^{\Fu,\rl}_{ik}(m_{u,i}^{2})
+m_{u,k}\Sigma^{\Fu,\rr}_{ik}(m_{u,i}^{2})
+m_{u,k}\Sigma^{\Fu,\rr}_{ik}(m_{u,k}^{2})
\biggr] \; V_{kj} \nl
&{}-V_{ik}\frac{1}{m_{d,k}^{2}-m_{d,j}^{2}} 
\biggl[
m_{d,k}^{2}\Sigma ^{d,\rL}_{kj}(m_{d,k}^{2})
+m_{d,j}^{2}\Sigma ^{d,\rL}_{kj}(m_{d,j}^{2})
+m_{d,k}m_{d,j}(\Sigma ^{d,\rR}_{kj}(m_{d,k}^{2})
+\Sigma ^{d,\rR}_{kj}(m_{d,j}^{2}))
\nl&\qquad{}
+m_{d,k}\Sigma^{\Fd,\rl}_{kj}(m_{d,k}^2)
+m_{d,k}\Sigma^{\Fd,\rl}_{kj}(m_{d,j}^2)
+m_{d,j}\Sigma^{\Fd,\rr}_{kj}(m_{d,k}^2)
+m_{d,j}\Sigma^{\Fd,\rr}_{kj}(m_{d,j}^2)
\biggr]\Biggr\}.
\end{align}

It remains to fix the charge renormalization constant $\de Z_e$.  The
renormalized vertex function for the $\gamma\Ffbar\Ff$ vertex reads
\beq
\Gamma^{\FA\Ffbar\Ff}_{\ren,ij,\mu }(k,\bar{p},p)=-e Q_{f}\gamma_{\mu }\de_{ij}
+e\Lambda^{\FA\Ffbar\Ff}_{\ren,ij,\mu }(k,\bar{p},p).
\label{eq:RCE1}
\eeq
For equal ($i=j$) OS external fermions the higher-order contribution
can be decomposed as ($k=-\bar{p}-p$)
\begin{align}\label{eq:RCE2}
\Lambda^{\FA\Ffbar\Ff}_{\ren,ii,\mu }(k,\bar{p},p)={}&
\biggl(
 \gamma _{\mu }\Lambda^{f}_{\ren,\mathrm{V}}(k^{2})
-\gamma _{\mu }\gamma _{5}\Lambda^{f}_{\ren,\mathrm{A}}(k^{2}) +
\frac{(p-\bar{p})_{\mu}}{2\Mf}\Lambda^{f}_{\ren,\mathrm{S}}(k^{2})-
\frac{(p+\bar{p})_{\mu}}{2\Mf}\gamma_{5}\Lambda^{f}_{\ren,\mathrm{P}}(k^{2})
\biggr).
\end{align}
Expressing the renormalized quantities in terms of the unrenormalized ones
and the counterterms, using the renormalization condition \refeqf{eq:RCE}
for arbitrary fermions, and making use of the Gordon identities and the
Ward identity
\begin{align}
\bar{u}(p)\Lambda^{\FA\Ffbar\Ff}_{ii,\mu }(0,-p,p)u(p)={}&
{-\Qf}\bar{u}(p)\left[\frac{\partial}{\partial
    p^\mu}\Si^{\Ffbar\Ff}_{ii}(-p,p)\right]u(p)
-a_f\bar{u}(p)\ga_\mu(1-\ga_5)u(p)\frac{\SIAZ(0)}{\MZ^2},
\label{eq:WIchargeSM}
\end{align}
where $\Si^{\Ffbar\Ff}_{ii}(-p,p)$ is the higher-order contribution to
$\Gamma^{\Ffbar\Ff}_{ii}(-p,p)$, one finally finds
\cite{Denner:1991kt,Bohm:2001yx}
\beq
\label{eq:DZE}
\de Z_e = \frac{\delta e}{e} = -\frac{1}{2}\delta Z_{\FA\FA}-
\frac{\sw}{\cw}\frac{1}{2}\delta Z_{\FZ\FA}
=\left.\frac{1}{2}\frac{\partial\SIA(k^{2})}{\partial k^{2}}
\right\vert_{k^{2}=0}-\frac{\sw}{\cw}\frac{\SIAZ(0)}{\MZ^{2}}.
\eeq
The result \refeqf{eq:DZE} is independent of the fermion species,
reflecting electric charge universality.  Consequently, the
analogue of \refeq{eq:RCE} holds for arbitrary fermions~$f$.

The Ward identity \refeqf{eq:WIchargeSM} is known to hold in the EWSM
at the one-loop level.  
In \citeres{Bohm:1986rj,Hollik:1988ii,Jegerlehner:1991dq,Denner:1991kt},
\refeq{eq:DZE} was derived via explicit one-loop calculation of self-energy
and vertex diagrams.  In \refapp{se:charge_WI} we provide a derivation of
\refeq{eq:DZE} from Slavnov--Taylor and Lee identities valid at the
one-loop level. While this derivation proceeds in an arbitrary 't~Hooft
gauge, it can easily be generalized to other gauges. It holds as well for
extensions of the SM that do not modify its gauge structure. Moreover, the
derivation presented in \refapp{se:charge_WI} may serve as a basis for a
generalization of the identity \refeqf{eq:DZE} to higher orders.


In the OS scheme the {\em weak mixing angle} is a derived quantity,
\ie merely an abbreviation.
Following Sirlin \cite{Sirlin:1980nh}, it is generally defined as
\beq \label{eq:swren}
\sin^{2}\theta_{\rw} = \sw^{2} = 1-\cw^2 = 1-\frac{\MW^{2}}{\MZ^{2}},
\eeq
in terms of the physical, renormalized gauge-boson masses. This
definition is independent of a specific process and valid to all
orders of perturbation theory. We note that the weak mixing angle is
not a directly measurable quantity and (in higher orders) could be
defined in different ways. In particular, this {\em OS weak
  mixing angle} differs from the {\em effective weak mixing} angle
defined from the OS fermion--Z-boson couplings 
that is often used in the analysis of LEP1/SLC data.
 
Since the auxiliary parameters $\sw^2$ and $\cw^2$ frequently
appear, it is useful to introduce the corresponding renormalization constants 
\beq
c_{0,\rw}^2 = \cw^2 + \delta \cw^2 , \qquad
s_{0,\rw}^2 = \sw^2 + \delta \sw^2 .
\eeq
Because of \refeq{eq:swren} these are directly related to the
renormalization constants for the gauge-boson masses, \ie  to one-loop order
\beq
\label{eq:Dsw}
\frac{\delta \cw^{2}}{\cw^{2}} ={}
\frac{\delta \MW^{2}}{\MW^{2}}-\frac{\delta \MZ^{2}}{\MZ^{2}} =
\rRe\left(\frac{\SIW(\MW^{2})}{\MW^{2}}
-\frac{\SIZ(\MZ^{2})}{\MZ^{2}}\right) ,
\qquad
\frac{\delta \sw^{2}}{\sw^{2}} ={} -\frac{\cw^{2}}{\sw^{2}}
\frac{\delta \cw^{2}}{\cw^{2}}.
\eeq

\subsubsection{Renormalization of the unphysical sector}
\label{se:ren_unphys}

The renormalization of the unphysical fields and of the gauge
parameters does not affect $S$-matrix elements. Therefore, the choice
of the corresponding renormalization conditions is a pure matter of
convenience. 

An appropriate field renormalization of the would-be Goldstone-boson and
the Faddeev--Popov fields allows us to render vertex and Green functions
involving these fields finite. A simple way to achieve this is to
require that the corresponding propagators have residue equal to one
at the poles and that the mixing between the  Faddeev--Popov ghosts corresponding
to the photon and Z-boson fields vanishes for $k^2=0$ and $k^2=\xi_\FZ\MZ^2$.

It can be shown that in linear gauges such as the 't~Hooft--Feynman
gauge, the gauge-fixing term need not be
renormalized~\cite{tHooft:1972qbu,Lee:1973rb}.  We therefore require
that the gauge-fixing Lagrangian has the form of \refeq{eq:Lfix} with
\refeq{eq:gf} in terms of renormalized fields and parameters.  This
renormalized form of the gauge-fixing Lagrangian can be translated
back into an expression in terms of unrenormalized quantities by
inverting the renormalization transformation of the parameters and
fields given in \refse{se:rtsm}.  Bare and renormalized gauge-fixing
Lagrangians can even be brought into identical functional form by
introducing bare and renormalized sets of gauge parameters and
adjusting the transformation between the two sets accordingly.
Finally, from the bare gauge-fixing functionals the Faddeev--Popov
Lagrangian is derived as usual and expressed in terms of renormalized
fields.

The renormalization of the unphysical sector can be arranged in such a
way that the Slavnov--Taylor identities still hold in terms of
renormalized fields and parameters. 
To this end, it is sufficient to
renormalize the sources of the BRS-transformed fields in the
generating functionals of renormalized Green functions appropriately.
More details can be found in
\citeres{Aoki:1982ed,Bohm:1986rj,Bohm:2001yx}.

\subsection{Renormalization within the background-field method}
\label{se:ren_bfm}

In the {\em background-field formalism} for the SM (see \refse{se:bfm}) the
parameters and quantum fields are renormalized in the same way as in the
conventional formalism, given in \refeqs{eq:Pren}--\refeqf{eq:rcxi}. In
fact, the renormalization of the quantum fields is irrelevant for
$S$-matrix elements and back\-ground-field vertex functions, since
field-renormalization counterterms related to internal lines in Feynman
diagrams generally cancel. In one-loop calculations, renormalization
constants for quantum fields do not even appear.  On the other hand, the
renormalization of the background fields is required for finiteness of the
background-field vertex functions and the validity of the background-field
Ward identities for renormalized vertex functions,
\begin{align}
\FWhat_{0}^{\pm}  ={}& Z_{\FWhat}^{1/2} \FWhat^{\pm}
  = \left(1+\frac{1}{2}\delta Z_{\FWhat}\right) \FWhat^{\pm}  ,\nl
\left(\barr{l} \FZhat_{0} \\ \FAhat_{0} \earr \right)  ={}&
\left(\barr{ll} Z_{\FZhat\FZhat}^{1/2} & Z_{\FZhat\FAhat}^{1/2}  \\[1ex]
                Z_{\FAhat\FZhat}^{1/2} & Z_{\FAhat\FAhat}^{1/2}
      \earr
\right)
\left(\barr{l} \FZhat \\ \FAhat \earr \right)   =
\left(\barr{cc} 1 + \frac{1}{2}\delta Z_{\FZhat\FZhat} &
\frac{1}{2}\delta Z_{\FZhat\FAhat} \\ [1ex]
\frac{1}{2}\delta Z_{\FAhat\FZhat}  & 1 + \frac{1}{2}\delta
Z_{\FAhat\FAhat}
\earr \right)
\left(\barr{l} \FZhat \\[1ex] \FAhat \earr \right)   ,\nl
\phihat_{0}^\pm ={}& Z_{\phihat}^{1/2} \phihat^\pm
 = \left(1+\frac{1}{2}\delta Z_{\phihat}\right) \phihat^\pm,
\qquad
\FShat_{0} ={} Z_{\FShat}^{1/2} \FShat
 = \left(1+\frac{1}{2}\delta Z_{\FShat}\right) \FShat, \qquad
\FShat = \FHhat,\chihat.
\qquad
\label{eq:renconsts2}
\end{align}

The renormalization constants in \refeq{eq:renconsts2} can be fixed in
such a way that the renormalized effective action is invariant under
background-field gauge transformations. This implies relations between
the renormalization constants for the background fields and the
parameter renormalization constants \cite{Denner:1994xt}. At the
one-loop level, these read
\begin{align}
\label{eq:delZB}
\delta Z_{\FAhat\FAhat} ={}& {- 2} \delta Z_e, \qquad
\delta Z_{\FZhat\FAhat} = 0, \qquad
\delta Z_{\FAhat\FZhat} = 2 \frac{\cw}{\sw}
    \frac{\delta \cw ^2}{\cw ^2}  ,\nl
\delta Z_{\FZhat\FZhat} ={}& {- 2} \delta Z_e -
    \frac{\cw ^2 - \sw ^2}{\sw^2} \frac{\delta \cw ^2}{\cw ^2} , \qquad
\delta Z_{\FWhat} = - 2 \delta Z_e -
    \frac{\cw ^2}{\sw^2} \frac{\delta \cw ^2}{\cw ^2}  ,\nl
\delta Z_{\FHhat} ={}& \delta Z_{\chihat} = \delta Z_{\phihat} 
      = - 2 \delta Z_e -
        \frac{\cw ^2}{\sw^2} \frac{\delta \cw ^2}{\cw ^2} +
        \frac{\delta \MW^2}{\MW^2} + 2\frac{\Delta \varv}{\varv},
\end{align}
with $\delta \cw ^2$ from \refeq{eq:Dsw}. The explicit term 
\beq
2\frac{\Delta \varv}{\varv} = 2\frac{T^{\FHhat}}{\MH^2\varv}
 = \frac{eT^{\FHhat}}{\sw\MW\MH^2}
\eeq
in $\delta Z_{\FHhat}$ is only present in the FJTS, and the explicit
tadpole $T^{\FHhat}$ appearing there has been replaced by the shift of the
vev upon using \refeqs{eq:RCT}  and \refeqf{eq:def_Delta_v}.
  
Although the expressions of the scalar field renormalization constants
$\delta Z_{\FHhat}=\delta Z_{\chihat} = \delta Z_{\phihat}$ in terms
of the parameter renormalization constants differ between the FJTS and
the PRTS, $\delta Z_{\FHhat}$ in the two tadpole schemes is the same;
the $\De\varv$ term results from tadpole contributions and is
implicitly contained in $\de\MW^2$ in the PRTS.  The relations
\refeqf{eq:delZB} express the field renormalization constants of all
gauge bosons and scalars completely in terms of the renormalization
constants of the electric charge and the particle masses.  Note, in
particular, that we get the same relation as in QED between the charge
and the background-photon-field renormalization constant,
\beq
\label{eq:dZeBFM}
 Z_{\FAhat\FAhat} = (Z_e)^{-2}.
\eeq
In the background-field formalism, this relation holds to all orders
of perturbation theory.  It has been employed to calculate the
two-loop counterterm for the renormalization of the electric charge in
the SM \cite{Degrassi:2003rw}.  In addition, the field renormalization
constants of the fermions obey
\beq
\delta Z^{\FF,\rL} =
\delta Z^{\FF_+,\rL} = \delta Z^{\FF_-,\rL},
\label{eq:zferm}
\eeq
\ie the renormalization constants for the two left-handed fermions in
a doublet $\FF=(\FF_+,\FF_-)^\rT$ must be equal (as in the minimal
renormalization scheme of \citere{Bohm:1986rj}).

As the field renormalization constants are fixed by \refeq{eq:delZB},
in general the propagators acquire residues different from one, and
different fields can mix on shell.  This is similar to the minimal
OS scheme~\cite{Bohm:1986rj} of the conventional formalism.
Therefore, when calculating $S$-matrix elements, one has to introduce
(UV-finite) wave-function renormalization constants, which are
explicitly given in \citere{Denner:1996gb} for the gauge fields.
However, just as in QED, the OS definition of the electric
charge together with gauge invariance automatically fixes the residue
of the photon propagator to one. Moreover, the photon--\PZ-boson mixing
vanishes for OS photons as a consequence of \refeq{eq:gaAZ1} and
the analyticity of the 2-point functions.

The gauge-invariance relations among the renormalizations constants
can also be derived by demanding that
the form of the BFM Ward identities discussed in \refse{se:bfm} 
remains the same after renormalization~\cite{Denner:1994xt}. 
In this context, we emphasize that the self-energies in
\citere{Denner:1994xt} contain only 1PI contributions, so that
the form of Eqs.~(29)--(32) of that reference is changed 
in the notation of this review by some tadpole contributions,
while Eqs.~(25)--(28) of \citere{Denner:1994xt} formally remain the same.
To illustrate the derivation of some of the relations \refeqf{eq:delZB}
among the BFM renormalization constants in our conventions, 
we introduce the BFM self-energies
$\Sigma$ in the gauge-boson--Goldstone-boson sector
analogous to our definitions in the conventional formalism as follows,
\begin{align}
\hat\Gamma^{\FWhat^+\FWhat^-}_{\ren,\rL}(k^2) ={}& 
\MW^{2}-\Sigma^{\FWhat}_{\ren,\rL}(k^{2})
=\MW^{2}-\Sigma^{\FWhat}_{\rL}(k^{2})
+\left(\delta Z_{\FWhat}  + \frac{\delta \MW^2}{\MW^2}  
\right)\MW^2, 
\nl
\hat\Gamma^{\FZhat\FZhat}_{\ren,\rL}(k^2) ={}& 
\MZ^2 -\Sigma^{\FZhat\FZhat}_{\ren,\rL}(k^{2})
=\MZ^2 -\Sigma^{\FZhat\FZhat}_{\rL}(k^{2})
+\left( \delta Z_{\FZhat\FZhat}  + \frac{\delta \MZ^2}{\MZ^2}
\right)\MZ^2,
\nl
\hat\Gamma^{\FWhat^\pm\phihat^\mp}_{\ren,\mu}(k,-k) ={}& 
\pm k_\mu \left[ \MW+\Sigma^{\FWhat\phihat}_{\ren}(k^2)\right]
=\pm k_\mu \left[ \MW+\Sigma^{\FWhat\phihat}(k^2)
+\frac{1}{2}\left(\de Z_{\phihat}
+\de Z_{\FWhat}+\frac{\de\MW^2}{\MW^2} \right)\MW
\right], 
\nl
\hat\Gamma^{\FZhat\chihat}_{\ren,\mu}(k,-k) ={}& 
k_\mu \left[ \ri\MZ+\Sigma^{\FZhat\chihat}_{\ren}(k^2)\right]
=k_\mu \left[ \ri\MZ+\Sigma^{\FZhat\chihat}(k^2)
+\frac{1}{2}\left(\de Z_{\chihat}
+\de Z_{\FZhat\FZhat}+\frac{\de\MZ^2}{\MZ^2} \right)\ri\MZ
\right],
\nl
\hat\Gamma^{\phihat^+\phihat^-}_{\ren}(k,-k) 
={}&\hat\Gamma^{\phihat^+\phihat^-}_{\ren}(k^2) 
= k^{2} + \Sigma^{\phihat}_{\ren}(k^{2})
= k^{2} + \Sigma^{\phihat}(k^{2})+\de Z_{\phihat}k^2,
\nl
\hat\Gamma^{\chihat\chihat}_{\ren} (k,-k)
={}&\hat\Gamma^{\chihat\chihat}_{\ren} (k^2)
= k^{2} + \Sigma^{\chihat}_{\ren}(k^{2})
= k^{2} + \Sigma^{\chihat}(k^{2})+\de Z_{\chihat}k^2.
\end{align}
Recall that the tree-level contributions to the vertex functions
differ from the conventional ones, since the BFM vertex functions
do not include gauge-fixing terms.
As already pointed out in \refse{se:bfWI},
the unrenormalized vertex functions including tadpole contributions,
as encoded in the tree-level parts plus the 
$\Sigma$ functions, obey \refeq{eq:BFMWI2ptfctfull}
with $r_{\FHhat}$ as in \refeq{eq:rHhatFJTS} or
$r_{\FHhat}=1$ in the FJTS or PRTS, respectively.
Expressed in terms of self-energies $\Sigma$, the identities are given by
\begin{alignat}{3}
& \Si^{\FWhat}_{\rL}(k^2) + \MW \Si^{\FWhat\phihat}(k^2)
+ \MW^2\frac{\De\varv}{\varv} =0 , 
&\qquad&
k^2 \Si^{\FWhat\phihat}(k^2) - \MW \Si^{\phihat}(k^2) 
- \MW k^2\frac{\De\varv}{\varv} =0 , 
\nn\\
& \Si^{\FZhat\FZhat}_{\rL}(k^2) - \ri \MZ \Si^{\FZhat\chihat}(k^2)
+ \MZ^2\frac{\De\varv}{\varv} =0 , 
&&
k^2 \Si^{\FZhat\chihat}(k^2) - \ri \MZ
\Si^{\chihat}(k^2) 
- \ri\MZ k^2\frac{\De\varv}{\varv} =0 , 
\end{alignat}
with the $\De\varv$ terms only contributing in the FJTS.
Demanding that these relations carry over to
renormalized quantities in the form
\begin{alignat}{3}
& \Si^{\FWhat}_{\ren,\rL}(k^2) + \MW \Si^{\FWhat\phihat}_{\ren}(k^2)
=0, 
&\qquad&
k^2 \Si^{\FWhat\phihat}_{\ren}(k^2) - \MW \Si^{\phihat}_{\ren}(k^2) 
=0 , 
\nn\\
& \Si^{\FZhat\FZhat}_{\ren,\rL}(k^2) - \ri \MZ \Si^{\FZhat\chihat}_{\ren}(k^2)
=0, 
&&
k^2 \Si^{\FZhat\chihat}_{\ren}(k^2) - \ri \MZ
\Si^{\chihat}_{\ren}(k^2) 
= 0  
\end{alignat}
is a non-trivial requirement, which has the unique solution 
\beq
\de Z_{\phihat} = \de Z_{\FWhat}+\frac{\de\MW^2}{\MW^2}+\frac{2\De\varv}{\varv},
\qquad
\de Z_{\chihat} = \de Z_{\FZhat\FZhat}+\frac{\de\MZ^2}{\MZ^2}
+\frac{2\De\varv}{\varv}.
\label{eq:dZphidZchiBFM}
\eeq
The derivation of these relations, which are part of \refeq{eq:delZB},
is valid both in the FJTS and the PRTS.
The relations \refeqf{eq:delZB} and \refeqf{eq:dZphidZchiBFM} can be
alternatively derived by requiring that the renormalization transformations
applied to gauge-invariant terms of the bare Lagrangian result in
equivalent terms expressed by renormalized parameters and fields up to
global factors.  
Taking into account the expressions of
\refeq{eq:delZB} for $\de Z_{\FWhat}$ and $\de Z_{\FZhat\FZhat}$, the
relations \refeqf{eq:dZphidZchiBFM} are compatible with the requirement
$\delta Z_{\FHhat}=\delta Z_{\chihat} = \delta Z_{\phihat}$, which
expresses rigid invariance, and lead to the last relation of
\refeqf{eq:delZB}.

As a consequence of the relations between the renormalization
constants, the counterterm vertices of the background fields have a
much simpler structure than the ones in the conventional formalism.
In fact, all vertices originating from a single gauge-invariant
term in the Lagrangian acquire the same renormalization constants.
The explicit form of the counterterm vertices at one-loop order can be
found in \refapp{app:FR_BFM}.

\subsection{Renormalization of QCD}
\label{se:ren_qcd}

When QCD is included as described in \refse{se:QCD}, the
renormalization has to be extended. Additional
counterterms for the strong coupling constant $\gs$ and the gluon
fields $\FG$ are introduced according to
\begin{equation}
g_{0,\mathrm{s}} ={} (1+\DZgs)\gs, \qquad
\FG_0 ={} Z_{\FG}^{1/2} \FG = \left(1+\frac{1}{2}\DZG\right) \FG.
\end{equation}
Moreover, the counterterms introduced in \refeqs{eq:Pren} and
\refeqf{eq:field_ren}, in particular those for the quark masses and
quark fields receive additional contributions from the strong
interaction. In order to render all Green functions finite, field
renormalization constants for the Faddeev--Popov ghosts of the strong
interaction and for the gauge parameter of QCD have to be introduced
in analogy to the strategy outlined in \refse{se:ren_unphys}.

Although the gluon field renormalization constant $\delta Z_{\FG}$ is
often defined in the $\MSbar$ scheme in the QCD literature, it can
also be fixed conveniently by an OS renormalization condition in
analogy to the second condition of \refeq{eq:RCM},
\beq\label{eq:RC_QCD}
\lim_{k^2\to0} \frac{1}{k^2}
\left. \Gamma^{\FG\FG}_{\ren,\mu\nu} (-k,k) \,\veps^{\nu }(k)
\right\vert_{k^{2}= 0} ={} -\veps_{\mu }(k),
\eeq
leading to 
\beq
\delta Z_{\FG} ={} \left.
-\frac{\partial\Sigma_{\rT}^{\FG}(k^{2})}{\partial k^{2}}
\right\vert_{k^{2}=0} .
\eeq
Since 2-point functions at momentum transfer zero do not develop
absorptive parts, it is not necessary to include the $\rRe$ operation.
The corresponding explicit result for $\delta Z_{\FG}$ can, e.g., be
found in \citere{Beenakker:2002nc}.

The strong coupling $\gs$ is usually fixed in the $\MSbar$ scheme, which is
tied to dimensional regularization (DR) with $D=4-2\epsilon$ dimensions.  In
this scheme, the renormalization constant contains only the UV divergence
along with related finite constants in the form
\beq\label{eq:Delta}
\Delta = \frac{2}{4-D} + \ln 4\pi - \gamma_{\mathrm{E}},
\eeq
where $\gamma_{\mathrm{E}}=0.577\ldots$ is the Euler--Mascheroni
constant.  In the $\MSbar$ scheme, the renormalization constant $\de
Z_{\gs}$ can be determined from any vertex function that involves the
strong coupling in leading order (LO), a convenient choice being the
quark--antiquark--gluon coupling.  A common definition is
\beq
\de Z_{\gs} = -\frac{\alphas(\muR^2)}{4\pi}\left[
\left(\frac{11}{2}-\frac{N_{\mathrm{f}}}{3}\right)
\left(\Delta +\ln\frac{\mu^2}{\muR^2} \right)
-\frac{1}{3}\sum_{F}\left(\De+\ln\frac{\mu^2}{m_F^2}\right)
\right]
\eeq
in the {\em $N_\mathrm{f}$-flavour scheme}.  Here $\mu$ is the mass
parameter of DR (see \refse{se:regul}) and
$\muR$ the renormalization scale for the strong coupling.
Furthermore, $N_\mathrm{f}$ is the number of active (light) flavours,
and $F$ runs over the inactive (heavy) flavours. The contribution of
active flavours (and gluons) is renormalized in the $\MSbar$ scheme,
while the one from inactive flavours, \ie heavy degrees of freedom
like the top quark, is subtracted at zero momentum transfer.  As a
consequence, the contribution of heavy flavours disappears (decouples)
from the renormalized gluon self-energy if the corresponding masses
tend to infinity.
Most of the higher-order calculations for high-energy scattering
processes at the LHC are carried out in the {\em 5-flavour scheme},
where $N_\mathrm{f}=5$ and $\sum_{F}$ comprises only the top-quark
contribution.

In the BFM, background-field gauge invariance implies the relation
\beq
Z_{\FGhat}= (Z_{\gs})^{-2}
\eeq
to all orders in perturbation theory. Consequently, $\de Z_{\gs}$ can
be obtained from the gluon self-energy alone (without any vertex
corrections) in the BFM \cite{Abbott:1980hw}.

\subsection{Techniques for electroweak one-loop calculations}
\label{se:loop_techniques}

In this section we give an overview of the techniques used for
calculating EW one-loop corrections. We first discuss the
regularization of soft and collinear singularities based on
DR and mass regularization (MR).  Then we
sketch the traditional tensor-integral reduction method that goes back
to Brown and Feynman \cite{Brown:1952eu} and Passarino and Veltman
\cite{Passarino:1978jh} and the reduction at the integrand level
introduced by Ossola, Pittau, and Papadopoulos \cite{Ossola:2006us}.
Finally, we discuss codes for tensor and scalar one-loop integrals and
make some remarks on purely numerical methods for the calculation of NLO
corrections.

\subsubsection{Dimensional versus mass regularization}
\label{se:regul}

Perturbative calculations in relativistic quantum field theories lead
to divergent integrals. To allow for a proper treatment of
perturbative quantities a {\em regularization} procedure is needed.
This amounts to a modification of the theory so that possibly
divergent expressions become well defined and that in a suitable limit
the original (divergent) theory is recovered. Once renormalization has
been performed and virtual and real corrections are properly combined,
all singularities cancel, and the limit to the original theory is
well defined. The final results are independent of the regularization
procedure.

Quantum field theories involve both UV and infrared (IR)
divergences. The former result from regions of the loop integrals
where the integration momentum tends to infinity, the latter originate
from regions in phase space where loop momenta or momenta of external
particles become small or collinear to each other. 

For the regularization of UV and IR singularities various
regularization schemes exist. Naive regularization schemes based on
momentum cutoffs violate Lorentz invariance and gauge invariance and
are therefore not used in practical calculations. Lattice
regularization in space--time is very useful for the numerical
calculation of non-perturbative quantities but too complicated for
perturbative calculations.

The method of choice for perturbative calculations in relativistic
quantum field theories and gauge theories is {\it dimensional
  regularization}, because it is convenient and respects Lorentz and
gauge invariance.  It was introduced by Bollini and Giambiagi
\cite{Bollini:1972ui} and 't~Hooft and Veltman~\cite{tHooft:1972tcz}.
For a precise definition and thorough discussion we refer to Section~4
of the book of Collins~\cite{Collins:1984xc}.

In DR, calculations are performed in $D$ instead of four dimensions.
Since the loop integrals converge in suitably chosen numbers of
dimensions (small dimensions for UV singularities, large dimensions
for IR singularities), the usual calculational rules for integrals,
such as linearity, translational and rotational invariance, and
scaling properties can be used.  The structure of the loop integrals
allows for a continuation to arbitrary complex values of $D$ so that
integrals involving UV and IR singularities can be
regularized and the limit $D\to4$ can be taken.%
\footnote{For integrals that are both UV and IR divergent, this
  analytic continuation is subtle, since no value of $D$ exists for
  which those integrals are finite (for proper arguments, see, \eg,
  \citere{Leibbrandt:1975dj}).}  
The divergences manifest themselves as poles at integer values of $D$.
Changing the space--time dimension of an integral also changes its
mass dimension.  This effect gets compensated by multiplying each
integral $\int \rd^D q$ over a 4-momentum $q$ with a factor $(2\pi
\mu)^{4-D}$, where $\mu$ has the dimension of a mass. In one-loop
calculations the poles appear always in the combination $\Delta$ as
defined in \refeq{eq:Delta}.

DR is the de facto standard for regularizing UV singularities in gauge
theories. It is the basis for the $\MSbar$ renormalization scheme (\cf
\refse{se:input_msbar}), where the renormalization constants just
consist of terms of the form \refeqf{eq:Delta} involving the UV
divergence along with some universal finite terms.  
This renormalization scheme is
widely used in QCD, where the masses of the light quarks are often
neglected.  DR regularizes also the soft and collinear singularities
related to massless gluons and quarks.  Since UV and IR singularities
result from different parts of the integration region, they can be
discriminated by introducing formally different regulator parameters
$\Delta_{\mathrm{UV}}$ and $\Delta_{\mathrm{IR}}$ corresponding to
different numbers of dimensions, $D_{\mathrm{UV}}$ and
$D_{\mathrm{IR}}$.
In practice, this separation can be done in different ways.  Since IR
divergences are tied to specific kinematical configurations such as
OS conditions or massless limits, the UV divergences can be
extracted from a given integral upon first considering the integral
with general off-shell external momenta, determining the divergent
parts, which are all of UV origin, and finally specializing the result
to the actual kinematic situation.  Alternatively, the $1/\eps$ poles
of UV origin can be obtained by regularizing all IR singularities by
small mass parameters, so that no $1/\eps$ poles of IR origin exist.%
\footnote{To this end, it is even enough to give all internal massless
  particles some infinitesimal regulator mass~$\lambda$, even though
  this procedure does not correspond to a consistent IR regularization
  of gauge theories.  More details on translating mass singularities
  between different regularization schemes can,\eg, be found in
  \citere{Dittmaier:2003bc}.}
At the one-loop level, no standard scalar integral is both IR and UV
singular. Thus, when expressing amplitudes by the standard scalar
integrals, the separation of IR and UV singularities becomes trivial.

DR can be used in the EW theory in the same way as in QCD. However, in
QED and the EWSM, IR singularities have been traditionally
regularized with infinitesimal photon masses and small fermion masses,
\ie in {\em mass regularization}.  In contrast to the non-abelian
theories, the photon mass does not ruin the renormalizability of a
$\U(1)$ gauge theory (see, for instance, \citere{Piguet:1974tp} or
\citere{Collins:1984xc}, Section 12.9) and can be used to regularize
soft singularities without violating Ward identities. Furthermore, the
small lepton masses are well-defined physical parameters, which offer
a physical regularization of corresponding collinear singularities in
QED and in the SM. DR of UV singularities can be combined with MR of
IR singularities, providing a natural separation of the two kinds of
singularities. The basic loop and phase-space integrals depend on the
chosen regularization, but are available for one-loop calculations in
both regularization schemes and are of similar complexity
\cite{Dittmaier:2003bc,Denner:2010tr}.  More details on IR
singularities, their regularization and evaluation can be found in
\refse{se:real}.

Finally, the choice of the regularization scheme is a matter of
convenience. For hadron-collider processes, the partonic cross
sections have to be combined with parton distributions. Since the
latter are usually defined in the $\MSbar$ scheme, their combination
with perturbative matrix elements is more convenient in DR. If on the
other hand final-state fermions and photons are not recombined in the
collinear regions, as it may be relevant for muons, then EW
corrections become sensitive to the masses of the fermions, so that MR
of the corresponding singularities is appropriate. In any case,
results can be translated between regularization schemes, and different
regularization methods can be combined with the due care.

Further regularization schemes, including in particular 4-dimensional
methods, are discussed in the review~\cite{Gnendiger:2017pys}.

\subsubsection{The \texorpdfstring{$\gamma_5$}{gamma-5} problem}
\label{se:gamma5}

While DR preserves Lorentz and gauge invariance, it does not preserve
chiral invariance. This is due to the fact that the matrix
\beq
\gamma_5=\ri\ga^0\ga^1\ga^2\ga^3
\label{eq:ga5}
\eeq
appearing in chiral projectors $\om_\pm=(1\pm\ga^5)/2$ is a genuinely
$4$-dimensional object. The definition \refeqf{eq:ga5} is consistent
in $D$ dimensions, but breaks $D$-dimensional Lorentz invariance, since
the first four  dimensions are distinguished. From the existence of the
axial anomaly it is clear that no fully Lorentz-invariant
generalization of the Dirac algebra exists. On an algebraic level, it
is impossible to define a $D$-dimensional $\ga$ algebra with
\beq
\{\ga^\mu,\ga^\mu\} = \ga^\mu\ga^\nu + \ga^\nu\ga^\mu  
= 2g^{\mu\nu}, \quad \mu=0,\ldots,D-1
\label{eq:diracalgebra}
\eeq
that obeys both
the anticommutation relation 
\beq\label{eq:anticomm}
\{\ga^\mu,\ga_5\} = \ga^\mu\ga_5 + \ga_5\ga^\mu = 0
\eeq
and the trace condition
\beq
\Tr[\ga^\mu\ga^\nu\ga^\rho\ga^\si\ga_5\} =-4\ri\eps^{\mu\nu\rho\si},
\label{eq:trace_cond_4d}
\eeq
or a $D$-dimensional generalization thereof.  In a theory with chiral
fields, as the SM, a gauge-invariant DR can
only be obtained by giving up either anticommutativity
\refeqf{eq:anticomm} or the trace condition \refeqf{eq:trace_cond_4d}
\cite{Jegerlehner:2000dz}. 

In the literature, different prescriptions for the treatment of
$\ga_5$ within DR have been proposed:
\begin{myitemize}
\item In the {\it 't~Hooft--Veltman prescription}
  \cite{tHooft:1972tcz,Akyeampong:1973xi,Breitenlohner:1977hr,Bonneau:1980yb}
  the 4-dimensional trace condition \refeqf{eq:trace_cond_4d} is
  preserved.  Dirac matrices are split into 4-dimensional
  ($\hat\ga^\mu$) and $(D-4)$-dimensional ($\check\ga^\mu$) parts,
  $\ga^\mu=\hat\ga^\mu+\check\ga^\mu$, obeying the
  (anti)com\-mu\-ta\-tion relations
\beq
\{\hat\ga^\mu,\ga_5\} = 0, \qquad
[\check\ga^\mu,\ga_5] = 0, \qquad
\{\check\ga^\mu,\ga_5\} \ne 0.
\eeq
This prescription is algebraically consistent, but violates gauge
invariance, \ie modifies the Ward identities. These can be restored
upon adding extra $(D-4)$-dimensional counterterms.  Results on those
counterterms (even beyond one loop) as well as a technical variant to
treat $\gamma_5$ in the 't~Hooft--Veltman scheme that avoids the split
of $D$-dimensional indices into $(D-4)$-dimensional and 4-dimensional
ones in practical calculations can be found in \citere{Larin:1993tq}.

\item In {\it naive dimensional regularization (NDR)}
  \cite{Chanowitz:1979zu}, which can be safely used at the one-loop level,
  the anticommutation relations
  \refeqf{eq:anticomm} as well as the trace condition
  \refeqf{eq:trace_cond_4d} are kept.  The algebraic inconsistency of
  this scheme leads to ambiguities in the results for Feynman diagrams
  that are related to the Adler--Bell--Jackiw anomaly
  \cite{Adler:1969gk,Bell:1969ts}.  All diagrams without closed
  fermion loops and contributions to closed fermion loops with an even
  number of $\gamma_5$ can be calculated consistently in $D$
  dimensions.  Contributions to closed fermion loops exhibiting traces
  involving an odd number of $\gamma_5$ cannot be obtained uniquely by
  dimensional continuation.  In theories with a chiral anomaly,
  the anomalous terms in the usual conventions are recovered upon
  requiring Bose symmetry and vector-current conservation
\cite{Adler:1969gk,Adler:1970,Jegerlehner:2000dz}.
\item Kreimer, K\"orner and Schilcher (KKS) proposed a
  scheme~\cite{Kreimer:1989ke,Korner:1991sx,Kreimer:1993bh} with
  anticommuting $\ga_5$ by giving up the cyclicity of Dirac traces.
  Instead, traces in different diagrams have to be written down using
  the same {\it reading point}, defining the starting point of the
  $\gamma$-string in a Dirac trace.  In addition, specific rules have
  to be used to calculate Dirac traces involving an odd number of
  $\ga_5$.  As a consequence all ambiguities disappear, and the
  anomalies in the usual conventions are obtained.  Beyond one loop,
  the detailed application is sometimes under debate.  In practice,
  the consistency of the scheme can be checked by verifying the
  relevant Ward identities before loop integration.
\end{myitemize}

In theories without chiral anomalies, such as the SM, traces with an
odd number of $\ga_5$ matrices can be calculated using NDR while
ensuring that all potentially anomalous terms cancel. At the one-loop
level the ambiguities concern fermion-loop diagrams with three or four
external vector bosons. The ambiguous terms are UV-finite, polynomial
(rational) terms involving $\eps^{\mu\nu\rho\si}$ and are linear in
the external momenta for the 3-point functions and independent of the
external momenta for the 4-point functions. In a consistent
calculation these rational terms cancel when summing over complete
fermion generations. To ensure the cancellation in practical
calculations, the same analytical expressions have to be used for all
diagrams involving (linearly UV-divergent)
triangular fermion loops, and the diagrams with
clockwise and counterclockwise fermion flow have to be related
appropriately. It is crucial that Dirac traces with reversed order of
Dirac matrices are calculated in the same way as the traces with the
original order. This can be ensured by starting the traces at the same
vertex and performing the same algebraic manipulations. It can also be
guaranteed by requiring Bose symmetry or in case of charged vector
bosons the symmetry under the interchange of their momenta,
Lorentz indices, and couplings. The rules of the KKS scheme also ensure
cancellation of the ambiguities.  The NDR scheme has been used in the
majority of one-loop calculations in the past, including
\citeres{Lemoine:1979pm,Fleischer:1980ub,Aoki:1980ix,Bardin:1981sv,%
Fleischer:1982af,Bohm:1986rj,Denner:1991kt,Denner:2000bj,Denner:2005es}.

In modern recursive approaches (see \refse{se:automation}), the
numerators of the loop diagrams are calculated in four dimensions, and
the extra rational terms are calculated with dedicated Feynman
rules~\cite{Ossola:2008xq}.  Only the latter depend on the treatment
of $\ga_5$.  Feynman rules for the rational parts of the EW one-loop
amplitudes have been calculated in the KKS
scheme~\cite{Garzelli:2010qm,Shao:2011tg}. No rational terms
proportional to $\eps$~tensors show up, confirming the absence of
rational terms associated with anomalies in the SM and the equivalence
of the NDR and KKS schemes at the one-loop level.

\subsubsection{Tensor-integral reduction}
\label{se:ti_reduction}

\providecommand{\TNtilde}{{\tilde{T}^N}}
\providecommand{\sst}{\scriptstyle}
\providecommand{\debar}{\bar\delta}

A general one-loop amplitude $\de\cM$ can be written as a linear combination of
one-loop tensor integrals $T_{(j)}^{N_j,\mu_1\ldots\mu_{r_j}}$ with tensor
coefficients $c^{(j,r_j,N_j)}_{\mu_1\ldots\mu_{r_j}}$ and a
contribution $\cM_{\mathrm{ct}}$ from
counterterms, \ie
\beq
\de\cM = \sum_j c^{(j,r_j,N_j)}_{\mu_1\ldots\mu_{r_j}}
T_{(j)}^{N_j,\mu_1\ldots\mu_{r_j}}
+ \cM_{\mathrm{ct}},
\label{eq:M_oneloop}
\eeq
where the Lorentz indices and tensors have to be understood as
$D$-dimensional objects in DR.  
The indices $j$ characterize the different tensor integrals, and $N_j$
denotes the number of propagators of these integrals.  In the
't~Hooft--Feynman gauge the rank of the tensor integrals is bounded by
$r_j\le N_j$ in renormalizable theories.  The counterterms cancel the
UV singularities present in the tensor integrals which manifest
themselves as poles in $\deltaDR=(4-D)/2$ [\cf \refeq{eq:Delta}].

If the Lorentz indices are only allowed to take 4-dimensional values,
as is the case if the amplitudes are calculated numerically in
4~dimensions, finite {\em rational terms of type
  $R_2$}~\cite{Ossola:2008xq} are missed in the amplitude calculation
(see \refse{se:opp}), since algebraic terms $f(D)$ depending on
$D=4-2\eps$ produce finite rational terms when they multiply $1/\eps$
poles, \ie $f(D)/\eps=f(4)/\eps+ \text{finite rational terms}$.  If $f(D)$
appears as $f(4)$ from the beginning, the missing rational terms have
to be supplemented by hand, and the matrix element
\refeqf{eq:M_oneloop} takes the form
\beq
\de\cM = \sum_j \hat{c}^{(j,r_j,N_j)}_{\mu_1\ldots\mu_{r_j}}
T_{(j)}^{N_j,\mu_1\ldots\mu_{r_j}}
+ \cM_{\mathrm{ct}} + \cM_{\mathrm{R_2}}  ,
\label{eq:M_oneloop_R2}
\eeq
where the caret on the tensor coefficient
$\hat{c}^{(j,r_j,N_j)}_{\mu_1\ldots\mu_{r_j}}$ indicates that it
involves only 4-dimensional components.

In \citeres{Binoth:2006hk,Bredenstein:2008zb} it was demonstrated that
rational terms of type $R_2$ of IR origin cancel in unrenormalized
scattering amplitudes and only result from wave-function
renormalization constants. Thus, only the rational terms of type $R_2$
of UV origin have to be calculated.
Since the UV-divergent terms of one-loop integrals are always
polynomial in internal masses and external momenta (see, e.g.,
App.~A in \citere{Denner:2005nn} for explicit results) and are not
subject to specific kinematical configurations (in contrast to
IR singularities), the calculation of rational terms of type $R_2$ 
is simple and easy to automate (see \refse{se:opp}d). 

Tensor integrals have the general form \cite{Denner:2005nn}
\beq
\label{eq:tensorint}
T^{N,\mu_{1}\ldots\mu_{P}}(p_{1},\ldots,p_{N-1},m_{0},\ldots,m_{N-1})=
\displaystyle{\frac{(2\pi\mu)^{4-D}}{\ri\pi^{2}}\int \rd^{D}q\,
\frac{q^{\mu_{1}}\cdots q^{\mu_{P}}}
{\PD_0\PD_1\ldots \PD_{N-1}}}
\eeq
with the denominator factors
\beq \label{eq:D0Di}
\PD_{k}= (q+p_{k})^{2}-m_{k}^{2}+\ri\eps, \qquad k=0,\ldots,N-1 ,
\qquad p_0=0,
\eeq
where $\ri\eps$ $(\eps>0)$ is an infinitesimally small imaginary
part.%
\footnote{Note that the $\eps$ in the propagator denominators and the
  $\eps$ that parametrizes the deviation of the dimension from 4 are
  different, although we use the same symbol. The meaning should be
  clear from the context.}  Here $N$ denotes the number of propagators
in the loop, and $P$ the rank of the tensor integral. For $P=0$, \ie
rank zero, \refeq{eq:tensorint} defines the scalar $N$-point integral
$T^N_0$.  Following the notation of \citere{'tHooft:1978xw}, we set
$T^{1}= A,$ $T^{2}= B,$ $T^{3}= C,$ $T^{4}= D$, $T^{5}= E,$ $T^{6}=
F$.

We use the conventions of
\citeres{Denner:1991kt,Denner:2002ii,Denner:2005nn} to decompose the
tensor integrals into Lorentz-covariant structures.  For tensor
integrals up to rank five the decompositions read
\begin{align}
T^{N,\mu}={}&\sum_{i_1=1}^{N-1} p_{i_1}^{\mu}T^N_{i_1}, \qquad
T^{N,\mu\nu}=\sum_{i_1,i_2=1}^{N-1} p_{i_1}^{\mu}p_{i_2}^{\nu}T^N_{i_1i_2}
+g^{\mu\nu}T^N_{00},\nl
T^{N,\mu\nu\rho}={}&\sum_{i_1,i_2,i_3=1}^{N-1} p_{i_1}^{\mu}p_{i_2}^{\nu}p_{i_3}^{\rho}T^N_{i_1i_2i_3}
+\sum_{i_1=1}^{N-1}\{g p\}_{i_1}^{\mu\nu\rho} T^N_{00i_1}, 
\nl
T^{N,\mu\nu\rho\si} ={}& 
\sum_{i_1,i_2,i_3,i_4=1}^{N-1} p_{i_1}^{\mu}p_{i_2}^{\nu}p_{i_3}^{\rho}p_{i_4}^\si T^N_{i_1i_2i_3i_4}
+\sum_{i_1,i_2=1}^{N-1}
\{g pp\}_{i_1i_2}^{\mu\nu\rho\si}T^N_{00i_1i_2}
+\{g g\}^{\mu\nu\rho\si} T^N_{0000},\nl
T^{N,\mu\nu\rho\si\tau} ={}& 
\sum_{i_1,i_2,i_3,i_4,i_5=1}^{N-1}
p_{i_1}^{\mu}p_{i_2}^{\nu}p_{i_3}^{\rho}p_{i_4}^\si p_{i_5}^\tau T^N_{i_1
i_2i_3i_4i_5}
+\sum_{i_1,i_2,i_3=1}^{N-1}
\{g ppp\}_{i_1i_2i_3}^{\mu\nu\rho\si\tau}
T^N_{00 i_1 i_2 i_3}
+\sum_{i_1=1}^{N-1} 
\{g g p\}_{i_1}^{\mu\nu\rho\si\tau} 
T^N_{0000i_1}.
\label{eq:TI_decomposition}
\end{align}
Because of the symmetry of the tensor $T^{N}_{\mu_{1}\ldots\mu_{P}}$
all coefficients $T^N_{i_1\ldots i_P}$ are symmetric under permutation
of all indices. 
The symmetrized tensors in \refeq{eq:TI_decomposition} read
\begin{align}
\{g p\}_{i_1}^{\mu\nu\rho} ={}& 
g^{\mu\nu}p_{i_1}^{\rho}+g^{\nu\rho}p_{i_1}^{\mu}+g^{\rho\mu}p_{i_1}^{\nu},
\nl[.3em]
\{g pp \}^{\mu\nu\rho\si}_{i_1 i_2} ={}& 
 g^{\mu\nu}p_{i_1}^{\rho}p_{i_2}^{\si}
+g^{\mu\rho}p_{i_1}^{\si}p_{i_2}^{\nu}
+g^{\mu\si}p_{i_1}^{\nu}p_{i_2}^{\rho}
+g^{\nu\rho}p_{i_1}^{\si}p_{i_2}^{\mu}
+g^{\rho\si}p_{i_1}^{\nu}p_{i_2}^{\mu}
+g^{\si\nu}p_{i_1}^{\rho}p_{i_2}^{\mu},
\nl[.3em]
\{gg\}^{\mu\nu\rho\si} ={}& g^{\mu\nu}g^{\rho\si}+g^{\nu\rho}g^{\mu\si}
+g^{\rho\mu}g^{\nu\si},
\label{eq:covbrace}
\end{align}
with obvious generalizations to other tensors like $\{gppp\}$,
\ie the curly brackets denote symmetrization so that all
non-equivalent permutations of the Lorentz indices of metric tensors
and a generic momentum are included. The decomposition of general
tensor integrals can be found in \citere{Denner:2005nn}. The following
presentation consistently sticks to the conventions introduced there.

Tensor-integral reduction allows us to reduce the tensor integrals to
those with less denominators or lower rank. Thus, in the reduction of
the tensor integral $T^{N+1}_{\mu_1\ldots\mu_P}$, tensor integrals
appear that are obtained by omitting the $k$th denominator $D_k$; such
integrals are denoted $T^{N}_{\mu_1\ldots\mu_P}(k)$.  In the Lorentz
decomposition of $T^{N}_{\mu_1\ldots\mu_P}(k)$, $k=1,\dots,N$, shifted
indices appear which are denoted as
\beq\label{eq:defji}
 (i_r)_k=\left\{\barr{lll} i_r & \mathrm{\ for\ }&  k>i_r,\\
                     i_r-1 & \mathrm{\ for\ }& k < i_r.\earr \right.
\eeq
When cancelling the first denominator $D_0$ the resulting tensor
integrals are not in the standard form, but can be expressed in terms
of standard integrals by shifting the integration momentum. Choosing
the shift $q\to q-p_1$, the following $N$-point integrals appear:
\beq
\label{Dshifted}
\tilde {T}^{N,\mu_1\ldots\mu_P}(0)
={}
\frac{(2\pi\mu)^{(4-D)}}{\ri\pi^{2}}\int\rd^{D}q\,
\frac{q^{\mu_1}\cdots q^{\mu_P}}{\PDs_1 \cdots \PDs_N},
\qquad
\PDs_{k}={} (q+p_{k}-p_{1})^{2}-m_{k}^{2}+\ri\eps, \qquad k=1,\ldots,N .
\eeq
The scalar integral $T^N_{0}= T^N$ and the tensor
coefficients $T^N_{00}$, $T^N_{0000}$, $\ldots$ are invariant under this
shift. The other coefficients of $T^{N}_{\mu_1\ldots\mu_P}(k)$ can be
recursively obtained as
\begin{align}\label{eq:TNaux}
T^N_{\sst\underbrace{\mathclap{\sst{0\ldots0}}}_{2n} i_{2n+1}\ldots i_{P}}(0) ={}& 
\TNtilde_{\underbrace{\sst 0\ldots0}_{2n} i_{2n+1}-1,\ldots, i_P-1}(0), \quad
i_{2n+1},\ldots,i_P>1, \nl
T^N_{\underbrace{\sst 0\ldots0}_{2n} 1 i_{2n+2} \ldots i_P}(0) ={}& 
-T^N_{\underbrace{\sst 0\ldots0}_{2n} i_{2n+2}\ldots i_P}(0)
-\sum_{r=2}^{N} T^N_{\underbrace{\sst 0\ldots0}_{2n} r i_{2n+2} \ldots
  i_P}(0), \quad i_{2n+2},\ldots,i_P>0.
\end{align}


We also use the notation $\debar_{ij} = 1-\de_{ij}$, \ie $\sum_i
\debar_{ij}(\dots) = \sum_{i\ne j}(\dots)$, and employ the caret
``$\hat{\phantom{x}}$'' to indicate indices that are omitted, \ie
\beq
T^N_{i_1\ldots\hat i_r \ldots i_P} = T^N_{i_1\ldots i_{r-1}  i_{r+1}
  \ldots i_P}.
\eeq

\myparagraph{(a) Reduction of tensor integrals up to $N\le4$ according
  to Passarino--Veltman} 

All one-loop tensor integrals can be algebraically reduced to one-loop
standard scalar integrals. The standard procedure is known as
{\it Passarino--Veltman reduction}~\cite{Passarino:1978jh}, which proceeds
via contracting the tensor integrals \refeqf{eq:tensorint} with the
external momenta $p_k^\mu$ and the metric tensor $g^{\mu\nu}$. Using
\begin{align}
2p_k q ={}& D_k-D_0-f_k , 
\qquad  f_k=p_k^2-m_k^2+m_0^2,\nn\\
q^2 ={}& D_0 + m_0^2,  
\label{contrtensor}
\end{align}
all scalar products with an integration momentum in the numerator can be
cancelled against propagator denominators $D_k$, resulting in tensor
integrals with lower rank and/or lower number of propagators.
Inserting the Lorentz decompositions of the tensor integrals
\refeqf{eq:TI_decomposition} results in systems of linear equations for
the tensor-integral coefficients. These equations have the 
solutions~\cite{Denner:1991kt}
\begin{align}\label{eq:PVrecursionD1}
T^N_{00i_3\ldots i_P} ={}& \frac{1}{2(3+P-N)}\biggl[
 -2(D-4)T^N_{00i_3\ldots i_P} 
+ T^{N-1}_{i_3\ldots i_P}(0) 
+ 2m_0^2  T^N_{i_3\ldots i_P}  
+ \sum_{n=1}^{N-1} f_n T^N_{ni_3\ldots i_P} \biggr],
\\\label{eq:PVrecursionD2}
 T^N_{i_1\ldots i_P} ={}&  \sum_{n=1}^{N-1}\left(Z^{(N-1)}\right)^{-1}_{i_1n}
\left(  T^{N-1}_{(i_2)_n\ldots (i_P)_n}(n)
\debar_{ni_2}\ldots \debar_{ni_P}
-  T^{N-1}_{i_2\ldots i_P}(0) - f_n  T^{N}_{i_2\ldots i_P} -  2 \sum_{r=2}^P
\de_{ni_r}  T^{N}_{00i_2\ldots \hat i_r\ldots i_P}\right), 
\quad i_1\ne0.
\end{align}
The matrix $\left(Z^{(N)}\right)^{-1}$ is the inverse of the Gram matrix
\beq\label{eq:matrixZ}
Z^{(N)} = \left(\barr{ccc}
2 p_1 p_1     & \ldots & 2 p_1 p_{N} \\
\vdots        & \ddots & \vdots      \\
2 p_{N} p_1 & \ldots & 2 p_{N}p_{N} 
\end{array} \right).
\eeq
The relations \refeqf{eq:PVrecursionD1} and \refeqf{eq:PVrecursionD2}
determine $T^N_{i_1\ldots i_P}$ in terms of $T^N_{i_1\ldots i_{P-1}}$,
$T^N_{i_1\ldots i_{P-2}}$, $T^{N-1}_{i_1\ldots i_{P-1}}$, and
$T^{N-1}_{i_1\ldots i_{P-2}}$. Using these relations recursively, all
coefficients of $N$-point functions can be expressed in terms of
$(N-1)$-point functions and the scalar $N$-point function $T^N_0$. The
finite polynomial quantities $(D-4)T^N_{00i_3\ldots i_P}$, the
{\em rational terms of type $R_1$}
(defined in \refse{se:opp}d), can easily be derived by exploiting
\refeqs{eq:PVrecursionD1} and \refeqf{eq:PVrecursionD2} for the
UV-singular parts and using those for the scalar 1-point and 2-point
functions as input; explicit results for tensors up to rank~7 are
summarized in \citere{Denner:2005nn}.  
Rational terms of type $R_1$ do not result from IR divergences, since
$T^N_{00i_3\ldots i_P}$ do not involve IR singularities
\cite{Denner:2005nn}. 
More explicit formulas for the
Passarino--Veltman reduction for all tensor functions up to rank~5 are,
\eg, given in Appendix~B of \citere{Denner:2002ii}.

Equation \refeqf{eq:PVrecursionD2} becomes numerically unstable if the
Gram matrix $Z^{(N-1)}$ is nearly singular, \ie if its determinant,
the Gram determinant $\Delta^{(N-1)}=\det(Z^{(N-1)})$, is close to
zero. In this case different iterative solutions of the reduction
equations that avoid those instabilities
are possible, as described in detail in
\citere{Denner:2005nn}. This reference contains also an alternative
Passarino--Veltman reduction that is numerically more stable in
specific phase-space regions.

Different variants of Passarino--Veltman reduction can be found in the
literature. Van Oldenborgh and Vermaseren constructed a different tensor basis
that concentrates some of the numerical instabilities into a
number of determinants \cite{vanOldenborgh:1989wn}. Ezawa \etal
performed the reduction using an orthonormal tensor basis
\cite{Ezawa:1990dh}.  A reduction in Feynman-parameter space, which is
equivalent to the Passarino--Veltman scheme, is used in the GRACE
package \cite{Belanger:2003sd}.

Alternative tensor reduction schemes have been
developed using different sets of master integrals. Davydychev could
relate the coefficients of one-loop tensor integrals to scalar
integrals in a different number of space--time dimensions
\cite{Davydychev:1991va}, and Tarasov found recursion relations
between those integrals \cite{Tarasov:1996br}. These methods have been
further elaborated by different groups
\cite{Bern:1992em,Binoth:1999sp,Duplancic:2003tv,Giele:2004iy,Giele:2004ub}.
In this approach all one-loop tensor integrals can be reduced to
finite 4-point integrals in $(D+2)$ dimensions and divergent 3-point
and 2-point integrals in $D$ dimensions.  
These basic integrals are typically reduced to the usual scalar
integrals or, in particular for small Gram determinants, calculated by
numerical integration \cite{Binoth:2005ff}.

\myparagraph{(b) Reduction of tensor integrals with $N\ge5$}

Owing to the 4-dimensionality of space--time further reduction
formulas exist for scalar and tensor integrals with $N\ge5$.  It was
realized already in the 1960s by Melrose \cite{Melrose:1965kb} that
scalar integrals with more than four lines in the loop, \ie 5-point
and higher-point scalar integrals, can be reduced to scalar integrals
with fewer internal propagators in four dimensions. These methods were
subsequently extended and improved by several authors
\cite{vanOldenborgh:1989wn,Campbell:1996zw,Davydychev:1991va,%
  Bern:1992em,Binoth:1999sp,Duplancic:2003tv,vanNeerven:1983vr,%
  Denner:1991kt,Denner:2002ii,Belanger:2002ik,Denner:2005nn,%
  Fleischer:2010sq,Fleischer:2011hc} and generalized to DR
in \citeres{Bern:1992em,Beenakker:2002nc,Dittmaier:2003bc}.  In
\citere{Denner:2002ii}, a method for the reduction of 5-point
integrals that completely avoids inverse leading Gram determinants was
worked out. Later, a similar reduction was found that
reduces 5-point tensor integrals to 4-point integrals with rank
reduced by one \cite{Binoth:2005ff,Denner:2005nn}.

We here present the results for the reduction of $(N\ge5)$-point
integrals using the conventions of \citere{Denner:2005nn}. These can
be directly written in terms of tensor integrals without the need to
introduce Lorentz decompositions.  For $N\ge6$ the reduction formulas
read
\begin{align}\label{eq:Fredtensor3}
T^N_{\mu \mu_1\ldots\mu_{P-1}} ={}&
\sum_{n=1}^5 
\smash{\sum_{\substack{m=1\\m\ne k}}^5} 
\left(M_{(k)}^{-1}\right)_{nm} \, p_{m,\mu}
\Bigl[T^{N-1}_{\mu_1\ldots\mu_{P-1}}(n)- T^{N-1}_{\mu_1\ldots\mu_{P-1}}(0)\Bigr]
+T^N_{\al \mu_1\ldots\mu_{P-1}}(g^\al_\mu-{\gfour}^{\al}_\mu), \quad
N\ge6,
\end{align}
which express the $N$-point tensor integral of rank~$P$ in terms of
six $(N-1)$-point tensor integrals of rank $(P-1)$. The matrix
$M_{(k)}^{-1}$ is the inverse of 
\beq\arraycolsep 6pt
M_{(k)}=
\left(
\barr{ccc@{\hspace{\arraycolsep}\;}c}
2p_{1}p_1 & \ldots & \;2p_{1}p_{5} \\
 \vdots     & \ddots     &\;\vdots     \\
 2p_{k-1}p_1 & \ldots & \;2p_{k-1}p_{5} \\
 f_1       & \ldots & \;f_5 \\
 2p_{k+1}p_1 & \ldots & \;2p_{k+1}p_{5} \\
 \vdots     & \ddots     &\;\vdots     \\
2p_{5}p_{1} & \ldots &\; 2p_{5}p_5
\earr\right)\,.
\eeq
The index $k$ can be chosen arbitrarily, and for $N>6$ any five
momenta $p_i$
can be selected out of the $(N-1)$ available momenta of the $N$-point
function.  The last term in \refeq{eq:Fredtensor3}, which involves the
4-dimensional metric tensor ${\gfour}^{\al}_\mu$ gives rise to
rational terms and only contributes if $T^N_{\al \mu_1\ldots\mu_{P-1}}$ is
singular. For UV singularities this is the case if $P\ge2N-4$, which is
not needed in renormalizable theories, where $P\le N$.  IR (soft and
collinear) singularities of $T^N_{\al \mu_1\ldots\mu_{P-1}}$ only appear
in contributions that are proportional to some momentum $p_{i,\al}$ and
thus drop out in the last term of \refeq{eq:Fredtensor3}. Therefore, the terms
involving $(g-\gfour)$ in \refeq{eq:Fredtensor3} can be omitted for
$P\le2N-5$. Further variants of the reduction \refeqf{eq:Fredtensor3} can be
found in \citere{Denner:2005nn}.

Equation \refeqf{eq:Fredtensor3} is not applicable to the reduction of
scalar functions. These can be reduced via
\begin{align}\label{eq:Fredf}
 T^N_{\mu_1\ldots\mu_P} ={}& - \sum_{n=0}^5
 \,\eta_n \,T^{N-1}_{\mu_1\ldots\mu_P}(n),
\quad \eta_n = \frac{\det(Y^{(5)}_n)}{\det(Y^{(5)})}.
\end{align}
The modified Cayley matrix $Y^{(5)}=(Y^{(5)}_{ij})$, $i,j=0,\ldots,5$ is defined by
\beq
Y^{(5)}_{ij}=m_i^2+m_j^2-(p_i-p_j)^2
\eeq 
in terms of the momenta and masses of five different propagators of
the $N$-point function, and the matrix $Y^{(5)}_n$ is obtained from $Y^{(5)}$ by
replacing all entries in the $n$th column by~1.  For the scalar
integral $T^N_0$, this result is identical with the one of
\citere{Melrose:1965kb}. For $P>0$, it provides an alternative, though
less efficient reduction for the tensor integrals of rank $N\ge6$
\cite{Denner:1991kt,Denner:2005nn}.

The formulas for the reduction of 5-point functions are somewhat
more complicated. The relation analogous to \refeq{eq:Fredtensor3}
reads \cite{Denner:2005nn}
\begin{align}\label{eq:Eredtensor1}
T^5_{\mu \mu_1\ldots\mu_{P-1}} ={}&
\sum_{n,m=1}^4  \left(X^{(4)}\right)^{-1}_{nm}p_{m,\mu}
\Bigl[T^4_{\mu_1\ldots\mu_{P-1}}(n)- T^4_{\mu_1\ldots\mu_{P-1}}(0)\Bigr]
+\sum_{n=1}^4  \left(X^{(4)}\right)^{-1}_{0n}\Bigl[-p_{n,\mu}
T^4_{\mu_1\ldots\mu_{P-1}}(0)+
\mathcal{T}^4_{\mu\mu_1\ldots\mu_{P-1}}(n)\Bigr]
\nl&{}
-\sum_{n=1}^4 \mathcal{T}^4_{\al\mu_1\ldots\mu_{P-1}}(n) 
\sum_{m,l=1}^4  2p_{m}^{\al}p_{l,\mu}
\left[\left(X^{(4)}\right)^{-1}_{0n}\left(X^{(4)}\right)^{-1}_{ml}
-\left(X^{(4)}\right)^{-1}_{0l}\left(X^{(4)}\right)^{-1}_{mn}\right]
+2m_0^2\left(X^{(4)}\right)^{-1}_{00}(g_\mu^\al-{\gfour}_\mu^\al)T^5_{\al\mu_1\ldots\mu_{P-1}}
\nl&{}
+2\sum_{n=1}^4 \left(X^{(4)}\right)^{-1}_{0n}
\Bigl[p_{n,\mu}(g^{\al\be}-{\gfour}^{\al\be})
- p_{n}^{\be} (g_\mu^\al-{\gfour}_\mu^\al)\Bigr]T^5_{\al\be \mu_1\ldots\mu_{P-1}},
\hspace{2em}
\end{align}
which expresses the $5$-point function of rank $P$ in terms of five
4-point functions of rank $(P-1)$ and contributions of the
corresponding 4-point functions of rank $P$ involving metric tensors
in their covariant decomposition as well as rational terms.  The
matrix $X^{(4)}$ reads
\beq\label{eq:defYmod}
X^{(4)}=\left(\barr{cccc}
2m_0^2 & f_1 & \ldots & f_{4} \\
f_1    & 2p_1p_1 & \ldots & 2p_1p_{4} \\
\vdots & \vdots  & \ddots & \vdots \\
f_{4}    & 2p_{4}p_1 & \ldots & 2p_{4}p_{4} 
\earr\right), 
\eeq
and 
\beq
\mathcal{T}^4_{\al\mu_1\ldots\mu_{P-1}}(i) =
\left.\left[T^4_{\al\mu_1\ldots\mu_{P-1}}(i)-T^4_{\al\mu_1\ldots\mu_{P-1}}(0)\right]\right|_{\text{contributions
with~}g_{\alpha\mu_r}} , 
\quad i=1,\dots,4.
\eeq
The rational terms proportional to $(g_\mu^\al-{\gfour}_\mu^\al)$ only
contribute if $T^5_{\al\be \mu_1\ldots\mu_{P-1}} $ is UV~divergent,
\ie for $P\ge5$. The rational terms for $P=5$ are given in
\citere{Denner:2005nn}, and the case $P>5$ does not appear in
renormalizable theories and renormalizable gauges.

The scalar 5-point function can be reduced to scalar 4-point functions
with the relation
\begin{align}\label{eq:oldEfinal}
T^5_{\mu_1\ldots\mu_P} ={}& - \sum_{n=0}^4
\eta_n \,T^4_{\mu_1\ldots\mu_P}(n)
+\sum_{n,m=1}^4
\left[\left(X^{(4)}\right)^{-1}_{00}\left(X^{(4)}\right)^{-1}_{nm}
-\left(X^{(4)}\right)^{-1}_{n0}\left(X^{(4)}\right)^{-1}_{0m}\right]
\,2p_{m}^{\al}
\mathcal{T}^4_{\al\mu_1\ldots\mu_P}(n)
\nn\\
& {}
+2T^5_{\al\be\mu_1\ldots\mu_P} \left(X^{(4)}\right)^{-1}_{00} \left(g^{\al\be}-\gfour^{\al\be}\right).
\end{align}
The $\eta_n$ are defined analogously to those in \refeq{eq:Fredf} from
the 5-dimensional modified Cayley matrix $Y^{(4)}=(Y^{(4)}_{ij})$,
$i,j=0,\ldots4$.  For the scalar case, $P=0$, the result is identical
with the one of \citere{Melrose:1965kb}. For $P>0$,
\refeq{eq:oldEfinal} provides an alternative, though less efficient,
reduction for the tensor integrals. Explicit expressions for the
rational terms for $P=4$ can be found in
\citeres{Denner:2002ii,Denner:2005nn}.

Reduction formulas for tensor integrals can be directly used in
reduction algorithms for one-loop amplitudes, \ie tensor components
can be employed instead of tensor coefficients for a basis of
covariants.  This approach is, \eg, implemented in \Recola
\cite{Actis:2012qn,Actis:2016mpe}.  Avoiding the introduction of
tensor coefficients allows for a speedup of the calculations. On the
other hand, these reduction formulas can be transferred to those for
the tensor coefficients upon inserting the Lorentz decompositions of
the tensor integrals.

\myparagraph{(c) Libraries for tensor-integral  reduction}

There are various software packages for tensor-integral reduction on the
market: 
\begin{myitemize}
\item {\sc LoopTools} \cite{Hahn:1998yk} is a package for the
  evaluation of scalar and tensor one-loop integrals based on the {\sc
    FF} package by van Oldenborgh \cite{vanOldenborgh:1990yc}. It
  provides the actual numerical implementations of the functions
  appearing in {\sc FormCalc} output \cite{Hahn:2016ebn}.
  UV~singularities are regularized dimensionally, IR singularities are
  regularized either dimensionally or in MR. Complex
  masses are supported using the implementation of \citere{Nhung:2009pm}.
\item {\sc
    Golem95C}~\cite{Binoth:2008uq,Cullen:2011kv,Guillet:2013msa} is a
  program for the numerical evaluation of scalar and tensor one-loop
  integrals which supports the use of complex masses. It performs an
  alternative reduction to a set of basis integrals involving 4-point
  functions in 6 dimensions, which are IR and UV finite, UV-divergent
  4-point functions in $D + 4$ dimensions, and various 2-point and
  3-point functions as defined in
  \citeres{Binoth:2005ff,Binoth:1999sp}.
\item The Fortran package \Collier
  \cite{Denner:2014gla,Denner:2016kdg} provides a numerically stable
  reduction of tensor integrals to scalar integrals based on
  conventional methods and expansions in regions with small
  kinematical determinants \cite{Denner:2005nn} and simultaneously
  calculates the scalar integrals. It supports DR and MR
  of IR singularities and complex masses.
\item
The C++ package {\sc PJFry} \cite{Fleischer:2011zz} encodes a
numerical reduction of one-loop tensor integrals in terms of scalar
1- to 4-point functions, which have to be supplied by an external
library.
\item
{\sc PackageX} is a Mathematica package for the analytic calculation
and symbolic manipulation of one-loop Feynman integrals in
relativistic quantum field theory \cite{Patel:2015tea,Patel:2016fam}.
It generates analytic expressions for dimensionally regulated tensor
integrals with up to four distinct propagators and arbitrarily high
rank.
\end{myitemize}

\subsubsection{Reduction of tensor integrals at the integrand level}
\label{se:integrand_reduction}

In \citeres{delAguila:2004nf,Pittau:2004bc} an alternative method was
proposed that allows for recursively reducing tensor integrals to
tensor integrals of lower rank and integrals with fewer propagator
denominators.  Therewith, high-rank tensor integrals can be
recursively expressed in terms of lower-rank ones at the integrand
level. Here, we sketch a reformulation of this method as described in
\citere{Buccioni:2017yxi}.  The idea of the algorithm is to express
products of two integration momenta in the numerators of the tensor
integrals by terms involving at most one integration momentum in
combination with denominators of the tensor integrals.  In four
space--time dimensions, the basic identity reads
\beq \qfour^\mu \qfour^\nu
 = \left[ A^{\mu\nu}_{-1} + A^{\mu\nu}_{0} \Dfour_0  \right]
+ \left[ B^{\mu\nu}_{-1,\lambda}+
 \sum\limits_{j=0}^{3} B^{\mu\nu}_{j,\lambda} \Dfour_j \right]\qfour^{\lambda}{}.
\label{eq:qmuqnuredfinal}
\eeq
The rank-1 polynomial on the r.h.s.\ is a linear combination of four
4-dimensional loop denominators, $\Dfour_0,\dots,\Dfour_3$, and the
corresponding tensor coefficients, $A_j^{\mu\nu}$ and
$B_{j,\lambda}^{\mu\nu}$, depend only on the three external momenta
$p_1,p_2,p_3$ corresponding to the loop-propagator denominators
$\Dfour_1,\Dfour_2,\Dfour_3$ as defined in \refeq{eq:D0Di} but for
4-dimensional loop momenta $\qfour$.
%
The coefficients of the loop denominators
$\Dfour_0,\Dfour_1,\Dfour_2,\Dfour_3$ are labelled with indices
$j=0,\dots,3$, while $j=-1$ is used for the constant parts.

The identity~\refeqf{eq:qmuqnuredfinal} provides an exact
representation of $\qfour^\mu \qfour^\nu$ in terms of 4-dimensional
loop denominators, but can be easily generalized to $D$-dimensional
denominators by substituting
\beq
\Dfour_j =  \DD_{j}-\qDmf^2, \quad j=0,1,2,3\,,
\label{eq:DiDibar}
\eeq
where quantities with a check denote $(D-4)$-dimensional objects, \eg
$\qD^2=\qfour^2+\qDmf^2$.  The terms of the form
$\DD_{i}-\Dfour_i=\qDmf^2$ account for the mismatch between the
$D$-dimensional and 4-dimensional parts of loop denominators.  The
corresponding tensor integrals with $(D-4)$-dimensional $\qDmf^{2}$
terms in the numerator give rise to finite terms
\cite{delAguila:2004nf,Ossola:2008xq}, so-called rational terms of
type $R_1$ (see \refse{se:opp}d).

The reduction identities of~\citere{delAguila:2004nf}
are based on a decomposition of the 4-dimensional loop momentum,
\beq
\qfour^\mu=\sum\limits_{i=1}^{4} c_i l_i^\mu, 
\label{eq:qdecomb}
\eeq
in a basis $l_1,\dots,l_4$, formed by 
massless momenta $l_i$, which span two orthogonal planes, 
\begin{align} 
l_i^2=0,\quad i=1,2,3,4, \qquad
l_{j} l_{k} =0{}, \quad j=1,2,\quad k=3,4. 
\label{eq:redbasisprop}
\end{align}
This reduction basis $\{l_i\}$ is constructed from two external momenta
$p_1,p_2$ which appear in the propagators $D_1$, $D_2$. The momenta
$l_{1},l_{2}$ lie in the plane spanned by $p_1$ and $p_2$,
\begin{align}
l^{\mu}_{1} = p^{\mu}_{1} - \alpha_{1} p^{\mu}_{2},  \qquad 
l^{\mu}_{2} = p^{\mu}_{2} - \alpha_{2} p^{\mu}_{1}. 
\label{eq:basismom12}
\end{align}
The momenta $l_3$, $l_4$ are defined as
\begin{align}
l^{\mu}_{3} ={}& \bar{\varv}(l_{1}) \gamma^{\mu} \om_- u(l_{2}),\qquad
l^{\mu}_{4} ={} \bar{\varv}(l_{2}) \gamma^{\mu} \om_- u(l_{1}),
\label{eq:basismom34}
\end{align}
with the massless Dirac spinors $u$ and $\bar{\varv}$,
and thus lie in the plane orthogonal to $p_1$, $p_2$.
The basis is normalized according to
\begin{equation}
\kappa = 2(l_{1} l_{2}) = -\frac{1}{2}(l_{3} l_{4}){},
\label{eq:basisnorm}
\end{equation}
and the coefficients $\alpha_{1,2}$ in \refeq{eq:basismom12} read
\begin{equation}
\alpha_{i} = \frac{p^{2}_{i}}{p_{1} p_{2} +\sgn(p_1p_2) \sqrt{\Delta}}, 
\label{eq:al12def}
\end{equation}
where $\Delta$ is the negative of the rank-2 Gram determinant,
\begin{equation}
\Delta =  (p_{1} p_{2})^2 - p^2_{1} p^2_{2}= -\det(p_1 p_2),
\label{eq:Gram12}
\end{equation}
and the sign in front of  the positive square root is chosen such that 
$\alpha_i$ behaves smoothly in the limits $p_i^2\to0$.
The Gram determinant is related to the normalization factor $\kappa$
via
\begin{equation}
\kappa = \frac{4\Delta}{p_{1} p_{2} +\sgn(p_1p_2) \sqrt{\Delta}}.
\label{eq:gammafactor}
\end{equation}
%

Rewriting \refeq{eq:qmuqnuredfinal} in the more generic form
\beq 
\qfour^\mu \qfour^\nu = 
\sum\limits_{j=-1}^{3} \left( A^{\mu\nu}_j+
B^{\mu\nu}_{j,\lambda}\qfour^{\lambda}\right) \Dfour_j,
\qquad \Dfour_{-1}=1,
\eeq
the reduction coefficients $A^{\mu\nu}_j$, $B^{\mu\nu}_{j,\lambda}$
are expressed in terms of the vectors $l_i$ as follows,
\begin{align}
\label{eq:ABmunutensors4}
A_{-1}^{\mu \nu} ={}& m_{0}^2 A_{0}^{\mu \nu},
\qquad
A_{0}^{\mu \nu} ={}  \frac{1}{4 \kappa} \left(\alpha L_{33}^{\mu \nu} + \frac{1}{\alpha} L^{\mu \nu}_{44} - L^{\mu \nu}_{34}\right),
\qquad
A_{1,2,3}^{\mu \nu} =0,
\nonumber\\
B^{\mu \nu}_{-1,\lambda} ={}& \sum_{i=1}^3 f_{i0} B^{\mu \nu}_{i,\lambda},
\qquad
B^{\mu \nu}_{0,\lambda} = -\sum_{i=1}^3 B^{\mu \nu}_{i,\lambda},
\nonumber\\
B^{\mu \nu}_{1,\lambda} = {}& \frac{1}{4 \kappa^2} 
\Bigg[
\frac{2(p_{3} r_{2})}{p_{3} l_{3}} 
\left( L_{33}^{\mu \nu} l_{4,\lambda} + \frac{1}{\alpha} L_{44}^{\mu \nu} l_{3,\lambda} 
\right) 
- \Bigl( r_{2}^{\mu} L^{\nu}_{34,\lambda} + r_{2}^{\nu} L^{\mu}_{34,\lambda} \Bigr)\Biggr]  
+ \frac{1}{\kappa}\Bigl(r_{2}^{\mu} \delta^{\nu}_{\lambda}- A_0^{\mu \nu} r_{2,\lambda} \Bigr)
\, ,\nonumber \\[2mm]
B^{\mu \nu}_{2,\lambda} ={}& B^{\mu \nu}_{1,\lambda}\big|_{r_1\leftrightarrow r_2},
\qquad
B^{\mu \nu}_{3,\lambda} ={}  - \frac{1}{4 \kappa (p_{3} l_{3})} 
\left( L_{33}^{\mu \nu} l_{4,\lambda} + \frac{1}{\alpha} L_{44}^{\mu \nu} l_{3,\lambda} \right) \, ,
\end{align}
with
\begin{align}
L^{\mu \nu}_{33}  = l_{3}^{\mu} l_{3}^{\nu},\qquad
L^{\mu \nu}_{44}  = l_{4}^{\mu} l_{4}^{\nu},\qquad
L^{\mu \nu}_{34}  = l_{3}^{\mu} l_{4}^{\nu} + l_{4}^{\mu} l_{3}^{\nu}, \qquad\,
\alpha = \frac{p_3  l_{4}}{p_3  l_{3}}\, ,
\label{eq:alphadef}
\end{align}
and
\begin{equation}
r^{\mu}_{1} = l^{\mu}_{1} - \alpha_{1} l^{\mu}_{2},\qquad
r^{\mu}_{2} =l^{\mu}_{2} -\alpha_{2} l^{\mu}_{1}, \qquad
f_{i0}=m_{i}^2-m_{0}^2-p_{i}^2{}.
\end{equation}
Equation \refeqf{eq:qmuqnuredfinal} is applicable to any integrand with $N\ge
4$ loop propagators,
\begin{align}
\int\rd^D\qD\,
\frac{\qfour^\mu \qfour^\nu S(\qfour)}{\DD_{0}\cdots \DD_{N-1}}={}&
\int\rd^D\qD\,
\frac{\left(A^{\mu\nu}_{-1}+B^{\mu\nu}_{-1,\lambda}\,\qfour^\lambda\right) S(\qfour)}{\DD_{0}\cdots \DD_{N-1}}
+\sum_{j=0}^3
\int\rd^D\qD\,
\frac{\left(A^{\mu\nu}_{j}+B^{\mu\nu}_{j,\lambda}\,\qfour^\lambda\right) S(\qfour)}{\DD_{0}\cdots
\DD_{j-1}\DD_{j+1} \cdots\DD_{N-1}}
{}+ \text{rational terms},
\label{eq:4pointintredA}
\end{align}
where $S(\qfour)$ is an arbitrary polynomial in the integration
momentum.  
The relation \refeqf{eq:4pointintredA} allows for reducing tensor
integrals of arbitrary high rank to simpler tensor integrals. In the
on-the-fly reduction approach of
\citeres{Buccioni:2017yxi,Buccioni:2019sur}, sketched at the end of
\refse{se:Openloops}, it is applied to incomplete loop integrals
during the recursive construction and avoids the calculation of tensor
integrals and the manipulation of tensors of rank higher than~2.

For 3-point integrals of rank $P\le 3$ a similar relation can be
derived \cite{delAguila:2004nf,Buccioni:2017yxi}:
\begin{align}
\int\rd^D\qD\,
\frac{\qfour^\mu \qfour^\nu S(\qfour)}{\DD_{0}\cdots \DD_{2}}={}&
\int\rd^D\qD\,
\frac{\left(A^{\mu\nu}_{-1}+B^{\mu\nu}_{-1,\lambda}\,\qfour^\lambda\right) S(\qfour)}{\DD_{0}\cdots \DD_{2}}
+\sum_{j=0}^2
\int\rd^D\qD\,
\frac{\left(A^{\mu\nu}_{j}+B^{\mu\nu}_{j,\lambda}\,\qfour^\lambda\right) S(\qfour)}{\DD_{0}\cdots 
\DD_{j-1}\DD_{j+1}\cdots\DD_{2}}
{}+ \text{rational terms}.
\label{eq:3pointintredA}
\end{align}
Here $S(\qfour)=S+S_\rho \qfour^\rho$ is an arbitrary rank-1 polynomial, the
sum over $j$ runs only from 0 to 2, and the
tensors $A_j^{\mu\nu}$ and  $B_{j,\lambda}^{\mu\nu}$ are obtained
from~\refeq{eq:ABmunutensors4} through the trivial replacements
\begin{align}
L^{\mu \nu}_{33} \to 0, \qquad
L^{\mu \nu}_{44} \to 0.
\label{eq:triangleOFRrepl}
\end{align}
The identity \refeqf{eq:3pointintredA} allows for a reduction of
3-point tensor integrals of ranks~2 and 3 to rank-1 integrals.

While similar methods can be found for the reduction of 2-point
functions, these do not provide advantages with respect to
conventional reduction methods.
Formulae for the reduction of rank-1 integrals at the integrand
level can be found in \citere{delAguila:2004nf}.

The algorithm described above has also been used in
\citere{vanHameren:2005ed}.  As tensor-integral reduction, it suffers
from the appearance of inverse Gram determinants.  A systematic way of
avoiding Gram-determinant instabilities in this reduction algorithm
was presented in \citeres{Buccioni:2017yxi,Buccioni:2019sur}.

\subsubsection{The OPP method}
\label{se:opp}

The {\it reduction at the integrand level} or {\it OPP method}
according to its authors Ossola, Papadopoulos, and
Pittau~\cite{Ossola:2006us} is based on the fact that any
one-loop amplitude can be expressed in terms of scalar 1-, 2-, 3-, and
4-point functions and that, as shown in \citere{delAguila:2004nf},
already at the integrand level 
the dependence of numerators on loop-integration momenta can be
expressed in terms of propagator denominators.

\myparagraph{(a) Coefficients of scalar functions}
The integrand of any $N$-point one-loop amplitude with loop momentum
$q$ has the form%
\beq
\label{eq:OLintegrand}
A(\qD)= \frac{\ND(\qD)}{\DD_{0}\DD_{1}\cdots \DD_{N-1}}\,,\qquad
\DD_{i} = ({\qD} + p_i)^2-m_i^2\,,\qquad p_0 \ne 0\,,
\eeq
and the choice $p_0 \ne 0$ allows for a completely symmetric treatment
of all denominators $\DD_{i}$.  The OPP method starts from the
related quantity
\beq
\label{eq:OPPintegrand}
\Afour(\qD)= \frac{\Nfour(\qfour)}{\DD_{0}\DD_{1}\cdots \DD_{N-1}}\,,
\eeq
with the purely 4-dimensional numerator $\Nfour(\qfour)$.  According to
\citere{Ossola:2006us}, this numerator can be cast into the form
\begin{align}
\label{eq:OPPdecomp}
\Nfour(\qfour) ={}&
\sum_{i_0 < i_1 < i_2 < i_3}^{N-1}
\left[
          d( i_0 i_1 i_2 i_3 ) +
     \tilde{d}(\qfour;i_0 i_1 i_2 i_3)
\right]
\prod_{i \ne i_0, i_1, i_2, i_3}^{N-1} \Dfour_{i} 
+ \sum_{i_0 < i_1 < i_2 }^{N-1}
\left[
          c( i_0 i_1 i_2) +
     \tilde{c}(\qfour;i_0 i_1 i_2)
\right]
\prod_{i \ne i_0, i_1, i_2}^{N-1} \Dfour_{i} \nl
     &+
\sum_{i_0 < i_1 }^{N-1}
\left[
          b(i_0 i_1) +
     \tilde{b}(\qfour;i_0 i_1)
\right]
\prod_{i \ne i_0, i_1}^{N-1} \Dfour_{i} 
+ \sum_{i_0}^{N-1}
\left[
          a(i_0) +
     \tilde{a}(\qfour;i_0)
\right]
\prod_{i \ne i_0}^{N-1} \Dfour_{i} 
+ \tilde{P}(\qfour)
\prod_{i}^{N-1} \Dfour_{i}\,. 
\end{align}
The quantity $d(i_0 i_1 i_2 i_3)$ is the coefficient of the scalar
4-point function with the denominator $D_{i_0} D_{i_1} D_{i_2} D_{i_3}$. 
Analogously, $c(i_0 i_1 i_2)$, $b(i_0 i_1)$, and $a(i_0)$
are the coefficients of all possible scalar 3-point, 2-point, and
1-point functions, respectively.

\myparagraph{(b) Spurious terms}
The {\em spurious terms} $\tilde{d}$, $\tilde{c}$, $\tilde{b}$,
$\tilde{a}$, and $\tilde{P}$ still depend on $\qfour$, but are required
to vanish upon integration over $\rd^D \qD$.  The complete set of
spurious terms in renormalizable theories (and renormalizable gauges)
was constructed in \citere{Ossola:2006us} in terms of the
massless momenta $l_i$ introduced in \refse{se:integrand_reduction}.
The actual form of the spurious terms depends on the maximal rank of the
tensors formed 
from loop momenta in the amplitude. The 4-point-like spurious terms
involve only one coefficient,
\beq
\tilde{d}(\qfour;0 1 2 3) = \tilde{d}(0 1 2 3) \, (\qfour+p_0) n,
\eeq
where 
\beq
n^\mu \propto \eps^{\mu\nu\rho\si} l_{1\nu} l_{2\rho} 
(p_{3}-p_{0})_\si, 
\eeq
is a four-vector orthogonal to all momenta $p_i-p_0$ ($i=1,2,3$),
and $l_1,l_2$ are constructed in the same way from $p_1-p_0$ and $p_2-p_0$
as described in the previous section for $p_1$ and $p_2$.
The 3-point-like spurious terms are of the form
\beq\label{eq:C_spurious}
\tilde{c}(\qfour;0 1 2) =\sum_{j=1}^{j_{\max}}
\left\{\tilde{c}_{1j}(012)[(\qfour+p_0) l_3]^j 
+ \tilde{c}_{2j}(012)[(\qfour+p_0) l_4]^j \right\}
\eeq
with constants $\tilde{c}_{1j}(012)$, $\tilde{c}_{2j}(012)$,
where $j_{\max}=3$ in renormalizable theories. The number of constant
coefficients of spurious terms is eight for the 2-point and four for
the 1-point functions in renormalizable theories. The spurious term
$\tilde{P}(q)$ vanishes in renormalizable theories.

\myparagraph{(c) Determination of coefficients}
All integration momenta $\qfour$ in the numerator $\Nfour(\qfour)$ of
\refeq{eq:OPPdecomp} are 4-dimensional. Equation \refeqf{eq:OPPdecomp}
is thus directly applicable to the purely 4-dimensional terms, which
are usually the most difficult to compute. To this end, we can set
$\DD_i\to \Dfour_i$ and $\qD\to \qfour$ in \refeq{eq:OPPintegrand}.

Since the standard scalar 1-, 2-, 3-, 4-point functions are known,
\refeq{eq:OPPdecomp} reduces the problem of calculating $\Afour(\qD)$
to the algebraic problem of extracting all possible coefficients in
\refeq{eq:OPPdecomp} by computing $\Nfour(\qfour)$ a sufficient number
of times, for different values of $\qfour$, and then inverting the
system of equations.  This procedure can be directly performed {\em at
  the amplitude level}, 
provided that the sum of all Feynman diagrams
is known.  Note, however, that using the tensor decomposition
\refeq{eq:M_oneloop_R2} is crucial in minimizing the evaluation time
of the many evaluations of $N(q)$ that the OPP method requires.

The resulting system of algebraic equations is in general pretty
large, and various strategies have been proposed to solve it
efficiently.  The original solution suggested in \citere{Ossola:2006us}
consists in using particular choices of $\qfour$ so that, systematically,
4, 3, 2, or 1 among all possible denominators $\Dfour_{i}$ vanish.
Then, the system of equations becomes block-triangular, and one can solve
first for the coefficients of all possible 4-point functions, then for
those of all 3-point functions, and so on.

For the determination of the coefficients of the 4-point function
with denominators $\Dfour_0,\dots,\Dfour_3$
one determines $\qfour$ in such a way that
\beq\label{eq:fixD4}
\Dfour_0=\Dfour_1=\Dfour_2=\Dfour_3=0,
\eeq
resulting in two complex solutions $\qfour_0^\pm$. Since
$(\qfour_0^++p_0)n=-(\qfour_0^-+p_0)n$, one finds in terms of these solutions,
\begin{align}
d(0123) ={}& \frac{1}{2}\left[\frac{N(\qfour_0^+)}{\prod_{i=4}^{N-1} D_i(\qfour_0^+)}
+\frac{N(\qfour_0^-)}{\prod_{i=4}^{N-1} D_i(\qfour_0^-)}\right],\notag\\
\tilde{d}(0123) ={}& \frac{1}{2(\qfour_0^++p_0)n}\left[\frac{N(\qfour_0^+)}{\prod_{i=4}^{N-1} D_i(\qfour_0^+)}
-\frac{N(\qfour_0^-)}{\prod_{i=4}^{N-1} D_i(\qfour_0^-)}\right].
\end{align}

The coefficients of the 3-point functions are determined in a similar
way after subtracting the known contributions of the 4-point
functions from the integrand. When selecting $\qfour$ in such a way that
\beq\label{eq:fixD3}
\Dfour_{0}= \Dfour_{1}= \Dfour_{2}= 0, \qquad \Dfour_{i} \ne 0, \quad i \ne 0,1,2,
\eeq
\refeq{eq:OPPdecomp} becomes
\beq
\Nfour(\qfour) - \sum_{i_3>2}[d(012i_3)
           + \tilde{d}(\qfour;012i_3)]\prod_{i \ne 0,1,2,i_3}\Dfour_{i}(\qfour) 
\,=\,{} R^\prime(\qfour) \prod_{i \ne 0,1,2} \Dfour_{i}(\qfour)
\,=\, [c(012) + \tilde{c}(\qfour;012)]\prod_{i \ne 0,1,2} \Dfour_{i}(\qfour)\,,
\eeq         
and one can extract $c(012)$ together with all the six coefficients 
$\tilde{c}_{ij}(012)$ of \refeq{eq:C_spurious} by
computing $R^\prime(\qfour)$ at seven different $\qfour$'s that fulfil
\refeq{eq:fixD3}. In fact, there is an infinite number of complex
solutions of \refeq{eq:fixD3}.

Once the coefficients of the 3-point contributions are known, those for
the 2-point and subsequently for the 1-point functions can be
determined in an analogous manner.

The determination of the coefficients was improved in
\citere{Mastrolia:2008jb} by making use of the polynomial structure of
 the integrand when evaluated at values of the loop momentum
fulfilling multiple cut conditions. The coefficients can be
efficiently extracted by means of a projection technique based on
discrete Fourier transform.  A further improvement was introduced
in \citere{Mastrolia:2012bu} exploiting the asymptotic behaviour of
the integrand decomposition.  It is based on the systematic
application of a Laurent expansion to the integrand decomposition
which is implemented through polynomial division.  

\myparagraph{(d) Rational terms}
\label{test}
 
For the calculation of the rational
terms one has to use \refeq{eq:OPPdecomp} in $D$ dimensions. The
mismatch between the $D$-dimensional denominators and the
4-dimensional inverse propagators in \refeq{eq:OPPdecomp} can be dealt
with by using \refeq{eq:DiDibar}.  The resulting terms depending on
$\qDmf^2$ are the {\em rational terms of type $R_1$} defined as
\cite{Ossola:2008xq},
\beq
R_1 = \frac{1}{(2\pi)^D}\int\rd^D \qD  
\frac{f(\qDmf^2,\qfour)}{\DD_{0}\DD_{1}\cdots \DD_{N-1}},
\eeq
where $f(\qDmf^2,\qfour)$ is some polynomial in $\qDmf^2$
with $f(0,\qfour)=0$.
Rational terms of type $R_1$ arise from the $(D-4)$-dimensional part
of the denominators.  They can be determined through a mass shift
$m_i^2 \to m_i^2 - \qDmf^2$ (to be applied also for $m_i=0$) in all
the coefficients of \refeq{eq:OPPdecomp}.  This procedure is formally
equivalent to the application of $D$-dimensional cuts in the
generalized unitarity framework.  In this way all coefficients of the
OPP expansion start depending on $\qDmf^2$. The coefficients of the
various powers of $\qDmf^2$, obtained through this mass shift, are the
coefficients of the extra integrals introduced in
\citere{delAguila:2004nf}. In renormalizable theories the only
possible contributions to those rational terms come from the extra
scalar integrals
\begin{align} \label{eq:rat_int} 
\int \rd^D \qD\,
\frac{\qDmf^4}{\DD_{i}\DD_{j}\DD_{k}\DD_{l}} ={}& {- \frac{\ri \pi^2}{6}} +
\cal{O}(\deltaDR)\,,\notag\nl 
\int \rd^D \qD\,
\frac{\qDmf^2}{\DD_{i}\DD_{j}\DD_{k}}       ={}& {- \frac{\ri \pi^2}{2}} +
\cal{O}(\deltaDR)\,,\nl 
\int \rd^D \qD\,
\frac{\qDmf^2}{\DD_{i}\DD_{j}}             ={}& {- \frac{\ri \pi^2}{2}}
\left[m_i^2+m_j^2-\frac{(p_i-p_j)^2}{3} \right]   +
\cal{O}(\deltaDR)\,. 
\end{align} 
Despite the fact that these integrals involve powers of
$(D-4)$-dimensional momenta in the numerator they give rise to finite
contributions for $D\to4$.  The $\qDmf^2$ dependence of the 2- and
3-point coefficients can be written as
\beq
b(ij;\qDmf^2) ={} b(ij)+ \qDmf^2 b^{(2)}(ij), 
\qquad
c(ijk;\qDmf^2) ={} c(ijk)+ \qDmf^2 c^{(2)}(ijk).
\eeq
The results for the contributions of the 4-point functions
are slightly more complicated and involve $(\qDmf^2)^2$ 
terms~\cite{Ossola:2007ax,Ossola:2008xq}.

To determine the $R_1$ terms, \ie the coefficients of the integrals
\refeqf{eq:rat_int}, in practice in the OPP method, the fits of the
coefficients are redone for different values of $\qDmf^2$, once
the 4-dimensional coefficients have been determined.  
An alternative approach to compute the $R_1$ terms
\cite{Ossola:2007bb} only requires the knowledge of the coefficients
in \refeq{eq:OPPdecomp} without the need to refit them at different
values of $\qDmf^2$. It is based on rewriting the denominators in
\refeq{eq:OPPintegrand} using
\beq
\frac{1}{D_i} = \frac{Z_i}{\Dfour_i}, \quad 
Z_i = 1 - \frac{\qDmf^2}{D_i}.
\eeq
While in this formulation 
more integrals than those in \refeq{eq:rat_int} contribute to
the rational parts, these are all classified and calculated in
\citere{Ossola:2007bb}.
When using tensor reduction techniques as described in \refse{se:ti_reduction},
the rational terms of type $R_1$ are obtained directly by the
$D$-dimensional tensor reduction without extra effort.

Rational terms of type $R_1$ arise only in the presence of
$1/(D-4)$ poles of UV origin and can be easily identified by means of
simple power counting.  The fact that rational terms of type $R_1$ of
IR origin do not occur at the one-loop level follows from the
observation that no such terms result from the Passarino--Veltman
reduction of one-loop integrals \cite{Denner:2005nn}.

The difference between \refeq{eq:OLintegrand} and
\refeq{eq:OPPintegrand} leads to the {\em rational terms of type
  $R_2$}.  The $D$-dimensional numerator function $\ND(\qD)$ can be
split into its 4-dimensional part $\Nfour(\qfour)$ plus a
$(D-4)$-dimensional part
\beq
\ND(\qD)=  \Nfour(\qfour) + \NDmf(\qDmf^2,\qfour,\deltaDR).
\eeq
The $R_2$ terms induced by the $(D-4)$-dimensional
part $\NDmf(\qDmf^2,\qfour,\deltaDR)$ are given by~\cite{Ossola:2008xq}
\beq
\label{eq:r2def}
R_2 =\frac{1}{(2\pi)^D}\int\rd^D \qD\, 
\frac{\NDmf(\qDmf^2,\qfour,\deltaDR)}{\DD_{0}\DD_{1}\cdots \DD_{N-1}}\,.
\eeq

As already pointed out in \refse{se:ti_reduction}, only the rational 
terms of type $R_2$ of UV origin remain in unrenormalized
scattering amplitudes, while $R_2$ terms of IR origin can only
originate from wave-function renormalization 
constants~\cite{Binoth:2006hk,Bredenstein:2008zb}.

Based on this observation, it was shown in \citere{Ossola:2008xq} that
the rational terms of type $R_2$ can be obtained from
a tree-level amplitude with a finite, universal set of
theory-dependent Feynman rules, with at most quartic vertices in
renormalizable theories.  Besides these new Feynman rules, this
approach does not introduce additional complications, and its
computational complexity is negligible compared to the one for the
4-dimensional parts. Therefore, it is the most widely used method
nowadays in integrand reduction and recursive loop calculation
techniques.  The Feynman rules for the $R_2$ terms have been
determined for QCD~\cite{Draggiotis:2009yb}, for the EWSM
\cite{Garzelli:2009is,Garzelli:2010qm,Shao:2011tg}, for a generic
theory containing scalars, vectors, and fermions~\cite{Pittau:2011qp},
as well as for specific BSM theories \cite{Shao:2012ja,Page:2013xla}.

While the calculation of rational terms via dedicated Feynman rules is
widely employed, different methods for the calculation of the rational
terms have been used in the QCD literature, in particular in the
context of generalized unitarity methods. An early method
\cite{Bern:2005cq} relies on OS 
recursion relations and a
unitarity-factorization bootstrap \cite{Bern:1997sc}. An alternative
is provided by $D$-dimensional unitarity methods
\cite{Anastasiou:2006jv,Anastasiou:2006gt,Giele:2008ve}. Here the
rational terms are determined in a similar manner as the 4-dimensional
terms in the OPP formalism by evaluating the amplitude in different
integer space--time dimensions such as $D=6$ and $D=8$.  The
calculations in additional dimensions can be avoided by using
appropriate mass shifts and exchanging the $D$-dimensional cuts by
4-dimensional massive cuts \cite{Badger:2008cm}.
 

\myparagraph{(e) Libraries for OPP reduction}

The OPP reduction algorithm has been implemented in different
numerical codes. The original implementation by Ossola, Papadopoulos
and Pittau is {\sc CutTools}~\cite{Ossola:2007ax}, which was
subsequently improved 
by the libraries {\sc Samurai}~\cite{Mastrolia:2010nb} and {\sc
  Ninja}~\cite{Peraro:2014cba}. {\sc Samurai} extends the OPP approach
to accommodate an implementation of the generalized $D$-dimensional
unitarity-cut technique and uses a polynomial interpolation
exploiting the discrete Fourier transform \cite{Mastrolia:2008jb}.
{\sc Ninja} implements an alternative reduction algorithm based on the
systematic application of Laurent series expansion
\cite{Mastrolia:2012bu}.

\myparagraph{(f) Generalized unitarity methods} 

The OPP reduction technique is closely related to {\em generalized
  unitarity methods}.  While unitarity can be verified by using
ordinary cuts of Feynman diagrams, generalized unitarity is based on
multiple cuts that do not necessarily correspond to physical
intermediate particle states.  These are defined by the requirement
that a subset of loop denominators vanishes, \ie that the
corresponding momenta are on shell [for examples see \refeqs{eq:fixD4}
and \refeqf{eq:fixD3}].  In \citere{Ellis:2007br} it was shown that
the coefficients of the master integrals in the OPP reduction can be
expressed in terms of tree-level amplitudes with complex external
momenta. The use of cut equations in $D$~dimensions corresponds to the
$D$-dimensional unitarity method \cite{Giele:2008ve}. While
generalized unitarity methods have been applied in NLO calculations
within QCD, they have not been generalized to the calculation of EW
NLO corrections yet.

\subsubsection{Scalar one-loop integrals}

Apart from purely numerical approaches, all methods for the
calculation of one-loop amplitudes eventually need results for the
one-loop scalar integrals $A_0$, $B_0$, $C_0$, and $D_0$. 

The classic paper with analytic expressions for general one-loop
scalar 1-, 2-, 3-, and 4-point integrals is due to 't~Hooft and
Veltman \cite{'tHooft:1978xw}. For the regular 1-, 2-, 3-point
integrals compact explicit expressions are given for arbitrary
internal masses and arbitrary physical momenta. For the regular scalar
4-point function results in terms of 24 or 48 dilogarithms were
presented for real internal masses.  A compact expression with 16
dilogarithms for the general scalar 4-point function with real masses
was provided in \citere{Denner:1991qq}.  The general result for the
regular box integral with real masses in $D$~dimensions was 
calculated in \citere{Fleischer:2003rm}.

In the EWSM and in particular in QCD, many particles are massless, so that
soft and/or collinear singularities appear in the scalar integrals.
Explicit results for soft-divergent scalar 3-point and 4-point integrals
in MR were published in \citere{Beenakker:1988jr}.
Results for scalar box integrals with vanishing internal masses 
were calculated in
\citeres{Bern:1993kr,Duplancic:2000sk,Duplancic:2002dh}.  A
compilation of mass-singular 3-point integrals in DR and MR
can be found in the appendix of
\citere{Dittmaier:2003bc}. Finally, all scalar one-loop integrals
required for calculations in QCD have been compiled in
\citere{Ellis:2007qk} (see also \citere{Denner:2010tr}).

For calculations with unstable particles loop integrals with complex
masses are needed. For 1-, 2-, 3-point functions, the results of
\citere{'tHooft:1978xw} can be directly used. This reference contains
also a result for the regular 4-point function with complex masses in
the physical region in terms of 108 dilogarithms. Based on this
result, a code for the scalar 4-point function with complex masses was
presented in \citere{Nhung:2009pm}.  More compact results for the
4-point function with complex masses, together with results for all
IR-singular 4-point functions both in MR and in
DR were worked out in \citere{Denner:2010tr}.

A novel approach for the computation of one-loop scalar 3- and 4-point
functions that proceeds in terms of quantities driving the algebraic
reduction methods, like the Gram and Cayley determinants, was recently
proposed in \citeres{Guillet:2018cdm,Guillet:2018skp,Guillet:2018fsc}.

Various codes are available for the numerical evaluation of one-loop
scalar integrals:
\begin{myitemize}
\item {\sc LoopTools} \cite{Hahn:1998yk} is a package for the
  evaluation of scalar and tensor one-loop integrals based on the {\sc
    FF} package by van Oldenborgh \cite{vanOldenborgh:1990yc}.  Like the
  {\sc FF} package, {\sc LoopTools} can handle one-loop integrals with
  complex masses up to 3-point functions. For the calculation of
  4-point functions with complex masses the code of
  \citere{Nhung:2009pm} has been integrated.
  UV~singularities are regularized dimensionally, IR singularities are
  regularized either dimensionally or in MR.
\item 
{\sc Golem95C}~\cite{Binoth:2008uq,Cullen:2011kv,Guillet:2013msa}
  is a program for the numerical evaluation of
  scalar and tensor one-loop integrals which supports the use of
  complex masses.
\item {\sc OneLOop}~\cite{vanHameren:2010cp} is a program to evaluate the
  one-loop scalar 1-point, 2-point, 3-point, and 4-point functions, for all
  kinematical configurations relevant for collider physics, and for complex
  internal squared masses based on the formulas of \citere{Nhung:2009pm}.
  It deals with all UV and IR divergences within DR.
  Furthermore, it provides routines to evaluate these
  functions using straightforward numerical integration.
\item {\sc QCDLoop 2.0}~\cite{Carrazza:2016gav}, a C{+}+
  reimplementation and extension of the Fortran library 
  {\sc QCDLoop}~\cite{Ellis:2007qk}, provides all finite and singular
  one-loop integrals in DR with at least one
  vanishing internal mass.  It allows for complex masses based on a
  generalization of the results of \citere{Denner:1991qq} and offers the
  possibility to switch from double to quadruple precision on the fly.
\item \Collier \cite{Denner:2016kdg} is a Fortran library for the numerical
  evaluation of one-loop scalar and tensor integrals appearing in
  perturbative relativistic quantum field theory. While UV singularities
  are treated in DR, soft and collinear
  singularities can be handled in both DR and MR.
  Complex masses are fully supported based on the results of
  \citere{Denner:2010tr} for the 4-point function and of
  \citere{'tHooft:1978xw} for lower-point functions.
\end{myitemize}

\subsubsection{Purely numerical methods}

Besides the methods described above that rely on algebraic reduction
(even if performed numerically) and numerical evaluation of scalar
integrals there are alternative approaches that aim at calculating the
loop integrals directly via numerical integration. While all these
methods are presently of minor importance in actual one-loop
calculations, they may turn out to be useful when generalized to higher
orders. For an overview of the methods for the numerical evaluation of
multi-loop integrals we refer to \citere{Freitas:2016sty}.

As far as the one-loop level is concerned, different numerical
approaches exist.  The method of \citere{Ferroglia:2002mz}
uses the Bernstein--Tkachov theorem
\cite{Bernshtein1972,Tkachov:1996wh} together with a dedicated
treatment of phase-space singularities to evaluate arbitrary one-loop
Feynman-parameter integrals via numerical integration.  This approach
has been used to calculate the two-loop EW corrections to Higgs
production in gluon fusion \cite{Actis:2008ug} and the Higgs decays to
photons and gluons \cite{Passarino:2007fp,Actis:2008ts}.

Other numerical methods have so far only been used for QCD
calculations so that we mention them only briefly.
A general method using subtraction techniques for the numerical
calculation of one-loop QCD matrix elements has been proposed in
\citere{Nagy:2003qn}.
It consists in performing a simultaneous numerical integration over
the phase space and the loop momentum of NLO amplitudes and has been
further investigated by various groups. 
For details and references we refer to the contribution of Seth and
Weinzierl in \citere{Bendavid:2018nar}.  These methods have been
proven to work for multi-jet production in $\Pep\Pem$ annihilation
\cite{Becker:2010ng,Becker:2011vg,Goetz:2014lla}.
A related approach is based on the {\em loop--tree duality}
\cite{Catani:2008xa}, which links loop graphs to tree graphs.  The
loop integrals are transformed into phase-space integrals and are
evaluated numerically along with the phase-space integral over of the
external partons of the process under consideration.  Consequently,
there is only one step of numerical integration involved in the
computation of any particular integrated cross section or
distribution.  For more details and references we
refer to \citere{Capatti:2019ypt} and the contribution of Chachamis
\etal in \citere{Bendavid:2018nar}.

\subsection{Automation and tools}
\label{se:automation}

Compared to QCD, the Lagrangian of the EWSM involves many terms and
gives rise to a large set of Feynman rules, rendering calculations
of EW corrections more tedious.  Therefore, already very early efforts
towards the automation of the calculation of EW corrections have been
undertaken.%
\footnote{We do not restrict the term {\em automation} to the complete
  calculation of observables without user intervention, but use it
  also for generic tools that perform part of this task.}
For the pioneering calculations of EW corrections to
fermion-pair production \cite{Passarino:1978jh,Consoli:1979xw} and
W-pair production \cite{Lemoine:1979pm} in $\Pep\Pem$ annihilation and
Higgs decays \cite{Fleischer:1980ub} the computer program 
{\sc Schoonschip}~\cite{Strubbe:1974vj}
for the symbolic evaluation of algebraic expressions, 
originally developed by Veltman, was used.
This code was later replaced by the still very popular symbolic
manipulation system {\FORM} by Vermaseren
\cite{vanOldenborgh:1990wj,Ruijl:2017dtg}.
  
\myparagraph{(a) Traditional methods} 
The original strategy for the calculation of EW corrections has been
to interfere the one-loop contributions
to the matrix elements with the LO matrix element and to algebraically
reduce the expressions to tensor integrals and coefficient functions or
form factors (see \refse{se:ti_reduction}).  The tensor integrals 
are then numerically reduced with Passarino--Veltman reduction
\cite{Passarino:1978jh} to scalar integrals \cite{'tHooft:1978xw} which in
turn are evaluated numerically based on explicit representations in form
of logarithms and dilogarithms. For simple processes the reduction to
scalar integrals can also be done algebraically. Computer algebra
packages that have been developed along these lines are {\sc FeynCalc}
\cite{Mertig:1990an,Shtabovenko:2016sxi} and {\sc FormCalc}
\cite{Hahn:1998yk,Hahn:2016ebn}. Note, however, that these have been 
continuously improved over time to include state-of-the-art methods.

When considering processes with more external particles it has been
realized that methods calculating the helicity amplitudes are more
efficient than those calculating directly squared amplitudes, since
the computational complexity scales only linearly with the number of
Feynman diagrams instead of quadratically.  While, on the one hand,
methods for the analytic calculation of helicity amplitudes have been
developed (\cf \citeres{Berends:1984gf,Dittmaier:1998nn} and
references therein), on the other hand, tools for the direct numerical
evaluation have become popular.  One of these codes is {\sc
  HELAS}~\cite{Murayama:1992gi}, designed for the calculation of
helicity amplitudes for arbitrary tree-level Feynman diagrams with
sequences of calls of Fortran subroutines.  {\sc HELAS} can be used to
recycle substructures of Feynman diagrams and is, thus, already much
better than a naive diagram-by-diagram calculation.  {\sc HELAS} was
combined with the {\sc Madgraph} \cite{Stelzer:1994ta} Feynman-diagram
generator to generate and calculate tree-level amplitudes in QCD, the
EWSM, and beyond.  Starting from {\sc Madgraph5}
\cite{Alwall:2014hca}, the HELAS library was replaced by {\sc ALOHA}
\cite{deAquino:2011ub}, which automatically generates the {\sc HELAS}
library for any quantum field theory based on Feynman rules provided
in the {\it Universal FeynRules Output (UFO)} format
\cite{Degrande:2011ua}.  This procedure is, however, still based on
Feynman diagrams so that the complexity increases factorially with
increasing number of external particles.

\myparagraph{(b) Feynman diagram generation}

In the traditional methods, the calculation of matrix elements starts
with the generation of Feynman diagrams. To this end, various programs
are available, including:
\begin{myitemize}
\item {\sc FeynArts~3} \cite{Hahn:2000kx}, an extension of {\sc
    FeynArts~1} \cite{Kublbeck:1990xc}, is a Mathematica package for the
  generation and visualization of Feynman diagrams and amplitudes.  
  It generates diagrams at generic or specific levels for particle
  insertions, employs user-definable model files,
  supports supersymmetric models, and provides publication-quality
  Feynman diagrams in Postscript or LATEX.
\item The Fortran program {\sc QGRAF} \cite{Nogueira:1991ex} generates
  Feynman diagrams and represents them by symbolic expressions.
A graphical representation can be obtained with the Feynman diagram
analyzer {\sc DIANA} \cite{Tentyukov:1999is}, which also translates
the output of {\sc QGRAF} into a source code for analytical or
numerical evaluations.
\end{myitemize}
The Mathematica package {\sc FeynRules}
\cite{Christensen:2008py,Alloul:2013bka} derives the Feynman rules for
a particle physics model from a list of fields and parameters, and a
corresponding Lagrangian. The output can be given in the {\it Universal
FeynRules Output (UFO)} format \cite{Degrande:2011ua}, a generic form
suitable for the translation to any Feynman diagram calculation
program.  Translation interfaces exist for {\sc CalcHEP/CompHEP}
\cite{Belyaev:2012qa}, {\sc Feyn\-Arts} \cite{Hahn:2000kx}, \Gosam
\cite{Cullen:2011ac,Cullen:2014yla}, {\sc MadGraph/MadEvent}
\cite{Alwall:2011uj}, {\sc Sherpa} \cite{Gleisberg:2008ta}, 
and {\sc Whizard} \cite{Moretti:2001zz,Kilian:2007gr,Christensen:2010wz}. A
tool similar to {\sc FeynRules} is the program {\sc LanHEP}
\cite{Semenov:2014rea}.  Starting from a Lagrangian it derives Feynman
rules and writes them in terms of physical fields and independent
parameters in the form of {\sc CompHep} \cite{Boos:2004kh} or {\sc
  CalcHEP} \cite {Belyaev:2012qa} model files, or in the {\sc
  Feyn\-Arts} \cite{Hahn:2000kx} or UFO formats
\cite{Degrande:2011ua}. The program can also generate one-loop
counterterms in the {\sc FeynArts} format.

\myparagraph{(c) Modern recursive methods} Berends and Giele
\cite{Berends:1987me} had shown that tree-level gluon amplitudes can
be calculated via recursion relations with a computational complexity
growing only exponentially. This recursive algorithm was generalized
to calculate tree-level matrix elements in the complete SM in the
codes {\sc ALPHA} \cite{Caravaglios:1995cd} and {\sc HELAC}
\cite{Kanaki:2000ey} based on Dyson--Schwinger equations
\cite{Dyson:1949ha,Schwinger:1951ex,Schwinger:1951hq}.

The calculation of one-loop amplitudes, in particular in QCD, has been
boosted with the advent of generalized unitarity techniques
\cite{Bern:1994zx,Bern:1994cg,Britto:2004nc,Ellis:2007br,Ellis:2008ir}
some ten years ago, which allow for expressing loop amplitudes in
terms of (generalized) tree-level amplitudes.  
The related OPP method (\cf \refse{se:opp}) has been more or less
directly implemented in generic codes. In parallel, powerful
automated tools have been developed based on improvements of
traditional techniques.
Van Hameren \cite{vanHameren:2009vq} proposed
to generalize the recursive construction of tree-level amplitudes to
the coefficient functions of tensor integrals in one-loop
calculations.

Presently there is a variety of tools available for the automated
calculation of NLO QCD corrections.
For generic processes the following codes exist:
\begin{myitemize}
\item
In \Gosam
\cite{Cullen:2011ac,Cullen:2014yla} the amplitudes are constructed
from Feynman diagrams and evaluated using $D$-dimensional reduction at
the integrand level,
based on {\sc Ninja} and {\sc Samurai}, or alternatively with tensor reduction using
the formalism described in \citere{Binoth:2005ff}, based on {\sc Golem95C}.
\item
{\sc HELAC-1LOOP}~\cite{vanHameren:2009dr}, which is part of the 
{\sc HELAC-NLO} Monte Carlo framework \cite{Bevilacqua:2011xh}, uses
the OPP method \cite{Ossola:2006us,Ossola:2008xq,Draggiotis:2009yb}
for the reduction of one-loop amplitudes at the integrand level,
specifically the implementation {\sc CutTools} \cite{Ossola:2007ax}.
\item 
  The OpenLoops generator \cite{Buccioni:2019sur} computes matrix
  elements
  by applying a numerical recursion analogous to the one of
  \citere{vanHameren:2009vq} to colour-stripped cut-opened loop
  diagrams. In OpenLoops1 \cite{Cascioli:2011va} the reduction to scalar
  integrals is done at the integral or integrand level using external
  tools, while OpenLoops2 \cite{Buccioni:2019sur} implements the
  {\it on-the-fly reduction} technique of \citere{Buccioni:2017yxi}.  
\item The algorithm developed for OpenLoops1 is also adopted in the
  code \MGNLO \cite{Alwall:2014hca}.
\item
The matrix-element generator \Recola \cite{Actis:2012qn,Actis:2016mpe}
generates arbitrary one-loop amplitudes 
based on an implementation of the recursive algorithm of
\citere{vanHameren:2009vq} for amplitudes with colour structures.
\end{myitemize}
Other codes are limited to specific process classes:
\begin{myitemize}
\item
The program {\sc Blackhat}~\cite{Berger:2008sj} implements the
unitarity method and 
OS recursion to construct one-loop
amplitudes in a numerical approach. As input, it uses compact analytic
formulae for tree-level amplitudes for 4-dimensional helicity
states. 
The program has been used to calculate NLO QCD corrections to
vector-boson plus jets production with up to 5~jets
\cite{Bern:2013gka,Anger:2017glm}.
\item {\sc NJet}~\cite{Badger:2012pg} is a library for multi-parton
  one-loop matrix elements in massless QCD. It is based on the
  generalized unitarity program {\sc NGluon}~\cite{Badger:2010nx} and
  uses {\sc QCDLoop} \cite{Ellis:2007qk} and {\sc FF}
  \cite{vanOldenborgh:1990yc} for evaluating scalar integrals.
  It was used to calculate the NLO QCD corrections to 5-jet production
  at the LHC \cite{Badger:2013yda}.
\end{myitemize}

\myparagraph{(d) Automated tools for NLO EW corrections}
The automation of the calculation of EW corrections is presently still
ongoing.  The methods are basically the same as for QCD corrections
with some necessary modifications. The complexity increases due to the
mixing of QCD and EW corrections, the increasing number of
contributions, and the presence of more and very different mass
scales.  Additional complications arise from the chiral structure of
the EW interactions and the instability of many SM particles. Since
the SM particles decay via EW interactions, the instability can be
counted as an NLO EW effect (\cf\refse{se:unstable}).  

The following tools exist for the calculation of NLO EW corrections:
\begin{myitemize}
\item {\sc FormCalc} \cite{Hahn:1998yk,Hahn:2016ebn} is a Mathematica
  package for the calculation of tree-level and one-loop Feynman
  diagrams. It is based on the conventional approach and speeds up the
  calculation by the automatic isolation of subexpressions.  {\sc
    FormCalc} has been used for a large number of applications within
  but notably also beyond the SM. 
  The emphasis of {\sc FormCalc} is on flexibility and not so much on
  the calculation of high-multiplicity processes.
  
\item {\Gosam} \cite{Cullen:2011ac,Cullen:2014yla} was extended to
  allow for the calculation of NLO EW corrections \cite{Chiesa:2015mya}.
  It is based on algebraic generation of $D$-dimensional integrals derived
  from Feynman diagrams generated with {\sc QGRAF} \cite{Nogueira:1991ex}
  and {\FORM} \cite{Ruijl:2017dtg}. 
  The amplitudes are evaluated using $D$-dimensional
  reduction at integrand level
  \cite{Ossola:2006us,Mastrolia:2008jb,Ellis:2008ir}, which is available
  through two different reduction procedures and libraries, {\sc Samurai}
  \cite{Mastrolia:2010nb} and {\sc Ninja} \cite{Peraro:2014cba}, and {\sc
    Golem95C} 
 \cite{Binoth:2008uq,Cullen:2011kv,Guillet:2013msa} is
  used as the default rescue system.
  \Gosam proceeds purely algebraically, grouping diagrams with
  identical loop topologies and performing colour projections at the
  symbolic level, and writing optimized numerical code with the help
  of \FORM. While this purely analytical route for code generation is
  comparably slow, it offers great flexibility in the manipulation of
  the output.  The scalar one-loop integrals can be evaluated using
  {\sc OneLOop} \cite{vanHameren:2010cp}, {\sc Golem95C}
  \cite{Binoth:2008uq,Cullen:2011kv}, or {\sc QCDLoop}
  \cite{Ellis:2007qk}.
  
\item {\sc MadLoop} \cite{Alwall:2014hca} has demonstrated that it can
  calculate the EW corrections to arbitrary $2\to2$ and $2\to3$
  processes in the SM within the framework of
  \MGNLO~\cite{Frixione:2015zaa,Frederix:2018nkq}, but its technical
  capabilities are not restricted to those multiplicities.  It is
  capable of calculating EW corrections also in models beyond the SM
  and is publicly available. More details on the \MGNLO implementation
  can be found in \refse{se:Madgraph}.
  
\item \NLOX \cite{Honeywell:2018fcl} provides 
  QCD and EW one-loop corrections to SM processes at the squared-amplitude
  level for all possible coupling-power combinations in the strong
  and EW couplings and for processes with up to six external
  particles.  It is based on a Feynman-diagrammatic approach,
  utilizing {\sc QGRAF}, {\sc FORM}, and {\sc Python}, to algebraically
  generate C++ code for the Born and virtual contributions. The
  one-loop tensor coefficients are calculated recursively at run time
  through an integrated C++ library based on the tensor-reduction
  methods of \citeres{Passarino:1978jh,Denner:2005nn}. The scalar
  one-loop integrals are evaluated by either using {\sc OneLOop}
  \cite{vanHameren:2010cp} or {\sc QCDLoop} \cite{Ellis:2007qk}.  The
  code has been used to calculate one-loop corrections to selected
  processes such as QCD and EW corrections to $\PZ + \Pb$ production
  \cite{Figueroa:2018chn}.
  
\item With {\Openloops} \cite{Cascioli:2011va}, EW and QCD corrections
  have been calculated for vector-boson plus 1,2,3-jet production
  \cite{Kallweit:2014xda,Kallweit:2015dum},
  $\Pp\Pp\to\Pep\mu^-\nu_\Pe\nu_{\bar\mu}/\Pep\Pem\nu\bar{\nu}+X$
  \cite{Kallweit:2017khh}, 
and $\Pp\Pp\to\mu^+\mu^-\Pep\nu_\Pe\Pj\Pj+X$
\cite{Denner:2019tmn}.  Moreover, the NLO QCD and EW corrections to
$\Pp\Pp\to\PH\Pl\Pl(\Pj)/\PH\nu\Pl(\Pj)+X$ \cite{Granata:2017iod} have
been combined with QCD+QED parton shower via the {\sc POWHEG BOX RES}
framework \cite{Jezo:2015aia}, an extension of the {\sc POWHEG BOX}
\cite{Alioli:2010xd},
and the processes $\Pp\Pp\to\Pt\bar\Pt+0,1\,$jets have been studied with
multi-jet merging at NLO QCD and EW~\cite{Gutschow:2018tuk}.
OpenLoops process libraries for many processes are publicly available.
More details on \Openloops are given in \refse{se:Openloops}.
  
\item With the code \Recola \cite{Actis:2012qn,Actis:2016mpe} several
  cutting-edge calculations have been performed such as the EW
  corrections to the processes
  $\Pp\Pp\to\Pep\nu_\Pe\mu^-\bar{\nu}_\mu\Pb\bar\Pb+X$
  \cite{Denner:2016jyo},
  $\Pp\Pp\to\Pep\nu_\Pe\mu^-\bar{\nu}_\mu\Pb\bar\Pb\PH+X$
  \cite{Denner:2016wet},
  $\Pp\Pp\to\mu^\pm\nu_\mu\Pe^\pm\nu_\Pe\Pj\Pj+X$
  \cite{Biedermann:2016yds,Biedermann:2017bss,Chiesa:2019ulk},
  $\Pp\Pp\to\mu^+\mu^-\Pep\nu_\Pe\Pj\Pj+X$ \cite{Denner:2019tmn}, and
  $\Pp\Pp\to\Pl_1^-\bar{\nu}_{\Pl_1}\Pl_2^-\bar{\nu}_{\Pl_2}\Pl_3^+{\nu}_{\Pl_3}+X$
  \cite{Schonherr:2018jva}.  \Recola is fully public and allows for
  the calculation of EW corrections also in models beyond the SM for
  which \Recola model files are available
  \cite{Denner:2016etu,Denner:2018opp}. More details on \Recola are
  provided in \refse{se:Recola}.
\end{myitemize}
A performance comparison of the different one-loop providers was
presented in \citere{Degrande:2018neu}.

After a sketch of the conventional approach for the calculation of
one-loop matrix elements, in the following we give some more details
on the implementation of the general one-loop matrix element providers
\Recola, \Openloops, and \MGNLO, using widely the notation of the
original literature. We note that many NLO EW
calculations, such as the first (EW) NLO calculation for a $2\to4$
process, $\Pep\Pem\to4\,$fermions \cite{Denner:2005es,Denner:2005fg},
have been carried out with non-public semi-automated tools.

\subsubsection{Conventional approach}
\label{se:conv_ampl_red}

We begin by describing the conventional technique for the calculation
of one-loop matrix elements. More details for processes without
coloured particles can be found in \citere{Denner:1991kt}; for a
more detailed discussion of the treatment of colour we refer to
\citeres{Bredenstein:2010rs,Denner:2012yc}. Upon inserting the Lorentz
decompositions \refeqf{eq:TI_decomposition} of the one-loop tensor
integrals, an arbitrary one-loop matrix element with $E$ external
particles with momenta $p_i$ and polarizations
$\la_i$, $i=1,\ldots,E$, \refeq{eq:M_oneloop} can be written as
\beq
\de\cM_{\la_1\ldots\la_E}(\{p_k\}) 
=\sum_a \cC_a
\de\cA_{a,\la_1\ldots\la_E}(\{p_k\}) 
=\sum_i \sum_a \cC_a
\cA_{i,\la_1\ldots\la_E}(\{p_k\}) F_{i,a}(\{p_k p_l\}),
\label{eq:M_oneloop_standard}
\eeq
where $\de\cA_{a}$ are colour-stripped amplitudes and the colour
structures $\cC_a$ form a basis in colour space for the process under
consideration (colour indices are suppressed).  The {\em standard
  matrix elements} $\cA_i$ carry the dependence on polarizations of
the external particles. They are simple, purely kinematical,
model-independent functions of the external momenta.  The
Lorentz-invariant form factors $F_{i,a}$, on the other hand, are
independent of the polarizations and depend only on scalar products of
external momenta. They are linear functions of the scalar coefficients
of the tensor integrals $T^N$ and the counterterms and contain all
dynamical information.  The tree-level amplitude
$\cM_{0,\la_1\ldots\la_E}(\{p_k\})$ has the same structure as
\refeq{eq:M_oneloop_standard}, however, with typically fewer
non-vanishing and much simpler form factors without integrals,
\beq
\cM_{0,\la_1\ldots\la_E}(\{p_k\})
=\sum_a \cC_a
\cA_{0,a,\la_1\ldots\la_E}(\{p_k\})
=\sum_i \sum_a \cC_a
\cA_{i,\la_1\ldots\la_E}(\{p_k\}) F_{0,i,a}(\{p_k p_l\}).
\label{eq:M_0_standard}
\eeq

\begin{sloppypar}
The LO scattering probability density $\mathcal{W}_0$ and the corresponding
virtual one-loop correction 
$\de\mathcal{W}$ are obtained after summing/averaging over
polarizations and colours,
\begin{align}
\mathcal{W}_0={}&\frac{1}{N_{\rs,\rc}}\sum_{\text{pol,col}}|\cM_{0}^2|
=\sumbar_{\text{pol,col}}|\cM_{0}^2|
=\sumbar_{\text{pol}}\sum_{a,b} \cA_{0,a}^*\cA_{0,b}\sumbar_{\text{col}}\cC^*_a\cC_b, \nn\\
\de\mathcal{W}={}&\frac{1}{N_{\rs,\rc}}\sum_{\text{pol,col}}2\Re\left[\cM_{0}^*\de\cM\right]
=\sumbar_{\text{pol,col}}2\Re\left[\cM_{0}^*\de\cM\right]
=\sumbar_{\text{pol}}\sum_{a,b} 2\Re\left[\cA_{0,a}^*\de\cA_{b}\right]\sumbar_{\text{col}}\cC^*_a\cC_b,
\label{eq:M2_ol}
\end{align}
where $N_{\rs,\rc}$ is the number of combinations of colour and spin degrees of
freedom in the initial state. The corresponding average is denoted by
a bar over the summation symbol. The sum over external colours can be
carried out once and for all at the level of the colour basis,
\beq
\cK_{ab}=\sumbar_{\text{col}}\cC^*_a\cC_b.
\eeq
The representation \refeqf{eq:M_oneloop_standard}
allows us to write the one-loop correction to the scattering
probability in the form
\beq
\de\mathcal{W}=2\Re \sum_{i,b}\left(\sumbar_{\text{pol}}\sum_{a}\cK_{ab}
\cA_{0,a}^* \cA_i\right)F_{i,b},
\label{eq:m2_ol}
\eeq
\ie the one-loop contribution is a linear function of the loop integrals.
The colour--Born interference terms 
$\sumbar_{\text{pol}}\sum_{a}{\cK}_{ab} \cA_{0,a}^*\cA_i$ can be
precomputed without knowledge of the one-loop contributions. For the
calculation of the polarization sum only the basis of standard matrix
elements has to be known. Thus, for the calculation of polarization-summed
squared matrix elements the expensive evaluation of the form factors
$F_{i,b}$ has to be performed only once per phase-space point, while the
polarization and colour sums inside the brackets of \refeq{eq:m2_ol} can be
performed independently before the actual evaluation of the loop integrals.
\end{sloppypar}

Individual Feynman diagrams~$d$ that do not involve quartic gluon
couplings consist of a single colour-stripped amplitude $\cA_a^{(d)}$
multiplied by a single colour factor $\cC_a^{(d)}$, \ie the colour structure
factorizes. Diagrams with quartic gluon couplings can be decomposed in
subamplitudes with factorizing colour part. Therefore, in a
diagram-by-diagram approach~\cite{Bredenstein:2010rs,Denner:2012yc},
also the sum over~$b$ in \refeq{eq:m2_ol} is absent in the
contributions of (sub)diagrams, rendering the
colour treatment even more efficient.
Writing the contribution of~$d$ to the
form factor as $F_{i,b}^{(d)}=\cC_b^{(d)} f_{i}^{(d)}$
and likewise the contribution of some Born diagram~$d_0$ to the
LO form factor as $F_{0,i,b}^{(d_0)}=\cC_{0,b}^{(d_0)} f_{0,i}^{(d_0)}$,
the contribution of~$d$ to the probability $\de\mathcal{W}$ can be
arranged as follows,
\beq
\de\mathcal{W}^{(d)} = 2\Re \sum_i f_{i}^{(d)}
\sum_{d_0} 
\left( \sum_{a,b} (\cC_{0,a}^{(d_0)})^* \cK_{ab} \cC_{b}^{(d)} \right)
\left[ \sum_j \sumbar_{\text{pol}} (f_{0,j}^{(d_0)})^* \cA_j^* \cA_i \right].
\label{eq:m2_ol2}
\eeq
The colour sum within round brackets neither depends on phase space,
nor on polarizations, nor on the basis of standard matrix elements (indices
$i,j$), so that this sum can be precomputed even before generating
phase-space points.  The sum over $j$ and polarizations neither depends
on colour, nor on the specific (sub)diagram~$d$ and can, thus, be
precomputed after phase-space generation before summing over the
(sub)diagrams~$d$. These optimizations speed up the numerical
evaluation considerably, since only one set of CPU-expensive
form factors $f_{i}^{(d)}$ has to be calculated for all colour and
polarization structures, which are in this sense treated simultaneously.

The form factors $F_{i,b}$ or $f_{i}^{(d)}$ are reduced to scalar one-loop
integrals either algebraically or numerically. In the traditional methods,
$\de\cM$ is derived from Feynman diagrams, and the resulting expression is
simplified algebraically. Finally, the so-obtained algebraic expressions
are evaluated numerically.  

For loop-induced processes, \ie those with vanishing $\cM_{0}$,
the  scattering probability density is obtained as
\beq\label{eq:M2_li}
\mathcal{W}=\frac{1}{N_{\rs,\rc}}\sum_{\text{pol,col}}|\de\cM|^2
=\sumbar_{\text{pol,col}}|\de\cM|^2.
\eeq
This quantity depends quadratically on the loop amplitude, implying that
optimizations somewhat different from those of \refeq{eq:m2_ol2} may be
required.  Notably, tensor-integral reduction is superior to the OPP method
in this case (see \refse{se:Madgraph}).

\subsubsection{\Recola}
\label{se:Recola}

Modern methods for the evaluation of one-loop corrections use
recursive techniques. These are consistently applied in \Recola
\cite{Actis:2012qn,Actis:2016mpe}. In order to explain the underlying
algorithm, we start with the recursive construction of tree-level
amplitudes \cite{Actis:2012qn}. The basic building block is the
{\em tree-level off-shell current},
\beq
w_\be(P,\{n\})
\ =\ \vcenter{\hbox{
\raisebox{-18pt}{\includegraphics[page=3]{diagrams.pdf}}}} ,
\label{current}
\eeq
which is the sum of all Feynman diagrams with an external off-shell
leg $P$ and $n$ external 
OS legs with corresponding wave
functions. The dot indicates the off-shell part of the current, and
the corresponding leg includes the propagator for the particle $P$ for
$n\ge2$. For $n=1$, the off-shell current coincides with the wave
function (spinor or polarization vector) of the particle $P$. The
multi-index $\beta$ represents the Lorentz index for a vector and the
spinor index for a spinor as well as possible colour indices.

{\sloppypar In a theory with tri- and quadri-linear couplings only,
  like renormalizable quantum field theories, the off-shell currents
  can be constructed using the Dyson--Schwinger equations
  \cite{Dyson:1949ha,Schwinger:1951ex,Schwinger:1951hq} recursively,
\beq
\raisebox{-18pt}{\includegraphics[page=3]{diagrams.pdf}}
\
=
\
\sum_{\substack{\{i\},\{j\}\\ i+j=n}}\;
\sum_{P_i,P_j}\;
\raisebox{-26pt}{\includegraphics[page=4]{diagrams.pdf}}
\
+
\;
\sum_{\substack{\{i\},\{j\},\{k\} \\i+j+k=n}}\;
\sum_{P_{\!i},P_{\!j},P_{\!k}}\;
\raisebox{-36pt}{\includegraphics[page=5]{diagrams.pdf}}
\;.
\label{eq:recursive}
\eeq
Each term in the sums on the r.h.s.\ is obtained by multiplication of the generating
currents $w_\ga(P_{i},\{i\})$, $w_\de(P_{j},\{j\})$ [and
$w_\al(P_{k},\{k\})$]
with the interaction vertex, indicated by the small box, and the
propagator of $P$, marked by the thick line. The sums run over all
partitions of the set of $n$ external particles into subsets $\{i\}$,
$\{j\}$ [and $\{k\}$] with $i+j\:({}+k)=n$.  More explicitly, for
the terms involving triple vertices this amounts to}
\beq
w_\be(P,\{n\})\big|_{\mathrm{triple\,vertex}} 
= \sum_{\substack{\{i\},\{j\}\\ i+j=n}}\;
\sum_{P_i,P_j}\;
w_\ga(P_i,\{i\})\,w_\de(P_j,\{j\})\,\mathcal{P}^P_{\be\al}(p_P,m_P)
\,\mathcal{V}_{PP_iP_j}^{\al\ga\de}(p_P,p_i,p_j),
\label{eq:recursion_step_tree}
\eeq
where $\mathcal{V}_{PP_iP_j}^{\al\ga\de}(p_P,p_i,p_j)$ describes the vertex
connecting the lines $P$, $P_i$, $P_j$ with momenta $p_P$, $p_i$, $p_j$,
and $\mathcal{P}^P(p_P,m_P)$ is the propagator of the particle $P$. Since
all quantities depend only on the momenta, colours, and helicities of the
external particles, the recursion can be implemented as simple matrix
multiplication on numerical matrices.  Working with off-shell currents
rather than Feynman diagrams allows one to avoid recomputing identical
subgraphs contributing to different diagrams.  The generalization to
theories with elementary couplings of more than four fields is
straightforward.  The recursive evaluation of the currents begins with the
external currents ($n=1$), which, for colourless particles of a given
polarization $\lambda$ and momentum $p$, are given by the corresponding
wave functions: \vspace{-1ex}
\begin{equation}
\raisebox{-17pt}{\includegraphics[page=6]{diagrams.pdf}}
= \,
u_\lambda(p),
\qquad
\raisebox{-17pt}{\includegraphics[page=7]{diagrams.pdf}}
= \,
\bar{\varv}_\lambda(p),
\qquad
\raisebox{-17pt}{\includegraphics[page=8]{diagrams.pdf}}
= \;
\veps_\lambda(p),
\qquad
\raisebox{-17pt}{\includegraphics[page=9]{diagrams.pdf}}
= \;
1\,.
\end{equation}
For coloured external particles the colour indices have to be added.

Finally, the $S$-matrix element with $n+1$ external particles is obtained by 
multiplying the off-shell current  $w_\be(P,\{n\})$ with the inverse
propagator and the appropriate wave function of particle~$P$,
\beq
\cM \;=
\raisebox{-18pt}{\includegraphics[page=3]{diagrams.pdf}}
\times
\left(\raisebox{-13pt}{\includegraphics[page=10]{diagrams.pdf}}\right)^{-1}_{\be\ga}
\times
\raisebox{-13pt}{\includegraphics[page=11]{diagrams.pdf}}.
\label{eq:last step}
\eeq
In the code, the truncation of the external line is performed by
discarding the propagator of the $(n+1)$th external line,
since the propagator for an external OS line does not exist.  
To construct $S$-matrix elements for physical processes, incoming and
outgoing particles can be interchanged upon using crossing symmetry.

Following an idea of van Hameren \cite{vanHameren:2009vq} the
recursive calculation of tree amplitudes can be generalized to the
calculation of the coefficients
$\hat{c}^{(j,r_j,N_j)}_{\mu_1\ldots\mu_{r_j}}$ of the tensor integrals
in \refeq{eq:M_oneloop_R2} at one loop.  Since the numerical
calculation proceeds in four dimensions,
the rational terms of type $R_2$~\cite{Ossola:2008xq}
have to be supplemented. While the integration momenta in the
numerator of the loop integrals are 4-dimensional, those in the
propagator denominators are $D$-dimensional. As in \refse{se:opp} we
indicate $4$-dimensional loop momenta with carets.
Upon cutting one of the {\em loop lines}, each loop diagram with $E$
external lines can be mapped to a tree-level diagram with $E+2$
external legs.  The correspondence between loop and tree-level
diagrams can be made unique via defining precisely which loop line has
to be cut \cite{Actis:2012qn}. For the resulting tree-level amplitudes
with two external loop lines a recursion relation similar to
\refeq{eq:recursive} can be formulated.  One of the loop lines is
identified with the off-shell line $P$, the second loop line is part
of the 
OS lines $n$ but marked as a loop line. The basic
building blocks involving a loop line are called {\em loop off-shell
currents}. In contrast to the tree-level off-shell currents they are
not pure numerical spinors or vectors but depend on the integration
momentum and on the momenta and masses of the propagators that enter
their construction.  In the 't~Hooft--Feynman gauge the product of the
vertex and the propagator in the generalization of
\refeq{eq:recursion_step_tree} has the following dependence on the
integration momentum $q$:
\beq
\mathcal{P}^P_{\be\al}(q,\qfour,p_P,m_P)
\,\mathcal{V}_{P P_i P_j}^{\al\ga\de}(\qfour,p_P,p_i,p_j) =
\frac{a^{\ga\de}_{\be,\mu}(p_i,p_j) \qfour^\mu + b^{\ga\de}_{\be}(p_i,p_j)}{(q+p_P)^2-m_P^2}.
\label{eq:loop_vertexprop}
\eeq
After $l$ recursion steps, the loop off-shell current takes the form
\beq
w_{\be}^{(l)}(P,\{n\},q,\qfour,\{p_h\},\{m_h\}) =
\sum_{k=0}^l
\frac{d^{(k,l)}_{\be,\mu_1\cdots\mu_k}(P,\{n\},\{p_h\},\{m_h\})\qfour^{\mu_1}\cdots \qfour^{\mu_k}}
     {\prod_{h={1}}^l[(q+p_h)^2-m_h^2]}.
\label{eq:wl}
\eeq
A generalization of the algorithm to arbitrary gauges or
non-renormalizable theories is straightforward and amounts to
including terms with higher powers of integration momenta in
\refeqs{eq:loop_vertexprop} and \refeqf{eq:wl}.

While the loop off-shell currents obey a recursion relation like
\refeq{eq:recursive}, this cannot be used straightforwardly to compute
them numerically since they depend on the loop momentum. However, the
recursion relation can be turned into a recursion relation for the
coefficients $d^{(k,l)}_{\be,\mu_1\cdots\mu_k}$, which can be
evaluated numerically. At the end of the recursion, the coefficients
$d^{(k,l)}_{\be,\mu_1\cdots\mu_k}$ determine the tensor coefficients
$\hat c^{(j,r_j,N_j)}_{\mu_1\ldots\mu_{r_j}}$ in \refeq{eq:M_oneloop_R2},
where each set $\{p_h,m_h\}$ corresponds to an index $j$. The loops
are closed by forming traces of the free spinor or Lorentz indices of
the two corresponding loop lines.

In \Recola, the colour matrices are included in the recursive
numerical treatment, \ie\ \Recola directly calculates coloured
amplitudes. In order to optimize the colour treatment, \Recola uses
the colour-flow representation
\cite{tHooft:1973alw,Kanaki:2000ms,Maltoni:2002mq}, where the
conventional 8 gluon fields $G_{\mu}^{A}$ are replaced by a traceless
$3\times 3$ matrix $({\cal G}_{\mu})^{i}_{j} =
\frac{1}{\sqrt{2}}\,G_{\mu}^{A}(\lambda^{A})^{i}_{j}$.  Quarks and
antiquarks carry the usual colour index $i=1,{2},3$, while gluons get
a pair of indices $i,j=1,{2},3$. The Gell-Mann matrices
{$\lambda^{A}$} and the structure constants in the Feynman rules are
thus substituted by combinations of Kronecker deltas. Moreover,
instead of colour-dressed amplitudes, \Recola introduces {\em
  structure-dressed} amplitudes, where each current gets an explicit
{\em colour structure} corresponding to a product of Kronecker deltas
in colour indices (for more details see \citere{Actis:2012qn}).
In the code, the colour structures are
represented by integer numbers in a binary notation. Basing the
reduction on these colour structures instead of using directly
colour-dressed amplitudes reduces the number of currents and renders
the recursive construction more efficient.

Rational terms and counterterms are included via additional tree-level
Feynman rules
\cite{Ossola:2008xq,Draggiotis:2009yb,Garzelli:2009is,Shao:2011tg}.
For the calculation of the tensor integrals \Recola uses the \Collier
library \cite{Denner:2016kdg}. No other external libraries are
required.  

\Recola calculates loop-induced processes based on the same recursive
algorithm for the one-loop matrix elements~\cite{Actis:2016mpe}.

The enhanced version \RecolaTwo \cite{Denner:2017wsf} allows for the
calculation of EW corrections in models beyond the SM for which
\Recola model files are available
\cite{Denner:2016etu,Denner:2018opp}. Such model files, which include
counterterms and rational terms, can be generated in a semi-automatic
way using the tool {\sc Rept1l} \cite{Denner:2017vms}.

\subsubsection{\Openloops}
\label{se:Openloops}

In {\Openloops} \cite{Cascioli:2011va} one-loop amplitudes are
generated via a numerical algorithm for the recursive construction of
Feynman diagrams. Colour sums are performed by exploiting the
factorization of individual (sub)diagrams $d$ into colour factors and
colour-stripped amplitudes discussed at the end of \refse{se:conv_ampl_red},
\beq
\cM^{(d)} = \cC^{(d)} \cA^{(d)}.
\eeq
The algebraic reduction of the colour factors to a standard basis
$\{\cC_a\}$ permits one 
to encode all colour sums in the matrix
${\cK}_{ab}=\sumbar_{\mathrm{col}}\cC^*_a\cC_b$, which needs to be
computed only once per process.
Tree amplitudes in OpenLoops are computed through a numerical
recursion of Dyson--Schwinger type [see \refeq{eq:recursion_step_tree}]
for individual colour-stripped tree diagrams.
In contrast to \Recola, instead of
off-shell currents, individual topologies are constructed.


In {\Openloops} the construction of one-loop diagrams proceeds as
follows: A colour-stripped $N$-point one-loop diagram is an ordered
set $\cI_N=\{i_1,\ldots,i_N\}$ of $N$ subtrees $i_n$, called segments,
connected by loop propagators:
\beq
\de\cA^{(d)} = \int\rd^D q \frac{\cN(\cI_N;\qfour)}{\PD_0\PD_1\ldots
  \PD_{N-1}} =\;
\vcenter{\hbox{{\includegraphics[page=17]{diagrams.pdf}}}} .
\eeq
The ordered set $\{i_1,\ldots,i_N\}$ defines the topology of this
particular Feynman diagram. The denominators $\PD_i$ are the ones of
\refeq{eq:D0Di}.  All other contributions from loop propagators,
vertices, and external subtrees are summarized in the numerator, which
(in the 't~Hooft--Feynman gauge) is a polynomial of degree $R\le N$ in
the 4-dimensional part $\qfour$ of the loop momentum,
\beq
\cN(\cI_N;\qfour)=\sum_{r=0}^R\cN_{\mu_1\ldots\mu_r}(\cI_N)\,\qfour^{\mu_1}\cdots
  \qfour^{\mu_r}.
\label{eq:loop-numerator}
\eeq
Ambiguities related to possible shifts in the loop momentum
are eliminated by setting $p_0=0$, singling
out the $\PD_0$ propagator.  Cutting the loop at this propagator and
removing the denominators results in the {\em open loop}
\beq
\cN_\al^\be(\cI_N;\qfour)
=\sum_{r=0}^R\cN^\be_{\mu_1\ldots\mu_r;\al}(\cI_N)\,\qfour^{\mu_1}\cdots
  \qfour^{\mu_r},
\eeq
where the indices $\al,\be$ belong to the ends of the cut line.
The original polynomial \refeqf{eq:loop-numerator} is obtained upon
taking the trace with respect to $\al,\be$.   Open loops are
constructed with an algorithm similar to the numerical recursion of
\citere{vanHameren:2009vq}.
 While the one-loop off-shell currents as used in
\Recola correspond to complete amplitudes, the open loops correspond
to colour-stripped Feynman diagrams. The efficiency of the open-loop
recursion is increased by using parts of pre-computed open loops for
the construction of more involved open loops, whenever the same open
loop occurs in more than one diagram.  Finally, the
contribution of a one-loop diagram to the unpolarized transition
probability results in a linear combination
\beq
\de\mathcal{W}^{(d)}=
\Re\left[\sum_{r=0}^R\de\mathcal{W}^{(d)}_{\mu_1\ldots\mu_r} T^{N,\mu_1\ldots\mu_r}\right]
\eeq
with the helicity and colour-summed coefficients
\beq
\de\mathcal{W}^{(d)}_{\mu_1\ldots\mu_r} =
2\sumbar_{\text{hel}}\left(\sumbar_{\text{col}}\cM_{0}^*\cC^{(d)}\right)
\cN_{\mu_1\ldots\mu_r}(\cI_N).
\eeq

The open-loop algorithm can be interfaced to both OPP and
tensor-integral reduction. For tensor-integral reduction the
\Collier~\cite{Denner:2016kdg} library is used, while OPP reduction is
performed with {\sc CutTools}~\cite{Ossola:2007ax}. 
The $R_2$ rational
terms are restored via process-independent counterterms
\cite{Draggiotis:2009yb}.

More recently, \Openloops2 was introduced
\cite{Buccioni:2017yxi,Buccioni:2019sur}.  It implements a new method,
dubbed {\em on-the-fly reduction,} which unifies the construction of
loop amplitudes and their reduction in a single recursive algorithm.
This is achieved by factorizing the loop amplitude into segments and
interleaving segment multiplications with reduction operations at the
integrand level. The latter are based on the algorithm by del Aguila
and Pittau \cite{delAguila:2004nf} described in
\refse{se:integrand_reduction}.  
Using this method, objects with
tensor rank higher than two are avoided throughout the calculation.
While the complexity due to the tensor rank is kept low, the creation
of pinched topologies potentially leads to a huge proliferation of
open loops to be processed. This problem is solved by on-the-fly
merging of pinched open loops with open loops of the same topology and
the same undressed segments. Here, a segment ${\cal S}_i$ consists of
a subtree that involves a certain set of external particles connected
to the $i$th loop vertex and the adjacent loop propagator $D_i$. An
undressed segment refers to a segment that has not yet been used in
the calculation of the numerator.  The final rank-0 and rank-1 tensor
integrals are reduced to scalar integrals with $N\le4$ using OPP
reduction for $N\ge5$, integrand-level identities for $N=4$, and
Passarino--Veltman reduction for $N\le3$
\cite{delAguila:2004nf,Buccioni:2017yxi}.

Exploiting the factorized structure of open loops in a systematic way,
besides tensor reduction also helicity summation and diagram merging
are performed on the fly during open-loop recursion. This approach
reduces the complexity of intermediate results and leads to
improvements in CPU efficiency. Numerical instabilities can be avoided
by isolating them in triangle topologies, expressing the tensor
integrals in terms of scalar integrals and expanding the scalar
integrals in the limit of small Gram determinants, similar to the
expansions described in App.~B of \citere{Denner:2005nn},
but carried out to arbitrary precision in the expansion
parameter~\cite{Buccioni:2019sur}.
The algorithm is implemented in double and quadruple precision. Moreover,
it is equipped with a hybrid precision system that avoids residual
instabilities in a CPU-efficient way by restricting the usage of
quadruple precision to specific reduction steps.

\subsubsection{\sc Madgraph}
\label{se:Madgraph}

The construction of tree-level amplitudes in \MGNLO
\cite{Alwall:2014hca} is based on Feynman diagrams, helicity
amplitudes, and colour decomposition.  Employing colour decompositions
\cite{Maltoni:2002mq}, the colour matrix appearing in the squared
amplitude is computed automatically once and for all and then stored
in memory, similarly as in {\Openloops}. \MGNLO has its own
diagram-generation algorithm {\sc Madgraph5} \cite{Alwall:2011uj} that
needs as an input the Feynman rules corresponding to the Lagrangian of
a given theory. The information on the Feynman rules is typically
provided by {\sc FeynRules} \cite{Christensen:2008py} in the UFO
format \cite{Degrande:2011ua}.  With the information from UFO, the
dedicated routines that actually perform the computation of the
elementary blocks that enter helicity amplitudes are built by a
modified version of {\sc ALOHA} \cite{deAquino:2011ub}.

Loop amplitudes are calculated with the module {\sc MadLoop5}
\cite{Hirschi:2011pa}.  It is based on an independent implementation
of the recursive procedure used in {\Openloops}.  Like \Openloops,
{\sc MadLoop5} is based on subamplitudes $\cA_{0,h,b}$ with definite
helicities that possess a single colour factor $\cC_b$, so that the
full tree-level amplitude $\cM_{0,h}$ can be written as
\beq
\cM_{0,h}(\{p_k\}) = \sum_b \cC_b \cA_{0,h,b},
\eeq
where the index $h=\{\la_k\}$ indicates the dependence on
helicities.
The contributions to the subamplitudes $\cA_{0,h,b}$ are in one-to-one
correspondence with the amplitudes of the constituting Feynman
diagrams, except for those involving vertices featuring multiple
colour factors (such as 4-gluon vertices); in this case individual
diagrams contribute to several $\cA_{0,h,b}$.  The one-loop amplitude
is decomposed in the same way and can be written as
\beq
\de\cM_h = \sum_{t} \sum_{l\in t} \cC_l\int\rd^D q \frac{\cN^{(t)}_{h,l}(\qfour)}{\PD^{(t)}_0\PD^{(t)}_1\ldots \PD^{(t)}_{N_t-1}}.
\eeq
The expression is organized as a sum over sets of {\em topologies}
$t\,$ that are characterized by a specific product 
${\PD^{(t)}_0\cdots\PD^{(t)}_{N_t-1}}$ of propagator denominators.

The one-loop contribution to the unpolarized transition probability
can be written as
\beq
\de\mathcal{W}=
\sum_{t}\int\rd^D q\frac{1}{\PD^{(t)}_0\cdots\PD^{(t)}_{N_t-1}}\sumbar_h 
\sum_{l\in t}\sum_b\sumbar_{\mathrm{col}}(\cC_b^*\cC_l)  \cA_{0,h,b}^* \cN^{(t)}_{h,l}(\qfour)
\label{eq:M2_oneloop_MG}
\eeq
with
\beq
 \cN^{(t)}_{h,l}(\qfour)=\sum_{r=0}^{r_t}
 \hat{c}^{(t,r,N_t)}_{h,l;\mu_1\ldots\mu_{r}}
\qfour^{\mu_1}\cdots\qfour^{\mu_{r}}.
\label{eq:numerator_decomp}
\eeq
\Eref[b]{eq:M2_oneloop_MG} like \refeq{eq:m2_ol2} is optimized to
reduce the number of evaluations of loop integrals.  As in
{\Openloops}, the representation \refeqf{eq:M2_oneloop_MG} can be used
both for OPP reduction and for tensor-integral reduction.  For OPP,
after determination of the coefficients
$\hat{c}^{(t,r,N_t)}_{h,l;\mu_1\ldots\mu_{r}}$ the calculation of
the numerators can be performed upon using
\refeq{eq:numerator_decomp}. When inserting
\refeq{eq:numerator_decomp} into \refeq{eq:M2_oneloop_MG} the tensor
integrals emerge.  In fact, {\sc MadLoop5} supports the possibility to
switch between the OPP and tensor-integral reduction methods.  To this
end, it uses {\sc Ninja}~\cite{Peraro:2014cba}, {\sc
  Samurai}~\cite{Mastrolia:2010nb}, and {\sc
  CutTools}~\cite{Ossola:2007ax} or the internal tensor-integral
library {\sc IREGI}, {\sc PJFry}~\cite{Fleischer:2010sq}, {\sc
  Golem95C}~\cite{Cullen:2011kv}, and
\Collier~\cite{Denner:2014gla,Denner:2016kdg}, respectively.  {\sc
  MadLoop5} dynamically switches between different integral reduction
codes based on user-defined preferences and allows for dynamic
switching between double and quadruple precision. This ensures a
robust stability rescue mechanism as well as a reliable measure of the
numerical uncertainty of the result.

As in \Recola and {\sc Openloops}, the counterterms and rational terms
of type $R_2$ have to be added to the result of the loop-integration
procedure. Both are cast into the form of a tree-level-like amplitude
times the Born amplitude. The Feynman rules for the counterterms and
rational terms of type $R_2$ have to be provided as a set of
instructions in the UFO file. 
The implementation of such Feynman rules is automated in {\sc
  FeynRules} via {\sc NLOCT} \cite{Degrande:2014vpa} and similarly in
\RecolaTwo \cite{Denner:2017wsf} via {\sc Rept1l} \cite{Denner:2017vms}.

\MGNLO also offers an automated implementation of
cross-section computation and event generation for loop-induced processes
\cite{Hirschi:2015iia}. The unpolarized transition probability reads
in this case
\beq
\mathcal{W}=
\sumbar_{h}\sum_{l_1}\sum_{l_2}
\left[\int\rd^D q\frac{1}{\PD^{(l_1)}_0\cdots\PD^{(l_1)}_{N_{l_1}-1}} \cN_{h,l_1}(\qfour)\right]^* 
\left[\int\rd^D q\frac{1}{\PD^{(l_2)}_0\cdots\PD^{(l_2)}_{N_{l_2}-1}} \cN_{h,l_2}(\qfour)\right] 
\sumbar_{\mathrm{col}}(\cC_{l_1}^*\cC_{l_2}).  
\label{eq:M2_li_MG}
\eeq
The sums over $l_1$ and $l_2$ run over the subamplitudes corresponding
to single colour factors each. The quadratic scaling with the number
of diagrams is problematic for more complicated processes. To
circumvent this problem, the colour factors $\cC_i$ are projected into
the colour-flow basis \cite{Maltoni:2002mq}. The corresponding number
of basis vectors grows only power-like, reducing the computational
cost considerably.  This is similar to the approach in \Openloops and
the use of colour structures in \Recola.  The inability to perform
integrand reduction at the squared-matrix-element level leads to a
crucial difference between tensor-integral reduction and the OPP
reduction methods. While the OPP reduction operates on the
helicity-dependent numerators \refeqf{eq:numerator_decomp}, the tensor
integral reduction is based on helicity-independent tensor integrals.
Consequently, the number of independent OPP reductions for a kinematic
configuration is necessarily $L\times H\times C$, where $L$ is the
number of independent loop integrals, $H$ the number of helicities,
and $C$ the number of independent colour configurations.  On the other
hand, the number of independent tensor-integral reductions per
phase-space point is only $L$.

In \citere{Frederix:2018nkq} the EW corrections are quoted for
the integrated cross sections with minimal selection cuts for many
$2\to2$ and $2\to3$ processes at the $13\TeV$ LHC.
\begin{table}
\centerline{
\begin{tabular}{lr@{$\,\,\pm\,\,$}lr@{$\,\,\pm\,\,$}lr@{$\,\,\pm\,\,$}l}
\hline\\[-1.5ex]
Final state $F$
&\multicolumn{2}{c}{$\sigma_\mathrm{LO}$ [pb]} &
\multicolumn{2}{c}{$\sigma_\mathrm{NLO~EW}$ [pb]} &  \multicolumn{2}{c}{$\de_{\EW}$ [\%]}\\
\hline\\[-1.5ex]
$\Pep\Pne$ & 
$5.2498 $&$ 0.0005\,\cdot 10^{3}$ & $5.2113 $&$ 0.0006\,\cdot 10^{3}$ & $-0.73 $&$ 0.01$\\
$\Pep\Pne \Pj$ & 
$9.1468 $&$ 0.0012\,\cdot 10^{2}$ & $9.0449 $&$ 0.0014\,\cdot 10^{2}$ & $-1.11 $&$ 0.02$\\
$\Pep\Pne \Pj\Pj$ & 
$3.1562 $&$ 0.0003\,\cdot 10^{2}$ & $3.0985 $&$ 0.0005\,\cdot 10^{2}$ & $-1.83 $&$ 0.02$\\
$\Pep \Pem$ & 
$7.5367 $&$ 0.0008\,\cdot 10^{2}$ & $7.4997 $&$ 0.0010\,\cdot 10^{2}$ & $-0.49 $&$ 0.02$\\
$\Pep \Pem\Pj$ & 
$1.5059 $&$ 0.0001\,\cdot 10^{2}$ & $1.4909 $&$ 0.0002\,\cdot 10^{2}$ & $-1.00 $&$ 0.02$\\
$\Pep \Pem\Pj\Pj$ & 
$5.1424 $&$ 0.0004\,\cdot 10^{1}$ & $5.0410 $&$ 0.0007\,\cdot 10^{1}$ & $-1.97 $&$ 0.02$\\
$\Pep \Pem \mu^+ \mu^-$ & 
$1.2750 $&$ 0.0000\,\cdot 10^{-2}$ & $1.2083 $&$ 0.0001\,\cdot 10^{-2}$ & $-5.23 $&$ 0.01$\\
$\Pep\Pne \mu^- \bar{\nu}_{\mu}$ & 
 $5.1144 $&$ 0.0007\,\cdot 10^{-1}$ & $5.3019 $&$ 0.0009\,\cdot 10^{-1}$ & {$+3.67 $}&$ 0.02$\\
$\PH \Pep\Pne$ & 
$6.7643 $&$ 0.0001\,\cdot 10^{-2}$ & $6.4914 $&$ 0.0012\,\cdot 10^{-2}$ & $-4.03 $&$ 0.02$\\
$\PH \Pep  \Pem$ & 
$1.4554 $&$ 0.0001\,\cdot 10^{-2}$ & $1.3700 $&$ 0.0002\,\cdot 10^{-2}$ & $-5.87 $&$ 0.02$\\
$\PH \Pj \Pj$ & 
$2.8268 $&$ 0.0002\,\cdot 10^{0}$ & $2.7075 $&$ 0.0003\,\cdot 10^{0}$ & $-4.22 $&$ 0.01$\\
$\PWp\PWm\PWp$ & 
$8.2874 $&$ 0.0004\,\cdot 10^{-2}$ & $8.8017 $&$ 0.0012\,\cdot 10^{-2}$ & {$+6.21 $}&$ 0.02$\\
$\PZ \PZ \PWp$ & 
$1.9874 $&$ 0.0001\,\cdot 10^{-2}$ & $2.0189 $&$ 0.0003\,\cdot 10^{-2}$ & $+1.58 $&$ 0.02$\\
$\PZ\PZ\PZ$ & 
$1.0761 $&$ 0.0001\,\cdot 10^{-2}$ & $0.9741 $&$ 0.0001\,\cdot 10^{-2}$ & $-9.47 $&$ 0.02$\\
$\PH\PZ\PZ$ & 
 $2.1005 $&$ 0.0003\,\cdot 10^{-3}$ & $1.9155 $&$ 0.0003\,\cdot 10^{-3}$ & $-8.81 $&$ 0.02$\\
$\PH\PZ\PWp$ & 
 $2.4408 $&$ 0.0000\,\cdot 10^{-3}$ & $2.4809 $&$ 0.0005\,\cdot 10^{-3}$ & $+1.64 $&$ 0.02$\\
$\PH\PH\PWp$ & 
 $2.7827 $&$ 0.0001\,\cdot 10^{-4}$ & $2.4259 $&$ 0.0027\,\cdot
 10^{-4}$ & {$-12.82$} &$ 0.10$\\
$\PH\PH\PZ$ & 
$2.6914 $&$ 0.0003\,\cdot 10^{-4}$ & $2.3926 $&$ 0.0003\,\cdot 10^{-4}$ & ${-11.10} $&$ 0.02$\\
$\Pt \bar{\Pt} \PWp$ & 
 $2.4119 $&$ 0.0003\,\cdot 10^{-1}$ & $2.3025 $&$ 0.0003\,\cdot 10^{-1}$ & $-4.54 $&$ 0.02$\\
$\Pt \bar{\Pt} \PZ$ & 
$5.0456 $&$ 0.0006\,\cdot 10^{-1}$ & $5.0033 $&$ 0.0007\,\cdot 10^{-1}$ & $-0.84 $&$ 0.02$\\
$\Pt \bar{\Pt} \PH$ & 
$3.4480 $&$ 0.0004\,\cdot 10^{-1}$ & $3.5102 $&$ 0.0005\,\cdot 10^{-1}$ & $+1.81 $&$ 0.02$\\
$\Pt \bar{\Pt} \Pj$ & 
$3.0277 $&$ 0.0003\,\cdot 10^{2}$ & $2.9683 $&$ 0.0004\,\cdot 10^{2}$ & $-1.96 $&$ 0.02$\\
$\Pj\Pj\Pj$ & 
$7.9639 $&$ 0.0010\,\cdot 10^{6}$ & $7.9472 $&$ 0.0011\,\cdot 10^{6}$ & $-0.21 $&$ 0.02$\\
$\Pt \Pj$ & 
$1.0613 $&$ 0.0001\,\cdot 10^{2}$ & $1.0539 $&$ 0.0001\,\cdot 10^{2}$ & $-0.70 $&$ 0.02$\\
\hline
\end{tabular}}
\caption{Cross sections at LO, $\si_{\mathrm{LO}}$, and including NLO EW
  corrections, $\si_{\mathrm{NLO~EW}}$ ($=\si_{\mathrm{NLO}_2}$
  as defined in \refse{se:EW_QCD_mixing}),  as well
  as relative EW corrections, $\de_{\EW}=\si_{\mathrm{NLO~EW}}/\si_{\mathrm{LO}}-1$, 
  for a variety of scattering
  processes $\Pp\Pp\to F+X$ 
  at the LHC with $\sqrt{s}=13\TeV$ (taken from \citere{Frederix:2018nkq}).}
\label{tab:EWC_Madgraph}
\end{table}
The results, which are reproduced in \refta{tab:EWC_Madgraph},
demonstrate, on the one hand, the power of the automated tools, on the
other hand, they illustrate the spectrum of the size of the EW
corrections (in the $\GF$ scheme) for inclusive cross sections at the
LHC.  The EW corrections tend to increase with the number of
final-state particles and are larger for bosons in the final state
than for fermions.  Owing to the virtual contributions the EW
corrections are mostly negative.  While they are below 5\% in size for
the majority of processes, there are some notable exceptions. The
positive corrections, for instance for $\Pp\Pp \to \Pep\Pne \mu^-
\bar{\nu}_{\mu}+X$ (see also
\citeres{Baglio:2013toa,Billoni:2013aba,Biedermann:2016guo}), $\Pp\Pp
\to \PWp\PWm\PWp+X$ (see also
\citeres{Yong-Bai:2016sal,Dittmaier:2017bnh,Schonherr:2018jva,Dittmaier:2019twg}), or
$\Pp\Pp \to \PH\PZ\PWp+X$ result from contributions of 
photon--quark-induced processes that are enhanced by the presence of the
$\FA\FW^+\FW^-$ coupling and by quasi-soft--collinear $\PW$-boson
emission from 
photons.  It should, however, be realized
that photon--quark-induced contributions imply an additional jet in
the final state, so that their impact is often shadowed by QCD
corrections and drastically reduced if jet vetoes are applied.  For
some processes for charged-particle production via neutral initial
states, such as $\PW\PW$ production, even partonic $\ga\ga$ channels
can produce significant positive contributions. Typically, they appear
at high partonic scattering energies, but in some cases they can even lead to
significant corrections to integrated cross sections, as observed for
$\Pp\Pp \to \PWp\PWm\PZ+X$ \cite{Nhung:2013jta} (not included in
\refta{tab:EWC_Madgraph}).  For a further discussion of possible
patterns in EW corrections we refer to dedicated articles on specific
processes and to \citere{Frederix:2018nkq}.  Note that the size of EW
corrections to integrated cross sections may depend strongly on the
imposed cuts and the corrections are typically significantly larger in
differential distributions, most notably in high-energy tails.

{\providecommand{\LO}{\mathrm{LO}} 
\providecommand{\NLO}{\mathrm{NLO}}

As discussed in \refse{se:EW_QCD_mixing}, for general processes
contributions of different orders in the strong and EW couplings
appear at LO and at NLO. While one naively expects that the size of
the contributions is set by the corresponding coupling powers, this
is not always the case. In \refta{tab:tt_Madgraph}, we reproduce some
results from \MGNLO on $\Pt\bar{\Pt}$ production processes at the LHC
\cite{Frederix:2018nkq} using the notation introduced in
\citeres{Frixione:2015zaa,Frederix:2016ost,Frederix:2018nkq} and
\refse{se:EW_QCD_mixing}.  For inclusive rates all contributions apart
from the $\LO_1$ and $\NLO_1$ ones are small, with the exception of
the $\NLO_3$ term and, to a smaller extent, of the $\NLO_2$ one in
$\Pt\bar\Pt \PW^+$ production. The $+12\%$ correction to the $\LO_1$
cross section owing to the $\NLO_3$ contribution results from the
opening of a $\Pt\PW$ scattering channel, as noted in
\citeres{Dror:2015nkp,Frederix:2017wme}.
\begin{table}
\begin{center}
\begin{small}
\begin{tabular}{lr@{$\,\,\pm\,\,$}lr@{$\,\,\pm\,\,$}lr@{$\,\,\pm\,\,$}lr@{$\,\,\pm\,\,$}lr@{$\,\,\pm\,\,$}l}
\toprule
 & \multicolumn{2}{c}{$\Pp\Pp \to \Pt \bar{\Pt}+X$}
 & \multicolumn{2}{c}{$\Pp\Pp \to \Pt \bar{\Pt} \PZ+X$}
 & \multicolumn{2}{c}{$\Pp\Pp \to \Pt \bar{\Pt} \PW^+ +X$} 
\\
\midrule
$\LO_1$ & \multicolumn{2}{c}{$4.3803 \pm 0.0005\,\cdot 10^{2}$~pb} &
\multicolumn{2}{c}{$5.0463 \pm 0.0003\,\cdot 10^{-1}$~pb} &
\multicolumn{2}{c}{$2.4116 \pm 0.0001\,\cdot 10^{-1}$~pb} 
\\
$\LO_2$ & $+0.405 $&$ 0.001$~\% & $-0.691 $&$ 0.001$~\% & $+0.000 $&$
0.000$~\% 
\\
$\LO_3$ & $+0.630 $&$ 0.001$~\% & $+2.259 $&$ 0.001$~\% & $+0.962 $&$
0.000$~\% 
\\
$\NLO_1$ & $+46.164 $&$ 0.022$~\% & $+44.809 $&$ 0.028$~\% & $+49.504
$&$ 0.015$~\% 
\\
$\NLO_2$ & $-1.075 $&$ 0.003$~\% & $-0.846 $&$ 0.004$~\% & $-4.541 $&$
0.003$~\% 
\\
$\NLO_3$ & $+0.552 $&$ 0.002$~\% & $+0.845 $&$ 0.003$~\% & $+12.242
$&$ 0.014$~\% 
\\
$\NLO_4$ & $+0.005 $&$ 0.000$~\% & $-0.082 $&$ 0.000$~\% & $+0.017 $&$
0.003$~\% 
\\
\bottomrule
\end{tabular}
\caption{\label{tab:blobs} Cross sections for $\Pp\Pp\to\Pt\bar\Pt+X$,
  $\Pt\bar\Pt\PZ+X$, and $\Pt\bar\Pt\PWp+X$. Besides the leading $\LO_1$
  contribution in pb, the subleading LO and NLO contributions are
  given as percentage fractions of $\LO_1$ (taken from
  \citere{Frederix:2018nkq}).}
\label{tab:tt_Madgraph}
\end{small}
\end{center}
\end{table}
}

Another example, where the hierarchy of the NLO corrections differs
from the naive expectation is vector-boson scattering.  While the NLO
QCD corrections are comparably small for these processes, the NLO EW
corrections dominate the NLO corrections. More details can
be found at the end of \refse{se:ewlogs@1loop} and in
\citeres{Biedermann:2016yds,Biedermann:2017bss,Denner:2019tmn}.

\providecommand{\M}{{\cal{M}}}
\providecommand{\sub}{{\mathrm{sub}}}
\providecommand{\real}{{\mathrm{real}}}
\providecommand{\gsub}{g^{(\sub)}}
\providecommand{\bgsub}{{\bar g}^{(\sub)}}
\providecommand{\Gsub}{G^{(\sub)}}
\providecommand{\cGsub}{{\cal G}^{(\sub)}}
\providecommand{\bcGsub}{\bar{\cal G}^{(\sub)}}
\providecommand{\ssub}{\sigma^{(\sub)}}
\providecommand{\gin}{g^{\mathrm{(in)}}}
\providecommand{\gout}{g^{\mathrm{(out)}}}
\providecommand{\goutin}{g^{\mathrm{(out/in)}}}
\providecommand{\Hsub}{H}
\providecommand{\cHsub}{{\cal H}}
\providecommand{\bcHsub}{\bar{\cal H}}

\section{Electroweak radiative \texorpdfstring{corrections---real}
{corrections - real} emission effects}
\label{se:real}

\subsection{Infrared divergences in real electroweak corrections}
\label{se:IRrealEW}

In gauge theories involving massless gauge bosons or massive
gauge bosons in their high-energy limit, virtual corrections to
particle scattering processes in general contain mass singularities
originating from the exchange of soft or collinear massless or light particles.
In \refse{se:ewc@he} we discuss those virtual effects for energies well
above the EW scale in some detail.
In the following, we focus on energies of the order of the EW scale and
refer to the occurring soft and/or collinear 
mass singularities---following the usual terminology in QCD---globally
as infrared~(IR) singularities.
In one-loop corrections, {\it soft} singularities arise
from the exchange of a massless gauge boson (gluon or photon),
while {\it collinear} singularities occur whenever a massless
external particle (gluon, photon, or light fermion in the massless limit)
splits into two massless lines in a loop diagram~\cite{Kinoshita:1962ur}.
These IR divergences cancel in predictions for observables after
combining {\it virtual} corrections with {\it real} corrections
that involve the emission of additional massless particles.
At NLO, the real emission of one extra particle is relevant. 
The counterpart to soft virtual corrections is the emission 
{\it (bremsstrahlung)} of one
massless gauge boson (gluon or photon).
Within QED, summing virtual corrections and photonic
bremsstrahlung corrections already leads to IR-finite corrections
to all perturbative orders---this is the statement of the
{\it Bloch--Nordsieck theorem}~\cite{Bloch:1937pw}.
In the limit of a small fermion
mass $m_f$, those IR-finite corrections 
involve logarithmic terms of the form $\ln(m_f/Q)$, where
$Q$ is any kinematical quantity such as the scattering energy.
Such logarithms originate from collinear particle exchange in loops and
from the collinear splitting of fermions and photons in real emission
corrections.

Collinear singularities require special attention and treatment
for several reasons: Firstly, whenever quarks or gluons are involved in 
those singularities, the incomplete cancellation of IR singularities
is a sign of non-perturbative physics of strong interactions at low
energy scales. This is also the case if the singularities are formally
arising as logarithms of light quark masses, because those masses do
not represent perturbatively well-defined quantities. 
In such cases, the definition of observables either has to be refined
in such a way that IR singularities systematically cancel (like in jets),
or the IR-singular non-perturbative part has to be isolated properly
and eventually extracted from experimental data (like in
parton distribution or fragmentation functions).
To achieve this cancellation, the {\it Ki\-no\-shi\-ta--Lee--Nauen\-berg 
theorem (KLN)}~\cite{Kinoshita:1962ur,Lee:1964is}
is crucial, which predicts the cancellation of singularities in
observables that are sufficiently inclusive w.r.t.\ energy-degenerate
particle configurations.
The second type of collinear singularities concerns leptons and
photons. The corresponding corrections can be calculated perturbatively,
because lepton masses are well-defined parameters and measured to
good precision, but at present-day collider energies the mentioned
logarithms lead to corrections of the form $[\alpha\ln(\Ml/Q)]^n$
at the $n$-loop level. Since those corrections can get very large
at NLO ($n=1$), at least their dominating effects in higher orders 
should be known. 
A third reason why collinear singularities require some special 
treatment is of technical nature. Even if physical observables
are free of IR~singularities, such divergences appear in
different ingredients of the calculation, so that suitable
techniques for their treatment are required.

In the following, we first review the structure of soft and
collinear singularities appearing in {\it real} NLO EW corrections, 
then turn to frequently used techniques 
to isolate and calculate them, and subsequently discuss their absorption
in parton distribution and fragmentation functions of hadronic systems.
Finally, we add some brief discussion of enhanced EW corrections
beyond NLO in lepton--photon systems
that are connected to higher-order collinear singularities.

\subsubsection{Soft singularities}

In QED, the asymptotic form of soft-photon singularities
as well as their cancellation was worked out in the
classic paper of Yennie, Frautschi, and Suura~\cite{Yennie:1961ad}.
The results carry over to the full SM and even apply to all
electrically charged particles (such as W~bosons), because soft
photons are blind to the spin of charged particles.
A soft photon (outgoing momentum $k$)
only couples to the {\it eikonal current} $J^\mu_{\mathrm{eik}}$
generated by the charges $Q_n$ of all particles 
(momenta $p_n$) taking part in the considered process,
\begin{align}\label{eq:eikonal_current}
J^\mu_{\mathrm{eik}} = -\sum_n \sigma_n Q_n e\,\frac{p_n^\mu}{p_n k}.
\end{align}
The momenta $p_n$ might be incoming or outgoing.
The sign factors $\sigma_n=\pm1$ are determined by the physical
charge flow in the process.
We define $\sigma_n=+1$ for an
incoming particle and outgoing antiparticle, and $\sigma_n=-1$ for
an outgoing particle and incoming antiparticle.
Charge conservation for some process, thus, implies
\begin{align}
\sum_n Q_n \sigma_n = 0.
\label{eq:Qcons}
\end{align}

In the calculation of NLO EW corrections, one-photon bremsstrahlung
is relevant. Denoting the amplitudes for the bremsstrahlung 
process and for the underlying hard process without photon emission by
$\M_1$ and $\M_0$, respectively, the asymptotic behaviour 
of $|\M_1|^2$, summed over photon polarizations $\lambda_\gamma$,
in the soft-photon limit $k\to0$ is given by
\begin{align}
\sum_{\lambda_\gamma} |\M_1|^2 \;\asymp{k\to 0}\;
- J^\mu_{\mathrm{eik}} J_{\mathrm{eik},\mu}^* \, |\M_0|^2
= -\sum_{n,n'} Q_n \sigma_n Q_{n'} \sigma_{n'} e^2
\frac{p_n p_{n'}}{(p_n k)(p_{n'}k)} |\M_0|^2.
\label{eq:softphotonlimit}
\end{align}
Integrating this squared amplitude over the soft-photon phase
space, leads to a logarithmic singularity which can be
either regularized by an infinitesimal photon mass $m_\gamma$
or by dimensional regularization (DR) in $D=4-2\eps$ dimensions,
where the singularity shows up as $1/\epsilon$ pole.
As long as all charged particles have non-vanishing masses,
there is a simple universal correspondence between the 
soft divergences in the two regularizations schemes,
\begin{align}
\ln m_\gamma^2 \;\leftrightarrow\;
\frac{(4\pi\mu^2)^\epsilon\Gamma(1+\epsilon)}{\epsilon}
+ {\cal O}(\epsilon)
= \frac{(4\pi\mu^2)^\epsilon}{\epsilon\Gamma(1-\epsilon)}
+ {\cal O}(\epsilon),
\end{align}
where $\mu$ is the arbitrary reference mass scale of DR.  If at least
one radiating particle is truly massless, DR should be used; the
interplay between soft and collinear singularities then leads to
$1/\epsilon^2$~poles.  The correspondence between such IR-singular
terms in DR and the corresponding logarithms of small mass parameters
is more complicated and is discussed in
\refse{se:techniques4realcorrs} for light fermions.

\subsubsection{Collinear singularities}

\begin{figure}
\centering
\subfloat[$ab\to f\gamma+X$]{\includegraphics{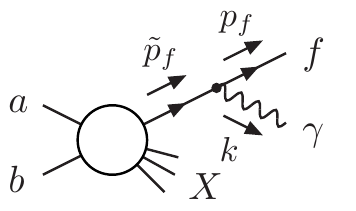}\label{fig:ab2fg}}
\qquad
\subfloat[$ab\to f\bar f+X$]{\includegraphics{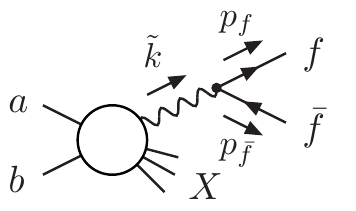}\label{fig:ab2ff}}
\\
\centering
\subfloat[$fb\to \gamma+X$]{\includegraphics{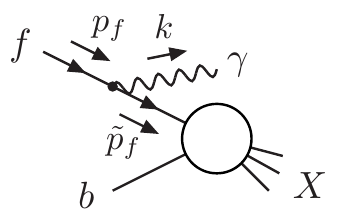}\label{fig:fa2g}}%
\qquad
\subfloat[$\gamma b\to f+X$]{\includegraphics{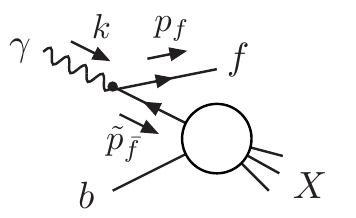}\label{fig:ga2f}}%
\qquad
\subfloat[$fb\to f+X$]{\includegraphics{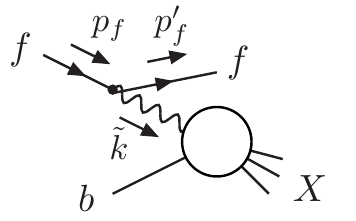}\label{fig:fa2f}}%
\caption{Structural diagrams illustrating the various 
fermion--photon splittings leading to 
collinear singularities in scattering processes.}
\label{fig:collsing}
\end{figure}

In the SM at experimentally relevant energy scales, collinear
singularities involving photons only appear in connection with light
fermions (W~bosons should be treated with their full mass dependence),
so that we restrict our discussion to fermions in the following.
Figure~\ref{fig:collsing} depicts the various cases in which such
collinear singularities can appear in scattering processes.
In the first two cases, the collinear splittings happen in the final state,
\ie after the hard scattering process. In the remaining three cases,
an initial-state particle is splitting, so that only a reduced momentum
flows into the hard scattering process.
For later convenience we introduce the complete set of relevant
splitting functions $P_{ab}(z)$ already here,
\begin{alignat}{5}
  \label{eq:pff-pfa-paf-paa}
  \Pff(z)   &{}= \left(\frac{1+z^2}{1-z}\right)_+, \qquad&
  \Pffunreg(z)   &{}= \frac{1+z^2}{1-z}, \nl
  P_{f \ga}(z)&{}= z^2 + (1-z)^2, \qquad&
  P_{\ga f}(z)&{}= \frac{1 + (1-z)^2}{z}, \qquad&
  P_{\ga\ga}(z)&{}= -\frac{2}{3}\delta(1-z),
\end{alignat}
which determine the probability to find parton $a$ in parton $b$.
Note, in particular, the two different variants $\Pff$ and
$\Pffunreg$, which are related by the 
$(\cdots)_+$ distribution, defined by
\begin{align}
\int_0^1 \rd x\, \bigl[f(x)\bigr]_+ \,g(x) \,=
\int_0^1 \rd x\, f(x) \, \bigl[g(x)-g(1)\bigr],
\label{eq:plusdist}
\end{align}
with $g$ representing some test function.
The kinematical region that develops the collinear singularity
for two external light particles of momenta $p_1$ and $p_2$ is
characterized by ${\cal O}(p_1 p_2)={\cal O}(m_f^2)\ll Q^2$, i.e.\
the squared mass of the fermion $f$ participating in the splitting and
the virtuality $(p_1+p_2)^2$ of the splitting propagator are much smaller
than the typical 
(squared) energy scale $Q^2$ of the hard process.
Using the momentum assignments defined in \reffi{fig:collsing},
the unpolarized squared matrix elements (spin-averaged for the initial
and spin-summed for the final states) $\langle|\M_1^{\cdots}|^2\rangle$ asymptotically
behave as follows in the collinear limits,
\begin{align}
\label{eq:ab2fgasymp}
\Bigl\langle |\M_1^{ab\to f\gamma X}(p_f,k)|^2 \Bigr\rangle
\;\asymp{p_f k\to 0}\; {}& \Qf^2 e^2 \, h_{f\gamma}(p_f,k) \; 
\Bigl\langle |\M_0^{ab\to f X}(\tilde p_f)|^2 \Bigr\rangle,
\\[.3em]
\label{eq:ab2ffasymp}
\Bigl\langle |\M_1^{ab\to f\bar f X}(p_f,p_{\bar f})|^2 \Bigr\rangle
\;\asymp{p_f p_{\bar f}\to 0}\; {}& \NCf \Qf^2 e^2 \, h_{f\bar f}^{\mu\nu}(p_f,p_{\bar f}) \; 
\Bigl\langle 
T_{0,\mu}^{ab\to \gamma X}(\tilde k)
\left(T_{0,\nu}^{ab\to \gamma X}(\tilde k)\right)^*
\Bigr\rangle,
\\[.3em]
\label{eq:fb2gasymp}
\Bigl\langle |\M_1^{fb\to \gamma X}(p_f,k)|^2 \Bigr\rangle
\;\asymp{p_f k\to 0}\; {}& \Qf^2 e^2 \, h^{f\gamma}(p_f,k) \; 
\Bigl\langle |\M_0^{fb\to X}(\tilde p_f)|^2 \Bigr\rangle,
\\[.3em]
\label{eq:gb2fasymp}
\Bigl\langle |\M_1^{\gamma b\to f X}(k,p_f)|^2 \Bigr\rangle
\;\asymp{p_f k\to 0}\; {}& \Qf^2 e^2 \, h^{\gamma f}(k,p_f) \; 
\Bigl\langle |\M_0^{\bar fb\to X}(\tilde p_{\bar f})|^2 \Bigr\rangle,
\\[.3em]
\label{eq:fb2fasymp}
\Bigl\langle |\M_1^{fb\to f X}(p_f,p'_f)|^2 \Bigr\rangle
\;\asymp{p_f p'_f\to 0}\; {}& \NCf \Qf^2 e^2 \, h^{ff,\mu\nu}(p_f,p'_f) \; 
\Bigl\langle T_{0,\mu}^{\gamma b\to X}(\tilde k)
\left(T_{0,\nu}^{\gamma b\to X}(\tilde k)\right)^* \Bigr\rangle,
\end{align}
where $\NCf$ is the colour multiplicity of $f$ (\ie $1$ or $3$ if $f$
is a lepton or a quark, respectively).  Here
$\langle|\M_0^{\cdots}|^2\rangle$ are the unpolarized squared matrix
elements of the hard scattering processes, and $T_{0,\mu}^{\cdots}$
the amplitudes of the hard processes with photon polarization vector
$\veps^\mu(\tilde k)$ truncated,
\begin{align}
\M_{0}^{ab\to \gamma X}(\tilde k) = 
\veps^\mu(\tilde k)^* \, T_{0,\mu}^{ab\to \gamma X}(\tilde k),
\qquad
\M_{0}^{\gamma b\to X}(\tilde k) = 
\veps^\mu(\tilde k) \, T_{0,\mu}^{\gamma b\to X}(\tilde k).
\label{eq:M0T0}
\end{align}
All squared matrix elements $\langle|\M_k^{\cdots}|^2\rangle$
are colour summed, but
summed over outgoing and averaged over incoming polarizations,
where the number of polarizations for fermions and the photon 
are $2$ and $D-2=2(1-\epsilon)$, respectively.%
\footnote{The spin average on the right-hand side of
  \refeq{eq:fb2fasymp} is for the incoming fermion $f$ and not yet for the
  incoming photon, \ie it yields a factor~$1/2$.}
If the incoming fermion turns into a photon or vice versa,
this leads to additional $\epsilon$-terms in the 
radiator functions
$h^{\cdots}_{\cdots}(p_1,p_2)$ which are given by
\begin{align}
h_{f\gamma}(p_f,k) 
={} & \frac{1}{p_f k}
\left[ \Pffunreg(z)-\epsilon(1-z)-\frac{m_f^2}{p_f k} \right],
\label{eq:hab2fg}
\nn\\[.3em]
h_{f\bar f}^{\mu\nu}(p_f,p_{\bar f}) 
={} & \frac{2}{(p_f+p_{\bar f})^2}
\left[ -g^{\mu\nu}+4z(1-z)\frac{k_\perp^\mu k_\perp^\nu}{k_\perp^2-m_f^2}
\right],
\nn\\[.3em]
h^{f\gamma}(p_f,k) ={} & \frac{1}{x(p_f k)}
\left[ \Pffunreg(x)-\epsilon(1-x)-\frac{x m_f^2}{p_f k} \right],
\nn\\[.3em]
h^{\gamma f}(k,p_f)
={} & \frac{1}{x(kp_f)} \,
\left[P_{f\ga}(x)-\frac{1}{1-\epsilon}
\left(2\epsilon x(1-x) -\frac{xm_f^2}{kp_f}\right)\right],
\nn\\[.3em]
h^{ff,\mu\nu}(p_f,p'_f)
={} & \frac{-2}{(p_f-p'_f)^2} 
\left[ -g^{\mu\nu}
-\frac{4(1-x)}{x^2}\frac{k_\perp^\mu k_\perp^\nu}{k_\perp^2-x^2 m_f^2} \right].
\end{align}
Here and in the following, we give results for unpolarized particles
that are both valid in DR and mass regularization (MR), which are
obtained upon combining the results given in
\citeres{Dittmaier:1999mb,Dittmaier:2008md,Catani:2002hc}.  Results
valid for polarized hard particles (in MR) can be found in
\citeres{Dittmaier:1999mb,Dittmaier:2008md}.  With the momentum
assignments in $h_{ii'}(p_1,p_2)$ and $h^{aa'}(p_1,p_2)$, the
dimensionless variables $z$ and $x$ are given by
\begin{equation}
z = \frac{p_1^0}{p_1^0+p_2^0}, \qquad
x = \frac{p_1^0-p_2^0}{p_1^0},
\label{eq:zx}
\end{equation}
\ie $z$ is the energy fraction taken by particle $i$ in a final-state
splitting $\tilde{i}\to i+i'$ and $x$ is the energy fraction
carried into the hard process by particle $\tilde{a}$ resulting
from the splitting $a\to a'+\tilde{a}$.  The momentum $k_\perp$ is
the orthogonal component of the momenta $p_1$ and $p_2$ in their {\it
  Sudakov parametrization} around some collinear axis~$\tilde{p}$
($\tilde{p} k_\perp=0$).  Following \citere{Catani:2002hc}
(Sect.~4.2), a parametrization appropriate for a final-state splitting
is given by
\beq
p_1^\mu {}= z \tilde{p}^\mu + k_\perp^\mu
- \frac{k_\perp^2+ z^2 m^2- m_1^2}{z}\frac{n^\mu}{2\tilde{p}n}, 
\qquad
p_2^\mu {}= (1-z) \tilde{p}^\mu - k_\perp^\mu
- \frac{k_\perp^2+(1-z)^2m^2 - m_2^2}{1-z}\frac{n^\mu}{2\tilde{p}n},
\eeq
with the squared masses $p_1^2=m_1^2$, $p_2^2=m_2^2$, $\tilde{p}^2=m^2$,
and $n^\mu$ denoting an auxiliary light-like vector 
($n^2=n k_\perp=0$). Depending on the specific splitting, the masses
$m_1$, $m_2$, and $m$ are either $m_f$ or zero.
Owing to
\begin{equation}
(p_1 + p_2)^2 = -\frac{k_\perp^2}{z(1-z)}
+\frac{m_1^2}{z} +\frac{m_2^2}{1-z},
\end{equation}
we have ${\cal O}\bigl((p_1+p_2)^2\bigr) = 
{\cal O}\bigl(k_\perp^2\bigr)={\cal O}\bigl(m_f^2\bigr)$ in the collinear limit.
A Sudakov parametrization for an initial-state splitting can be obtained
analogously.

The momenta $\tilde p_f$, $\tilde p_{\bar f}$, $\tilde k$ in the hard
matrix elements $\M_0$, $T_0$ (\ref{eq:ab2fgasymp})--(\ref{eq:fb2fasymp}) 
correspond to the collinear axes, i.e.\ to $\tilde{p}$ in the case of a final-state
splitting. Note that those momenta have to fulfil the corresponding
OS conditions (\ie they are not identical with $p_1\pm p_2$ as
suggested by momentum conservation in \reffi{fig:collsing}), otherwise
the hard matrix element would be ill-defined and in general gauge
dependent.

Let us further inspect the terms of the form 
$\big\langle k_\perp^\mu T_{0,\mu} k_\perp^\nu (T_{0,\nu})^* \big\rangle$
in the asymptotics of $\langle |\M_1|^2\rangle$ for splittings
with an external photon in the hard scattering process,
Eqs.~(\ref{eq:ab2ffasymp}) and (\ref{eq:fb2fasymp}).
These terms express correlations between the polarization of the
hard photon and the azimuthal angle $\phi$ of the nearly collinear momenta
around the collinear axis.
Assuming inclusiveness w.r.t.\ $\phi$ in the final state, 
$\langle |\M_1|^2\rangle$ can be averaged over $\phi$
(which is equivalent to a spin average of the splitting particle), so that the
asymptotics of Eqs.~(\ref{eq:ab2ffasymp}) and (\ref{eq:fb2fasymp}) 
further simplifies and shows factorization from the squared
hard matrix elements $\langle |\M_0|^2\rangle$,
\begin{align}
\label{eq:ab2ffasymp2}
\Bigl\langle |\M_1^{ab\to f\bar f X}(p_f,p_{\bar f})|^2 \Bigr\rangle_\phi
\;\asymp{p_f p_{\bar f}\to 0}\; {}& \NCf \Qf^2 e^2 \, h_{f\bar f}(p_f,p_{\bar f}) \; 
\Bigl\langle |\M_0^{ab\to \gamma X}(\tilde k)|^2 \Bigr\rangle,
\\[.3em]
\label{eq:fb2fasymp2}
\Bigl\langle |\M_1^{fb\to f X}(p_f,p'_f)|^2 \Bigr\rangle_\phi
\;\asymp{p_f p'_f\to 0}\; {}& \NCf \Qf^2 e^2 \, h^{ff}(p_f,p'_f) \; 
\Bigl\langle |\M_0^{\gamma b\to X}(\tilde k)|^2 \Bigr\rangle,
\end{align}
where the azimuthal average is indicated as $\langle\cdots\rangle_\phi$.
The radiator functions simplify to
\begin{align}
h_{f\bar f}(p_f,p_{\bar f}) &{}= \frac{2}{(p_f+p_{\bar f})^2}
\left[ P_{f\ga}(z)+\frac{2}{1-\epsilon}\left(-\epsilon z(1-z)
+\frac{m_f^2}{(p_f+p_{\bar f})^2} \right)\right],
\nn\\[.3em]
h^{ff}(p_f,p'_f) &{}= \frac{-2}{x(p_f-p'_f)^2}
\left[ P_{\ga f}(x) -\epsilon x
+\frac{2x m_f^2}{(p_f-p'_f)^2} \right].
\end{align}
For the other splittings, $\langle|\M_1|^2\rangle$ and
$\langle|\M_1|^2\rangle_\phi$ are identical.
Integrating the asymptotic forms of all
$\langle |\M_1|^2\rangle_\phi$ over small emission angles in the
collinear regions produces mass-singular corrections proportional
to $\alpha\ln(m_f/Q)P_{ab}(\xi)$ with $\xi=x,z$.

In the following we describe two different techniques how the
singular contributions to real NLO corrections originating from soft or
collinear regions can be extracted from the full phase-space integral,
integrated with regularization parameters,
and eventually combined with the remaining non-singular contributions.

\subsection{Techniques for calculating real electroweak corrections}
\label{se:techniques4realcorrs}

An analytical calculation of real radiative corrections is typically
only aimed at for fully inclusive quantities, such as total cross
sections or decay widths (if they are simple enough), while real
corrections to differential quantities are mostly evaluated
numerically.  Phase-space cuts and event-selection procedures such as
jet algorithms render an analytical treatment in general impossible.
Apart from feasibility, flexibility is another motivation for the
numerical approach, because analytical calculations---if possible at
all---are tied to idealized setups, while numerical phase-space
integrations can deliver many differential distributions
simultaneously and for arbitrary (physically reasonable) event
selections.  On the other hand, it is desirable to obtain the
IR-singular contributions originating from soft and/or collinear
phase-space regions in analytic form, in order to properly cancel the
singularities against their counterparts in the virtual corrections.
This means that we need flexible, general procedures that allow for a
numerical evaluation of real-emission contributions without any
regulators in the regular phase-space regions, but provide an
analytical treatment of the phase-space integration at least over the
subspaces containing the IR-singular contributions in the presence of
IR regulators, such as phase spaces with $D\ne4$ dimensions or small
mass parameters.  Among such procedures, two fundamentally different
approaches are used in applications: {\it phase-space slicing} and
{\it subtraction methods}.

In {\it slicing} approaches, the singular regions are cut away from
phase space in the numerical integration and treated separately.
Employing general factorization properties of squared amplitudes in
the soft or collinear regions, the singular integrations can be
carried out analytically.  In the limit of small technical slicing cut
parameters the sum of the two contributions reproduces the full
phase-space integral.  There is a trade-off between residual technical
cut dependences and numerical integration errors which increase with
decreasing slicing cuts; in practice, one is forced to search for a
plateau in the integrated result within integration errors by varying
the slicing cut parameters.  Technically, mostly two different slicing
approaches are in use: {\it
  one-cutoff}~\cite{Giele:1991vf,Giele:1993dj,Keller:1998tf} and {\it
  two-cutoff slicing} (see, for instance,
\citeres{Baur:1998kt,Harris:2001sx}).  One-cutoff slicing excludes the
IR-singular regions by cuts $s_{ij}>s_{\min}$ on the invariant masses
$s_{ij}$ of particle pairs $ij$ that can develop an IR~singularity, so
that the separation between regular and singular regions is ruled by a
single parameter $s_{\min}\ll Q^2$, which is small w.r.t.\ any
relevant scale $Q^2$ in the considered process.  By construction, this
method is particularly appealing in QCD calculations that employ
colour-ordered amplitudes. Without colour ordering, the separation of
soft and collinear singularities involves subtleties in disentangling
the overlap between different cut conditions $s_{ij}>s_{\min}$, as
discussed in \citere{Basso:2015gca} in detail.  Since there is no
analogue to colour ordering for electric charges, one-cutoff slicing
does not seem to be the method of choice for real EW corrections.
Two-cutoff slicing, which is frequently used in NLO EW calculations,
separates soft and collinear regions by two different types of cuts.
Singularities from soft-photon (or gluon) emission are excluded by a
cut $E_\gamma>\Delta E$ on the photon energy $E_\gamma$ with $\Delta
E\ll Q$, and collinear singularities are excluded by angular cuts
$\theta_{ij}>\Delta\theta$ ($\Delta\theta\ll1$), where $\theta_{ij}$
is the angle between the directions of two particles that can develop
a collinear singularity.  Note that all individual cross-section
contributions have to be calculated in the same frame of reference,
since the cut procedure is based on energies and angles, which are not
Lorentz invariant.  Finally, we mention another type of a slicing
procedure, called (by some misleading naming) {\it jettiness
  subtraction}~\cite{Gaunt:2015pea}, which employs the event-shape
variable {\it jettiness}~\cite{Stewart:2010tn} to separate IR-singular
regions. To our knowledge, jettiness subtraction was not yet used in
EW higher-order calculations.

{\it Subtraction techniques} are based on the idea of subtracting a
simple auxiliary function from the singular integrand and adding this
contribution again.  The auxiliary function has to be chosen in such a
way that it cancels all singularities of the original integrand so
that the phase-space integration of the difference can be performed
numerically, even over the singular regions of the original integrand.
In this difference the original matrix element can be evaluated
without regulators for soft or collinear singularities.  The auxiliary
function has to be simple enough so that it can be integrated over the
singular regions (ideally analytically)
with the help of regulators, before the
subtracted contribution is added again.  This singular analytical
integration can be done once and for all in a process-independent way
because of the general factorization properties of squared amplitudes
in the singular regions.  At NLO several subtraction variants have
been proposed and worked out in the literature, first within
QCD~\cite{Ellis:1980wv,Frixione:1995ms,Catani:1996jh,Catani:1996vz,Phaf:2001gc,Catani:2002hc,Frederix:2009yq}
and subsequently for photonic corrections~\cite{Dittmaier:1999mb,%
Dittmaier:2008md,Basso:2015gca,Schonherr:2017qcj,Frederix:2018nkq}.
The most frequently used variants are {\it FKS subtraction} and 
{\it dipole subtraction}, which have their roots in the papers of
Frixione, Kunszt, Signer~\cite{Frixione:1995ms} and 
Catani, Seymour~\cite{Catani:1996jh,Catani:1996vz}, respectively.
While dipole subtraction for EW corrections, including its formulation
in MR, is described in the literature very 
explicitly~\cite{Dittmaier:1999mb,Dittmaier:2008md,Basso:2015gca,%
  Schonherr:2017qcj}.  FKS subtraction for EW
corrections~\cite{Frederix:2018nkq} mostly follows the literal
translation of its formulation in QCD. The application of FKS
subtraction to the EW corrections to Drell--Yan processes is, for
example, described in \citeres{Barze:2012tt,Muck:2016pko} in some
detail.  At NNLO, several
approaches~\cite{GehrmannDeRidder:2005cm,Catani:2007vq,Czakon:2010td,%
  Boughezal:2011jf,Currie:2013vh,Cacciari:2015jma,Magnea:2018hab},
some of them based on numerical procedures, have been suggested and
successfully used in QCD calculations. In principle, the latter
procedures could be used in the calculation of real EW corrections as
well, the only caveat might be the fact that MR is
not supported in most cases.  On the other hand, a generalization of
the FKS~method~\cite{Frixione:1995ms} to real NNLO QED corrections
(and beyond) with truly massive charged particles, \ie in the case
where only soft IR singularities exist, was proposed in
\citere{Engel:2019nfw} recently.  We will not consider the issue of
real corrections beyond NLO in this review.

By experience, subtraction techniques are often superior to slicing
approaches in the sense that integration errors in cross-section
predictions are typically much smaller if a comparable amount of
statistics is used in the numerics within two comparable calculations.
On the other hand, subtraction techniques produce events in Monte
Carlo integrations with (even unbounded) negative weights, because the
difference between the original integrand and the auxiliary function
is not positive definite, even if the original integrand is. This
problem is tricky to handle in Monte Carlo event generation and less
delicate in slicing approaches.

In the following, we describe two very popular techniques for handling 
IR singularities in real NLO EW corrections in quite some detail: 
{\it two-cutoff slicing} and {\it dipole subtraction}. 
To keep the presentation transparent, we restrict it to the case
of IR (soft and/or collinear) singularities appearing in processes involving
light, unpolarized fermions only. 
For an account of massive or polarized particles, 
and for a description of the other techniques,
we refer to the original papers quoted above.
We do not consider the separation of mass singularities related to
particle masses at the EW scale, which \eg appear for
W-boson bremsstrahlung at energies way above the EW scale.
Those issues are still under discussion and corresponding methods
under construction.

\subsubsection{Two-cutoff slicing}
\label{se:twocutoffslicing}

As already indicated, two-cutoff slicing employs two different types of 
phase-space cuts to separate IR~singularities.
Soft singularities from soft-photon (or gluon) emission are excluded
from the full photon emission phase space upon demanding 
$E_{\gamma}>\Delta E$, where $E_\gamma=k_0$ is the energy of the emitted photon
of momentum~$k$.
The other IR-singular splittings described in \refse{se:IRrealEW}, which
do not involve real photons in the final state, do not lead to soft
singularities.
Collinear singularities are separated by the 
cuts $\theta_{ij}>\Delta\theta$ on the emission angles $\theta_{ij}$
between the two particles $i,j$ whose collinear splitting leads to a singularity.
In total, the cross-section contribution of the real NLO EW corrections 
is split into three parts,
\begin{align}
\rd\sigma^{\real} = 
\rd\sigma^{\hard}\big|_{E_\gamma>\Delta E,\,\mathrm{all}\,\theta_{ij}>\Delta\theta} +
\rd\sigma^{\soft}\big|_{E_\gamma<\Delta E} +
\sum_{\mathrm{pairs}\,ij}\rd\sigma^{\coll}_{ij}\big|_{E_\gamma>\Delta E,\,\theta_{ij}<\Delta\theta} \,.
\end{align}
In the {\it hard} contribution 
$\rd\sigma^{\hard}$ the dependence on the cut parameters
$\Delta E$ and $\Delta\theta$ emerges from the numerical phase-space
integration which extends rather deeply into the IR~region owing to
$\Delta E\ll Q$ and $\Delta\theta\ll 1$ and is, thus, rather CPU expensive 
in practice.

The {\it soft} contribution 
$\rd\sigma^{\soft}$ corresponding to the cross section for a process
$ab\to X+\gamma$ involves only an integration over the
soft-photon region, which can be separated from the full 
phase-space integral according to
\begin{align}
\int_{E_\gamma<\Delta E} \rd\Phi_{ab\to X+\gamma} \;=\;
\left. \int \rd\Phi_{ab\to X} \, \mu^{4-D}
\int_{E_\gamma<\Delta E} 
\frac{\rd^{D-1} \bf k}{2E_{\gamma}(2\pi)^{D-1}}
\,\right|_{E_\gamma=\sqrt{|{\bf k}|^2+m_\gamma^2}} .
\label{eq:softPS}
\end{align}
Here we have left open whether we want to apply DR with $D\ne4$ or MR
with an
infinitesimal photon mass $m_\gamma$.  Using the asymptotic behaviour
(\ref{eq:softphotonlimit}) of the squared matrix element $|\M_1|^2$
for a process involving external charged particles with charges $Q_n$
and momenta $p_n$, the soft contribution fully factorizes from the
differential LO cross section $\rd\sigma^{\LO}$,
\begin{equation}
\rd\sigma^{\soft} = \delta^{\soft}\,\rd\sigma^{\LO}\,,
\end{equation}
and the soft-photon factor can be calculated to
\begin{align}
\delta^{\soft} &{}= 
-\frac{\alpha}{4\pi^2} \sum_{n,n'}
\sigma_n Q_n \sigma_{n'} Q_{n'} \,
(2\pi\mu)^{4-D} \int_{E_{\gamma}< \Delta E}
\frac{\rd^{D-1} \bf k}{E_{\gamma}}
\left.\frac{p_n p_{n'}}{(p_n k)(p_{n'}k)} 
\,\right|_{E_\gamma=\sqrt{|{\bf k}|^2+m_\gamma^2}}
\nonumber\\
&{}= 
\frac{\alpha}{2 \pi} \sum_{\substack{n,n'\\n<n'} }
\sigma_n Q_n \sigma_{n'} Q_{n'}
\left[ I_{nn}+I_{n'n'}-2I_{nn'} \right],
\end{align}
where we have used charge conservation (\ref{eq:Qcons})
and introduced the basic integrals
\begin{align}
I_{nn'}=\frac{(2\pi\mu)^{4-D}}{2\pi}
\int_{E_{\gamma}< \Delta E}
\left. \frac{\rd^{D-1} {\bf k}}{E_{\gamma}} \frac{p_n p_{n'}}{(p_n k) (p_{n'} k)}
\,\right|_{E_\gamma=\sqrt{|{\bf k}|^2+m_\gamma^2}},
\end{align}
in which $p_n^0,p_{n'}^0>0$ is assumed.
Explicit expressions for the integrals $I_{nn'}$ for $D=4$ but arbitrary
non-vanishing mass parameters $p_n^2=m_n^2$, $p_{n'}^2=m_{n'}^2$
can be found in \citeres{tHooft:1978jhc,Denner:1991kt}.
Here we just give the solutions for small masses $m_n, m_{n'}$, which still
obey the hierarchy $m_\gamma\ll m_n,m_{n'}\ll Q$,
\begin{align}
I_{nn}\big|_{\MR} &{}=
2\ln\left(\frac{2\Delta E}{m_\gamma}\right)
+ 2\ln\left(\frac{m_n}{2p_n^0}\right),
\\
I_{nn'}\big|_{\MR} &{}=
2\ln\left(\frac{2\Delta E}{m_\gamma}\right)
\ln\left(\frac{s_{nn'}}{m_n m_{n'}}\right)
-\ln^2\left(\frac{m_n}{2p^0_n}\right)
-\ln^2\left(\frac{m_{n'}}{2p^0_{n'}}\right)
-\frac{\pi^2}{3}
-\Li_2\left(1-\frac{4 p^0_n p^0_{n'}}{s_{nn'}}\right),
\end{align}
with $s_{nn'}=2p_n p_{n'}\gg m_n^2,m_{n'}^2$.  
In DR with $p_n^2=p_{n'}^2=k^2=0$ these
integrals read
\begin{align}
I_{nn}\big|_{\DR} ={}& 0,
\\
I_{nn'}\big|_{\DR} 
={}& \frac{(4\pi)^\epsilon}{\epsilon^2\Gamma(1-\epsilon)}
\left[ 1 +\epsilon\ln\left(\frac{\mu^2}{4\Delta E^2}\right)
+\eps\ln\left(\frac{4 p^0_n p^0_{n'}}{s_{nn'}}\right)
+\frac{\epsilon^2}{2}\ln^2\left(\frac{\mu^2}{4\Delta E^2}\right)
\right.
\nn\\&{} \left.
{} +\eps^2 \ln\left(\frac{\mu^2}{4\Delta E^2}\right)
           \ln\left(\frac{4 p^0_n p^0_{n'}}{s_{nn'}}\right)
-\eps^2\Li_2\left(1-\frac{4 p^0_n p^0_{n'}}{s_{nn'}}\right)
-\epsilon^2\frac{\pi^2}{6} \right]
+{\cal O}(\epsilon).
\end{align}

The integration over the {\it collinear} phase-space regions is
straightforward as well, although the necessary phase-space factorization
is somewhat more complicated than \refeq{eq:softPS}.
One way of carrying out the integrals is to employ the phase-space
factorization of the dipole subtraction formalism (see next section)
and restricting it to the collinear regions, as \eg described in
\citere{Dittmaier:2008md}.
The resulting cross-section contributions 
(in the same order as shown in \reffi{fig:collsing})
can be written as
\begin{align}
\int \rd\sigma^{\coll}_{ab\to f\gamma X}(p_f,k)
={}& \frac{ \Qf^2\alpha}{2\pi}
\int \rd\sigma^{\LO}_{ab\to f X}(\tilde p_f) \,
\int_0^1 \rd z\,
\left\{ H_{f\gamma}(\tilde p_f^0) \,\de(1-z)
+ \left[\bar{\cal H}_{f\gamma}(\tilde p_f^0,z)\right]_+ \right\}
\nn\\[.3em]
& {}\times
\Theta_{\mathrm{cut}}\Bigl(p_f=z\tilde p_f,k=(1-z)\tilde p_f\Bigr),
\label{eq:sli-ab2fg}
\\[.5em]
\int \rd\sigma^{\coll}_{ab\to f\bar fX}(p_f,p_{\bar f})
={}& \NCf\, \frac{\Qf^2\alpha}{2\pi} \,
\int \rd\sigma^{\LO}_{ab\to \gamma X}(\tilde k) \,
\int_0^1 \rd z\,
\left\{ H_{f\bar f}(\tilde k^0) \, \de(1-z) 
+ \left[\bar{\cal H}_{f\bar f}(\tilde k^0,z)\right]_+ \right\}
\nn\\
& {}\times
\Theta_{\mathrm{cut}}\Bigl(p_f=z\tilde k,
p_{\bar f}=(1-z)\tilde k\Bigr),
\label{eq:sli-ab2ff}
\\[.5em]
\int \rd\sigma^{\coll}_{fb\to \gamma X}(p_f,k)
={}& \frac{ \Qf^2\alpha}{2\pi}
\int_0^1 \rd x
\int \rd\sigma^{\LO}_{fb\to X}(\tilde p_f=x p_f) \,
\left\{ H^{f\gamma}(p_f^0) \de(1-x)
+ \left[{\cal H}^{f\gamma}(p_f^0,x)\right]_+ \right\},
\label{eq:sli-fb2g}
\\[.5em]
\int \rd\sigma^{\coll}_{\gamma b\to fX}(k,p_f)
={}& \NCf\,
\frac{\Qf^2\alpha}{2\pi} 
\int_0^1\rd x
\int \rd\sigma^{\LO}_{\bar fb\to X}(\tilde p_{\bar f}=xk)\,
{\cal H}^{\ga f}(k^0,x),
\label{eq:sli-gb2f}
\\[.5em]
\int \rd\sigma^{\coll}_{fb\to fX}(p_f,p'_f)
={}& \frac{\Qf^2\alpha}{2\pi} \int_0^1\rd x
\int\rd\sigma^{\LO}_{\gamma b\to X}(\tilde k=xp_f) \,
{\cal H}^{ff}(p_f^0,x), 
\label{eq:sli-fb2f}
\end{align}
where the momenta $p_1$ and $p_2$ that get collinear are indicated
as arguments in $\rd\sigma^{\coll}(p_1,p_2)$ on the l.h.s.\
of the equations.
The relations between $p_{1,2}$ and the momentum $\tilde p$
playing the role of $p_1\pm p_2$ 
in the hard LO cross section $\rd\sigma^{\LO}(\tilde p)$ 
are made explicit on the r.h.s., where the step functions
$\Theta_{\mathrm{cut}}$ (being $1$ or $0$)
show which momenta are subject to possible phase-space cuts
in the case of final-state splittings (for initial-state splittings
the collinear particles are assumed to escape in the beam pipe).
Note that the hard cross section does not depend on the splitting
variable~$z$ for final-state splittings (first two equations).
For initial-state splittings, however,
it depends on the variable $x$, since the hard
kinematics is initiated by the reduced incoming momentum $xp_1$
and the CM frame of the hard scattering is boosted in the direction 
opposite to ${\bf p}_1$ by the velocity $(1-x)/(1+x)$ in the CM
frame of the two original incoming momenta.
The cross-section contributions take the form of convolutions 
of the hard LO cross sections
over the splitting variables~$z$ and $x$ with radiator functions
$\bar{\cal H}$, ${\cal H}$, and $H$ describing the radiation pattern.
In some of the cases it is appropriate to separate the
{\it endpoint contributions,} which correspond to the LO
kinematics, from the continuum contributions to the convolution
by employing $(\cdots)_+$ distributions. 
For the splittings $f\to f\gamma$ with real photon emission,
this procedure isolates the soft-singular contributions in the endpoint
contributions, which facilitates the cancellation of soft singularities.
The continuum parts $\bar{\cal H}$, ${\cal H}$ of the radiator functions, 
thus do not depend on $\Delta E$. 

Before moving on to the explicit results on the radiator function
$\bar{\cal H}_{f\gamma}$, etc., we want to emphasize a subtlety in the
use of the $(\cdots)_+$ distributions introduced above.
Definition (\ref{eq:plusdist}) does not make the fact explicit that
in general some kinematical variable $Q$ (such as $\tilde p_f^0$ etc.)
appears inside the
$(\cdots)_+$ symbol in addition to the integration variable~$x$, i.e.\
we have to evaluate a distribution like $[f(x,Q)]_+$, where
$Q=Q(\tilde\Phi(x))$ is a function on the phase space $\tilde\Phi(x)$ of the
hard scattering process.
In the analytical calculation of the endpoint contribution proportional
to $\delta(1-x)$,
the value of $Q$ is kept fixed in the integration over $x$.
In the numerical evaluation, the integral appears in the form
\begin{align}
& \int_0^1\rd x \int \rd \tilde\Phi(x)\,
\bigl[f\bigl(x,Q(\tilde\Phi(x))\bigr)\bigr]_+ \,\big|\M\bigl(\tilde\Phi(x)\bigr)\big|^2 \nl
&{}= \int_0^1\rd x \left[ \int \rd \tilde\Phi(x)\, f\bigl(x,Q(\tilde\Phi(x))\bigr) \, \big|\M\bigl(\tilde\Phi(x)\bigr)\big|^2
- \int \rd \tilde\Phi(1)\, f\bigl(x,Q(\tilde\Phi(1))\bigr) \,
\big|\M\bigl(\tilde\Phi(1)\bigr)\big|^2\right],
\end{align}
\ie
the variable~$Q$ inside $[f(x,Q)]_+$ has to be equal to
the $Q$~value in the corresponding matrix element $|\M\bigl(\tilde\Phi\bigr)|^2$
(see also related discussion at the end of Sect.~5.2.3 and in App.~B of
\citere{Catani:2002hc}).

In MR, \ie for small fermion masses
$m_f\ll \tilde p^0$, the radiator functions read
\begin{align}
\bar{\cal H}_{f\gamma}(\tilde p_f^0,z)\big|_{\MR} ={}&
\Pffunreg(z)\left[2\ln\left(\frac{\tilde p_f^0\De\theta}{m_f}\right)
+2\ln z -1\right] + 1-z,
\label{eq:Hcsli-ab2fg}
\\[.5em]
\bar{\cal H}_{f\bar f}(\tilde k^0,z)\big|_{\MR} ={}& 2P_{f\ga}(z) \left[
\ln\left(\frac{\tilde k^0\De\theta}{m_f}\right) 
+\ln z + \ln(1-z) \right] + 2z(1-z),
\label{eq:Hcsli-ab2ff}
\\[.5em]
{\cal H}^{f\gamma}(p_f^0,x)\big|_{\MR} ={}&
\Pffunreg(x)\left[2\ln\left(\frac{p_f^0\De\theta}{m_f}\right)-1\right] + 1-x,
\label{eq:Hcsli-fb2g}
\\[.5em]
{\cal H}^{\ga f}(k^0,x)\big|_{\MR} ={}& 2P_{f \ga}(x) \left[
\ln\left(\frac{k^0\De\theta}{m_f}\right) 
+\ln(1-x) \right] + 2x(1-x),
\label{eq:Hcsli-gb2f}
\\[.5em]
{\cal H}^{ff}(p_f^0,x)\big|_{\MR} ={}& P_{\ga f}(x) \left[ 
2\ln\left(\frac{p_f^0\De\theta}{m_f}\right)
+2\ln(1-x) -2\ln x -1 \right] + x.
\label{eq:Hcsli-fb2f}
\end{align}
The corresponding endpoint parts are given by
\begin{align}
H_{f\gamma}(\tilde p_f^0)\big|_{\MR} ={}&
-\left[2\ln\left(\frac{\De E}{\tilde p_f^0}\right)+\frac{3}{2}\right]
\left[2\ln\left(\frac{\tilde p_f^0\De\theta}{m_f}\right)-1\right]
+ 3 - \frac{2\pi^2}{3},
\label{eq:Hsli-ab2fg}
\\[.5em]
H_{f\bar f}(\tilde k^0)\big|_{\MR} ={}&
\frac{4}{3} \ln\left(\frac{\tilde k^0\De\theta}{m_f}\right)
-\frac{23}{9},
\label{eq:Hsli-ab2ff}
\\[.5em]
H^{f\gamma}(p_f^0)\big|_{\MR} ={}&
-\left[2\ln\left(\frac{\De E}{p_f^0}\right)+\frac{3}{2}\right]
\left[2\ln\left(\frac{p_f^0\De\theta}{m_f}\right)-1\right]
+ \frac{1}{2}.
\label{eq:Hsli-fb2g}
\end{align}

The radiator functions in DR can be
obtained by substituting the mass-singular terms appropriately.
For all splittings that do not involve soft singularities,
the logarithm $\ln m_f$ turns into the usual $1/\epsilon$ pole according to
the correspondence
\begin{align}
\ln m_f^2 \;\leftrightarrow\; 
\frac{(4\pi\mu^2)^\epsilon}{\epsilon\Gamma(1-\epsilon)} 
= \frac{1}{\epsilon\Gamma(1-\epsilon)} + \ln(4\pi)+\ln\mu^2 + {\cal O}(\epsilon)
= \Delta + \ln\mu^2 + {\cal O}(\epsilon),
\label{eq:collsingmassdim}
\end{align}
with the same formal definition of $\Delta$ as for UV singularities
in \refeq{eq:Delta},
but in general some finite terms change as well.  For the case
at hand, the correspondence \refeqf{eq:collsingmassdim} is exact
including finite terms for $\bar{\cal H}_{f\bar f}$ in
\refeq{eq:Hcsli-ab2ff}, ${\cal H}^{\ga f}$ in \refeq{eq:Hcsli-gb2f},
and $H_{f\bar f}$ in \refeq{eq:Hsli-ab2ff}.  For ${\cal H}^{ff}$ of
\refeq{eq:Hcsli-fb2f}, the transition to DR changes also some finite
terms:
\begin{align}
{\cal H}^{ff}(p_f^0,x)\big|_{\DR} ={}& P_{\ga f}(x) \left[ 
-\frac{(4\pi)^\epsilon}{\epsilon\Gamma(1-\epsilon)}+
2\ln\left(\frac{p_f^0\De\theta}{\mu}\right)
+2\ln(1-x) \right] + x.
\label{eq:Hcsli-fb2fdimreg}
\end{align}
For the soft-singular $f\to f\gamma$ splittings with real-photon
emission the translation to DR is somewhat more complicated owing to
the interplay between the soft and collinear singularities.  Those
radiator functions can, for instance, be easily obtained in DR upon
integrating the factorization formulae of \citere{Catani:1996jh} with
the slicing cuts (see also \citeres{Harris:2001sx,Beenakker:2002nc}).
The results for the continuum parts read
\begin{align}
\bar{\cal H}_{f\gamma}(\tilde p_f^0,z)\big|_{\DR} ={}&
\Pffunreg(z)\left[-\frac{(4\pi)^\epsilon}{\epsilon\Gamma(1-\epsilon)}+
2\ln\left(\frac{\tilde p_f^0\De\theta}{\mu}\right)
+2\ln(1-z)+2\ln z\right] + 1-z,
\label{eq:Hcsli-ab2fg-dimreg}
\\[.5em]
{\cal H}^{f\gamma}(p_f^0,x)\big|_{\DR} ={}&
\Pffunreg(x)\left[
-\frac{(4\pi)^\epsilon}{\epsilon\Gamma(1-\epsilon)}+
2\ln\left(\frac{p_f^0\De\theta}{\mu}\right)
+2\ln(1-x)\right] + 1-x,
\label{eq:Hcsli-fb2g-dimreg}
\end{align}
and the corresponding (soft-singular) endpoint parts are given by
\begin{align}
H_{f\gamma}(\tilde p_f^0)\big|_{\DR} ={}&
\left[2\ln\left(\frac{\De E}{\tilde p_f^0}\right)+\frac{3}{2}\right]
\left[\frac{(4\pi)^\epsilon}{\epsilon\Gamma(1-\epsilon)}
+2\ln\left(\frac{\mu}{\tilde p_f^0\De\theta}\right)\right]
- \frac{2\pi^2}{3}
+ \frac{13}{2} - 2\ln^2\left(\frac{\De E}{\tilde p_f^0}\right),
\label{eq:Hsli-ab2fg-dimreg}
\\[.5em]
H^{f\gamma}(p_f^0)\big|_{\DR} ={}&
\left[2\ln\left(\frac{\De E}{p_f^0}\right)+\frac{3}{2}\right]
\left[
\frac{(4\pi)^\epsilon}{\epsilon\Gamma(1-\epsilon)}
+2\ln\left(\frac{\mu}{p_f^0\De\theta}\right) \right]
+ 4 - 2\ln^2\left(\frac{\De E}{p_f^0}\right).
\label{eq:Hsli-fb2g-dimreg}
\end{align}

We end our discussion of two-cutoff slicing by considering the 
convolutions (\ref{eq:sli-ab2fg})--(\ref{eq:sli-fb2f}) in more detail.
For the final-state splittings (\ref{eq:sli-ab2fg}) and 
(\ref{eq:sli-ab2ff}), the integrations over $z$ do not involve
the hard cross section $\rd\sigma^{\LO}$, but only depend
on the event selection encoded in $\Theta_{\mathrm{cut}}$.
If the two collinear particles are not resolved, but clustered
into a single ``quasiparticle'' (\eg as part of a jet or
electromagnetic shower), only the sum of the two momenta enters
the event selection, and 
$\Theta_{\mathrm{cut}}\equiv1$ for all $z$.
In this case, the continuum parts $\bar{\cal H}$ integrate to zero
because of their appearance inside $(\cdots)_+$ distributions.
This observation is in line with the statement of the KLN theorem
which predicts the cancellation of all mass-singular contributions 
related to the considered final-state splitting if all energy-degenerate
configurations of the two collinear particles are integrated over.
For the $\gamma\to f\bar f$ splitting, the singularity in the collinear 
endpoint part (\ref{eq:Hsli-ab2ff}) cancels against its counterpart in the
one-loop correction to the photon production process $ab\to\gamma+X$;
for the $f\to f\gamma$ splitting the cancellation occurs between the
endpoint part (\ref{eq:Hsli-ab2fg}), corresponding soft-photon 
contributions in $\rd\sigma^{\soft}$,
and one-loop corrections involving photons coupling
to the external fermion line.
Observables in which collinear final-state singularities cancel
are called {\it collinear-safe}, and {\it non-collinear-safe} otherwise.
This issue frequently occurs in the identification of charged
leptons which copiously radiate photons in their direction of flight. 
We further elaborate on this point in \refse{se:lepton-photon}.

In contrast to collinear singularities from final-state splittings,
those originating from initial-state splittings do not cancel
in cross-section predictions, as can be seen explicitly in the
convolutions (\ref{eq:sli-fb2g})--(\ref{eq:sli-fb2f}) by the
fact that the hard cross sections depend on the convolution variable~$x$.
We discuss the role of those singularities in the description 
of hadronic collisions in the QCD-improved parton model in \refse{se:PDF-QED},
and in the predictions for leptonic collisions in \refse{se:lepton-photon}.

\subsubsection{Dipole subtraction}
\label{se:dipsub}

{\it Dipole subtraction} is one of the standard methods to treat IR
singularities in both NLO QCD and EW calculations.  It was
continuously extended since its original formulation for massless QCD
in \citeres{Catani:1996jh,Catani:1996vz}.  The step towards the
inclusion of massive particles was first done for photonic
corrections~\cite{Dittmaier:1999mb} and subsequently for QCD with
massive quarks~\cite{Phaf:2001gc,Catani:2002hc}.  For a comprehensive
treatment of NLO EW corrections to scattering processes, further
generalizations towards the treatment of non-collinear-safe
observables and all possible mass-singular fermion--photon splittings
were described in \citere{Dittmaier:2008md}; the modification for
decay processes was described in \citere{Basso:2015gca}.  Dipole
subtraction has been successfully automated for generic applications
in NLO calculations in various
variants~\cite{Frederix:2008hu,Hasegawa:2009tx,Czakon:2009ss,%
Frederix:2010cj,Gehrmann:2010ry,Schonherr:2017qcj} 
and is part of several multipurpose Monte Carlo generators.%
\footnote{The automation of FKS subtraction~\cite{Frixione:1995ms} has
  been accomplished as well~\cite{Frederix:2009yq,Frederix:2018nkq} and
  represents the underlying IR subtraction method of the
  \MGNLO~\cite{Alwall:2014hca} and 
{\sc POWHEG}~\cite{Nason:2004rx,Frixione:2007vw} Monte Carlo programs.}
The origin of the name {\it dipole subtraction} roots in the
construction of the subtraction function which resembles the
singularity structure of the squared matrix element $\langle
|\M_1|^2\rangle$ of real-emission corrections.  In the EW case, photon
emission amplitudes possess the most complicated IR structure, since
they are the only ones with soft {\it and} collinear singularities. In
detail, each light charged particle with momentum $p_n$ participating
in some bremsstrahlung process defines the axis of some collinear
cone around ${\bf p}_n$, in which potential photon emission is ruled
by the factorization formulae (\ref{eq:ab2fgasymp}) and
(\ref{eq:fb2gasymp}).  Constructing a subtraction function
$|\M_\sub|^2$ from a sum over those radiating particles seems natural,
but the fact that all collinear cones for photon radiation meet in the
soft region of small photon momentum complicates the construction.
The requirement that all contributions resembling the asymptotics in
the collinear cones conspire to reproduce the soft asymptotics given
in \refeq{eq:softphotonlimit} is a rather non-trivial constraint. 
Dipole subtraction employs partial fractioning
\begin{align}
\label{eq:partfracphotonoffres}
\frac{p_n p_{n'}}{(p_n k)(p_{n'}k)}
= \frac{p_n p_{n'}}{p_n k+p_{n'}k}\, \frac{1}{p_n k} 
+ \frac{p_n p_{n'}}{p_n k+p_{n'}k}\, \frac{1}{p_{n'}k} 
\end{align}
to the individual terms with $n\ne n'$ in \refeq{eq:softphotonlimit}, so that
each of the two terms on the r.h.s.\ can be attributed to
a single collinear cone.
Splitting the contributions with $n=n'$ with the help of \refeq{eq:Qcons}
expressing charge conservation, the asymptotic form (\ref{eq:softphotonlimit})
of the soft correction can be written as a sum over all
(ordered) charged-particle pairs $nn'$ with $n\ne n'$ as follows
\begin{align}
-\sum_{n,n'} Q_n \sigma_n Q_{n'} \sigma_{n'} e^2\,
\frac{p_{n'} p_n}{(p_{n'} k)(p_n k)}
&{}= -\sum_{\substack{n,n'\\ n\neq n'}} Q_n \sigma_n Q_{n'} \sigma_{n'} e^2\,
\frac{p_{n'} p_n}{(p_{n'} k)(p_n k)}
-\sum_n Q_n^2\si_n^2  e^2\,\frac{m_n^2}{(p_n k)^2}
\nn\\[.3em]
&{}= -\sum_{\substack{n,n'\\ n\neq n'}} Q_n \sigma_n Q_{n'} \sigma_{n'} e^2\,
\frac{1}{p_n k} \left[
\frac{2(p_n p_{n'})}{p_n k+p_{n'} k}
-\frac{m_n^2}{p_n k} \right].
\end{align}
In this double sum, particle $n$ defines a particular collinear cone,
while $n'$ is needed to balance charge and to get the soft limit in
the correct way. 
The subtraction function, thus, can be constructed from a sum over all
ordered pairs $nn'$ called {\it dipoles}, where $n$ is called {\it emitter} 
and $n'$ {\it spectator}.

Finally, we mention another role of the spectator in the construction
of $|\M_\sub|^2$.  In all collinear limits
(\ref{eq:ab2fgasymp})--(\ref{eq:fb2fasymp}), the original matrix
element $\M_1$ on the l.h.s.\ and the hard LO matrix elements $\M_0$
(or partial matrix element $T_0^\mu$) on the r.h.s.\ are defined on
different phase spaces. If $\M_1$ involves $N+1$ particles, then
$\M_0$ involves only $N$ particles where the two particles $i,i'$
getting collinear in $\M_1$ are merged to one ``quasiparticle''
$\tilde{i}$ with momentum $\tilde p_{i}$. To define
$|\M_\sub|^2$ on the phase space of $\M_1$, we, thus, have to find
appropriate phase-space mappings from the $(N+1)$-particle phase space
$\Phi_1$ 
with emitter~$i$ and spectator~$j$
to the $N$-particle phase space $\tilde\Phi_{0,ij}$ with the
property that $p_i\pm p'_i$ (the sign depending on the splitting type)
asymptotically approaches $\tilde p_{i}$ in the collinear limit $p_i
p'_i\to0$. Moreover, the consistency of the $N$-particle matrix elements
requires that the mappings respect overall momentum conservation and
all OS conditions including $\tilde p_{i}^2=m_{\tilde i}^2$.  These
constraints on the mappings can only be respected if more than the two
momenta $p_i$ and $p'_i$ of $\Phi_1$ are distorted.  In the
construction described below, in most cases it is sufficient to modify
the spectator momentum in addition, leaving all other momenta
unchanged.

The construction of $|\M_\sub|^2$ resembling the collinear asymptotics
of the remaining splittings that are not of bremsstrahlung type
actually do not require a construction via a sum over all
emitter--spectator pairs.  For each collinear configuration
represented by an emitter, it is sufficient to choose one spectator
particle from the initial or final state.

In the following, we give a brief account of all dipole subtraction
functions needed in NLO EW calculations for scattering processes with
external light, unpolarized, charged fermions, following closely
\citeres{Dittmaier:1999mb,Dittmaier:2008md}.  For the treatment of
massive or polarized radiating particles we refer to those papers.
Again, we support both MR and DR in the results.  For a process
$ab\to\gamma+X$ of bremsstrahlung type, the subtraction function reads
\begin{align}
|\M_\sub^{ab\to \gamma X}(\Phi_1)|^2 =
-\sum_{\substack{f,f'\\ f\neq f'}} \Qf \sigma_f Q_{f'} \sigma_{f'} e^2 \,
\gsub_{ff'}(p_f,p_{f'},k) \,
\Bigl\langle |\M_0^{ab\to X}(\tilde\Phi_{0,ff'})|^2 \Bigr\rangle,
\label{eq:m2bremsub}
\end{align}
with $f$ and $f'$ denoting all emitter and spectator fermions, respectively.
For processes not of bremsstrahlung type, no soft singularities occur, and
the subtraction function can be obtained as a single sum over
all external particles of the hard scattering process that can undergo a
collinear splitting. Nevertheless, a spectator parton is required for all
splittings to preserve momentum conservation and mass-shell conditions,
but it is possible to select any other external particle as spectator.
The corresponding contributions to the subtraction function can be constructed
from the following building blocks, 
\begin{align}
\label{eq:ab2ffsub}
|\M_{\sub,j}^{ab\to f\bar f X}(\Phi_1)|^2 
={}& \NCf \Qf^2 e^2 \, h_{f\bar f,j}^{\mu\nu}(p_f,p_{\bar f},p_j) \; 
\Bigl\langle 
T_{0,\mu}^{ab\to \gamma X}(\tilde\Phi_{0,fj})
\left(T_{0,\nu}^{ab\to \gamma X}(\tilde\Phi_{0,fj})\right)^*
\Bigr\rangle,
\\[.3em]
\label{eq:gb2fsub}
|\M_{\sub,j}^{\gamma b\to f X}(\Phi_1)|^2
={}& \Qf^2 e^2 \, h^{\gamma f}_j(k,p_f,p_j) \; 
\Bigl\langle |\M_0^{\bar fb\to X}(\tilde\Phi_{0,\gamma j})|^2 \Bigr\rangle,
\\[.3em]
\label{eq:fb2fsub}
|\M_{\sub,j}^{fb\to f X}(\Phi_1)|^2 
={}& \NCf \Qf^2 e^2 \, h^{ff,\mu\nu}_j(p_f,p'_f,p_j) \; 
\Bigl\langle T_{0,\mu}^{\gamma b\to X}(\tilde\Phi_{0,fj})
\left(T_{0,\nu}^{\gamma b\to X}(\tilde\Phi_{0,fj})\right)^* \Bigr\rangle,
\end{align}
where each of the terms stands for one specific splitting type.  Here
we specifically chose a spectator particle~$j$ from the final state.
Choosing an initial-state spectator $a$ does not change the form of
those definitions, but only the index position; by convention, the
lower spectator index~$j$ just turns into an upper index~$a$ in the
subtraction kernels~$h_{\cdots}^{\cdots}$.

Once the full subtraction function is constructed, the difference
$\langle |\M_1| \rangle^2-|\M_\sub|^2$ can be numerically integrated over the full
real-emission phase space $\Phi_1$ without any regulators, because
$\langle |\M_1| \rangle^2\sim |\M_\sub|^2$ in all IR-singular limits by construction.
To obtain the complete real corrections, we have to add back the
integral of $|\M_\sub|^2$ over $\Phi_1$, which contains the IR singularities.
If $|\M_\sub|^2$ is chosen sufficiently simple, the integration over the
degrees of freedom in $\Phi_1$ leading to the singularities can be done
analytically in some regularization scheme. To this end, the 
phase-space measure $\rd\Phi_1$ is split into the one of each hard
subspace, $\rd\tilde\Phi_{0,{ij}}$, and an effective one-particle part
$[\rd p']$ parametrizing the splitting process. 
The full cross-section contribution of the real corrections 
in the dipole subtraction approach can, thus, be schematically summarized as follows,
\begin{align}
\int\rd\sigma^{\real} = \frac{1}{2s} \int\rd\Phi_1\, \Bigl\langle |\M_1|^2 \Bigr\rangle =
\int\rd\sigma^{\real-\sub} + \int\rd\sigma^{\sub}
\end{align}
with
\begin{align}
\int\rd\sigma^{\real-\sub} &{}=
\frac{1}{2s} \int\rd\Phi_1\, \left(\Bigl\langle |\M_1|^2\Bigr\rangle -|\M_\sub|^2\right),
\label{eq:XSrealsub}
\\ 
\int\rd\sigma^{\sub} &{}=
\frac{1}{2s} \int\rd\tilde\Phi_0\,\otimes\left(\int [\rd p'] \, |\M_\sub|^2\right),
\label{eq:XSsub}
\end{align}
where $1/(2s)$ is the flux factor for massless incoming particles with
CM energy $\sqrt{s}$.  The symbol~$\otimes$ indicates that the
phase-space factorization is not necessarily an ordinary product, but
might still involve convolutions or summations.  The detailed
prescription to evaluate the r.h.s.\ of \refeq{eq:XSsub} depends on
the emitter--spectator configuration and will become clear in the
results below.  Details on the construction of the whole formalism can
be found in the original
papers~\cite{Catani:1996vz,Dittmaier:1999mb,Catani:2002hc,Dittmaier:2008md};
in this review we just outline the basic concepts and quote some final
results, following the original notation, but omitting unnecessary
clutter as much as possible.

We now go through the explicit construction of the subtraction kernels
$\gsub_{\cdots}$, $h_{\cdots}^{\cdots}$, and their integrated
counterparts for the four different cases with emitter and spectator
from the final or initial state.
In detail, the functions $\gsub_{\cdots}$, $h_{\cdots}^{\cdots}$ coincide
with the respective quantities of \citeres{Dittmaier:1999mb,Dittmaier:2008md}
in the limit of light fermions. In all cases but one, those
results are directly related to the QCD results of
\citeres{Catani:1996vz,Catani:2002hc}, from which at least the results
in DR can be obtained by an appropriate substitution {\it (abelianization)}
of the QCD couplings by its QED counterparts, as described at the
end of this section.%
\footnote{The only exception, in which the definitions of
\citeres{Dittmaier:1999mb,Dittmaier:2008md} and
\citeres{Catani:1996vz,Catani:2002hc} differ in the massless limit,
is the bremsstrahlung case of final--final type.
To be precise, the factor $(1-y)^{-1}$ in the function
$\gsub_{ij}$ of \refeq{eq:gsubij} is absent in the counterparts
given in \citeres{Catani:1996vz,Catani:2002hc}, which leads to
different non-singular terms in the corresponding integrated
dipole functions. Marking the quantities corresponding to the
definition of \citeres{Catani:1996vz,Catani:2002hc} by CS,
those are related to the quantities of this work by
$\gsub_{ij}\big|_{\mathrm{CS}} = (1-y)\gsub_{ij}$,
$\bcGsub_{ij}\big|_{\mathrm{CS}} = 
\bcGsub_{ij}+2\ln(1-z)/z+1+z$,
$\Gsub_{ij}\big|_{\mathrm{CS}} = \Gsub_{ij}-\pi^2/3+3/2$,
which modifies the relations \refeqf{eq:gsubij},
\refeqf{eq:bcGsubij}, \refeqf{eq:Gsubij}, and
\refeqf{eq:bcGsubijDR}.}

\begin{figure}
\centering
\subfloat[FS emitter and FS spectator]{\label{fig:FFdipoles}
\hspace*{.5em}
\includegraphics[scale=.7]{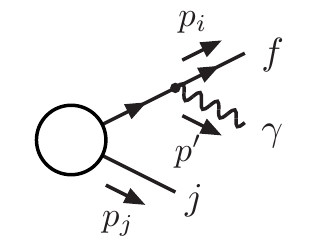}
\includegraphics[scale=.7]{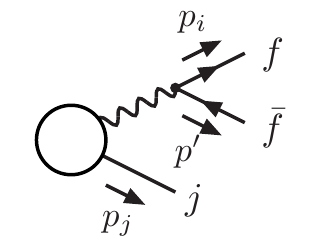}}
\hspace{4em}
\subfloat[IS emitter and FS spectator]{\label{fig:IFdipoles}
\includegraphics[scale=.7]{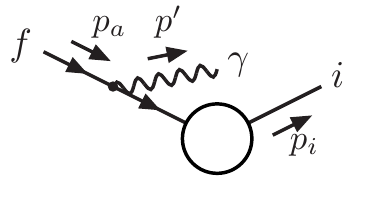}
\hspace*{-.5em}
\includegraphics[scale=.7]{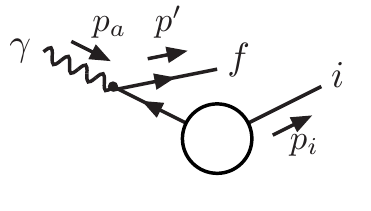}
\hspace*{-.5em}
\includegraphics[scale=.7]{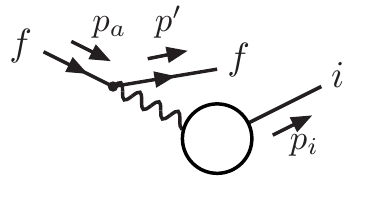}}
\\
\centering
\subfloat[FS emitter and IS spectator]{\label{fig:FIdipoles}
\includegraphics[scale=.7]{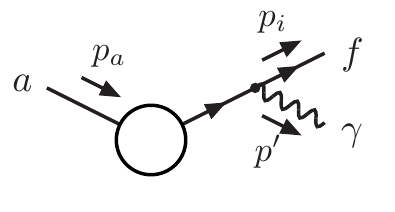}
\includegraphics[scale=.7]{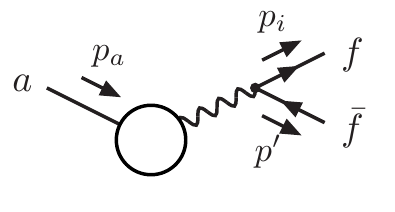}}
\hspace{4em}
\subfloat[IS emitter and IS spectator]{\label{fig:IIdipoles}
\includegraphics[scale=.7]{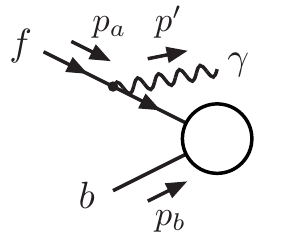}
\includegraphics[scale=.7]{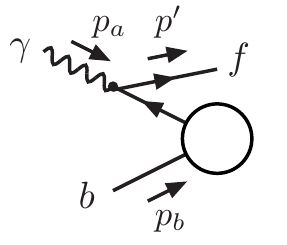}
\includegraphics[scale=.7]{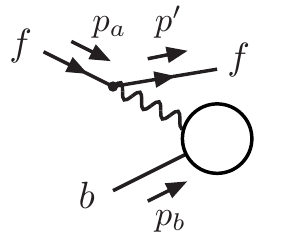}
\hspace*{1.5em}
}
\caption{Diagrams illustrating the EW dipole terms with 
emitters and spectators in the final state (FS) or initial state~(IS).}
\label{fig:dipoles}
\end{figure}

\myparagraph{(a) Final-state emitter and final-state spectator} 
The subtraction kernels $\gsub_{ij}$ and $h_{f\bar f,j}^{\mu\nu}$
with emitter $i=f$ and spectator $j$ in the final state
are graphically illustrated in \reffi{fig:FFdipoles}
and defined by
\begin{align}
\label{eq:gsubij}
\gsub_{ij}(p_i,p_j,k=p') &{}=
\frac{1}{(p_i p')(1-y)} \biggl[
\frac{2}{1-z(1-y)}-1-z \biggr],
\\[.3em]
h_{f\bar f,j}^{\mu\nu}(p_f=p_i,p_{\bar f}=p',p_j) &{}=
\frac{1}{p_i p'} \left[-g^{\mu\nu}-\frac{2}{p_i p'} \,
\bigl(z p_i-\bar zp'\bigr)^\mu\, \bigl(z p_i-\bar zp'\bigr)^\nu \right],
\end{align}
where the variables $y$ and $z$ are given by
\begin{align}
y = \frac{p_i p'}{p_i p_j + p_i p' + p_j p'}, \qquad
z = 1-\bar z = \frac{p_i p_j}{p_i p_j + p_j p'}.
\label{eq:yijzij}
\end{align}
It is straightforward to check that the functions $\gsub_{ij}$ 
and $h_{f\bar f,j}^{\mu\nu}$ have the required 
asymptotic behaviour in the soft ($p'\to0$) and collinear ($p_i p'\to0$) limits,
in which $y\to0$ and $z$ plays the role of the splitting variable~$z$
defined in \refeq{eq:zx}.
Note that we did not include explicit mass terms or terms of ${\cal O}(\epsilon)$
in the form of $\gsub_{ij}$ and $h_{f\bar f,j}^{\mu\nu}$ given above, because
those terms do not contribute in the (IR-finite) phase-space integral of the
difference $\langle |\M_1| \rangle^2-|\M_\sub|^2$.
Note, however,
that those mass and $\epsilon$~terms are essential in the full construction and
integration of the subtraction function
over the singular degrees of freedom in~$[\rd p']$.

For the evaluation of $|\M_0^{ab\to X}(\tilde\Phi_{0,ij})|^2$
and $T_{0,\mu}^{ab\to \gamma X}(\tilde\Phi_{0,fj})$
we have to define the mapping $\tilde\Phi_{0,ij}$ from $\Phi_1$ to $\Phi_0$. 
For $i$ and $j$ both in the final state, it is possible
to redefine only the momenta $p_i$, $p'$, and $p_j$,
leaving all other momenta $k_n$ in the process unaffected, $\tilde k_n=k_n$.
The momenta $\tilde p_i$ and $\tilde p_j$ are chosen as
\begin{equation}
\tilde p_i^\mu = p_i^\mu + p^{\prime\mu} - \frac{y}{1-y} p_j^\mu,
\qquad
\tilde p_j^\mu = \frac{1}{1-y} p_j^\mu,
\label{eq:tmomij0}
\end{equation}
where $\tilde p_i$ is the momentum of the quasiparticle $\tilde{i}$
replacing the collinear particle pair $i,i'$.
Obviously, the new momenta have the required behaviour
$\tilde p_i^\mu \to p_i^\mu + p^{\prime\mu}$ and $\tilde p_j^\mu \to p_j^\mu$
in the IR limits, and the momentum
\begin{equation}
P_{ij}=p_i+p_j+p'=\tilde p_i+\tilde p_j
\end{equation}
is left invariant by the phase-space mapping. 
In particular, the 3-particle invariant mass
$P_{ij}^2\ge 0$ is not affected by the mapping.
The above definitions comprise all ingredients for the evaluation of the
$ij$ contribution to the difference
$\langle |\M_1| \rangle^2-|\M_\sub|^2$.

The second step in the subtraction procedure consists in adding back
the contribution $\rd\sigma^{\sub}$ of \refeq{eq:XSsub},
analytically integrated over $[\rd p']$.
This integration implicitly involves an integration
over the azimuthal angle $\phi$ 
around the collinear axis, so that $h_{f\bar f,j}^{\mu\nu}$ can be replaced
by its $\phi$-averaged form, as explained already in our discussion of the
asymptotics in the collinear limit in the transition from
\refeq{eq:ab2ffasymp} to \refeq{eq:ab2ffasymp2}.
The singular integration with regulators is quite
complicated in general, but has to be done only once owing to the process
independence of the integration over the singular degrees of freedom
in $[\rd p']$.
We therefore refer to the original papers
for the details and just quote the final results,
\begin{align}
\int\rd\sigma^{\sub,ij}_{ab\to f\gamma X}
={}& {-}\frac{\alpha}{2\pi} Q_i\sigma_i Q_j\sigma_j  \,
\int\rd\sigma^{\LO}_{ab\to fX}(\tilde\Phi_{0,ij})
\int_0^1 \rd z\,
\left\{ \Gsub_{ij}(P_{ij}^2) \de(1-z)
+ \left[\bcGsub_{ij}(P_{ij}^2,z)\right]_+ \right\}
\nn\\[.3em]
& {}\times
\Theta_{\mathrm{cut}}\Bigl(p_i=z\tilde p_i,p'=(1-z)\tilde p_i,\tilde p_j\Bigr),
\label{eq:sigmasubij}
\\[.5em]
\int\rd\sigma^{\sub,fj}_{ab\to f\bar fX}
={}& \NCf\, \frac{\Qf^2\alpha}{2\pi} \,
\int\rd\sigma^{\LO}_{ab\to\gamma X}(\tilde\Phi_{0,fj})
\int_0^1 \rd z\, 
\left\{ \Hsub_{f\bar f,j}(P_{ij}^2) \, \de(1-z)
+ \left[\bcHsub_{f\bar f,j}(P_{ij}^2,z)\right]_+ \right\}
\nn\\[.3em]
& {}\times
\Theta_{\mathrm{cut}}\Bigl(p_f=z\tilde p_i,
p_{\bar f}=(1-z)\tilde p_i,\tilde p_j\Bigr).
\end{align}
The arguments of the cut functions $\Theta_{\mathrm{cut}}$ again
indicate which momenta are subject to phase-space cuts and, in
particular, show how the collinear momentum of the splitting particle
is shared by the two particles resulting from the splitting.  As
explained already in \refse{se:twocutoffslicing}, this momentum
assignment is essential in the calculation of non-collinear-safe
observables, where cuts (or histogram boundaries in differential
distributions) restrict the $z$~integration in a non-trivial way, so
that the continuum parts quantified by $\bcGsub_{ij}$ and
$\bcHsub_{f\bar f,j}$ do not integrate to zero in general.  In this
context we emphasize that the cut functions $\Theta_{\mathrm{cut}}$
are implicitly also part of the differential subtraction functions
(\ref{eq:m2bremsub}) and (\ref{eq:ab2ffsub}), \ie in the subtraction
terms the momenta in the arguments of $\Theta_{\mathrm{cut}}$ are
subject to the phase-space cuts, not the original momenta $p_i$, $p'$,
and $p_j$ of $\Phi_1$.  The treatment of non-collinear-safe
observables is described in detail in \citere{Dittmaier:2008md}.

In MR, the integrated subtraction kernels $\bcGsub_{ij}$,
$\Gsub_{ij}$, $\bcHsub_{f\bar f,j}$, and $\Hsub_{f\bar f,j}$ are
explicitly given by
\begin{align}
\bcGsub_{ij}(P^2,z)\big|_{\MR} &{}=
\Pffunreg(z)\,\left[ 
\ln\left(\frac{P^2}{m_i^2}\right) + \ln z
-1\right]
+(1+z)\ln(1-z) + 1-z,
\label{eq:bcGsubij}
\\
\Gsub_{ij}(P^2)\big|_{\MR} &{}= {\cal L}(P^2,m_i^2)\big|_{\MR} - \frac{\pi^2}{3} + \frac{3}{2},
\label{eq:Gsubij}
\\
\bcHsub_{f\bar f,j}(P^2,z)\big|_{\MR} &{}= P_{f\ga}(z) \left[
\ln\left(\frac{P^2}{m_f^2}\right) +\ln z + \ln(1-z)
-1\right] + 2z(1-z),
\\
\Hsub_{f\bar f,j}(P^2)\big|_{\MR} &{}=
\frac{2}{3}\ln\left(\frac{P^2}{m_f^2}\right) - \frac{16}{9},
\end{align}
where we have introduced the auxiliary function
\begin{align}
{\cal L}(P^2,m^2)\big|_{\MR} =
\ln\left(\frac{m^2}{P^2}\right)
\ln\left(\frac{m_\gamma^2}{P^2}\right)
+ \ln\left(\frac{m_\gamma^2}{P^2}\right)
- \frac{1}{2}\ln^2\left(\frac{m^2}{P^2}\right)
+ \frac{1}{2}\ln\left(\frac{m^2}{P^2}\right).
\label{eq:Lmassreg}
\end{align}

For the subtraction kernels $\bcHsub_{f\bar f,j}$ and $\Hsub_{f\bar
  f,j}$ of the $\gamma\to f\bar f$ splitting, the transition from MR
to DR follows the simple correspondence
(\ref{eq:collsingmassdim}) including finite terms.  For $\bcGsub_{ij}$
the result in DR reads
\begin{align}
\label{eq:bcGsubijDR}
\bcGsub_{ij}(P^2,z)\big|_{\DR} ={}&
\Pffunreg(z)\,\left[ 
-\frac{(4\pi)^\epsilon}{\epsilon\Gamma(1-\epsilon)}
+ \ln\left(\frac{P^2}{\mu^2}\right) 
+\ln z+2\ln(1-z) \right]
+(1+z)\ln(1-z) + 1-z.
\end{align}
The result for $\Gsub_{ij}$ takes the same form as in
\refeq{eq:Gsubij}, but the auxiliary function ${\cal L}$ in DR is
given by (see, e.g., appendix of \citere{Basso:2015gca})
\begin{align}
{\cal L}(P^2,0)\big|_{\DR} &{}= 
\Gamma(1+\epsilon)
\left(\frac{4\pi\mu^2}{P^2}\right)^\epsilon\,
\left( \frac{1}{\epsilon^2} +\frac{3}{2\epsilon} \right) +2
\nn\\
&{}= 
\frac{(4\pi)^\epsilon}{\Gamma(1-\epsilon)} 
\left[ \frac{1}{\epsilon^2} +\frac{3}{2\epsilon} 
+\frac{1}{\epsilon}\ln\left(\frac{\mu^2}{P^2}\right)
\right] 
+ \frac{1}{2}\ln^2\left(\frac{\mu^2}{P^2}\right)
+ \frac{3}{2}\ln\left(\frac{\mu^2}{P^2}\right) + \frac{\pi^2}{6}+ 2.
\label{eq:Ldimreg}
\end{align}

\myparagraph{(b) Initial-state emitter and final-state spectator}
The subtraction kernels $\gsub_{ai}$, $h^{\ga f}_{i}$, and $h_{i}^{ff,\mu\nu}$ 
with an initial-state emitter $a$ and a final-state spectator $i$,
are graphically illustrated in \reffi{fig:IFdipoles}
and defined by
\begin{align}
\gsub_{ai}(p_a,p_i,k=p') &{}=
\frac{1}{(p_a p')x} \biggl[ \frac{2}{2-x-z}-1-x \biggr],
\\[.3em]
h^{\ga f}_{i}(k=p_a,p_f=p',p_i) &{}= \frac{P_{f\ga}(x)}{(p_a p')x},
\\[.3em]
h_{i}^{ff,\mu\nu}(p_f=p_a,p'_f=p',p_i) &{}=
\frac{1}{p_a p'} \Biggl[-g^{\mu\nu}
+\frac{2}{(p_a p')x^2 z} \,
\bigl(z p'-\bar{z}p_i\bigr)^\mu \,
\bigl(z p'-\bar{z}p_i\bigr)^\nu
\Biggr],
\end{align}
with the kinematical variables
\begin{align}
x = \frac{p_a p_i + p_a p' - p_i p'}{p_a p_i + p_a p'}, \qquad
z = 1-\bar z=\frac{p_a p_i}{p_a p_i + p_a p'}.
\label{eq:xiazia}
\end{align}
In the soft ($p'\to0$) and collinear ($p_a p'\to0$) limits, $z\to1$
and $x$ plays the role of the splitting variable~$x$ defined in
\refeq{eq:zx}.  The momentum mapping from $\Phi_1$ to the reduced
phase space $\tilde\Phi_{0,ia}$ is given by
\begin{equation}
\tilde p_i^\mu = p_i^\mu+p^{\prime\mu}-(1-x)p_a^\mu, \qquad
\tilde p_a^\mu = x p_a^\mu,
\label{eq:tpitpa0}
\end{equation}
with all other momenta of the process left unchanged.
This momentum redefinition changes the initial state,
so that the original CM frame and the one of the hard process
(with $xp_a$ as incoming momentum) are related by a boost along
the beam axis with the relative velocity $(1-x)/(1+x)$.
The OS relations $\tilde p_i^2=\tilde p_a^2=0$ 
as well as momentum conservation,
\begin{equation}
P_{ia}=p_i+p'-p_a=\tilde p_i-\tilde p_a,
\label{eq:Pia}
\end{equation}
are easily seen to hold, and the virtuality $P_{ia}^2<0$ is left
unchanged by the momentum mapping.
Together with the hard scattering amplitudes $\M_0$
(or partial amplitudes $T^\mu_0$),
these ingredients define the considered contributions to
$|\M_\sub|^2$ completely, so that the contribution to
$\rd\sigma^{\real-\sub}$ of
\refeq{eq:XSrealsub} can be evaluated numerically.

In order to evaluate the contribution to $\rd\sigma^{\sub}$ of
\refeq{eq:XSsub}, the subtraction kernels have to be
analytically integrated over the $[\rd p']$.
In view of this, the same comments as made for the previous
case of final-state emitter and spectator apply, up to one point.
Now the hard LO cross section depends on $x$, so that the
$[\rd p']$~integration cannot be carried out completely analytically.
The process-dependent $x$~integration has to be left for a
numerical treatment. 
Since the $x$~integration is soft singular near its endpoint at
$x\to1$ for the $f\to f\gamma$ splitting with a real photon, 
it is convenient to split off this endpoint contribution
by a $(\cdots)_+$ prescription.
The resulting cross-section contributions read
\begin{align}
\int\rd\sigma^{\sub,ai}_{fb\to \gamma X}
={}& 
{-}\frac{\alpha}{2\pi} Q_a\sigma_a Q_i\sigma_i  
\int_0^1\rd x
\int\rd\sigma^{\LO}_{fb\to X}\bigl(\tilde\Phi_{0,ai}(x)\bigr) \,
\left\{ \left[\cGsub_{ai}(P_{ia}^2,x)\right]_+
+\delta(1-x)\, \Gsub_{ai}(P_{ia}^2) \right\},
\\[.5em]
\int\rd\sigma^{\sub,\gamma i}_{\gamma b\to fX}
={}& \NCf\,
\frac{\Qf^2\alpha}{2\pi} \int_0^1\rd x\,
\int\rd\sigma^{\LO}_{\bar fb\to X}\bigl(\tilde\Phi_{0,ai}(x)\bigr) \,
\cHsub^{\ga f}_{i}(P_{ia}^2,x), 
\\[.5em]
\int\rd\sigma^{\sub,fi}_{fb\to fX} ={}&
\frac{\Qf^2\alpha}{2\pi} \int_0^1\rd x
\int\rd\sigma_{\gamma b\to X}\bigl(\tilde\Phi_{0,ai}(x)\bigr) \,
\cHsub^{ff}_{i}(P_{ia}^2,x),
\end{align}
with the mass-regularized integrated subtraction kernels
\begin{align}
\cGsub_{ai}(P^2,x)\big|_{\MR} &{}= \Pffunreg(x)
\left[\ln\left(\frac{-P^2}{m_f^2}\right) -\ln x -1\right]
-\frac{2\ln(2-x)}{1-x}+(1+x)\ln(1-x) + 1-x,
\label{eq:cGsubai}
\\[.5em]
\Gsub_{ai}(P^2)\big|_{\MR} &{}=
{\cal L}(-P^2,m_f^2)\big|_{\MR} + \frac{\pi^2}{6} - 1,
\label{eq:Gsubai}
\\[.5em]
\cHsub^{\ga f}_{i}(P^2,x)\big|_{\MR} &{}= P_{f\ga}(x) \left[
\ln\left(\frac{-P^2}{m_f^2} \right) +\ln(1-x) - \ln x \right]
     + 2x(1-x),
\\[.5em]
\cHsub^{ff}_i(P^2,x)\big|_{\MR} &{}= P_{\ga f}(x) \left[
\ln\left(\frac{-P^2}{m_f^2}\right) +\ln(1-x) - 3\ln x -1\right] + x.
\end{align}

The subtraction kernel $\cHsub^{\ga f}_{i}$ translates into DR
according to \refeq{eq:collsingmassdim} including finite terms.  In
DR, the kernels $\cGsub_{ai}$ and $\cHsub^{ff}_i$ read
\begin{align}
\cGsub_{ai}(P^2,x)\big|_{\DR} ={}& \Pffunreg(x)
\left[
-\frac{(4\pi)^\epsilon}{\epsilon\Gamma(1-\epsilon)}
+\ln\left(\frac{-P^2}{\mu^2}\right)
-\ln x+2\ln(1-x) \right]
-\frac{2\ln(2-x)}{1-x}+(1+x)\ln(1-x) + 1-x,
\\
\cHsub^{ff}_i(P^2,x)\big|_{\DR} ={}& P_{\ga f}(x) \left[
-\frac{(4\pi)^\epsilon}{\epsilon\Gamma(1-\epsilon)}
+\ln\left(\frac{-P^2}{\mu^2}\right) +\ln(1-x) - \ln x \right] + x,
\end{align}
and $\Gsub_{ai}$ is evaluated as in \refeq{eq:Gsubai}, but with ${\cal L}$ given in
\refeq{eq:Ldimreg}.

\myparagraph{(c) Final-state emitter and initial-state spectator}
The subtraction kernels $\gsub_{ia}$ and $h_{f\bar f}^{a,\mu\nu}$
with a final-state emitter $i$ and an initial-state spectator $a$
are graphically illustrated in \reffi{fig:FIdipoles}
and defined by
\begin{align}
\gsub_{ia}(p_i,p_a,k=p') &{}=
\frac{1}{(p_i p')x} \left[ \frac{2}{2-x-z}-1-z \right],
\\[.3em]
h_{f\bar f}^{a,\mu\nu}(p_f=p_i,p_{\bar f}=p',p_a) &{}=
\frac{1}{p_i p'} \left[-g^{\mu\nu}-\frac{2}{p_i p'} \,
\bigl(z p_i-\bar zp'\bigr)^\mu \,\bigl(z p_i-\bar zp'\bigr)^\nu \right],
\end{align}
with the kinematical variables defined in \refeq{eq:xiazia}.
The whole kinematical construction of new momenta $\tilde p_a$ and $\tilde p_i$
is the same as in the previous case, since only the roles of $a$ and $i$
as emitter or spectator are interchanged, \ie Eqs.~(\ref{eq:tpitpa0})
and (\ref{eq:Pia}) remain valid.
The interpretation of the variables $x$ and $z$, however, changes.
In the soft ($p'\to0$) and collinear ($p_i p'\to0$)
limits with $i$ as emitter, $x\to1$ and $z$ plays the role of the splitting
variable defined in \refeq{eq:zx}.

The contributions to $\rd\sigma^{\sub}$ of \refeq{eq:XSsub}
can be worked out in a way similar to the previous case where
the roles of emitter and spectator are interchanged.
The results take the form of a convolution over $x$, and for
the $f\to f\gamma$ splitting with a real photon the soft-singular
endpoint contribution is again separated by a $(\cdots)_+$ prescription.
In order to cover the case of non-collinear-safe observables,
however, a new feature arises, since the information on both the
variable~$x$ and the
splitting variable~$z$, which rules the momentum share in the 
configuration of the collinear particle pair, should be made
accessible in the numerical evaluation.
However, the $z$~integration is soft singular at its endpoint
at $z\to1$, which renders a plain numerical integration over $z$
rather inconvenient. The solution to this problem is again to
introduce a $(\cdots)_+$ prescription in the $z$~integration as well.
The resulting distribution $(\cdots)_+^{(x,z)}$ for the double integral
is defined as
\begin{align}
\int_0^1\rd x\int_0^1\rd z\,\Bigl[f(x,z)\Bigr]_+^{(x,z)} g(x,z)
= \int_0^1\rd x\int_0^1\rd z\,f(x,z) \,
\bigl[g(x,z)-g(x,1)-g(1,z)+g(1,1)\bigr],
\end{align}
where $g(x,z)$ is 
some test function of two independent variables.
Further details on the introduction and the evaluation of this
distribution can be found in \citere{Dittmaier:2008md}.
Using this prescription, the cross-section contributions
to $\rd\sigma^{\sub}$ can be written as
\begin{align}
\int\rd\sigma^{\sub,ia}_{ab\to f\gamma X}
={}& 
{-}\frac{\alpha}{2\pi} Q_a\sigma_a Q_i\sigma_i  
\int_0^1\rd x
\int\rd\sigma^{\LO}_{ab\to fX}\bigl(\tilde\Phi_{0,ia}(x)\bigr) 
\int_0^1\rd z \,
\biggl\{ \left[\bgsub_{ia}(x,z)\right]_+^{(x,z)}
\nn\\[.3em] & {} 
+\delta(1-z)\, \left[\cGsub_{ia}(P_{ia}^2,x)\right]_+
+\delta(1-x)\, \left[\bcGsub_{ia}(P_{ia}^2,z)\right]_+ 
+\delta(1-x)\, \delta(1-z)\,\Gsub_{ia}(P_{ia}^2) 
\biggr\}
\nn\\[.3em] & {} \times
\,
\Theta_{\mathrm{cut}}\Bigl(p_f=z\tilde p_i(x),k=(1-z)\tilde p_i(x)\}\Bigr)\,,
\label{eq:sigmasubia}
\\[.5em]
\int\rd\sigma^{\sub,fa}_{ab\to f\bar fX}
={}& \NCf\, \frac{\Qf^2 \alpha}{2\pi}  \,
\int_0^1\rd x\,
\int\rd\sigma^{\LO}_{ab\to \gamma X}\bigl(\tilde\Phi_{0,ia}(x)\bigr)
\int_0^1 \rd z\,
\biggl\{ \, \left[\bar h^a_{f\bar f}(x,z)\right]^{(x,z)}_+
\nn\\[.3em]
& \hspace*{0em} {} 
+ \de(1-z)\, \left[\cHsub^a_{f\bar f}(P_{ia}^2,x)\right]_+
+ \de(1-x)\, \left[\bcHsub^a_{f\bar f}(P_{ia}^2,z)\right]_+ 
+ \de(1-x)\, \de(1-z)\, \Hsub^a_{f\bar f}(P_{ia}^2)
\biggr\}
\nn\\[.3em]
& {} \times
\Theta_{\mathrm{cut}}\Bigl(p_f=z\tilde p_i(x),p_{\bar f}=(1-z)\tilde p_i(x)\}\Bigr),
\end{align}
where the arguments of the step function $\Theta_{\mathrm{cut}}$
again indicate which momenta resulting from the final-state splitting
are subject to phase-space cuts.

Using MR, 
the various subtraction kernels are given by
\begin{align}
\bgsub_{ia}(x,z)\big|_{\MR} ={}&
\frac{1}{1-x}\left(\frac{2}{2-x-z}-1-z\right),
\\
\cGsub_{ia}(P^2,x)\big|_{\MR}  ={}&
\frac{1}{1-x}\left[2\ln\biggl(\frac{2-x}{1-x}\biggr)-\frac{3}{2}\right],
\\
\bcGsub_{ia}(P^2,z)\big|_{\MR} ={}&
\Pffunreg(z)\left[\ln\left(\frac{-P^2}{m_f^2}\right)+\ln z-1\right]
-\frac{2\ln(2-z)}{1-z}+(1+z)\ln(1-z)+ 1-z,
\label{eq:bcGsubia}
\\
\Gsub_{ia}(P_{ia}^2)\big|_{\MR} ={}&
{\cal L}(-P^2,m_f^2)\big|_{\MR}
- \frac{\pi^2}{2} + \frac{3}{2},
\label{eq:Gsubia}
\\
\bar h^a_{f\bar f}(x,z)\big|_{\MR} ={}& \frac{x}{1-x}P_{f\ga}(z),
\\
\cHsub^a_{f\bar f}(P^2,x)\big|_{\MR} ={}& \frac{2x}{3(1-x)},
\\
\bcHsub^a_{f\bar f}(P^2,z)\big|_{\MR} ={}& P_{f\ga}(z)
\left[\ln\left(\frac{-P^2}{m_f^2}\right) +\ln z+\ln(1-z)
-1\right] + 2z(1-z),
\\
\Hsub^a_{f\bar f}(P^2)\big|_{\MR} ={}& \frac{2}{3} \ln\left(\frac{-P^2}{m_f^2}\right)
-\frac{16}{9}.
\end{align}
In the transition to DR, the non-singular functions remain unchanged.
The mass-singular integrated $\gamma\to f\bar f$ splitting kernels
$\bcHsub^a_{f\bar f}$ and $\Hsub^a_{f\bar f}$  translate into DR
according to \refeq{eq:collsingmassdim}.  For the mass-singular $f\to
f\gamma$ splitting kernels the translation is more complicated; in DR
$\bcGsub_{ia}$ reads
\begin{align}
\bcGsub_{ia}(P^2,z)\big|_{\DR} ={}&
\Pffunreg(z)\left[
-\frac{(4\pi)^\epsilon}{\epsilon\Gamma(1-\epsilon)}
+\ln\left(\frac{-P^2}{\mu^2}\right)
+\ln z+2\ln(1-z)\right]
-\frac{2\ln(2-z)}{1-z}+(1+z)\ln(1-z)+ 1-z,
\end{align}
and $\Gsub_{ia}$ has to be evaluated 
as in \refeq{eq:Gsubia}
but with ${\cal L}$ given in \refeq{eq:Ldimreg}.

\myparagraph{(d) Initial-state emitter and initial-state spectator}
The subtraction kernels $\gsub_{ab}$, $h^{\ga f,a}$, and $h^{ff,a,\mu\nu}$
with emitter $a$ and spectator $b$ in the initial state
are graphically illustrated in \reffi{fig:IIdipoles}
and defined by
\begin{align}
\gsub_{ab}(p_a,p_b,k=p') &{}=
\frac{1}{(p_a p')x} \biggl[ \frac{2}{1-x}-1-x \biggr],
\\[.3em]
h^{\ga f,b}(k=p_a,p_f=p',p_b) &{}= \frac{P_{f\ga}(x)}{(p_a p')x},
\\[.3em]
h^{ff,b,\mu\nu}(p_f=p_a,p'_f=p',p_b) &{}=
\frac{1}{p_a p'} \Biggl[-g^{\mu\nu}
+\frac{2(1-x)}{(p' p_b)x^2 y} \,
\bigl(p'-yp_b\bigr)^\mu \, \bigl(p'-yp_b\bigr)^\nu
\Biggr],
\end{align}
with the kinematical variables
\begin{align}
x = \frac{p_a p_b-p_a p'-p_b p'}{p_a p_b}, \qquad
y = \frac{p_a p'}{p_a p_b}.
\end{align}
In the soft $(p'\to0)$ and collinear $(p_a p'\to0)$ limits, $y\to0$ and $x$
takes over the role of the splitting variable of \refeq{eq:xiazia}.
If both emitter and spectator are from the initial state, a mere
modification of the emitter and spectator momenta in the mapping $\Phi_1\to\tilde\Phi_0$
would change the direction of the
beam axis, so that it is more appropriate to modify only the emitter
momentum $p_a$ and all outgoing momenta $k_n$, while keeping the spectator
momentum $p_b$ fixed.
Defining the new incoming momenta according to
\begin{equation}
\tilde p_a^\mu = x p_a^\mu, \qquad
\tilde p_b^\mu = p_b^\mu, \qquad
\end{equation}
the total outgoing momentum for the hard scattering process (\ie 
without $p'$ from the splitting) before and after the momentum mapping is
\begin{align}
P_{ab} = p_a + p_b - p' = \sum_n k_n, \qquad
\tilde P_{ab}^\mu = x_{ab} p_a^\mu + p_b^\mu = \sum_n \tilde k_n,
\end{align}
respectively. Obviously, the OS conditions $\tilde p_a^2=\tilde p_b^2=0$ are fulfilled.
Moreover, by construction the invariant mass squared of the produced particles, 
$P_{ab}^2>0$, does not change by the momentum mapping,
$P_{ab}^2=\tilde P_{ab}^2$, so that the mapping of all $k_n$ to $\tilde k_n$
can be accomplished by a Lorentz transformation
\begin{equation}
\tilde k_n^\mu = \Lambda^\mu_{\phantom{\mu}\nu} k_n^\nu
\end{equation}
with 
\begin{equation}
\Lambda^\mu_{\phantom{\mu}\nu} = g^\mu_{\phantom{\mu}\nu}
-\frac{(P_{ab}+\tilde P_{ab})^\mu(P_{ab}+\tilde P_{ab})_\nu}
{P_{ab}^2+P_{ab}\tilde P_{ab}}
+\frac{2\tilde P_{ab}^\mu P_{ab,\nu}}{P_{ab}^2},
\label{eq:LT}
\end{equation}
and all outgoing particles from the hard process stay on their
mass shell, $\tilde k_n^2=k_n^2$.
This completes the construction of the respective contributions
to the subtraction function $|\M_\sub|^2$ needed for the 
numerical evaluation of \refeq{eq:XSrealsub}.

The contributions to $\rd\sigma^{\sub}$ of \refeq{eq:XSsub}
can be worked out without further complications and
take the form of one-dimensional convolutions over $x$,
and the soft-singular endpoint contribution 
of the $f\to f\gamma$ splitting with a real photon 
is again separated by a $(\cdots)_+$ prescription.
The corresponding results are
\begin{align}
\int\rd\sigma^{\sub,ab}_{fb\to \gamma X}
={}& 
{-}\frac{\alpha}{2\pi} Q_a\sigma_a Q_b\sigma_b  
\int_0^1\rd x
\int\rd\sigma^{\LO}_{fb\to X}\bigl(\tilde\Phi_{0,ab}(x)\bigr) \,
\left\{ \left[\cGsub_{ab}(s,x)\right]_+
+\delta(1-x)\, \Gsub_{ab}(s) \right\},
\\[.5em]
\int\rd\sigma^{\sub,\gamma b}_{\gamma b\to fX}
={}& \NCf\,
\frac{\Qf^2\alpha}{2\pi} \int_0^1\rd x
\int\rd\sigma^{\LO}_{\bar fb\to X}(\tilde\Phi_{0,ab}(x)\bigr) \,
\cHsub^{\ga f,b}(s,x),
\\[.5em]
\int\rd\sigma^{\sub,fb}_{fb\to fX} ={}&
\frac{\Qf^2\alpha}{2\pi} \int_0^1\rd x
\int\rd\sigma^{\LO}_{\gamma b\to X}\bigl(\tilde\Phi_{0,ab}(x)\bigr)\,
\cHsub^{ff,b}(s,x),
\end{align}
where $s=(p_a+p_b)^2$ is the squared CM energy in the original process.

In MR the corresponding integrated subtraction
kernels are given by
\begin{align}
\cGsub_{ab}(s,x)\big|_{\MR} &{}=
\Pffunreg(x)\left[\ln\left(\frac{s}{m_f^2}\right)-1\right] + 1-x,
\label{eq:cGsubab}
\\
\Gsub_{ab}(s)\big|_{\MR} &{}= {\cal L}(s,m_f^2)\big|_{\MR} -\frac{\pi^2}{3} + 2, 
\label{eq:Gsubab}
\\
\cHsub^{\ga f,b}(s,x)\big|_{\MR} &{}= P_{f \ga}(x) \left[
\ln\left(\frac{s}{m_f^2}\right) +2\ln(1-x)
\right] + 2x(1-x),
\\
\cHsub^{ff,b}(s,x)\big|_{\MR} &{}= P_{\ga f}(x) \left[
\ln\left(\frac{s}{m_f^2}\right) +2\ln(1-x)-2\ln x -1
\right] + x.
\end{align}

In DR, $\cHsub^{\ga f,b}$ 
can be obtained by the correspondence (\ref{eq:collsingmassdim}) including finite terms,
while $\cGsub_{ab}$ and $\cHsub^{ff,b}$ are given by
\begin{align}
\cGsub_{ab}(s,x)\big|_{\DR} &{}=
\Pffunreg(x)\left[
-\frac{(4\pi)^\epsilon}{\epsilon\Gamma(1-\epsilon)}
+\ln\left(\frac{s}{\mu^2}\right) +2\ln(1-x) \right] + 1-x,
\\
\cHsub^{ff,b}(s,x)\big|_{\DR} &{}= P_{\ga f}(x) \left[
-\frac{(4\pi)^\epsilon}{\epsilon\Gamma(1-\epsilon)}
+\ln\left(\frac{s}{\mu^2}\right) +2\ln(1-x)
\right] + x,
\end{align}
and $\Gsub_{ab}$ has to be evaluated as in \refeq{eq:Gsubab},
but with the dimensionally
regularized version of ${\cal L}$ given in \refeq{eq:Ldimreg}.

\vspace{1em}

We end this brief introduction to dipole subtraction by mentioning a
byproduct of the formalism. The subtraction function $|\M_\sub|^2$
resembles all IR~singularities in the squared real-emission
amplitudes, so that $\rd\sigma^{\sub}$ of \refeq{eq:XSsub} comprises
all IR-singular real corrections to cross sections.  As stated by the
KLN theorem, all IR-singular (virtual + real) corrections to
differential cross sections cancel for processes with neutral
initial-state particles if final-state particles are treated in a
collinear-safe way.  This implies that the IR-singular contributions
of the virtual corrections $\rd\sigma^{\virt}$ are the same as in
$-\rd\sigma^{\sub}$ for processes with neutral initial-state
particles.  The generalization of this statement on
$\rd\sigma^{\virt}$ to processes with charged initial-state particles
can be obtained via {\it crossing}, \ie for a given process first
consider all particles outgoing to derive the virtual IR~singularities
and then apply crossing relations to the result in order to restore
the original initial state.  This strategy was used in
\citere{Catani:2000ef} to predict the complete IR~structure of virtual
NLO QCD corrections in DR or MR for any generic process.  It is
straightforward to transfer this result to NLO EW corrections by
``abelianization''.  Using the notation of \citere{Catani:2000ef} for
QCD amplitudes, the mass-singular NLO EW corrections can be obtained
by the substitutions
\begin{align}
g_{\mathrm{s}} \to e, \quad
\alphas \to \alpha, \quad
{\bf T}_j\cdot {\bf T}_k \to Q_j\sigma_j Q_k\sigma_k, \quad
C_{\mathrm{A}}\to0, \quad
C_{\mathrm{F}}\to \Qf^2, \quad 
T_{\mathrm{R}}\to \NCf\Qf^2.
\end{align}
Note, however, that this procedure only produces all mass-singular terms
that have a counterpart in real-emission processes. 
For instance, potentially mass-singular contributions resulting from electric
charge renormalization are not concerned.

\subsection{Electromagnetic corrections to parton distribution functions}
\label{se:PDF-QED}

\subsubsection{Factorization of photonic initial-state singularities}

In quantum field theory, the cancellation of IR divergences is generally
ruled by the
KLN theorem~\cite{Kinoshita:1962ur,Lee:1964is}.
The key feature for the cancellation is the necessary level of inclusiveness
in the particle configuration w.r.t.\ the degeneracy in energy.
For soft singularities, the degeneracy concerns states with different
numbers of soft massless particles, \ie gluons or photons.
For NLO EW corrections this means that soft singularities 
always cancel between virtual EW and real photonic corrections
in physical (IR-finite) observables, a statement that is analogous to the
Bloch--Nordsieck theorem of QED~\cite{Bloch:1937pw}.
For collinear singularities, a complete compensation only takes place
if the collinear configurations of massless (or light) particles or partons
are treated inclusively in the event selection.
Phrased in more simple words, distributing the total momentum of the
collinear many-particle state to the individual collinear particles differently
should not shift the considered event over any phase-space cut or
histogram boundary in a cross-section calculation. 

According to the parton model (see standard textbooks such as 
\citeres{Ellis:1991qj,Sterman:1994ce,Peskin:1995ev,Bohm:2001yx,Schwartz:2013pla}),
a parton (quark, gluon, etc.) inside a hadron~$h$ of momentum~$p$
is characterized by its flavour~$a$,
its momentum~$xp$, with $x$ denoting the momentum fraction of the parton,
and by its parton distribution function (PDF) 
$f^{(h)}_a(x)$, which---naively interpreted---is the probability density
to find~$a$ inside~$h$ with momentum fraction~$x$.
In the case of collinear initial-state radiation, where the parton momentum~$xp$
is further reduced by some factor $\xi$, the process is effectively initiated
by the momentum $\xi xp$ with a weight depending on $\xi$ non-trivially.
This implies that IR singularities from collinear initial-state radiation 
remain after summing virtual and real corrections, since inclusiveness
w.r.t.\ all $\xi$~values is missing.
This flaw of the {\it naive parton model} is cured within the {\it QCD-improved
parton model} by {\it factorization}, which absorbs the uncancelled collinear
singularities into redefined PDFs. 
In other words, the effects of collinear initial-state radiation are considered
as a property of the hadron rather than of the hard scattering process.
The proof of the universality (process independence)
of the collinear singularities, however, is non-trivial; 
a detailed discussion
of this issue can, for instance, be found in \citere{Collins:2011zzd}.
The PDF redefinition effectively separates the long-distance
effects caused by collinear radiation from the short-distance 
physics going on in the hard
scattering process, a procedure that introduces the so-called 
{\it factorization scale}~$\muF$, which is widely ambiguous.
After their redefinition, the PDFs inherit a dependence on $\muF$,
which is controlled by the famous 
Dokshitzer--Gribov--Lipatov--Altarelli--Parisi (DGLAP) 
equations~\cite{Gribov:1972ri,Altarelli:1977zs,Dokshitzer:1977sg}.
Likewise the hard partonic scattering process receives some explicit
$\muF$~dependence from the subtraction of the collinear singularities,
but physically observable hadronic cross sections $\sigma$ must not depend on
$\muF$. In practice, some $\muF$~dependence, however, remains in predictions
owing to the necessary truncation of the perturbative series on which
the calculation of the hard partonic cross section
$\hat\sigma$ is based. Such residual scale dependences
are part of the overall theoretical uncertainty of the prediction.
We exemplify the calculation of a hadronic cross section from the
scattering of two hadrons $h_1$ and $h_2$ (\eg protons at the LHC):
\begin{align}
\int \rd\sigma_{h_1h_2}(p_1,p_2) = \sum_{a,b} 
\int_0^1\rd x_1 \int_0^1\rd x_2\,
f^{(h_1)}_a(x_1,\muF^2) \,f^{(h_2)}_b(x_2,\muF^2) 
\int \rd\hat\sigma_{ab}(x_1p_1,x_2p_2,\muF^2),
\label{eq:hadXS}
\end{align}
where $a$ and $b$ are all relevant partons in $h_1$ and $h_2$, respectively.
If $h$ is not explicitly written in $f^{(h)}_a(x,\muF^2)$ in the 
following, $h=\Pp={}$proton is assumed by default.

The calculation of EW corrections to hadronic scattering processes requires
the generalization of the concept of QCD factorization and PDF redefinition 
to QED, \ie to the inclusion of the photon as parton.
Fortunately, this generalization is straightforward. 
The generic size of the collinear QED singularities is controlled by
$\alpha$ instead of $\alphas$, so that the corresponding QED-driven PDF evolution effects, 
factorization-scale dependences, etc., are typically much smaller than in QCD.
Another important difference is due to the fact that photons do not self-interact. 

Since we have worked out the general form of NLO EW collinear singularities
for all possible splittings involving light charged fermions and photons 
in the previous section, we can immediately identify the contributions of
QED singularities from collinear initial-state splittings to cross sections.
Starting from a hadronic cross section in the form 
\refeqf{eq:hadXS}, but calculated from {\it bare PDFs} $f^{(h)}_{a}(x)$,
the following PDF redefinition for (anti)quarks and photons
removes all QED initial-state singularities from $\sigma_{h_1 h_2}$,
\begin{align}
  \label{eq:qpdf-redef-dreg}
  f^{(h)}_{q/\bar q}(x) \;\rightarrow\; 
  f^{(h)}_{q/\bar q}(x,\mu_{\mathrm{F}}^2)
  &{}- \frac{\alpha\,\Qq^2}{2\pi}  \int^1_x \frac{\rd \xi}{\xi}\,
  f^{(h)}_{q/\bar q}\left(\frac{x}{\xi},\mu_{\mathrm{F}}^2\right) 
  \left\{
  -\frac{(4\pi)^\epsilon}{\Gamma(1-\epsilon)}\,
  \biggl(\frac{\mu^2}{\mu_{\mathrm{F}}^2}\biggr)^\epsilon \,
  \frac{1}{\epsilon} \, \Pff(\xi)
  + C_{ff}(\xi) \right\} 
\nl
  & {}-\NCq \frac{\alpha\,\Qq^2}{2\pi} \int^1_x \frac{\rd \xi}{\xi} \,
  f^{(h)}_\ga\left(\frac{x}{\xi},\mu_{\mathrm{F}}^2\right) 
  \left\{
  -\frac{(4\pi)^\epsilon}{\Gamma(1-\epsilon)}\,
  \biggl(\frac{\mu^2}{\mu_{\mathrm{F}}^2}\biggr)^\epsilon \,
  \frac{1}{\epsilon} \,
  P_{f\gamma}(\xi) + C_{f\gamma}(\xi) \right\}, 
\\[.5em]
  f^{(h)}_\ga(x) \;\rightarrow\; 
  f^{(h)}_\ga(x,\mu_{\mathrm{F}}^2) 
  &{}- \NCq\frac{\alpha}{2\pi}
  \int^1_x \frac{\rd \xi}{\xi} \,
  f^{(h)}_\ga\left(\frac{x}{\xi},\mu_{\mathrm{F}}^2\right) 
  \sum_{a=q} Q_a^2
  \left\{
  -\frac{(4\pi)^\epsilon}{\Gamma(1-\epsilon)}\,
  \biggl(\frac{\mu^2}{\mu_{\mathrm{F}}^2}\biggr)^\epsilon \,
  \frac{1}{\epsilon} \,
  P_{\gamma\gamma}(\xi) + C_{\gamma\gamma}(\xi) \right\}
\nn\\
  &{}- \frac{\alpha}{2\pi}
  \sum_{a=q,\bar{q}} Q_a^2\int^1_x \frac{\rd \xi}{\xi} \,
  f^{(h)}_a\left(\frac{x}{\xi},\mu_{\mathrm{F}}^2\right)
  \left\{
  -\frac{(4\pi)^\epsilon}{\Gamma(1-\epsilon)}\,
  \biggl(\frac{\mu^2}{\mu_{\mathrm{F}}^2}\biggr)^\epsilon \,
  \frac{1}{\epsilon} \,
  P_{\gamma f}(\xi) + C_{\gamma f}(\xi) \right\}.
  \label{eq:apdf-redef-dreg}
\end{align}
The gluon PDF is not redefined at NLO EW, since it does not have any EW coupling.
The splitting functions $P_{ab}$ are the ones introduced in \refeq{eq:pff-pfa-paf-paa}.
Up to the different coupling factors, they are the same as for the 
quark--gluon system, with the only exception of $P_{\ga\ga}(\xi)$, which
is entirely due to some ``hard contact term'' $\propto\delta(1-\xi)$.
This term is needed in the sum rules
\begin{align}
\int_0^1\rd \xi\, \Pff(\xi) =0,
\quad
\int_0^1\rd \xi\, \xi\left[ \Pff(\xi) +P_{\gamma f}(\xi)\right] =0,
\quad
\int_0^1\rd \xi\, \xi\left[ 2P_{f\gamma}(\xi) +P_{\gamma\gamma}(\xi)\right] =0,
\end{align}
which guarantee charge and momentum conservation when
$\muF$ is changed.

Note that the replacements \refeqf{eq:qpdf-redef-dreg} and
\refeqf{eq:apdf-redef-dreg} are to be applied in the strict NLO sense,
\ie after the PDF replacements only linear contributions of relative
order ${\cal O}(\alpha)$ are kept, while ${\cal O}(\alpha^2)$ terms
are dropped.  This effectively leads to an additional NLO
contribution---also known as {\it collinear counterterm}---to the
partonic cross section, cancelling the initial-state collinear
singularities of the virtual one-loop and real emission corrections.
The full NLO partonic cross section, thus, is given by the sum
\begin{equation}
\rd\hat\sigma_{ab} = 
\rd\hat\sigma_{ab}^{0} +  
\rd\hat\sigma_{ab}^{\virt} + 
\rd\hat\sigma_{ab}^{\real} + 
\rd\hat\sigma_{ab}^{\collCT},
\end{equation}
where $\rd\hat\sigma_{ab}^{0}$, $\rd\hat\sigma_{ab}^{\virt}$,
$\rd\hat\sigma_{ab}^{\real}$, and $\rd\hat\sigma_{ab}^{\collCT}$
refer to the lowest-order, virtual one-loop, real emission, and collinear
counterterm contributions, respectively.

The PDF redefinitions (\ref{eq:qpdf-redef-dreg}) and (\ref{eq:apdf-redef-dreg})
are formulated in DR, which corresponds to the common
standard in QCD. Recall that $\mu$ is the reference mass scale of 
DR without any physical meaning; it is neither
identical to the renormalization scale $\muR$ nor to $\muF$.
Many NLO EW calculations, however, employ MR with small fermion
masses as regulators for mass singularities. Regularizing initial-state
collinear singularities with small quark masses $m_q$, the PDF redefinitions read
\begin{align}
  \label{eq:qpdf-redef-mreg}
  f^{(h)}_{q/\bar q}(x) \;\rightarrow\; 
  f^{(h)}_{q/\bar q}(x,\mu_{\mathrm{F}}^2)
  &{}- \frac{\alpha\,\Qq^2}{2\pi}  \int^1_x \frac{\rd \xi}{\xi}\,
  f^{(h)}_{q/\bar q}\left(\frac{x}{\xi},\mu_{\mathrm{F}}^2\right) 
  \left\{\ln\biggl(\frac{\mu_{\mathrm{F}}^2}{m_q^2}\biggr)
  \Pff(\xi) - \left[\Pffunreg(\xi)\;(2\ln(1-\xi)+1)\right]_+
  + C_{ff}(\xi) \right\} 
\nl
  & {}-\NCq \frac{\alpha\,\Qq^2}{2\pi} \int^1_x \frac{\rd \xi}{\xi} \,
  f^{(h)}_\ga\left(\frac{x}{\xi},\mu_{\mathrm{F}}^2\right) 
  \left\{\ln\biggl(\frac{\mu_{\mathrm{F}}^2}{m_q^2}\biggr) \, P_{f\gamma}(\xi) +
  C_{f\gamma}(\xi) \right\}, 
\\[.5em]
  f^{(h)}_\ga(x) \;\rightarrow\; 
  f^{(h)}_\ga(x,\mu_{\mathrm{F}}^2) 
  &{}- \NCq\frac{\alpha}{2\pi}
  \int^1_x \frac{\rd \xi}{\xi} \,
  f^{(h)}_\ga\left(\frac{x}{\xi},\mu_{\mathrm{F}}^2\right) 
  \sum_{a=q} Q_a^2
  \left\{
  \ln\biggl(\frac{\mu_{\mathrm{F}}^2}{m_a^2}\biggr) \,
  P_{\gamma\gamma}(\xi) + C_{\gamma\gamma}(\xi) \right\}
\nn \\
  &{}- \frac{\alpha}{2\pi}
  \sum_{a=q,\bar{q}} Q_a^2\int^1_x \frac{\rd \xi}{\xi} \,
  f^{(h)}_a\left(\frac{x}{\xi},\mu_{\mathrm{F}}^2\right)
\left\{\ln\biggl(\frac{\mu_{\mathrm{F}}^2}{m_a^2}\biggr) \,
  P_{\gamma f}(\xi)
  - P_{\ga f}(\xi)\,(2\ln \xi+1)
  + C_{\gamma f}(\xi) \right\}.
  \label{eq:apdf-redef-mreg}
\end{align}
Obviously the standard collinear divergence $\Delta+\ln(\mu^2)$
in DR corresponds to the mass-singular term $\ln m_q^2$, as expressed
by the correspondence (\ref{eq:collsingmassdim}), but it should be
realized that the correspondence involves non-trivial IR-finite
contributions for the parts involving the splitting functions $\Pff$
and $P_{\gamma f}$.  The precise form of those terms can be obtained
by comparing the collinear singularities in the functions ${\cal
  H}^{f\gamma}$ and ${\cal H}^{ff}$ in MR, \ie \refeqs{eq:Hcsli-fb2g}
and \refeqf{eq:Hcsli-fb2f}, 
with their counterparts in DR,
\refeqs{eq:Hcsli-fb2g-dimreg} and \refeqf{eq:Hcsli-fb2fdimreg}.  The
differences between the functions in the two regularization schemes
can be calculated either taking the results for two-cutoff slicing
given in \refse{se:twocutoffslicing} or the ones for dipole
subtraction, where ${\cal H}^{f\gamma}$ is replaced by ${\cal
  G}^{ab}$, given in \refse{se:dipsub}.  Translating the PDF
redefinitions between regularization schemes in this way guarantees
that the {\it coefficient functions} $C_{ab}(\xi)$, which define the
{\it factorization scheme}, are identical in the two regularization
schemes.  Nowadays, practically all hadronic cross sections are
calculated within the {\it $\MSbar$ factorization scheme}, which
corresponds to
\begin{equation}
C_{ab}^{\MSbar}(\xi) = 0.
\end{equation}
The {\it DIS scheme} 
\cite{Altarelli:1978id,Ellis:1991qj}, which absorbs all
corrections to the structure function $F_2$ of deep-inelastic
electron--proton scattering into the PDFs, is hardly in use anymore.

In the PDF redefinitions at NLO EW, we did not yet make the precise
form of the electromagnetic coupling~$\alpha$ explicit.  Taking the
formalism of the QCD-improved parton model literally, the running
coupling $\alpha(\muF^2)$ as dictated by the employed factorization
scheme should be used.  In NLO EW calculations (or in higher EW
orders), however, the choice of $\alpha$ is typically adjusted to the
specific process, in order to minimize universal corrections of higher
order (\cf\refse{se:input_schemes}).  Note that this is not a
contradiction.  It simply means that the explicit $\alpha$ appearing
in Eqs.~(\ref{eq:qpdf-redef-dreg}), (\ref{eq:apdf-redef-dreg}),
(\ref{eq:qpdf-redef-mreg}), and (\ref{eq:apdf-redef-mreg}) has to be
identified with the $\alpha$ used to calculate all matrix elements;
otherwise the cancellation of the initial-state collinear
singularities would be incomplete. The mismatch between the $\alpha$
used in the PDF evolution and in the matrix elements only concerns
effects of NNLO EW or higher. {\it Leading} universal corrections
beyond NLO EW can still be controlled.  We will come back to the issue
of schemes to set $\alpha$ in the discussion of photon-induced
processes below.

Finally, we comment on the role of charged leptons $\Pl$ in the
context of mass singularities from collinear splittings in the
partonic initial state.  Initial-state photons can split according to
$\gamma\to\Pl\bar\Pl^*/\bar\Pl\Pl^*$, where $\bar\Pl^*/\Pl^*$
initiates the hard scattering process. This leads to collinear
singularities if the outgoing lepton $\Pl/\bar\Pl$ gets lost in the
beam direction.  If such initial-state singularities are absorbed into
the photon PDF, the DGLAP formalism demands also PDFs for charged
leptons \cite{Bertone:2015lqa}.  Since the introduction and evolution
of lepton PDFs indirectly proceeds via the photon PDF, which is
already suppressed, PDFs of charged leptons are usually not
considered.  Moreover, mass singularities of leptons can be treated
perturbatively.

\subsubsection{QED-corrected PDFs and photon distribution function}
\label{se:QED-PDFS}

The inclusion of EW ${\cal O}(\alpha)$ corrections in the DGLAP
evolution has an influence on all
PDFs~\cite{Spiesberger:1994dm,Roth:2004ti,Bertone:2013vaa}, but owing
to the hierarchy $\alpha\ll\alphas$ the impact is quite small. For
$\muF\sim\MW$, which is typical for many LHC processes, quark and
antiquark PDFs are changed at the level of $0.3\%$~$(1\%)$ for $x\lsim
0.1$~$(0.4)$. This influences also the gluon PDF via the momentum sum
rule for the proton.  Moreover, including the photon in the list of
partons, a photon distribution function appears.  The latter is
similar to the gluon PDF in shape, but relatively suppressed by
roughly a factor of $\langle \Qq^2\rangle\alpha/\alphas\sim 10^{-2}$
for not too high $x$ values, where $\langle \Qq^2\rangle$ stands for
some average squared quark charge. It should also be realized that the
photon PDF actually consists of both inelastic and elastic components,
where the latter corresponds to the situation when the proton (or the
considered hadron) does not break up in the scattering process.

The first global PDF set including QED corrections was
MRST2004qed~\cite{Martin:2004dh} where the photon PDF was based on
some simple modelling and a fit to deep-inelastic $\Pe\Pp$ scattering
data; an error estimate in a later version of this PDF set confirmed
the initial expectation that the photon PDF was good within roughly
$20\%$.  About ten years later, the NNPDF group supported QED
corrections and photon PDFs in the NNPDF23qed~\cite{Ball:2013hta} and
NNPDF30qed~\cite{Ball:2014uwa} PDF sets, which were derived from
experimental data (DIS, Drell--Yan-like W/Z production).  Since
photon-induced contributions are generically small in standard candle
processes used in PDF fits, the so-obtained photon PDF involved very
large uncertainties with up to $100\%$ in the high-$x$ range.  The
photon PDF of the CT14qed PDF set~\cite{Schmidt:2015zda}, on the other
hand, combined model assumptions and experimental input from
$\Pe\Pp\rightarrow\Pe\gamma+X$ data, achieving an accuracy at the
level of $10{-}20\%$.  The situation was drastically improved in 2016
with the advent of the LUXqed photon
PDF~\cite{Manohar:2016nzj,Manohar:2017eqh}, which was derived by
exploiting the observation that hadronic collisions mediated by
only virtual photons can be equivalently described by using a photon
PDF or by the parametrization of the hadronic tensor in terms of the
structure functions $F_2$ and $F_{\mathrm{L}}$.  In this way, it is
possible to derive a relation between the photon PDF
$f_\gamma(x,\mu^2)$ and the structure functions,
\begin{align}
{x f_{\gamma}(x,\mu^2)} ={}&
\frac{1}{2\pi\alpha(\mu^2)}
\int_x^1 \frac{\rd \xi}{\xi} \, \left\{ \,
\int^{\mu^2/(1-\xi)}_{x^2m_\Pp^2/(1-\xi)}
\frac{\rd Q^2}{Q^2} \, \left(\alpha(Q^2)\right)^2
\, \Biggl[ \Biggl( \xi P_{\gamma f}(\xi)+\frac{2x^2m_\Pp^2}{Q^2}\Biggr)\,
{F_2\biggl(\frac{x}{\xi},Q^2\biggr)}\right.
\nonumber\\
&
\hspace*{8em}\left.
{}-\xi^2 \, {F_\rL\biggl(\frac{x}{\xi},Q^2 \biggr)} \Biggr]
-\left(\alpha(\mu^2)\right)^2\, \xi^2 \,{F_2\biggl(\frac{x}{\xi},\mu^2\biggr)} \right\},
\hspace{2em}
\label{eq:gammaPDF}
\end{align}
where $m_\Pp$ is the proton mass.
Based on \refeq{eq:gammaPDF},
$f_{\gamma}(x,\mu^2)$ can be numerically evaluated from data on
$F_2(x,Q^2)$ and $F_{\mathrm{L}}(x,Q^2)$, leading to an extremely
accurate result for the photon PDF.
\begin{figure}[t]
\centering
\raisebox{2.5em}{\includegraphics[scale=1]{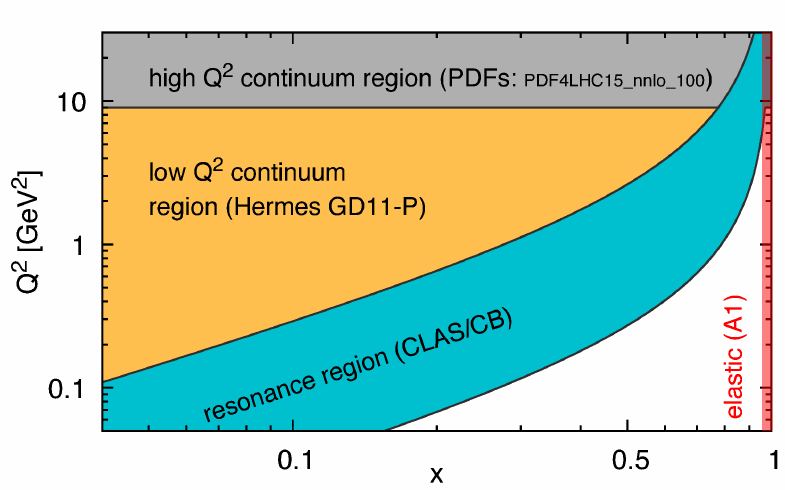}}
\hspace{3em}
\includegraphics[scale=.8]{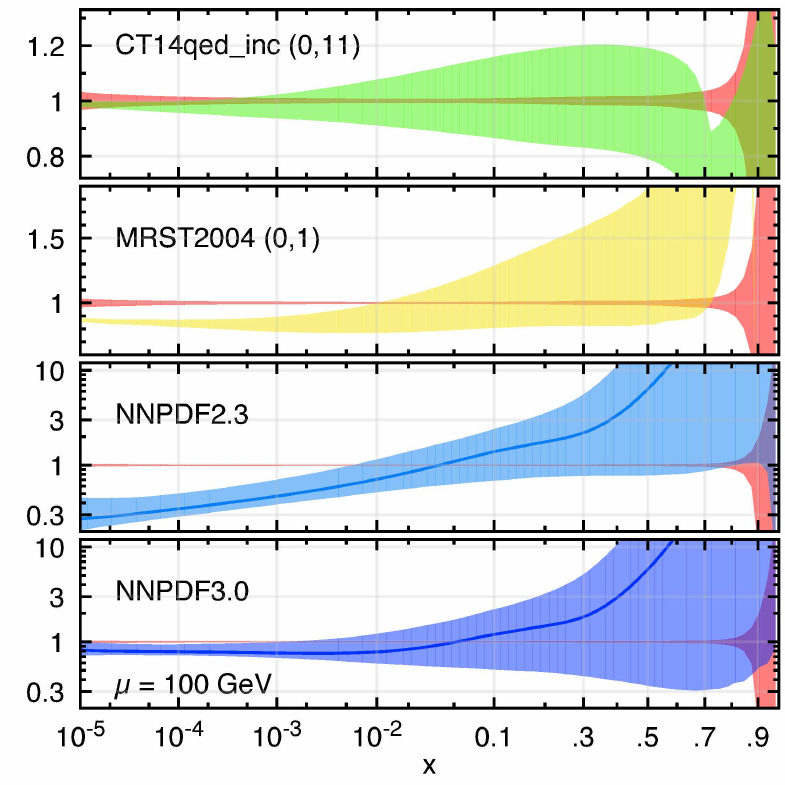}
\vspace*{-.5em}
\caption{Left: breakup of the $(x,Q^2)$ plane in terms of the
$F_2(x,Q^2)$ and $F_\rL(x,Q^2)$ data used in \protect\refeq{eq:gammaPDF}.
Right: ratio of photon PDFs from some common PDF sets (with uncertainty bands)
to the LUXqed photon PDF (uncertainty band in red).
(Taken from \citere{Manohar:2016nzj}.)
}
\label{fig:photonPDF}
\end{figure}
The l.h.s.\ of Fig.~\ref{fig:photonPDF} illustrates the coverage of
the $(x,Q^2)$ plane by data from different experiments.  Note that the
region at $x=1$ contains the contribution from elastic scattering,
where the proton does not break up in the collision.  The r.h.s.,
finally, shows the comparison of the mentioned determinations of the
photon PDF, normalized to LUXqed, with respective error bands.  The
LUXqed photon PDF is good within $1{-}2\%$ in the typical $x$~range of
LHC physics and, thus, even the best known of all PDFs.

Other approaches to extract the photon PDF from DIS and LHC data
were described in \citeres{Harland-Lang:2016apc,Giuli:2017oii}.
Nowadays, several PDF sets, which are publically available via
LHAPDF~\cite{Buckley:2014ana}, include QED corrections and a photon PDF.

\subsubsection{Photon-induced processes}

The inclusion of the photon in the set of partons leads to so-called
photon-induced processes, \ie partonic channels with photons in the
initial state.  At NLO EW, photon-induced channels always result as
crossed counterparts of photonic bremsstrahlung corrections.  For
instance, quark-initiated $qq$, $q\bar q$, $\bar q\bar q$ channels
always receive (real) ${\cal O}(\alpha)$ corrections from $q\gamma$
and/or $\bar q\gamma$ scattering, where the additional $q$ or $\bar q$
in the final state leads to an additional jet with respect to the LO
signature, similar to real NLO QCD corrections.  For specific final
states with charged particles, but without net electric charge, there
is also a contribution from $\gamma\gamma$ scattering with LO
kinematics and without additional partons in the final state.  This,
for instance, happens in the case for $\mu^+\mu^-$ or $\PW^+\PW^-$
production.

If photonic channels already exist in LO predictions, the redefinition
of the photon PDF becomes relevant at NLO EW.  At this point, special
attention has to be paid to the term $\propto
P_{\gamma\gamma}(\xi)\sum_{a=q} Q_a^2 \ln m_a^2$ in
\refeq{eq:apdf-redef-mreg}, or its counterpart in
\refeq{eq:apdf-redef-dreg}.%
\footnote{In early calculations of photon-induced processes, those
  terms were often disregarded, a procedure that was acceptable,
  because early QED-corrected PDFs did not make use of a consistent
  QED factorization scheme in all ingredients of the PDF
  determination.  Present-day QED-corrected PDF sets, however, employ
  $\MSbar$ factorization consistently, so that also all terms in the
  PDF redefinition have to be taken into account properly.}  Owing to
$P_{\gamma\gamma}(\xi)\propto\delta(1-\xi)$, the corresponding
contribution to the NLO cross section is proportional to the
photon-induced LO cross section and has some impact on the issue of
the EW input-parameter scheme for photon-induced processes, as pointed
out in \citeres{Harland-Lang:2016lhw,Kallweit:2017khh}.  In detail,
for processes with one or two incoming photons the partonic
cross-section contribution is given by
\begin{align}
\rd\hat\sigma^{\collCT}_{a\gamma} = {}& 
\left[\Dalphahad(\muF^2)+\dots\right]
\rd\hat\sigma^{0}_{a\gamma},\qquad
\rd\hat\sigma^{\collCT}_{\gamma\gamma} = {} 
2\left[\Dalphahad(\muF^2)+\dots\right]
\rd\hat\sigma^{0}_{\gamma\gamma},
\end{align}
where $a$ denotes an incoming parton different from a photon, and the
dots stand for non-singular constant terms.  The quantity
$\Delta\alpha(Q^2)$, which is defined in
\refse{se:input_reco}, controls the running of the electromagnetic
coupling $\alpha$, and the subscript ``had'' indicates that only
light-quark loops (all but the top-quark loop) are taken into account
in $\Dalphahad(Q^2)$.  Recall that $\Dalphahad(Q^2)$ receives
non-perturbative effects from all hadronic resonances of mass
$\sqrt{Q^2}$ lying between the Thomson limit at $\sqrt{Q^2}=0$ and some high
value of $\sqrt{Q^2}$, which results in the appearance of the light quark
masses in the perturbative result \refeqf{eq:Delta_alpha_ons} for
$\Delta\alpha(Q^2)$.  As discussed in \refse{se:input_reco}, a sound
way to account for those effects in perturbation theory is provided by
their absorption into an appropriately defined input value for
$\alpha$, \ie by a properly chosen input-parameter scheme.  For an NLO
EW amplitude with an initial-state photon, there are four sources for
terms involving $\Dalphahad(Q^2)$: (i)~the charge renormalization
constant $\delta Z_e$, (ii)~the photon wave-function renormalization
constant $\delta Z_{AA}$, (iii)~the hadronic contribution to
$-\Delta\alpha(\MZ^2)$ in the $\alpha(\MZ^2)$ or the $\GF$ scheme, and
(iv)~the contribution from the photon PDF redefinition discussed
above.  For the charge factor $e$ of each initial-state photon the
$\Delta\alpha$ parts in (i) and (ii) cancel in the sum $2\delta
Z_e+\delta Z_{AA}$, so that the mass-singular contribution
$\Dalphahad(\muF^2)$ of type~(iv) should be compensated by a term
$-\Delta\alpha(\MZ^2)$ of type~(iii) in order to avoid large
logarithmic corrections sensitive to non-perturbative physics.  This
means that the electromagnetic coupling for initial-state photons in
hadronic collisions should be renormalized in some input-parameter
scheme with $\alpha$ defined at some high momentum transfer, such as
in the $\alpha(\MZ^2)$ or the $\GF$ scheme. The fact that mass
logarithms of leptons still survive in the difference
$\Dalphahad(\muF^2)-\Delta\alpha(\MZ^2)$ is not harmful, since they
can be safely calculated in perturbation theory.

For processes where the diagrams of the $q\gamma$-induced corrections
have direct counterparts in the $q\Pg$ channels of the NLO QCD
corrections, the $q\gamma$ channels typically contribute at the
percent level to the hadronic cross section, or even less, which is a
direct consequence of the suppression of the photon PDF w.r.t.\ gluon
PDF by roughly $10^{-2}$.  A similar statement holds for
$\gamma\gamma$ channels whenever similar $\Pg\gamma$ or $\Pg\Pg$
channels exist.  More significant contributions from photon-induced
processes can arise if the photonic channels involve diagrams without
QCD counterparts, \ie diagrams where the photon couples to
colour-neutral charged particles like muons or W~bosons.  Typical
examples for enhanced $\gamma\gamma$~contributions are the channels
$\gamma\gamma\to\Pl^+\Pl^-$~\cite{CarloniCalame:2007cd,Dittmaier:2009cr,Boughezal:2013cwa}
and
$\gamma\gamma\to\PW^+\PW^-$~\cite{Bierweiler:2012kw,Bierweiler:2013dja,Baglio:2013toa,%
Billoni:2013aba,Biedermann:2016guo},
where the $\gamma\gamma$ channels comprise more than $10\%$ of the differential
cross sections in certain regions of phase space.
An extreme case for large $q\gamma$ contributions is provided by
$\PW\PW\PW$~production at the 
LHC~\cite{Yong-Bai:2016sal,Dittmaier:2017bnh,Schonherr:2018jva,Dittmaier:2019twg},
where photon-induced channels deliver $5{-}10\%$ of the integrated cross section,
strongly depending on the event selection (in particular on possible jet vetoes).

\subsection{\texorpdfstring{Photon--jet}{Photon-jet} systems}
\label{se:photon-jet}

\begin{sloppypar}
The perturbative treatment of 
final-state configurations with collinear photons
and light quarks in general leads to mass singularities in the quark
masses, similar to quark--gluon systems in QCD.  In principle those
singularities could be removed from predictions by consistently
considering the photon as another QCD parton, which takes part in all
clustering procedures of jet algorithms just like gluons and quarks.
This procedure, however, would merge all processes with hard photons
in the final state with corresponding processes with jet activity, \ie
this procedure is blind to any processes with tagged hard photon production.
Since processes with hard photons are certainly interesting to
analyze, we need a procedure to separate photon from jet production.
\end{sloppypar}

In the following we distinguish two different collinear
photon--quark configurations which deserve some special 
treatment:
\begin{myitemize}
\item Quark-to-photon splittings $q\to q\gamma$: This subprocess (see
  \reffi{fig:ab2fg} with $f$ being a quark) intertwines QCD and EW
  corrections in predictions for the production of a specific final
  state $F$ in association with a hard jet or a hard photon.
  Specifically, the real corrections to $F+\mathrm{jet}$ and
  $F+\gamma$ production both involve $F+\mathrm{jet}+\gamma$ final
  states, which contribute to the NLO EW corrections to
  $F+\mathrm{jet}$ production and to the NLO QCD corrections to
  $F+\gamma$ production at the same time.  A consistent isolation of
  $F+\mathrm{jet}$ from $F+\gamma$ signatures at full (QCD+EW) NLO
  accuracy requires a photon--jet separation that guarantees a proper
  cancellation of all IR (soft and collinear) singularities on the
  theory side and that is experimentally feasible at the particle
  level.
\item Photon-to-quark splittings $\gamma\to q\bar q$: Low-virtuality
  photons creating quark--antiquark pairs (see \reffi{fig:ab2ff} with
  $f$ representing a quark) lead to hadronic final states.  The
  perturbative treatment of this splitting leads to mass singularities
  in cross-section predictions that would be cancelled against loop
  corrections to the underlying hard process with a photon in the
  final state instead of the $q\bar q$~pair.  Assuming that hadronic
  activity can be experimentally distinguished from a hard photon, we
  need a procedure to treat the effect of the low-virtuality $q\bar
  q$~pair in a non-perturbative way.
\end{myitemize}

\subsubsection{\texorpdfstring{Photon--jet}{Photon-jet} separation}

Since calorimetric information is decisive for the reconstruction of
jets and photons, it is natural to take the electromagnetic energy
fraction inside some hadronic/electromagnetic shower as criterion to
decide whether it is called a jet or a photon.  Note that this
criterion, however, necessarily leads to an incomplete cancellation of
collinear singularities for photon emission off (anti)quarks.  This
can be explicitly seen by inspecting the mass-singular contributions
from photon radiation off final-state quarks, as spelled out for
two-cutoff slicing in \refeq{eq:sli-ab2fg} with \refeq{eq:Hcsli-ab2fg}
and for dipole subtraction in Eqs.~(\ref{eq:sigmasubij}) and
(\ref{eq:sigmasubia}) with Eqs.~(\ref{eq:bcGsubij}) and
(\ref{eq:bcGsubia}).  Separating jets from photons via their energy
flow, cuts into the integration over the splitting variable $z$, so
that the cancellation of collinear final-state singularities is
destroyed, which would require the integration over the
full $z$-range.  Two different methods have been suggested in the
literature to cope with this situation: One method introduces a
phenomenological {\it quark-to-photon fragmentation
  function}~\cite{Glover:1993xc} that absorbs the collinear
singularity in the same way as PDFs absorb collinear singularities
from the initial state.  Another method, known as {\it Frixione
  isolation}~\cite{Frixione:1998jh}, shifts the complete collinear
singularity to the jet side upon designing a specially adapted
definition of the allowed hadronic energy fraction in a photonic
shower:

\myparagraph{(a) Quark-to-photon fragmentation function}
The quark-to-photon fragmentation function
$D_{q\rightarrow\gamma}(z_\gamma)$ describes the probability of a
quark fragmenting into a jet containing a photon carrying the fraction
$z_\gamma$ of the total jet energy.  Since fragmentation is a
long-distance process, it cannot be calculated entirely in
perturbation theory, but receives two types of contributions: (i) the
perturbatively calculable radiation of a photon off a quark, which
contains a collinear divergence;
(ii) the non-perturbative production of
a photon during the hadronization of the quark,
which is described by a bare non-perturbative fragmentation function
$D^{{\rm bare}}_{q\rightarrow\gamma}(z_\gamma)$.
The latter contributes to the photon-emission cross section as
\begin{equation}
\rd \sigma^{{\rm frag}}_{ab\to\gamma X}(z_{\mathrm{cut}}) =
\rd \sigma^{\LO}_{ab\to qX}
\int_{z_{\mathrm{cut}}}^1 \rd z_\gamma\,
D^{{\rm bare}}_{q\rightarrow\gamma}(z_\gamma),
\end{equation}
where $z_{\mathrm{cut}}$ is the smallest photon energy fraction
required in the collinear photon--quark system to be identified as a
photon.  The perturbative and non-perturbative (bare) contributions to
the fragmentation function are sensitive to the soft dynamics inside
the quark jet and can a priori not be separated from each other in a
unique way.  Since the IR singularity present in the perturbative
contribution must be balanced by a divergent piece in the bare
fragmentation function, $D^{{\rm
    bare}}_{q\rightarrow\gamma}(z_\gamma)$ can be decomposed at a
factorization scale $\mu_{\mathrm{F}}$ into a universal divergent
piece and a phenomenological contribution
$D_{q\rightarrow\gamma}(z_\gamma,\mu_{\mathrm{F}}^2)$, which has to be
taken from experiment.  The precise form of this separation, including
the choice of the IR-finite contributions in the perturbative part,
determines the {\it factorization scheme} in which
$D_{q\rightarrow\gamma}$ is defined.  The $\overline{\mathrm{MS}}$
factorization scheme, which removes only the singular parts as defined
in DR from $D^{{\rm bare}}_{q\rightarrow\gamma}$, is the most common
choice \cite{Glover:1993xc}:
\begin{equation}
D_{q\rightarrow\gamma}^{\mathrm{bare}}(z_\gamma)\Big|_{\DR} =
\frac{\alpha \Qq^2}{2\pi}
\frac{(4\pi)^\epsilon}{\Gamma(1-\epsilon)}
\left(\frac{\mu^2}{\mu_{\mathrm{F}}^2}\right)^\epsilon
\frac{1}{\epsilon}
\Pffunreg(1-z_\gamma) 
+ D_{q\rightarrow\gamma}(z_\gamma,\mu_{\mathrm{F}}^2).
\label{eq:massfactDR}
\end{equation}
The respective result for MR with a small quark mass $m_q$,
which corresponds to the same phenomenological contribution
$D_{q\rightarrow\gamma}$, reads~\cite{Denner:2010ia}
\begin{equation}
D_{q\rightarrow\gamma}^{\mathrm{bare}}(z_\gamma)\Big|_{\MR} =
\frac{\alpha \Qq^2}{2\pi}
\left[\ln\left( \frac{m_q^2}{\muF^2}z_\gamma^2 \right)+1 \right]
\Pffunreg(1-z_\gamma)
+D_{q\rightarrow\gamma}(z_\gamma,\muF^2).
\label{eq:massfactMR}
\end{equation}

In \citeres{Glover:1993xc,Glover:1994he}
a method was proposed how to measure
$D_{q\rightarrow\gamma}(z_\gamma,\mu_{\mathrm{F}}^2)$ upon analyzing the process
$\Pe^+\Pe^-\rightarrow n\,\mathrm{jets}+\mathrm{photon}$.
The key feature of the proposed method is the democratic clustering of
both hadrons and photons into jets, while keeping track of the
photonic energy fraction in the jet, i.e.\
technically one has to deal with identified
particles in the final state that lead to non-collinear-safe observables.
The combination of the various ingredients in such a calculation
based on one-cutoff phase-space slicing or
dipole subtraction is described in \citeres{Glover:1993xc,Glover:1994he} and
\cite{Denner:2010ia}, respectively.
The first determination of $D_{q \to \gamma}(z,\muF^2)$ 
was performed by the ALEPH
collaboration~\cite{Buskulic:1995au} using the ansatz
\begin{equation}
D_{q\rightarrow\gamma}^{\mathrm{ALEPH}}(z_\gamma,\mu_{\mathrm{F}}^2)=
\frac{\alpha \Qq^2}{2\pi}
\left[
\Pffunreg(1-z_\gamma)\ln\left(\frac{\mu_{\mathrm{F}}^2}{\mu_0^2}
\frac{1}{(1-z_\gamma)^2}\right)+C
\right],
\end{equation}
with two fitting parameters, $C=-12.1$ and $\mu_0=0.22\GeV$.

\myparagraph{(b) Frixione isolation}
This procedure defines isolated hard photons by the requirement
that only soft partons can become collinear to the photon.
In detail, the total transverse energy $\sum_i E_{\rT,i}$ of all
partons $i$ with small distances
$R_{i\gamma}=\sqrt{(\eta_i-\eta_\gamma)^2+\Delta\phi_{i\gamma}^2}$
to the photon
in the pseudo-rapidity--azimuthal-angle plane is required to
go to zero with the maximally allowed distance $R_{i\gamma}$,
i.e.\
\begin{equation}
\sum_i E_{\rT,i}\theta(\delta-R_{i\gamma}) \le {\cal X}(\delta)
\qquad \mbox{for all} \quad \delta\le\delta_0,
\label{eq:frixione}
\end{equation}
where $\delta_0$ is a measure for the size of the cone around the
photon, in which the criterion is used.  The sum on the l.h.s.\ of
\refeq{eq:frixione} receives contributions only from hadrons with a
distance to the photon smaller or equal to $\delta$.  In
\refeq{eq:frixione}, ${\cal X}(\delta)$ is an appropriate function
with ${\cal X}(\delta)\to0$ for $\delta\to0$.  Specifically, the form
\begin{equation}
{\cal X}(\delta) = E_{\rT,\gamma}\epsilon_\gamma
\left(\frac{1-\cos\delta}{1-\cos\delta_0}\right)^n
\label{eq:frixione_chi}
\end{equation}
is suggested in \citere{Frixione:1998jh} for a photon of transverse
energy $E_{\rT,\gamma}$, where the two parameters $\epsilon_\gamma$
and $n$ can be taken to be $1$.   While we here
reproduce the definitions for a hadron collider, for lepton colliders the
transverse energies should be replaced by the ordinary energies in
\refeqs{eq:frixione} and \refeqf{eq:frixione_chi}. The condition
\refeqf{eq:frixione} excludes any hard jet activity collinear to the
photon, but still takes into account soft jet activity at a
sufficiently inclusive level to guarantee the proper cancellation of
IR singularities when calculating NLO QCD corrections to $F+\gamma$
production with $\gamma$ being the isolated photon.  Taking the
inverse procedure to define $F+\mathrm{jet}$ production, \ie
interpreting $F+\mathrm{jet}+\gamma$ production as photonic EW
correction to $F+\mathrm{jet}$ if condition (\ref{eq:frixione}) is not
fulfilled, formally shifts the complete contribution of the
quark-to-photon fragmentation function to $F+\mathrm{jet}$ production.
Whether this means that $F+\gamma$ production is really insensitive to
non-perturbative corrections is not proven and sometimes under debate.
Moreover, the implementation of condition (\ref{eq:frixione}) at the
experimental level raises issues, in particular concerning shower
effects on the hard photon kinematics and the realizability at the
apex of the cone.

\vspace{1em}

At the LHC, a prominent example for photon--jet separation is given by
$\PW/\PZ$ production in association with a jet or a photon, where the
NLO EW corrections to $\PW/\PZ+\mathrm{jet}$ production and the NLO
QCD corrections to $\PW/\PZ+\gamma$ production overlap.  For
$\PW/\PZ+\mathrm{jet}$ and $\PW/\PZ+\gamma$ production, NLO QCD+EW
results obtained with the quark-to-photon fragmentation function were
discussed in \citeres{Denner:2009gj,Denner:2011vu,Denner:2012ts} and
\citeres{Denner:2014bna,Denner:2015fca}; NLO QCD results based on
Frixione isolation can, for instance, be obtained with the program
MCFM~\cite{Campbell:2011bn}.  Comparing results on $\PZ+\gamma$
production obtained with the two separation procedures, reveals
differences at the level of $\sim1\%$ only~\cite{Denner:2015fca}, if
the separation parameters of the
two methods are matched as far as possible.%
\footnote{Using $R_{\gamma \mathrm{jet}}\sim R_{0}$
for the critical events,
the two parameters $ z_{\mathrm{cut}}$ and $\epsilon_\gamma$ can be related by
$z_{\mathrm{cut}}\approx 1/(1+\epsilon_\gamma)$.}
Note, however, that it is not possible to make generic, process-independent
statements on such differences.

\subsubsection{Photon-to-jet conversion}

The phase-space integral over squared amplitudes that involve some
$\gamma^* \to \Pq\bar\Pq$ splitting process contains a mass singularity
for light quarks~$\Pq$, originating from the collinear region
(\ie low-virtuality photons).
As discussed in \refse{se:IRrealEW}, the structure of this
singularity is universal in the sense that the underlying squared
matrix elements factorize into a universal radiator function
and the square of the hard matrix element of the underlying
process with a real photon instead of the $\Pq\bar\Pq$ pair.  
Note, however, that the physical final state is still a jet, or at least
some hadronic activity, emerging from the photon initiating the
splitting. 
Perturbatively,
the mass-singular cross-section contribution
can be calculated in a straightforward way, \eg via
two-cutoff slicing or dipole subtraction, as outlined in 
\refse{se:techniques4realcorrs}.
The singular contributions show up as $1/\epsilon$ poles in
DR or as logarithms $\ln m_q$ in small
quark masses $m_q$ in MR.
Either way, the resulting singular contribution
is not yet described in a physically meaningful way, since the
splitting contains non-perturbative contributions,
which have to be taken from experiment.

The non-perturbative cross-section contribution can be combined
with the perturbative part by means of a {\it photon-to-quark conversion
function} 
$D^{\mathrm{bare}}_{\gamma\to\mathrm{jet}}$
similar to the concept of fragmentation functions for 
identified-particle production~\cite{Denner:2019zfp},
\begin{align}
\rd\sigma^{\mathrm{conv}}_{ab\to \mathrm{jet}+X} &{}=
\rd\sigma^{\mathrm{LO}}_{ab\to \gamma X} \, \int_0^1\rd z\, 
D^{\mathrm{bare}}_{\gamma\to\mathrm{jet}}(z).
\end{align}
Here $D^{\mathrm{bare}}_{\gamma\to\mathrm{jet}}(z)$ is the {\it bare}
$\gamma\to\mathrm{jet}$
conversion function, which depends on the variable $z$ describing
the fraction of the photon momentum $\tilde k$ transferred to one of the
jets \mbox{($p_{\mathrm{jet}}=z\tilde k$)}. The bare conversion function
contains singular contributions so that the sum of the 
conversion part $\rd\sigma^{\mathrm{conv}}$ and the remaining 
perturbative cross-section contribution is non-singular. 
The extraction of the singular contribution from
$D^{\mathrm{bare}}_{\gamma\to\mathrm{jet}}(z)$ at some factorization scale
$\mu_{\mathrm{F}}$ requires a factorization scheme, for which we
take the $\MSbar$ scheme following common practice.
In DR, $D^{\mathrm{bare}}_{\gamma\to\mathrm{jet}}(z)$
is decomposed into a singular and a phenomenological part
$D_{\gamma\to\mathrm{jet}}(z,\mu_{\mathrm{F}}^2)$ as follows,
\begin{align}
\label{eq:Dbar_dreg}
D^{\mathrm{bare}}_{\gamma\to\mathrm{jet}}(z)\Big|_{\DR} 
&{}= 
\sum_q \NCq\, \frac{\Qq^2\alpha}{2\pi} \, 
\frac{(4\pi)^\epsilon}{\Gamma(1-\eps)} 
\left(\frac{\mu^2}{\mu_{\mathrm{F}}^2}\right)^\eps 
\frac{1}{\epsilon}
P_{f\ga}(z) 
+ D_{\gamma\to\mathrm{jet}}(z,\mu_{\mathrm{F}}^2).
\end{align}
In MR, this splitting reads
\begin{align}
D^{\mathrm{bare}}_{\gamma\to\mathrm{jet}}(z)\Big|_{\MR} 
&{}= 
\sum_q \NCq\, \frac{\Qq^2\alpha}{2\pi} \,
\ln\left(\frac{m_q^2}{\mu_{\mathrm{F}}^2}\right) 
P_{f\ga}(z) 
+ D_{\gamma\to\mathrm{jet}}(z,\mu_{\mathrm{F}}^2),
\label{eq:Dbar_mreg}
\end{align}
where the finite non-perturbative part
$D_{\gamma\to\mathrm{jet}}(z,\mu_{\mathrm{F}}^2)$ is the same in the two 
versions.

The non-perturbative contributions to
$D_{\gamma\to\mathrm{jet}}(z,\mu_{\mathrm{F}}^2)$ have to be extracted
from experimental data.  Ideally, this information would come from an
accurate differential measurement of a jet production cross section
(with low jet invariant mass) and of its corresponding prompt-photon
counterpart, \ie experimental information that is not available at
present.

In \citere{Denner:2019zfp} it was shown that at least the inclusive
$z$-integral over $D_{\gamma\to\mathrm{jet}}(z,\mu_{\mathrm{F}}^2)$ can
be obtained from a dispersion integral for the $R$~ratio of the cross
sections for $\Pe^+\Pe^-\to\mathrm{hadrons}/\mu^+\mu^-$.  This
dispersion integral, in turn, can be tied to the quantity
$\Dalphahad(\MZ^2)$, which is fitted to experimental data (see also
\refse{se:input_reco}).  Based on this feature, it is possible to
predict the following form of $D_{\gamma\to
  \mathrm{jet}}(z,\mu_{\mathrm{F}}^2)$,
\begin{align}
D_{\gamma\to \mathrm{jet}}(z,\mu_{\mathrm{F}}^2) = \Dalphahad(\MZ^2)
+ \sum_q\NCq\frac{\Qq^2\alpha}{2\pi} \,
\left[\ln\left(\frac{\mu_{\mathrm{F}}^2}{\MZ^2}\right)+\frac{5}{3}\right]\, P_{f\ga}(z),
\end{align}
which is valid up to $z$-dependent terms that integrate to zero.
Here the sum over $q$ runs over all quarks but the top quark.
Since mostly the inclusive integral over $z$ is needed in predictions
for cross sections, and since the impact of $D_{\gamma\to \mathrm{jet}}(z)$
is quite small in general, this result should be sufficient for all
phenomenological purposes.

\subsection{Radiation effects in \texorpdfstring{lepton--photon}
{lepton-photon} systems}
\label{se:lepton-photon}

\subsubsection{Prominent features of radiation effects}

Collinear photon radiation off massive fermions~$f$ (mass $m_f$,
electric charge $\Qf$) or collinear photons splitting into massive
fermion--antifermion pairs~$f\bar f$ lead to real radiative
corrections $\propto \Qf^2\alpha\ln(m_f^2/Q^2)$ that are enhanced by
mass logarithms if a typical energy scale~$Q$ of the underlying
process is much larger than the fermion mass $m_f$, \ie 
$Q\gg m_f$, a fact
that is already known for a long time \cite{Chen:1974wv}.  In contrast
to light quarks, whose masses are perturbatively not well defined, the
masses of charged leptons~$\Pl$ are physical and measured to very high
precision.  Leptonic mass-singular corrections are, thus,
perturbatively calculable and are often dominating EW corrections to
cross sections involving charged leptons in the initial or final
state.

At NLO, the methods outlined in \refse{se:techniques4realcorrs} are
suitable to calculate those corrections. Technically, it is not
favourable to employ the full lepton mass dependence in real-emission
amplitudes.  Rather, it is more appropriate to neglect the lepton
masses wherever possible and to keep them only as regulators in the
mass-singular logarithms. The lepton masses can be neglected in all
real-emission amplitudes, and non-vanishing masses are only required
in the phase-space integration over the regions of collinear emission
and splittings, which are isolated via slicing or subtraction
techniques.  This procedure significantly increases efficiency, speed,
and stability in the phase-space integration of cross sections.

Phenomenologically, the most important situations in which
collinear enhancements in corrections occur are:
\myparagraph{(a) Photonic initial-state radiation (ISR):} 
Since the
  factorization property of collinear photon emission off light
  fermions is universal, the corresponding mass-singular corrections
  originating from each incoming charged-lepton line take the form of
  a convolution of the underlying LO cross section $\sigma^{\LO}$ with
  a universal {\it structure function} $\Gamma^{\LL}_{\Pl\Pl}(x,Q^2)$,
  where ``LL'' indicates the leading-logarithmic approximation.  For a
  process with two incoming charged leptons of momenta $p_1$ and $p_2$
  this convolution is given by
\begin{align}
  \int \rd\sigma^{\LL,\ISR} =
  \int^1_0 \rd x_1 \int^1_0 \rd x_2 \,
  \Gamma_{\Pl\Pl}^{\LL}(x_1,Q^2) \, \Gamma_{\Pl\Pl}^{\LL}(x_2,Q^2) 
  \int \rd\sigma^{\LO}(x_1 p_1,x_2 p_2),
\label{eq:ISRconv}
\end{align}
where the integration variables $x_k$ are the momentum fractions of
the leptons initiating the hard scattering process.  From the analysis
of the collinear limit in \refse{se:techniques4realcorrs} we can
directly read off the form of $\Gamma_{\Pl\Pl}^{\LL}(x,Q^2)$ at NLO,%
\footnote{The mass-singular logarithms of $\Gamma_{\Pl\Pl}^{\LL}(x,Q^2)$
can, e.g., be read from the sum of the continuum parts
$[\cGsub_{ai}(P^2,x)|_{\MR}]_+$ and
$[\cGsub_{ab}(s,x)|_{\MR}]_+$ of dipole subtraction with a massive
initial-state emitter, as given in
\refeqs{eq:cGsubai} and \refeqf{eq:cGsubab},
since the sum of the respective endpoint parts
$\Gsub_{ai}(P^2,x)|_{\MR}$ and
$\Gsub_{ab}(s,x)|_{\MR}$ cancels against the corresponding 
mass singularities of the virtual corrections by virtue of the KLN theorem.}
\begin{align}
  \Gamma_{\Pl\Pl}^{\LL}(x,Q^2) 
&{}= \delta(1-x) +
  \frac{\beta_\Pl}{4} \Pff(x)
+\,{\cal O}(\alpha^2)
\label{eq:GammaLL}
\\
  &{}= \left.
  \delta(1-x)\left[ 1+
    \frac{\beta_\Pl}{4} \left(\frac{3}{2}+2\ln\eps\right)\right]
  + \frac{\beta_\Pl}{4}\theta(1-x-\eps)\Pffunreg(x)\right|_{\eps\to 0+}
+\,{\cal O}(\alpha^2),
\end{align}
where the difference between the two forms is only of technical nature.
The large logarithmic enhancement is contained in the parameter
\begin{align}
  \beta_\Pl = \frac{2Q_\Pl^2\alpha}{\pi} \left(\ln\frac{Q^2}{\Ml^2}-1\right),
\label{eq:betal}
\end{align}
where the non-logarithmic term is motivated from the soft-photon limit
\cite{Yennie:1961ad}.
The generic energy scale $Q$ of the process is not uniquely
determined in LL approximation; in a full NLO EW correction
the dependence on $Q$ completely cancels against the remaining
NLO EW contributions, from which the LL contribution of
$\sigma^{\LL,\ISR}$ has to be subtracted to avoid double counting.
If not the full NLO EW corrections are included in the prediction,
or more generally, if the accuracy in $\sigma^{\LL,\ISR}$
goes beyond the order of the remaining completely calculated
corrections, some residual $Q$ dependence remains.
For lepton--lepton annihilation
processes at a centre-of-mass energy $\sqrt{s}$, 
the scale $Q^2=s=(p_1+p_2)^2$ often is a reasonable choice, and
varying $Q$ by some numerical factor illustrates part of the
theoretical uncertainty from missing higher-order corrections.
 
Apart from the enhancement by the large correction factor
$\beta_\Pl$, ISR is distinguished by the fact that it 
effectively reduces the energy that is available for the
hard scattering process after ISR. 
Neglecting $\Ml$ in the kinematics, the squared CM energy~$s$
is reduced to $x_1 x_2 s$ for the ISR contribution
in \refeq{eq:ISRconv}.
The reduction of the CM energy in 
$\rd\sigma^{\LO}(x_1 p_1,x_2 p_2)$ can have a large impact
if $\sigma^{\LO}$ has a strong energy dependence below the 
nominal scattering energy~$\sqrt{s}$.
The most prominent case is an $s$-channel resonance
with mass smaller than $\sqrt{s}$ which can be reached by an ISR reduction
of the available scattering energy---a phenomenon known as
{\it radiative return}.
This effect, for example, leads to a significant distortion
of the Z-boson line 
shape~\cite{Berends:1987bg,Hollik:1988ii,Bardin:1989di,Bardin:1989qr,%
Montagna:1993py,Bardin:1997xq,Montagna:1998kp,Bardin:1999gt,Bardin:1999yd}
as observed at LEP1~\cite{ALEPH:2005ab} via
$\Pep\Pem\to\gamma^*/\PZ\to f\bar f$ and to radiative corrections of
${\cal O}(100\%)$ well above the Z-boson
resonance~\cite{Boudjema:1996qg}, as illustrated on the l.h.s.\ of 
\reffi{fig:QEDpeaksdistortions}.
\bfi
\centering
{\includegraphics[bb=0 10 560 537,scale=.35]{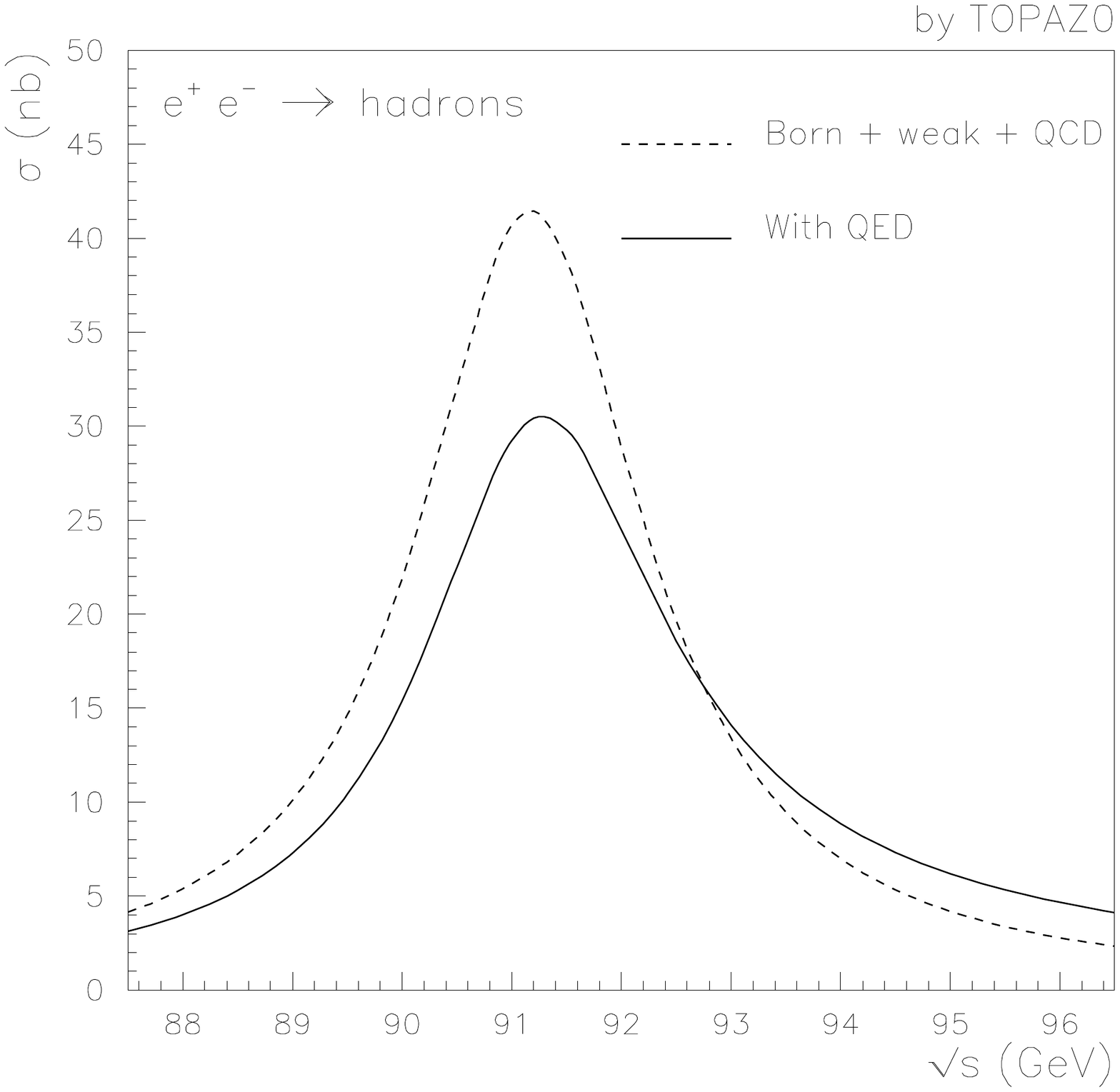}
\hspace*{2em}
\begin{picture}(220,180)
\put(0,10){\includegraphics[bb=55 262 571 633,scale=.44]{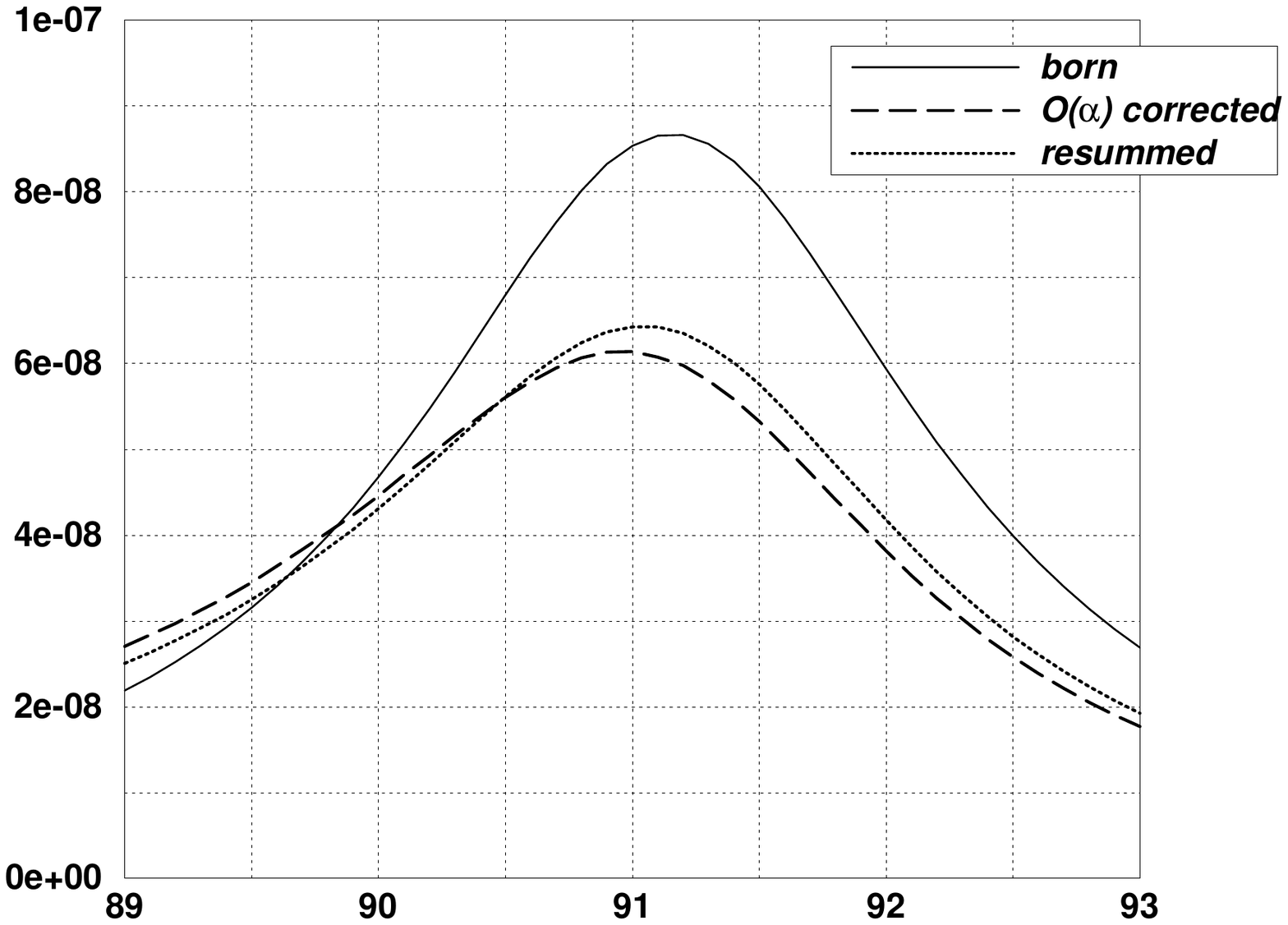}}
\put(160,4){\footnotesize $M_{\Pep\Pem}[\mathrm{GeV}]$}
\put(0,176){\footnotesize $\rd^2\sigma/\rd M_{\Pep\Pem}^2/\rd M_{\nu_\tau\bar\nu_\tau}^2[\mathrm{pb/GeV^4}]$ \qquad $\Pep\Pem\to\PZ\PZ\to\Pep\Pem\nu_\tau\bar\nu_\tau$}
\end{picture}
}
\caption{Resonance distortion due to ISR (left)
and FSR (right). Left:
Cross-section prediction for $\Pep\Pem\to\gamma^*/\PZ\to\mathrm{hadrons}$ 
as function of the CM energy $\sqrt{s}$ from
{\sc TOPAZ0}~\cite{Montagna:1993ai,Montagna:1995ja}, in which ``with QED''
refers to photonic ISR (taken from \citere{Montagna:1998sp}).
Right: Differential cross section for 
$\Pep\Pem\to\PZ\PZ\to\Pep\Pem\nu_\tau\bar\nu_\tau$ in the invariant
mass $M_{\Pep\Pem}$ of the final-state $\Pep\Pem$ pair
for $M_{\nu_\tau\bar\nu_\tau}=\MZ$, in which
``${\cal O}(\alpha)$ corrected'' refers to photonic FSR
and ``resummed'' to the corresponding effect of soft-photon
resummation (taken from \citere{Beenakker:1998cu}).}
\label{fig:QEDpeaksdistortions}
\efi
W-boson pair production near the pair production threshold at
$\sqrt{s}=2\MW$ provides another example where ISR corrections 
of ${\cal O}(10{-}20\%)$ are the by far dominating EW higher-order effects
(see also \refse{se:multires}), which have to
be controlled with high precision in order to achieve a sound
confrontation between theory and experiment.

Obviously, a pure NLO treatment of the logarithmically enhanced
ISR corrections is often insufficient. 
We consider the
inclusion of logarithmically enhanced radiation effects beyond NLO further below.
For $\Pep\Pem$ physics at the Z-boson resonance,
$\Pep\Pem\to\PZ/\gamma^*\to f\bar f$,
even NNLO QED effects beyond the logarithmic approximation are
necessary to adequately describe the high-precision measurements
carried out at LEP1 and planned for future $\Pep\Pem$ colliders.
The account of NNLO corrections is, however, beyond the scope of this
review (see \citeres{Blumlein:2019srk,Blumlein:2019pqb} for a
recent recalculation and some corrections to the original 
work~\cite{Berends:1987ab} and references in 
\citeres{Altarelli:1989hv,Bardin:1997xq,Bardin:1999ak,Bardin:1999gt}).

\myparagraph{(b) Photonic final-state radiation (FSR):}
At the LL level, the universal factorization property of collinear
photon emission is identical for initial- and final-state fermions.
Nevertheless there are characteristic differences in the impact of the
corresponding corrections on cross sections which are mainly due to
the different momentum flow: ISR photons reduce the effective
scattering energy of the hard process, but FSR photons do not.  The LL
approximation for the inclusion of FSR off a single lepton $\Pl$
schematically reads
\begin{align}
  \int \rd\sigma^{\LL,\FSR(\Pl)} =
  \int^1_0 \rd z \,
  \Gamma_{\Pl\Pl}^{\LL}(z,Q^2) 
  \int \rd\sigma^{\LO}(\hat p_\Pl) \, 
\Theta_{\mathrm{cut}}(p_\Pl=z \hat p_\Pl,k=(1-z)\hat p_\Pl),
\label{eq:FSRconv}
\end{align}
where the structure function $\Gamma_{\Pl\Pl}^{\LL}(z,Q^2)$ is the 
same as for ISR in LL approximation and
$\hat p_\Pl$ is the hard lepton momentum before FSR.
The theta function $\Theta_{\mathrm{cut}}$ indicates that
the collinear lepton and photon momenta
$p_\Pl$ and $k$ are relevant for the event selection
that defines the measurable cross section.
Note that only $\Gamma_{\Pl\Pl}^{\LL}$ and $\Theta_{\mathrm{cut}}$ 
depend on the momentum fraction~$z$ which controls the momentum
share of lepton and photon after FSR, while the
LO cross section $\rd\sigma^{\LO}$ is $z$~independent.
The scale~$Q$ plays the same role as for ISR discussed above.

The KLN theorem~\cite{Kinoshita:1962ur,Lee:1964is} guarantees 
that the mass-singular FSR corrections to cross sections cancel if
collinear lepton--photon systems are treated fully inclusively,
\ie if the $z$~integral in \refeq{eq:FSRconv} is carried out
over the full domain with $\Theta_{\mathrm{cut}}\equiv1$.
A total cross section, defined without any phase-space cuts, 
is an obvious observable of this kind.
The NLO approximation of $\Gamma_{\Pl\Pl}^{\LL}$ given in
\refeq{eq:GammaLL} makes this property of FSR very explicit by
the appearance of the $(\cdots)_+$~distribution.
Observables that are sufficiently inclusive so that
FSR mass singularities cancel 
are called {\it collinear safe}, as already discussed in 
\refse{se:twocutoffslicing}.

In the presence of phase-space cuts and in differential cross
sections, in general, mass-singular contributions survive, leading to
enhanced FSR effects.  The r.h.s.\ of \reffi{fig:QEDpeaksdistortions}
illustrates those FSR effects in the $\Pep\Pem$ invariant-mass
distribution of the process
$\Pep\Pem\to\PZ\PZ\to\Pep\Pem\nu_\tau\bar\nu_\tau$, where collinear
FSR effectively reduces the invariant mass after the resonant
production of the $\Pep\Pem$ pair, leading to large positive
corrections for $M_{\Pep\Pem}\lsim\MZ$ known as {\it radiative tail}.
Examples of this kind also show up in hadron-collider observables for
processes involving final-state leptons, which are copiously produced
in any process of W- or Z-boson production.  Considering, e.g., the
distribution in the $\Pl^+\Pl^-$~invariant-mass $M_{\Pl^+\Pl^-}$ from
Drell--Yan-like Z-boson production (see in particular
\refse{se:NLOEW-DY}), the spectrum receives FSR corrections of
several tens of percent for $M_{\Pl^+\Pl^-}$ values below the
Z~resonance peak at $M_{\Pl^+\Pl^-}=\MZ$.

For differential observables the level of inclusiveness necessary for
collinear safety can be restored by a procedure known as {\it photon
  recombination}, which treats collinear lepton--photon systems as one
quasi-particle.  This procedure is similar to the application of a jet
algorithm in QCD.  For final-state electrons, photon recombination
automatically is involved in their experimental reconstruction from
electromagnetic showers detected in calorimeters.  Muons, on the other
hand, can be observed as {\it bare} leptons from their tracks in the
muon chambers, but in order to reduce large FSR corrections, observed
muons are sometimes also reconstructed as {\it dressed} muons via
photon recombination, as e.g.\ described in \citere{Aad:2011gj} for an
ATLAS analysis.  Working with dressed leptons, where mass-singular FSR
effects cancel, has the advantage that the resulting cross section
does not depend on the mass (and thus on the flavour) of the charged
lepton, \ie the reconstructed lepton looks universal (at least for
electrons and muons).
\myparagraph{(c) Photon processes in $\Pe\Pe$ collisions:}
The complementary process to photonic ISR with subsequent
$\Pe\Pe$ scattering is the process where the radiated photon
instead of the $\Pe^\pm$ is initiating the hard scattering 
process (see \reffi{fig:fa2f} with $f$ representing an $\Pe^\pm$).
Similar to the LL approximation for ISR and FSR,
the universal factorization of the collinear singularity in
the $\Pl\to\Pl\gamma^*$ splitting can be employed to 
calculate the hard photon scattering contribution in 
leptonic collisions---a procedure known as
{\em Weizs\"acker--Williams} or {\it equivalent photon approximation
(EPA)}~\cite{vonWeizsacker:1934nji,Williams:1934ad}.
For the $\gamma\gamma$~contribution to a high-energy
$\Pep\Pem$ scattering cross section, the EPA reads
\begin{align}
  \int \rd\sigma^{\EPA}_{\Pe\Pe} =
  \int^1_0 \rd x_1 \int^1_0 \rd x_2 \,
  \Gamma_{\gamma\Pl}^{\LL}(x_1,Q^2) \, \Gamma_{\gamma\Pl}^{\LL}(x_2,Q^2) 
  \int \rd\sigma^{\LO}_{\gamma\gamma}(x_1 p_1,x_2 p_2),
\label{eq:EPAconv}
\end{align}
with $\sigma^{\LO}_{\gamma\gamma}$ denoting the LO cross section for
hard $\gamma\gamma$ scattering into any targeted final state.  The
leading contribution to the corresponding structure function
$\Gamma_{\gamma\Pl}^{\LL}$, which describes the probability to find a
photon with momentum fraction~$x$ in a lepton, can be read off from
the results of \refse{se:techniques4realcorrs},%
\footnote{The result for $\Gamma_{\gamma\Pl}^{\LL}(x_1,Q^2)$ with 
$Q^2 = (1-x)(p_i^0)^2\De\theta^2$ can be directly read from
${\cal H}^{ff}(p_f^0,x)\big|_{\MR}$ as given in the slicing approach in
\refeq{eq:Hcsli-fb2f}.}
\begin{equation}
  \Gamma_{\gamma\Pl}^{\LL}(x,Q^2) {}= 
  \frac{Q_\Pl^2\alpha}{2\pi} \left[
P_{\gamma f}(x) \ln\left(\frac{Q^2(1-x)}{\Ml^2 x^2}\right)
-\frac{2(1-x)}{x} \right]
+\,{\cal O}(\alpha^2).
\label{eq:GammaAL}
\end{equation}
Here we have included finite terms beyond LL approximation
in such a way that choosing 
$Q^2 = (1-x)E^2\De\theta^2$
reproduces the cross section for incoming leptons with 
energy $E$ and outgoing leptons with 
scattering angle from $0$ to 
$\De\theta\gg\Ml/E$~\cite{Aurenche:1996mz}
in the centre-of-mass frame of the process.

Since the physics of photon--photon collisions is more characterized
by the strong rather than the EW interaction, we will not consider it
any further in this review and instead refer to
\citeres{Aurenche:1996mz,Accomando:1997wt,AguilarSaavedra:2001rg} for
more details.

\subsubsection{QED structure functions to higher orders}
\label{se:SF}

As already indicated in the phenomenological examples mentioned
in the previous section, radiative corrections from collinear
photon emission off leptons can be rather large, so that 
NLO EW precision for those corrections often
is not sufficient to achieve percent accuracy (or beyond).
In such cases, at least the dominant ISR or FSR corrections 
beyond NLO should be better taken into account. 
Since those corrections are tied to IR singularities of a system
of light fermions and photons, many field-theoretical statements of
perturbative QCD on systems of light quarks and gluons can be
adapted to the QED case. Although many simplifications occur
in the transition from non-abelian QCD to abelian QED, some care
has to be taken, because leptons and photons are not confined, i.e.\
in the QED case observables are typically not based on objects
at the same level of inclusiveness as jets in QCD.

In QCD, {\it IR evolution equations} can be used to resum
IR-singular contributions to cross sections, the most famous
example being the DGLAP equations for PDFs.
Generically, those equations are consistency relations expressing
the fact that physical observables cannot depend on some
{\it factorization scale} $\muF$ which is introduced in the
theory to separate {\it soft} 
(\ie IR-sensitive) and {\it hard}
interaction effects. Both soft and hard contributions to an observable
depend on $\muF$, but only the effects from the hard interactions 
are perturbatively calculable. 
The soft effects are absorbed into some non-perturbative
quantity like a PDF, which in most cases is fitted to experimental data.
In spite of its non-perturbative nature, however, the $\muF$ dependence
of the quantity comprising the soft, IR-sensitive physics
is controlled by perturbation theory via 
IR evolution equations.
Thorough treatments of these concepts and of the formal construction
can be found in many text books on quantum field theory, such as
\citeres{Collins:1984xc,Sterman:1994ce,Peskin:1995ev,Bohm:2001yx,Collins:2011zzd,Schwartz:2013pla}.

In the following we sketch the application of IR evolution equations
to the LL QED structure functions $\Gamma_{ab}^{\LL}(x,\muF^2)$ as
introduced in \refeq{eq:ISRconv} for photonic ISR.  These functions
are the precise counterparts to the PDFs of QCD partons, but now
formulated for a system of leptons and photons. We first consider the
case of one lepton flavour and a photon and comment on the
generalization to more leptons below. In detail,
$\Gamma_{ab}^{\LL}(x,\muF^2)$ is the generalized (\ie not necessarily
positive) probability density to find the lepton, antilepton, or
photon $a=\Pl,\bar\Pl,\gamma$ with momentum fraction $x$ inside some
lepton, antilepton, or photon $b=\Pl,\bar\Pl,\gamma$.  Identifying
$\Gamma_{ab}^{\LL}(x,\muF^2)$ with the parton density
$f^{(b)}_a(x,\mu_{\mathrm{F}}^2)$ of \refeq{eq:qpdf-redef-dreg} or
\refeq{eq:qpdf-redef-mreg}, upon taking the derivative w.r.t.\ 
$\ln\muF^2$ we get the evolution equations
\begin{align}
\frac{\partial}{\partial\ln\muF^2}
\begin{pmatrix} \,
\Gamma_{lb}^{\LL}(x,\muF^2) \,\\[.5em] 
\Gamma_{\bar lb}^{\LL}(x,\muF^2) \,\\[.5em] 
\Gamma_{\gamma b}^{\LL}(x,\muF^2) 
\end{pmatrix}
&{}=
\frac{\Ql^2\alpha(0)}{2\pi} \int_x^1\frac{\rd \xi}{\xi} \,
\begin{pmatrix} \,
\Pff\left(\frac{x}{\xi}\right) & 0
& P_{f\gamma}\left(\frac{x}{\xi}\right) 
\\[.5em]
0& \Pff\left(\frac{x}{\xi}\right)
& P_{f\gamma}\left(\frac{x}{\xi}\right) 
\\[.5em]
P_{\gamma f}\left(\frac{x}{\xi}\right) & 
P_{\gamma f}\left(\frac{x}{\xi}\right) & 
P_{\gamma\gamma}\left(\frac{x}{\xi}\right) \,
\end{pmatrix} \,
\begin{pmatrix} \,
\Gamma_{lb}^{\LL}(\xi,\muF^2) \,\\[.5em] 
\Gamma_{\bar lb}^{\LL}(\xi,\muF^2) \,\\[.5em] 
\Gamma_{\gamma b}^{\LL}(\xi,\muF^2) \end{pmatrix},
\label{se:GammaQEDevolve}
\end{align}
which are valid in LL accuracy with the splitting functions
given in \refeq{eq:pff-pfa-paf-paa}.

At LO the structure functions are given by
\begin{equation}
\Gamma_{ab}^{\LL}(x,\muF^2)\big|_{\alpha^0} = \delta_{ab}\,\delta(1-x).
\label{eq:GammaQEDLO}
\end{equation}
This simply states that without any interaction a lepton stays a lepton,
an antilepton stays an antilepton,
and a photon stays a photon.
Given $\Gamma_{ab}^{\LL}$ at LO, the evolution equations can be
solved iteratively by integrating them over $\ln\muF^2$.
The upper limit in this integration is set by the target scale $\muF^2$,
the lower limit is set by the IR mass scale of the chosen IR regularization
scheme. Since lepton masses are physically meaningful, we make use
of MR where the lower limit in the integration
over $\ln\muF^2$ is given by $\ln\Ml^2$.
In LL approximation, the IR evolution equations do not
contain any additional information beyond the coefficients of the
logarithms $\left[\alpha\ln(\muF^2/\Ml^2)\right]^n$.
Integrating \refeq{se:GammaQEDevolve} and 
inserting \refeq{eq:GammaQEDLO} on the r.h.s.\ leads to
\begin{align}
\begin{pmatrix} \,
\Gamma_{ll}^{\LL} & \Gamma_{l\bar l}^{\LL} &\Gamma_{l\gamma}^{\LL}
\,\\[.5em] 
\Gamma_{\bar ll}^{\LL} & \Gamma_{\bar l\bar l}^{\LL} &\Gamma_{\bar l\gamma}^{\LL}
\,\\[.5em] 
\Gamma_{\gamma l}^{\LL} & \Gamma_{\gamma\bar l}^{\LL} & \Gamma_{\gamma\gamma}^{\LL}
\end{pmatrix}(x,\muF^2) 
\;=\;{}& \delta(1-x)\,{\mathbb{1}}_3
+ \frac{\Ql^2\alpha(0)}{2\pi} 
\int_{\ln\Ml^2}^{\ln\muF^2} \rd\ln\mu^2 \,
\int_x^1\frac{\rd \xi}{\xi} \,
\nn\\*
\label{eq:GammaQEDiterate}
& {} \,\times{}
\begin{pmatrix} \,
\Pff\left(\frac{x}{\xi}\right) & 0
& P_{f\gamma}\left(\frac{x}{\xi}\right) 
\\[.5em]
0& \Pff\left(\frac{x}{\xi}\right) 
& P_{f\gamma}\left(\frac{x}{\xi}\right) 
\\[.5em]
P_{\gamma f}\left(\frac{x}{\xi}\right) &
P_{\gamma f}\left(\frac{x}{\xi}\right) &
P_{\gamma\gamma}\left(\frac{x}{\xi}\right) \,
\end{pmatrix} \,
\begin{pmatrix} \,
\Gamma_{ll}^{\LL} & \Gamma_{l\bar l}^{\LL} &\Gamma_{l\gamma}^{\LL}
\,\\[.5em] 
\Gamma_{\bar ll}^{\LL} & \Gamma_{\bar l\bar l}^{\LL} &\Gamma_{\bar l\gamma}^{\LL}
\,\\[.5em] 
\Gamma_{\gamma l}^{\LL} & \Gamma_{\gamma\bar l}^{\LL} & \Gamma_{\gamma\gamma}^{\LL}
\end{pmatrix}(\xi,\mu^2) 
+{\cal O}(\alpha^2)
\\[.5em]
\;=\;{}& \delta(1-x)\,{\mathbb{1}}_3
+ L_l\,
\begin{pmatrix} \,
\Pff\left(x\right) & 0
& P_{f\gamma}\left(x\right) 
\\[.5em]
0 & \Pff\left(x\right) 
& P_{f\gamma}\left(x\right) 
\\[.5em]
P_{\gamma f}\left(x\right) & P_{\gamma f}\left(x\right) &
P_{\gamma\gamma}\left(x\right) \,
\end{pmatrix} 
+{\cal O}(\alpha^2)
\\[.5em]
\;=\;{}& 
\begin{pmatrix} \,
{ \includegraphics[scale=.8]{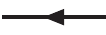} }
+
{ \includegraphics[scale=.8]{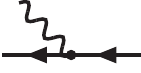} } \,
& 0 &
{ \includegraphics[scale=.8]{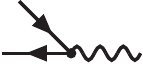} }
\\[.3em]
0 &
{ \includegraphics[scale=.8]{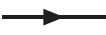} }
+
{ \includegraphics[scale=.8]{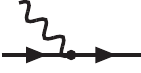} } \,
&
{ \includegraphics[scale=.8]{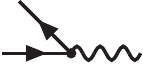} }
\\[.3em]
{ \includegraphics[scale=.8]{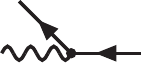} }
&
{ \includegraphics[scale=.8]{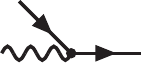} }
&
{ \includegraphics[scale=.8]{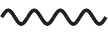} }
+
{ \includegraphics[scale=.8]{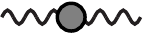} } \,
\end{pmatrix} 
\label{eq:GammaQEDNLO}
\end{align}
with
\begin{align}
\label{eq:QEDlog}
L_l = \frac{\Ql^2\alpha(0)}{2\pi} \,
\ln\left(\frac{\muF^2}{\Ml^2}\right).
\end{align}
This, in particular, confirms the NLO results for
$\Gamma_{ll}^{\LL}(x,q^2)$ and $\Gamma_{\gamma\Pl}^{\LL}(x,q^2)$ of
the previous section up to terms that are beyond the LL approximation.
The diagrammatic illustration in the last line shows the individual
contributions to each structure function $\Gamma_{ab}^{\LL}$ with $b$
coming in from the right, $a$ going out on the left.  For more than
one lepton species, each lepton and antilepton just behaves as $\Pl$
and $\bar\Pl$ at NLO, respectively; 
in $\Gamma_{\gamma\gamma}^{\LL}$ the sum over all lepton types
has to be taken.

Iterating this procedure by inserting the results for the structure
functions into the r.h.s.\ of \refeq{eq:GammaQEDiterate},
higher-order LL contributions to $\Gamma_{ab}^{\LL}$ can be
calculated.  The NNLO contribution is still easy to calculate and
reads
\begin{align}
\Gamma_{ll}^{\LL}(x,\muF^2)\big|_{\alpha^2} 
={}& \frac{L_l^2}{2}\,
\left\{ \Pffunreg(x)\left[4\ln(1-x)+3\right]
-\frac{1+3x^2}{1-x}\ln x-2+2x \right\}_+ 
& { \includegraphics[scale=.8]{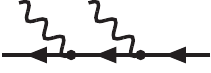} } 
\nn\\*
& {}+ \frac{L_l^2}{2}\,
\left\{ \frac{1-x}{3x}(4+7x+4x^2) + 2(1+x)\ln x \right\},
& { \includegraphics[scale=.8]{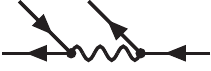} } 
\label{eq:GammallQEDNNLO}
\\
\Gamma_{l\bar l}^{\LL}(x,\muF^2)\big|_{\alpha^2} 
={}& \frac{L_l^2}{2}\,
\left\{ \frac{1-x}{3x}(4+7x+4x^2) + 2(1+x)\ln x \right\},
& { \includegraphics[scale=.8]{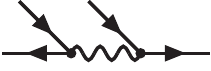} } 
\\
\Gamma_{l\gamma}^{\LL}(x,\muF^2) \,\Big|_{\alpha^2}
={}& \frac{L_l^2}{2}\,
\left\{ 2P_{f\gamma}(x)\ln(1-x)-(1-2x+4x^2)\ln x-\frac{1}{2}+2x \right\}
& { \includegraphics[scale=.8]{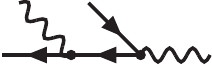} } 
\nn\\*
& {} -  \frac{L_l^2}{3}\,
P_{f\gamma}(x),
& { \includegraphics[scale=.8]{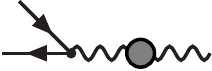} } 
\\
\Gamma_{\bar ll}^{\LL}(x,\muF^2) \,\Big|_{\alpha^2}
={}& \Gamma_{l\bar l}^{\LL}(x,\muF^2) \,\Big|_{\alpha^2},
& { \includegraphics[scale=.8]{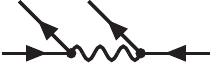} } 
\\
\Gamma_{\bar l\bar l}^{\LL}(x,\muF^2) \,\Big|_{\alpha^2}
={}& \Gamma_{ll}^{\LL}(x,\muF^2) \,\Big|_{\alpha^2},
& \hspace{-4em}
{ \includegraphics[scale=.8]{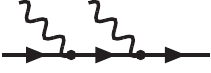} } 
+ { \includegraphics[scale=.8]{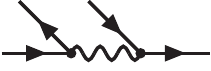} } 
\\
\Gamma_{\bar l\gamma}^{\LL}(x,\muF^2) \,\Big|_{\alpha^2}
={}& \Gamma_{l\gamma}^{\LL}(x,\muF^2) \,\Big|_{\alpha^2},
& \hspace*{-4em}
{ \includegraphics[scale=.8]{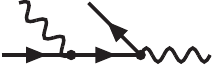} } 
+ { \includegraphics[scale=.8]{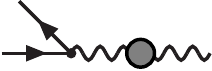} } 
\\
\Gamma_{\gamma l}^{\LL}(x,\muF^2) \,\Big|_{\alpha^2}
={}& \frac{L_l^2}{2}\,
\left\{2P_{\gamma f}(x)\,\ln(1-x) +(2-x)\ln x+2-\frac{x\vphantom{1}}{2} \right\}
& { \includegraphics[scale=.8]{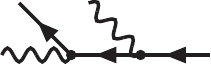} } 
\nn\\*
& {} - \frac{L_l^2}{3}\,
P_{\gamma f}(x),
& { \includegraphics[scale=.8]{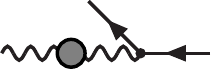} } 
\\
\Gamma_{\gamma\bar l}^{\LL}(x,\muF^2) \,\Big|_{\alpha^2}
={}&
\Gamma_{\gamma l}^{\LL}(x,\muF^2) \,\Big|_{\alpha^2},
& \hspace{-4em}
{ \includegraphics[scale=.8]{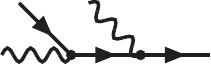} } 
+ { \includegraphics[scale=.8]{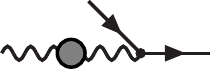} } 
\\
\Gamma_{\gamma\gamma}^{\LL}(x,\muF^2) \,\Big|_{\alpha^2}
={}& {L_l^2}\,
\left\{ \frac{1-x}{3x}(4+7x+4x^2) + 2(1+x)\ln x \right\}
& \hspace{-4em}
{ \includegraphics[scale=.8]{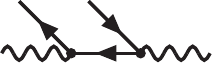} } 
+ { \includegraphics[scale=.8]{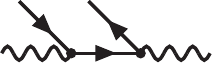} } 
\nn\\*
& {} -  \frac{L_l^2}{3}\,
P_{\gamma\gamma}(x).
& { \includegraphics[scale=.8]{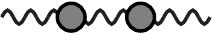} } 
\label{eq:GammaQEDNNLO}
\end{align}
The diagrams illustrate the splitting processes contributing to
$\Gamma_{ab}^{\LL}$ with $b$ coming in from the right, $a$ going
out on the left, and the diagonal lines showing the particles
being radiated off.
In the case of more than one lepton type, the structural diagrams
determine by charge and flavour
conservation which types of processes can take place.
Note that at NNLO, some new structure functions
appear, such as $\Gamma_{ll'}^{\LL}$ for $l\ne l'$
or $\Gamma_{l\bar l'}^{\LL}$. In $\Gamma_{\gamma\gamma}^{\LL}$
the sum over all possible lepton types has to be taken for each
$P_{\gamma\gamma}$ insertion, and all possible $\Pl\bar \Pl$ and $\bar\Pl\Pl$
emissions have to be taken into account.

Instead of working out higher iterations explicitly, we
concentrate on the important special cases of genuine
photonic ISR and FSR, i.e.\
on the phenomenological situation
in which any further (anti)lepton emission during ISR or FSR
(or hadronic activity via quarks)
can be vetoed in the event selection. 
In this case, only the splitting $\Pl\to\Pl\gamma$ 
(splitting function $P_{ff}$) with real~$\gamma$ emission
is relevant, \ie diagrammatically the initial- or final-state
lepton line goes into the hard process without conversion to
intermediate photons [first term in \refeq{eq:GammallQEDNNLO}].
Explicit analytical results are known in the soft-photon
region ($x\to1$) to all orders in $\alpha$~\cite{Gribov:1972ri} 
and for arbitrary
$x$~values even up to 
order~$\alpha^5$~\cite{Montagna:1991ku,Cacciari:1992pz,Arbuzov:1999cq,%
Blumlein:2004bs,Blumlein:2007kx}.
Here we just quote a frequently used form of the 
LL structure function which combines the full LL correction
up to order $\alpha^3$ with the LL soft-photon exponentiation
and is sufficient for many practical purposes~\cite{Beenakker:1996kt},
\begin{align}
\Gamma_{ll}^{\LL,\gamma\mbox{\scriptsize -rad}}(x,\muF^2) ={}&
    \frac{\exp\left(-\frac{1}{2}\beta_\Pl\gamma_{\rE} +
        \frac{3}{8}\beta_\Pl\right)}
{\Gamma\left(1+\frac{1}{2}\beta_\Pl\right)}
    \frac{\beta_\Pl}{2} (1-x)^{\frac{\beta_\Pl}{2}-1} - \frac{\beta_\Pl}{4}(1+x)
\nn\\
&  {} - \frac{\beta_\Pl^2}{32} \Biggl\{ \frac{1+3x^2}{1-x}\ln x
    + 4(1+x)\ln(1-x) + 5 + x \Biggr\}
\nn\\
&  {} - \frac{\beta_\Pl^3}{384}\Biggl\{
      (1+x)\left[6\Li_2(x)+12\ln^2(1-x)-3\pi^2\right]
+\frac{1}{1-x}\Biggl[ \frac{3}{2}(1+8x+3x^2)\ln x
\nn\\& \quad\quad {}
{}+6(x+5)(1-x)\ln(1-x)
+12(1+x^2)\ln x\ln(1-x)-\frac{1}{2}(1+7x^2)\ln^2\! x
\nn\\
& \quad\quad  {}
+\frac{1}{4}(39-24x-15x^2)\Biggr] \Biggr\},
\end{align}
where $\beta_\Pl$ is defined in \refeq{eq:betal} 
and $\gamma_{\rE}$ denotes the Euler--Mascheroni constant.

Finally, we recall that the full series of powers $L_l^n$ with
$L_l$ given in \refeq{eq:QEDlog} does not comprise all
logarithmically enhanced QED corrections, but defines the
{\it leading logarithmic} (LL) accuracy. 
The {\it next-to-leading logarithmic} (NLL) order scales like $\alpha
L_l^{n-1}$ and involves one logarithmic enhancement factor less than the LL
level in each given order $\alpha^n$.
The QED structure functions $\Gamma^{\mathrm{NLL}}$ were
calculated recently in \citeres{Frixione:2019lga,Bertone:2019hks}.

\subsubsection{QED structure functions versus parton showers}
\label{se:SFandPS}

While analytical results for structure functions are quite easy to
implement in predictions, they also have some shortcomings, which
are due to the fact that the kinematics of the emitted particles
is integrated over up to their total energy fraction $(1-x)$.
In particular, all information on transverse momenta from
nearly collinear splittings is lost, and no information on the
momentum share between emitted particles from multiple emissions is kept.
In this respect, a probabilistic parton-shower approach is doing better.
Parton showers are based on the universal factorization properties
of emission processes as well, and as such also limited to some
(leading or subleading) logarithmic approximation,
but they provide the fully differential information on all
emitted particles, which can be important if decaying particles
have to be reconstructed to very high precision.
For this reason,
multi-purpose event generators typically employ QED parton showers
to account for photonic radiation effects, as described in the
program descriptions of 
{\sc Herwig}~\cite{Seymour:1991xa,Hamilton:2006xz},
{\sc Pythia}~\cite{Sjostrand:2014zea}, and
{\sc Sherpa}~\cite{Schonherr:2008av,Hoeche:2009xc}
in more detail.
A standalone QED parton shower is, e.g., provided by the
program {\sc Photos}~\cite{Barberio:1990ms,Barberio:1993qi,Golonka:2005pn}, which
can be linked to any Monte Carlo program providing cross-section 
predictions.
Finally, applications and further developments of
QED parton showers have been described in the context of predictions for
specific processes, such as for Drell--Yan-like processes with the
Monte Carlo generators 
{\sc Horace}~\cite{CarloniCalame:2003ux,CarloniCalame:2005vc,CarloniCalame:2006zq,%
CarloniCalame:2016ouw},
{\sc Wgrad}~\cite{Bernaciak:2012hj}, and
{\sc Winhac}~\cite{Placzek:2003zg}.

In complete analogy to higher-order QCD calculations, a consistent
combination of NLO (or even higher-order) calculations with parton
showers is non-trivial and has to carefully avoid double-counting of
fixed-order contributions, in order not to destroy NLO (or higher)
accuracy. Note that this combination is quite simple if higher-order
QED effects are included via LL structure functions, where
double-counting can be avoided easily due to the analytic knowledge of
all higher-order effects.  For a consistent matching of NLO EW
corrections with QED parton showers mostly the 
{\sc POWHEG}~\cite{Nason:2004rx,Frixione:2007vw} approach was employed,
as, for instance, described for Drell--Yan production in
\citeres{Bernaciak:2012hj,CarloniCalame:2016ouw,Muck:2016pko}.  For
many important processes with leptons analyzed at the LHC with high
precision, such as EW diboson production, this step is still to be
done.

More generally, also the W and Z bosons should be included in PDFs and
EW parton showers. While this will become relevant at very-high-energy
colliders, it is presently work in progress.  PDFs for transverse and
longitudinal EW gauge bosons were computed in terms of deep-inelastic
scattering structure functions, following the LUXqed method to
determine the photon PDF \cite{Fornal:2018znf}.  LO evolution of
parton fragmentation functions for all the SM fermions and bosons were
studied in \citere{Bauer:2018xag}.  The inclusion of the emission of W
and Z bosons in a traditional QCD + QED shower was investigated in
\citere{Christiansen:2014kba} and the emission of collinear W-bosons
in jets in \citere{Krauss:2014yaa}.  EW collinear splitting functions
for the complete SM were derived and implemented in a comprehensive,
practical EW showering scheme in \citere{Chen:2016wkt}.

\section{Electroweak radiative \texorpdfstring{corrections---general}
{corrections - general} features}
\label{se:general}

\subsection{Input-parameter schemes}
\label{se:input_schemes}

In order to perform consistent higher-order calculations in the SM, a
set of input parameters together with the corresponding numerical
values has to be specified. It is crucial to stick to a set of
independent parameters in order to guarantee gauge independence and
consistency in the results.  Moreover, the input-parameter set should
be chosen in a phenomenologically sensible way, \ie precisely known
and well-defined parameters should be preferred.  Since this is the
case for the electromagnetic coupling and many of the masses, these
are often used as input parameters.  Such input-parameter schemes that
are suited for on-shell (OS) renormalization schemes are presented in
\refse{se:input_reco}.  In \refse{se:input_msbar} we discuss $\MSbar$
renormalization schemes, which are sometimes used as an alternative to
OS renormalization schemes and require different or at least converted
input parameters.

\subsubsection{The \texorpdfstring{$\alpha(0)$, $\alpha(\MZ^2)$,
    and $\GF$}{alpha(0), alpha(MZ), and GF)} schemes}
\label{se:input_reco} 

An obvious choice for the input parameters of the SM consists of 
the electromagnetic coupling $\alpha=e^2/(4\pi)$,
the strong coupling $\alphas=\gs^2/(4\pi)$,
the weak gauge-boson masses $\MW$ and $\MZ$,
the Higgs-boson mass $\MH$,
the fermion masses $\Mf$, and the CKM matrix $V$.

In the EW sector of the SM, the masses of the particles are
conveniently defined as {\it pole masses}, fixed by the locations of
the particle poles in the respective propagators. Since the
non-perturbative strong interaction at the scale of the quark masses
renders the definition of pole masses problematic, for quarks it is
often useful to switch to a running mass at some appropriate scale
\cite{Tarrach:1980up,Brock:1993sz,Tanabashi:2018oca}.  Properly
defined observables and their predictions should be insensitive to the
perturbatively problematic light quark masses, as discussed below in
more detail.  The Yukawa couplings do not represent independent
parameters, but are fixed by the fermion masses and the other EW input
parameters.  Disturbing the relation between Yukawa couplings and
fermion masses, violates EW gauge invariance and can lead to
inconsistent and wrong predictions, especially in the calculation of
EW corrections.

As discussed in \refse{se:renCKM}, the definition of the CKM matrix in
the presence of EW corrections is a non-trivial task.  However, apart
from applications in flavour physics, for high-energy scattering the
approximation of taking all quarks 
other than the top quark and possibly the bottom quark massless
and ignoring mixing with the third generation is an
appropriate procedure.  In this approximation, the CKM matrix can only
become relevant in charged-current processes that are not democratic
w.r.t.\ quark flavours so that the unitarity relations
$\sum_{k}V_{ik}V^*_{kj}=\de_{ij}$ are disturbed. This is, for
instance, the case in charged-current quark--antiquark-annihilation
channels, such as Drell--Yan-like W-boson production, where it leads
to global factors $|V_{ij}|^2$ in partonic cross sections with
$q_i\bar q_j$ or $q_j\bar q_i$ initial states.  This statement holds
at the level of NLO EW corrections as well because of the mass
degeneracy in the first two quark generations where the mixing takes
place.

For the boson masses $\MW$, $\MZ$, and $\MH$, real OS masses are
usually employed. Details about different schemes that become relevant
if the instability of those bosons matters are discussed in
\refse{se:unstable}.  Here, we emphasize that the weak mixing angle
$\theta_{\rw}$ is not an independent input parameter. The most common
choice in EW physics follows the OS prescription
\cite{Sirlin:1980nh} which defines $\cw=\cos\theta_{\rw}=\MW/\MZ$ and
$\sw=\sqrt{1-\cw^2}$ via the OS W- and Z-boson masses.  Taking
$\sw$ as independent parameter in addition to $\MW$ and $\MZ$, e.g.\ 
by setting it to some ad hoc value or to the sine of the effective
weak mixing angle measured at the Z~pole, in general breaks gauge
invariance, destroys gauge cancellations, and can lead to totally
wrong results, even at LO.

For the electromagnetic coupling constant $\alpha$ basically three
different input values are used: the fine-structure constant
$\alpha(0)\approx1/137$ ({\em $\alpha(0)$~scheme}), the effective
value $\alpha(\MZ^2)\approx1/129$ ({\em $\alpha(\MZ^2)$~scheme}),
where $\alpha(0)$ is evolved via renormalization-group equations from
$Q^2=0$
to the Z~pole, and an effective value derived
from the Fermi constant $\GF$ leading to
$\alpha_{\GF}=\sqrt{2}\GF\MW^2(1-\MW^2/\MZ^2)/\pi\approx1/132$,
defining the so-called {\em $\GF$~scheme}. The various choices for
$\alpha$ differ by $2{-}6\%$ and represent an important part of the
{\it input-parameter scheme}.  In practice, the most appropriate
scheme depends on the nature of the process under consideration. In
any case, it is crucial that a common coupling factor $\alpha^n$ is
used in complete gauge-invariant subsets of diagrams, otherwise
important consistency relations (compensation of divergences,
unitarity cancellations, etc.)  are destroyed. Note that this does not
necessarily mean that there is only one value of $\alpha$, but that
all factors $\alpha$ have to be global factors to gauge-invariant
pieces of amplitudes.

In the following we describe the input-parameter schemes 
appropriate for a contribution to a cross section (or squared matrix
element) whose LO contribution is proportional to a fixed order
$\alpha_{\mathrm{s}}^m\alpha^n$, \ie NLO EW contributions scale like
$\alpha_{\mathrm{s}}^m\alpha^{n+1}$.  Corrections to LO contributions
scaling with different powers $m$ or $n$ belong to disjoint
gauge-invariant contributions, which can be treated
independently.  The standard QED definition of $\alpha$ (see
\refse{se:charge_ren}) employs an
OS renormalization condition in the Thomson limit (photon
momentum transfer $Q^\mu=0$), leading to the renormalized value
$\alpha=\alpha(0)$ of the $\alpha(0)$~scheme.  
When applied to processes at energies of the scale of the EW gauge
bosons or above, this scheme leads to large logarithmic corrections
involving the ratios of the light fermion masses and the gauge-boson
masses. It turns out that these logarithmic corrections are related to
the running of the electromagnetic coupling from low ($Q^2=0$) to high
($|Q^2|\gtrsim\MW^2$) 
scales $Q^2$. Thus, they can be
absorbed upon using a suitable input scheme for $\alpha$.%
\footnote{There are additional logarithmic corrections involving
  fermion masses. These originate from collinear singularities that
  can be linked to the external lines of the process under
  consideration. For (anti)quarks they are absorbed in parton
  distribution functions or fragmentation functions, for leptons
  those contributions can be safely calculated in perturbation theory
  and, if needed, supplemented by leading corrections beyond NLO.
  These issues are discussed in \refse{se:real}.}

Let us first consider a process without external photons in the
initial or final state.  In the $\alpha(0)$~scheme with OS charge
renormalization, each of the $n$ EW couplings leads to a relative
correction $2\delta Z_e$ to the cross section.  The charge
renormalization constant $\delta Z_e$ contains mass-singular terms of
the form $\alpha\ln \Mf$ from each light fermion $f$ that remain
uncancelled in the EW corrections.  These terms are contained in the
quantity
\begin{equation}\label{eq:Delta_alpha_ons}
  \Delta\alpha(\MZ^2) =
  \PIA_{\Pf\ne \Pt}(0) - \Re\left\{\PIA_{f\ne \Pt}(\MZ^2)\right\}
=
\frac{\alpha(0)}{3\pi}\sum_{f\ne\mathrm{t}} \NCf \Qf^2
\left[\ln\frac{\MZ^2}{\Mf^2}-\frac{5}{3}\right]
+{\cal O}\left(m_f^2/\MZ^2\right),
\end{equation}
with $\Pi^{\FA\FA}_{\Pf\ne \Pt}$ denoting the photonic vacuum
polarization induced by all fermions $\Pf$ (with charge $\Qf$) other
than the top quark (see also \citere{Denner:1991kt}), and
$\NCl=1$ and $\NCq=3$ are the colour multiplicities
for leptons and quarks, respectively.  The vacuum polarization is
related to the transverse part of the photon self-energy defined in
\refse{se453eforc} according to,
\beq
\Pi^{\FA\FA}(Q^2) = \frac{\SIAA(Q^2)}{Q^2}.
\eeq
The correction $\Delta\alpha(\MZ^2)\approx6\%$ quantifies the running of $\alpha$
from $Q^2=0$ to the high scale $Q^2=\MZ^2$ induced by vacuum polarization effects
of the light fermions,
\begin{equation}
\alpha(\MZ^2)=\frac{\alpha(0)}{1-\Delta\alpha(\MZ^2)},
\end{equation}
a quantity that is non-perturbative, as signalled by its sensitivity
to the light-quark masses.  The appearance of $\Delta\alpha(\MZ^2)$ in
the denominator results from a resummation of the large logarithms
\cite{Marciano:1979yg,Sirlin:1983ys}. The numerical value for the
hadronic contribution $\Dalphahad(\MZ^2)$ to
$\Delta\alpha(\MZ^2)$ and thus to $\alpha(\MZ^2)$ is obtained from an
experimental analysis of $\Pep\Pem$ annihilation into hadrons and
hadronic $\tau$~decays at low energies below the $\PZ$-boson
resonance, combined with theoretical arguments using dispersion
relations~\cite{Eidelman:1995ny}.  Eliminating $\alpha(0)$ in favour
of $\alpha(\MZ^2)$ in the LO prediction, \ie turning to the
$\alpha(\MZ^2)$~scheme, effectively subtracts the
$\Delta\alpha(\MZ^2)$ terms from the EW corrections and thus cancels
all light-fermion logarithms resulting from charge renormalization in
the EW corrections.  This cancellation happens at each loop order in
$\alpha$, \ie employing the $\alpha(\MZ^2)$~scheme resums the dominant
effects from the running of $\alpha$ and, at the same time, removes
the (perturbatively unpleasant) light-quark masses, which should have
been taken from a fit to $\Re\Pi^{\FA\FA}_{f\ne
  \Pt}(Q^2)-\Pi^{\FA\FA}_{f\ne \Pt}(0)$ otherwise.  For high-energy
processes without external photons the $\alpha(\MZ^2)$~scheme is
appropriate from this point of view.  Practically, the
$\alpha(\MZ^2)$~scheme can be implemented by 
using $\alpha(\MZ^2)$ as electromagnetic coupling and
replacing the charge
renormalization constant \refeqf{eq:DZE} by
\begin{align}  \label{eq:DZE_alz}
\DZe\big|_{\alpha(\MZ^2)} ={}&
\frac{1}{2}\PIA(0)-\frac{\sw}{\cw}\frac{\SIAZ(0)}{\MZ^{2}}
-\frac{1}{2}\Delta\al(\MZ^2).
\end{align}
Note that the same value of $\alpha=\alpha(\MZ^2)$ should be used
in all terms of \refeq{eq:DZE_alz} to guarantee the complete
cancellation of the fermion-mass logarithms.
While all terms involving logarithms of fermion masses are absorbed
into $\Delta\alpha(\MZ^2)$ or $\alpha(\MZ^2)$ (and parton distribution and
fragmentation functions), a non-logarithmic dependence on the small
quark masses remains. These contributions are suppressed as
$m_\Pq^2/E^2$ and thus negligible at sufficiently high energies $E$.

For processes with $l$ external photons ($l\le n$), however, the
relative EW corrections contain $l$~times the photonic wave-function
renormalization constant $\de Z_{AA}$, which exactly cancels the
light-fermion mass logarithms appearing in $2\de Z_e$. This statement
expresses the fact that external, \ie real, photons effectively couple
with the scale $Q^2=0$.  Consequently, the coupling factor $\alpha^n$
in the LO cross section should be parametrized as
$\alpha(0)^l\alpha(\MZ^2)^{n-l}$, and the corresponding
$\Delta\alpha(\MZ^2)$ terms should be subtracted only $(n-l)$ times
from the EW correction, in order to absorb the large effects from
$\Delta\alpha(\MZ^2)$ into the LO prediction.  For $l=n$, this scaling
corresponds to the pure $\alpha(0)$~scheme, but for $l<n$ to a mixed
scheme. Thus, if the appropriate definition of $\alpha$ is used, large
logarithms are avoided and the hadronic vacuum polarization is not
required as input. On the other hand, $\alpha(\MZ^2)$ is needed as input
parameter in this case.  This is valid if photons truly act as
external states, either in the preparation of initial states or in
their detection. Exceptions are initial-state photons in
photon-induced processes in hadronic collisions (see
\refse{se:QED-PDFS}) and final-state photons as parts of
electromagnetic showers of photons and leptons (see
\refse{se:SFandPS}).

The $\GF$~scheme, finally, offers the possibility to absorb some
significant universal corrections connected with the renormalization
of the weak mixing angle into LO contributions. At NLO, the $\GF$ and
$\alpha(0)$~schemes are related according to
\begin{equation}
\alpha_{\GF}=\frac{\sqrt{2}\GF\MW^2}{\pi}
\left(1-\frac{\MW^2}{\MZ^2}\right)
=\alpha(0)\left(1+\Delta r^{(1)}\right) + {\cal O}(\alpha^3),
\label{eq:dr}
\end{equation}
where $\Delta r^{(1)}$ is the NLO EW correction to muon
decay~\cite{Sirlin:1980nh,Marciano:1980pb,Hollik:1988ii,Denner:1991kt},
which can be written as
\begin{equation}
\Delta r^{(1)} ={} \Pi^{AA}(0) -\frac{c_{W}^{2}}{s_{W}^{2}}
\left(\frac{\Sigma^{ZZ}_{\rT}(M_{Z}^{2})}{M_{Z}^{2}}
-\frac{\Sigma^{W}_{\rT}(M_{W}^{2})}{M_{W}^{2}}\right)
+\frac{\Sigma^{W}_{\rT}(0)-\Sigma^{W}_{\rT}(M_{W}^{2})}{M_{W}^{2}} 
+2\frac{c_{W}}{s_{W}}\frac{\Sigma^{AZ}_{\rT}(0)}{M_{Z}^{2}}
+\frac{\alpha(0) }{4\pi s_{W}^{2}}
\left(6 +  \frac{7-4s_{W}^{2}}{2s_{W}^{2}}\ln c_{W}^{2}\right).
\end{equation}
The last term as well as part of the term involving
$\Sigma^{AZ}_{\rT}(0)$ result from vertex and box corrections.  The
quantity $\Delta r^{(1)}$ can be decomposed according to
\beq
\Delta r^{(1)}=\Delta\alpha(\MZ^2)-\Delta\rho^{(1)}\cw^2/\sw^2
+\Delta r_{\mathrm{rem}}
\eeq
with $\Delta\alpha(\MZ^2)$ from \refeq{eq:Delta_alpha_ons}, the
universal (top-mass-enhanced) correction to the $\rho$~parameter
\cite{Ross:1975fq,Veltman:1977kh,Chanowitz:1978uj}
\beq
\Delta\rho^{(1)}
=\frac{3\alpha(0)\Mt^2}{16\pi\sw^2\MW^2}
\eeq
and a small remainder $\Delta r_{\mathrm{rem}}$.  In view of the
running of $\alpha$, the $\GF$~scheme corresponds to an $\alpha$ at
the EW scale similar to the $\alpha(\MZ^2)$~scheme, since $\Delta
r^{(1)}$ contains exactly one unit of $\Delta\alpha(\MZ^2)$. The
$\GF$~scheme is thus similar to the $\alpha(\MZ^2)$~scheme as far as
the running of $\alpha$ is concerned, \ie it is preferable over the
$\alpha(0)$~scheme except for the case of external photons.  The
choice between the $\GF$ or $\alpha(\MZ^2)$~scheme is driven by the
appearance of $\sw$ in the EW couplings. Whenever $\sw$ (or $\cw$) is
involved in an EW coupling, the corresponding EW correction receives a
contribution from $\Delta\rho^{(1)}$ according to
$\sw^2\to\sw^2+\Delta\rho^{(1)}\cw^2$ originating from the OS
renormalization of the weak mixing angle.  Using \refeq{eq:dr}, it is
easy to see that the combination $\alpha_{\GF}/\sw^2$, which corresponds
to the $\SU(2)_\rw$ gauge coupling, does not receive this universal
correction, since the $\Delta\rho^{(1)}$ terms from $\Delta r^{(1)}$
and the correction associated with $\sw^2$ cancel. In other words, in
the $\GF$~scheme the leading correction to the $\rho$~parameter is
absorbed into the LO $\SU(2)_\rw$ gauge coupling.  This statement
holds also at the two-loop level~\cite{Consoli:1989fg}.  The
$\GF$~scheme is thus most appropriate for describing couplings of
W~bosons. For Z~bosons this scheme absorbs at least part of the
$\Delta\rho^{(1)}$ corrections because of additional $\cw$ factors in
the coupling from the weak mixing, while the scheme is actually not
appropriate for photonic couplings.  However, also here it should be
kept in mind that a fixed scheme with a global definition of couplings
has to be employed within gauge-invariant subsets of diagrams.  In
most cases, it is advisable to use the $\GF$~scheme for couplings that
involve weak bosons, although the gauge-invariant subsets of diagrams,
in general, also contain internal photons.  Practically, the
$\GF$~~scheme can be implemented by substituting the charge
renormalization constant \refeqf{eq:DZE} by
\begin{align}  \label{eq:DZE_alGF}
\DZe\big|_{\GF} ={}& 
\frac{1}{2}\PIA(0)-\frac{\sw}{\cw}\frac{\SIAZ(0)}{\MZ^{2}}
-\frac{1}{2}\Delta r^{(1)},
\end{align}
and the r.h.s.\ should involve a common factor of $\alpha=\alpha_{\GF}$.


\subsubsection{\texorpdfstring{\MSbar}{MSbar}~scheme and running couplings}
\label{se:input_msbar}

Besides the OS scheme, the $\MSbar$ scheme is sometimes used for
the renormalization of the EW gauge couplings. While masses could also
be renormalized in the $\MSbar$ scheme, a hybrid scheme where masses
are defined on shell and couplings via $\MSbar$ subtraction is more
customary \cite{Fanchiotti:1992tu} and used for instance by the
Particle Data Group \cite{Tanabashi:2018oca}.  In this section we
follow the convention of \citeres{Fanchiotti:1992tu,Tanabashi:2018oca}
to denote $\MSbar$ parameters with a caret
(not to be confused with the carets used in the BFM above).

Inspecting the UV divergences and the top-quark contributions in the
OS charge renormalization constant \refeqf{eq:DZE},%
\footnote{Explicit expressions for the 1PI contributions to the
one-loop self-energies can, \eg, be found in \citere{Denner:1991kt}.}
we obtain
the charge renormalization constant in the $\MSbar$ scheme as
\beq
\de Z_{e}(\muR^2) 
= -\frac{\alphams(\muR^2)}{4\pi}
\left[
\left(\frac{7}{2}-\frac{2}{3}\sum_{f} \NCf \Qf^2\right)
\left(\Delta +\ln\frac{\mu^2}{\muR^2} \right)
-2\theta\left(\Mt^2-\muR^2\right)\Qt^2\ln\frac{\muR^2}{\Mt^2}
\right],
\label{eq:dZems}
\eeq
where $\alphams = \hat{e}^2/(4\pi)$ with the renormalized coupling $\hat{e}$
in the \MSbar\ scheme, $\sum_{f} \NCf \Qf^2=8$, $\mu$ is the
mass parameter of DR, and $\muR$ the renormalization scale. Here the
contributions of all degrees of freedom lighter than $\muR$ are
renormalized in the $\MSbar$ scheme, while the top-quark contribution
is subtracted at zero momentum transfer for $\Mt^2>\muR^2$ to ensure
decoupling.

The OS and $\MSbar$ couplings are related via the bare coupling by
\beq
(1+\de Z_e)e = e_0 = 
\left[1+\de Z_{e}(\muR^2)\right]\ems(\muR^2).
\eeq
This can be solved for the running $\MSbar$ coupling, yielding
\beq\label{eq:running_alpha_MSbar}
\alphams(\muR^2) = \frac{\alpha(0)}{1-\Delta\alphams(\muR^2)}
\eeq   
up to two-loop terms with
\begin{align}\label{eq:Delta_alpha_MSbar}
\Delta\alphams(\muR^2) 
={} 2\left[\de Z_e - \de Z_{e}(\muR^2)\right] 
={}& \Pi^{\FA\FA}_{\Pq\ne\Pt}(0) - 
\Re\Pi^{\FA\FA}_{\Pq\ne\Pt}(\MZ^2)
+ \frac{\alpha(0)}{\pi}\Biggl[
\sum_{\Pq\ne\Pt}\Qq^2\left(\frac{5}{3}-\ln\frac{\MZ^2}{\muR^2}\right)
\notag\\ &{}
- \frac{1}{3}\sum_{\Pl}\Ql^2\ln\frac{\Ml^2}{\muR^2}
- \Qt^2\ln\frac{\Mt^2}{\muR^2}\theta(\muR^2-\Mt^2)
+\frac{7}{4}\ln\frac{\MW^2}{\muR^2}-\frac{1}{6}\Biggr].
\end{align}
Here we have substituted
\beq
 \Pi^{\FA\FA}_{\Pq\ne\Pt}(0) = 
\left[ \Pi^{\FA\FA}_{\Pq\ne\Pt}(0) - 
\Re\Pi^{\FA\FA}_{\Pq\ne\Pt}(\MZ^2)\right]
+\Re\Pi^{\FA\FA}_{\Pq\ne\Pt}(\MZ^2)
\eeq
and used the perturbative one-loop expressions for the contributions
of the leptons and the top quark as well as for the contribution
$\Re\Pi^{\FA\FA}_{\Pq\ne\Pt}(\MZ^2)$ of the light quarks to the vacuum
polarization in the limit $\muR^2,\MZ^2\gg\Mf^2$.  The difference
$\left[ \Pi^{\FA\FA}_{\Pq\ne\Pt}(0) -
  \Re\Pi^{\FA\FA}_{\Pq\ne\Pt}(\MZ^2)\right]$ can be determined
experimentally as described in \refse{se:input_reco}.  Since the top
quark is decoupled for $\muR^2<\Mt^2$, \refeq{eq:running_alpha_MSbar}
allows us to compute the running coupling $\alphams(\muR^2)$
independently of $\Mt$ in this case. A similar procedure can be
applied to other unknown heavy particles.

The difference between \refeqs{eq:Delta_alpha_MSbar} and
\refeqf{eq:Delta_alpha_ons} for $\muR^2=\MZ^2$ is given by
\beq\label{eq:Delta_alpha_diff}
\Delta\alphams(\MZ^2) - \Delta\alpha(\MZ^2) = 
\frac{\alpha(0)}{\pi}\left[\frac{100}{27}-\frac{1}{6}
  -\frac{7}{4}\ln\frac{\MZ^2}{\MW^2}\right].
\eeq
Contributions of higher-order corrections to
\refeqs{eq:Delta_alpha_ons} and \refeqf{eq:Delta_alpha_MSbar} or
\refeqf{eq:Delta_alpha_diff} can be found in
\citeres{Fanchiotti:1992tu,Tanabashi:2018oca} and references
therein.\footnote{The quantity $\Delta\alphams(\MZ^2)$ is called
  $(\alpha/\pi)\Delta_\gamma$ in \citere{Fanchiotti:1992tu}.}

In the OS scheme, the weak mixing angle is defined according to
\refeq{eq:swren} \cite{Sirlin:1980nh}.  Alternatively, it can be
defined in the $\MSbar$ scheme
\cite{Marciano:1979yg,Sirlin:1989uf,Marciano:1991ix,Fanchiotti:1992tu,Sirlin:2012mh},
which is useful for comparing with predictions of grand unification.
In this scheme, the weak mixing angle is related to the running
electromagnetic $\ems^2(\muR^2)$ and $\SU(2)_\rw$ gauge coupling
$\gms_2(\muR^2)$ as follows,
\beq
\swms^2(\muR^2) \equiv
\sin^2\hat{\theta}_\rw(\muR^2) = \frac{\ems^2(\muR^2)}{\gms_2^2(\muR^2)}.
\eeq
If $\swms^2(\muR^2)$ is used as free parameter in a calculation, only
one of the masses $\MW$ and $\MZ$ can be treated as a free
parameter, while the other one plays the role of a derived parameter.
Otherwise the parameter set would be overdetermined leading to potential
gauge-invariance violation in predictions.

The pure $\MSbar$ counterterm $\de\swms^2$ can be
determined from any definition of the weak mixing angle. Extracting
the UV-divergent part of the OS definition \refeqf{eq:Dsw} yields
\begin{align}
\de\swms^2(\muR^2) = {}&\cwms^2\left[
\frac{\SIZZ(\MZ^2)}{\MZ^2} - \frac{\SIW(\MW^2)}{\MW^2} \right]
_{\text{UV}} 
= {} \frac{\hat\al(\muR^2)}{4\pi}\left[\frac{11}{3}\swms^2+\frac{19}{6}\right]
\left(\De+\ln\frac{\mu^2}{\muR^2}\right),
\label{eq:swms1}
\end{align}
where the UV-divergent part involves all singular terms in the form
$\De+\ln({\mu^2}/{\muR^2})$.
The running $\swms^2(\muR^2)$ satisfies the renormalization-group
equation
\beq
\muR^2\frac{\partial}{\partial \muR^2} \swms^2(\muR^2) = 
-\muR^2\frac{\partial}{\partial \muR^2} \de\swms^2(\muR^2) = 
\frac{\hat\al(\muR^2)}{4\pi}
\left[\frac{11}{3}\swms^2+\frac{19}{6}\right] + \text{higher orders}.
\label{eq:swms2}
\eeq

The running of $\swms^2(\muR^2)$, as determined by \refeqs{eq:swms1}
and \refeqf{eq:swms2}, receives contributions from bosonic and
fermionic loops, with equal contributions from all three fermion
generations. With this definition, however, the effects of 
a heavy top quark do in general not decouple in EW corrections.
To decouple heavy-top-quark contributions several modifications
in the definition of $\swms^2(\muR^2)$ have been proposed in the
literature.
A standard choice~\cite{Fanchiotti:1992tu} is to use the
neutral-current interaction of the \PZ~boson with a massless fermion
which has the general form \cite{Sarantakos:1982bp,Sirlin:2012mh}
\begin{align}
\langle 0|J_\FZ^\mu|\bar f(\bar p)f(p)\rangle 
= {}& V_f(q^2)\, \vbar(\bar p) \gamma^\mu
\left[\If\om_- - \Qf\kappams_f(q^2)\swms^2(q^2)\right]u(p) 
\notag\\ = {}& 
V_f(q^2) \,\vbar(\bar p) \gamma^\mu
\left[\If\om_- - \Qf\kappa_f(q^2)\sw^2\right]u(p),
\end{align}
where $V_f(q^2)$, $\kappams_f(q^2)$ and its OS counterpart
$\kappa_f(q^2)$ are EW form factors and 
$q^2=(\bar p+p)^2$ is the invariant mass of
the \PZ~boson. The form factors $\kappams_f(q^2)$ or
$\kappa_f(q^2)$ get contributions from vertex corrections and from
the photon--\PZ-boson mixing energy, and the $\MSbar$ counterterm can
be written as \cite{Sirlin:1989uf}
\beq
\de\swms^2(\muR^2) =
\frac{\cwms\swms}{\MZ^2}\left[\SIAZ(\MZ^2)+\SIAZ(0)\right]_{\text{UV}}.
\label{eq:swms3}
\eeq
Decoupling of the top quark for $\muR^2<\Mt^2$
(and analogously for other heavy particles) can be
incorporated by including the full top-quark contribution to 
$\SIAZ(q^2)/q^2$ in \refeq{eq:swms3} for $q^2\to0$
instead of using $\SIAZ(\MZ^2)\bigr|_{\text{UV}}/\MZ^2$.
This leads to the following counterterm~\cite{Marciano:1991ix},
\begin{align}
\de\swms^2(\muR^2) = {}&
\frac{\al}{4\pi}\left(\Delta+\ln\frac{\mu^2}{\muR^2}\right)
\left[\frac{11}{3}\swms^2+\frac{19}{6}\right] 
+\frac{\al}{\pi}\ln\frac{\Mt}{\muR}
\left[\frac{1}{3}-\frac{8}{9}\swms^2\right] \theta(\Mt^2-\muR^2).
\label{eq:swms4}
\end{align}
For higher-order corrections to $\de\swms^2(\muR^2)$ we refer to
\citeres{Fanchiotti:1992tu,Tanabashi:2018oca} and references therein.
The scale $\muR$ is conveniently chosen to be $\MZ$ for many EW processes.

The derivation of the \MSbar\ renormalization constants,
including their modifications for decoupling a heavy top quark,
is formally significantly simplified if the background-field 
vertex functions are used, owing to the connection between 
gauge-field and gauge-coupling renormalization constants 
as given in \refeq{eq:delZB} for the EWSM.
The divergences of the gauge-coupling renormalization constants
can, thus, be obtained directly from the derivatives of the
relevant BFM self-energies at zero momentum transfer,
\beq
\de Z_e(\muR^2) = \frac{1}{2}
\Sigma^{\prime\FAhat\FAhat}_{\rT}(0)\big|_{\mathrm{\overline{UV}}},
\qquad
\de\swms^2(\muR^2) = 
\cwms\swms\frac{\Sigma^{\FAhat\FZhat}_{\rT}(\MZ^2)\bigr|_{\mathrm{\overline{UV}}}}{\MZ^2}
= \cwms\swms\Sigma^{\prime\FAhat\FZhat}_{\rT}(0)\big|_{\mathrm{\overline{UV}}},
\eeq
where $\overline{\mathrm{UV}}$ here means to include besides all
UV-divergent parts the full top-quark loops to ensure decoupling.
In the last equation we used the fact that
$\Sigma^{\FAhat\FZhat}_{\rT}(q^2)$ vanishes for $q^2=0$ and that its
divergent part is proportional to $q^2$.
This procedure directly reproduces the results given
in \refeqs{eq:dZems} and \refeqf{eq:swms4}.

Finally, we point out that the relation between the muon decay constant
and the SM parameters somewhat changes in the $\MSbar$ 
scheme~\cite{Fanchiotti:1989wv},
\beq
\sqrt{2}\GF = \frac{\pi\al(0)}{\MW^2\swms^2(\MZ^2)(1-\Delta \hat{r}_\FW)}
\eeq
with the modified EW correction $\Delta \hat{r}_\FW$.
Moreover, the modified relation between the weak mixing angle and 
the OS gauge-boson masses leads to a non-trivial 
$\rho$~parameter~\cite{Degrassi:1990tu},
\beq
\hat\rho = \frac{\cw^2}{\cwms^2(\MZ^2)} = \frac{\MW^2}{\cwms^2(\MZ^2)\MZ^2}.
\eeq 

\subsection{The structure of NLO electroweak corrections}

\subsubsection{EW and QCD corrections for general processes}
\label{se:EW_QCD_mixing}

The LO of simple processes receives contributions of a single power in
the strong and EW couplings, \ie the LO cross section is proportional
to $\alphas^n\alpha^m$ with unique $n$ and $m$. In this case, the NLO
corrections can be trivially separated into NLO QCD corrections and
NLO EW corrections of the orders $\alphas^{n+1}\alpha^m$ and
$\alphas^n\alpha^{m+1}$, respectively. Since both contributions are
characterized by complete contributions to physical matrix elements in
the corresponding orders of the couplings, they are individually gauge
invariant.

For general processes, however, the LO matrix element gets
contributions of different orders in the strong and EW couplings.
While the different sizes of the strong and EW coupling constants
suggest a hierarchy of contributions, this is not always respected in
specific processes.  Moreover, in some cases a formally subleading
contribution is of specific interest in view of a particular physical
ingredient.

\begin{figure}
\centerline{\includegraphics[scale=.9,page=16]{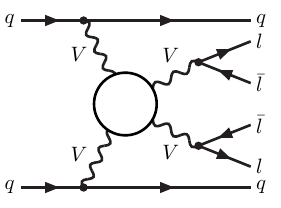}
\qquad\raisebox{-.5em}
{\includegraphics[scale=1,page=21]{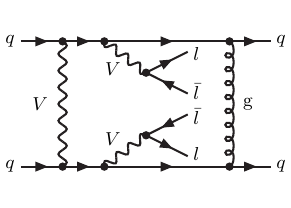}}}
\vspace{-1em}
\caption{Structure of vector-boson scattering at hadron colliders
  (left) and an example for a 
  loop diagram for $\Pq\Pq\to\Pq\Pq\Pl\bar\Pl\bar\Pl\Pl$ (right).}
\label{fig:vbsww}
\end{figure}

An example in this respect is provided by vector-boson scattering
(VBS), \ie processes of the form
$\Pp\Pp\to2\,\mathrm{jets}+4\,$leptons${}+X$.  The LO matrix element
receives purely EW contributions of order $\order{e^6}$ as well as
QCD-induced contributions of order $\order{\gs^2e^4}$. While the
former involve the physically interesting VBS diagrams
(\cf\reffi{fig:vbsww} left), the latter dominate the complete cross
section for most final states.
The LO cross section, derived from squared matrix elements,
then receives contributions of three different
orders, the EW contributions of $\order{\alpha^6}$, the QCD-induced
contributions of $\order{\alpha^4\alphas^2}$, and the interference
contributions of $\order{\alpha^5\alphas}$.

\begin{figure}
\centerline{\includegraphics[page=20]{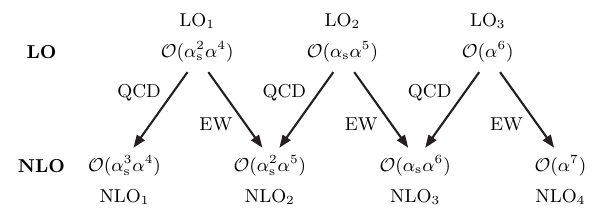}}
\vspace{-.5em}
\caption{Contributing orders at LO and NLO for the processes
  $\Pp\Pp\to2\,\mathrm{jets}+4\,$leptons${}+X$.} 
\label{fig:allorders}
\end{figure}

This tower of contributions proliferates at higher orders as
illustrated in \reffi{fig:allorders}.
There are QCD corrections of relative order $\mathcal{O}{(\alphas)}$
and EW corrections of relative order $\mathcal{O}{(\alpha)}$ to each
LO contribution, leading to four different NLO contributions:
$\mathcal{O}{(\alpha^{7})}$, $\mathcal{O}{(\alphas\alpha^{6})}$,
$\mathcal{O}{(\alphas^{2}\alpha^{5})}$, and
$\mathcal{O}{(\alphas^{3}\alpha^{4})}$.  The order
$\mathcal{O}{(\alpha^{7})}$ contributions are the NLO EW corrections
to the EW-induced LO processes
\cite{Biedermann:2016yds,Denner:2019tmn}.  The order
$\mathcal{O}{(\alpha_{\rm s}^{3}\alpha^{4})}$ contributions furnish
the QCD corrections to the QCD-induced process
\cite{Melia:2010bm,Melia:2011dw,Greiner:2012im,Campanario:2013qba,Campanario:2013gea}.
For the orders $\mathcal{O}{(\alphas\alpha^{6})}$ and
$\mathcal{O}{(\alphas^{2}\alpha^{5})}$, a simple separation of the
EW-induced process and the QCD-induced process is not possible
anymore, and only the complete orders are gauge independent.  In fact,
there are loop diagrams, like the one in the right of
\reffi{fig:vbsww}, that cannot be attributed to one of these
processes. Thus, the order $\mathcal{O}{(\alphas\alpha^{6})}$ contains
QCD corrections to the VBS process as well as EW corrections to the LO
interference.  The QCD corrections were originally computed in
the VBS approximation in
\citeres{Jager:2006cp,Bozzi:2007ur,Jager:2009xx,Jager:2011ms,Denner:2012dz,Campanario:2013gea,Baglio:2014uba},
where the $s$-channel diagrams as well as the interference of $t$- and
$u$-channel diagrams are neglected.  In this approximation, the
interferences of the LO VBS and QCD-induced contributions are
neglected, and the order $\mathcal{O}{(\alphas\alpha^{6})}$ contains
only QCD corrections.  Similarly to the
$\mathcal{O}{(\alphas\alpha^{6})}$, also the order
$\mathcal{O}{(\alphas^{2}\alpha^{5})}$ contains EW corrections to the
QCD-induced contribution as well as QCD corrections to the LO
interference.

Other examples where a tower of contributions appears at LO and NLO
are dijet production \cite{Dittmaier:2012kx,Frederix:2016ost} and
top--antitop production processes \cite{Frederix:2018nkq}.

The treatment of simultaneous expansions in the two coupling constants
$\alphas$ and $\alpha$ in \MGNLO is discussed in
\citere{Alwall:2014hca}. In
\citeres{Frixione:2015zaa,Frederix:2016ost,Frederix:2018nkq} the
notation $\LO_x$ and $\NLO_x$ for the different contributions was
introduced. Here $\LO_1$ stands for the LO contribution with highest
power in the strong coupling, say the contribution of order
$\alphas^n\alpha^m$. Then, $\LO_x$ indicates the contribution of order
$\alphas^{n+1-x}\alpha^{m+x-1}$ and $\NLO_x$ the one of order
$\alphas^{n+2-x}\alpha^{m+x-1}$. For the process
$\Pp\Pp\to\Pt\bar\Pt\Pt\bar\Pt+X$ \cite{Frederix:2017wme} five
different coupling orders show up at LO ($\LO_1$, \ldots, $\LO_5$),
leading to six different coupling combinations at NLO ($\NLO_1$,
\ldots, $\NLO_6$).

\subsubsection{Disentangling weak and electromagnetic corrections}
\label{se:split_ew_qed}

For phenomenological applications and the understanding of the
structure of the EW corrections it is useful to split NLO amplitudes
of a given order $\alphas^n\alpha^m$ further into gauge-invariant
parts as far as possible. We restrict the following discussion to the
one-loop level.

For arbitrary processes, the subset of diagrams (loop diagrams and
counterterms) that involve a closed fermion loop form a
gauge-invariant subset, often called {\em fermionic} NLO corrections.
This is evident from the fact that adding further fermion generations
to the SM (in the absence of generation mixing) would not violate its
gauge invariance.  The remaining NLO corrections are called {\em
  bosonic}.

It is often desirable to split the bosonic NLO corrections further
into an {\em electromagnetic} (or photonic) part and a purely {\em
  weak} part.  In a gauge-invariant way this is possible for processes
that proceed in LO via neutral-current interactions only, but not for
processes that involve W~bosons (and charged would-be Goldstone
bosons) at LO. This fact can, for instance, be seen as follows.
Removing the W~bosons and charged would-be Goldstone bosons from the
SM Lagrangian results is a Lagrangian for a spontaneously broken
$\U(1)_{\mathrm{em}}\times \U(1)_Z$ gauge theory with the same fermion
content, the same electromagnetic couplings, the same neutral-current
couplings, and the same neutral Higgs couplings as in the SM. Here,
$\U(1)_{\mathrm{em}}$ refers to the electromagnetic gauge group with
the photon as massless gauge boson and $\U(1)_Z$ to an abelian gauge
symmetry with the Z~boson as gauge boson and the associated weak
charges of the (left- and right-handed) fermions and neutral scalars
($\FH,\chi$) given by
\begin{equation}
Q_Z = \frac{I^3_{\rw}}{\cw\sw} - \frac{\sw}{\cw}Q.
\end{equation}
The resulting Lagrangian defines a consistent and renormalizable
field theory and, thus, isolates gauge-invariant subsets of SM
contributions to arbitrary gauge-invariant quantities. In particular,
it provides a gauge-invariant subset of the full SM corrections for
processes involving only fermions, photons, Z~bosons, and neutral
scalar bosons.
Since the $\U(1)_{\mathrm{em}}\times \U(1)_Z$ gauge group is abelian
and the photon couples only to fermions, the photon--fermion couplings
can be modified arbitrarily within this theory (while keeping all
other couplings fixed) without destroying its gauge invariance.  Thus,
gauge-invariant subsets can be further classified according to
independent powers of the fermion--photon couplings.  In particular,
for processes involving no charged scalar or vector bosons at LO, this
implies that the bosonic corrections of the full SM can be split into
three gauge-invariant subsets: 1) contributions resulting from photon
exchange between fermions (\ie the QED corrections), 2) corrections
resulting from Z-, $\chi$-, or Higgs-boson exchange, and 3)
corrections that involve W- or $\phi$-boson exchange.
The photonic corrections can be further split
into smaller gauge-invariant subsets by considering the coefficients
of the global charge factors $Q_1Q_2$ of different fermions.

In practice, the gauge-invariant decomposition of the bosonic NLO EW
corrections $\Delta\sigma_{\rm NLO}$ into a purely weak part
$\Delta\sigma_{\rm NLO}^{\rm weak}$ and a photonic part
$\Delta\sigma_{\rm NLO}^{\rm phot}$ can be performed as follows for a
process that does not involve charged-current interactions at LO.  The
virtual photonic part is defined as the set of all diagrams with at
least one photon in the loop coupling to the (charged) fermion lines.
These diagrams are obtained upon inserting one additional photon line
into a LO diagram and do not contain W bosons.  The weak contribution
is then the set of all remaining bosonic one-loop diagrams, which may
also involve W~bosons.  The contributions to the renormalization
constants have to be split accordingly.  

\begin{figure}
\centerline{\includegraphics[page=22,scale=0.9]{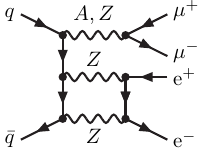}
\quad\includegraphics[page=23,scale=0.9]{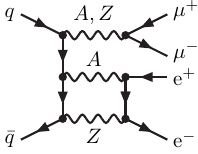}
\quad\includegraphics[page=24,scale=0.9]{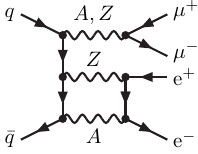}
\quad\includegraphics[page=25,scale=0.9]{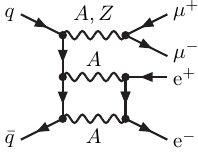}}
\caption{Illustration of the splitting of EW corrections into purely
  weak (first diagram) and photonic (2nd-4th diagrams) contributions
  for a sample process that proceeds via neutral-current interactions at tree level.}
\label{fig:weakPhotonicSplitting}
\end{figure}

As an example we consider the production of two different
lepton--antilepton pairs in proton--proton scattering
\cite{Biedermann:2016lvg}, assuming massless leptons and massless
incoming quarks. Out of the eight diagrams shown in
\reffi{fig:weakPhotonicSplitting} only the first pair with exclusively
Z~bosons in the loop belongs to the weak corrections while the three
other pairs with one or two photons in the loop are part of the
electromagnetic corrections.
Note that the criterion for the splitting considers only the vector
bosons in the loop, while it does not refer to the tree-level part of
the diagram.  The contributions to the field renormalization constants
of the fermions are decomposed in an analogous manner.  Note that the
photon-exchange contributions to the renormalization constants for the
W-boson mass and the weak mixing angle count as weak corrections.
Since only loops with internal photons 
coupling to fermions lead to soft and collinear
divergences, the purely weak contribution is IR finite. The
IR singularities of the virtual photonic corrections are
cancelled by the real photonic corrections, which are purely photonic
corrections.

\subsection{Goldstone-boson equivalence theorem}
\label{se:gbet}

An important consequence of the Slavnov--Taylor identities in
spontaneously broken gauge theories is the {\it Gold\-stone-boson
  equivalence theorem}
\cite{Cornwall:1974km,Chanowitz:1985hj,Gounaris:1986cr}. It states
that the amplitudes for reactions involving high-energetic,
longitudinal vector bosons are asymptotically proportional to the
amplitudes where those vector bosons are replaced by their associated
would-be Goldstone bosons.  The practical virtue of the
Goldstone-boson equivalence theorem is twofold. First of all, it
facilitates the calculation of cross sections for reactions with
longitudinal vector bosons at high energies, as the amplitudes for
external scalars are much easier to evaluate.  On the other hand, it
might allow us to derive information on the mechanism of spontaneous
symmetry breaking from the experimental study of longitudinal vector
bosons.

The Goldstone-boson equivalence theorem can be derived starting from
the Slavnov--Taylor identities \refeqf{eq:sti_c+phys} and
\refeqf{eq:sti_nc+phys} (see for instance
\citere{He:1993yd,Bohm:2001yx}). For the case of one external
\PW~boson the corresponding equation \refeqf{eq:sti_c+phys} reads
\beq\label{eq:WIWphiphys}
0 = \Bigl\langle T (\partial^{\mu }W^{\mp}_{\mu} 
\pm \ri \MW \xi'_\FW \phi^{\mp}) 
\prod_l \Psi_{I_l}^{\text{on{-}shell,phys.}}\Bigr\rangle.
\eeq
This identity holds for unrenormalized quantities, but also for
renormalized quantities in the renormalization scheme, where the
linear gauge-fixing functions are not renormalized.

In order to obtain relations for $S$-matrix elements, one truncates
the external legs of the considered longitudinal vector bosons and
transforms to momentum space.  In \citere{Yao:1988aj} it was realized
that the truncation leads to non-trivial proportionality factors
originating from higher-order corrections.  The resulting relation for a
\PW~boson can be written as~\cite{Bohm:2001yx}
\begin{equation}\label{eq:WIET1}
k^\nu \Bigl\langle T \underline{W}^{\pm}_{\nu } 
\prod_l \Psi_{I_l}^{\text{on{-}shell,phys.}}\Bigr\rangle
=  \pm \MW A^\FW(k^2) \Bigl\langle T \underline{\phi}^{\pm}
\prod_l \Psi_{I_l}^{\text{on{-}shell,phys.}}\Bigr\rangle,
\end{equation}
where $\langle T\underline{\FW}^\pm(k)\ldots\rangle$ and $\langle
T\underline{\phi}^\pm(k)\ldots\rangle$ denote Green functions with the
$\FW$ and $\phi$ legs truncated, respectively.  We recall 
our conventions from \refapp{se:conventions} that the
fields in \refeq{eq:WIWphiphys} are outgoing, but the underlined field
labels for the truncated lines in \refeq{eq:WIET1} correspond to
incoming fields; the momentum $k$ is incoming.  Analogous equations
can be derived for the \PZ-boson sector.

At LO the correction factors $A^V(k^2)$ are equal to one; at higher orders
they depend on the gauge and the renormalization of the unphysical sector
\cite{He:1993yd}.  To derive explicit expressions for $A^V(k^2)$ (see,
e.g., \citere{Bohm:2001yx}), the matrix-valued propagator for the external
vector--scalar system of $V$ and its corresponding Goldstone field~$S$ 
is truncated from the full (reducible, connected) Green functions; for the
\PZ-boson this system includes the mixing to the photon as well.
Subsequently $A^V(k^2)$ can be directly read off in terms of propagator
functions~$G^{ab}$ of the $VS$~system and expressed in terms of 2-point
vertex functions~$\Gamma^{ab}$, which result from the inverse of the matrix
$(G^{ab})$.  Using the Slavnov--Taylor identities~\refeqf{eq:WIVV} for the
propagators~$G^{ab}$, 
helps to bring the factors $A^V(k^2)$ into a simple final form.  In
terms of unrenormalized $\Gamma^{ab}$ in the 't~Hooft gauge, the
$A^V(k^2)$ read
\begin{align}\label{eq:AWAZ_unren}
A^\FW(k^2) ={}& 
\frac{k^2/\xi_\FW + \Gamma^{\FW\FW}_{\rL}(k^2)}{\MW\left(\MW +\Gamma^{\FW\phi}(k^2)\right)}
= r_{\FH}
\frac{k^2/\xi_\FW + \Gamma^{\FW\FW}_{\mathrm{1PI},\rL}(k^2)}{\MW\left(\MW +\Gamma^{\FW\phi}_{\mathrm{1PI}}(k^2)\right)}
,\nl
A^\FZ(k^2) ={}&
\frac{k^2/\xi_\FZ + \Gamma^{\FZ\FZ}_{\rL}(k^2)}{\MZ\left(\MZ - \ri\Gamma^{\FZ\chi}(k^2)\right)}
= r_{\FH}
\frac{k^2/\xi_\FZ + \Gamma^{\FZ\FZ}_{\mathrm{1PI},\rL}(k^2)}{\MZ\left(\MZ - \ri\Gamma^{\FZ\chi}_{\mathrm{1PI}}(k^2)\right)},
\end{align}
where the factor $r_{\FH}$ depends on the tadpole scheme (see
\refse{se:tadpoles}):
\begin{align}
r_{\FH}^{\mathrm{PRTS}} = 1, \qquad 
r_{\FH}^{\mathrm{FJTS}} = 1+\frac{T^{\FH}}{\MH^2\varv}.
\end{align}
This difference originates from the fact that the unrenormalized
vertex functions are derived from the bare Lagrangian and do include
besides the explicit loop contributions also contributions from
tadpole counterterms, as already pointed out in
\refeq{eq:definition_se} for the 2-point vertex functions and the
corresponding definition of the self-energies $\Sigma$.
The Lorentz decompositions of the vector--vector 2-point functions 
are given in
\refeq{eq:Lorentz_SE}, those for the vector--scalar 2-point functions
read
\begin{align}\label{eq:GWphi}
\Gamma^{\FW^\pm\phi^\mp}_{\mu}(k,-k) ={} \pm k_\mu \Gamma^{\FW\phi}(k^2),
\qquad
\Gamma^{\FZ\chi}_{\mu}(k,-k) ={} k_\mu \Gamma^{\FZ\chi}(k^2).
\end{align}
Note that the first equality in \refeq{eq:AWAZ_unren} for each
$A^V(k^2)$ holds to all orders, but the second equality 
is restricted to the one-loop approximation with respect to the 
tadpole contribution $T^{\FH}$. 
%
Splitting the vertex functions into LO contributions and
self-energies $\Sigma$ according to
\beq
\Gamma^{VV}_{\rL}(k^2)=M_V^2-\frac{k^2}{\xi_V}-\Sigma^{VV}_{\rL},\qquad
\Gamma^{\FW\phi}(k^2) = \Sigma^{\FW\phi}(k^2),\qquad 
\Gamma^{\FZ\chi}(k^2) = \Sigma^{\FZ\chi}(k^2),
\eeq
the unrenormalized correction factors at the one-loop level read
\begin{align}\label{eq:AWAZ_oneloop}
A^\FW(k^2)  ={}&  
1 -\frac{\Sigma^{\FW\FW}_{\rL}(k^2)}{\MW^2} 
                 - \frac{\Sigma^{\FW\phi}(k^2)}{\MW}
= r_{\FH} -\frac{\Sigma^{\FW\FW}_{\mathrm{1PI},\rL}(k^2)}{\MW^2} 
                 - \frac{\Sigma^{\FW\phi}_{\mathrm{1PI}}(k^2)}{\MW},
\nl
A^\FZ(k^2)  ={}&  
1 -\frac{\Sigma^{\FZ\FZ}_{\rL}(k^2)}{\MZ^2} 
+\ri\frac{\Sigma^{\FZ\chi}(k^2)}{\MZ}
= r_{\FH} -\frac{\Sigma^{\FZ\FZ}_{\mathrm{1PI},\rL}(k^2)}{\MZ^2} 
+\ri\frac{\Sigma^{\FZ\chi}_{\mathrm{1PI}}(k^2)}{\MZ}.
\end{align}

In the renormalization scheme where the gauge-fixing functions are not
renormalized, the renormalized correction factors $A_\ren^V(k^2)$ are given
by
\begin{align}\label{eq:AWAZ}
A_\ren^\FW(k^2) ={}& 
\frac{k^2/\xi_\FW + \Gamma^{\FW\FW}_{\ren,\rL}(k^2)}
{\MW\left(\MW +\Gamma^{\FW\phi}_{\ren}(k^2)\right)}
,\qquad
A_\ren^\FZ(k^2) ={}
\frac{k^2/\xi_\FZ + \Gamma^{\FZ\FZ}_{\ren,\rL}(k^2)}{\MZ\left(\MZ - \ri\Gamma^{\FZ\chi}_\ren(k^2)\right)}.
\end{align}
Inserting the decomposition of the renormalized vertex functions
into one-loop self-energies and counterterms, we get
\begin{align}\label{eq:AWAZ2}
A_\ren^\FW(k^2) ={}& 
A^\FW(k^2) + \frac{1}{2}\DZW - \frac{1}{2}\de Z_\phi  + \frac{\DMWS}{2\MW^2}, \qquad
A_\ren^\FZ(k^2) ={}
A^\FZ(k^2) + \frac{1}{2}\DZZ  - \frac{1}{2}\de Z_\chi + \frac{\DMZS}{2\MZ^2}
\end{align}
in one-loop approximation, where $\de Z_\phi$ and $\de Z_\chi$ are
field renormalization constants for the would-be Goldstone fields that
render Green functions involving these fields as well as the factors
$A_\ren^V(k^2)$ finite.
Note that the tadpole counterterms are already contained in the
unrenormalized functions $A^V(k^2)$.  The renormalized correction
factors are independent of the tadpole scheme, since all tadpole
contributions cancel in the sum of the unrenormalized factors
$A^V(k^2)$ and the contributions of the mass counterterms $\de M_V^2$.

Upon expressing the external momentum of the vector boson that appears
explicitly in the identity \refeq{eq:WIET1} by the corresponding
longitudinal polarization vector $\veps_{\rL}^\mu(k)$ using
\begin{equation}\label{eq:polvecdecomp}
\veps_{\rL}^\mu(k) = \frac{k^\mu}{\MW} + \cO\left(\frac{\MW}{k^0}\right),
\end{equation}       
for the case of a single external longitudinal \PW~boson,
one arrives at
\begin{align}\label{eq:EToneW}
\langle\ldots|S|W_\rL^\pm(k)\ldots\rangle ={}& 
{\pm} A^\FW_\ren(\MW^2)\,
\langle T\underline{\phi}^\pm(k)\ldots\rangle
+ \cO\left(\frac{\MW}{k^0}\right),
\end{align}
and similarly for a single  external longitudinal \PZ~boson
\begin{align}\label{eq:EToneZ}
\langle\ldots|S|Z_\rL(k)\ldots\rangle ={}&
\ri  A^\FZ_\ren(\MZ^2)\,
\langle T\underline{\chi}(k)\ldots\rangle
+ \cO\left(\frac{\MZ}{k^0}\right),
\end{align}
where the left-hand sides represent $S$-matrix elements. 
All quantities are renormalized in the complete OS scheme as defined
in \refse{se:rcsm}, so that no wave-function renormalization
corrections for the external vector bosons appear, because all
corrections to the propagator residues and all mixing effects of OS
particles are absorbed by the field renormalization.  The
$\cO(\MV/k^0)$ terms in \refeqs{eq:EToneW} and \refeqf{eq:EToneZ}
describe mass-suppressed terms relative to the leading canonical mass
dimension of the inspected Green function, i.e.\ if the explicit terms
on the r.h.s.\ of \refeqs{eq:EToneW} and \refeqf{eq:EToneZ} are
already mass suppressed, the r.h.s.\ does not predict the leading term
explicitly.

The relations \refeqf{eq:EToneW} and \refeqf{eq:EToneZ} can be
generalized to $S$-matrix elements with more longitudinal vector
bosons, \eg for $n$ charged \PW bosons:
\begin{align}\label{eq:ETW2}
\Bigl\langle 
\ldots\big|\,S\,\big|
\Bigl(\prod_{n}W_\rL^\pm(k_{n})\Bigr)\ldots\Bigr\rangle
=
\Bigl(\prod_n \Bigl[\pm A^\FW_\ren(\MW^2)\Bigr] \Bigr) \,
\Bigl\langle T \Bigl(\prod_n
\underline{\phi}^{\pm}(k_n) \Bigr)
\ldots\Bigr\rangle
+ \cO\left(\frac{\MW}{k^0_n}\right).
\end{align}
The inclusion of longitudinal $\PZ$ bosons is obtained via an obvious
generalization. The relations \refeqf{eq:ETW2} and their generalization
constitute the  Goldstone-boson equivalence theorem.
It relates $S$-matrix elements involving longitudinal gauge bosons to
matrix elements involving the corresponding would-be Goldstone-boson
fields in the high-energy limit.

Finally, we briefly consider the derivation of the 
Goldstone-boson equivalence theorem within the BFM, which was
discussed in detail in \citere{Denner:1996gb}.
The major result of that paper
is that the renormalized correction factors reduce to one,
\begin{align}\label{eq:AVrenBFM}
A_\ren^{\FVhat}(k^2)\big|_{\mathrm{BFM}} = 1,
\quad \FVhat=\FWhat,\FZhat,
\end{align}
in all orders of perturbation theory if the BFM renormalization
scheme~\cite{Denner:1994xt} is employed that preserves all BFM
Ward identities for renormalized vertex functions,
as outlined in \refse{se:ren_bfm}.
In \citere{Denner:1996gb}, this fact was proven within the
PRTS, but the all-order result \refeqf{eq:AVrenBFM} does not depend on
the tadpole treatment.
It should, however, be realized that the final form 
of \refeqs{eq:EToneW}--\refeqf{eq:ETW2} for $S$-matrix elements
receives another UV-finite correction originating from the
fact that the employed BFM renormalization scheme does not
normalize the residues of the $W$ and $Z$ propagators to one.
In detail, each factor $A_\ren^{\FVhat}(k^2)$ has to be replaced by
the wave-function correction factor $R_{\FVhat}^{1/2}$, where
$R_{\FVhat}$ in one-loop approximation is given by
\begin{align}
R_{\FVhat} = 1-\left.\rRe\frac{\partial\Sigma_{\ren,\rT}^{\hat V\hat V}(k^2)}{\partial k^2}
\right\vert_{k^2=M_V^2}.
\end{align}
The generalization of this factor beyond one loop is described
in \citere{Denner:1996gb} as well.

\subsection{Electroweak corrections at high energies}
\label{se:ewc@he}

At energies that are large compared to the masses of the EW gauge
bosons, the EW corrections can become large due to the appearance of
large logarithms that result from the virtual exchange of soft and/or
collinear massive weak gauge bosons
\cite{Kuroda:1990wn,Degrassi:1992ue,Beenakker:1993tt,Denner:1995jv,
  Denner:1996ug,Beccaria:1998qe,Fadin:1999bq,Kuhn:1999nn,Ciafaloni:1999ub}.
The leading contributions from the soft--collinear limit are known as
{\em Sudakov logarithms} \cite{Sudakov:1954sw}.  At NLO, the leading
and subleading terms are of the form $(\alpha/\sw^2)\ln^2(Q^2/\MW^2)$
and $(\alpha/\sw^2)\ln(Q^2/\MW^2)$, respectively, where $Q$ denotes
the energy scale of the hard scattering process, which is typically
determined by the centre-of-mass energy $\sqrt{s}$ of the partonic
process. In QED and QCD the corresponding double logarithms resulting
from virtual photons and gluons are cancelled against the related
real-emission corrections, and the remaining single logarithms can be
absorbed into parton distribution and fragmentation functions.
However, since the masses of the W and Z~bosons provide a physical
cutoff and since the EW charges are not confined, the radiated
real W or Z~bosons can be experimentally reconstructed to a large
extent. Consequently, real massive vector-boson radiation needs not be
included in the definition of cross sections, and typically only a
small fraction that remains unresolved compensates for part of the
virtual corrections \cite{Baur:2006sn}.  Thus, at high scales
$Q\gg\MW$, which are accessible at the LHC and future colliders, the
Sudakov logarithms can produce large corrections. Since the real
corrections that would cancel the enhanced virtual corrections in
fully inclusive quantities result from separate process classes (with
their own squared matrix elements $|{\cal M}|^2$) and are thus
positive, the latter are negative.  In the TeV range these negative
corrections typically amount to tens of percent rendering the EW
corrections very significant.\footnote{The fact that EW corrections at
  energies above the EW scale are dominated by large single and double
  logarithms has been observed in the 1990s in various calculations.
  See, for instance,
  \citeres{Kuroda:1990wn,Degrassi:1992ue,Beenakker:1993tt,Denner:1995jv,%
   Denner:1996ug,Beccaria:1998qe}.}

The {\it Sudakov regime} is characterized by the situation that all
invariants $s_{ij}=(p_i+p_j)^2$ formed from pairs of external
particles' four-momenta $p_{i}$ (taken all incoming) become large
($|s_{ij}|\gg\MW^2$).  In this limit, the large logarithms are
associated with soft and/or collinear singularities arising if the
masses are small compared to the relevant energies.  Since these
singularities are associated with external particles of the scattering
processes and thus universal \cite{Kinoshita:1962ur}, this allows us 
to derive general process-independent results for the EW high-energy
logarithms.  We note that this applies to processes that are not mass
suppressed, \ie the corresponding matrix elements scale with the
scattering energy according to their mass dimension.  For
mass-suppressed processes, like Higgs production in vector-boson
fusion or Higgs-strahlung off a vector boson, the EW logarithms may
have a different structure.

The structure of EW corrections in the Sudakov regime has been investigated
in detail at ${\cal O}(\alpha)$ and beyond by several groups
(see e.g.\
\citeres{Fadin:1999bq,Kuhn:1999nn,Ciafaloni:1999ub,Ciafaloni:2000df,%
Hori:2000tm, Denner:2000jv,Denner:2001gw,Melles:2001ye,Melles:2001dh,%
Beenakker:2001kf, Denner:2003wi, Jantzen:2005xi,Jantzen:2005az,%
Denner:2006jr} and references therein).  As described for example in
\citeres{Denner:2000jv,Denner:2001gw,Denner:2003wi,Denner:2006jr}, the
leading EW logarithmic corrections, which are enhanced by large
factors $L=\ln(s_{ij}/\MW^2)$, can be divided into an
$\SU(2)_\rw\times\U(1)_Y$-symmetric part, an electromagnetic part
related to the running of EW couplings below $\MW$, and a subleading
part induced by the mass difference between $\PW$ and \PZ~bosons. The
leading (Sudakov) logarithms $\propto(\alpha L^2)^n$ of
electromagnetic origin cancel between virtual and real (soft) photonic
bremsstrahlung corrections, so that the only source of leading
logarithms is the symmetric EW part, which can be characterized by
comprising \PW~bosons, \PZ~bosons, and photons of a common mass $\MW$.
These leading EW Sudakov corrections can be obtained to all orders
from the respective NLO result via exponentiation \cite{Fadin:1999bq}.
For the subleading EW high-energy logarithms, corresponding
resummations are not rigorously proven, but the corrections are
expected to obey {\em IR evolution equations}
\cite{Kuhn:1999nn,Kuhn:2001hz}, a statement that is backed by explicit
two-loop calculations
\cite{Denner:2003wi,Denner:2006jr,Denner:2008yn}.

A general result for the large logarithmic contributions at one loop
was derived in \citeres{Denner:2000jv,Denner:2001gw}, and a
general method for the resummation of these logarithms was 
developed in \citeres{Chiu:2007yn,Chiu:2008vv} based on {\it
  soft--collinear effective
  theory}~\cite{Bauer:2000ew,Bauer:2000yr,Bauer:2001ct,Bauer:2001yt}.
Owing to these results, the resummed EW corrections to all hard
scattering processes that are not mass suppressed and do not involve
kinematic scales of the order of $\MW$, such as processes with intermediate
resonances, are known explicitly at next-to-leading logarithmic (NLL) order
\cite{Chiu:2009mg,Chiu:2009ft,Fuhrer:2010eu} and can be incorporated
into LHC cross-section calculations.

We note that the EW logarithmic corrections are independent of the
quark-mixing matrix, because in the high-energy limit, the masses of
the light quarks are irrelevant, so that the quark-mixing matrix can
be replaced by the unit matrix. Consequently, quark mixing only enters
the underlying LO matrix elements, while the relative correction
factors to these do not depend on it.

\subsubsection{General results for one-loop electroweak logarithmic corrections}
\label{se:ewlogs@1loop}

At the one-loop level, the single and double EW logarithms 
result from soft and/or collinear singularities and, thus, have a
universal form. General results for processes that are not mass
suppressed were presented in \citere{Denner:2000jv}, and the
factorization of these logarithms in the EWSM was proven in
\citere{Denner:2001gw}.  These results were derived in the
't~Hooft--Feynman gauge in the broken phase of the EW theory, \ie with
fields in the physical basis. Upon choosing the mass scale $\mu$ of
DR of the order of the energy of the scattering process, only
mass-singular logarithms of the form $\ln(\mu^2/M^2)$ or $\ln(s/M^2)$
are large, where $M$ is of the order of $\MW$.  Field renormalization
is fixed in such a way that no separate wave-function renormalization
is required (see \refse{se:rcsm}), and for parameter renormalization
the OS scheme is adopted. The fact that the polarization vector of
longitudinal gauge bosons involves the gauge-boson mass in the
denominator complicates the extraction of the logarithmic corrections
in the massless limit. This problem is circumvented by using the
Goldstone-boson equivalence theorem
\cite{Cornwall:1974km,Chanowitz:1985hj,Gounaris:1986cr} taking into
account correction factors from higher-order
contributions~\cite{Yao:1988aj}, as discussed in \refse{se:gbet}.  All
external particles are considered as on shell and incoming. Outgoing
particles are obtained via crossing symmetry.

In this setup, EW logarithms are due to  the following contributions:
\begin{myitemize}
\item Double-logarithmic contributions result from loop diagrams where
  soft--collinear gauge bosons are exchanged between pairs of
  external legs (see \reffi{fig:Ldiags} left). These contributions can
  be obtained within the {\em eikonal approximation}, \ie in the
  approximation where the momenta of the soft--collinear gauge bosons
  can be neglected everywhere but in the singular propagators. In
  fact, the eikonal approximation includes already all contributions
  associated with soft singularities, \ie singularities related to the
  vanishing of a gauge-boson loop momentum. The soft contributions
  associated with single external lines are treated separately
  together with the collinear contributions in the field
  renormalization constants.
\begin{figure}
\centerline{\includegraphics[page=12]{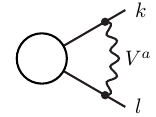}\qquad\qquad
\includegraphics[page=13]{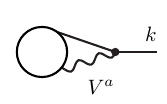}}
\caption{The types of Feynman diagrams leading to double-logarithmic
  corrections (left) and single-logarithmic corrections (right).}
\label{fig:Ldiags}
\end{figure}

\item Single-logarithmic contributions result from the emission of
  virtual collinear gauge bosons from external lines
  \cite{Kinoshita:1962ur} (see \reffi{fig:Ldiags} right).  These
  contributions are extracted in the collinear limit by means of Ward
  identities and are found to factorize into the LO amplitude times
  collinear factors \cite{Denner:2001gw}.

\item
Additional single-logarithmic contributions originating from soft or
collinear regions are contained in the field renormalization
constants.

\item Single-logarithmic contributions of UV origin show up in the
  parameter renormalization constants, \ie the renormalization of the
  electromagnetic coupling, the weak mixing angle, the Yukawa
  couplings, and the scalar self-coupling. The corresponding
  logarithms are controlled by the renormalization-group equations and
  associated to the running of the couplings.
\end{myitemize}

In the following, we summarize the results of \citere{Denner:2000jv}
on the structure of EW high-energy logarithms at NLO. We concentrate
on logarithms that involve the ratio of the high-energy scale
$\sqrt{s}$ and a generic EW mass for which we choose $\MW$, since
these logarithms give rise to the most important corrections at high
energies for many processes. These EW logarithms could directly be
derived in the symmetric basis with a subsequent transformation to the
physical basis. This is, however, not the case for the electromagnetic
logarithms and terms involving $\ln(\MW/\MZ)$, which can be found in
\citere{Denner:2000jv} as well, but are not reproduced here.
While the former cancel to a large extent against similar logarithms
in the real electromagnetic corrections, the latter are of the order
of the finite terms. We do not consider enhanced logarithms from real
corrections in this section either.

The LO matrix element for the generic process
\beq \label{process}
\varphi_{i_1}(p_1)\dots \varphi_{i_n}(p_n)\rightarrow 0
\eeq
is denoted as $\cM_0^{i_1 \ldots i_n}(p_1,\ldots, p_n)$.  The
particles (or antiparticles) $\varphi_{i_k}$ correspond to the
components of the various multiplets $\varphi$ present in the SM.
Chiral fermions and antifermions are represented by $\Ff^\kappa$ and
$\Ffbar^\kappa$, respectively, with the chirality $\kappa=\rR,\rL$.
The gauge bosons are denoted by $\FV^a=\FA,\FZ,\FWpm$, and can be
transversely (T) or longitudinally (L) polarized.  The components
$\Phi_i$ of the scalar doublet correspond to the physical Higgs
particle $\FH$ and the would-be Goldstone bosons $\chi,\phi^\pm$,
which are used to describe the longitudinally polarized massive gauge
bosons $\FZ_\rL$ and $\FW^\pm_\rL$ with help of the Goldstone-boson
equivalence theorem (see \refse{se:gbet}).  In logarithmic
approximation, the correction $\delta \cM$ to a matrix element takes
the form
\beq \label{LAfactorization} 
\delta \cM^{i_1 \ldots i_n}(p_1,\ldots, p_n)=
\cM_0^{i'_1 \ldots i'_n}(p_1, \ldots, p_n)\,\delta_{i'_1i_1 \ldots i'_ni_n},
\end{equation}
\ie it can be written as a matrix product of the LO matrix element and
a relative correction $\de$. The latter is split into various
contributions according to their origin:
\beq
\delta=\de^{\LSC}+\de^{\SSC}+\de^{\LC}+\de^\PR.
\eeq
The leading and subleading soft--collinear logarithms are denoted by
$\de^\LSC$ and $\de^\SSC$, respectively, the collinear logarithms by
$\de^\LC$, and the logarithms resulting from parameter renormalization
by~$\de^\PR$. Note that in \refeq{LAfactorization} and in the
following equations, sums over $i'_1 \ldots i'_n$ are understood.

Considering the diagrams of the form shown in the left of
\reffi{fig:Ldiags}, using the eikonal approximation and neglecting all
terms proportional to masses, the soft--collinear one-loop corrections
read
\begin{align} \label{eikonalappA}
\delta^{\mathrm{eik}} \cM^{i_1 \ldots i_n}
={}&\sum_{\substack{k,l=1\\l< k}}^n \sum_{\FV^a=\FA,\FZ,\FW^\pm} 
(-4\ri e^2) (p_kp_l) I^{\FV^a}_{i'_k i_k}(k) I^{\bar{\FV}^a}_{i'_l
  i_l}(l) 
\cM_0^{i_1 \ldots i'_k \ldots i'_l \ldots i_n} \notag\nl
&\qquad \times\int\frac{\rd^4 q}{(2\pi)^4} 
\frac{1}{(q^2-M_{\FV^a}^2)[(q+p_k)^2-m_{k'}^2][(q-p_l)^2-m_{l'}^2]}
\notag\nl
={}&\sum_{\substack{k,l=1\\l< k}}^n
  \sum_{\FV^a=\FA,\FZ,\FW^\pm} \frac{e^2}{4\pi^2}
  (p_kp_l) I^{\FV^a}_{i'_k i_k}(k) I^{\bar{\FV}^a}_{i'_l i_l}(l) \cM_0^{i_1 \ldots i'_k
    \ldots i'_l \ldots i_n}
C_0(p_k,-p_l,M_{V^a}^2,m_{k'}^2,m_{l'}^2) \notag\nl
\sim{}&\frac{1}{2}\sum_{\substack{k,l=1\\l\ne k}}^n\sum_{\FV^a=\FA,\FZ,\FW^\pm}
I^{\FV^a}_{i'_k i_k}(k) I^{\bar{\FV}^a}_{i'_l i_l}(l) \cM_0^{i_1 \ldots
  i'_k \ldots i'_l \ldots i_n}
\frac{\al}{4\pi}\left[
\ln^2\frac{s_{kl}}{M_{\FV^a}^2}
-\de_{\FV^a\FA}\ln^2\frac{m_k^2}{m_\ga^2}\right].
\end{align}
While in the last line all potentially large logarithms in the Sudakov
regime are kept, in the following we only retain terms involving
logarithms of ratios of large energy scales $s_{kl}=(p_k+p_l)^2$ and
the W-boson mass squared.
The couplings of the external fields $\varphi_i$ to the gauge bosons
$\FV^a$ in the eikonal approximation are denoted by $\ri e
I_{ii'}^{\FV^a}(\varphi)$. To be precise, these couplings correspond
to the $\FV^a\bar\varphi_i\varphi_{i'}$ vertex, where all fields are
considered incoming, and $\bar\varphi$ denotes the charge conjugate of
$\varphi$.  The couplings of antiparticles are related to the couplings
of the particles as
\beq
\ri e I^{\FVbar^a}_{ii'}(\bar\varphi) = \left(\ri e I^{\FV^a}_{ii'}(\varphi)\right)^*.
\eeq
Furthermore, $C_0$ is the scalar 3-point function as defined in
\refse{se:ti_reduction}. In the last step in \refeq{eikonalappA} the
high-energy expansion of the scalar integral \cite{Roth:1996pd} is
inserted with the (infinitesimal) photon mass $m_\ga$ as regulator.

The single collinear logarithms result from the type of diagrams shown
on the right of \reffi{fig:Ldiags} and from the field renormalization
constants, which are calculated separately.  In order to avoid double
counting, the contributions of the field renormalization constants and
those in the soft--collinear corrections have to be subtracted as
illustrated here:
\begin{align}
\sum_{\FV^a=\FA,\FZ,\FW^\pm} & \left\{
\vcenter{\hbox{\includegraphics[page=13]{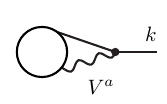}}}
- \vcenter{\hbox{\includegraphics[page=14]{diagrams.pdf}}} \right.
\notag\\ & \left.
- \sum_{l\neq k}\left[
\vcenter{\hbox{\includegraphics[page=12]{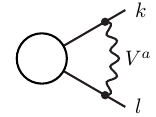}}}
\right]_{\mathrm{eik.~appr.}}\right\}_{\mathrm{coll.}}
=\de^\coll(k)
\vcenter{\hbox{\includegraphics[page=15]{diagrams.pdf}}}.
\end{align}
Using Ward identities resulting from the BRS invariance of the
complete EWSM, the factorization of the collinear logarithms is
proven, and the collinear factors $\de^\coll(k)$ are determined for
the various fields of the EWSM in \citere{Denner:2001gw}.

\myparagraph{(a) Leading soft--collinear contributions}
The leading soft--collinear contributions are the double-logarithmic
contributions, arising from \refeq{eikonalappA}
with the different kinematical invariants replaced by one high-energy
scale $s$.  Upon using $\SU(2)_\rw\times\U(1)_Y$ invariance, 
\beq\label{eq:gi}
0=\ri e \sum_{k=1}^n I^{\FV^a}_{i'_k i_k}(k) \cM^{i_1 \ldots i'_k \ldots i_n},
\eeq
the leading soft--collinear
contributions can be cast into a single sum over external legs,
\beq \label{SCsum}
\de^{\LSC} \cM^{i_1 \ldots i_n} =\sum_{k=1}^n \delta^\LSC_{i'_k i_k}(k)
\cM_0^{i_1 \ldots i'_k\ldots i_n},
\end{equation}
with the correction factors
\beq \label{deSC} 
\de^\LSC_{i'_k i_k}(k)=- \frac{1}{2} C^{\ew}_{i'_k i_k}(k)L(s) ,
\eeq
involving the double logarithm
\beq
L(s)=\frac{\al}{4\pi}\ln^2\frac{s}{\MW^2}.
\label{eq:Ls}
\eeq
The EW Casimir operator $ C^{\ew}$ is defined from the
eikonal couplings $I^{\FV^a}(\varphi)_{ii'}$ of the external fields
$\varphi_i$ to the gauge bosons $V^a$  as
\begin{equation}\label{CasimirEW} 
\cew=\sum_{\FV^a=A,Z,W^\pm} I^{\FV^a}I^{\bar{\FV}^a}
=\frac{1}{\cw^2}\left(\frac{Y_\rw}{2}\right)^2+\frac{1}{\sw^2}I_\rw(I_\rw+1)
\end{equation}
with the weak hypercharge $Y_\rw$ and the weak isospin $I_\rw$ defined in
\refse{se:symmetric_lagrangian}.  For most of the fields, $\cew$ is
diagonal and given by:
\beq
\begin{array}{lcccccccc}
\varphi & \FW^\pm, \FW^3 & \FB & \Phi,\FH,\chi,\phi^\pm &
\Fnu^\rL,\Fnubar^\rL, \Fl^\rL,\Flbar^\rL &  \Fl^\rR,\Flbar^\rR &
\Fu^\rL,\Fubar^\rL, \Fd^\rL,\Fdbar^\rL &  \Fu^\rR,\Fubar^\rR &
\Fd^\rR,\Fdbar^\rR \\\midrule
C^{\ew} & \frac{2}{\sw^2} & 0 &  \frac{1+2\cw^2}{4\sw^2\cw^2} &
\frac{1+2\cw^2}{4\sw^2\cw^2} &  \frac{1}{\cw^2} &
\frac{1+26\cw^2}{36\sw^2\cw^2} &  \frac{4}{9\cw^2} & \frac{1}{9\cw^2} 
\end{array}
\eeq
For the neutral gauge bosons, $\cew$ becomes non-diagonal owing to the
mixing between photon and Z~boson, and the Casimir matrix entries read
\beq
C^{\ew}_{\FA\FA} = 2, \qquad 
C^{\ew}_{\FA\FZ} = C^{\ew}_{\FZ\FA} = -2\frac{\cw}{\sw}, \qquad
C^{\ew}_{\FZ\FZ} = 2 \frac{\cw^2}{\sw^2}.
\eeq

\myparagraph{(b) Subleading soft--collinear contributions} The
subleading soft--collinear contributions result from
\refeq{eikonalappA} after subtracting the leading double-logarithmic
contributions. Upon using
\beq
\ln^2\frac{s_{ij}}{M^2} = \ln^2\frac{s}{M^2}
+2  \ln\frac{s}{M^2} \ln\frac{s_{ij}}{s} + \ln^2\frac{s_{ij}}{s},
\eeq
the angular-dependent part of the form $\ln\frac{s}{M^2}
\ln\frac{s_{ij}}{s}$ furnishes the subleading soft--collinear
contributions, which can be written as 
\beq \label{SScorr}
\de^\SSC \cM^{i_1 \ldots i_n} =
\sum_{\substack{k,l=1\\l< k}}^n\delta^{\SSC}_{i'_k i_k i'_l i_l}(k,l)
\cM_0^{i_1\ldots i'_k\ldots i'_l\ldots i_n},
\eeq
with
\beq \label{subdl1} 
\delta^{\SSC}_{i'_k i_k i'_l i_l}(k,l)=
2l(s)\ln\frac{s_{k l}}{s} \sum_{\FV^a=\FA,\FZ,\FW^\pm} I_{i'_k i_k}^{\FV^a}(k)I_{i'_l i_l}^{\FVbar^a}(l)
\eeq
and the single logarithm
\beq
l(s)=\frac{\al}{4\pi}\ln\frac{s}{\MW^2}.
\label{eq:ls}
\eeq
In the physical basis the couplings are given by
\beq
I^A=-Q,\qquad I^Z=\frac{I_{\rw}^{3}-\sw^2 Q}{\sw\cw},\qquad
I^\pm=\frac{I_{\rw}^\pm}{\sw}=\frac{I_{\rw}^1\pm\ri I_{\rw}^2}{\sqrt{2}\sw}
\eeq
with the generators of electric charge and weak isospin as defined in
\refse{se:symmetric_lagrangian}.  Specifically, for gauge bosons,
charged would-be Goldstone bosons, and fermions the diagonal couplings
to photons and $\PZ$~bosons are obtained to:
\beq\label{eq:neutral_eikonal_couplings}
\begin{array}{lcccccccccc}
\varphi & \FW^\pm & \FA,\FZ & \phi^\pm &
\Fnu^\rL,\Fnubar^\rL &  \Fl^\rL,\Flbar^\rL &  \Fl^\rR,\Flbar^\rR &
\Fu^\rL,\Fubar^\rL & \Fd^\rL,\Fdbar^\rL &  \Fu^\rR,\Fubar^\rR &
\Fd^\rR,\Fdbar^\rR \\\midrule
I^\FA(\varphi) & \mp1 & 0 &  \mp1 &  
0 &  \pm 1 & \pm1 & \mp\frac{2}{3} & \pm\frac{1}{3} &  
\mp\frac{2}{3} & \pm\frac{1}{3}
\\\midrule
I^\FZ(\varphi) & \pm\frac{\cw}{\sw} & 0 &  
 \pm\frac{1-2\sw^2}{2\cw\sw} & 
\pm\frac{1}{2\cw\sw} & \mp\frac{1-2\sw^2}{2\cw\sw} & \pm\frac{\sw}{\cw} &
\pm\frac{3-4\sw^2}{6\cw\sw} & \mp\frac{3-2\sw^2}{6\cw\sw} &
\mp\frac{2\sw}{3\cw} & \pm\frac{\sw}{3\cw} 
\end{array}
\eeq
The non-vanishing couplings of the gauge bosons to the charged gauge
bosons are given by
\beq
I^{\FW^\pm}_{\FA\FW^\mp} = -I^{\FW^\pm}_{\FW^\pm\FA} = \mp1,\qquad
I^{\FW^\pm}_{\FZ\FW^\mp} = -I^{\FW^\pm}_{\FW^\pm\FZ} = \pm\frac{\cw}{\sw},
\eeq
and the non-vanishing couplings of left-handed fermions to the $\FW^\pm$
bosons by
\begin{align}
  I^{\FW^+}_{\Fu\Fd}(\FQ^\rL) ={}& {-}
  I^{\FW^+}_{\Fdbar\Fubar}(\FQbar^\rL) = I^{\FW^-}_{\Fd\Fu}(\FQ^\rL) =
  - I^{\FW^-}_{\Fubar\Fdbar}(\FQbar^\rL) =
  \frac{1}{\sqrt{2}\sw},\notag\nl I^{\FW^+}_{\Fnu\Fl}(\FL^\rL) ={}& {-}
  I^{\FW^+}_{\Flbar\Fnubar}(\FLbar^\rL) = I^{\FW^-}_{\Fl\Fnu}(\FL^\rL) =
  - I^{\FW^-}_{\Fnubar\Flbar}(\FLbar^\rL) = \frac{1}{\sqrt{2}\sw},
\end{align}
while those to right-handed fermions vanish.  In the physical basis,
the couplings of the neutral scalar fields to the gauge bosons become
non-diagonal. All couplings of neutral scalars to the photon field
vanish, while the non-vanishing couplings to the Z-boson field read
\beq \label{ZHcoup}
I^\FZ_{\FH\chi}=-I^\FZ_{\chi \FH}=\frac{-\ri}{2\sw\cw}.
\eeq
The non-vanishing eikonal
couplings of the scalar fields to the charged W~bosons read
\beq\label{eq:IWSS}
I^{\FW^\pm}_{\FH\phi^\mp} = -I^{\FW^\pm}_{\phi^\pm\FH} =
\mp\frac{1}{2\sw},\qquad
I^{\FW^\pm}_{\chi\phi^\mp} = -I^{\FW^\pm}_{\phi^\pm\chi} =
-\frac{\ri}{2\sw}.
\eeq

\begin{sloppypar}
\myparagraph{(c) Collinear and soft single logarithms}
The complete single-logarithmic contributions
originating from soft or collinear regions
can be written as a sum over the external legs,
\beq\label{subllogfact}
\de^{\LC} \cM^{i_1 \ldots i_n} =\sum_{k=1}^n \delta^\LC_{i'_k i_k}(k)
\cM_0^{i_1 \ldots i'_k \ldots i_n}
\eeq
with
\beq\label{subllogfact2}
 \delta^\LC_{i'_k i_k}(k)=\left.\delta^\coll_{i'_k i_k}(k)+\frac{1}{2}\delta Z^{\varphi}_{i'_k i_k}\right|_{\mu^2=s},
\eeq
where $\delta Z^{\varphi}$ are the field renormalization constants in
the OS renormalization scheme [see \refeqs{eq:FR_bos} and
\refeqf{eq:CTF}] in logarithmic approximation.  The collinear factors
$\de^\coll(k)$ and the corrections $\de^\LC(k)$ depend on the quantum
numbers of the external fields $\varphi_{i_k}$.
\end{sloppypar}

The correction factor for the fermions is diagonal and reads
\beq \label{deccfer}
\de^{\LC}_{ff}(f^\kappa)=\left[\frac{3}{2} \cew_{f^\kappa} -\frac{1}{8\sw^2}\left((1+\delta_{\kappa \rR})\frac{m_{f}^2}{\MW^2}+\delta_{\kappa \rL}\frac{m_{\tilde{f}}^2}{\MW^2}\right)\right]l(s),
\eeq
where $\tilde{f}$ denotes the isospin partner of the fermion $\Ff$,
$\kappa=\rL,\rR$, and $\delta_{\kappa \rL}$, $\delta_{\kappa \rR}$ are
Kronecker deltas. The fermion-mass-dependent Yukawa contributions are
only relevant for top and bottom quarks.

While the correction factor for the charged transverse gauge bosons is
diagonal
\beq \label{deccWT}
\delta^\LC_{W^\pm W^\pm}(\FV_{\rT})=\frac{1}{2}\bew_{\FW}l(s),
\eeq
those for the neutral transverse gauge bosons involve non-diagonal
terms,
\beq \label{deccVVT} 
\de^\LC_{\FA\FA}(\FV_{\rT})= \frac{1}{2}\bew_{\FA\FA}l(s), \qquad
\de^\LC_{\FZ\FZ}(\FV_{\rT})= \frac{1}{2}\bew_{\FZ\FZ}l(s), \qquad
\de^\LC_{\FA\FZ}(\FV_{\rT})= \bew_{\FA\FZ}l(s), \qquad
\de^\LC_{\FZ\FA}(\FV_{\rT})= 0. 
\eeq
The coefficients $\bew$ are proportional to the one-loop coefficients
in the $\beta$ functions of the EW couplings and explicitly read 
\beq \label{betarelations}
\bew_{\FW}={}\frac{19}{6\sw^2},\qquad
\bew_{\FA\FA}=-\frac{11}{3},\qquad
\bew_{\FA\FZ}=-\frac{19+22\sw^2}{6\sw\cw},\qquad
\bew_{\FZ\FZ}={}\frac{19-38\sw^2-22\sw^4}{6\sw^2\cw^2}.
\eeq

The correction factors for the longitudinal gauge bosons are extracted
with the help of the Goldstone-boson equivalence theorem taking into
account higher-order corrections \cite{Yao:1988aj}
\beq \label{longeq:coll} 
\de^\LC_{\FW^\pm\FW^\pm}(\FV_\rL)=  \de^\LC_{\FZ\FZ}(\FV_\rL)=
\left[2\cew_\Phi-\frac{\NCt}{4\sw^2}\frac{\Mt^2}{\MW^2}\right] l(s),
\eeq
where the $\Mt^2$ term accounts for the enhanced coupling to the top quark.

Finally, the correction factor for the Higgs boson reads
\beq
\de^{\LC}_{\FH\FH}(\Phi)= \left[2\cew_\Phi-\frac{\NCt}{4\sw^2}\frac{\Mt^2}{\MW^2}\right]l(s).
\eeq

\myparagraph{(d) Logarithms from parameter renormalization}

Additional EW logarithms result from the renormalization of
dimensionless parameters at scales small compared to $\mu^2=s$.  We
here only give contributions related to the running of the couplings
from $\MW^2$ to $s$.  Further contributions from the running of the
electromagnetic coupling from zero to $\MW^2$ result from the
renormalization of the electric charge.  In the $\alpha(0)$~scheme,
those contributions appear as $\De\al(\MW^2)$ terms, as explicitly
given in \citere{Denner:2000jv}; in other EW input-parameter schemes
those terms are different (cf.~\refse{se:input_schemes}).

While neglecting
light fermion masses, it is convenient to
choose the four dimensionless parameters
\beq
e, \qquad \cw^2, \qquad g_\Ft= \frac{e}{\sqrt{2}\sw}\frac{\Mt}{\MW}, 
\qquad \la=\frac{e^2}{2\sw^2}\frac{\MH^2}{\MW^2}
\eeq
for the renormalization transformation.
Then, the UV logarithms are obtained from the LO matrix element
$\cM_0=\cM_0(e,\cw^2,g_\Ft,\la)$ in the high-energy limit by
\beq\label{eq:PRlogs}
\de^\PR \cM = \left. \frac{\de\cM_0}{\de e}\de e
+ \frac{\de\cM_0}{\de\cw^2}\de\cw^2
+ \frac{\de\cM_0}{\de g_\Ft}\de g_{\Ft}
+ \frac{\de\cM_0}{\de\la}\de\la
\, \right|_{\mu^2=s}.
\eeq
In the case of processes with longitudinal gauge bosons, these
logarithms should be determined from the matrix elements for external
Goldstone bosons according to the equivalence theorem. Note that the
correction factors $A_\FV(k^2)$ in \refeqs{eq:EToneW} and
\refeqf{eq:EToneZ} do not give rise to logarithmic corrections.

The leading-logarithmic contributions to the counterterms read
for the charge renormalization
\beq \label{chargerenorm}
\de Z_e = -\frac{1}{2}\bew_{AA}l(\mu^2),
\eeq
for the renormalization of the weak mixing angle
\beq \label{weinbergrenorm}
\frac{\delta \cw^2}{\cw^2}=\frac{\sw}{\cw}\bew_{AZ}l(\mu^2),
\eeq
for the renormalization of the top-quark Yukawa coupling
\begin{align}
\frac{\de g_\Ft}{g_\Ft} 
={}&
\left[-\frac{3}{4\sw^2}-\frac{3}{8\sw^2\cw^2}+\frac{3}{2\cw^2}\Qt-\frac{3}{\cw^2}\Qt^2
+\frac{3+2\NCt}{8\sw^2}\frac{\Mt^2}{\MW^2}\right]l(\mu^2)
=
\left[-\frac{17+10\cw^2}{24\cw^2\sw^2}
+\frac{9}{8\sw^2}\frac{\Mt^2}{\MW^2}\right]l(\mu^2),
\end{align}
and for the renormalization of the scalar self-coupling
\beq
\frac{\de\la}{\la} = 
\frac{3}{2\sw^2}\left[\frac{\MW^2}{\MH^2}\left(2+\frac{1}{\cw^4}\right)
-\left(2+\frac{1}{\cw^2}\right) + \frac{\MH^2}{\MW^2}\right]l(\mu^2)
+\frac{\NCt}{\sw^2}\frac{\Mt^2}{\MW^2}\left(1-2\frac{\Mt^2}{\MH^2}\right)l(\mu^2).
\eeq
We note that the counterterms for the dimensionless parameters are
independent of the tadpole scheme.

While the collinear and soft logarithms involve $\beta$-function
coefficients $\bew$ with a positive sign for external transverse gauge
bosons, the logarithms from the renormalization of the electromagnetic
coupling contain such terms with negative sign leading to partial
cancellation of single EW logarithms for processes with external
transverse gauge bosons. For the particular case of processes with
exactly one external fermion--antifermion pair and only external
transverse gauge bosons (but no other external fields) all these
contributions cancel. This feature is a consequence of Ward
identities.

\myparagraph{(e) Application to scattering processes} The formulae
given above are applicable to arbitrary scattering processes subject
to the condition that all kinematic invariants are large compared to
$\MW^2$. This is, however, not directly the case for processes that
involve the production of unstable particles with their subsequent
decays. The above results can nevertheless be applied in such cases
within the pole approximation (\cf \refse{se:pole_scheme}), since then
the conditions are fulfilled for the production subprocesses while no
large logarithms appear in the corrections to decay subprocesses where
the scale is set by the mass of the decaying
resonance~\cite{Accomando:2001fn}.

An interesting physical example is the scattering of EW vector bosons,
which occurs as 
subprocess in partonic processes such as
$\Pq\Pq\to\Pq\Pq\Pl\bar\Pl\Pl\bar\Pl$. In this case, the double-pole
approximation for the produced gauge bosons can be combined
\cite{Accomando:2006hq} with the effective vector-boson approximation
\cite{Dawson:1984gx,Kane:1984bb,Lindfors:1985yp} for the incoming
vector bosons, and the dominant contributions result from diagrams
with the generic form shown in the left of \reffi{fig:vbsww} 
on page \pageref{fig:vbsww}.
While the effective vector-boson approximation yields only a crude
approximation for the matrix elements in the very-high-energy limit
\cite{Kuss:1995yv,Accomando:2006hq}, it provides a reasonable basis
for an approximation for the relative corrections.  Within this
approach, the logarithmic approximation can, for instance, be applied
to the subprocess $\PWp\PWp\to\PWp\PWp$, while the emission of the
W~bosons from the quark lines and the decay of the W~bosons do not
give rise to large EW logarithms. Disregarding the angular-dependent
subleading soft--collinear corrections, but including the leading
soft--collinear corrections, the single soft and collinear logarithms
for the dominant transverse polarization of the scattering W~bosons,
as well as the EW logarithms from parameter renormalization for the
dominant transverse vector bosons, one finds a simple approximation
for the corrections to the cross section for the process
$\Pp\Pp\to\mu^+\nu_\mu\Pep\nu_\Pe\Pj\Pj+X$ \cite{Biedermann:2016yds}
\begin{equation}\label{eq:WWWW_log_corr}
 \sigma_{\rm LL, T} = \sigma_{\rm LO} \left[1  -\frac{\alpha}{4\pi} 4
   C^{\rm ew}_{\PW} \ln^2\frac{Q^2}{\MW^2}
+\frac{\alpha}{4\pi} 2 b^{\rm ew}_{\PW} \ln \frac{Q^2}{\MW^2} \right],
\end{equation}
where $\sigma_{\rm LO}$ is the full LO cross section.  Using $\langle
M_{4 \Pl} \rangle\sim 390\GeV$ as a typical scale $Q$ for the
vector-boson scattering subprocess, which can be determined from a LO
calculation, leads to an EW correction of about $-16\%$ in remarkably
good agreement with the full calculation of the NLO EW corrections
\cite{Biedermann:2016yds,Biedermann:2017bss}.  Using $Q= M_{4 \Pl}$ on
an event-by-event basis results in a correction of about $-15\%$.
Since the self-interaction of EW gauge bosons is only due to the
$\SU(2)_\rw$ interaction, the result \refeqf{eq:WWWW_log_corr} is
valid for arbitrary scattering processes of EW vector bosons. This can
be derived either in the symmetric phase of the EW theory or using the
results given above together with the fact that LO matrix elements for
$VV\to VV$ scattering in the high-energy limit differ only by factors
$-\cw/\sw$ if a photon field is replaced by a Z-boson field. While the
effective scale $\langle M_{4 \Pl} \rangle$ depends somewhat on the
considered process and the event selection, the EW corrections are
expected to be of the same order of magnitude for all scattering
processes of EW gauge bosons.  This has been confirmed for
$\PW\PZ\to\PW\PZ$ scattering, where relative EW corrections of $-16\%$
were found in nice agreement with the approximation
\refeqf{eq:WWWW_log_corr} \cite{Denner:2019tmn}.

The angular-dependent subleading soft--collinear corrections, on
the other hand, depend on the considered scattering process $VV\to VV$,
and the corresponding correction factor to the integrated cross
section requires a non-trivial integration over the scattering angle.
The correction factors for the differential cross sections can be
easily constructed from the results of \citere{Accomando:2006hq}
[see also \refeqs{SScorr}--\refeqf{eq:IWSS}]. They
depend on the polarization of the vector bosons and on the various LO
matrix elements with different intermediate vector bosons. Using the
high-energy relations between the LO matrix elements, the result for
the dominant contribution of purely transverse $\PWp\PWp$ scattering
reads 
\beq\label{eq:WWWW_SSC_corr_TTTT}
 \rd\sigma_{\SSC,\rT} = \rd\sigma_{\rm LO} \frac{\alpha}{\pi\sw^2} 
 2\ln\left(\frac{Q^2}{\MW^2}\right)
\left[\ln\frac{s_{12}}{Q^2}
-\left(1-\frac{s_{13}}{s_{12}}\right)\ln\frac{s_{13}}{Q^2}
-\left(1-\frac{s_{23}}{s_{12}}\right)\ln\frac{s_{23}}{Q^2}
\right],
\eeq
where $s_{ij}$ are the Mandelstam variables of the vector-boson
scattering subprocess.  When integrating over the phase space, the
angular-dependent corrections \refeqf{eq:WWWW_SSC_corr_TTTT}
average out to a large extent such that the resulting contribution
is only of the order of one percent.

The generic results of \citere{Denner:2000jv} were used to calculate
the one-loop logarithmic corrections to various processes including
vector-boson pair production at the LHC
\cite{Accomando:2001fn,Accomando:2004de}, vector-boson scattering in
electron--positron annihilation \cite{Accomando:2006hq}, and $V+\Pj$
production at the LHC \cite{Kuhn:2004em,Kuhn:2005gv,Kuhn:2007qc}.  In
\refse{se:EWlogspract} below, we consider the evaluation of the
leading EW logarithms to the Drell--Yan-like W- and Z-boson
production.  The results of \citere{Denner:2000jv} have been
implemented in the event generator ALPGEN \cite{Mangano:2002ea} and
applied to the production of jets with missing energy
\cite{Chiesa:2013yma}.

\subsubsection{Resummation of EW double-logarithmic corrections}
\label{se:evol_eqs}

Since the EW high-energy logarithmic corrections can grow to several
tens of percent in the TeV range for typical scattering processes, EW
logarithms beyond NLO can become relevant as well. These can be taken
into account via appropriate resummation methods. The resummation of
the leading EW logarithms (double soft--collinear logarithms) was
studied in a pioneering paper by Fadin \etal \cite{Fadin:1999bq}.
There, the {\em IR evolution equations} \cite{Kirschner:1982qf}, describing
the dependence of amplitudes on some IR cutoff $\muIR$ of the virtual
particle transverse momenta, were used to perform the resummation. In
the following we basically follow the arguments of
\citere{Fadin:1999bq}.

We start with a generic non-abelian gauge theory and consider an
arbitrary amplitude, where all invariants $s_{ij}$ are large and of
the same order $|s_{ij}|\sim s$. We extract the virtual particle with
the smallest value of transverse momentum $|\qperp|$ in such a way
that the transverse momenta $|\qperp'|$ of the other virtual particles
are much bigger,
\beq
{\qperp'}\!^2\gg\qperp^2\gg\muIR^2.
\eeq
For the other particles, $\qperp^2$ plays the role of the initial IR
cutoff $\muIR^2$. The Sudakov double logarithms result from the
exchange of soft--collinear gauge bosons (see \reffi{fig:Ldiags} left
on page~\pageref{fig:Ldiags}).  In this case, the integral over the
momentum $q$ of the soft virtual boson with smallest $\qperp$ can be
extracted based on the non-abelian generalization of the Gribov
theorem \cite{Gribov:1966hs,Lipatov:1988ii,DelDuca:1990gz}. This
leads, via a generalization of the one-loop results
\refeqf{eikonalappA}, to the IR evolution equation
\begin{align} \label{IRevolutionLL}
\cM^{i_1 \ldots i_n}&(p_1,\ldots,p_n,\muIR^2) ={}
\cM_{0}^{i_1 \ldots i_n}(p_1,\ldots,p_n)
+\frac{\ri}{2}g^2
\sum_{\substack{k,l=1\\l\ne k}}^n \sum_a
 T^{a}_{i'_k i_k}(k) T^{a}_{i'_l i_l}(l) \\
& \times\int_{s>\qperp^2>\muIR^2}
\frac{\rd^4 q}{(2\pi)^4} 
\frac{-4p_k p_l}{(q^2-M^2+\ri\eps)[(q+p_k)^2-m_k^2][(q-p_l)^2-m_l^2]}
\cM^{i_1 \ldots i'_k\ldots i'_l\ldots i_n}(p_1,\ldots,p_n,\qperp^2),
\notag
\end{align}
where $M$ is the mass of the exchanged gauge boson,
the external particles are on their mass shell ($p_l^2=m_l^2$,
$p_k^2=m_k^2$).
In the double logarithmic approximation $m_l$ and $m_k$ are only needed
for photon exchange, where they are equal, while they are negligible
for $\PW$- or $\PZ$-boson exchange.
Moreover, $g$ denotes the non-abelian gauge coupling and $T^a(l)$ the
representation of the generators of the gauge group corresponding to
the external particle~$l$.  In \refeq{IRevolutionLL}, $\qperp$
represents the component of the gauge-boson 3-momentum $\mathbf{q}$
transverse to the particles $k$ and $l$ coupling to this boson,
defined in the CM frame of the two momenta $p_k,p_l$.  In the limit of
small masses $m_k,m_l$ and small virtuality $q^2$, it can be expressed
in Lorentz-invariant form according to
\beq
\qperp^2=2\min_{k\ne l} \left|\frac{(q p_l)(q p_k)}{p_k p_l}\right|.
\eeq
Equation \refeqf{IRevolutionLL} is valid in a covariant gauge for the
gauge boson with momentum $q$, but can be written in a gauge-invariant way
upon including the term with $k=l$ in the sum (which does not give a
double-logarithmic term).

In order to evaluate the integral in \refeq{IRevolutionLL}, 
the Sudakov parametrization 
\beq
q=-x \left(p_k-\frac{m_k^2}{2p_k p_l}p_l\right)
+ y \left(p_l-\frac{m_l^2}{2p_k p_l}p_k\right) 
+ q_\perp, \quad p_k q_\perp = p_l q_\perp =0, 
\eeq
is used, where here and in the following terms of order $m_k^4$,
$m_l^4$, $m_k^2m_l^2$, which are irrelevant in double-logarithmic
approximation, are neglected.  
In the CM frame of $p_k,p_l$, we have $q_\perp^0=0$, so that
$q_\perp^2=-\qperp^2<0$.
The integral in \refeq{IRevolutionLL}
can be rewritten as
\begin{align}\label{eq:soft-collinear-integral}
I = {} &\int_{s>\qperp^2>\muIR^2}\frac{\rd^4 q}{(2\pi)^4} 
\frac{ -4p_k p_l}{(q^2-M^2+\ri\eps)[(q+p_k)^2-m_k^2][(q-p_l)^2-m_l^2]}
\notag\\
\sim{}& {-(p_kp_l)} |p_kp_l| \frac{4\pi}{(2\pi)^4}
\int_{\muIR^2}^s\rd \qperp^2\int\rd x \int\rd y \,
\Bigl[-2xy(p_kp_l)-\qperp^2-M^2+\ri\eps\Bigr]^{-1}
\notag\\
&\times\Bigl[2(1-x)y(p_kp_l)-\qperp^2- x m_k^2 \Bigr]^{-1}
\Bigl[2x(1-y)(p_kp_l)-\qperp^2- y m_l^2 \Bigr]^{-1},
\end{align}
where the asymptotic equality $\sim$ here and in the following refers
to double-logarithmic approximation.  To evaluate $I$, it is
convenient to apply a {\it sector decomposition} which separates the
two types of collinear singularities w.r.t.\ $p_k$ or $p_l$ by
assuming a hierarchy between $|x|$ and $|y|$,
\beq
I = I_{|x|>|y|} + I_{|x|<|y|}.
\eeq
In the following we evaluate $I_{|x|>|y|}$, which contains the
collinear singularity w.r.t.\ $p_k$, and get $I_{|x|<|y|}$ from
$I_{|x|>|y|}$ upon interchanging $k$ and $l$.  Using
\beq\label{eq:PVdec}
\frac{1}{2x y(p_k p_l)+\qperp^2+M^2-\ri\eps}
=\ri\pi\,\delta\left(2x y(p_k p_l)+\qperp^2+M^2\right)
+\mathrm{PV}\frac{1}{2x y(p_k p_l)+\qperp^2+M^2}
\eeq
for the gauge-boson propagator, only the part with the $\delta$
function but not the principle-value (PV) part contributes in the
double-logarithmic approximation.
Performing the $y$~integration in the $|x|>|y|$ sector of
\refeq{eq:soft-collinear-integral} with the help of the $\de$~function
in \refeq{eq:PVdec}, results in
\beq
I_{|x|>|y|}\sim{} -\frac{\ri\pi^2}{(2\pi)^4}
\int_{\muIR^2}^s\rd \qperp^2 
\int_{x^2>\frac{\qperp^2+M^2}{|2p_k p_l|}}\rd x \,
\frac{1}{ \qperp^2+M^2(1-x)+ x^2 m_k^2} \,
\frac{4|x|(p_kp_l)^2 }{4x^2(p_kp_l)^2 + 2x(p_kp_l)M^2+ m_l^2(\qperp^2+M^2)}.
\eeq
The restriction on the $x$-integration originates from the combination
of the requirements $|x|>|y|$ and that the argument of the $\de$
function in \refeq{eq:PVdec} has to vanish.  Owing to this restriction
the mass term $m_l^2$ in the integral never regularizes a singularity,
so that we can set $m_l=0$ in the integral $I_{|x|>|y|}$.
Substituting $\qperp^2\to\qperp^2-M^2$, we get
\begin{align}
I_{|x|>|y|}&{}\sim 
-\frac{\ri\pi^2}{(2\pi)^4}
\int_{\muIR^2+M^2}^s\rd \qperp^2 
\int_{x^2>\frac{\qperp^2}{|2p_k p_l|}}\rd x \,
\frac{1}{ \qperp^2-x M^2+ x^2 m_k^2} \,
\frac{2(p_kp_l) }{2|x|(p_kp_l) + \sgn(x)M^2}.
\nn\\
&{}\sim
-\frac{\ri\pi^2}{(2\pi)^4}
\int_{\max\{\muIR^2,M^2\}}^s\rd \qperp^2 
\int_{x^2>\qperp^2/s}\frac{\rd x}{|x|} \,
\frac{1}{ \qperp^2+ x^2 m_k^2},
\end{align}
where the second form is valid, because the $M^2$ terms in the integrand of
the first form do not regularize a singularity. Moreover, we have
modified the integration boundaries in a way that does not change
the integral in double-logarithmic approximation.
Finally, we translate the $m_k^2$ term in the integrand into an
effective upper bound of the $x$-integration:
For $m_k=0$, the upper limit of $x$ can be set to any number
of $\ord(1)$ that does not lead to an enhanced logarithm;
for $m_k\ne0$, the $m_k^2$ term truncates the logarithmic contribution
for $x^2\gsim\qperp^2/m_k^2$.
Thus, the double
logarithmic contribution from the sector $|x|>|y|$ is obtained as
\begin{align}
I_{|x|>|y|} \sim{}& 
-\frac{2\ri\pi^2}{(2\pi)^4}
\int_{\max\{\muIR^2,M^2\}}^{s} \frac{\rd\qperp^2}{\qperp^2} \,
\int_{|\qperp|/\sqrt{s}}^{\min\{1,|\qperp|/m_k\}}\frac{\rd x}{x} 
= -\frac{\ri\pi^2}{(2\pi)^4}
\int_{\max\{\muIR^2,M^2\}}^{s} \frac{\rd\qperp^2}{\qperp^2} \,
 \ln\frac{s}{\max\{\qperp^2,m_k^2\}} 
\nl
\equiv {}&
-\frac{\ri\pi^2}{(2\pi)^4} J(\muIR^2,M^2,m_k^2),
\end{align}
where a factor 2 results from combining regions of positive and
negative $x$.  Adding the contribution of the sector $|x|<|y|$, the
full result for $I$ in double-logarithmic approximation is given by
\beq
I\sim -\frac{\ri\pi^2}{(2\pi)^4}\left[ J(\muIR^2,M^2,m_k^2) +
  J(\muIR^2,M^2,m_l^2)\right]
\eeq
with 
\begin{align}
J(\muIR^2,M^2,m^2)=
\left\{
\begin{array}{ll}
\frac{1}{2}\ln^2\frac{s}{\max\{\muIR^2,M^2\}} &
\text{for} \quad \max\{\muIR^2,M^2\}>m^2, \\[.5em]
\frac{1}{2}\ln^2\frac{s}{m^2} +
\ln\frac{s}{m^2}\ln\frac{m^2}{\max\{\muIR^2,M^2\}}\quad &
\text{for} \quad m^2>\max\{\muIR^2,M^2\}.
\end{array}\right.
\end{align}

Each contribution for $|x|> |y|$ or $|y|< |x|$ depends on the mass and
momentum of only one of the external legs. Using the above results
as well as charge conservation
\beq
\sum_k T^a_{i'_k i_k}(k)\cM_{0}^{i_1 \ldots i'_k\ldots  i_n}(p_1,\ldots,p_n)=0, 
\eeq
the IR evolution equation \refeqf{IRevolutionLL} can be transformed to
\begin{align} \label{IRevolutionLL2}
\cM^{i_1 \ldots i_n}&(p_1,\ldots,p_n,\muIR^2)=
\cM_{0}^{i_1 \ldots i_n}(p_1,\ldots,p_n)
\notag\\
& -\frac{g^2}{(4\pi)^2} \sum_{k=1}^n
\int_{\max\{\muIR^2,M^2\}}^{s} \frac{\rd\qperp^2}{\qperp^2} \,
 \ln\frac{s}{\max\{\qperp^2,m_k^2\}}
  \,
C_{i'_ki_k}(k) \cM^{i_1\ldots i'_k\ldots i_n}(p_1,\ldots,p_n,\qperp^2),
\end{align}
with the Casimir operator $C_{i'_li_l}(l) =\sum_a
[T^a(l)T^a(l)]_{i'_li_l}$ of the generic non-abelian gauge group.

The differential form of the evolution equation then reads
\beq \label{IRevolutionLLexp}
\frac{\partial\cM^{i_1 \ldots i_n}(p_1,\ldots,p_n,\muIR^2)}{\partial \ln(\muIR^2)}=
 -\frac{g^2}{(4\pi)^2}\sum_{k=1}^n C_{i'_k i_k}(k)
\frac{\partial J(\muIR^2,M^2,m_k^2)}{\partial\ln(\muIR^2)}
\prod_{\substack{l=1\\l\ne k}}^n \de_{i'_l i_l}
\cM^{i'_1 \ldots i'_n}(p_1,\ldots,p_n,\muIR^2),
\eeq
with
\beq
\frac{\partial J(\muIR^2,M^2,m_k^2)}{\partial\ln(\muIR^2)}
= -\theta(\muIR^2-M^2) \ln\frac{s}{\max\{\muIR^2,m_k^2\}}.
\eeq

Since there are  no large logarithms for large cutoff scale
$\muIR^2=s$, we have the initial condition
\beq
\cM^{i_1 \ldots i_n}(p_1,\ldots,p_n,\muIR^2=s) =
\cM_0^{i_1 \ldots i_n}(p_1,\ldots,p_n).
\eeq
Using the fact that for unbroken gauge theories the Casimir operators
are diagonal, 
$C_{i'_k i_k}(k) = C(k) \de_{i'_k i_k}$,
the evolution equation \refeqf{IRevolutionLLexp} can be
solved as
\beq
\cM^{i_1 \ldots i_n}(p_1,\ldots,p_n,\muIR^2) =
\cM_0^{i_1 \ldots i_n}(p_1,\ldots,p_n)
\exp\left(-\frac{g^2}{(4\pi)^2} \sum_{k=1}^n
C(k)  J(\muIR^2,M^2,m_k^2)
 \right),
\eeq
\ie the Sudakov double logarithms exponentiate in the non-abelian case
in the same way as in the abelian case.

The method based on the IR evolution equations  is also applicable to
broken gauge theories like the EWSM. To double-logarithmic accuracy
all heavy masses can be considered equal,
\beq
\MZ\sim\MW\sim\MH\sim\Mt\sim M,
\eeq
and the energy should be much larger, $\sqrt{s}\gg M$.  For
$\sqrt{s} > \muIR >  M$, effects of spontaneous symmetry breaking and
gauge-boson masses can be neglected, and the evolution equation can be
studied in the unbroken phase.  If the equation is formulated in the
broken phase, the photon contributions have to be taken into account
in order not to violate gauge invariance.  For $\muIR\ll M$ only the
photon exchange drives the running in $\muIR$ as in QED.

If we assume for simplicity that all charged particles have masses
$m_k\lsim M$, the IR evolution equation reads for $\sqrt{s}>\muIR> M$
\beq \label{IRevolutionLL_SM}
\frac{\partial\cM^{i_1 \ldots i_n}(p_1,\ldots,p_n,\muIR^2)}{\partial
  \ln(\muIR^2)}=
\frac{e^2}{(4\pi)^2}\ln\frac{s}{\muIR^2} \sum_{k=1}^n
C^{\ew}_{i'_k i_k}(k)
\cM^{i_1 \ldots i'_k\ldots i_n}(p_1,\ldots,p_n,\muIR^2)
\eeq
with the EW Casimir operator defined in \refeq{CasimirEW}.
In the symmetric basis, \ie diagonal EW Casimir operators,
the solution of \refeq{IRevolutionLL_SM} is given by
\beq\label{Solution_IRevolutionLL_SM}
\cM^{i_1 \ldots i_n}(p_1,\ldots,p_n,\muIR^2) =
\cM_0^{i_1 \ldots i_n}(p_1,\ldots,p_n)
\exp\left[
-\frac{e^2}{2(4\pi)^2}\ln^2\frac{s}{\muIR^2} \sum_{k=1}^n
C^{\ew}(k)
\right].
\eeq
The corresponding results in the physical basis can be easily obtained
by transforming the matrix elements or by using the matrices for the
Casimir operators in the exponential function.

Choosing the cutoff $\muIR$ in the region $\muIR < M$, only the
photon contribution remains.  In order to carry over the above results
to this new situation, the IR regulator mass previously called $M$ is
now the infinitesimal photon mass $m_\gamma\ll m_k$.
The corresponding IR evolution equation then reads
\beq \label{IRevolutionLL_QED}
\frac{\partial\cM^{i_1 \ldots i_n}(p_1,\ldots,p_n,\muIR^2)}{\partial
  \ln(\muIR^2)}=
\cM^{i_1 \ldots i_n}(p_1,\ldots,p_n,\muIR^2)
\frac{e^2}{(4\pi)^2} \sum_{k=1}^n
Q_k^2
\ln\frac{s}{\max\{\muIR^2,m_k^2\}}.
\eeq
The appropriate initial condition is given by
\refeq{Solution_IRevolutionLL_SM} evaluated at the matching point
$\muIR=M$, and the solution is
\begin{align}\label{Solution_IRevolutionLL_QED}
\cM^{i_1 \ldots i_n}(p_1,\ldots,p_n,\muIR^2) ={}&{}
\cM_0^{i_1 \ldots i_n}(p_1,\ldots,p_n)
\exp\left[
-\frac{e^2}{2(4\pi)^2}\ln^2\frac{s}{M^2} \sum_{k=1}^n
C^{\ew}(k)
\right]
\notag\\
&\times\exp\left[
-\frac{e^2}{(4\pi)^2} \sum_{k=1}^n
Q_k^2\left(
J(\muIR^2,m_\ga^2,m_k^2) - J(M^2,m_\ga^2,m_k^2)
\right)
\right],
\end{align}
which takes the following explicit form in the region
$m_\gamma< \muIR<m_k < M$,
\begin{align}
\cM^{i_1 \ldots i_n}(p_1,\ldots,p_n,\muIR^2) 
={}&{}
\cM_0^{i_1 \ldots i_n}(p_1,\ldots,p_n)
\exp\left[
-\frac{e^2}{2(4\pi)^2}\ln^2\frac{s}{M^2} \sum_{k=1}^n
C^{\ew}(k)
\right]
\notag\\
&\times\exp\left[
-\frac{e^2}{(4\pi)^2} \sum_{k=1}^n
Q_k^2\left(
\ln\frac{s}{m_k^2}\ln\frac{M^2}{\muIR^2}
-\frac{1}{2}\ln^2\frac{m_k^2}{M^2}
\right)
\right],
\end{align}

The results \refeqf{Solution_IRevolutionLL_SM} and
\refeqf{Solution_IRevolutionLL_QED} are applicable for processes
involving chiral fermions, transverse gauge bosons, and Higgs bosons,
provided that all invariants of order $s$ are large compared to $M^2$.
For longitudinal gauge bosons, the equivalence theorem (see
\refse{se:gbet})has to be used.  When expanded to one-loop accuracy,
the results of \citere{Denner:2000jv} including the electromagnetic
logarithms are reproduced upon substituting $m_\gamma$ (called
$\lambda$ in \citere{Denner:2000jv}) for $\muIR$.  It has been shown
for specific processes that the exponentiation of the EW double
logarithms is in agreement with explicit two-loop calculations of
these contributions \cite{Melles:2000ed,Hori:2000tm,Beenakker:2001kf}.

IR evolution equations have also been used to sum subleading EW
Sudakov logarithms (see
\citeres{Jantzen:2005xi,Melles:2001ye,Melles:2001dh} and references
therein). To this end, the evolution equations have been taken over
from QCD \cite{Collins:1980ih,Sen:1981sd,Sen:1982bt} and applied to
the EWSM.  
This approach has been extended to the 
NLL and next-to-next-to-leading logarithmic (N$^2$LL) approximation
\cite{Kuhn:1999nn,Kuhn:2001hz}.  Starting with the N$^3$LL
approximation, the corrections become sensitive to the details of the
gauge-boson mass generation \cite{Jantzen:2005xi}.  This method has
been applied to sum the EW logarithms to the vector form factor and
neutral-current 4-fermion amplitudes at the NLL \cite{Kuhn:1999nn},
N$^2$LL \cite{Kuhn:2001hz}, and N$^3$LL level \cite{Jantzen:2005xi},
and for W-pair production at the ILC and LHC at the N$^2$LL level
\cite{Kuhn:2007ca,Kuhn:2011mh}.
Various explicit two-loop calculations have been performed to verify
the resummation of the subleading logarithms
\cite{Denner:2003wi,Denner:2006jr,Denner:2008yn}.

\subsubsection{EW logarithmic corrections from Soft--Collinear
  Effective Theory}

The computation of EW corrections in the high-energy domain can be
greatly simplified by using effective field theory (EFT) methods,
specifically {\em Soft--Collinear Effective field Theory (SCET)}
\cite{Chiu:2007yn,Chiu:2008vv,Chiu:2009mg}. In the SCET approach
\cite{Bauer:2000ew,Bauer:2000yr,Bauer:2001ct,Bauer:2001yt}
factorization of physics at different scales is assumed. 
As a result, the hard scattering amplitude can be written as a product
of process-independent one-particle collinear functions depending on
external particle energies and a universal soft function depending
only on external particle directions \cite{Chiu:2009mg}.

\newcommand{\lL}{\mathsf{L}}
\newcommand{\lM}{\mathsf{L_M}}
\newcommand{\lQ}{\mathsf{L_Q}}
\newcommand{\lp}{\mathsf{L_p}}
\newcommand{\vev}[1]{\left\langle #1 \right \rangle}
\newcommand{\scetew}{$\text{SCET}_{\text{EW}}$}
\newcommand{\mul}{\mu_\mathrm{l}}
\newcommand{\muh}{\mu_\mathrm{h}}

The treatment of EW corrections within the SCET approach (\scetew)
\cite{Chiu:2007yn,Chiu:2007dg,Chiu:2008vv,Chiu:2009mg,Chiu:2009ft,Fuhrer:2010eu}
is summarized in \citere{Manohar:2014vxa} as follows:

\begin{myitemize}
  
\item At a high scale $\muh$ of order $\sqrt{s}$, the scattering
  amplitudes are matched onto $\SU(3)_{\rc} \times \SU(2)_{\rw} \times
  \U(1)_Y$ gauge-invariant local operators $O_k$ with Wilson
  coefficients $C_k$ which can be computed perturbatively in terms of
  a power series in the three gauge coupling constants 
  $\alpha_i(\mu^2_h)$ of the SM. As an example, for $\Pg(p_1)+\Pg(p_2) \to
  \Pq(p_3)+\overline \Pq(p_4)$ the operators read
\beq
O_1 =  \bar q_4 q_3 \FG_2^A \FG_1^A, 
\qquad
O_2 =  d^{ABC} \bar q_4 T^C q_3 \FG_2^A \FG_1^B,
\qquad
O_3 = \ri f^{ABC} \bar q_4 T^C q_3 \FG_2^A \FG_1^B,
\label{eq:ops}
\eeq
which represent the possible colour structures of the amplitude. The
subscripts $1,2,3,4$ label the different external particles.

\item The Wilson coefficients $C_i$ are evolved using
  renormalization-group
  equations (RGEs) down to a low scale $\mul$ of order $\MW$. The
  anomalous dimensions can be computed in the unbroken $\SU(3)_{\rc} \times
  \SU(2)_{\rw} \times \U(1)_Y$ theory.
  
\item At the scale $\mul$, the fields of the EW gauge bosons W and Z,
  the Higgs field, and the top-quark field are integrated out. This
  calculation must be performed in the broken theory.  A single
  gauge-invariant operator breaks up into different components,
  because the weak gauge symmetry is broken. For example, each
  of the operators $O_i$ in \refeq{eq:ops} splits into an
  $\SU(3)_{\rc}$-invariant $\Pg\Pg \to \Pt \bar\Pt$ and $\Pg\Pg \to \Pb
  \bar\Pb$ operator, if $\Pq$ is an EW doublet $(\Pt_\rL,\Pb_\rL)$.
 
\item The operators in the theory below $\mul$ are used to
  compute the scattering cross sections.

\end{myitemize}
As final result, the scattering amplitudes $\mathcal{M}$ can be
written in resummed form
\begin{align}
\mathcal{M} &=  \exp \left[D_C(\mul,\lM,\bar{n}p) \right] d_{\mathrm{S}}(\mul,\lM)
\,P\exp \left[\int_{\muh}^{\mul} \frac{\rd \mu}{\mu}
  \gamma(\mu) \right] 
C(\muh,\lQ).
\label{eq:m}
\end{align}
The operator $P$ denotes path ordering so that the values of
$\mu$ increase from left to right. The ingredients in \refeq{eq:m},
for which explicit formulas are given in \citere{Chiu:2009ft}, are
discussed in the following.

The {\it high-scale matching} $C(\muh,\lQ)$ is an $n$-dimensional
column vector with a perturbative expansion in $\alpha_i(\mu^2_h)$,
with $i=1,2,3$ being the respective $\U(1)_Y$, $\SU_{\rw}(2)$, and
$\SU(3)_{\rc}$ gauge couplings.  For the example \refeqf{eq:ops}, we
have $n=3$, since there are 3 gauge-invariant amplitudes.  The
high-scale matching depends on $\lQ=\ln s/\muh^2$, which is not a
large logarithm if one chooses $\muh^2 \sim s$, and is computed from
graphs of the full theory, setting all small scales such as the
gauge-boson and fermion masses to zero.  It can be calculated in the
unbroken theory without worrying about mass effects and EW mixing. In
an NLO calculation, the high-scale matching corresponds to the finite
non-logarithmic corrections in the high-energy limit.  The high-scale
matching does not obey the factorization structure of the EFT
amplitude and cannot be calculated using EFT results. For processes
including a small number of external particles it is known
\cite{Chiu:2008vv}.

The {\it SCET anomalous dimension}
$\gamma(\mu)$ determines the running
of the amplitude between the high scale $\muh\sim\sqrt{s}$ and the low scale
$\mul\sim\MW$. It is an $n \times n$ anomalous-dimension matrix which
can be written as the sum of a collinear and a soft part,
\begin{align}
\gamma(\mu) &= \gamma_{\mathrm{C}}(\mu,\bar{n}p)+ \gamma_{\mathrm{S}}(\mu,\{n\}),
\end{align}
where the collinear part is proportional to a unit matrix in colour space,
\begin{align}\label{eq:gammaC}
\gamma_{\mathrm{C}}(\mu,\bar{n}p)&= \mathbb{1} \sum_r \left[A_r(\mu) \ln \frac{2E_r}{\mu}+B_r(\mu) \right],
\end{align}
and linear in $\ln (\bar n_r p_r)=\ln (2E_r)$ to all orders in
perturbation theory~\cite{Manohar:2003vb,Chiu:2009mg}, with $E_r$
denoting the energy of parton~$r$. For incoming particles,
$n_r=(1,\mathbf{n}_r)$ and $\bar{n}_r=(1,-\mathbf{n}_r)$, where
$\mathbf{n}_r$, with $|\mathbf{n}_r|=1$, points into the direction of
motion of parton $r$; for outgoing particles $n_r=-(1,\mathbf{n}_r)$
and $\bar{n}_r=-(1,-\mathbf{n}_r)$. The sum over $r$ comprises all
partons in the scattering process, and $A_r(\mu)$, the {\it cusp
  anomalous dimension}, and $B_r(\mu)$, the {\it non-cusp anomalous
  dimension}, have a perturbative expansion in $\alpha_i(\mu^2)$. The
one-loop expressions for the SM fields are given in Table 1 of
\citere{Chiu:2009ft}. They are in one-to-one correspondence with the
corresponding results of \citere{Denner:2000jv}.  The results for
$\gamma_{\mathrm{C}}$ in Table 1 of \citere{Chiu:2009ft} summarize the
leading soft--collinear contributions [terms containing
$\lp=\ln(\bar{n}p/\mu)$], the collinear factors $\de^\coll$ (explicit
constant terms $\propto T\cdot T$), and the contributions from field
renormalization constants $\gamma_\phi$ from \citere{Denner:2000jv}.
At one-loop order, the soft part reads
\begin{equation}\label{eq:gamma_soft}
\gamma_{\mathrm{S}}(\mu,\{n\}) = -\sum_{i=1}^3 
\sum_{\mathrm{pairs}\,r,s}
\frac{\alpha_i(\mu^2)}{\pi} 
T_r^{(i)} \cdot T_s^{(i)}  \ln \frac{-n_r  n_s - \ri\eps}{2},
\end{equation}
where the sum is over all parton pairs $r,s$,
and $T_r^{(i)}$ is the gauge generator for the
$i$th gauge group acting on parton $r$.  The soft part
$\ga_{\mathrm{S}}$ corresponds to the subleading soft--collinear
contributions of \citere{Denner:2000jv}.  The anomalous dimension
$\gamma(\mu)$ is independent of the low-energy scales and can be
computed in the unbroken gauge theory. The $\bar{n}$ dependence 
(Lorentz-frame dependence) cancels between the collinear and soft
functions.  The renormalization-group evolution governed by the SCET
anomalous dimension sums the Sudakov double logarithms which provide
the dominant effect of the EW corrections.

The {\it low-scale matching} consists of a collinear part
$D_{\mathrm{C}}$ and a soft part $d_{\mathrm{S}}$. The soft part
$d_{\mathrm{S}}$ is an $m \times n$ matrix, where $m$ is the number of
amplitudes produced after $\SU(2)_{\rw} \times \U(1)_Y$ breaking. In
$\Pg\Pg \to q \bar q$, if $q$ is an EW doublet of left-handed quarks
$(\Pt_\rL,\Pb_\rL)$, 
the operators  in \refeq{eq:ops} give rise 
to $m=6$ operators after $\SU(2)_{\rw} \times
\U(1)_Y$ breaking, where $\bar q_4 q_3 = \bar \Pt_4 \Pt_3$ or $\bar
q_4 q_3 = \bar \Pb_4 \Pb_3$.  If $q$ in \refeq{eq:ops} is an EW
singlet, such as $\Pb_\rR$ or $\Pt_\rR$, then $m=3$.  The collinear
matching $D_{\mathrm{C}}$ is proportional to an $m \times m$ unit
matrix and given by
\begin{align}
\left[D_{\mathrm{C}}(\mul,\bar{n}p,\lM)\right]_{ii} &= 
\sum_r \left[ J_r (\mul,\lM)\ln \frac{2E_r}{\mul} + H_r (\mul,\lM) \right].
\label{DC}
\end{align}
The sum over $r$ includes all particles in the operator $O_i$ produced
after EW symmetry breaking, and $D_{\mathrm{C}}$ is linear in $\ln
(\bar np)$ to all orders in perturbation
theory~\cite{Manohar:2003vb,Chiu:2009mg}. The collinear contributions
$D_{\mathrm{C}}$ [and also $\gamma_{\mathrm{C}}$ in \refeq{eq:gammaC}]
are process independent.  The soft part
$d_{\mathrm{S}}(\mul,\{n\},\lM)$ and the functions $J_r$ and $H_r$ of
the collinear part have an expansion in
$\alpha_{\rs,\rw,\text{em}}(\mul)$, \ie in the couplings of the broken
theory, and depend on $\mul$ and on masses of the order of the EW
scale via dimensionless ratios such as $\MW/\MZ$ and $\lM=\ln
(\MZ/\mul)$.
The one-loop contribution to the soft part has the structure of
\refeq{eq:gamma_soft} with an additional logarithm of the gauge-boson
mass.
These logarithms are small if one chooses $\mul \sim
\MZ$.  The low-scale matching has to be computed in the broken EW
theory. It translates the amplitudes for the fields of the unbroken
theory to those of the broken theory \cite{Chiu:2009ft}.  Since all
light lepton and quark masses can be neglected, the quark-mixing
matrix does not enter the SCET computation. Once the EW corrections
are computed, one can introduce the quark-mixing matrix by
transforming to the mass-eigenstate basis.

The exponent in \refeq{eq:m} contains at most a double (Sudakov)
logarithm resulting from integrating the $A_r(\mu)$ terms in the
collinear anomalous dimension.  The high-scale matching is by two
powers of logarithms less important than the leading Sudakov
logarithms.  The low-scale matching contains a single-logarithmic
term.  This is a new feature of \scetew\ with respect to
SCET$_\mathrm{QCD}$ first pointed out in \citere{Chiu:2007yn}.  One
can show that the low-scale matching contains at most a single
logarithm to all orders in perturbation
theory~\cite{Chiu:2007yn,Chiu:2009ft}.  As a consequence, resummed
perturbation theory remains valid even at high energy, because
$\alpha_i^k \ln(s/\MW^2) \ll 1$ for sufficiently large~$k$.

The logarithmic term in the matching \refeqf{DC} is needed for a
proper factorization of scales. A typical Sudakov double-logarithmic
term at one loop has the form (suppressing coupling factors)
\begin{align}
 \ln^2 \frac{Q^2}{M^2} &=\ln^2 \frac{Q^2}{\muh^2} + \left[\ln^2
   \frac{Q^2}{\mul^2}-\ln^2 \frac{Q^2}{\muh^2} \right]
+\left[  \ln^2  \frac{M^2}{\mul^2} - 2 \ln  \frac{Q^2}{\mul^2} \ln \frac{M^2}{\mul^2} \right].
\end{align}
The first term on the r.h.s.\ belongs to the high-scale matching $C$,
the second term arises from integrating the $\ln (Q^2/\mu^2)$
anomalous dimension from $\muh$ to $\mul$, and the third term enters
the low-scale matching $D_{\mathrm{C}}$. The existence of the
logarithmic term in the low-scale matching also follows from the
consistency condition that the theory is independent of $\mul$.  Since
changes in the running between $\muh$ and $\mul$ contain a single
logarithm from the anomalous dimension, there must be a single
logarithm in the matching.

The resummed EW corrections can be grouped into LL, NLL, etc., in the
usual way; the precise definition for \scetew\ can be found in
\citere{Chiu:2009mg}. The LL series is determined by the one-loop cusp
anomalous dimension $A_r(\mu)$. The NLL series is determined by the
two-loop cusp anomalous dimension and the one-loop values of
$B_r(\mu)$ and $J_r(\mu,\lM)$. All terms needed for an NLL computation
are known, so that \emph{all} processes can be computed to resummed
NLL order.  In \citeres{Chiu:2009ft,Fuhrer:2010eu} the one-loop
$d_{\mathrm{S}}$ and $C$ terms were computed for all $2 \to 2$
processes.  The three-loop cusp anomalous dimension $A_r(\mu)$ and the
two-loop non-cusp anomalous dimension $B_r(\mu)$ are available in the
literature, except for the scalar Higgs contributions, which are
numerically small. The two-loop contribution to $D_{\mathrm{C}}$ has
not been calculated yet. The NNLL results are known, with the
exception of these terms.

Based on the \scetew\ results, one can compute radiative corrections
in the EFT to all partonic hard scattering processes at the LHC with
an arbitrary number of external particles, if the high-scale matching
is known.  These calculations are valid in the Sudakov region and
neglect power corrections of the form $M^2_{\PW,\PZ}/s$ and $\Mf^2/s$
in the one-loop and higher-order radiative corrections.  In this
regard, we expect that the results do not apply to processes that are
mass-suppressed at LO.  For the Sudakov form factor, the
mass-suppressed one-loop corrections are $2\%$ for $\MW/Q\sim 1$ and
decrease fast with increasing $Q$ \cite{Chiu:2009mg}. How this
translates to general processes is not obvious. In this respect it
might be interesting to note that for Higgs production processes in
$\Pep\Pem$ annihilation and at hadron colliders bosonic EW corrections
at the level of $5\%$ have been found that could not be attributed to
a known source of enhanced corrections
\cite{Denner:2003yg,Ciccolini:2003jy,Ciccolini:2007ec,Figy:2010ct,Denner:2011id}.


\subsubsection{EW logarithmic corrections for practical calculations}
\label{se:EWlogspract}

The detailed knowledge of the tower of EW high-energy
logarithms is important for a deeper understanding of
EW dynamics, but making use of it in predictions is a subject
that deserves care:
\begin{myitemize}
\item It is certainly advisable to make use of full NLO EW
  corrections, \ie without applying expansions for high energies,
  whenever possible for a given process.  Non-logarithmic corrections
  are process specific and typically amount to some percent, depending
  on the kinematical domain and the process under consideration.
  After including finite, non-logarithmic terms in the high-energy
  approximation,
  power-suppressed terms
  are still missing.  A safe assessment of the quality of high-energy
  approximations requires the comparison to full NLO results
  \cite{Beenakker:1993tt,Kuhn:2004em,Kuhn:2005gv,Accomando:2005ra}.
\item Beyond NLO, the knowledge of higher-order EW logarithms can be
  very useful to improve pure NLO predictions.  However,
  particular care has to be taken if the EW logarithms show large
  cancellations between leading and subleading terms, as for instance
  observed in the case of neutral-current fermion--antifermion
  scattering~\cite{Jantzen:2005xi}.  If the full tower of logarithms
  of a fixed perturbative order is not known, it is not clear to which
  accuracy truncated towers approximate the full correction. However,
  the known part of the tower can at least deliver estimates for the
  size of missing corrections and be used in the assessment of
  theoretical uncertainties.
\item Since the EW high-energy logarithmic corrections are associated
  with virtual soft and/or collinear weak-boson or photon exchange,
  they all have counterparts in real weak-boson or photon-emission
  processes which can partially cancel the large negative virtual
  corrections.  This cancellation is incomplete
  \cite{Ciafaloni:2000df}, since $\SU(2)_\rw$ doublets are in general
  not treated inclusively in EW corrections---a fact that is by some
  abuse
of language called {\it Bloch--Nordsieck violation}.%
\footnote{The Bloch--Nordsieck theorem \cite{Bloch:1937pw} simply does
  not apply to non-abelian gauge theories.}  To which extent the
cancellation occurs depends on the experimental capabilities to
separate final states with or without weak bosons or photons.
Logarithmic approximations, as implemented in the {\sc Pythia}
shower~\cite{Christiansen:2014kba} can deliver first estimates, but
solid predictions have to be based on complete matrix elements.  The
general issue and specific examples were discussed in
\citeres{Ciafaloni:2006qu,Baur:2006sn,Chiu:2009mg,Bell:2010gi,Stirling:2012ak,Chiesa:2013yma,Manohar:2014vxa,Bauer:2016kkv}.
For instance, the numerical analysis~\cite{Baur:2006sn} of
neutral-current Drell--Yan production demonstrates the effect of real
weak-boson emission in the distributions in the transverse lepton
momentum $p_{\mathrm{T},\Pl}$ and in the invariant mass $M_{ll}$ of
the lepton pair.  At the LHC, at $M_{ll}=2\TeV$ the EW corrections are
reduced from about $-11\%$ to $-8\%$ by weak-boson emission. At
$p_{\mathrm{T},\Pl}=1\TeV$ the corresponding reduction from about
$-10\%$ to $-3\%$ is somewhat larger.
A framework to perform an all-order resummation of EW logarithms for
inclusive scattering processes at energies much above the EW scale was
 developed in \citere{Manohar:2018kfx}.
\end{myitemize}

While the analytical structure of EW corrections was studied in the
literature in detail in the Sudakov regime, there is only little
knowledge on EW corrections beyond NLO in more general kinematical
situations where not all invariants $s_{ij}$ are large. Note that
there are many cross sections that are in fact not dominated by the
Sudakov regime in the high-energy limit, including all processes that
are dominated by $t$-channel diagrams. For example, unless
specifically designed cuts are applied, reactions like W-boson pair
production via $\Pep\Pem$, $\Pp\Pp$, or $\gamma\gamma$ collisions
receive sizeable contributions from the {\it Regge limit}, where the
Mandelstam variable $t$ remains small while $s$ gets large.  Moreover,
it often depends on the specific observable which regime is probed in
high-energy tails of kinematical distributions.  Taking
neutral-current Drell--Yan processes (see, e.g.,
\citeres{Dittmaier:2001ay,Dittmaier:2009cr}) and dijet
production~\cite{Dittmaier:2012kx} at the LHC as examples,
differential distributions in the transverse momenta of the produced
leptons or jets probe the Sudakov regime in the high-momentum tails.
On the other hand, the dilepton or dijet invariant-mass distributions
of these processes are not dominated by this regime at high scales, so
that the EW high-energy logarithms derived in the Sudakov regime do
not approximate the EW corrections well in those observables.

In \citere{Brensing:2007qm} the full EW corrections for the process
$\Pp\Pp\to \Pl^+ \nu_\Pl+X$, where $\Pl\/$ is a massless lepton, were
compared to the double-logarithmic 
approximation for the LHC with $\sqrt{s}=14\TeV$.
In \refta{tab:EWC_pplnu} we show the LO cross section for different
cuts on the transverse mass $M_{\rT,\Pl\nu_\Pl}$ of the produced W~boson.
\begin{table}\small
\centerline{\begin{tabular}{c|cccccc}
\hline
 $M_{\rT,\Pl \nu_\Pl}/\mathrm{GeV}$ &
 $50{-}\infty$ & $100{-}\infty$ & $200{-}\infty$ &
 $500{-}\infty$ & $1000{-}\infty$ & $2000{-}\infty$\\  \hline
$\sigma_0/\mathrm{pb}$ & $ \! 4495.7(2)\! $ & $ \! 27.589(2)\! $ & $ \! 1.7906(1)\! $ & $ \!
0.084697(4)\! $ & $ \! 0.0065222(4)\! $ & $ \! 0.00027322(1)\! $ \\ \hline
$\de^{\mu^+\nu_\mu}_{\Pq \bar{\Pq}}/\mathrm{\%}$ &         $  -2.9(1) $ & $  -5.2(1) $ & $  -8.1(1) $ & $  -14.8(1) $ &
$ -22.6(1) $ & $  -33.2(1) $ \\
$\de^{\mathrm{rec}}_{\Pq \bar{\Pq}}/\mathrm{\%}$ &         $  -1.8(1) $ & $  -3.5(1) $ & $  -6.5(1) $ & $  -12.7(1) $ &
$ -20.0(1) $ & $  -29.6(1) $ \\ \hline
{$\de^{(1)}_{\mathrm{EWdlog}}/\mathrm{\%}$} &               {$  0.0005 $} & {$  0.5 $} & {$  -1.9 $} &
{$ -9.5 $} & {$ -18.5 $} & {$ -29.7 $} \\
{$\de^{(1)}_{\mathrm{EWslog}}/\mathrm{\%}$} &               {$  0.008 $} & {$  0.9 $} & {$  2.3 $} &
{$ 3.8 $} & {$ 4.8 $} & {$ 5.9 $} \\
$\de^{(2)}_{\mathrm{EWdlog}}/\mathrm{\%}$ &         $  -0.0002 $ & $  -0.023 $ & $  -0.082 $ & $  0.21 $ & $  1.3 $ & $
3.8 $ \\ \hline
 \end{tabular}}
\caption{
Integrated LO cross sections for $\Pp\Pp\to \Pl^+ \nu_\Pl+X$ at the LHC
for $\sqrt{s}=14\TeV$ for different ranges in the transverse mass
$M_{\rT,\Pl\nu_\Pl}$ and corresponding relative corrections (results
taken from \citere{Brensing:2007qm} and extended).}
\label{tab:EWC_pplnu}
\vspace*{3ex}
\small
\centerline{\begin{tabular}{c|cccccc}
\hline$M_{\Pl \Pl}/\mathrm{GeV}$ & $50{-}\infty$ & $100{-}\infty$ & $200{-}\infty$
& $500{-}\infty$ & $1000{-}\infty$ & $2000{-}\infty$\\
\hline
$\sigma_0/\mathrm{pb} $ & $  738.733(6) $ & $  32.7236(3) $ & $  1.48479(1) $ & $  \!0.0809420(6) \! $ & $  \!0.00679953(3) \! $ & $ \!
0.000303744(1) \!  $ \\
$\delta^{\mathrm{rec}}_{\Pq\bar{\Pq},\mathrm{phot}}/\mathrm{\%} $ & $ -1.81 $ & $ -4.71 $ & $ -2.92 $ & $ -3.36 $ & $
-4.24 $ & $ -5.66 $ \\
$\delta_{\Pq\bar{\Pq},\mathrm{weak}}/\mathrm{\%} $ & $ -0.71 $ & $ -1.02 $ & $ -0.14 $ & $ -2.38 $ & $ -5.87 $ & $
-11.12 $ \\
\hline
{$\delta^{(1)}_{\mathrm{EWdlog}}/\mathrm{\%} $} & {$ 0.27 $} & {$ 0.54 $} & {$ -1.43 $} & {$
-7.93$} & {$ -15.52 $} & {$ -25.50 $} \\
$\delta^{(2)}_{\mathrm{EWdlog}}/\mathrm{\%} $ & $ -0.00046 $ & $ -0.0067 $ & $ -0.035 $ & $ 0.23 $ & $ 1.14 $ & $ 3.38
$ \\
\hline
\end{tabular}}
\caption{Integrated LO cross sections for $\Pp\Pp\to \Pl^+\Pl^- +X$ at
  the LHC for $\sqrt{s}=14\TeV$ for different ranges in the invariant
  mass $M_{\Pl\Pl}$ and corresponding relative corrections (results
  taken from \citere{Dittmaier:2009cr} and extended)).}
\label{tab:EWC_ppll}
\end{table}
We list the EW corrections for bare muon final states
$\de^{\mu^+\nu_\mu}_{\Pq \bar{\Pq}}$ and with photon recombination
applied $\de^{\mathrm{rec}}_{\Pq \bar{\Pq}}$, as well as the NLO
corrections due to EW double logarithms $\delta^{(1)}_{\mathrm{EWdlog}}$ 
and EW single logarithms $\delta^{(1)}_{\mathrm{EWslog}}$
as derived from Eqs.~\refeqf{SCsum}, \refeqf{SScorr},
and \refeqf{subllogfact}, \refeqf{eq:PRlogs} respectively,
\begin{align}
\label{eq:deltaEWdlog}
\delta^{(1)}_{\mathrm{EWdlog}} ={}&
{-C^{\EW}}_{\Fdbar\Fu\to\Fl^+\Fnu} \,L(\hat s)
+4l_{\Fdbar\Fu\to\Fl^+\Fnu}(\hat s,\hat t,\hat u) \,l(\hat s),
\\
\delta^{(1)}_{\mathrm{EWslog}} ={}&
 3 C^{\EW}_{\Fdbar\Fu\to\Fl^+\Fnu} \,l(\hat s)
+ p_{\Fdbar\Fu\to\Fl^+\Fnu} \,l(\hat s),
\label{eq:deltaEWslog}
\end{align}
with the large logarithms $L(\hat s)$ and $l(\hat s)$ defined in
\refeqs{eq:Ls} and \refeqf{eq:ls} and
\begin{align}
C^{\EW}_{\Fdbar\Fu\to\Fl^+\Fnu} &{}=
C^{\EW}(\Fd^\rL) + C^{\EW}(\Fu^\rL) + C^{\EW}(\Fl^{\rL}) + C^{\EW}(\Fnu^\rL) 
= \frac{3}{\sw^2}+\frac{5}{9\cw^2},
\nn\\ 
l_{\Fdbar\Fu\to\Fl^+\Fnu}(\hat s,\hat t,\hat u) &{}=
-\frac{1}{\sw^2}\ln\left(\frac{\hat t\,\hat u}{\hat s^2}\right)
+ \frac{1}{3\cw^2}\ln\left(\frac{\hat t}{\hat u}\right),
\nn\\
p_{\Fdbar\Fu\to\Fl^+\Fnu}(\hat s,\hat t,\hat u) &{}=
2\left(\frac{\cw}{\sw}\bew_{\FA\FZ} - \bew_{\FA\FA}\right)
= -\frac{19}{3\sw^2}.
\end{align}
Here, $\hat s,\hat t,\hat u$ are the usual Mandelstam variables of the
partonic process $\Fdbar\Fu\to\Fl^+\Fnu$, \ie $\hat
s=(p_{\Fdbar}+p_{\Fu})^2$, $\hat t=(p_{\Fdbar}-p_{\Fl^+})^2$, and
$\hat u=(p_{\Fdbar}-p_{\Fnu})^2$.  Moreover, we include an estimate
$\delta^{(2)}_{\mathrm{EWdlog}}$ of the NNLO EW logarithms based on the
first two leading two-loop logarithms obtained from
$\delta^{(1)}_{\mathrm{EWdlog}}$ by exponentiation,
\beq
\delta^{(2)}_{\mathrm{EWdlog}} =
C^{\EW}_{\Fdbar\Fu\to\Fl^+\Fnu} \,L(\hat s)
\left[\frac{1}{2}C^{\EW}_{\Fdbar\Fu\to\Fl^+\Fnu} \,L(\hat s)
-4l_{\Fdbar\Fu\to\Fl^+\Fnu}(\hat s,\hat t,\hat u) \,l(\hat s)\right].
\label{eq:deltaEWlog2}
\eeq
We observe that the relative NLO EW corrections are apparently well
reproduced for large $M_{\rT,\Pl\nu_\Pl}$ by
$\delta^{(1)}_{\mathrm{EWdlog}}$, i.e.\ by the leading and subleading
EW logarithms of types LSC and SSC.  However, the perfect match is
accidental, since the single EW logarithms
$\delta^{(1)}_{\mathrm{EWslog}}$ from collinear singularities and
parameter renormalization contribute at the level of $5\%$.
Logarithmic contributions from real radiation, which are not included
in $\delta^{(1)}_{\mathrm{EWdlog}}+\delta^{(1)}_{\mathrm{EWslog}}$,
and terms involving constants or pure angular-dependent 
logarithms are at the
level of a few percent (as also demonstrated for $\PW\ga$ production
at the LHC in \citere{Accomando:2005ra}). Thus, a generic uncertainty
of a few per cent should be attributed to an approximation based on
enhanced logarithmic contributions only. Within this uncertainty
margin, the logarithmic approximation works well for the considered
case.  This is due to the fact that demanding large values of
$M_{\rT,\Pl\nu_\Pl}$ by a cut enforces large Mandelstam invariants
$\hat s$, $|\hat t|$, and $|\hat u|$ in the underlying partonic
process, which corresponds to the Sudakov domain.

As a further example we consider the neutral current Drell--Yan
process $\Pp\Pp\to\Pl^+\Pl^- +X$ at $\sqrt{s}=14\TeV$ as
investigated in \citere{Dittmaier:2009cr}. The corresponding cross
section and EW corrections for different cuts on the lepton-pair
invariant mass $M_{\Pl\Pl}$ are shown in \refta{tab:EWC_ppll}.
The table shows the complete EW corrections, the gauge-invariant weak
corrections as well as the double-logarithmic 
EW corrections at NLO (types
LSC and SSC) and NNLO as defined in the previous case, \ie which are
calculated as in \refeqs{eq:deltaEWdlog} and \refeqf{eq:deltaEWlog2}.
However, for the partonic neutral-current process $\bar q^\si q^\si\to
\Fl^{+,\tau}\Fl^\tau$ the factors (see also \citere{Denner:2000jv})
\begin{align}
C^{\EW}_{\bar q^\si q^\si\to \Fl^{+,\tau}\Fl^\tau} ={} &
2C^{\EW}(\Fq^\si) + 2C^{\EW}(\Fl^{\tau}) 
= \frac{2}{\sw^2}\left[ I_{\rw,\Pq^\si}(I_{\rw,\Pq^\si}+1)
      + I_{\rw,\Pl^\tau}(I_{\rw,\Pl^\tau}+1) \right]
+ \frac{1}{2\cw^2}\left[ (Y_{\rw,\Pq^\si})^2
      + (Y_{\rw,\Pl^\tau})^2 \right],
\nn\\ 
l_{\bar q^\si q^\si\to \Fl^{+,\tau}\Fl^\tau}(\hat s,\hat t,\hat u) ={} &
2\left( \frac{I^3_{\rw,\Pq^\si}I^3_{\rw,\Pl^\tau}}{\sw^2}
       + \frac{Y_{\rw,\Pq^\si}Y_{\rw,\Pl^\tau}}{4\cw^2} \right)
\ln\left(\frac{\hat u}{\hat t}\right)
-\frac{2}{\sw^2}
\frac{4\cw^2 I^3_{\rw,\Pq^\si}I^3_{\rw,\Pl^\tau}}%
{4\cw^2 I^3_{\rw,\Pq^\si}I^3_{\rw,\Pl^\tau}
       + \sw^2 Y_{\rw,\Pq^\si}Y_{\rw,\Pl^\tau}}
\ln\left(\frac{|\hat r_{\si\tau}|}{\hat s}\right),
\nn\\& {}
\nn
\hat r_{\si\tau} = \left\{
\begin{array}{ll} \hat t, & I^3_{\rw,\Pq^\si}I^3_{\rw,\Pl^\tau}>0, \\[.3em] 
\hat u, &  I^3_{\rw,\Pq^\si}I^3_{\rw,\Pl^\tau}<0, 
\end{array}\right.
\\
p_{\bar q^\si q^\si\to \Fl^{+,\tau}\Fl^\tau} ={} &
2\left(
\frac{4\cw^4 I^3_{\rw,\Pq^\si}I^3_{\rw,\Pl^\tau}-\sw^4 Y_{\rw,\Pq^\si}Y_{\rw,\Pl^\tau}}%
{\cw\sw(4\cw^2 I^3_{\rw,\Pq^\si}I^3_{\rw,\Pl^\tau} + \sw^2
  Y_{\rw,\Pq^\si}Y_{\rw,\Pl^\tau})}\bew_{\FA\FZ} 
- \bew_{\FA\FA}\right),
\end{align}
depend on the chiralities $\si,\tau=\rR/\rL=+/-$.  In this case, the
EW logarithms fail to approximate the full EW corrections, as can
already be concluded from the significant discrepancy between the
double-logarithmic correction $\delta^{(1)}_{\mathrm{EWdlog}}$ and the
full correction.  The failure is a consequence of the different event
selection: The cut demanding large $M_{\Pl\Pl}$ values enforces only
large values of $\hat s$ (which is equal to $M_{\Pl\Pl}^2$ in LO
kinematics), but not large values of $|\hat t|$ and $|\hat u|$, which
would be required in the Sudakov regime. Instead, the cross section
still receives large contributions from regions where $\hat s$ is
large, but either $|\hat t|$ or $|\hat u|$ small, where the EW
corrections are less pronounced.

\providecommand{\cmps}{\mu^2_P}
\providecommand{\cmhs}{\mu^2_\PH}
\providecommand{\cmf}{\mu_\Pf}
\providecommand{\Gf}{\Ga_\Pf}
\providecommand{\cmt}{\mu_\Pt}
\providecommand{\cmws}{\mu^2_\PW}
\providecommand{\cmzs}{\mu^2_\PZ}
\providecommand{\cmw}{\mu_\PW}
\providecommand{\cmz}{\mu_\PZ}
\providecommand{\csw}{\sw}
\providecommand{\ccw}{\cw}
\providecommand{\cZ}{\mathcal{Z}}
\providecommand{\qqb}{q\bar q}
\providecommand{\llb}{l^- l^+}
\providecommand{\gqqA}{ \ensuremath{g_{qqA}}}
\providecommand{\gllA}{ \ensuremath{g_{llA}}}
\providecommand{\gffA}{ \ensuremath{g_{ffA}}}
\providecommand{\gqqZ}{ \ensuremath{g_{qqZ}}}
\providecommand{\gllZ}{ \ensuremath{g_{llZ}}}
\providecommand{\gffZ}{ \ensuremath{g_{ffZ}}}
\providecommand{\gqqV}{ \ensuremath{g_{qqV}}}
\providecommand{\gllV}{ \ensuremath{g_{llV}}}
\providecommand{\gffV}{ \ensuremath{g_{ffV}}}

\section{Issues with unstable particles}
\label{se:unstable}

\subsection{Unstable particles and resonances}

Since most of the elementary degrees of freedom (fermions, Higgs and gauge bosons)
of the SM are unstable, the appearance of unstable particles in high-energy
particle reactions is more the rule than an exception. 
Apart from rare exceptions, such as muons or B~mesons, 
unstable states do not exist long enough to leave directly accessible traces
in detectors, but rather appear as resonances that can only be reconstructed
from their decay products. 
The lifetime $\tau_P$ of such a resonance $P$ can, thus, 
not be determined in a time-of-flight
measurement, but only indirectly from its total decay width $\Gamma_P=1/\tau_P$,
which is accessible via the width of the resonance if the experimental
resolution in the reconstruction is good enough.%
\footnote{When restoring Planck's constant $\hbar$, which equals one
  in the natural units that we use, the relation between width and
  lifetime reads  $\tau_P=\hbar/\GP$.}
In an ideal situation, $\GP$ is obtained from the Breit--Wigner-like
resonance in the invariant-mass distribution of its decay products, but also 
other distributions (\eg transverse-mass or transverse-momentum distributions)
might be useful.

Theoretically, resonance processes lead to complications in standard
perturbation theory which requires stable asymptotic states in
$S$-matrix elements. 
Strictly speaking, an unstable particle $P$ (mass $\MP$) 
should only appear as intermediate state, \ie on internal
lines in Feynman diagrams. However, there it might happen that the
pole at invariant mass $p^2=\MP^2$ in a propagator
$\propto(p^2-\MP^2)^{-1}$ of $P$ is hit for some physical momentum transfer~$p$, 
leading to a singularity at any fixed order in perturbation theory.
This artificially introduced singularity is only lifted if at least
the imaginary parts of self-energy corrections near the singularity
are {\it Dyson summed}.
Denoting the renormalized self-energy $\Sigma_\ren(p^2)$ of $P$
graphically by a dark grey blob,%
\footnote{For a scalar field, $\Sigma$ is the usual self-energy as defined in
\refeq{eq:definition_se}, for a vector field $\Sigma$ stands for the
transverse part $\Sigma_\rT$, and for fermionic fields $\Sigma$ is effectively
formed by some combination of self-energies corresponding to the various
covariants spanning the 2-point function.} 
\beq
\Sigma_\ren(p^2) \;=\; 
\raisebox{-.45em}{\includegraphics[scale=.8]{diagrams/2pt-blob.pdf}}  \;=\;
\raisebox{-.45em}{\includegraphics[scale=.8]{diagrams/2pt-1loop.pdf}} \;+\;
\raisebox{-.45em}{\includegraphics[scale=.8]{diagrams/2pt-tad1.pdf}}  \;+\;
\raisebox{-.45em}{\includegraphics[scale=.8]{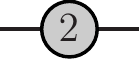}} \;+\;
\raisebox{-.45em}{\includegraphics[scale=.8]{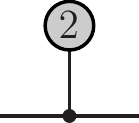}}  \;+\;
\raisebox{-.45em}{\includegraphics[scale=.8]{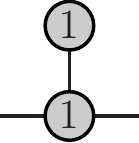}} \;+\; \dots,
\eeq
where the numbers represent the loop orders in the light grey 1PI blobs on the 
right-hand side,
the Dyson
summation modifies the LO propagator  $G_{0,P}(p^2)=\ri/(p^2-\MP^2)$ to
\begin{align}
\arraycolsep 4pt
\begin{array}{ccccccccc}
\raisebox{-.4em}{\includegraphics[scale=.8]{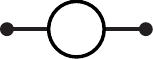}} &=&
\raisebox{-.4em}{\includegraphics[scale=.8]{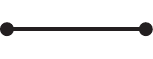}} &+&
\raisebox{-.4em}{\includegraphics[scale=.8]{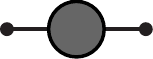}} &+& 
\raisebox{-.4em}{\includegraphics[scale=.8]{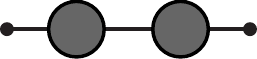}} &+& \dots
\\[.5em]
G_P(p^2) &=& \frac{\ri}{p^2-\MP^2} &+&
\frac{\ri}{p^2-\MP^2}\ri\Sigma_\ren(p^2)\frac{\ri}{p^2-\MP^2} &+&
\frac{\ri}{p^2-\MP^2}\ri\Sigma_\ren(p^2)\frac{\ri}{p^2-\MP^2} 
\ri\Sigma_\ren(p^2)\frac{\ri}{p^2-\MP^2} &+& \dots
\\[.5em]
&{}= \rlap{$\displaystyle\frac{\ri}{p^2-\MP^2+\Sigma_\ren(p^2)}$.}
\end{array}
\nn\\*[-1.8em]
\end{align}
The difference between a stable and an unstable particle of mass
$\MP>0$ rests in the fact that the pole of the propagator lies on the
positive real axis or strictly below it in the complex $p^2$ plane,
respectively.

The {\it optical theorem}, which is a consequence of the unitarity of
the $S$-matrix, yields the relation 
\beq\label{eq:optical_theorem}
\Im\Sigma_\ren(\MP^2) \,=\, \Im\left\{\,
\raisebox{-.45em}{\includegraphics[scale=.8]{diagrams/2pt-blob.pdf}}
\,\right\}\Big|_{p^2=\MP^2}
\,=\, \frac{1}{2}\sum_X 
\raisebox{-1.4em}{\includegraphics[scale=.8]{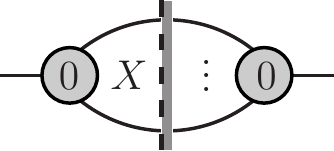}}
\,=\, \frac{1}{2} \sum_X \int\rd\Phi_X \, |\M_{P\to X}|^2
\,=\, \MP\GP 
\eeq
in one-loop approximation (see for instance
\citeres{Peskin:1995ev,Schwartz:2013pla}).  Since moreover
$\Re\Sigma_\ren(\MP^2)=0$ for the on-shell (OS)-renormalized mass
$\MP$, the propagator of an unstable particle behaves like
$[p^2-\MP^2+\ri\MP\GP]^{-1}$ with $\GP>0$ near the pole, so that its
singularity turns into the shape of a Breit--Wigner resonance after
squaring it.  The sign of $\GP$ is determined by causality, \ie
already fixed by Feynman's $\ri\eps$~prescription for propagators, and
guarantees that the intermediate resonance~$P$ exponentially decays in
forward time direction with the lifetime $\tau_P=1/\GP$.

In a quantum field theory with stable particles only, the unitarity of
the $S$-matrix can be expressed in terms of {\it Cutkosky's cutting
  rules}~\cite{Cutkosky:1960sp}, which relate imaginary parts of
Feynman integrals with corresponding {\it cut diagrams}.  Those cut
diagrams result from the original diagram by replacing the propagators
(with momentum~$p$) on the cut by {\it cut propagators} of the form
$2\pi\theta(p_0)\delta(p^2-\MP^2)$, which correspond to physical OS
intermediate states, and replacing the propagators and vertices on one
side of the cut by their complex conjugates.  Following common
practice, the complex-conjugated part of the cut diagram is indicated
in \refeq{eq:optical_theorem} by a shadow on the right side of the
cut.  Obviously, the cut relations require modification in the
presence of unstable particles, which cannot appear as intermediate
states with real mass values.  Already in 1963 Veltman showed for a
generic scalar field theory that unitarity is still respected and that
the cutting relations can be restored if cuts through intermediate
lines of unstable particles are omitted~\cite{Veltman:1963th}.
Veltman's proof, however, consistently makes use of Dyson-summed
propagators and all-order properties of amplitudes, so that the
arguments do not easily carry over to calculations in practice.
Unitarity within the {\it complex-mass scheme (CMS)}, 
which is presented in
\refse{se:CMS} below in detail, was investigated in
\citere{Denner:2014zga}.

Technically, the Dyson summation is straightforward, but conceptually
it bears some issues in combination with the truncation of the
perturbative series, which is necessary in practice for corrections
that are not of self-energy type, so that perturbative orders are only
partially included starting from some loop level.  However, consistency
relations, such as the gauge independence of $S$-matrix elements or
cutting relations expressing unitarity, are typically formulated order
by order in perturbation theory, so that those relations are in
general disturbed if some perturbative orders are not taken into
account completely.  Without taking special measures, this can totally
destroy the consistency and reliability of predictions.  Before
introducing and discussing solutions to this problem at tree and loop
level, we first look into the form of resonance propagators and naive
approaches that ignore issues connected with their embedding in full
matrix elements, in order to put into context simple approximations
and to clarify some details in the definition of mass and width of an
unstable particle.

\subsection{Narrow-width approximation and naive width schemes}

The simplest, but somewhat crude way to describe the production of an
unstable particle~$P$ is to employ the {\it narrow-width
  approximation (NWA)}, which treats $P$ as ``stable intermediate
state'', \ie the full process is decomposed into the production of an
OS particle $P$ and its decay to some final state $X$.  This
decomposition results from the limit $\GP\to0$ in the squared matrix
element of the full resonance process, where the squared propagator
with momentum transfer~$p$ behaves like
\begin{equation}
\frac{1}{|p^2-\MP^2+\ri \MP\GP|^2} 
\; \asymp{\GP\to0} \; \frac{\pi}{\MP\GP} \, \de(p^2-\MP^2).
\end{equation}
The $1/\GP$ factor on the r.h.s.\ is part of the well-known {\it branching ratio}
\begin{equation}
\mathrm{BR}_{P\to X} = \frac{\Gamma_{P\to X}}{\GP}, \qquad
\GP=\sum_X \Gamma_{P\to X},
\end{equation}
which emerges after the inclusive integration over the $P\to X$ decay phase space
and is the probability that $P$ chooses the final state $X$ in its decay.
The quantity $\Gamma_{P\to X}$ is the {\it partial decay width} for this channel,
which is calculable in perturbation theory similar to a cross section from the
$P\to X$ matrix elements.
This pattern repeats for each resonance if several unstable particles are produced.
By construction, the NWA fails to describe 
observables that resolve the invariant mass of the resonance, \ie it is restricted to
observables in which the resonances are integrated over. 

For unstable particles with spin, in general effects of spin correlations 
between production and decay or between various decaying particles appear
if cuts are imposed on the decay products, or if distributions in kinematical
variables of those are considered. 
The naive NWA, which employs unpolarized production cross sections and
decay widths, can be easily improved to include these correlations
by properly combining production and decay parts for definite polarization states
of the unstable particle.

The NWA is also useful in a first approximate calculation of
radiative corrections to cross sections for unstable-particle production.
The NWA can accommodate corrections to the production cross section and to the
relevant branching ratio, but neglects off-shell effects of 
${\cal O}(\GP/\MP)$ for each resonating particle $P$. 
These off-shell effects, which are due to the off-shell tails of the resonances
and {\it background diagrams} with fewer or no resonances,
have to be added to NWA predictions, e.g., to get full NLO accuracy
in cross-section predictions \cite{Tkachov:1998uy,Nekrasov:2007ta}.
Moreover, we note that naive estimates often underestimate the
theoretical uncertainties of the NWA, as \eg discussed in
\citeres{Kauer:2007zc,Uhlemann:2008pm}. Specifically,
in decays near some kinematical boundary or
in scenarios with cascade decays, which are quite common in non-standard models,
the NWA can even fail completely, as demonstrated in \citeres{Berdine:2007uv,Fuchs:2014ola}.

In view of EW physics at colliders, the most important resonances are
the ones of $\PW$ and $\PZ$~bosons, which decay into
lepton--antilepton or quark--antiquark pairs at LO. The decay widths
$\Ga_V$ ($V=\PW,\PZ$) formally count as ${\cal O}(\alpha)$ in the
electromagnetic coupling $\alpha$.  In numbers, we get ${\cal
  O}(\Ga_V/M_V)\sim3\%$, which is the expected order of relative
uncertainty of NWA predictions at LO, which start from the
approximation of OS 
$\PW/\PZ$~bosons in production and decay.
NLO corrections of relative order ${\cal O}(\alpha)$ do not only
comprise virtual one-loop and real emission corrections to OS
$\PW/\PZ$ production and decays, but also off-shell effects in LO
amplitudes.  Including the latter goes beyond a pure NWA calculation,
but nevertheless the NLO accuracy achieved by this improved
description~\cite{Nekrasov:2007ta}
of the resonance process is restricted to the resonance
region and to observables in which the resonances are not resolved,
\ie completely integrated over.

A detailed description of a resonance process, keeping the full
differential information of the kinematics of the decay products, has
to be based on complete matrix elements for the full process,
including both resonant and non-resonant diagrams.  Early LO
calculations based on full matrix elements proceeded in a minimalistic
way and merely modified propagators of unstable particles~$P$ upon
including an approximation for the imaginary part of the Dyson-summed
self-energies,
\begin{equation}
P_P(p^2) = \frac{1}{p^2-\MP^2+\ri \MP\GP(p^2)},
\label{eq:PropP}
\end{equation}
where $\GP(p^2)$ plays the role of the width of $P$ in the vicinity
of the resonance.
Two frequently used versions are:
\begin{myitemize}
\item
{\it Fixed width (FW)}: $\GP(p^2)=\GP=\mathrm{const.}$ \\
In this parametrization, the complex squared mass 
\begin{equation}
\mu_P^2 = \MP^2-\ri\MP\GP
\label{eq:muP}
\end{equation}
plays the role of the location of the pole in the propagator 
in the complex $p^2$~plane.
\item
{\it Running width (RW)}:
\begin{equation}
\GP(p^2)=\GP\times p^2/\MP^2\times\theta(p^2).
\label{eq:GPp2}
\end{equation}
The function $\GP(p^2)$ resembles the $p^2\!$ dependence of the
imaginary part of the one-loop self-energy of a vector particle $P$
that exclusively decays into pairs of massless fermions.  To good
approximation this applies to $\PW$ and $\PZ$~bosons.  The intention
in this scheme is to come closer to the off-shell behaviour of the
full propagator, in particular by including the factor $\theta(p^2)$
which switches off the imaginary part below the kinematical decay
threshold, as demanded by causality and unitarity.
The running-width scheme has, in particular, been used for the analysis of
the Z-boson resonance at LEP1/SLC 
(see \citere{ALEPH:2005ab} and references therein).
\end{myitemize}
We note, however, that none of these naive width schemes is satisfactory
even at LO, because both versions introduce a gauge dependence in
matrix elements. While the FW scheme for $\PW/\PZ$ nevertheless
delivers reasonable LO results by experience, 
(see, e.g., \citeres{Beenakker:1996kn,Denner:1999gp,Dittmaier:2002ap}),
the RW scheme
often fails completely, since the $p^2$ factor in the off-shell regions
is prone to enhance gauge-invariance-breaking terms by orders of
magnitude \cite{Kurihara:1994fz}.
We illustrate these facts in \refta{tab:LOXSwidthschemes}, where
we show the high-energy behaviour of some LO cross sections featuring
$\PW\PW$~production, $\PW\PW\gamma$~production, and $\PW\PW$~scattering
in $\Pep\Pem$~annihilation. 
\btab
\centering
\begin{tabular}{|l|r|r|r|r|}
\multicolumn{5}{l}{$\Pep\Pem\to{\Pu\bar\Pd\mu^-\bar{\nu}_\mu}$ 
\quad (no phase-space cuts applied)}
\\
\hline
$\sqrt{s}/\mathrm{GeV}$ &
\multicolumn{1}{c|}{$189$} &
\multicolumn{1}{c|}{$500$} &
\multicolumn{1}{c|}{$2000$} &
\multicolumn{1}{c|}{$10000$}\\
\hline
$\sigma($FW$)/\mathrm{fb}$
& $703.5(3)$ & $237.4(1)$ & $13.99(2)$ & $0.624(3)$
\\
$\sigma($RW$)/\mathrm{fb}$
& $703.4(3)$ & $238.9(1)$ & $34.39(3)$ & $498.8(1)$
\\
$\sigma($CMS$)/\mathrm{fb}$
& $703.1(3)$ & $237.3(1)$ & $13.98(2)$ & $0.624(3)$
\\\hline
\multicolumn{5}{l}{} \\
\multicolumn{5}{l}{$\Pep\Pem\to{\Pu\bar\Pd\mu^-\bar{\nu}_\mu+\gamma}$
\quad (separation cuts for $\gamma$ applied)}
\\ \hline
$\sqrt{s}/\mathrm{GeV}$ &
\multicolumn{1}{c|}{$189$} &
\multicolumn{1}{c|}{$500$} &
\multicolumn{1}{c|}{$2000$} &
\multicolumn{1}{c|}{$10000$}\\
\hline
$\sigma($FW$)/\mathrm{fb}$
& $224.0(4)$ & $83.4(3)$ & $6.98(5)$ & $0.457(6)$
\\\hline
$\sigma($RW$)/\mathrm{fb}$
& $224.6(4)$ & $84.2(3)$ & $19.2(1)$ & $368(6)$
\\\hline
$\sigma($CMS$)/\mathrm{fb}$
& $223.9(4)$ & $83.3(3)$ & $6.98(5)$ & $0.460(6)$
\\\hline
\multicolumn{5}{l}{} \\
\multicolumn{5}{l}{$\Pep\Pem\to\nu_\Pe\bar\nu_\Pe\mu^-\bar\nu_\mu\Pu\bar\Pd$
\quad (phase-space cuts applied)}
\\\hline
$\sqrt{s}/\mathrm{GeV}$ & 
\multicolumn{1}{c|}{$500$} & 
\multicolumn{1}{c|}{$800$} & 
\multicolumn{1}{c|}{$2000$} & 
\multicolumn{1}{c|}{$10000$}
\\ \hline \hline
$\sigma($FW$)/\mathrm{fb}$
& $1.633(1)$ & $4.105(4)$ & $11.74(2)$ & $26.38(6)$
\\ \hline
$\sigma($RW$)/\mathrm{fb}$
& $1.640(1)$ & $4.132(4)$ & $12.88(1)$ & $12965(12)$
\\ \hline
$\sigma($CMS$)/\mathrm{fb}$
& $1.633(1)$ & $4.104(3)$ & $11.73(1)$ & $26.39(6)$
\\\hline
\end{tabular}
\caption{Some LO predictions for cross sections 
based on different schemes to treat the EW gauge-boson resonances:
fixed-width (FW), running-width (RW), and complex-mass scheme (CMS).
(Results taken from \citeres{Denner:1999gp,Dittmaier:2002ap}).}
\label{tab:LOXSwidthschemes}
\etab
The FW and RW schemes are compared with the CMS, which fully respects
gauge-invariance requirements and is explained in detail in 
\refse{se:CMS} below. While the results from the FW and CMS
nicely agree, the RW scheme totally fails in the high-energy regime.

\subsection{The issue of gauge invariance}

Gauge invariance implies two crucial properties of amplitudes that are
required for consistency: independence of the gauge-fixing procedure
employed in the quantization of the gauge theory and the validity of
{\it Ward identities}. The first property obviously ensures that
amplitudes do not depend on gauge parameters, the second deserves
somewhat more explanation.  With Ward identities we mean the
Slavnov--Taylor identities discussed in \refse{se:wi} applied to
truncated Green functions with external lines that are either physical
states, would-be Goldstone bosons, or gauge-boson legs that are
contracted with their momenta.  Consider, for instance, an amplitude
${\cal M}$ with an outgoing gauge boson~$V$ of outgoing momentum $p$,
\ie
\beq
{\cal M} = \veps_V^\mu(p)^* \,T^V_\mu(p), \quad V=W^\pm,Z,\FA,
\eeq
where $\veps^\mu(p)^*$ is the polarization vector of $V$.  Then
electromagnetic $\U(1)_{\mathrm{em}}$ and $\SU(2)_{\rw}$ gauge
invariance imply the relations
\beq
p^\mu \,T_\mu^{\FW^\pm}(p) = \pm\MW T^{\phi^\pm}_\mu(p), \qquad
p^\mu \,T^\FZ_\mu(p) = -\ri\MZ T^\chi_\mu(p), \qquad
p^\mu \,T^\FA_\mu(p) = 0,
\label{eq:WIamp}
\eeq
which are equivalent to the relations \refeqf{eq:sti_c+phys} of
\refse{se:wi} or to \refeq{eq:WIET} within the BFM.  Disturbing these
relations can lead to unphysical behaviour of the corresponding
amplitudes.  Violating, for instance, the $\SU(2)_{\rw}$ relations for
$\PW$ and $\PZ$~bosons, has a direct impact on the longitudinal
polarizations owing to the Goldstone-boson equivalence theorem, which
was described in \refse{se:gbet}.  This can easily result in a totally
wrong high-energy behaviour of amplitudes and cross sections.

Note that the Ward identities \eqref{eq:WIamp} are not only relevant
in the case of external gauge bosons~$V$, but also for subamplitudes
in many-particle processes. The most important case arises in the
situation where an external fermion--antifermion current $\bar
u_{f_1}(p_1)\gamma^\mu\omega_\sigma \varv_{f_2}(p_2)$ becomes
asymptotically proportional to its total momentum~$p$, which is
$p=p_1+p_2$ if $p_1$ and $p_2$ are outgoing.  In the subamplitude
where the $f_1\bar f_2$ current is connected to the rest of the
diagram by a $V$~propagator, the Ward identities \eqref{eq:WIamp} are
crucial for the correct behaviour of the amplitude in this kinematic
limit.  For $\PW$ and $\PZ$~bosons this situation occurs, e.g., in the
high-energy limit of $V$~production and subsequent decay, which is
diagrammatically illustrated by
\beq
\raisebox{-1.8em}{\includegraphics[scale=.7]{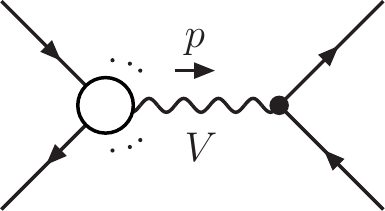}}
\;\sim\;\frac{\mathrm{const.}}{p^2-M_V^2+\ri M_V \Gamma_V}\;p^\mu\, T^V_\mu(p)
\quad \mbox{for } \;p^0\gg M_V,
\eeq
where the dots stand for any other produced particles.
It is the violation of the corresponding Ward identities for W~bosons that
leads to the bad high-energy behaviour of the cross sections shown in
\refta{tab:LOXSwidthschemes} of the previous section in the case of the RW
scheme. More such examples can be found in
\citere{Passarino:2000mt}.
For photons an analogous situation occurs, e.g., in forward scattering of light
fermions such as electrons,
\beq
\raisebox{-1.8em}{\includegraphics[scale=.7]{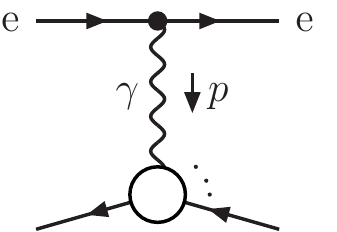}}
\;\sim\;\frac{\mathrm{const.}}{p^2}\;p^\mu\, T^\FA_\mu(p)
\quad \mbox{for } \;|p^2|\,\to\, {\cal O}(\Me^2)\;\ll\; (p^0)^2,
\eeq
where the validity of the Ward identity is required to guarantee a decent
behaviour of the amplitude in the limit of low photon virtuality~$p^2$.
In practice, this situation, for instance, occurs
in single-W production in $\Pep\Pem$
annihilation, as discussed in detail in
\citeres{Kurihara:1994fz,Argyres:1995ym,Beenakker:1996kn,Passarino:2000mt}.

\subsection{Mass and width of unstable particles}
\label{se:mass_width_unstableparticles}

Before we can describe schemes to treat resonances beyond LO, we have
to consider mass renormalization for unstable particles to properly
identify the definition of mass and width that parametrize the
resonances that are analyzed experimentally.  We compare the two most
common renormalization schemes: (i) the OS scheme that closely follows
the mass renormalization for stable particles described in
\refse{se:rcsm} and (ii) the scheme that introduces the complex pole
location $\mu_P^2$ of \refeq{eq:muP} to define a renormalized mass
$\MP$.  Running masses, as \eg defined in the \MSbar{} scheme, are not
suited to parametrize resonances and, thus, not considered in the
following.  We denote the previously used {\it bare mass} $M_{0,P}$ 
of the unstable particle and add the subscript OS to quantities
defined in the OS scheme.  The OS renormalization condition fixes
$\MPOS^2$ as the zero of the real part of the inverse $P$~propagator,
\beq
\MPOS^2 - M_{0,P}^2 + \Re\left\{\Sigma(\MPOS^2)\right\} = 0,
\label{eq:MPOScond}
\eeq
while the condition that identifies $\mu_P^2$ with the complex location
of the propagator pole reads
\beq
\mu_P^2 - M_{0,P}^2 + \Sigma(\mu_P^2) = 0,
\label{eq:muPcond}
\eeq
where $\Sigma(p^2)$ denotes the unrenormalized self-energy of~$P$.
The determination of OS renormalization constants works as for stable
particles, which is described in \refse{se453eforc};
the procedure for complex masses is described in \refse{se:CMS} below.
Here we are concerned with the relation between mass and width parameters
in the two schemes.
Eliminating the bare mass from the propagator factors in either scheme,
the corresponding leading resonance behaviour of $P_P(p^2)$ is given by
\begin{align}
P_{P,\,\OS}(p^2) &{}=
\frac{1}{p^2-\MPOS^2 + \Sigma(p^2) - \Re\{\Sigma(\MPOS^2)\}}
\;\asymp{p^2\to\MPOS^2}\;
\frac{1}{(p^2-\MPOS^2)[1+\Re\{\Sigma'(\MPOS^2)\}]
+ \ri\Im\{\Sigma(p^2)\} + \ldots},
\label{eq:PropOS}
\\
P_{P}(p^2) &{}=
\frac{1}{p^2-\mu_P^2+\Sigma(p^2)-\Sigma(\mu_P^2)}
\;\asymp{p^2\to\mu_P^2}\;
\frac{1}{(p^2-\mu_P^2)[1+\Sigma'(\mu_P^2)] + \ldots},
\label{eq:Proppole}
\end{align}
where the expansions in the denominators are correct up to ${\cal
  O}\left((p^2-\MPOS^2)^2\right)$ and ${\cal
  O}\left((p^2-\mu_P^2)^2\right)$, respectively.  Comparing
\refeq{eq:PropOS} near the resonance pole $p^2\sim\MPOS^2$ with the
standard form $1/(p^2-\MP^2+\ri\MP\GP)$, we can associate
\beq
\GPOS =
\frac{\Im\{\Sigma(\MPOS^2)\}}{\MPOS[1+\Re\{\Sigma'(\MPOS^2)\}]}
\label{eq:GPOS}
\eeq
with the width in the OS scheme.  A similar comparison of
\refeq{eq:Proppole} yields the {\it pole mass} $\MP$ and the
{\it pole width} $\GP$ in the scheme with
$\mu_P$ from \refeq{eq:muP} as renormalized parameter.
Since the pole location $\mu_P^2$ is an intrinsic property of the
$S$-matrix, the definition of $\MP$ and $\GP$ is 
gauge independent~\cite{Gambino:1999ai,Grassi:2001bz} in the sense
that the complex mass renormalization constant and the parametrization
of $S$-matrix elements in terms of $\mu_P^2$ are gauge independent.
On the other hand, the OS scheme involves gauge dependences starting from the
two-loop level~\cite{Sirlin:1991fd}.

For $\PW$ and $\PZ$~bosons the form of the propagator in the OS scheme
\eqref{eq:PropOS} is compatible with the form \refeq{eq:PropP} with
the running width \eqref{eq:GPp2} for massless decay products.  
Historically, the $\PW$ and $\PZ$ masses and widths were
experimentally determined at LEP, Tevatron, and the LHC in the OS
scheme, even though the (gauge-independent) pole definitions are
theoretically preferred.  Fortunately, the conversion from the OS to
the pole definitions can be easily obtained from
\refeqs{eq:muP}, \refeqf{eq:MPOScond}, \refeqf{eq:muPcond}, and
\refeqf{eq:GPOS} upon using the form \refeq{eq:GPp2} via an
order-by-order expansion of $\Sigma$, leading
to~\cite{Bardin:1988xt,Beenakker:1996kn} 
\begin{equation}
\MV = \frac{\MVOS}{\sqrt{1+\GVOS^2/\MVOS^2}},
\qquad
\GV = \frac{\GVOS}{\sqrt{1+\GVOS^2/\MVOS^2}}.
\label{eq:mvconversion}
\end{equation}
This implies
\begin{equation}
\MWOS-\MW\approx 27\MeV,
\qquad
\MZOS-\MZ\approx 34\MeV, 
\end{equation}
\ie the scheme differences are much larger than the corresponding 
experimental uncertainties on the $\PW$ and $\PZ$ masses, which are 
currently~\cite{Tanabashi:2018oca}
$12\MeV$ and $2.1\MeV$, respectively.

\subsection{Pole scheme and pole approximation}
\label{se:pole_scheme}

\subsubsection{Pole \texorpdfstring{scheme---general}{scheme-general} 
idea and subtleties}

The {\it pole scheme} is a prescription for 
a construction of gauge-independent cross sections including resonances,
exploiting the fact that both the location of the
$P$~propagator pole and its residue in amplitudes are gauge-independent
quantities.
The idea~\cite{Stuart:1991xk,Aeppli:1993rs}
is to first isolate the residue $R$ of each
resonance in the considered amplitude ${\cal M}$
and subsequently to introduce a finite decay width $\GP$
only in the gauge-independent resonant part, 
\begin{align}
{\cal M} &{}=
\frac{R(p^2)}{p^2-\MP^2} + N(p^2)
\;=\;
\frac{R(\MP^2)}{p^2-\MP^2}
+ \frac{R(p^2)-R(\MP^2)}{p^2-\MP^2}
+N(p^2)
\rightarrow\,
\underbrace{\frac{\tilde R(\mu_P^2)}{p^2-\mu_P^2}}_{\mbox{\scriptsize resonant}}
\;+\; \underbrace{\frac{R(p^2)-R(\MP^2)}{p^2-\MP^2}+\tilde N(p^2)}_{\mbox{\scriptsize non-resonant}}.
\label{eq:poleschemegeneric}
\end{align}
Here, $R(p^2)$ summarizes all contributions developing resonant parts
in the $P$~propagator with virtuality $p^2$, and $N(p^2)$ the remaining
non-resonant contributions.
The replacement, thus, dresses the propagator denominator 
$p^2-\MP^2$ with the contribution $\ri\MP\GP$ from the finite decay width,
\ie partially takes into account Dyson summation of propagator corrections.
Note, however, that in the resonance region $p^2\sim\MP^2$, the width
contribution $\ri\MP\GP$ becomes part of the leading term, although
$\GP$ accounts for a correction according to the counting in the
off-shell region. Thus, if radiative corrections are to be included,
the $\GP$ value inserted here has to include corrections one order higher
than the rest of the matrix-element calculation.
Moreover, double-counting has to be carefully avoided.
The tilde on the symbols $\tilde R$ and $\tilde N$ indicates that
there are some modifications of those terms in the course of replacing
$\MP^2\to\mu_P^2$, in order to avoid double counting.
If done carefully the prescription~\refeqf{eq:poleschemegeneric}
respects gauge invariance and can be used to make uniform
predictions in resonant and non-resonant phase-space regions.

Although the general idea seems quite simple and appealing,
we emphasize that \refeq{eq:poleschemegeneric} is a generic formula that
involves several subtleties in practice:
\begin{myitemize}
\item
There is some freedom in the actual implementation of the scheme,
because the resonant part of an amplitude is not uniquely determined by
the propagator structure alone, but depends on a specific phase-space
parameterization and in most cases also on the separation of
polarization-dependent parts (spinors, polarization vectors).
\item
Taking the pole prescription literally, the scattering amplitude
on the resonance pole (\ie with $\mu_P^2$ being complex) involves
matrix elements with complex kinematical variables, which is a subtle
issue~\cite{Passarino:2010qk}.%
\footnote{This {\it complex pole scheme} was, e.g., applied to
Higgs production via gluon fusion at the LHC in \citere{Goria:2011wa},
with particular emphasis on a heavy, \ie very broad, Higgs boson.}
For narrow resonances such as the weak gauge bosons
$V=\PW,\PZ$, or the SM Higgs boson, the complex kinematics can be avoided by suitable
expansions in $\Gamma_V/M_V$. 
\item
\begin{sloppypar}
Taking into account higher-order corrections, the resonance does not
appear as a pure pole structure in the amplitude, but rather as branch
point in the complex $p^2$ plane. The exchange of massless particles
such as photons or gluons with the resonance or between production and
decay subprocesses leads to correction terms involving factors like
\mbox{$\ln(p^2-\MP^2)$}. For such terms the concept of taking the residue,
$R(p^2)\to R(\mu_P^2)$, is ill-defined and has to be generalized. 
Since all non-analytic terms in $R(p^2)$ can be uniquely isolated, 
a possible generalization consists in the substitution 
$\ln(p^2-\MP^2)\to\ln(p^2-\mu_P^2)$ while we set $p^2\to \mu_P^2$ in all
regular terms.
\end{sloppypar}
\item
In the case of multiple resonances further generalizations are necessary.
Massless particle (\ie photon or gluon) exchange that involves more than one
resonance leads to deviations from the pole structure $K_P(p,\MP)^{-1}=(p^2-\MP^2)^{-1}$ 
for each resonance.
For the case of two resonances $P_1$ and $P_2$, for instance, in addition to
the pattern $K_{P_1}(p_1,M_{P_1})^{-1} K_{P_2}(p_2,M_{P_2})^{-1}$ also
doubly-resonant terms of the form
$[a_{11} K_{P_1}^2 + a_{12} K_{P_1}K_{P_2} + a_{22} K_{P_2}^2]^{-1}$ arise,
where $a_{ij}$ depend on further kinematical 
invariants~\cite{Melnikov:1995fx,Beenakker:1997ir,Denner:1997ia}.
Again, such terms should be uniquely isolated (in order to avoid gauge
dependences) and subsequently regularized by 
$K_{P_i}(p_i,M_{P_i})\to K_{P_i}(p_i,\mu_{P_i})$.
\item
If the pole scheme is applied to higher-order corrections, it should be kept
in mind that both virtual and real corrections have to be handled,
including a consistent matching of IR singularities.
On the side of the real corrections, this means that real-photon (or
real-gluon) emission might appear before, during, or after the resonant production of 
an unstable particle~$P$.
If $P$ is electrically (or colour neutral),
there is a gauge-invariant separation of radiation effects into 
emission from the production or decay processes in matrix elements, so
that the procedure outlined in \refeq{eq:poleschemegeneric} is applicable,
though being quite cumbersome in complicated cases.
If $P$ is charged, however, the problem of overlapping resonances occurs.
In this case, radiation off the resonating particle leads to propagator
products $K_P(p+k,\MP)^{-1} K_P(p,\MP)^{-1}$ with $k$ denoting the
momentum of the radiated photon (or gluon).
In principle, the partial fractioning 
\begin{align}
\frac{1}{K_P(p+k,\MP) K_P(p,\MP)} &{}\;=\;
\frac{1}{(2pk)K_P(p,\MP)} \;-\; \frac{1}{(2pk)K_P(p+k,\MP)}
\label{eq:partialfrac}
\\*[.3em]
\raisebox{-.8em}{\includegraphics[scale=.6]{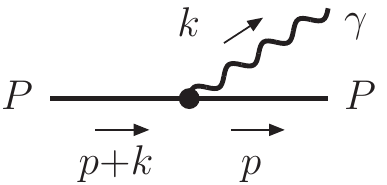}} 
\hspace*{2em} & \hspace*{2em} 
\raisebox{-.8em}{\includegraphics[scale=.6]{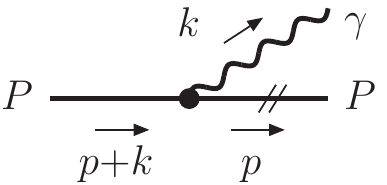}}
\hspace*{2em} 
\raisebox{-.8em}{\includegraphics[scale=.6]{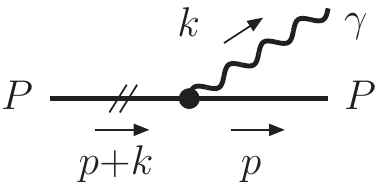}}
\nn
\end{align}
offers a separation into emission after or before the resonance,
where the double slash on a propagator line indicates
which momentum is set on its mass shell in the rest of the diagram 
(but not on the slashed line itself).
However, the $1/(2pk)$ factors on the r.h.s.\ induce spurious soft singularities in this
decomposition, because this factor mimics photon radiation off an 
OS state.
\end{myitemize}
To our knowledge, a full pole-scheme calculation 
including all non-resonant terms for multiple resonances
or for processes with charged resonances
has not been carried out in the literature yet, which is certainly
due to the subtleties and problems explained above.

Nevertheless, for cases where the pole scheme can be applied, it 
certainly offers various virtues.
For instance, it provides a well-defined separation between resonant
and non-resonant parts of a cross section, \ie in some sense
a definition of {\it signal} and {\it background}
for the production of a resonance. This scheme is, thus, well suited
for a parametrization of a resonance by so-called
{\it pseudo-observables}, such as total and partial decay widths,
peak cross sections, effective couplings, etc.
Moreover,
the consistent separation of signal and background contributions
is an ideal starting point to further improve the description of the
signal by higher-order corrections, a point that is particularly important
if the signal dominates over the background and, thus, deserves higher precision.

\subsubsection{Weak corrections to \texorpdfstring{Drell--Yan-like}
{Drell-Yan-like} Z-boson production in the pole scheme}
\label{se:DY-weak-PS}

As an example for the successful implementation of the pole scheme, we
consider the calculation of NLO weak corrections to \PZ-boson
production at hadron colliders, as presented in
\citere{Dittmaier:2009cr}.  Photonic corrections, which comprise all
contributions from virtual exchange and real emission of photons
between or off the external fermions, are separated in a
gauge-invariant way, following the procedure outlined in
\refse{se:split_ew_qed}.  At NLO the weak corrections, thus, comprise
all remaining virtual one-loop diagrams with W/Z- or Higgs-boson
exchange or closed fermion-loop corrections. The weak corrections do
not involve any real radiation effects, nor IR divergences.

In detail, we consider the partonic processes
$q(p_1)\bar q(p_2)\to\Pl^-(k_1)\Pl^+(k_2)$ with massless quarks $q$ and
leptons $\Pl$, the momenta of which are indicated in parentheses.
The corresponding LO diagrams and the generic
one-loop diagrams are shown in \reffi{fig:genericDYdiags}.
\bfi
  \centering
     \includegraphics[scale=1]{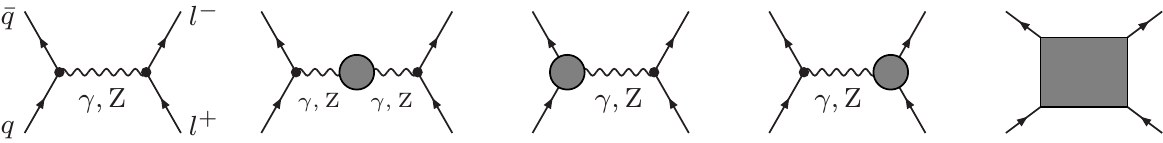}
\vspace*{-.5em}
  \caption{LO and generic one-loop diagrams for the process 
$q\bar q\to\Pl^-\Pl^+$, where the grey blobs and the grey box
stand for self-energy, vertex, and box corrections.}
\label{fig:genericDYdiags}
\efi
The LO matrix elements can be written in the generic way
\begin{equation}
\label{eq:DYm0}
{\cal M}_{\qqb}^{\mathrm{LO},\sigma\tau} =
-\frac{e^2}{\hat s} \sum_{V=\FA,\FZ} \gqqV^\sigma\, \gllV^\tau\,
\chi_V(\hat s)\,  \mathcal{A}^{\sigma \tau} 
\;=\; f_{\qqb}^{\mathrm{LO},\sigma\tau} \, \mathcal{A}^{\sigma \tau},
\end{equation}
where $\sigma=\pm$ and $\tau=\pm$ denote the chiralities on the quark
and lepton lines, respectively, and $\gffV^\si$ are the chiral
couplings of the fermions $f$ to the vector bosons $V$,
\beq
\gffA^\pm = -\Qf, \qquad
\gffZ^+ = -\frac{\sw}{\cw}\Qf, \qquad \gffZ^- =
\frac{\If-\sw^2\Qf}{\sw\cw}.
\eeq

The quantities
\begin{equation}
\label{eq:DYsme}
\mathcal{A}^{\sigma \tau}=
[ \bar{\varv}_q(p_2) \, \gamma^\mu \omega_\sigma\, u_q(p_1) ] \,
[ \bar{u}_l(k_1) \, \gamma_\mu \omega_\tau \, \varv_l(k_2) ],
\end{equation}
are {\it standard matrix elements} (with an obvious notation for Dirac spinors
$u_{q,l}, \varv_{q,l}$)
containing the spin information of the fermions, which are functions of
the usual Mandelstam variables
\begin{equation}
  \hat s = (p_1+p_2)^2, \qquad
  \hat t = (p_1-k_1)^2, \qquad
  \hat u = (p_1-k_2)^2. 
\end{equation}
The functions 
\begin{equation}
\chi_\FA(\hat s) = 1, \qquad
\chi_\FZ(\hat{s}) = \frac{\hat s}{\hat{s}-\mu_\PZ^2},
\end{equation}
describe the propagation of $V$ and contain the Z-boson resonance,
which is regularized by writing the complex mass $\mu_\PZ$ instead
of the real mass $\MZ$.

The NLO weak corrections receive contributions from self-energy,
vertex, and (irreducible) box corrections,
which are generically illustrated  in \reffi{fig:genericDYdiags}.
The weak correction factor $f^{\virt,\,\si\tau}_{\qqb,\weak}$
in the weak one-loop matrix element
\beq
{\cal M}_{\,\qqb}^{\weak,\,\sigma \tau} =
f^{\virt,\,\sigma \tau}_{\qqb,\weak}\, \mathcal{A}^{\sigma \tau},
\qquad
f^{\virt,\,\si\tau}_{\qqb,\weak} =
  f^{\mathrm{self},\,\si\tau}_{\qqb,\weak}(\hat s) +
  f^{\mathrm{vert},\,\si\tau}_{\qqb,\weak}(\hat s) +
  f^{\mathrm{box},\,\si\tau}_{\qqb,\weak}(\hat s,\hat t),
\label{eq:DYMweak}
\eeq
is decomposed into respective contributions accordingly.
The self-energy and vertex correction factors contain resonant
contributions, but depend only on $\hat s$; 
the box contribution depends both on $\hat s$ and $\hat t$, but does not
contain resonant parts. 
This means that only the former need modifications for the resonance
treatment in the pole scheme.
{}For the vertex corrections this procedure is very simple.
The contributions involving Z-boson exchange,
$f^{\mathrm{vert},\PZ,\,\si\tau}_{\qqb}$,
are modified as follows,
\begin{align}
f^{\mathrm{vert},\PZ,\,\si\tau}_{\qqb,\weak}(\hat s) &{}\,=\,
- e^2 \, \frac{\gqqZ^\sigma \gllZ^\tau}{\hat s - \MZ^2}
\left[F^\sigma_{\ren,qqZ,\weak}(\hat s) +F^\tau_{\ren,llZ,\weak}(\hat s)  \right]
\nl
&{}\,\to\, - e^2 \, \gqqZ^\sigma \gllZ^\tau
\left[ \frac{F^\sigma_{\ren,qqZ,\weak}(\MZ^2)
    +F^\tau_{\ren,llZ,\weak}(\MZ^2)}{\hat s - \mu_\PZ^2} \right.
\nl
&\qquad  {}
+ \left.\frac{F^\sigma_{\ren,qqZ,\weak}(\hat s)
    -F^\sigma_{\ren,qqZ,\weak}(\MZ^2)
  + F^\tau_{\ren,llZ,\weak}(\hat s)
    -F^\tau_{\ren,llZ,\weak}(\MZ^2)}{\hat s - \MZ^2}
\right] ,
\hspace{1em}
\end{align}
while the non-resonant contributions involving photon exchange
are kept unchanged. Off resonance the introduction of the finite Z-decay
width $\GZ$ in the denominator of the vertex corrections changes
the amplitude only in ${\cal O}(\alpha^2)$ relative to LO, \ie the
effect is beyond NLO.
The explicit form of the renormalized one-loop vertex form factors 
$F^\sigma_{\ren,ffV,\weak}(\hat s)$ can be found in \citere{Dittmaier:2009cr}.
The treatment of the self-energy corrections is somewhat more
involved, since it involves more propagator factors
$(\hat s-\MZ^2)^{-1}$ and care has to be taken that no double-counting is 
introduced in the modification of the sum of LO and NLO matrix elements,
\begin{align}
\label{eq:PSself}
{f_{\qqb}^{\mathrm{LO},\,\sigma \tau}
+f^{\mathrm{self},\,\sigma \tau}_{\qqb,\weak}}
&{}\,=\,
-e^2 \left\{ \frac{ \Qq \, Q_\Pl}{\hat{s}}
\left[ 1-\frac{\Sigma^{\FA\FA}_{\ren,\rT}(\hat s)}{\hat s}\right]
+ \frac{\gqqZ^\sigma\, \gllZ^\tau}{\hat{s}-M_{\PZ}^2}
\left[ 1-\frac{\Sigma^{\FZ\FZ}_{\ren,\rT}(\hat s)}{\hat s - \MZ^2}\right]
+ \frac{\Ql\, \gqqZ^\sigma +\Qq \,\gllZ^\tau}{\hat s}
\frac{\Sigma^{\FA\FZ}_{\ren,\rT}(\hat s)}{\hat s - \MZ^2}
\right\}
\nl
&{}\,=\, -e^2 \left\{ \frac{ \Qq \, Q_\Pl}{\hat{s}}
\left[ 1-\frac{\Sigma^{\FA\FA}_{\ren,\rT}(\hat s)}{\hat s}\right]
+\frac{\gqqZ^\sigma\, \gllZ^\tau}{\hat{s}-M_{\PZ}^2}
\left[1 - \frac{\Sigma^{\FZ\FZ}_{\ren,\rT}(\MZ^2)}{\hat s - \MZ^2} 
\right.  \right.
\nl
&\qquad\quad {}
- \left.\Sigma^{'\FZ\FZ}_{\ren,\rT}(\MZ^2) -\frac{\Sigma^{\FZ\FZ}_{\ren,\rT}(\hat s) -\Sigma^{\FZ\FZ}_{\ren,\rT}(\MZ^2) -(\hat s-\MZ^2)\Sigma^{'\FZ\FZ}_{\ren,\rT}(\MZ^2)}{\hat s - \MZ^2}
\right]
\nl
&\quad\qquad {}
+ \left.\frac{\Ql\, \gqqZ^\sigma +\Qq \,\gllZ^\tau}{\hat s}
\left[\frac{\Sigma^{\FA\FZ}_{\ren,\rT}(\MZ^2)}{\hat s - \MZ^2}
 + \frac{\Sigma^{\FA\FZ}_{\ren,\rT}(\hat s)-\Sigma^{\FA\FZ}_{\ren,\rT}(\MZ^2)}{\hat s - \MZ^2}  \right]\right\}
\nl
&{}\,\to\, - e^2 \left\{ \frac{ \Qq \, Q_\Pl}{\hat{s}}
\left[ 1-\frac{\Sigma^{\FA\FA}_{\ren,\rT}(\hat s)}{\hat s}\right] + \gqqZ^\sigma\, \gllZ^\tau \left[\frac{1-\Sigma^{'\FZ\FZ}_{\ren,\rT}(\MZ^2)}{\hat s-\mu_\PZ^2} 
\right.  \right.
\nl
&\qquad\quad {}
- \left.
\frac{\Sigma^{\FZ\FZ}_{\ren,\rT}(\hat s) -\Sigma^{\FZ\FZ}_{\ren,\rT}(\MZ^2) -(\hat s-\MZ^2)\Sigma^{'\FZ\FZ}_{\ren,\rT}(\MZ^2)}{(\hat s-\MZ^2)^2} \right]
\nn\\
&\qquad {}
+ \left(\Ql\, \gqqZ^\sigma +\Qq \,\gllZ^\tau \right) \left[
\frac{1}{\hat s - \mu_\PZ^2} \frac{\Sigma^{\FA\FZ}_{\ren,\rT}(\MZ^2)}{\MZ^2}  
+ \left.
\frac{1}{\hat s - \MZ^2} \Biggl(\frac{\Sigma^{\FA\FZ}_{\ren,\rT}(\hat s)}{\hat s} - \frac{\Sigma^{\FA\FZ}_{\ren,\rT}(\MZ^2)}{\MZ^2} \Biggr)  \right]\right\}.
\end{align}
Here we have
used the fact that in the OS renormalization scheme
the renormalized \PZ-boson self-energy fulfils
$\Re\Sigma^{\FZ\FZ}_{\ren,\rT}(\MZ^2) = 0$ and that the
resummed terms account for some imaginary parts via the optical
theorem  \refeqf{eq:optical_theorem},
$\Im\Sigma^{\FZ\FZ}_{\ren,\rT}(\MZ^2) = \MZ\GZ$, which holds
in \Oa.

Figure~\ref{fig:DYwidthschemes} shows some numerical results of
\citere{Dittmaier:2009cr} on the application of the pole scheme to the
LO and NLO contributions to the partonic scattering channels of the 
Drell--Yan process in the broad neighbourhood of the Z-boson resonance.
\bfi
\centering{%
{\includegraphics[viewport=80 409 423 651,clip,scale=.674]{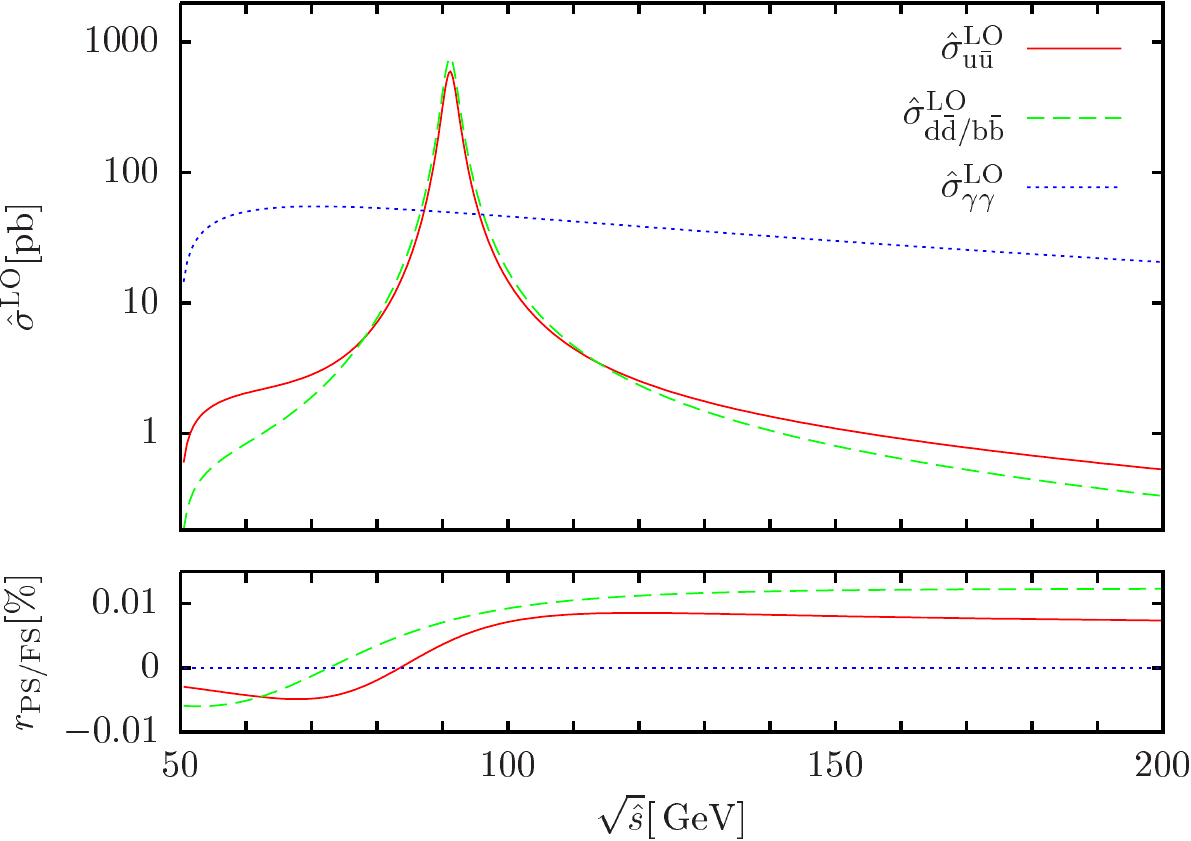}}
\hspace*{2em}
{\includegraphics[viewport=84 379 425 651,clip,scale=.6  ]{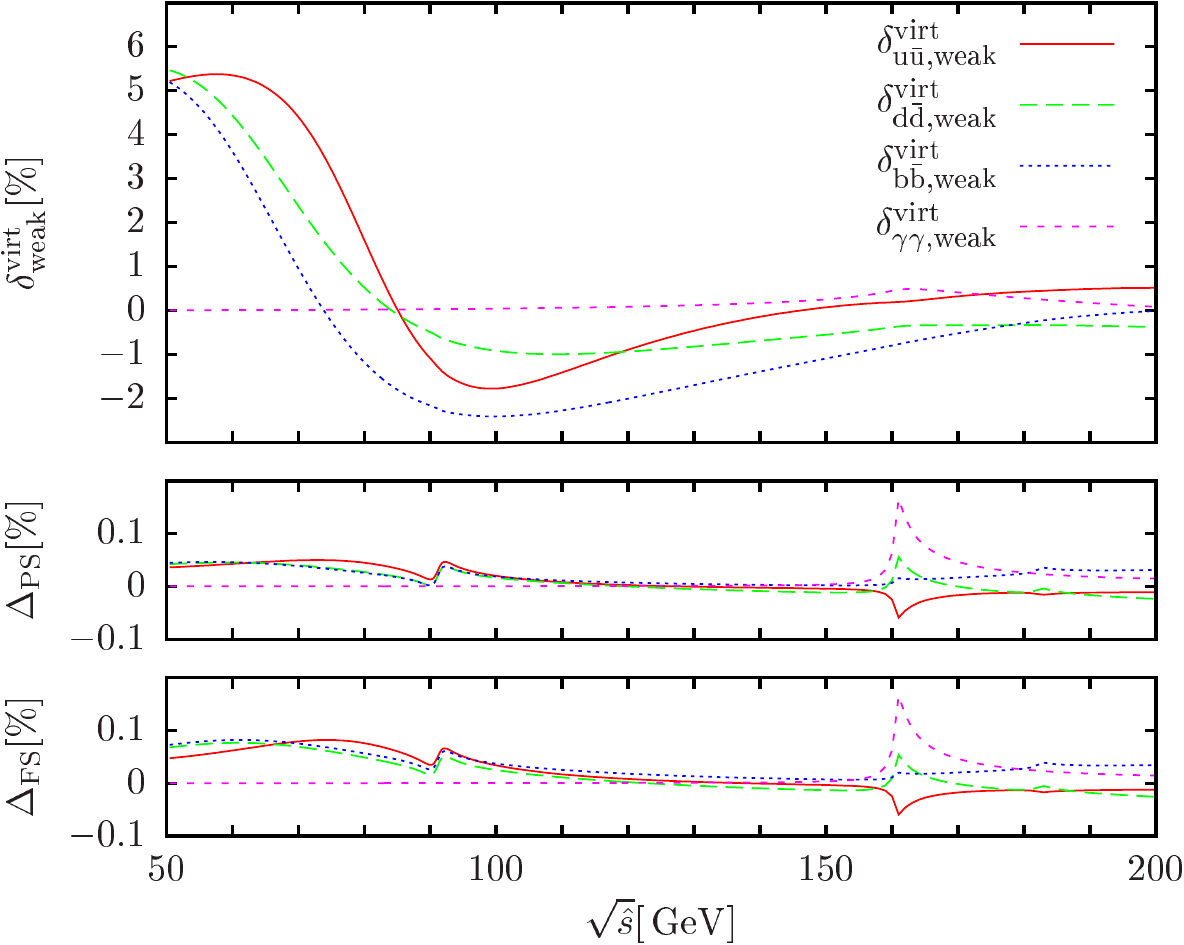}}
}
\caption{LO cross sections $\hat\sigma^{\LO}$
(top left) for the partonic processes 
$q\bar q/\gamma\gamma\to\Pl^-\Pl^+$ as function of the partonic CM energy 
$\sqrt{\hat s}$ and corresponding weak corrections $\delta^\mathrm{virt}_{\weak}$ 
(top right), indicating the partonic channels as subscripts.
The lower panels show the relative deviations
$r_{\mathrm{X}}=\hat \sigma^{\LO}|_{\mathrm{X}}\; /\; \hat
\sigma^{\LO}\big|_{\mathrm{CMS}} - 1$ of the LO cross sections
and the differences $\Delta_{\mathrm{X}}
= \delta^{\mathrm{virt}}_{\weak}\big|_{\mathrm{X}}
           -\delta^{\mathrm{virt}}_{\weak}\big|_{\mathrm{CMS}}$
between scheme X and the CMS for treating the Z~resonance, where
PS stands for pole scheme and FS for factorization scheme.
(Plots taken from \citere{Dittmaier:2009cr}.)}
\label{fig:DYwidthschemes}
\efi
The pole scheme (PS) results are compared to two further schemes that were
applied to uniformly describe the resonance processes in resonant and
non-resonant phase-space regions: the CMS
described in \refse{se:CMS} and a {\it factorization scheme (FS)} 
briefly sketched in \refse{se:unstableparts-other}.
The central result from this comparison is the excellent level of
agreement between NLO results from the various schemes, leaving
uncertainties from the resonance treatment that are well below the
impact of missing corrections beyond the NLO level.

\subsubsection{Pole approximation}

A {\it pole approximation (PA)}---in contrast to a full pole-scheme
calculation---results from a resonant amplitude defined in the pole
scheme upon neglecting non-resonant parts.  The general concept as
well as specific applications have been introduced and described by
several authors in different contexts (see, e.g.,
\citeres{Stuart:1991xk,Aeppli:1993rs,Fadin:1993dz,Melnikov:1995fx,Beenakker:1997ir,Denner:1997ia,Beenakker:1998gr,Denner:2000bj,Dittmaier:2014qza}).
Typically, the PA is applied to higher-order corrections where full
off-shell calculations are too difficult to carry out or not needed in
view of the aimed precision tag.  In this context, we recall the
interplay between different perturbative orders on and off resonance:
To achieve NLO precision in the resonance region of a cross section,
both off-shell and finite-width effects have to be included at least
in LO along with the NLO corrections on resonance, and likewise in
higher orders.  In the following, we describe the structure of NLO
calculations in PA, which are combined with complete LO calculations
including off-shell effects.  Such predictions deliver NLO accuracy
only in the neighbourhood of resonances, but only LO precision in
off-shell tails.  The corresponding inclusive cross sections
integrated over the resonances, however, possess NLO accuracy.  Let us
make this counting more quantitative for the case of the resonant
production of a weakly decaying particle~$P$, such as a W or a
Z~boson.  In this case, LO off-shell effects are of ${\cal
  O}(\GP/\MP)\sim{\cal O}(\alpha)$ relative to the leading resonant
contribution and, thus, of the same generic order as NLO EW
corrections on resonance.  In PA at NLO, the largest
missing EW contributions are the off-shell NLO EW effects which are of
the typical size $\alpha/(\pi\sw^2)\times\GP/\MP$ times some possible
enhancement factor.  Taking the latter at most to $\sim10$, still
offers an uncertainty estimate of $\lsim0.5\%$ for observables that
are dominated by resonant contributions, which is good enough for most
phenomenological purposes for a wide class of processes.  We will
further elaborate on the applicability, reliability, and limitations
of such NLO calculations in PA below.

In the simple generic form \refeqf{eq:poleschemegeneric} of a
resonance amplitude, the PA is defined by taking only the resonant
part $\tilde R(\mu_P^2)/(p^2-\mu_P^2)$ into account.  However, as
mentioned already above, whenever the resonance is involved in
(virtual or real) photon or gluon exchange, or if the resonance is
``bridged over'' by a photon or a gluon, this form is oversimplified.
In those cases, the resonance singularity is a branch point rather
than a pole, and resonant terms in addition to the ones proportional
to $\tilde R(\mu_P^2)$ arise.  More precisely, the $\tilde R(\mu_P^2)$
terms define the so-called {\it factorizable corrections}, which
include all resonant contributions that can be written as a product of
production and decay matrix elements with the intermediate particle
$P$ set on shell multiplied with the off-shell propagator
$(p^2-\mu_P^2)^{-1}$. The remaining resonant contributions define the
so-called {\it non-factorizable corrections}, where the
non-factorizability refers to the resonance structure which deviates
from the $(p^2-\mu_P^2)^{-1}$ factor.  The various types of
corrections appearing in an NLO calculation in PA for a generic
$2{\to}2$-particle process with a vector resonance~$V$ are illustrated
in \reffi{fig:genericPAdiags}.
\bfi
  \centering
  \begin{subfigure}[c]{0.32\linewidth}
    \centering
    \includegraphics[scale=.75]{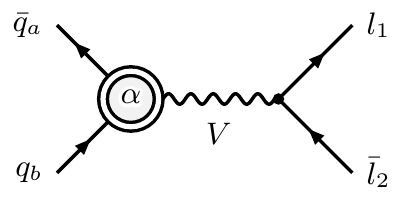}
    \vspace*{-1em}\subcaption{a}{}
  \end{subfigure}
  \begin{subfigure}[c]{0.32\linewidth}
    \centering
     \includegraphics[scale=.75]{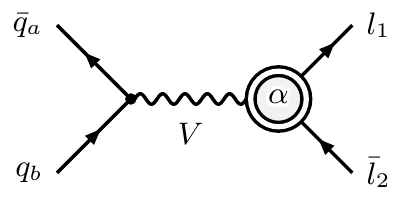}
     \vspace*{-1em}\subcaption{b}{}
  \end{subfigure}
  \begin{subfigure}[c]{0.32\linewidth}
    \centering
    \includegraphics[scale=.75]{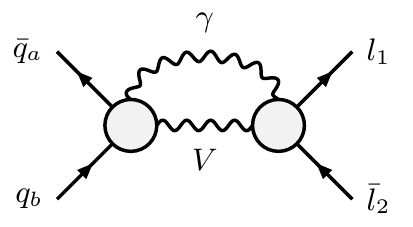}
     \vspace*{-1em}\subcaption{c}{}
  \end{subfigure}
\\[.5em]
  \centering
  \begin{subfigure}[c]{0.29\linewidth}
    \centering
    $\left\lvert\vphantom{\rule{1cm}{1cm}}
    \Vcenter{\includegraphics[scale=0.69]{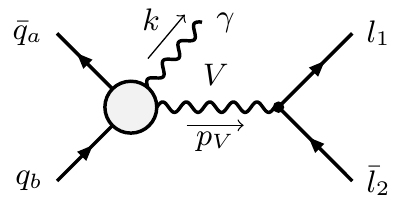}}
    \right\rvert^2$
    \subcaption{d}{}
  \end{subfigure}
  \begin{subfigure}[c]{0.29\linewidth}
    \centering
    $\left\lvert\vphantom{\rule{1cm}{1cm}}
    \Vcenter{\includegraphics[scale=0.69]{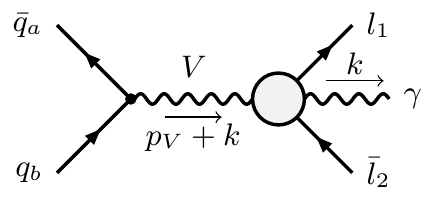}}
    \right\rvert^2$
    \subcaption{e}{}
  \end{subfigure}
  \begin{subfigure}[c]{0.4\linewidth}
    \centering
    \includegraphics[scale=0.69]{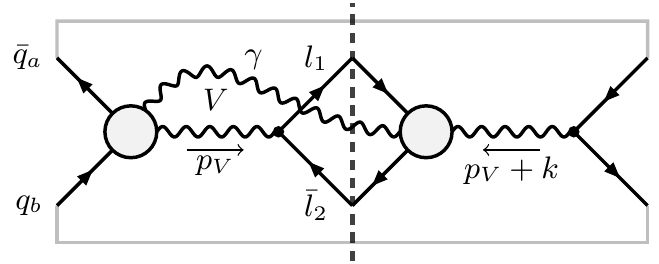}
    \subcaption{f}{}
  \end{subfigure}
  \caption{Generic diagrams for NLO EW corrections in PA for the
    resonance process $\bar q_a q_b\to V\to\Pl_1\bar\Pl_2$, 
    with $q$ and $\Pl$ denoting generic quarks and leptons: (a/b)
    virtual factorizable corrections to production/decay, (c) virtual
    non-factorizable corrections, (d/e) real factorizable corrections
    to production/decay, and (f) real non-factorizable corrections.
    The ``squared diagrams'' (d) and (e) and the
    ``interference diagram'' (f) are direct contributions to the
    squared amplitude $|{\cal M}|^2$.  The empty blobs stand for all
    relevant tree structures, the ones with ``$\alpha$'' for one-loop
    corrections of ${\cal O}(\alpha)$.  (Adapted from
    \citere{Dittmaier:2014qza}.)}
\label{fig:genericPAdiags}
\efi
We describe the characteristic features of factorizable and
non-factorizable corrections in the following.

The calculation of factorizable corrections can be done widely
independent for the production and decay subprocesses for specific
resonance patterns.  Some care has to be taken in the following
points:
\begin{myitemize}
\item
To properly take into account spin correlations between different subprocesses, 
the same polarization basis has to be taken
both in production and decay of each resonance.
\item The evaluation of the subamplitudes in the factorizable
  corrections requires kinematical variables or momenta that
  consistently respect the on-shellness of the resonance, in order to
  be theoretically consistent (gauge independence, etc.).  On the
  other hand, the factorizable contributions should be evaluated on
  the original off-shell phase space. This problem is solved by
  introducing an 
  {\it on-shell projection} of off-shell momenta,
  which deforms external momenta in such a way that the intermediate
  resonant particle~$P$ becomes on shell.  In the vicinity of the
  resonance, momenta are only changed by terms of ${\cal O}(\GP/\MP)$
  relative to their original momenta, \ie NLO corrections of ${\cal
    O}(\alpha)$ are formally changed at the NNLO level, so that NLO
  precision in the resonance region is maintained.  Comparing results
  obtained with different 
  OS projections, can, thus, help to assess
  theoretical uncertainties in NLO predictions in PA.
\item The factorizable corrections have the usual IR structure of OS
  production and decay of the resonating particles. Note that this IR
  structure in general is not the same as for the full off-shell
  amplitude, because additional IR divergences connected with a
  resonating charged particle are introduced by setting its momentum
  on shell.
\end{myitemize}

By definition, the non-factorizable corrections are the resonant 
remnants of the difference between the full off-shell amplitude and its
factorizable parts, which result from going on shell with the kinematics
of the resonating particle(s) in the production and decay subamplitudes.
This difference can be inspected graph by graph:
\begin{myitemize}
\item Loop diagrams with explicit resonance factors
  $(p^2-\mu_P^2)^{-1}$ originating from tree-like resonance
  propagators obviously contribute to the factorizable corrections.
  They also contribute to the non-factorizable corrections only if the
  process of setting the resonance momentum on shell in a subamplitude
  leads to an IR divergence (\cf
  \citeres{Denner:1997ia,Denner:2000bj}). This only happens in the
  case where a photon or gluon of a loop is attached to the resonance.
  In those cases, the
  non-factorizable contribution entirely stems from soft photon or
  gluon exchange with the resonance(s).  Formally the diagrams that
  contain both factorizable and non-factorizable parts can be split into the
  respective contributions via the partial fractioning given in
  \refeq{eq:partialfrac}.  This is illustrated in
  \reffi{fig:nonfactsplitDY}, where photon exchange between an
  initial-state fermion and a charged resonance ($V=\PW$) leads to
  factoriable and non-factorizable contributions which contribute to
  the corrections of types~(a) and (c) in \reffi{fig:genericPAdiags},
  respectively.
\bfi
  \centering
     \includegraphics[scale=.7]{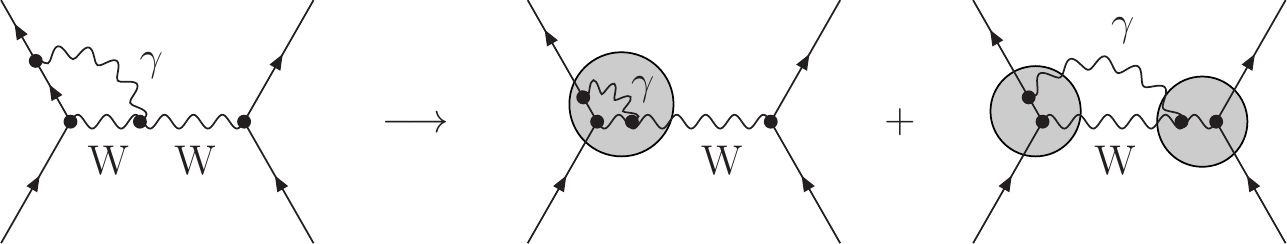}
\vspace*{-.5em}
  \caption{An example for the splitting of a diagram with photon exchange of
a resonance (left) into factorizable (middle) and non-factorizable (right) contributions.}
\label{fig:nonfactsplitDY}
\efi
\item
Loop diagrams with no tree-like resonance line, or more generally with less
tree-like resonance lines than the leading resonance pattern,
obviously do not possess factorizable parts, but still might deliver
non-factorizable resonant contributions.
Loop integrals can only deliver resonance enhancements if the loop
contains a propagator of the potentially resonating particle and if
the loop momentum corresponding to the resonance at the same time 
leads to an IR divergence. This is exactly the case if a soft photon or gluon
bridges over the resonance (or over several resonances),
\ie the corresponding diagrams lead only to contributions
of type~(c) of \reffi{fig:genericPAdiags}.
\item Real non-factorizable corrections are classified and constructed
  accordingly, but instead of considering graphs with photon exchange
  contributing to the amplitude ${\cal M}$, interference contributions
  of the type ${\cal M}_{\gamma,1} {\cal M}^*_{\gamma,2}$ have to be
  considered, where ${\cal M}_{\gamma,1}$ and ${\cal M}_{\gamma,2}$
  are contributions to the amplitude ${\cal M}_{\gamma}$ resulting
  from ${\cal M}$ by adding an extra photon emission.  Interference
  diagrams in which the photon in ${\cal M}_{\gamma,1}$ and ${\cal
    M}_{\gamma,2}$ is emitted from one and the same subprocess
  (production or decay) obviously lead to factorizable corrections to
  the correspondong subprocess; those contributions are of type~(d) or
  (e) of \reffi{fig:genericPAdiags}.  But diagrams with one of the
  photons coupling to the resonance and the other to an external leg
  or to a resonance lead to non-factorizable contributions as well.
  Again, this is due to the fact that the process of going on shell
  with the kinematics of the resonance in subamplitudes leads to new
  IR divergences in the factorizable contributions.  Finally, photon
  exchange between different production or decay subprocesses entirely
  contributes to non-factorizable corrections. 
  The generic structure of non-factorizable corrections
  is sketched in \reffi{fig:genericPAdiags}(f).
\end{myitemize}
Since the non-factorizable corrections are due to the exchange or
emission of soft massless particles, they can be evaluated using
modified eikonal currents [\cf\refeq{eq:eikonal_current} for their
original form] applied to LO matrix
elements~\cite{Denner:1997ia,Dittmaier:2015bfe}.  Ordinary eikonal
currents describe only the coupling of soft photons or gluons to
external OS states.  The modification necessary for the calculation of
non-factorizable corrections concerns the inclusion of the couplings
to the internal resonance lines and the corresponding momentum flow.
To this end, for a single resonance~$P$
the eikonal current $J^\mu_{\mathrm{eik}}$ for the
full process is decomposed into currents for the production and
decay subprocesses of $P$ according to
\beq
J^\mu_{\mathrm{eik}} = J^\mu_{\mathrm{eik,prod},P}
+ J^\mu_{\mathrm{eik,dec},P},
\eeq
where the radiation off $P$ contributes to $J^\mu_{\mathrm{eik,prod},P}$
and $J^\mu_{\mathrm{eik,dec},P}$. The two contributions result from splitting
all diagrams with photon radiation off $P$ via partial fractioning
the product of the two $P$ propagators before and after the resonance,
as shown in \refeq{eq:partfracphotonoffres}. Note that the two contributions
involve different $P$ momenta in the $P$~propagators. 
If $p$ is the $P$ momentum after photon emission off $P$,
$J^\mu_{\mathrm{eik,prod},P}$ involves a $P$~propagator with momentum~$p$
and $J^\mu_{\mathrm{eik,dec},P}$ one with momentum $p+k$.
The current $J^\mu_{\mathrm{eik,prod},P}$ is, thus, constructed
from all charged external states of the production process as
described in \refeq{eq:eikonal_current}, including
the resonance if it is charged.
The current $J^\mu_{\mathrm{eik,dec},P}$ is constructed analogously,
but receives an additional global factor
$(p^2-\MP^2)/[(p+k)^2-\MP^2]$ accounting for the momentum shift
in the resonance when the photon is radiated after the resonance has formed.
In spite of the modified eikonal factors, the factorization property
of soft-photon corrections remains, so that
non-factorizable corrections take the form of a differential, but 
universal correction factors to LO cross sections, which depend only
on the resonance pattern and the external states of the process.  On
the side of the virtual corrections, the loop integration leads to a
non-factorizable correction factor to the LO cross section.  On the
side of the real corrections, however, the non-factorizable
corrections depend on details of the event selection, \ie whether or
how ``semi-soft'' photons or gluons with energies of the size of the
width(s) of the resonance(s) are recombined into jets or dressed
leptons.

In spite of the restriction of its validity to resonance regions,
the PA has several virtues:
\begin{myitemize}
\item
In comparison to the application of the full pole scheme including
the calculation of non-resonant contributions, the PA is conceptually
much simpler to apply. Non-resonant diagrams can be omitted from the
beginning, reducing the number of graphs drastically. 
\item
Issues with artificially created 
IR singularities, which were mentioned for the pole
scheme, are conceptually solved by the systematic extraction and
evaluation of non-factorizable corrections.
By definition, the combination of the non-factorizable 
and factorizable corrections 
reproduces the IR structure of the full process in the neighbourhood
of the resonances.
Owing to the universal structure of the non-factorizable corrections
widely generic results for those exist in the 
literature~\cite{Melnikov:1995fx,Beenakker:1997bp,Beenakker:1997ir,%
Denner:1997ia,Accomando:2004de,Dittmaier:2015bfe}.
\item The PA, similar to the pole scheme, decomposes amplitudes into
  production and decay subprocesses and non-factorizable
  contributions.  As explained, the latter factorize from the
  underlying LO cross sections.  Predictions in PA, thus, are ideal as
  basis for {\it improved Born approximations} 
  to parametrize cross sections
  near resonances in terms of appropriate pseudo-observables, such as
  mass, width, and effective couplings of the resonating particle.
  Moreover, matrix elements in PA can be used to define cross sections
  for polarized unstable particle production
  \cite{Ballestrero:2017bxn,Ballestrero:2019qoy}.
\item
Finally, calculations in PA can be pushed to higher orders much more
easily than full off-shell calculations. Since off-shell contributions
of higher orders are further suppressed with respect to the dominating
resonant contribution, including higher-order corrections in PA often
is a reasonable first step towards higher-order predictions.
\end{myitemize}
Pole approximations for processes with one or two resonances were
worked out by several groups for many processes.
In the following we pick two prominent examples from EW gauge-boson
production where both results in PA and with full off-shell effects
are available.

\subsubsection{NLO EW corrections to  \texorpdfstring{Drell--Yan-like}
{Drell-Yan-like} Z-boson production in the pole approximation}
\label{se:NLOEW-DY}

Technically, the application of the PA to single-resonance processes
is rather simple. As an example, we consider again the neutral-current
Drell--Yan process of Z-boson production at the LHC following closely
\citere{Dittmaier:2014qza}, making use of the basic definitions given
in \refse{se:DY-weak-PS}, where we have described the calculation of
the purely weak corrections in the pole scheme.  Here, we take into
account the full NLO EW corrections, which consist of both weak
virtual contributions and photonic virtual and real contributions.

The structure of the matrix element ${\cal M}_{\,\qqb,\fact}^{\EW,\sigma \tau}$
of the factorizable virtual EW corrections can be obtained from 
\refeq{eq:DYMweak} by replacing the weak one-loop contributions with
the full EW one-loop contributions and performing the OS limit
$\hat s\to\MZ^2$ in the vertex form factors $F^\sigma_{\ren,ffZ,\EW}(\hat s)$
and the bosonic self-energies $\Sigma^{\FA\FZ}_{\ren,\rT}(\hat s)$ and
$\Sigma^{\FZ\FZ}_{\ren,\rT}(\hat s)$.
The irreducible box contributions do not exhibit an explicit resonance factor
$(\hat s - \mu_\PZ^2)^{-1}$ from a tree-like resonance propagator and
do not contribute to the factorizable corrections, which read
\begin{align}
{\cal M}_{\,\qqb,\fact}^{\EW,\sigma \tau} ={} &
f^{\virt,\,\sigma \tau}_{\qqb,\EW,\fact}\, \mathcal{A}^{\sigma \tau},
\label{eq:DYMPA}
\nn\\
f^{\virt,\,\sigma \tau}_{\qqb,\EW,\fact} ={} &
{- e^2} \, \gqqZ^\sigma \gllZ^\tau
\, \frac{F^\sigma_{\ren,qqZ,\EW}(\MZ^2)
    +F^\tau_{\ren,llZ,\EW}(\MZ^2)}{\hat s - \mu_\PZ^2} 
\nl
&{}
+ e^2 
\gqqZ^\sigma\, \gllZ^\tau 
\frac{\Sigma^{'\FZ\FZ}_{\ren,\rT}(\MZ^2)}{\hat s-\mu_\PZ^2} 
- e^2 \left(\Ql\, \gqqZ^\sigma +\Qq \,\gllZ^\tau \right) 
\frac{\Sigma^{\FA\FZ}_{\ren,\rT}(\MZ^2)}{(\hat s - \mu_\PZ^2)\MZ^2}.
\end{align}
Note that the self-energy contributions in the last line 
vanish in the complete OS renormalization scheme, underlining the simplicity
of the PA.
The standard matrix elements $\mathcal{A}^{\sigma \tau}$ can still be
evaluated on the off-shell phase space, where 
$\hat s=-\hat t-\hat u\ne\MZ^2$ off resonance, or alternatively
with the OS-projected kinematics, where $\hat s=-\hat t-\hat u=\MZ^2$.
In the latter case, the OS limit $\hat s\to\MZ^2$ is taken while keeping
the scattering angle in the partonic CM frame fixed.

Since the Z~boson is electrically neutral, there are no one-loop
graphs with photons coupling to the resonance, so that there are no graphs
that contribute to both factorizable and non-factorizable corrections. 
The non-factorizable corrections entirely result from box diagrams with
photon and Z-boson exchange between initial and final states (\cf
\reffi{fi:dy_nf})
\bfi
  \centerline{\includegraphics[bb=0 5 120 60,scale=.7]{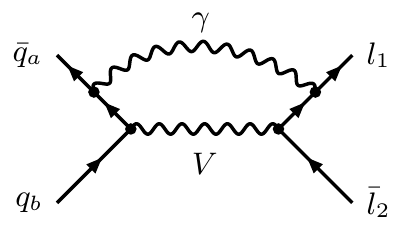}}
  \caption{Example for a
manifestly non-factorizable box diagram for Drell--Yan-like W- or Z-boson production.}
\label{fi:dy_nf}
\efi
and take the simple form
\beq
\label{eq:nf-virt}
{\cal M}_{\,\qqb,\nf}^{\EW,\sigma \tau} ={} 
\delta_{\nf}^{\qqb}\, 
{\cal M}_{\qqb,\mathrm{PA}}^{\mathrm{LO},\sigma\tau},
\qquad
\delta_{\nf}^{\qqb}
    ={}   \frac{\alpha}{\pi}
\Qq \Ql 
\Biggl\lbrace
\left[ \Delta - 2\ln\left(\frac{\mu_\PZ^2-\hat s}{\mu \MZ}\right) \right]
           \ln\left(\frac{\hat t}{\hat u}\right) 
    +\Li_{2}\left(1+\frac{\MZ^2}{\hat t}\right) 
    -\Li_{2}\left(1+\frac{\MZ^2}{\hat u}\right) 
    \Biggr\rbrace ,
\eeq
where the subscript PA in ${\cal
  M}_{\qqb,\mathrm{PA}}^{\mathrm{LO},\sigma\tau}$ indicates that the
non-resonant photon contribution is not included here.  In
$\delta_{\nf}^{\qqb}$, the divergent contribution $\Delta$, defined in
\refeq{eq:Delta}, originates from the soft IR divergence in DR.
Note that this IR singularity comes along with the non-analytic contribution
proportional to $\ln(\mu_\PZ^2-\hat s)$, which shows that $\hat s=\mu_\PZ^2$
is a branch point in the complex $\hat s$ plane.
As explained above, by construction the sum of factorizable and
non-factorizable corrections has the same IR structure as the full off-shell
EW one-loop amplitude.

For a full NLO prediction in PA, we still have to add the
real-photonic corrections, which can either be based on the full LO
prediction for $\qqb\to \Pl\bar\Pl\gamma$ or on a corresponding result
in PA.  Figure~\ref{fig:DY-PA} shows the
comparison~\cite{Dittmaier:2014qza} of the NLO EW corrections for
Z~production based on the PA for virtual and real corrections with
full off-shell results based on the CMS, described in \refse{se:CMS}.
\bfi
  \centering
  \begin{tabular}{r@{\hspace{.5cm}}r}
     \includegraphics[scale=0.63]{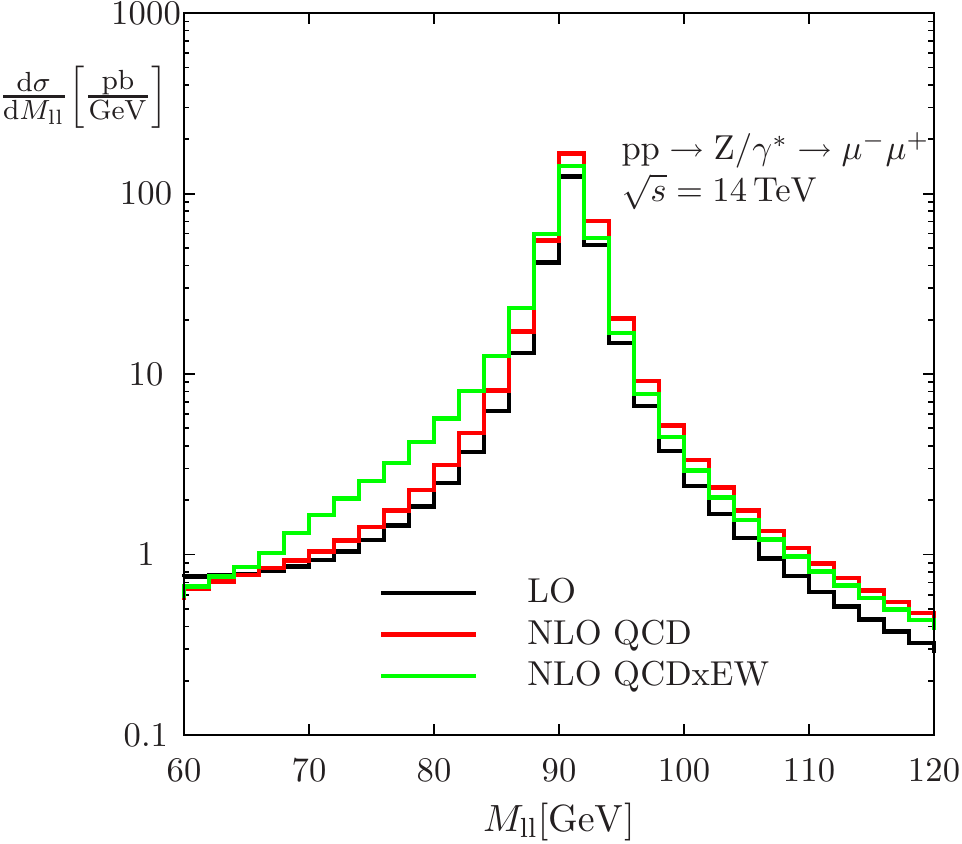}
    &\includegraphics[scale=0.63]{{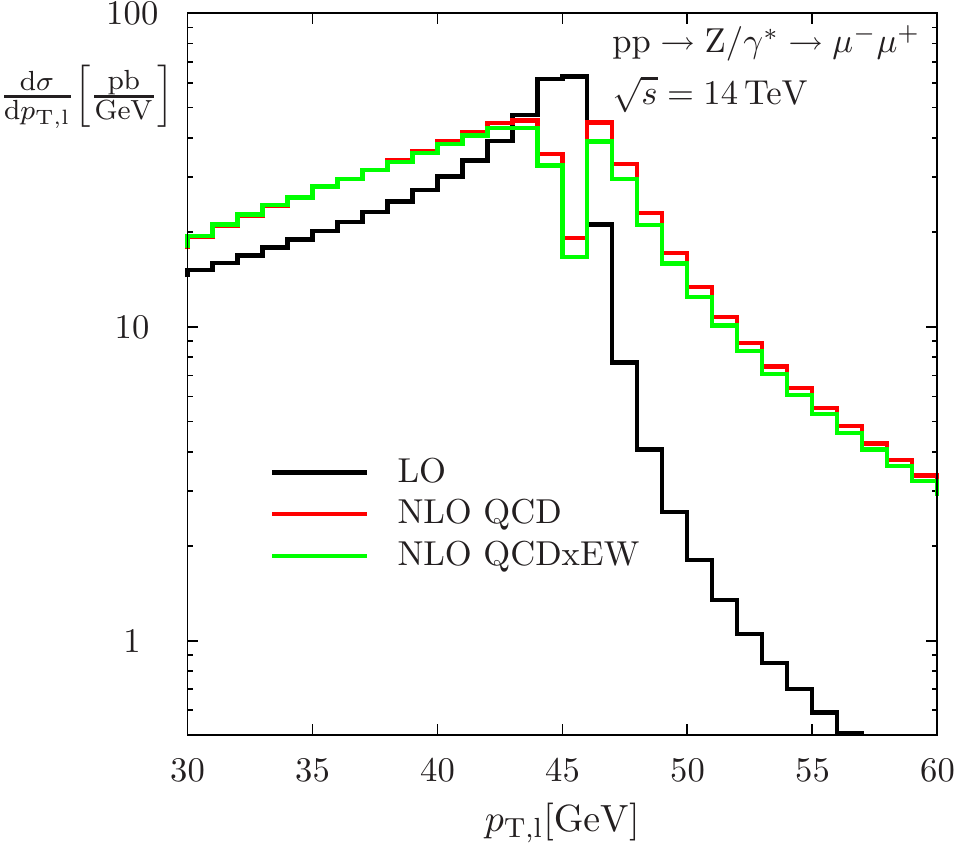}} \\[.5em]
     \includegraphics[scale=0.63]{{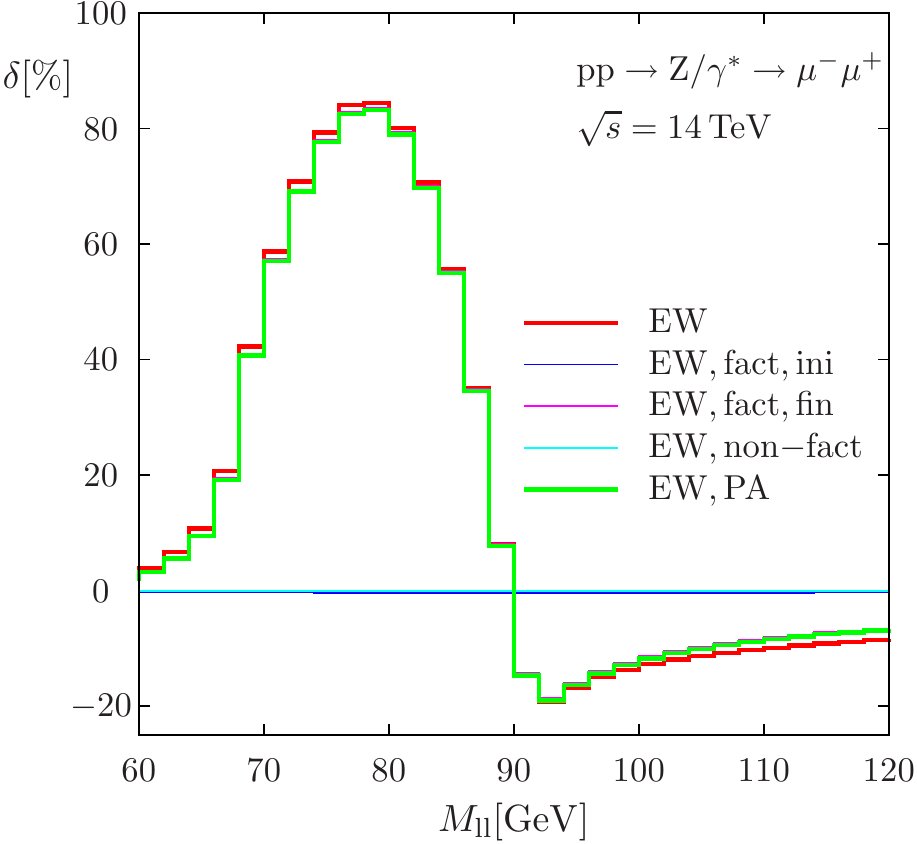}}
    &\includegraphics[scale=0.63]{{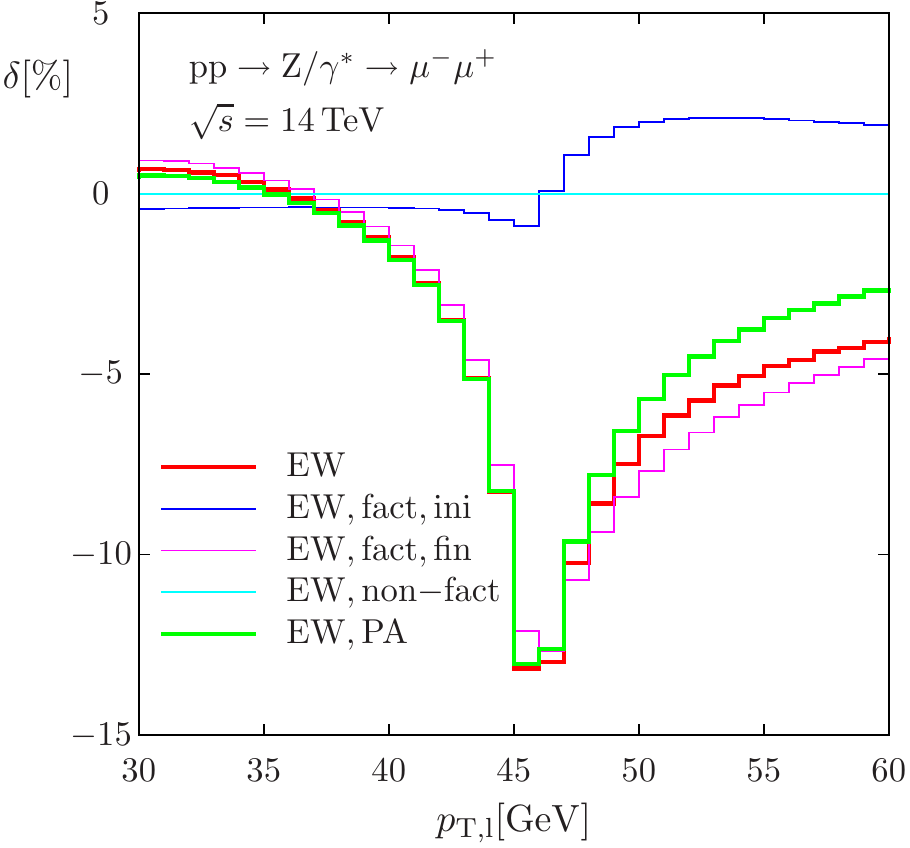}}
 \end{tabular}
  \caption{LO and NLO distributions in the invariant dilepton-mass $M_{\mathrm{ll}}$ 
and in the transverse-lepton-momentum $p_{\mathrm{T,l}}$ for Drell--Yan-like
Z~production $\Pp\Pp\to\PZ/\gamma^*\to\mu^-\mu^+ +X$ at the LHC (upper panels)
and corresponding relative EW corrections (lower panels), where the corrections
both show the full NLO EW results as well as the individual contributions (and their sum)
in pole approximation (PA).
(Plots taken from \citere{Dittmaier:2014qza}.)}
\label{fig:DY-PA}
\efi
The distribution in the dilepton invariant mass $M_{\Pl\Pl}$ shows the
resonance near $M_{\Pl\Pl}\sim\MZ$, but the distribution in the lepton
transverse momentum $p_{\rT,\Pl}$ is dominated by resonant
contributions for the whole range of $p_{\rT,\Pl}\lsim\MZ/2$.  The
results clearly show that the difference in the relative EW correction
between full off-shell result and PA is only some $0.1\%$ whenever the
resonance dominates the distribution.  In more detail, the plots also
show the individual contributions to the PA from (real+virtual)
factorizable corrections to production (``fact,ini'') and decay
(``fact,fin'') of the Z~boson and from the (real+virtual)
non-factorizable corrections (``non-fact''), which turn out to be
negligibly small.

More results and further details on the NLO EW corrections in PA,
including the corresponding charged-current process of W-boson
production, can be found in \citere{Dittmaier:2014qza}.  Moreover, the
generalization of the concept of the PA to NNLO QCD$\times$EW
corrections of ${\cal O}(\alphas\alpha)$ as well as the calculation of
the non-factorizable corrections at this order are given there.  The
factorizable ${\cal O}(\alphas\alpha)$ corrections can be further
decomposed into pure production or decay contributions and cross-talk
between production and decay. The latter are dominant and discussed in
\citere{Dittmaier:2015rxo}, while the pure decay corrections are
negligibly small~\cite{Dittmaier:2015rxo}.  The pure production
corrections are only known to NNLO QCD$\times$QED
yet~\cite{deFlorian:2018wcj,Delto:2019ewv}.

\subsubsection{Multiple resonances in pole approximation and
gauge-boson pair production in double-pole approximation}
\label{se:multires}

The PA concept can be applied to any resonance structure,
as long as resonances are kinematically well separated.%
\footnote{The necessary separation is, for instance, not given in
  processes with cascade decays such as $ab\to P_1+X_1\to
  P_2+X_2+X_1\to X_3+X_2+X_1$, where the mass gap $M_{P_1}-M_{P_2}$
  between the two resonances $P_1$ and $P_2$ is not large w.r.t.\ the
  widths $\Gamma_{P_1}$ and $\Gamma_{P_2}$.}  The factorizable
corrections are calculated as for single-resonance processes.  Merely
the OS projection of the kinematics becomes somewhat more complicated.
As indicated already above, the calculation of the non-factorizable
corrections, however, is more involved in the case of multiple
resonances.  For the class of processes with several ``parallel
resonances'' (\ie cascade decays are not included), generic results on
virtual one-loop EW non-factorizable corrections were given in
\citeres{Accomando:2001fn,Dittmaier:2015bfe}.  These results were
obtained by generalizing the corresponding corrections to EW
gauge-boson pair production processes first derived for W-pair
production~\cite{Melnikov:1995fx,Beenakker:1997bp,Beenakker:1997ir,Denner:1997ia}.
In the following, we briefly review some features of W-pair production
in $\Pep\Pem$ annihilation in {\it double-pole approximation (DPA)},
which was worked out for cross-section predictions for LEP2 in
different variants by various groups~\cite{Beenakker:1998gr,%
Jadach:1996hi,Jadach:1998tz,%
Denner:1999kn,Denner:2000bj,Denner:2002cg,%
Kurihara:2001um}
(see also \citere{Grunewald:2000ju}).
Finally, we show a comparison of the DPA with results from the full
off-shell calculation for 
$\Pep\Pem\to4\,$fermions~\cite{Denner:2005es,Denner:2005fg}.

Figure~\ref{fig:eeWWtree} shows the LO {\it signal diagrams} containing
the double-pole structure as well as a typical {\it background diagram}
with only one W~resonance, which are all needed to describe W-pair
production near the resonances at NLO accuracy.
Recall that the DPA is only applied to NLO corrections, since its
application at LO would introduce errors of the order of ${\cal
  O}(\GP/\MP)\sim{\cal O}(\alpha)$.
The generic diagram for the virtual factorizable corrections of the DPA as well as
a typical one-loop diagram beyond DPA are depicted in \reffi{fig:eeWWloop}.
\bfi
  \centering
     \includegraphics[scale=.8]{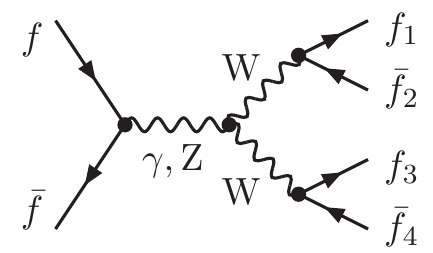}
     \includegraphics[scale=.8]{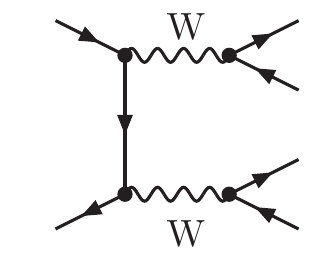}
     \includegraphics[scale=.8]{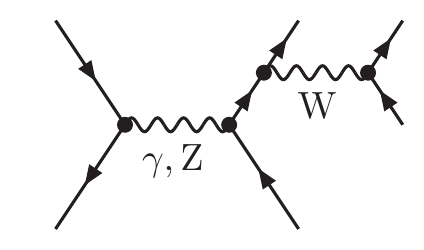}
\vspace*{-.5em}
  \caption{Doubly-resonant {\it signal diagrams} (left and middle) 
as well as an example for
a singly-resonant {\it background diagram} (right) for the 
charged-current 4-fermion production
process $f\bar f\to\PW\PW\to 4\,$fermions.}
\label{fig:eeWWtree}
\vspace{.3em}
  \centering
     \includegraphics[scale=.8]{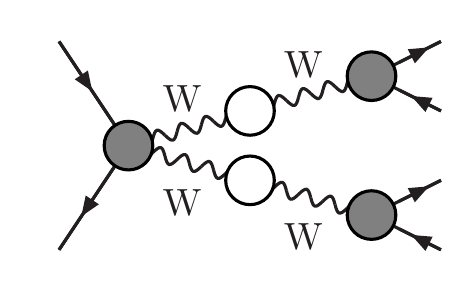}
\hspace*{1em}
     \includegraphics[scale=.8]{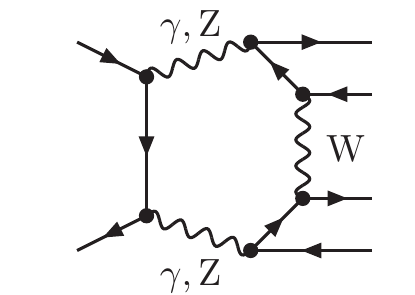}
\vspace*{-.5em}
  \caption{Structural diagram for the virtual factorizable corrections of the DPA (left)
as well as an example for an irreducible diagram type which is not included in the DPA (right)
for $f\bar f\to\PW\PW\to 4\,$fermions.}
\label{fig:eeWWloop}
\efi
Finally, Figure~\ref{fig:eeWWnf} illustrates the types of diagrams
contributing to the non-factorizable corrections in the DPA, where the
diagrams in the first line deliver both factorizable and non-factorizable parts,
while the diagrams in the second line exclusively contribute to the non-factorizable
corrections.
\begin{figure}[t]
  \centering
     \includegraphics[scale=.65]{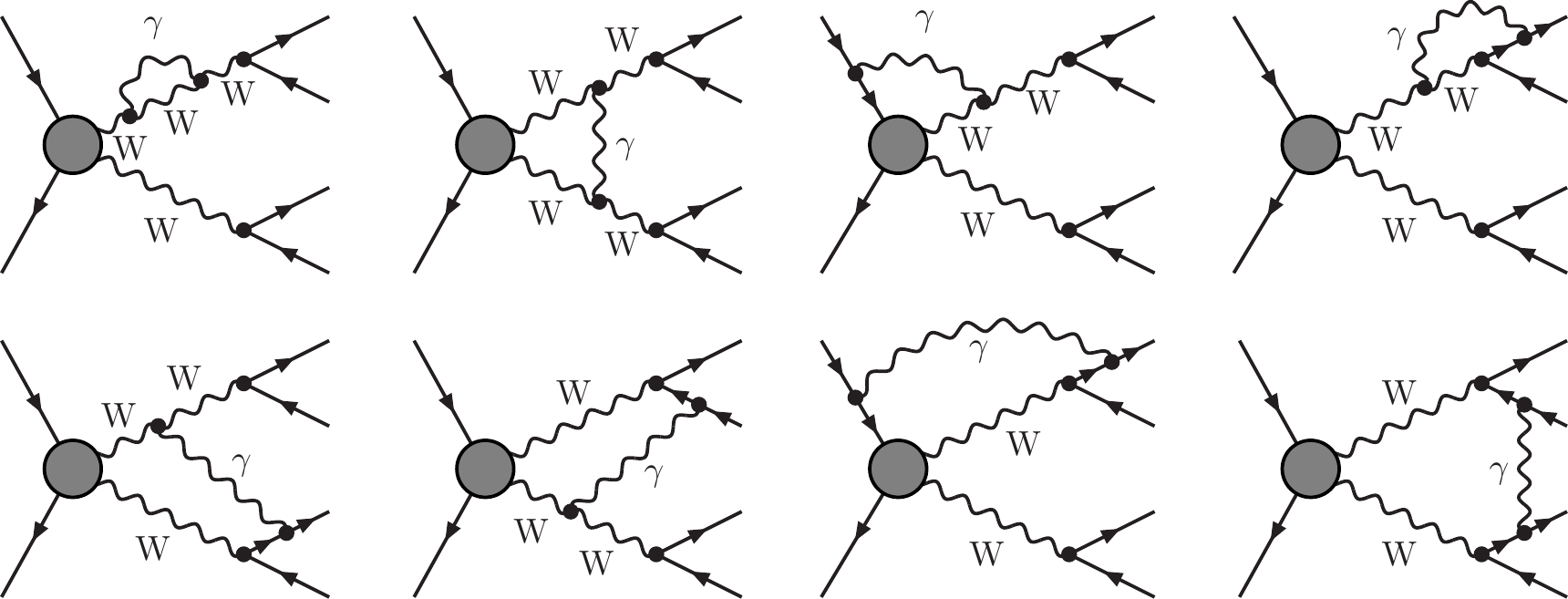}
\vspace*{-.5em}
  \caption{A representative set of diagrams for the virtual non-factorizable corrections
to the process $f\bar f\to\PW\PW\to 4\,$fermions,
where the grey blobs stand for any tree-like structures.}
\label{fig:eeWWnf}
\efi

NLO calculations in DPA can be set up in different ways as far as
the application of the DPA to virtual and/or real corrections is
concerned:
\begin{myitemize}
\item
If both virtual and real corrections are treated in DPA, it is possible to
separately discuss factorizable and non-factorizable corrections, which are
gauge independent each.
Both types receive IR-divergent virtual and real contributions, but the
two sums of virtual and real parts are each IR~finite. 
A numerical evaluation of the non-factorizable corrections alone,
as presented in \citeres{Beenakker:1997bp,Denner:1997ia}, reveals that
those are very small and phenomenologically unimportant.%
\footnote{This was also observed for Z-pair production at LEP2~\cite{Denner:1998rh}.}
This observation is in line with the known property that the sum of
virtual and real non-factorizable corrections cancels in observables that are
inclusive with respect to the resonance~\cite{Melnikov:1993np,Fadin:1993dz,Fadin:1993kt}, 
\ie if the invariant mass of the
resonating particle is fully integrated over the resonance region.
\item
An alternative ``hybrid version'' of the DPA was implemented in the
Monte Carlo generator {\sc RacoonWW}~\cite{Denner:1999kn,Denner:2000bj,Denner:2002cg}
for $\Pep\Pem\to\PW\PW\to4\,$fermions,
which applies the DPA only to the virtual corrections, but employs complete
LO matrix elements for $\Pep\Pem\to4f+\gamma$ in the real radiation part.
In this way, it was possible to check that the splitting of real
radiation off resonating W~bosons into production and decay contributions,
as formally done via the partial fractioning \eqref{eq:partialfrac} for
real photons, does not introduce additional uncertainties if the two W~resonances
overlap (\ie if the photon energy $k^0\sim\GW$).
\end{myitemize}
A detailed comparison~\cite{Grunewald:2000ju}
between the various results in DPA~\cite{Beenakker:1998gr,%
Jadach:1996hi,Jadach:1998tz,%
Denner:1999kn,Denner:2000bj,Denner:2002cg,%
Kurihara:2001um} as well as several uncertainty estimates within the
different approaches confirmed the validity of the DPA within a
precision margin of $\lsim0.5\%$ for W-pair production in the energy
range $170\GeV\lsim\sqrt{s}\lsim 300\GeV$, which in particular covers
the energy range of LEP2 where W-pair production was accessible within
good precision.  Note that the DPA necessarily breaks down near the
threshold energy $\sqrt{s}=2\MW$, because at least one of the two
W~bosons is shifted out of resonance below threshold. For
$\sqrt{s}\lsim 2\MW+n\GW$ with $n\sim2{-}3$ the calculation of the
W-pair cross section is based on an {\it improved Born approximation
(IBA)}~\cite{Dittmaier:1991np,Denner:2001zp}, which includes only
universal corrections such as leading-logarithmic initial-state
radiation, running-coupling effects, and the Coulomb singularity.

Finally, the reliability of the DPA for W-pair production was
ultimately checked after the completion of the full NLO EW corrections
to the charged-current processes
$\Pep\Pem\to4f$~\cite{Denner:2005es,Denner:2005fg} in the CMS,
which is described in \refse{se:CMS}.
Figure~\ref{fig:XSeeWW} shows a comparison of predictions for the inclusive
W-pair production cross section for CM energies ranging from the LEP2 energy
range up to $2\TeV$.
\bfi
  \centering
     \includegraphics[scale=.6]{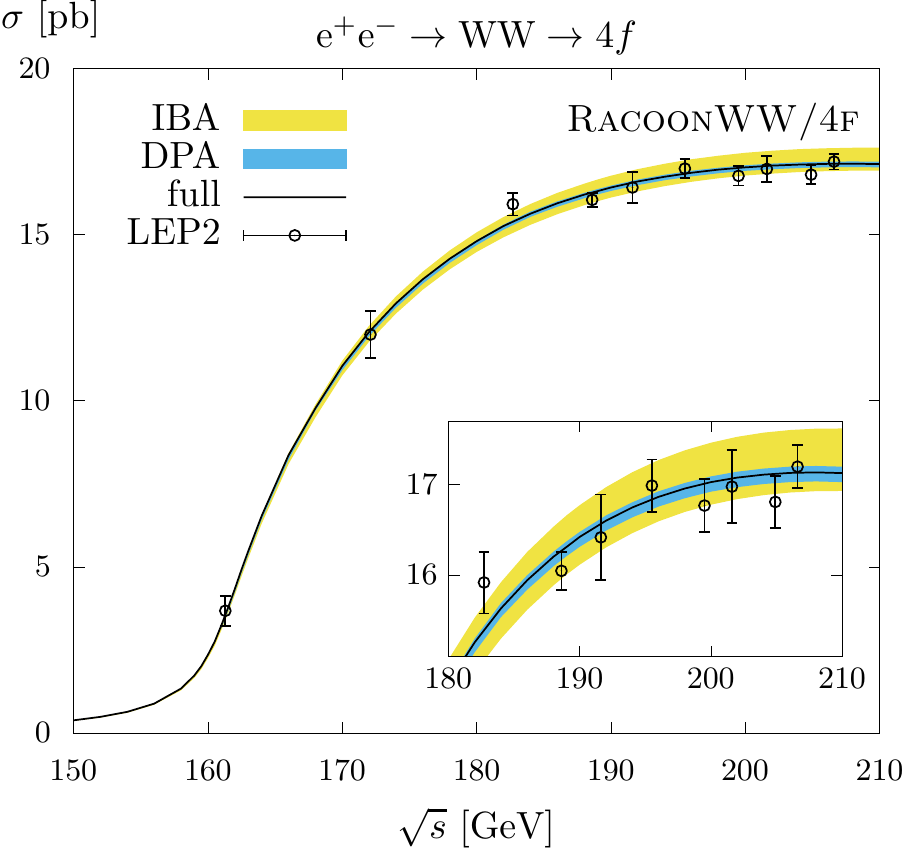}
\hspace*{1em}
     \includegraphics[scale=.6]{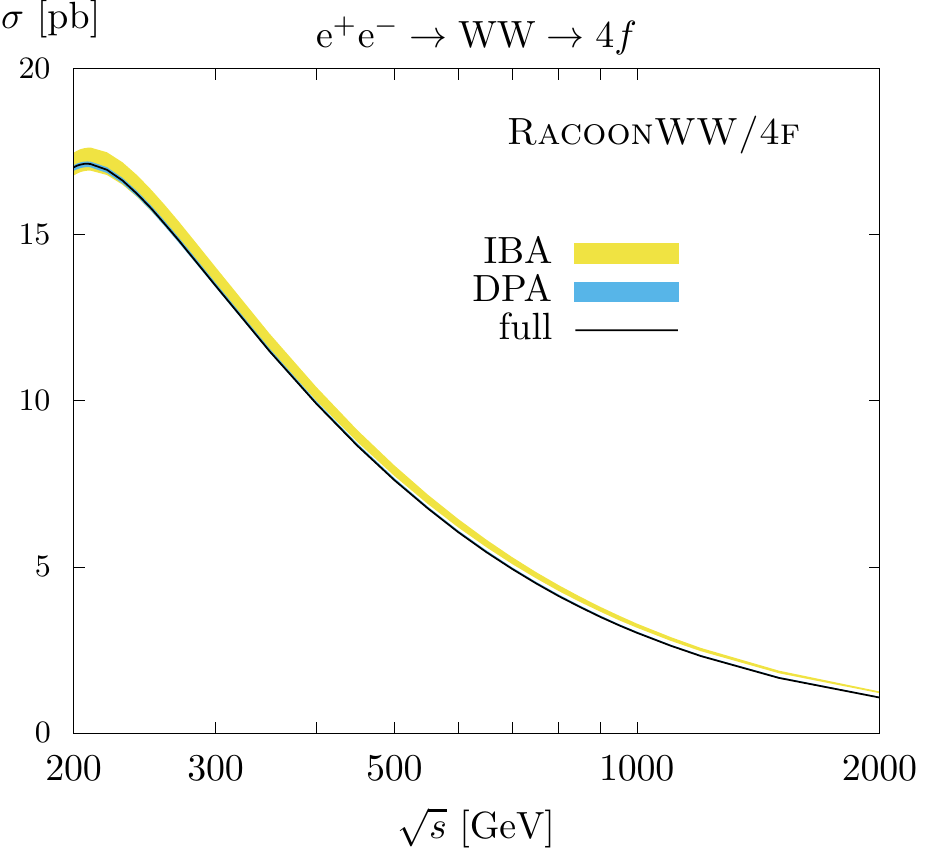} \\
     \includegraphics[scale=.6]{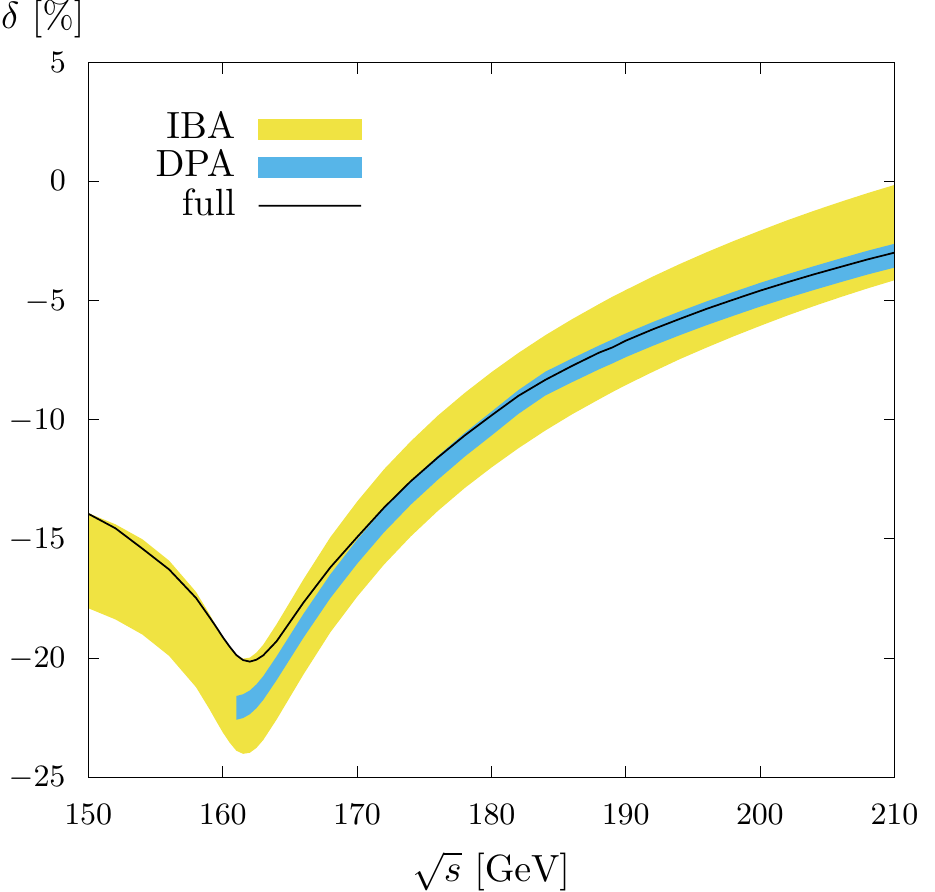}
\hspace*{1em}
     \includegraphics[scale=.6]{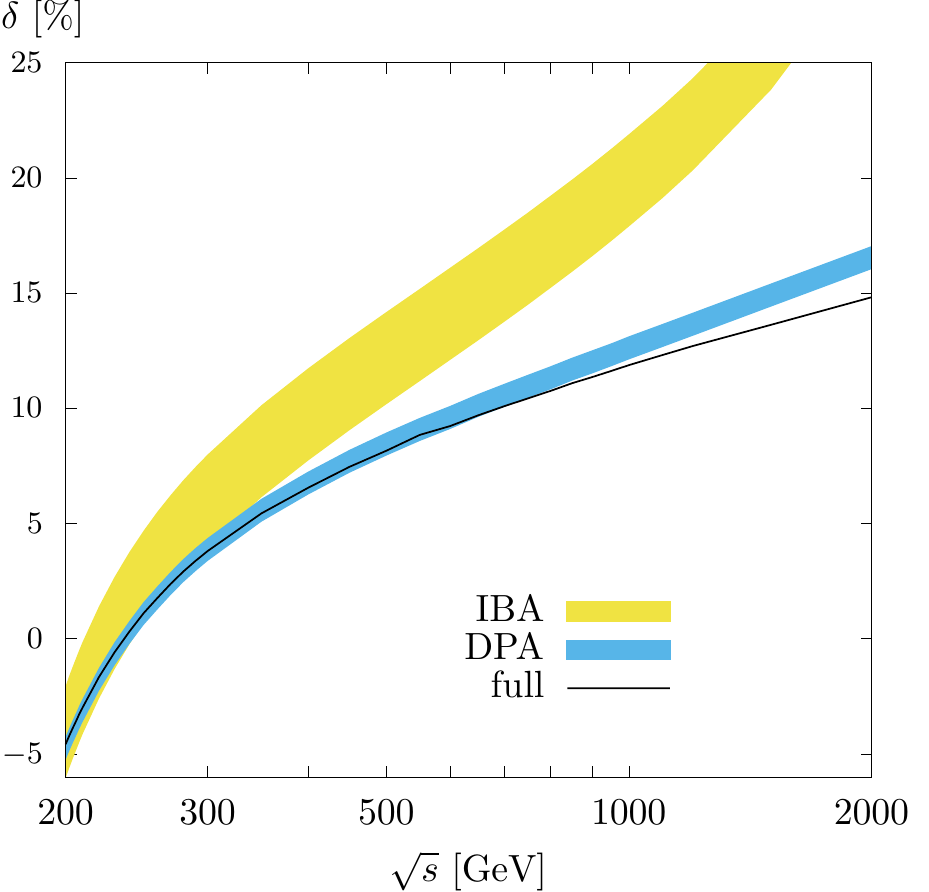} \\
\vspace*{-.5em}
  \caption{Inclusive W-pair production cross section in $\Pep\Pem$ annihilation
    as provided by {\sc RacoonWW/4f}, including NLO EW corrections in
    IBA, in DPA, and obtained from the ``full'' off-shell calculation
    in the CMS.  All predictions include leading higher-order effects
    from initial-state radiation. The IBA and DPA bands illustrate the
    uncertainty bands of $\pm2\%$ and $\pm0.5\%$, as assessed for the
    LEP2 energy range.  The cross-section measurements of
    LEP2~\cite{Schael:2013ita} are shown as data points.  (Results
    taken from \citeres{Denner:2005es,Denner:2005fg}.)}
\label{fig:XSeeWW}
\efi
The predictions include NLO EW corrections based on an IBA and the DPA
of {\sc RacoonWW} and the full off-shell calculation of
\citeres{Denner:2005es,Denner:2005fg} as implemented in the extension
{\sc Racoon4f} of {\sc RacoonWW}, as well as leading higher-order
effects from initial-state radiation in all versions, as described in
\refse{se:SF}.  In the LEP2 energy range, the IBA, the DPA, and the
full off-shell calculations nicely
agree within the assessed error margins.%
\footnote{Note that the central prediction of the IBA, as provided by
  {\sc RacoonWW}, includes also some fitted terms to come closer to
  the DPA, which accounts for non-universal NLO EW corrections.}  
Note
that a proper confrontation of LEP2 data~\cite{Schael:2013ita}, which
are included in the plot, with theory predictions in fact required the
inclusion of non-universal EW corrections, as provided by the DPA.
For energies above the LEP2 energy range, the error assessments of
$\pm2\%$ and $\pm0.5\%$ for IBA and DPA, respectively, start to fail.
For $\sqrt{s}\gsim300{-}400\GeV$, the onset of the enhanced negative
high-energy EW corrections from soft/collinear W/Z exchange
(\cf\refse{se:ewc@he}) can be observed, an effect that is not
accounted for by the IBA, but still by the DPA. Those corrections grow
to $\sim-10\%$ and more in the TeV range for the total cross sections
and are even larger in differential distributions.  The difference of
some percent between DPA and full off-shell result, on the other hand,
is due to the increasing contribution of background diagrams, which
are included in the DPA calculation only at LO (see last diagram in
\reffi{fig:eeWWtree}), but not at NLO (see second diagram in
\reffi{fig:eeWWloop}).  More details and results from this comparison
can be found in \citeres{Denner:2005es,Denner:2005fg}.

Predictions for EW gauge-boson production in DPA were worked out for
other channels and collider types as well, for instance for
WW~production at a $\gamma\gamma$
collider~\cite{Bredenstein:2004ef,Bredenstein:2005zk} and
WW~\cite{Billoni:2013aba} and WZ~\cite{Baglio:2018rcu} production at
the LHC.  Complete NLO EW corrections with fully off-shell gauge
bosons were presented for
WW~\cite{Biedermann:2016guo,Kallweit:2017khh},
WZ~\cite{Biedermann:2017oae}, and
ZZ~\cite{Biedermann:2016yvs,Biedermann:2016lvg} production at the LHC.

Recently, the concept of the pole approximation was also applied
to triple-W production at the LHC in \citere{Dittmaier:2019twg}, defining
a {\it triple pole approximation (TPA)}. 
Similar to the findings for W-pair production, the TPA is able to
reproduce the NLO corrections to the full off-shell $2\to6$ particle
reaction with W-boson decays whenever the W~resonances dominate the
cross section, as shown in \citere{Dittmaier:2019twg} for 
leptonically decaying W~bosons.

\subsection{Complex-mass scheme}
\label{se:CMS}

\subsubsection{LO procedure and general properties}

The {\it complex-mass scheme (CMS)} was introduced in
\citere{Denner:1999gp} for LO calculations with $\PW/\PZ$ resonances
and generalized to NLO in
\citeres{Denner:2005fg,Denner:2006ic} for arbitrary SM particles.%
\footnote{Here and in
  \citeres{Denner:1999gp,Denner:2005fg,Denner:2006ic} all external
  particles are assumed to be stable. Generalizations to external
  unstable particles have been discussed in the literature as well
  (see, e.g., \citere{Kniehl:2001ch,Espriu:2002xv,Bharucha:2012nx}).
  In this case, it is convenient to introduce different wave-function
  renormalization constants for incoming and outgoing
  particles~\cite{Espriu:2002xv,Bharucha:2012nx}.} 
This scheme is nowadays used in many state-of-the-art calculations and
implemented, for instance, in the tools
\MGNLO~\cite{Frederix:2018nkq}, {\Openloops2} \cite{Buccioni:2019sur},
and \Recola \cite{Actis:2012qn,Actis:2016mpe}.

In the complex-mass scheme the mass squared of each unstable particle~$P$
is consistently identified with 
the complex value $\mu_P^2$ as defined in \refeq{eq:muP},
not only in the $P$-propagators, but also in the couplings,
which therefore become complex.
In particular, couplings involve
a complex weak mixing angle fixed via 
\beq
\cw^2=1-\sw^2=\mu_\PW^2/\mu_\PZ^2.
\label{eq:complexthw}
\eeq
The CMS does not change the theory at all, but only
rearranges its perturbative expansion, 
so that in particular no double counting of terms occurs.
At LO, the above prescription already describes the full procedure.
At NLO, the scheme requires some changes in the NLO machinery, 
since the imaginary parts in the complex masses correspond to
higher-order contributions in the perturbative expansion in the
formalism for stable particles.
The necessary changes in the renormalization procedure are
described below, otherwise the usual perturbative calculus with Feynman
rules and counterterms works without modification.
Generalizations beyond NLO have not yet been described in the literature,
but should be straightforward.

LO and NLO calculations in the CMS have the following virtues:
\begin{myitemize}
\item All relations following from gauge invariance are respected,
  because the gauge-boson masses are modified only by an analytic
  continuation.  This concerns both the gauge independence in the
  parametrization of $S$-matrix elements in terms of renormalized
  input parameters and 
  the validity of all Ward or
  Slavnov--Taylor identities of the related Green functions.  This, in
  particular, implies that so-called unitarity cancellations within
  amplitudes remain intact, since those are a consequence of gauge
  invariance.
\item NLO calculations deliver uniform predictions with NLO accuracy
  everywhere in phase space, \ie both in resonant and non-resonant
  regions (if the widths of the unstable particles have NLO accuracy).
\item Amplitudes involve spurious, \ie artificially introduced,
  imaginary parts (in space-like propagators and complex couplings).
  As a consequence, amplitudes do not exactly obey the standard 
  cutting relations~\cite{Cutkosky:1960sp} which express unitarity,
  because these relations involve complex conjugation.  However, the
  spurious terms spoiling unitarity are
  unproblematic, since they are of (N)NLO in an
  (N)LO calculation, \ie of higher order, without any unnatural
  amplification, because unitarity cancellations are respected.  
  Modified cutting rules for the CMS were formulated, and the
  unitarity within this scheme investigated in
  \citere{Denner:2014zga}.
\end{myitemize}

\subsubsection{Complex renormalization}

Let us consider the modified renormalization procedure in detail in
't~Hooft--Feynman gauge, following closely its original introduction
presented in \citere{Denner:2005fg}.  Further below we comment on the
procedure in the framework of the background-field method.
In the following, we assume a unit quark-mixing matrix.

We start by considering the gauge-boson sector.  The squares of the
(real) bare $\PW$, $\PZ$ masses, $\MWb^2$ and $\MZb^2$, are split into
complex renormalized mass squares $\cmws$, $\cmzs$ and complex
counterterms $\de\cmws$, $\de\cmzs$,
\beq
\MWb^2=\cmws+\de\cmws, \qquad \MZb^2=\cmzs+\de\cmzs,
\eeq
and the field renormalization transformation is carried
out in the same way as in the OS renormalization for stable particles described in
\refse{se:ren_sm},
\beq
\FW_{0}^{\pm}   {}=  \left(1+{\frac{1}{2}}\delta \cZ_{\FW}\right) \FW^{\pm}, 
\qquad
\bpm \FZ_{0} \\ \FA_{0} \epm   {}= 
\bpm 1 + \frac{1}{2}\delta \cZ_{\FZ\FZ} & \frac{1}{2}\delta \cZ_{\FZ\FA} \\[1ex]
               \frac{1}{2}\delta \cZ_{\FA\FZ}  & 1 + \frac{1}{2}\delta \cZ_{\FA\FA}
\epm
\bpm \FZ \\[1ex] \FA \epm .
\eeq
To make the changes transparent, 
we denote the field renormalization constants in the
CMS by calligraphic letters.
The renormalized
transverse ($\rT$) gauge-boson self-energies now read
\begin{alignat}{3}
\label{eq:ren-se}
\Si^{\FW}_{\ren,\rT}(k^2) &{}= \Si^{\FW}_{\rT}(k^2) - \de\cmws 
+(k^2-\cmws)\de \cZ_{\FW}, \qquad&
\Si^{\FZ\FZ}_{\ren,\rT}(k^2) &{}= \Si^{\FZ\FZ}_{\rT}(k^2) - \de\cmzs 
+(k^2-\cmzs)\de \cZ_{\FZ\FZ}, \nl
\Si^{\FA\FA}_{\ren,\rT}(k^2) &{}= \Si^{\FA\FA}_{\rT}(k^2)
+ k^2 \de \cZ_{\FA\FA}, \qquad&
\Si^{\FA\FZ}_{\ren,\rT}(k^2) &{}= \Si^{\FA\FZ}_{\rT}(k^2) 
+ k^2 \frac{1}{2}\de \cZ_{\FA\FZ} +(k^2-\cmzs) \frac{1}{2}\de \cZ_{\FZ\FA}.
\end{alignat}
Note that the renormalized complex masses $\cmw$ and $\cmz$ are used 
everywhere, \ie also within the self-energies. 

The counterterms are fixed by generalizing the renormalization
conditions of the complete OS scheme~\cite{Denner:1991kt,Aoki:1980ix} 
for stable particles to
\begin{align}
\label{eq:ren-cond-CMS1}
\Si^{\FW}_{\ren,\rT}(\cmws) &{}= 0, \qquad 
\Si^{\FZ\FZ}_{\ren,\rT}(\cmzs) = 0, \\
\Si^{\FA\FZ}_{\ren,\rT}(0) &{}= 0, \qquad
\Si^{\FA\FZ}_{\ren,\rT}(\cmzs) = 0, \nl
\Si^{\prime \FW}_{\ren,\rT}(\cmws) &{}= 0, \qquad
\Si^{\prime \FZ\FZ}_{\ren,\rT}(\cmzs) = 0, \qquad
\Si^{\prime \FA\FA}_{\ren,\rT}(0) = 0,
\label{eq:ren-cond-CMS2}
\end{align}
where the prime on $\Si^{\prime V}_{\ren,\rT}(p^2)$ denotes
differentiation with respect to the argument~$p^2$.  The conditions
(\ref{eq:ren-cond-CMS1}) fix the mass counterterms in such a way that
the squared renormalized masses are equal to the location of the
corresponding propagator pole in the complex plane.  The conditions
\refeqf{eq:ren-cond-CMS2} fix the field renormalization constants.  In
contrast to OS renormalization for stable particles described in
\refse{se453eforc}, the imaginary parts of the self-energies are kept
in the renormalization conditions, so that both mass and field
renormalization constants become complex.  This, in particular,
implies that the renormalized Z-boson field is not real anymore and
that the renormalized $\FW^\pm$ fields are not complex conjugates of
each other.  Thus, the renormalized Lagrangian, \ie the Lagrangian in
terms of renormalized fields without counterterms, is not hermitean,
but the total Lagrangian (which is equal to the bare Lagrangian) of
course is.

The renormalization conditions (\ref{eq:ren-cond-CMS1}) and
(\ref{eq:ren-cond-CMS2}) have the solutions
\begin{alignat}{5}
\label{exact-complex-ren-const-mass}
\de\cmws &{}= \Si^{\FW}_{\rT}(\cmws), \qquad&
\de\cmzs &{}= \Si^{\FZ\FZ}_{\rT}(\cmzs), \\
\label{exact-complex-ren-const-field}
\de \cZ_{\FZ\FA} &{}= \frac{2}{\cmzs}\Si^{\FA\FZ}_{\rT}(0), \qquad &
\de \cZ_{\FA\FZ} &{}= -\frac{2}{\cmzs}\Si^{\FA\FZ}_{\rT}(\cmzs),
\qquad &\nl
\de \cZ_{\FW} &{}= - \Si^{\prime \FW}_{\rT}(\cmws), \qquad&
\de \cZ_{\FZ\FZ} &{}= -\Si^{\prime \FZ\FZ}_{\rT}(\cmzs), \qquad&
\de \cZ_{\FA\FA} = -\Si^{\prime \FA\FA}_{\rT}(0),
\end{alignat}
which require to calculate the self-energies for complex squared
momenta. 

Owing to its definition (\ref{eq:complexthw}), the renormalization
of the complex weak mixing angle is determined by
\beq\label{eq:ren-mixing-angle}
\frac{\de\csw}{\csw} = -\frac{\ccw^2}{\csw^2}\frac{\de\ccw}{\ccw}
=-\frac{\ccw^2}{2\csw^2}
\left(\frac{\de\cmws}{\cmws}-\frac{\de\cmzs}{\cmzs}\right).
\eeq

The electric charge is fixed in the OS scheme by requiring that
there are no higher-order corrections to the $\Pe\Pe\ga$ vertex in the
Thomson limit.  In the CMS this condition reads
\beq\label{eq:ren-charge}
\de Z_e = \frac{\de e}{e} = \frac{1}{2}\Si^{\prime \FA\FA}(0) -
  \frac{\sw}{\cw}\frac{\Si^{\FA\FZ}_{\rT}(0)}{\cmzs}.
\eeq
Because of the explicit  and implicit 
presence of the complex masses and couplings in the expressions
on the r.h.s.\ of \refeq{eq:ren-charge},
the charge renormalization constant and thus the renormalized charge
become complex. Since the imaginary part of the bare charge vanishes,
the imaginary part of the charge renormalization constant is directly
fixed by the imaginary part of self-energies. 
As further discussed in \refse{se:CMSinput} below,
the imaginary part of the renormalized charge is not relevant in a
one-loop calculation, but has to be taken into account at the two-loop level.

For the Higgs boson the renormalization proceeds along the same lines
as for the gauge bosons above.
The complex Higgs mass squared
\beq
\cmhs= \MH^2 - \ri\MH\GH = \MHb^2 - \de\cmhs
\eeq
is defined as the location of the zero in $k^2$ in the renormalized
Higgs-boson self-energy
\beq
\Si^{\FH}_\ren(k^2) = \Si^{\FH}(k^2) - \de\cmhs 
+(k^2-\cmhs)\de \cZ_{\FH}. 
\eeq
Fixing the Higgs field renormalization on shell, we obtain the
renormalization constants 
\beq \label{eq:complex-ren-const-Higgs-mass}
\de\cmhs = \Si^{\FH}(\cmhs), \qquad
\de \cZ_{\FH} = - \Si^{\prime \FH}(\cmhs).
\eeq

To complete the renormalization in the scalar sector, the field
renormalization for the would-be Goldstone bosons can simply be set
equal to the Higgs field renormalization constant $\de \cZ_H$, which
is sufficient to cancel all UV divergences in vertex functions.  In
the case of the CMS, the same tadpole schemes can be
used as discussed for the OS scheme in \refse{se:tadpoles}. In
particular, the tadpole counterterm $\de t$ can be introduced to
cancel explicitly occurring tadpole graphs, \ie we set $\de t=-T^\FH$,
where $\Gamma^H_{\mathrm{1PI}}(0)=T^\FH$ 
is the one-loop contribution to the 1-point vertex function for the Higgs
boson at one loop. The linear 't~Hooft gauge-fixing term need not be
renormalized.

Among the fermions of the SM, only very few appear 
as high-energy resonance (mainly the top quark and the $\tau$~lepton).
In the following we nevertheless formulate the CMS for a generic
unstable fermion~$f$ and comment on the treatment of stable fermions
further below.
For the unstable fermion~$f$, 
the complex mass and its renormalization constant are introduced via
\beq
\cmf^2= \Mf^2 - \ri \Mf\Gf, 
\qquad
\Mfb=\cmf+\de\cmf.
\eeq
The field renormalization constants become complex, but
as for the bosons the fermion fields $\Ff_i^\si$ $(\si=\rR,\rL)$
and their adjoint counterparts $\bar\Ff_i^\si$
both are rescaled by the same field renormalization factor, 
\beq
\label{eq:CMSffieldren}
\Ff_{0,i}^\si = \left(1+\frac{1}{2} \de \cZ^{\Ff,\si} \right) \Ff_i^\si,
\qquad
\bar\Ff_{0,i}^\si = \left(1+\frac{1}{2} \de \cZ^{\Ff,\si} \right) \bar\Ff_i^\si,
\eeq
so that $(\Ff_{i}^\si)^\dagger\ga_0$ and $\bar\Ff_{i}^\si$ are not
identical, but differ by spurious imaginary parts.
The renormalized fermion self-energy reads 
\beq
\Sigma^{\Ff}_\ren(p) ={}
  \left[\Sigma^{\Ff,\mathrm{R}}(p^2)+\de \cZ^{\Ff,\mathrm{R}}\right]\slashed{p}\omega_+
+ \left[\Sigma^{\Ff,\mathrm{L}}(p^2)+\de \cZ^{\Ff,\mathrm{L}}\right]\slashed{p}\omega_-
 {} + \cmf\left[ \Sigma^{\Ff,\mathrm{S}}(p^2) 
- {\frac{1}{2}}(\de \cZ^{\Ff,\mathrm{R}} + \de \cZ^{\Ff,\mathrm{L}})
-\frac{\de\cmf}{\cmf} \right].
\eeq
Generalizing the OS renormalization conditions to complex
renormalization as 
\begin{align}
\label{eq:complex-ren-const-Top-mass}
\de\cmf &{}= \frac{\cmf}{2}
\left[  \Sigma^{\Ff,\mathrm{R}}(\cmf^2)
       +\Sigma^{\Ff,\mathrm{L}}(\cmf^2)
      +2\Sigma^{\Ff,\mathrm{S}}(\cmf^2) \right],
\nn\\
\de \cZ^{\Ff,\sigma} &{}= -\Sigma^{\Ff,\sigma}(\cmf^2)
-\cmf^2\left[  \Sigma^{\prime \Ff,\mathrm{R}}(\cmf^2)
             +\Sigma^{\prime \Ff,\mathrm{L}}(\cmf^2)
            +2\Sigma^{\prime \Ff,\mathrm{S}}(\cmf^2) \right], \qquad
\sigma=\mathrm{R,L},
\hspace{2em}
\end{align}
fixes $\cmf^2$ as the location of the complex pole in the fermion
propagator.  

Typically, only a few of the fields considered above develop
resonances in interesting processes at a time, so that the 
complex renormalization is not needed and often also not wanted
for all unstable particles. Concerning such a mix of real and complex
renormalization conditions, the following points should be respected:
\begin{myitemize}
\item
The masses of all
initial- and final-state particles of a resonance process should
be taken real, \ie those states are considered as stable.
\item Even the mass and field renormalization constants of stable (or
  supposed to be stable) particles become complex quantities if at
  least one mass is renormalized according to the CMS, because the
  complex mass enters in general all self-energies that appear in the
  renormalization. The imaginary parts in all renormalization
  constants have to be kept in order to guarantee a full cancellation
  of UV divergences in amplitudes, \ie no real parts of self-energies
  should be taken in the calculation of renormalization constants.
  However, the absorptive parts resulting from the self-energies of
  fields that are treated as stable have to be discarded. Thus, if
  some unstable particles are treated as stable, the operator $\rRe$
  (see \refse{se:rcsm}) should be introduced with the 
  understanding that it only eliminates the absorptive parts of the 
  loop integrals appearing in all field and mass renormalization constants 
  of the particles that are treated as stable, while it does not 
  affect complex couplings and masses. 
  A detailed example with stable
  top quark and unstable vector bosons is discussed in
  \citere{Frederix:2018nkq}, while a description of the implementation
  in \Openloops2 is provided in \citere{Buccioni:2019sur}.
\item
One and the same particle should not appear as external stable state
and internal resonance within the same process, because their
simultaneous appearance would mean two different types of
renormalization for one and the same particle.
Any attempt to allow for this asymmetric treatment would necessarily
face issues with gauge invariance.
\item
Even if only the $\PW$ or only the $\PZ$~boson develops a resonance
in the considered process, it is advisable to treat both $\PW$ and $\PZ$~bosons
in the CMS, although complex renormalization for only one of those
would formally work as well. The reason for this advice is more of
practical nature, because the spurious imaginary parts in couplings,
which result from the complex weak mixing angle, are numerically smaller if
both $\cmw$ and $\cmz$ are used in \refeq{eq:complexthw} to calculate~$\cw$.
\end{myitemize}

\subsubsection{Simplified version of the complex renormalization}
\label{se:simplifiedcomplexren}

Renormalization in the CMS requires to calculate the self-energies for
complex squared momenta. This demands an analytic continuation of the
2-point functions entering the self-energies in the momentum variable
to the unphysical Riemann sheet (\cf\citere{Passarino:2010qk}). 
A method for the analytic continuation of 2-point functions
for arbitrary complex momentum variables and masses based on
trajectories in the complex plane was proposed in
\citere{Frederix:2018nkq}. Such a general method  becomes relevant in models
involving particles with large widths. In the SM, where the widths of
all unstable particles are small compared to the masses, this complication
 can be avoided by expanding the self-energies appearing
in the renormalization constants about real arguments in such a way
that one-loop accuracy is retained.

We schematically illustrate the procedure for a scalar resonance $P$
with pole mass $M_P$ and pole width $\Gamma_P$, \ie
the location of the complex pole in the propagator is
$\mu_P^2=M_P^2-\ri M_P\Gamma_P$.
The one-loop self-energy correction to an amplitude in the CMS is proportional to 
\beq\label{eq:selffactor}
f(k^2)=\frac{\Sigma(k^2)-\de\mu_P^2}{k^2-\mu_P^2}
+\de\cZ_P
=\frac{\Sigma(k^2)-\Sigma(\mu_P^2)}{k^2-\mu_P^2}
-\Sigma'(\mu_P^2),
\eeq
where we used the OS counterterms
\beq
\de\mu^2_P  = \Si(\cmps) ,\qquad
\de \cZ_{P} = - \Si^{\prime}(\cmps).
\eeq
Note that the pole at $k^2=\mu_P^2$ cancels exactly in
\refeq{eq:selffactor}, and $f(k^2)$ is well-behaved in the vicinity of
the resonance ($k^2\approx M_P^2$), where $k^2-\mu_P^2\approx \ri
M_P\Gamma_P$ is of one-loop order, as the width $\Gamma_P$.  When
approximating $f(k^2)$ we have to make sure that the approximation is
NLO correct in the vicinity of the resonance.

If the self-energy can be expanded as
\begin{align}
\label{eq:expansion}
\Sigma(\mu_P^2)&{}=\Sigma(M_P^2)+(\mu_P^2-M_P^2)
\Sigma'(M_P^2)\;+\; {\cal O}\left((\mu_P^2-M_P^2)^2\right)
\nl
&{}=\Sigma(M_P^2)-\ri M_P\Gamma_P
\Sigma'(M_P^2)\;+\; {\cal O}\left((M_P\Gamma_P)^2\right),
\end{align}
we can approximate the mass and wave-function renormalization
counterterms by
\beq
\label{eq:expansionCT}
\de\mu^2_P  = 
\Si(M_P^2) +  (\cmps-M_P^2) \Si^{\prime}(M_P^2) + \Oaaa,
\qquad
\de \cZ_{P} = - \Si^{\prime}(M_P^2) + \Oaa,
\eeq
and the resulting approximation
\begin{align}
\label{eq:selffactorappr}
f(k^2)&{}=
\frac{\Sigma(k^2)-\Sigma(M_P^2)-(\cmps-M_P^2) \Si^{\prime}(M_P^2)}{k^2-\mu_P^2}
-\Sigma'(M_P^2) + \Oaa\nl
&{}=
\frac{\Sigma(k^2)-\Sigma(M_P^2)-(k^2-M_P^2) \Si^{\prime}(M_P^2)}{k^2-\mu_P^2}
 + \Oaa
\end{align}
is correct at NLO accuracy.  The $\Oaa$ and $\Oaaa$
contributions in \refeqs{eq:expansionCT} and \refeqf{eq:selffactorappr}
result from products of terms $\Si=\Oa$ and $(\mu_P^2-M_P^2)=\Oa$ and
are UV finite by construction at the one-loop level.

While the expansion \eqref{eq:expansion} holds true for a neutral and
colourless field, it breaks down for charged or coloured fields in
the presence of photon or gluon exchange due to the contributions with
a branch point at $k^2=\mu_P^2$.
The corresponding types of one-loop diagrams are depicted in
\reffi{fig:photon+gluon-exchange}.
\bfi
  \centering
\includegraphics[scale=1]{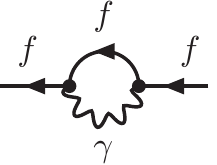} \hspace*{2em}
\includegraphics[scale=1]{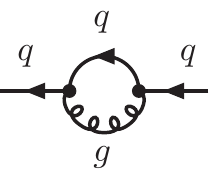} \hspace*{2em}
\includegraphics[scale=1]{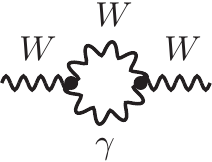}
\caption{Typical one-loop diagrams that lead to a non-analytic behaviour 
in the self-energies for charged fermions $\Ff$, quarks $\Fq$, and
W~bosons at OS momentum, where the
singularity in the virtuality $p^2$ becomes a logarithmic branch point.}
\label{fig:photon+gluon-exchange}
\efi
We explicitly see this by considering
\beq
\Sigma(k^2) = a\left(\frac{\mu_P^2}{k^2}-1\right)\ln\left(1-\frac{k^2}{\mu_P^2}\right) 
\;+\; \mbox{regular terms near $k^2\sim \mu_P^2$}
\eeq
with a constant $a$, which is the typical functional form
for a self-energy diagram with $P$ emitting and reabsorbing a
photon or gluon.
In this case, the difference between the exact self-energy and the
expansion is given by
\beq\label{eq:ccexpansion}
\left[\Sigma(M_P^2)-\ri M_P\Gamma_P \Sigma'(M_P^2)\right]
-\Sigma(\mu_P^2)
= \ri a \frac{\Gamma_P}{M_P}
\;+\; a\,{\cal O}(\Gamma_P^2 \ln\Gamma_P)
\eeq
in the limit $\Gamma_P\ll M_P$, \ie it is of two-loop order and thus
of one order higher than nominally required. Substituting
\refeq{eq:ccexpansion} into \refeq{eq:selffactorappr} shows that
$f(k^2)$ does not have one-loop accuracy anymore. The failure of the
expansion can be easily corrected by adding the missing term back to
the expanded counterterm:
\begin{align}
\label{eq:expansioncorr}
\Sigma(\mu_P^2)&{}=\Sigma(M_P^2)+(\mu_P^2-M_P^2)
\left[\Sigma'(M_P^2)+ \frac{a}{M_P^2}\right]\;+\; {\cal O}\left((\mu_P^2-M_P^2)^2\right)\nn\\
&{}=\Sigma(M_P^2)-\ri M_P\Gamma_P
\Sigma'(M_P^2) -\ri a\frac{\Gamma_P}{M_P}
\;+\; {\cal O}
\left((M_P\Gamma_P)^2\right).
\end{align}
In this way the counterterms can be consistently expressed in terms of
self-energies at real momentum arguments.

The described procedure can be applied to the CMS in
the SM as follows. The gauge-boson self-energies 
at the complex pole positions can be approximated as
\begin{align}\label{eq:exp_ons_SE}
\Si^{\FW}_{\rT}(\cmws) &{}= \Si^{\FW}_{\rT}(\MW^2) +  (\cmws-\MW^2)
\Si^{\prime \FW}_{\rT}(\MW^2)  + c^{\FW}_{\rT} + \Oaaa,\nl
\Si^{\FZ\FZ}_{\rT}(\cmzs) &{}= \Si^{\FZ\FZ}_{\rT}(\MZ^2) +  (\cmzs-\MZ^2)
\Si^{\prime \FZ\FZ}_{\rT}(\MZ^2) + \Oaaa,\nl
\frac{1}{\cmzs}\Si^{\FA\FZ}_{\rT}(\cmzs) &{}= 
\frac{1}{\cmzs}\Si^{\FA\FZ}_{\rT}(0)
+\frac{1}{\MZ^2}\Si^{\FA\FZ}_{\rT}(\MZ^2)  
-\frac{1}{\MZ^2}\Si^{\FA\FZ}_{\rT}(0) + \Oaa.
\end{align}
The constant
\beq
c^{\FW}_{\rT} \;=\; \frac{\ri\alpha}{\pi} \MW\GW
 \;=\; \frac{\alpha}{\pi}(\MW^2-\cmws)
\eeq
compensates for the failure of the expansion of the photon-exchange
diagram in the W-boson self-energy as described above.

By neglecting the  $\Oaa$  terms in the expansion of the mixing energy
and the $\Oaaa$ terms in the diagonal self-energies in \refeq{eq:exp_ons_SE}, we can 
replace Eqs.~\eqref{exact-complex-ren-const-mass} and 
\eqref{exact-complex-ren-const-field} by
\begin{alignat}{3}
\label{complex-ren-const-mass}
\de\cmws &{}={} 
\rlap{$\Si^{\FW}_{\rT}(\MW^2)+  (\cmws-\MW^2)
\Si^{\prime \FW}_{\rT}(\MW^2) + c^{\FW}_{\rT}$,} \nl 
\de\cmzs &{}={} 
\rlap{$\Si^{\FZ\FZ}_{\rT}(\MZ^2)+  (\cmzs-\MZ^2)
\Si^{\prime \FZ\FZ}_{\rT}(\MZ^2),$} 
\\
\label{complex-ren-const-field}
\de \cZ_{\FZ\FA} &{}= \frac{2}{\cmzs}\Si^{\FA\FZ}_{\rT}(0), \qquad&
\de \cZ_{\FA\FZ} &{}= -\frac{2}{\MZ^2}\Si^{\FA\FZ}_{\rT}(\MZ^2)
+\left(\frac{\cmzs}{\MZ^2}-1\right) \de \cZ_{\FZ\FA}
, \nl
\de \cZ_{\FW} &{}= - \Si^{\prime \FW}_{\rT}(\MW^2), \qquad&
\de \cZ_{\FZ\FZ} &{}= -\Si^{\prime \FZ\FZ}_{\rT}(\MZ^2).
\end{alignat}
The missing $\Oaa$ terms in $\de \cZ_{\FW}$, $\de \cZ_{\FZ\FZ}$, and
$\de \cZ_{\FA\FZ}$ do not influence our
results, since the gauge-boson field renormalization constants drop
out as there is no external unstable gauge boson allowed when using
the CMS, and the missing (finite) $\Oaaa$ terms in the mass
counterterms are beyond the accuracy of a one-loop calculation. The
counterterms \refeqf{complex-ren-const-mass} and
\refeqf{complex-ren-const-field} involve only functions
that appear also in the usual OS renormalization
scheme~\cite{Denner:1991kt,Aoki:1980ix}, but consistently take into
account the imaginary parts.

After inserting the counterterms \eqref{complex-ren-const-mass} and
\eqref{complex-ren-const-field} into
\refeq{eq:ren-se}, we can
rewrite the renormalized self-energies in the CMS as
\begin{align}
\label{eq:ren-se-cms}
\Si^{\FW}_{\ren,\rT}(k^2) &{}= \Si^{\FW}_{\rT}(k^2) - \de\MW^2
+(k^2-\MW^2)\de Z_{\FW} - c^{\FW}_{\rT}, \nl
\Si^{\FZ\FZ}_{\ren,\rT}(k^2) &{}= \Si^{\FZ\FZ}_{\rT}(k^2) - \de\MZ^2 
+(k^2-\MZ^2)\de Z_{\FZ\FZ}, \nl
\Si^{\FA\FA}_{\ren,\rT}(k^2) &{}= \Si^{\FA\FA}_{\rT}(k^2)
+ k^2 \de Z_{\FA\FA}, \nl
\Si^{\FA\FZ}_{\ren,\rT}(k^2) &{}= \Si^{\FA\FZ}_{\rT}(k^2) 
+ k^2 \frac{1}{2}\de Z_{\FA\FZ} +(k^2-\MZ^2) \frac{1}{2}\de Z_{\FZ\FA}
\end{align}
with 
\begin{alignat}{5}
\label{complex-ren-const-ons}
\de\MW^2 &{}= \Si^{\FW}_{\rT}(\MW^2), &            
\de\MZ^2 &{}= \Si^{\FZ\FZ}_{\rT}(\MZ^2), \nl
\de Z_{\FZ\FA} &{}= \frac{2}{\MZ^2}\Si^{\FA\FZ}_{\rT}(0), \qquad&            
\de Z_{\FA\FZ} &{}= -\frac{2}{\MZ^2}\Si^{\FA\FZ}_{\rT}(\MZ^2), \nl
\de Z_{\FW} &{}= - \Si^{\prime \FW}_{\rT}(\MW^2), \qquad &            
\de Z_{\FZ\FZ} &{}= -\Si^{\prime \FZ\FZ}_{\rT}(\MZ^2), &            
\de Z_{\FA\FA} &{}= -\Si^{\prime \FA\FA}_{\rT}(0).
\end{alignat}
Apart from the terms $ c^{\FW}_{\rT}$, \refeq{eq:ren-se-cms} with
\refeq{complex-ren-const-ons} have exactly the form of the
renormalized self-energies in the usual OS scheme, but without taking
the real parts of the counterterms.  While in the OS scheme for stable
particles the self-energies are calculated in terms of the real
renormalized masses $\MZ^2$ and $\MW^2$, in \refeqs{eq:ren-se-cms} and
(\ref{complex-ren-const-ons}) the self-energies are to be calculated
in terms of the complex internal masses $\cmzs$ and $\cmws$, although
with real squared momenta.  Note that this difference between usual OS
and complex renormalization also changes the form of the IR divergence
appearing in the W-field renormalization constant $\de Z_{\FW}$.  In
the former scheme, it appears as logarithm $\ln m_\gamma$ of an
infinitesimally small photon mass [or as the related $1/(4-D)$ pole in
DR]; in the latter, the W~width regularizes the singularity via $\ln\GW$.

The renormalization of  the complex weak mixing angle is given by
\refeq{eq:ren-mixing-angle} with mass counterterms from
\refeq{complex-ren-const-mass}. The renormalization of the 
electric charge stays the same as in 
\refeq{eq:ren-charge}.

For the (neutral and colourless) Higgs boson, the approximate
renormalization works as for the generic scalar $P$ discussed above.
The renormalization constants can be approximated as
\beq
\label{eq:exact-complex-ren-const-Higgs}
\de\cmhs = 
\Si^{\FH}(\MH^2) +  (\cmhs-\MH^2) \Si^{\prime \FH}(\MH^2) + \Oaaa,
\qquad
\de \cZ_{\FH} =  - \Si^{\prime \FH}(\MH^2) + \Oaa,
\eeq
so that the renormalized Higgs-boson self-energy up to finite $\Oaa$ terms
can be written as
\beq
\Si^{\FH}_\ren(k^2) = \Si^{\FH}(k^2) - \de\MH^2 +(k^2-\MH^2) \de Z_{\FH}
\eeq
with
\beq
\de\MH^2 = \Si^{\FH}(\MH^2),
\qquad
\de Z_{\FH} = - \Si^{\prime \FH}(\MH^2).
\eeq
For a SM Higgs boson with mass $\MH=125\GeV$, as determined by the
LHC experiments, the expansion parameter is $\GH/\MH\sim 3\times 10^{-5}$
and, thus, extremely small.
Note, however, that 
for large Higgs-boson masses ($\MH\gsim400\GeV$) the Higgs-boson width grows
drastically, so that the expansion of the mass counterterm
for $\GH/\MH\to0$ would not be justified anymore.

Expanding the self-energies appearing in the renormalization constants
\refeqf{eq:complex-ren-const-Top-mass} for an unstable fermion~$f$ about $\Mf^2$
and neglecting (UV-finite) terms of order $\Oaaa$ in the mass
counterterm and of order $\Oaa$ in the field renormalization constant,
the renormalized fermion self-energy can be expressed as
\begin{align}
\Sigma^{\Ff}_\ren(p) ={}& 
  \left\{\Sigma^{\Ff,\mathrm{R}}(p^2)+\de Z^{\Ff,\mathrm{R}}\right\}\slashed{p}\omega_+
+ \left\{\Sigma^{\Ff,\mathrm{L}}(p^2)+\de Z^{\Ff,\mathrm{L}}\right\}\slashed{p}\omega_-
\nn\\ & {}
+ \cmf\Biggl\{ \Sigma^{\Ff,\mathrm{S}}(p^2) 
- {\frac{1}{2}}(\de Z^{\Ff,\mathrm{R}} + \de Z^{\Ff,\mathrm{L}})
-\frac{\de \Mf}{ \Mf} 
+\ri \Mf\Gf \left[  {\frac{1}{2}}\Sigma^{\prime \Ff,\mathrm{R}}( \Mf^2)
             +{\frac{1}{2}}\Sigma^{\prime \Ff,\mathrm{L}}( \Mf^2)
            +\Sigma^{\prime \Ff,\mathrm{S}}( \Mf^2) \right]
-c^f \Biggr\}
\end{align}
with
\begin{align}
\de \Mf &{}= \frac{ \Mf}{2}
\left[  \Sigma^{\Ff,\mathrm{R}}( \Mf^2)
       +\Sigma^{\Ff,\mathrm{L}}( \Mf^2)
      +2\Sigma^{\Ff,\mathrm{S}}( \Mf^2) \right],
\nn\\
\de Z^{\Ff,\sigma} &{}= -\Sigma^{\Ff,\sigma}( \Mf^2)
- \Mf^2\left[  \Sigma^{\prime \Ff,\mathrm{R}}( \Mf^2)
             +\Sigma^{\prime \Ff,\mathrm{L}}( \Mf^2)
            +2\Sigma^{\prime \Ff,\mathrm{S}}( \Mf^2) \right], \qquad
\sigma=\mathrm{R,L}.
\hspace{2em}
\end{align}
Note that for QCD corrections, in case $f$ is a quark $q$,
the neglected terms are of order
$\ord{(\alphas^3)}$ or $\ord{(\alphas^2)}$, respectively.
The constant $c^f$ again originates from the
non-analytic terms from photon and gluon exchange, as explained above.
Taking EW and QCD corrections into account, it reads
\beq
c^\Ff = \frac{\ri\alpha \Qf^2}{\pi} \frac{\Gf}{ \Mf}
+ \frac{\ri\alpha_{\mathrm{s}}C_{\mathrm{F}}}{\pi} \frac{\Gf}{ \Mf} \delta_{\Ff\Fq}
\eeq
with the electric charge $\Qf$, the constant $C_{\mathrm{F}}=4/3$, and
$\delta_{\Ff\Fq}$ is $0$ or $1$ if $\Ff$ is a lepton or a quark,
respectively.

In summary, the calculation of self-energies with complex momentum
arguments can be avoided by carefully expanding these about real
values.  In case of charged or coloured particles, extra constants
must be added to the expanded self-energies in order not to spoil the
one-loop accuracy of the results. All the mass arguments of the
self-energies are complex, and no real parts should be taken.

\subsubsection{Input parameters to the complex-mass scheme}
\label{se:CMSinput}

In the above description we did not yet clarify how the width of an
unstable particle~$P$ should be chosen in the CMS. 
Note that the width $\GP$ is part of the renormalized parameter $\mu_P$,
but not an independent parameter of the theory. It is rather determined
by the theory and thus calculable,
either from the decay processes of $P$ or equivalently via unitarity cuts
from the imaginary part of its self-energy,
\beq
\MP\GP =\Im\left\{\Si(\MP^2-\ri\MP\GP )\right\}.
\label{eq:GPrelation}
\eeq
In principle, this equation can be iteratively solved for $\GP$,
but the result would only be of LO accuracy if $\Sigma$ is of NLO,
because imaginary parts of one-loop self-energies contain only
tree-level information about particle widths via unitarity cuts.
In order to obtain full NLO accuracy in the cross section near 
the resonance, however, NLO corrections to $\GP$ are required. 
This is obvious, because near resonance the offshellness of the
propagator with virtuality $k^2$ and the width part are of the same size,
$|k^2-\MP^2|={\cal O}(\MP\GP)$.
Equation~(\ref{eq:GPrelation}) should, thus, be solved for $\GP$ 
using self-energy contributions up to the two-loop level,
or alternatively $\GP$ can be calculated directly from decay amplitudes
at NLO.

This asymmetry in the loop level between input preparation and
matrix-element calculation in the CMS deserves further justification:
For each unstable particle mass, in the CMS we add and subtract the same
imaginary part in the Lagrangian. One of these terms provides the
imaginary part for the mass parameter and becomes part of the free
propagator, while the other becomes part of a counterterm vertex.
Thus, the first term is resummed, but the second is not.
Independently of the imaginary part that is added and subtracted,
this procedure does not spoil the algebraic relations that govern
gauge invariance, and unitarity cancellations 
(not to be confused with the full cut equations expressing unitarity)
are exactly respected.
In practice, this means that we can insert values for the
particle widths that are not directly related to the loop
order to which the amplitudes of the process are calculated. We
could even go beyond NLO in the calculation of the widths or
take an empirical value.
On the other hand, the validity of the  cut equations at NLO requires
to use the NLO decay widths \cite{Denner:2014zga}.

Note that this argument generalizes to all (independent) input
parameters, real or complex in the CMS: Changing the input parameter
or the corresponding counterterm by terms that affect amplitude
calculations beyond NLO precision, is a legal procedure in the sense
that it does not introduce any inconsistencies such as violating gauge
invariance or unitarity cancellations.

Similarly, the numerical setting of the electromagnetic coupling
deserves some care. Owing to the use of complex masses in the loop
corrections that enter the charge renormalization constant and the
fact that the bare elementary charge is real, the renormalized 
electromagnetic coupling $e$ actually becomes complex.
The imaginary part of $e$ is not a free input parameter, but
determined by the charge renormalization constant, \ie it
could be iteratively calculated.
Note, however, that the imaginary part of $e$ is entirely due to
spurious terms, since the charge renormalization constant only
involves self-energies at zero momentum transfer, which do not 
develop imaginary parts for real internal masses. The imaginary
parts in $e$ are, thus, of formal two-loop order.
Following the above argument, 
we conclude that it is consistent within NLO accuracy to
set the imaginary parts in $e$ to zero.
We also remark that using a complex electromagnetic coupling $e$
would disturb the cancellation of IR divergences between
virtual and real EW corrections, because those two types of corrections
only involve exactly the same power $e^n$ in the 
contributions to squared amplitudes
if $e$ is real, while the powers $e^{n_1} (e^*)^{n_2}$ for a complex $e$
would be different in spite of the same values of $n=n_1+n_2$.

The situation does not change much if the coupling $\alpha$ is not
fixed to the fine-structure constant $\alpha(0)$, but derived from a
high-energy value $\alpha(Q^2)$ or from the Fermi constant $\GF$ in
the $\GF$ scheme as described in \refse{se:input_reco}.  In the latter
case, the input value $\al_{\GF}$ should be calculated from the real
$\PW$ and $\PZ$ masses to avoid spurious terms of $\Oa$ in
$\al_{\GF}$. In this way the input conversion is done in the same way
as in the real OS scheme for stable particles, and the quantity $\De
r$, which modifies the charge renormalization constant $\de Z_e$, is
calculated from real input parameters as well.  Again, following the
above general argument about changing input parameters or
corresponding renormalization constants beyond NLO, it would also be
legal to calculate $\De r$ from complex masses, etc.\ (but keeping
$\al_{\GF}$ real), because the corresponding change in $\De r$ is of
two-loop order.  Other variants to get rid of the
complex phase of $\alpha$ are described in
\citeres{Frederix:2018nkq,Buccioni:2019sur}.

\subsubsection{Complex
  \texorpdfstring{renormalization---background-field
    method}{renormalization - background-field method}}

Using the background-field method, the gauge-boson field
renormalization constants can be determined in terms of the parameter
renormalization in such a way that Ward identities possess the same form
before and after renormalization \cite{Denner:1994xt},
as explained in \refse{se:bfWI}.
Real parameters have to be substituted by the corresponding complex
parameters everywhere when the complex renormalization is employed.
The complex parameter renormalization is fixed as above in
Eqs.~\eqref{exact-complex-ren-const-mass}, \eqref{eq:ren-mixing-angle},
\eqref{eq:ren-charge}, \eqref{eq:complex-ren-const-Higgs-mass}, and
\eqref{eq:complex-ren-const-Top-mass} or likewise by their 
simplified versions described in \refse{se:simplifiedcomplexren}.
Note that
$\Si^{\FA\FZ}_{\rT}(0)$ vanishes in the background-field method as a
consequence of the background-field gauge invariance of the effective
action, which in particular simplifies the charge renormalization
constant \eqref{eq:ren-charge}.

Since the gauge-boson field renormalization constants drop out in the
$S$-matrix elements without external gauge-boson fields, we can
also use the definitions in
\refeq{complex-ren-const-field} in the calculation of $S$-matrix
elements.

\subsection{Further schemes for unstable particles}
\label{se:unstableparts-other}

To our experience, pole scheme, pole approximation, and CMS
are the most frequently used methods to describe resonance processes beyond
the NWA, especially as far as the calculation of EW corrections is concerned.
There are also other methods to deal with unstable-particle effects
in resonance processes with different strengths and weaknesses which were
suggested and used in the literature. 
In the following we briefly sketch some of those alternative approaches,
without claiming to be exhaustive.

\myparagraph{(a) Factorization schemes} 
Different variants of factorizing resonance structures from amplitudes have
been proposed in the literature, but they all share the idea
to separate a simple resonance factor from complete (gauge-invariant) 
amplitudes or from gauge-invariant subsets of diagrams.
For pure LO predictions, this idea can be implemented easily: For each
potentially resonant propagator factor $(p^2-\MP^2)^{-1}$ of a
resonance~$P$, multiply the amplitude with the {\it fudge factor}
$f_P(p^2)=(p^2-\MP^2)/(p^2-\mu_P^2)$.  In the resonance region, this
factor restores the correct Breit--Wigner factor. On resonance the
non-resonant terms are put to zero and off resonance, where
$|p^2-\MP^2|\gg\MP\GP$, the amplitude is modified by a factor
$f_P=1+{\cal O}(\GP/\MP)$, which changes the result only at formal NLO
level.
LO results on W-pair production in $\Pep\Pem$ annihilation 
based on this scheme can, for instance, be found in
\citeres{Kurihara:1994fz,Argyres:1995ym}.

Beyond LO, however, it can be quite non-trivial to guarantee 
the aimed precision (\eg NLO) everywhere in phase space and to
match virtual and real corrections.
Simply modifying LO cross sections with fudge factors containing
decay widths for resonances in general introduces spurious ${\cal O}(\GP/\MP)$
terms destroying NLO accuracy.
The most simple example in which the method works perfectly well is the case
where the LO amplitude receives only contributions with a common resonance
structure without any non-resonant or subleading background diagrams.
In such cases, the relative correction to the LO amplitude does not involve
any resonance factors.
The Drell--Yan-like production of W~bosons at hadron colliders is such a 
fortunate case; the first calculation of NLO EW corrections to this
process was based on the factorization scheme~\cite{Dittmaier:2001ay}.
Owing to the non-resonant background by photon exchange, the application
of the factorization scheme to
Drell--Yan-like Z-boson production is already more complicated.
For NLO weak corrections such a calculation was described in
\citere{Dittmaier:2009cr}, and corresponding results were already
included in the comparison of schemes in \refse{se:DY-weak-PS} above.

\myparagraph{(b)
  Schemes based on resummation and the {\it fermion-loop scheme}}
  As discussed at the very beginning of this section, standard
  perturbation theory cannot describe particle resonances in any
  finite order, but requires at least some partial Dyson summation of
  self-energy contributions in the propagators of unstable particles.
  Although it would be straightforward to consistently construct all
  amplitudes from Dyson-summed propagators, the necessary truncation
  of the perturbative series when calculating irreducible vertex
  functions at some finite order in general invalidates this
  procedure, because consistency relations from gauge invariance and
  unitarity usually hold order by order and get violated in
  perturbative orders that are not completely taken into account.

In view of this situation, it seems natural to look for the technical possibility
to reconcile Dyson summation and constraints from gauge invariance and unitarity.
In the case of unstable particles that exclusively decay into fermion--antifermion
pairs at LO---a case that in particular includes W and Z~bosons---such 
a procedure is provided by the 
{\it fermion-loop scheme}~\cite{Argyres:1995ym,Beenakker:1996kn,Passarino:2000mt},
which was suggested and used to describe W- and Z-boson pair production processes
at LEP2.
In this scheme, all one-loop corrections induced by closed fermion loops are
Dyson summed to all orders, but no other corrections are taken into account.
This procedure obviously introduces all necessary finite-width effects in 
propagators of unstable particles decaying only into fermion--antifermion
pairs. Taking into account {\it all} closed fermion loops, \ie 
also in vertex corrections etc., however, preserves all Ward identities 
in amplitudes and subamplitudes even in the presence of resummed propagators
and does not introduce any dependence on gauge parameters.
This statement is quite non-trivial and originates from the fact that 
closed fermion loops always provide gauge-invariant subsets of
radiative corrections in each perturbative order
(see \refse{se:split_ew_qed}).%
\footnote{One possibility to establish the validity of the Ward
identities for amplitudes with propagators based on Dyson-summed
closed-fermion-loop corrections is to start from the background-field
Ward identities for connected Green functions established in 
\citere{Denner:1996gb} and to reduce the ${\cal O}(\alpha)$ corrections
to the closed fermion loops, which can be separated in a gauge-invariant way
and which are identical in the conventional and background-field approaches.}
Predictions based on the fermion-loop scheme, thus, provide fully consistent
cross sections at LO accuracy improved by the higher-order effects induced
by closed fermion loops, such as running effects in the electromagnetic coupling
or leading corrections in the $\rho$-parameter. 

Promoting those results to full NLO precision, however, is highly non-trivial
and has not yet been accomplished.
Some steps or field-theoretical statements of such a procedure are known though.
For example, in \citere{Denner:1996gb} it was shown that even the full 
${\cal O}(\alpha)$ one-loop corrections to amplitudes can be Dyson summed
without violating Ward identities if the field theory is quantized within the
background-field method (see \refse{se:bfm}), which is a straightforward
consequence of the gauge invariance of the background-field effective action.
This feature also lifts the restriction of the method to unstable
particles decaying only into fermion--antifermion pairs.%
\footnote{Recall that irreducible vertex functions in the BFM still
  depend on some quantum gauge parameter $\xiQ$, although BFM vertex
  functions fulfil the simple BFM Ward identities exactly. The
  dependence on $\xiQ$ drops out in complete amplitudes order by
  order. If Dyson-summed propagators are used, however, the
  cancellation of the $\xiQ$ dependence only happens within completely
  calculated orders, \ie a corresponding NLO calculation based on
  resummed propagators still shows some $\xiQ$ dependence of formal
  NNLO level.  If $\xiQ$ is not taken ``artificially large'', this
  residual dependence should not be too harmful. Since Ward identities
  are fulfilled exactly, there are no systematic enhancements of those
  artifacts by violation of unitarity cancellations.}

The most serious show-stopper in the Dyson-summation approach towards
complete higher orders seems to be the fact that the introduction of
widths via imaginary parts of self-energies lags behind by one
perturbative order.  Even taking into account the full set of ${\cal
  O}(\alpha)$ corrections in Dyson-summed amplitudes lacks the ${\cal
  O}(\alpha)$ corrections to total decay widths in propagator
denominators.  Moreover, the cancellation of IR singularities against
real-emission corrections might
require a non-trivial modification in such calculations.

\myparagraph{(c) Effective field theories} 
Effective field theories (EFTs) generically can describe physical systems consisting
of two or more components that are characterized by different energy scales
$\mu_i$ obeying some hierarchy $\mu_1\ll\mu_2\ll\dots$.
Field theories with one or more narrow resonances~$P$ with $\GP\ll\MP$
fulfil this requirement, rendering the formulation of EFTs
for unstable particles possible~\cite{Beneke:2003xh,Beneke:2004km,Hoang:2004tg}.
In the EFT approach to describe an unstable particle~$P$, an
effective Lagrangian is constructed that separates hard, collinear, and
soft degrees of freedom in the fields.
The splitting of modes corresponds to the different momentum regions in Feynman 
diagrams of the full theory as defined by the 
{\it strategy of regions}~\cite{Beneke:1997zp,Smirnov:2002pj}.
Particles with hard momenta (high-energetic, non-resonant, or massive particles of some
energy scale $|q^\mu|\sim\MP$) do not represent dynamical degrees of freedom in the EFT.
They are ``integrated out'', so that their effect is included through the Wilson coefficients 
of the operators describing $P$~production and decay. 
The soft momentum region, on the other hand,
accounts for off-shell modes of $P$ ($|p^2-\MP^2|\sim\MP\GP$)
and the exchange or emission of soft massless gauge bosons ($|q^\mu|\sim\GP$).
The domains of collinear momenta (with a non-trivial
hierarchy of its light-cone components in terms of $\GP$)
are needed to consistently include the interaction
of massless gauge bosons with light high-energetic particles (such as quarks or leptons).
The contributions of the different momentum regions involve artificial
UV and IR divergences which in their sum combine to a systematic expansion of
amplitudes in powers of $\GP/\MP$, similar to the result of a direct pole
expansion of Feynman diagrams. In particular, the leading contribution of the soft momentum region
corresponds to non-factorizable corrections in the PA.

In summary, the EFTs of \citeres{Beneke:2003xh,Beneke:2004km,Hoang:2004tg} 
deliver a field-theoretically elegant way to carry out pole expansions
owing to their formulation via effective actions.
Like the PA, their validity is restricted to the resonance region,
but they offer the combination with further expansions, \eg around thresholds,
and suggest better possibilities to carry out dedicated resummations
of higher-order effects.
The EFT shows limitations or complications if detailed
information on differential properties of observables is needed, since
experimental degrees of freedom do not correspond to the degrees of
freedom of the EFT.
For $\Pep\Pem$ collisions the EFT approach was, for instance,
used to evaluate NLO EW corrections to the W-pair production cross section
near the $\PW\PW$ threshold~\cite{Beneke:2007zg,Actis:2008rb}, 
including leading EW higher-order effects beyond NLO.

\section{Conclusions}
\label{se:conclusions}

After the discovery of a Higgs boson at the CERN LHC in 2012,
the full particle content of the SM of particle physics
is experimentally established. Moreover, all particle phenomena of the
strong and electroweak interactions observed at colliders are nicely
described by the SM without truly significant deviations.
In other words, any effects of physics beyond the 
SM---if accessible at all by current and future collider experiments---are 
small and subtle, putting the aspect of precision at the forefront
of experimental analyses and theory predictions. 
The inclusion of perturbative corrections in predictions
for present and future collider experiments
is at the heart of this task on the theory side.
Besides the indispensable corrections of the strong interaction, 
{\it electroweak radiative corrections} must be taken into account as
well. While their generic size is at the level of a few percent and
thus roughly comparable to the NNLO QCD corrections, they can be
enhanced via different mechanisms. The inclusion of NLO
electroweak corrections is becoming standard in the experimental analyses.

This review gives an overview of the most important,
established techniques that are presently used in the calculation of
electroweak corrections. While we have included some discussion on the
resummation of the leading logarithmic corrections in the high-energy
limit, in parton distribution functions, and structure functions
for photonic interactions, we
mainly focus on the methods for NLO electroweak calculations. In this
realm, a set of agreed-upon, well-developed, and efficient approaches
exists both for the calculation of virtual and real contributions. On
the other hand, the calculation of NNLO corrections, in particular for
the electroweak parts, is still in an exploratory phase.

The aim of this review is to provide a coherent introduction to the
well-established and most frequently used concepts and techniques for
the calculation of NLO electroweak corrections.  In detail, we have
spelled out the renormalization of the SM in the on-shell scheme and
discussed frequently used input-parameter schemes as well as the
structure of the NLO electroweak corrections. An overview over the
modern techniques for the calculation of electroweak one-loop
corrections and the available automated tools is given. As far as real
corrections are concerned, slicing and subtraction techniques for the
treatment of infrared singularities are described in some detail, and
applications to the electromagnetic corrections to parton distribution
functions, peculiarities of photon--jet systems, as well as real
radiation effects in lepton--photon systems are discussed.  Moreover,
a section is devoted to the treatment of unstable particles in quantum
field theory, in particular in connection with the calculation of NLO
electroweak corrections---a subject that is to our knowledge not
covered in such a coherent form in the literature yet.  Finally, the
appendix contains a complete list of Feynman rules with counterterms
for the SM both in the conventional and in the background-field
formalism, which might be helpful for practitioners.

While this review is primarily targeted to the description of the
basic concepts and
calculational techniques, we have included typical phenomenological
applications for illustration of generic features
whenever appropriate. 
A comprehensive account of the status of electroweak
corrections for collider processes is, however, way beyond the scope
of this review.
Moreover, extrapolating the great progress of recent years in the field,
such an account would be outdated very soon.
Regular compilations and updates of important precision calculations for
collider processes may be found in workshop proceedings such as
those of the bi-annual Les Houches workshop on 
{\it Physics at TeV Colliders}.

We end this review with a brief assessment of the current state of the
art concerning the calculation of electroweak corrections for
high-energy colliders and possible future directions.
With the present techniques, NLO calculations for processes with up to
six particles in the final state are achieved. We expect that
further refinements of the methods will allow us to push this limit even
somewhat further. A more pressing issue is, however, to provide
interfaces of NLO electroweak calculations with general-purpose
Monte Carlo generators including electromagnetic or even
electroweak parton showers, to
bring electroweak precision more directly into the analyses of
experimental data. Another line of development, but also beyond the 
reach of this review, concerns 
electroweak precision in the search for physics beyond the SM
which is currently pushed forward in two different directions.
On the one hand, electroweak corrections are calculated in various 
specific extensions of the SM, a task that is technically
straightforward, but requires great care in the formulation of
phenomenologically sound renormalization and input-parameter schemes. 
On the other hand, physics beyond the SM is searched for in the more
model-independent framework of effective field theories, which
poses new challenges in the renormalization procedure because of the
lack of renormalizability.
Currently, the field of electroweak precision calculations develops
rapidly in both directions.

Looking further ahead, potential future high-energy colliders
will implicate new theoretical challenges on the electroweak
precision frontier. 
If high-energy electron--positron colliders are realized, 
electroweak corrections at the NNLO level
and higher will be required. At present, such results are only
available for selected low-multiplicity processes, such as
pseudo-observables at the Z-boson resonance, but for future $\Pep\Pem$
colliders electroweak two-loop calculations have to become standard
for the most important processes.
In particular, this program requires
considerable advancements in the calculation of multi-loop integrals
with many scales, where the most promising road presently seems to be
numerical integration. 
If multi-TeV colliders, such as future circular hadron colliders, 
become reality, the higher scattering energies bring in further challenges.
Extrapolating the large electroweak
corrections appearing in the TeV range even to higher energies,
makes it clear that electroweak effects have to be controlled way
beyond NLO accuracy.
In particular, electroweak corrections from multiple W- and
Z-boson radiation will become more and more important, rendering it
more and more difficult to isolate different types of 
hard scattering processes. 
Eventually, we might face the problem of refining
the definition of appropriate observables.

In summary, looking back at the previous decades and looking ahead
to future challenges, the hunt for higher precision in collider
physics goes hand in hand with the need for a deeper understanding 
of quantum field theory, which is part of the charm that pushes us forward.


\appendix


\providecommand*\FRvskip{\vspace{-1.4em}}

\appendix

\section{Feynman rules}
\label{se:feynman_rules}

\newcommand{\dth}{\frac{e}{2s}\frac{\delta t}{\MW\MH^2}}

In this appendix we list the Feynman rules for the SM both in the
conventional and in the background-field formalism. 
We provide the complete set of propagators and vertices by listing
possible actual insertions for generic Feynman rules.
In the vertices all momenta and fields are defined as incoming.
For brevity we use the shorthand notation
\beq
c=\cw=\cos\theta_\rw, \qquad
s=\sw=\sin\theta_\rw.
\eeq

\begin{fleqn}
\section*{Feynman rules for QCD}

In the following we list the Feynman rules for QCD in the conventional
formalism in the $R_\xi$ gauge including the counterterms resulting
from the renormalization of parameters, gluon fields, and quark fields. We
do not include counterterms from the gauge-fixing term, which does not
need to be renormalized in linear gauges, and do not provide
counterterms for the vertices involving Faddeev--Popov fields as these
are not needed to render one-loop $S$-matrix elements finite.
\myparagraph{Propagators} 
\begin{myaitemize}
\item 
gluons ($A,B$ are colour indices of the adjoint representation)
\beq\label{FR:Gprop}
\vcenter{\hbox{\includegraphics[page=1]{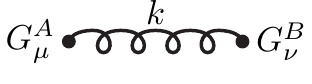}}} =
-\ri\left[\left(g_{\mu \nu }-\frac{k_{\mu }k_{\nu }}{k^{2}}\right)+
  \xi\frac{k_{\mu }k_{\nu }}{k^{2}}\right] \frac{\delta
  ^{AB}}{k^{2}+\ri\epsilon} 
\eeq

\item 
Faddeev--Popov ghosts
\beq\label{FR:uprop}
\vcenter{\hbox{\includegraphics[page=2]{FRQCDdiagrams.pdf}}}
 =\frac{\ri\delta^{AB}}{k^{2}+\ri\epsilon}
\eeq

\item 
quarks ($\al,\be$ are colour indices of the fundamental representation)
\beq\label{FR:qprop}
\vcenter{\hbox{\includegraphics[page=3]{FRQCDdiagrams.pdf}}}
 =\frac{\ri\de_{\al\be}(\slashed{k}+m_q)}{k^2-m_q^2+\ri\epsilon }
\eeq
\end{myaitemize}

\myparagraph{Vertices} 
\begin{myaitemize}
\item 
$\FG\FG$ vertex
\beq\label{FR:GG}
\vcenter{\hbox{\includegraphics[page=4]{FRQCDdiagrams.pdf}}}
= \ri \left(-g_{\mu\nu}k^2+ k_\mu k_\nu\right)\de^{AB}\DZG
\eeq 
\item 
$\Fqbar\Fq$ vertex
\beq\label{FR:FF}
\vcenter{\hbox{\includegraphics[page=5]{FRQCDdiagrams.pdf}}}
= \ri\Bigl[(\slashed{k}-m_q)\DZq - \de \Mq\Bigr]\de_{\al\be}
\eeq 

\item 
$GGGG$ vertex
\beq\label{FR:GGGG}
\vcenter{\hbox{\includegraphics[page=6]{FRQCDdiagrams.pdf}}}
=
-\ri \gs^{2}\Bigl(1+2\DZgs+2\DZG\Bigr)
\barr[t]{lll}
\left[f^{A_{1}A_{2}B}f^{A_{3}A_{4}B}(g_{\mu_{1}\mu_{3}}g_{\mu_{2}\mu_{4}}
-g_{\mu_{1}\mu_{4}}g_{\mu_{2}\mu_{3}})\right.\\
\left.\:{}+f^{A_{1}A_{3}B}f^{A_{4}A_{2}B}(g_{\mu_{1}\mu_{4}}g_{\mu_{3}\mu_{2}} -
g_{\mu_{1}\mu_{2}}g_{\mu_{3}\mu_{4}})\right.
\\
\left.
\:{}+f^{A_{1}A_{4}B}f^{A_{2}A_{3}B}(g_{\mu_{1}\mu_{2}}g_{\mu_{4}\mu_{3}} -
g_{\mu_{1}\mu_{3}}g_{\mu_{4}\mu_{2}})\right]
\earr
\eeq 

\item 
$GGG$ vertex
\beq
\vcenter{\hbox{\includegraphics[page=7]{FRQCDdiagrams.pdf}}}
=
\barr[t]{l}
\gs\Bigl(1+\DZgs+\frac{3}{2}\DZG\Bigr)
f^{A_{1}A_{2}A_{3}}
\left[g_{\mu_{1}\mu_{2}}(k_{1} -k_{2})_{\mu_{3} }
+ g_{\mu_{2}\mu_{3} }(k_{2}-k_{3})_{\mu_{1} } 
+g_{\mu_{3} \mu_{1} }(k_{3}-k_{1})_{\mu_{2}}\right]
\earr
\label{FR:GGG}
\raisetag{4ex}
\eeq

\item 
$G \Fqbar q$ vertex
\beq\label{FR:Gqq}
\vcenter{\hbox{\includegraphics[page=8]{FRQCDdiagrams.pdf}}}
=\textstyle
\ri \gs \Bigl(1+\DZgs+\frac{1}{2}\DZG + \DZq \Bigr)\gamma_\mu T^{A}_{\al\be}
\eeq

\item 
$\FG\FUbar\FU$ vertex
\beq\label{FR:Guu}
\vcenter{\hbox{\includegraphics[page=9]{FRQCDdiagrams.pdf}}}
=\textstyle
 -\gs f^{A_{1}A_{2}A_{3}}k_{2,\mu} 
\eeq

\end{myaitemize}

Next we list the Feynman rules for QCD in the background-field
formalism.  The Feynman rules involving only background fields are
identical to those in the conventional formalism given in
\refeqs{FR:Gprop}--\refeqf{FR:Guu} including the counterterms. 
Only the Feynman rules involving Faddeev--Popov fields or
exactly two quantum gluon fields differ from those of the conventional
formalism. In the quantum $R_\xi$ gauge these read:

\begin{myaitemize}
\item 
$\FGhat\FGhat\FG\FG$ vertex
\beq\label{FR:GhGhGhGh}
\vcenter{\hbox{\includegraphics[page=10]{FRQCDdiagrams.pdf}}}
\barr{lll}
=-\ri \gs^{2}
\Bigl[f^{A_{1}A_{2}B}f^{A_3 A_{4}B}
(g_{\mu_{1}\mu_{3}}g_{\mu_{2}\mu_{4}}
-g_{\mu_{1}\mu_{4}}g_{\mu_{2}\mu_{3}})\\[1ex]
\phantom{=-\ri \gs^2\Bigl[}
{}+f^{A_1A_{3}B}f^{A_{4}A_{2}B}
(g_{\mu_{1}\mu_{4}}g_{\mu_{3}\mu_{2}}
- g_{\mu_{1}\mu_{2}}g_{\mu_{3}\mu_{4}}
- g_{\mu_{1}\mu_{3}}g_{\mu_{2}\mu_{4}}/\xi)\\[1ex]
\phantom{=-\ri \gs^2\Bigl[}
{}+f^{A_{1}A_{4}B}f^{A_{2}A_{3}B}
(g_{\mu_{1}\mu_{2}}g_{\mu_{4}\mu_{3}} -
g_{\mu_{1}\mu_{3}}g_{\mu_{4}\mu_{2}}
+g_{\mu_{1}\mu_{4}}g_{\mu_{3}\mu_{2}}/\xi)\Bigr] 
\earr
\raisetag{3ex}
\eeq

\item 
$\FGhat\FG\FG$ vertex
\beq\label{FR:GhGhGh}
\vcenter{\hbox{\includegraphics[page=11]{FRQCDdiagrams.pdf}}}
\barr{l}
=\gs f^{A_{1}A_{2}A_{3}}  \Bigl[g_{\mu_1 \mu_2 }(2k_{1} +k_3(1-1/\xi))_{\mu_3} \\[1ex]
\phantom{+\gs f^{A_{1}A_{2}A_{3}}\Bigl[} +g_{\mu_2 \mu_3 }(k_{2}-k_{3})_{\mu_{1} } 
-g_{\mu_3 \mu_1 }(2k_{1}+k_2(1-1/\xi))_{\mu_{2}}\Bigr] 
\earr
\eeq

\item 
$\FGhat\FGhat\FUbar\FU$ vertex
\beq\label{FR:GhGhuu}
\vcenter{\hbox{\includegraphics[page=12]{FRQCDdiagrams.pdf}}}
= \ri \gs^2 (f^{A_{1}A_{3}B}f^{A_{2}A_4B}
 +f^{A_{1}A_{4}B}f^{A_2A_3B})
g_{\mu_1\mu_2} 
\eeq

\item 
$\FGhat G\FUbar\FU$ vertex
\beq\label{FR:GhGuu}
\vcenter{\hbox{\includegraphics[page=13]{FRQCDdiagrams.pdf}}}
= \ri \gs^2 f^{A_{1}A_{3}B}f^{A_{2}A_4B}g_{\mu_1\mu_2} 
\eeq

\item 
$\FGhat\FUbar\FU$ vertex
\beq\label{FR:Ghuu}
\vcenter{\hbox{\includegraphics[page=14]{FRQCDdiagrams.pdf}}}
\barr{l}
= -\gs f^{A_{1}A_{2}A_{3}}(k_2-k_3)_{\mu} 
\earr
\eeq

\end{myaitemize}

\section*{Feynman rules for the EWSM in the conventional formalism}

We provide the Feynman rules in a general 't~Hooft gauge with a
gauge-fixing term as defined in \refeqs{eq:gf} and \refeqf{eq:Lfix}
with $\xi'_\FW=\xi_\FW$, $\xi'_\FZ=\xi_\FZ$. In this gauge all
propagators are diagonal.  We include the one-loop contributions of
the counterterms from the renormalization of parameters and fields in
the physical basis, but do not include counterterms from the
renormalization of unphysical fields and counterterms due to the
gauge-fixing terms.  The tadpole counterterms $\de t$ are given both
in the Fleischer--Jegerlehner tadpole scheme (FJTS) 
\cite{Fleischer:1980ub,Krause:2016oke,Denner:2016etu} and in the
(PRTS) scheme of \citere{Denner:1991kt}, as described in \refse{se:tadpoles}.
The respective tadpole terms are called $\dtFJ$ or $\dtPR$; whenever
$\de t$ is written without subscript FJTS or PRTS, this term
contributes in both schemes.  We do not provide counterterms
for the vertices involving Faddeev--Popov fields, as these are not
needed for $S$-matrix elements at the one-loop level.

If the following Feynman rules are used in the CMS
described in \refse{se:CMS}, only minor modifications are required.
All parameters and renormalizations constants have to be substituted
by their complex counterparts.  In this context, one should recall
that spinors and corresponding adjoint spinors are rescaled by the
same field renormalization constants[\cf\refeq{eq:CMSffieldren}], so
that both $\delta Z^{\Ff,\si}$ and $(\delta Z^{\Ff,\si})^\dagger$ are
represented by $\de \cZ^{\Ff,\si}$. Recall that we assume a unit quark
mixing matrix in the CMS.

\myparagraph{Propagators} 
\begin{myaitemize}
\item 
gauge bosons $\FV = \FA$, $\FZ$, $\FW$  ($M_\FA = 0$):
\beq\label{FR:Vprop}
\vcenter{\hbox{\includegraphics[page=1]{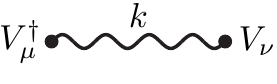}}}
= \frac{-\ri g_{\mu\nu}}{k^{2}-M_\FV^{2}+\ri\epsilon}
 +\frac{\ri(1-\xi_\FV)k_\mu k_\nu}{(k^{2}-M_\FV^{2}+\ri\epsilon)(k^{2}-\xi_\FV M_\FV^{2}+\ri\epsilon)}
\eeq

\item
Faddeev--Popov ghosts $\FU = u^{\FA}$, $u^{\FZ}$, $u^{\pm}$
($M_{u^{\FA}}=0$, $M_{u^{\FZ}} = \sqrt{\xi_\FZ}\MZ$, $M_{u^{\pm}} =
\sqrt{\xi_\FW}\MW$),
\beq
\vcenter{\hbox{\includegraphics[page=2]{FRSMconvdiagrams.pdf}}}
 = \frac{\ri}{k^{2}-M_{\FU}^{2}+\ri\epsilon} 
\eeq

\item
scalar fields $\FS = \FH$, $\chi$, $\phi$
($M_\chi = \sqrt{\xi_\FZ} \MZ$, $M_\phi = \sqrt{\xi_W} \MW$),
\beq
\vcenter{\hbox{\includegraphics[page=3]{FRSMconvdiagrams.pdf}}}
= \frac{\ri}{k^{2}-M_{S}^{2}+\ri\epsilon} 
\eeq

\item
fermion fields $\FF$,
\beq\label{FR:Fprop}
\vcenter{\hbox{\includegraphics[page=4]{FRSMconvdiagrams.pdf}}}
= \frac{\ri(\slashed{k} + m_{f,i})}{k^{2}-m_{f,i}^{2}+\ri\epsilon} 
\eeq
\end{myaitemize}

\myparagraph{Vertices and counterterms}

\begin{myaitemize}
\item tadpole
\beq
\vcenter{\hbox{\includegraphics[page=5]{FRSMconvdiagrams.pdf}}}
= \ri\delta t 
\eeq

\item $\FV\FV$ counterterm
\begin{align}
&\vcenter{\hbox{\includegraphics[page=6]{FRSMconvdiagrams.pdf}}}
= \ri \Bigl[(-g_{\mu\nu}k^2+ k_\mu k_\nu)C_{1} + g_{\mu\nu} C_{2} \Bigr]
\\
&\begin{array}[b]{lll}
\FV_1 \FV_2 & C_1 & C_2 \\\midrule
\FWp\FWm & \DZW & \DZW \MW^{2}+ \DMWS -\frac{e\MW}{s\MH^2}\dtFJ\\[1ex]
\FZ\FZ & \DZZ & \DZZ \MZ^{2}+ \DMZS  -\frac{e\MZ}{s c\MH^2}\dtFJ \\[1ex]
\FA\FZ & \frac{1}{2}\DZAZ + \frac{1}{2}\DZZA &
\frac{1}{2}\DZZA \MZ^{2} \\[1ex]
\FA\FA & \DZA & 0
\end{array}
\end{align}

\item $\FV\FS$ counterterm
\begin{align}
&\vcenter{\hbox{\includegraphics[page=7]{FRSMconvdiagrams.pdf}}}
= \ri k_\mu C
\\
&\begin{array}[b]{lll}
\FV \FS & C \\\midrule
\FW^\pm\phi^\mp & \pm\frac{1}{2}\left(\DZW + \frac{\DMWS}{\MW^2}
-\frac{e}{s\MH^2\MW}\dtFJ \right)\MW \\[1ex]
\FZ\chi & \ri \frac{1}{2}\left(\DZZ +\frac{\DMZS}{\MZ^2} 
-\frac{e}{s\MH^2\MW}\dtFJ \right)\MZ \\[1ex]
\FA\chi & \ri \frac{1}{2}\DZZA\MZ 
\end{array}
\end{align}

\item $\FS\FS$ counterterm
\begin{align}
&\vcenter{\hbox{\includegraphics[page=8]{FRSMconvdiagrams.pdf}}}
= \ri\Bigl[C_{1}k^{2} - C_{2} \Bigr]
\\
&\begin{array}[b]{lll}
\FV_1 \FV_2 & C_1 & C_2 \\
\midrule
\FH\FH & \DZH \quad &\DZH \MH^{2} + \DMHS  -\frac{3e}{2s\MW}\dtFJ \\
\barr{@{}l@{}} \chi \chi \\ \phi \phi \earr \left.\rule[-1.6ex]{0mm}{4ex}\right\}
& 0 \quad & -\frac{e}{2s\MW}\delta t 
\end{array}
\end{align}

\item $\FF\FFbar$ counterterm
\begin{align}
&\vcenter{\hbox{\includegraphics[page=9]{FRSMconvdiagrams.pdf}}}
= \ri\Bigl[\CL\slashed{k}\omega_{-} + \CR\slashed{k}\omega_{+}
- C_{\rS}^{-}\omega_{-} - C_{\rS}^{+}\omega_{+} \Bigr]
\\
&\begin{aligned}[b]
\CL ={}& \frac{1}{2}\left(\delta Z_{ij}^{f,\rL}+ \delta
Z_{ij}^{f,\rL\dagger}\right),
\qquad \CR = \frac{1}{2}\left(\delta Z_{ij}^{f,\rR}+ \delta
Z_{ij}^{f,\rR\dagger}\right)\\
C_{\rS}^{-} ={}& m_{f,i}\frac{1}{2}\delta Z_{ij}^{f,\rL}
+\frac{1}{2}\delta Z_{ij}^{f,R\dagger} m_{f,j} + \delta_{ij}\left(\delta
m_{f,i}-m_{f,i}\frac{e}{2s\MH^2\MW}\dtFJ\right)\\
C_{\rS}^{+} ={}& m_{f,i}\frac{1}{2}\delta Z_{ij}^{f,\rR}
+\frac{1}{2}\delta Z_{ij}^{f,\rL\dagger} m_{f,j} + \delta_{ij}\left(\delta
m_{f,i}-m_{f,i}\frac{e}{2s\MH^2\MW}\dtFJ\right) 
\end{aligned}
\end{align}

\item
$\FV\FV\FV\FV$ vertex:
\begin{align}
&\vcenter{\hbox{\includegraphics[page=10]{FRSMconvdiagrams.pdf}}}
= \ri e^{2}C \Bigl[2g_{\mu\nu}g_{\rho\sigma } -
g_{\mu \sigma }g_{\nu\rho} - g_{\mu\rho}g_{\nu\sigma}\Bigr]
\\
&\begin{array}[b]{ll}
\FV_{1} \FV_{2} \FV_{3} \FV_{4}  &  C  \\
\midrule
\FWp \FWp \FWm \FWm & \frac{1}{s^{2}}
\Bigl[1+2\DZe -2\frac{\delta s}{s}+ 2\DZW
\Bigr]\\[1em]
\FWp \FWm \FZ \FZ & -\frac{c^{2}}{s^{2}}
\Bigl[1 + 2\DZe - 2\frac{1}{c^{2}}\frac{\delta s}{s}
+ \DZW + \DZZ \Bigr]
+ \frac{c}{s}\DZAZ\\[1ex]
\FWp \FWm \FA \FZ & 
\frac{c}{s}
\Bigl[1 + 2\DZe -\frac{1}{c^{2}}\frac{\delta s}{s}
+ \DZW + \frac{1}{2}\DZZ + \frac{1}{2}\DZA \Bigr]
-\frac{1}{2}\DZAZ
-\frac{1}{2}\frac{c^{2}}{s^{2}}\DZZA 
\\[1ex]
\FWp \FWm \FA \FA & - \Bigl[1 + 2\DZe + \DZW + \DZA \Bigr]
+ \frac{c}{s}\DZZA 
\end{array}
\end{align}

\item
$\FV\FV\FV$ vertex:
\begin{align}
&\vcenter{\hbox{\includegraphics[page=11]{FRSMconvdiagrams.pdf}}}
= \ri eC \Bigl[g_{\mu \nu }(k_{1}-k_{2})_{\rho}
+g_{\nu\rho}(k_{2}-k_{3})_{\mu}
+g_{\rho\mu}(k_{3}-k_{1})_{\nu}\Bigr]
\\
&\begin{array}[b]{ll}
\FV_{1} \FV_{2} \FV_{3}  &  C  \\
\midrule
\FA \FWp \FWm &  1+\DZe+\DZW+\frac{1}{2}\DZA
-\frac{1}{2}\frac{c}{s}\DZZA\\[1ex]
\FZ \FWp \FWm &  -\frac{c}{s}(1 + \DZe
- \frac{1}{c^{2}}\frac{\delta s}{s}
+ \DZW + \frac{1}{2}\DZZ ) + \frac{1}{2}\DZAZ 
\end{array}
\end{align}

\item
$\FS\FS\FS\FS$ vertex:
\begin{align}
&\vcenter{\hbox{\includegraphics[page=12]{FRSMconvdiagrams.pdf}}}
= \ri e^{2}C
\\
&\begin{array}[b]{ll}
\FS_{1} \FS_{2} \FS_{3} \FS_{4}  &  C  \\
\midrule
\FH \FH \FH \FH & - \frac{3}{4s^{2}} \frac{\MH^{2}}{\MW^{2}}
\Bigl[1+2\DZe -2\frac{\delta s}{s}
+\frac{\DMHS}{\MH^{2}} + \frac{e}{2s\MH^2\MW}\dtPR -\frac{\delta \MW^{2}}{\MW^{2}}
+ 2\DZH \Bigr] \\[1ex]
\barr{@{}l@{}} \FH \FH \chi \chi \\ \FH \FH \phi \phi \earr \left.\rule[-
1.5ex]{0mm}{4ex}\right\}
& - \frac{1}{4s^{2}} \frac{\MH^{2}}{\MW^{2}}
\Bigl[1+2\DZe -2\frac{\delta s}{s}
+\frac{\DMHS}{\MH^{2}} + \frac{e}{2s\MH^2\MW}\dtPR -\frac{\delta \MW^{2}}{\MW^{2}}
+ \DZH \Bigr] \\[1ex]
\chi \chi \chi \chi & - \frac{3}{4s^{2}} \frac{\MH^{2}}{\MW^{2}}
\Bigl[1+2\DZe -2\frac{\delta s}{s}
+\frac{\DMHS}{\MH^{2}} + \frac{e}{2s\MH^2\MW}\dtPR -\frac{\delta \MW^{2}}{\MW^{2}}
\Bigr] \\[1ex]
\chi \chi \phi \phi & - \frac{1}{4s^{2}} \frac{\MH^{2}}{\MW^{2}}
\Bigl[1+2\DZe -2\frac{\delta s}{s}
+\frac{\DMHS}{\MH^{2}} + \frac{e}{2s\MH^2\MW}\dtPR -\frac{\delta \MW^{2}}{\MW^{2}}
\Bigr] \\[1ex]
\phi \phi \phi \phi & - \frac{1}{2s^{2}} \frac{\MH^{2}}{\MW^{2}}
\Bigl[1+2\DZe -2\frac{\delta s}{s}
+\frac{\DMHS}{\MH^{2}} + \frac{e}{2s\MH^2\MW}\dtPR -\frac{\delta \MW^{2}}{\MW^{2}}
\Bigr] 
\end{array}
\end{align}

\item
$\FS\FS\FS$ vertex:
\begin{align}
&\vcenter{\hbox{\includegraphics[page=13]{FRSMconvdiagrams.pdf}}}
= \ri eC
\\
&\begin{array}[b]{ll}
\FS_{1} \FS_{2} \FS_{3}   &  C  \\
\midrule
\FH \FH \FH &  - \frac{3}{2s} \frac{\MH^{2}}{\MW}
\Bigl[1+\DZe -\frac{\delta s}{s}
+\frac{\DMHS}{\MH^{2}} + \frac{e}{2s\MH^2\MW}(\dtPR-\dtFJ)
-\frac{1}{2}\frac{\delta \MW^{2}}{\MW^{2}}
+ \frac{3}{2}\DZH \Bigr] \\[1ex]
\barr{@{}l@{}} \FH\chi\chi \\ \FH\phi\phi \earr \left.
\rule[-1.5ex]{0mm}{4ex}\right\}
&  - \frac{1}{2s} \frac{\MH^{2}}{\MW}
\Bigl[1+\DZe -\frac{\delta s}{s}
+\frac{\DMHS}{\MH^{2}} + \frac{e}{2s\MH^2\MW}(\dtPR-\dtFJ)
-\frac{1}{2}\frac{\delta \MW^{2}}{\MW^{2}}
+ \frac{1}{2}\DZH \Bigr] 
\end{array}
\end{align}

\item
$\FV\FV\FS\FS$ vertex:
\begin{alignat}{2}
\mathrlap{
\vcenter{\hbox{\includegraphics[page=14]{FRSMconvdiagrams.pdf}}}
= \ri  e^{2}g_{\mu\nu}C
}
\\
&\FV_{1} \FV_{2} \FS_{1} \FS_2   &\quad&  C  \notag\\
\midrule
&\FWp \FWm \FH \FH &&\textstyle \frac{1}{2s^{2}}
\Bigl[1+2\DZe -2\frac{\delta s}{s}
+ \DZW + \DZH\Bigr] \notag\\[1ex]
&\barr{@{}l@{}} \FWp \FWm \chi \chi \\ \FWp \FWm \phi^{+} \phi^{-} \earr
\left.\rule[-1.5ex]{0mm}{4ex}\right\}
&&\textstyle   \frac{1}{2s^{2}}
\Bigl[1+2\DZe -2\frac{\delta s}{s}
+ \DZW \Bigr] \notag\\[1ex]
&\FZ \FZ \phi^{+} \phi^{-} &&\textstyle 
\frac{(s^{2}-c^{2})^{2}}{2s^{2}c^{2}}
\Bigl[1+2\DZe
+\frac{2}{(s^{2}-c^{2})c^{2}}\frac{\delta s}{s} + \DZZ \Bigr]
+\frac{s^{2}-c^{2}}{s c} \DZAZ \notag\\[1ex]
&\FZ\FA\phi^{+} \phi^{-} && \textstyle 
\frac{s^{2}-c^{2}}{s c}  \Bigl[1+2\DZe
+\frac{1}{(s^{2}-c^{2})c^{2}}\frac{\delta s}{s}
+ \frac{1}{2}\DZZ + \frac{1}{2}\DZA  \Bigr] 
+\frac{(s^{2}-c^{2})^{2}}{2s^{2}c^{2}}\frac{1}{2}\DZZA
+ \DZAZ  \notag\\
&\FA\FA\phi^{+} \phi^{-} &&\textstyle 
2\Bigl[1+2\DZe + \DZA \Bigr]
+\frac{s^{2}-c^{2}}{sc} \DZZA \notag\\[1ex]
&\FZ\FZ\FH\FH &&\textstyle   \frac{1}{2s^{2}c^{2}}
\Bigl[1+2\DZe
+ 2\frac{s^{2}-c^{2}}{c^{2}}\frac{\delta s}{s}
+ \DZZ+ \DZH \Bigr] \notag\\[1ex]
&\FZ\FZ\chi \chi &&\textstyle   \frac{1}{2s^{2}c^{2}}
\Bigl[1+2\DZe
+ 2\frac{s^{2}-c^{2}}{c^{2}}\frac{\delta s}{s}
+ \DZZ \Bigr] \notag\\[1ex]
&\barr{@{}l@{}} \FZ\FA\FH\FH\\ \FZ\FA\chi \chi \earr
\left.\vphantom{\barr{@{}l@{}} \FZ\FA\FH\FH\\ \FZ\FA\chi \chi \earr}
\rule[-1.5ex]{0mm}{4ex}\right\}
&&\textstyle    \frac{1}{2s^{2}c^{2}} \frac{1}{2}\DZZA \notag\\[1ex]
&\FW^{\pm} \FZ \phi^{\mp} \FH &&\textstyle
  {-\frac{1}{2c}}
\Bigl[1+2\DZe -\frac{\delta c}{c}
 + \frac{1}{2}\DZW + \frac{1}{2}\DZH
+ \frac{1}{2}\DZZ \Bigr]
- \frac{1}{2s}\frac{1}{2}\DZAZ \notag\\[1ex]
&\FW^{\pm} \FA \phi^{\mp} \FH &&\textstyle   {-\frac{1}{2s}}
\Bigl[1+2\DZe -\frac{\delta s}{s}
+ \frac{1}{2}\DZW + \frac{1}{2}\DZH
+ \frac{1}{2}\DZA \Bigr]
- \frac{1}{2c}\frac{1}{2}\DZZA \notag\\[1ex]
&\FW^{\pm} \FZ \phi^{\mp} \chi &&\textstyle   {\mp \frac{\ri}{2c}}
\Bigl[1+2\DZe -\frac{\delta c}{c}
+ \frac{1}{2}\DZW + \frac{1}{2}\DZZ \Bigr]
\mp \frac{\ri}{2s}\frac{1}{2}\DZAZ \notag\\[1ex]
&\FW^{\pm} \FA \phi^{\mp} \chi &&\textstyle    {\mp\frac{\ri}{2s}}
\Bigl[1+2\DZe -\frac{\delta s}{s}
+ \frac{1}{2}\DZW + \frac{1}{2}\DZA \Bigr]
\mp \frac{\ri}{2c}\frac{1}{2}\DZZA  
\raisetag{3ex}
\end{alignat}

\item
$\FV\FS\FS$ vertex:
\begin{align}
&\vcenter{\hbox{\includegraphics[page=15]{FRSMconvdiagrams.pdf}}}
= \ri eC(k_{1}-k_{2})_{\mu}
\\
\FRvskip
&\barr[b]{ll}
\FV \FS_{1} \FS_{2}   &  C  \\
\midrule
\FA \chi \FH &  - \frac{\ri}{2cs} \frac{1}{2}\DZZA\\[1ex]
\FZ \chi \FH &  - \frac{\ri}{2cs}
\Bigl[1+\DZe +\frac{s^{2}-c^{2}}{c^{2}}\frac{\delta s}{s}
+ \frac{1}{2}\DZH + \frac{1}{2}\DZZ\Bigr]\\[1ex]
\FA\phi^{+}\phi^{-} &  -\Bigl[1 + \DZe + \frac{1}{2}\DZA
+\frac{s^{2}-c^{2}}{2s c}\frac{1}{2}\DZZA\Bigr] \\[1ex]
\FZ \phi^{+} \phi^{-} &  - \frac{s^{2}-c^{2}}{2s c}
\Bigl[1+\DZe +\frac{1}{(s^{2}-c^{2})c^{2}}\frac{\delta s}{s}
+\frac{1}{2}\DZZ\Bigr]
- \frac{1}{2}\DZAZ \\[1ex]
\FW^{\pm} \phi^{\mp} \FH &  \mp \frac{1}{2s}
\Bigl[1+\DZe -\frac{\delta s}{s}
+ \frac{1}{2}\DZW + \frac{1}{2}\DZH \Bigr] \\[1ex]
\FW^{\pm} \phi^{\mp} \chi &  - \frac{\ri}{2s}
\Bigl[1+\DZe -\frac{\delta s}{s}
+ \frac{1}{2}\DZW\Bigr] 
\earr
\end{align}

\item
$\FS\FV\FV$ vertex:
\begin{align}
&\vcenter{\hbox{\includegraphics[page=16]{FRSMconvdiagrams.pdf}}}
= \ri eg_{\mu\nu}C
\\
&\barr[b]{ll}
\FS \FV_{1} \FV_{2}   &  C  \\
\midrule
\FH \FWp \FWm &   \MW\frac{1}{s}
\Bigl[1+\DZe -\frac{\delta s}{s}
+\frac{1}{2}\frac{\delta \MW^{2}}{\MW^{2}} - \frac{e}{2s\MH^2\MW}\dtFJ
+ \frac{1}{2}\DZH+ \DZW \Bigr] \\[1ex]
\FH\FZ\FZ&  \MW\frac{1}{s c^{2}}
\Bigl[1+\DZe +\frac{2s^{2}-c^{2}}{c^{2}}\frac{\delta s}{s}
+\frac{1}{2}\frac{\delta \MW^{2}}{\MW^{2}}- \frac{e}{2s\MH^2\MW}\dtFJ
+ \frac{1}{2}\DZH+ \DZZ \Bigr] \\[1ex]
\FH \FZ\FA&  \MW\frac{1}{s c^{2}} \frac{1}{2}\DZZA \\[1ex]
\phi^{\pm} \FW^{\mp} \FZ &  - \MW\frac{s}{c}
\Bigl[1+\DZe +\frac{1}{c^{2}}\frac{\delta s}{s}
+\frac{1}{2}\frac{\delta \MW^{2}}{\MW^{2}}- \frac{e}{2s\MH^2\MW}\dtFJ
+ \frac{1}{2}\DZW + \frac{1}{2}\DZZ \Bigr]
- \MW \frac{1}{2}\DZAZ \\[1ex]
\phi^{\pm} \FW^{\mp} \FA &  - \MW
\Bigl[1+\DZe +\frac{1}{2}\frac{\delta \MW^{2}}{\MW^{2}}- \frac{e}{2s\MH^2\MW}\dtFJ
+ \frac{1}{2}\DZW +  \frac{1}{2}\DZA \Bigr]
- \MW\frac{s}{c} \frac{1}{2}\DZZA  
\earr
\raisetag{3.7ex}
\end{align}

\item
$\FV\FFbar\FF$ vertex:
\begin{alignat}{2}
\mathrlap{
\vcenter{\hbox{\includegraphics[page=17]{FRSMconvdiagrams.pdf}}}
= \ri e\gamma_{\mu}(\CL\omega_{-} + \CR\omega_{+})
}
\\
& \FV\FFbar_{1} \FF_{2}   &\quad&  \CR,\CL  \notag\\
\midrule
&\FA \bar{f}_{i} f_{j} && \left\{
\barr{l}
\CR = -Q_{f}\Bigl[\delta_{ij}\Bigl(1 + \DZe +
\frac{1}{2}\DZA\Bigr)
+ \frac{1}{2}\left(\DZfR_{ij}+\DZfRd_{ij}\right)\Bigr]
+ \delta_{ij}g_{f}^{+}\frac{1}{2}\DZZA
\notag\\[1ex]
\CL = -Q_{f}\Bigl[\delta_{ij}\Bigl(1 + \DZe +
\frac{1}{2}\DZA\Bigr)
+ \frac{1}{2}\left(\DZfL_{ij}+\DZfLd_{ij}\right)\Bigr]
+ \delta_{ij}g_{f}^{-}\frac{1}{2}\DZZA
\earr \right.\notag\\[1ex]
&\FZ \bar{f}_{i} f_{j} && \left\{\barr{l}
\CR = g_{f}^{+}\Bigl[\delta_{ij}\Bigl(1 + \frac{\delta
g_{f}^{+}}{g_{f}^{+}}+\frac{1}{2}\DZZ\Bigr)
+ \frac{1}{2}\left(\DZfR_{ij}+\DZfRd_{ij}\right)\Bigr]
-\delta_{ij}Q_{f}\frac{1}{2}\DZAZ \notag\\[1ex]
\CL = g_{f}^{-}\Bigl[\delta_{ij}\Bigl(1 + \frac{\delta
g_{f}^{-}}{g_{f}^{-}}+\frac{1}{2}\DZZ\Bigr)
+ \frac{1}{2}\left(\DZfL_{ij}+\DZfLd_{ij}\right)\Bigr]
-\delta_{ij}Q_{f}\frac{1}{2}\DZAZ 
\earr \right.
\notag\\[1ex]
&\FWp \bar{u}_{i} d_{j} && 
\textstyle
\CR = 0,   \qquad
\CL = \frac{1}{\sqrt{2}s}\Bigl[V_{ij}\Bigl(1 + \DZe -
\frac{\delta s}{s}+\frac{1}{2}\DZW\Bigr) + \delta V_{ij} 
+\frac{1}{2}\sum_{k}\left(\DZuLd_{ik}V_{kj}+V_{ik}\DZdL_{kj}\right)\Bigr]
\notag\\[1ex]
&\FWm \bar{d}_{j} u_{i} && 
\textstyle
\CR = 0,   \qquad
\CL = \frac{1}{\sqrt{2}s}\Bigl[V_{ji}^\dagger\Bigl(1 + \DZe -
\frac{\delta s}{s}+\frac{1}{2}\DZW\Bigr) + \delta V_{ji}^\dagger 
+\frac{1}{2}\sum_{k}\left(\DZdLd_{jk}V_{ki}^\dagger
+V_{jk}^\dagger\DZuL_{ki}\right)\Bigr]
\notag\\[1ex]
&\FWp \bar{\nu}_{i} l_{j} &&
\textstyle
\CR = 0,   \qquad
\CL = \frac{1}{\sqrt{2}s}\delta_{ij}\Bigl[1 + \DZe -
\frac{\delta s}{s}+\frac{1}{2}\DZW
+\frac{1}{2}\left(\DZnLd_{ii}+\DZlL_{ii}\right)\Bigr]
\notag\\[1ex]
&\FWm \bar{l}_{j} \nu_{i} &&
\textstyle
\CR = 0,  \qquad
\CL = \frac{1}{\sqrt{2}s}\delta_{ij}\Bigl[1 + \DZe -
\frac{\delta s}{s}+\frac{1}{2}\DZW
+\frac{1}{2}\left(\DZlLd_{ii}+\DZnL_{ii}\right)\Bigr] 
\refstepcounter{equation}\tag{\theequation}
\end{alignat}
where
\beq \label{geZ}
\barr[b]{ll}
g_{f}^{+} = -\frac{s}{c} Q_{f}, &
\delta g_{f}^{+} = -\frac{s}{c} Q_{f}\Bigl[\DZe +
\frac{1}{c^{2}}\frac{\delta s}{s}\Bigr], \\[1.2em]
g_{f}^{-} =  \frac{I_{W,f}^{3}-s^{2}Q_{f}}{s c},\qquad &
\delta g_{f}^{-} =  \frac{I_{W,f}^{3}}{s c}\Bigl[\DZe +
\frac{s^{2}-c^{2}}{c^{2}}\frac{\delta s}{s}\Bigr] + \delta g_{f}^{+} .
\earr
\eeq
The vector and axial-vector couplings of the $Z$-boson are given by
\beq \label{vfaf}
\barr{l}
\varv_{f} = \frac{1}{2}(g_{f}^{-} + g_{f}^{+}) = \frac{I_{W,f}^{3}-
2s^{2}Q_{f}}{2s c}, \qquad
a_{f} = \frac{1}{2}(g_{f}^{-} - g_{f}^{+}) = \frac{I_{W,f}^{3}}{2s c}.
\earr
\eeq

\item
$\FS\FFbar\FF$ vertex:
\begin{alignat}{2}
\mathrlap{
\vcenter{\hbox{\includegraphics[page=18]{FRSMconvdiagrams.pdf}}}
= \ri e(\CL\omega_{-} + \CR\omega_{+})
}
\\
& S \FFbar_{1} \FF_{2}   &\quad&  \CR, \CL  \notag\\
\midrule
& \FH \bar{f}_{i} f_{j} && \left\{ \barr{l}
\CR = - \frac{1}{2s}\frac{1}{\MW}\Bigl[\delta_{ij}m_{f,i}
\Bigl(1 + \DZe -\frac{\delta s}{s} +
\frac{\delta m_{f,i}}{m_{f,i}} - \frac{1}{2}\frac{\delta
\MW^2}{\MW^2}+\frac{1}{2}\DZH\Bigr) 
{}+ \frac{1}{2}\Bigl(m_{f,i}\DZfR_{ij}+\DZfLd_{ij}m_{f,j}\Bigr)
\Bigr] \notag\\[1ex]
\CL = - \frac{1}{2s}\frac{1}{\MW}\Bigl[\delta_{ij}m_{f,i}
\Bigl(1 + \DZe -\frac{\delta s}{s} +
\frac{\delta m_{f,i}}{m_{f,i}} - \frac{1}{2}\frac{\delta
\MW^2}{\MW^2}+\frac{1}{2}\DZH\Bigr) 
{}+ \frac{1}{2}\Bigl(m_{f,i}\DZfL_{ij}+\DZfRd_{ij}m_{f,j}\Bigr)
\Bigr] \earr\right.\notag\\[1ex]
& \chi \bar{f}_{i} f_{j} && \left\{ \barr{l}
\CR = \ri\frac{1}{2s}2I_{W,f}^{3}\frac{1}{\MW}\Bigl[\delta_{ij}m_{f,i}
\Bigl(1 + \DZe -\frac{\delta s}{s} +
\frac{\delta m_{f,i}}{m_{f,i}} - \frac{1}{2}\frac{\delta \MW^2}{\MW^2}\Bigr)
{}+ \frac{1}{2}\Bigl(m_{f,i}\DZfR_{ij}+\DZfLd_{ij}m_{f,j}\Bigr)
\Bigr]
\\[1ex]
\CL = -\ri\frac{1}{2s}2I_{W,f}^{3}\frac{1}{\MW}\Bigl[\delta_{ij}m_{f,i}
\Bigl(1 + \DZe -\frac{\delta s}{s} +
\frac{\delta m_{f,i}}{m_{f,i}} - \frac{1}{2}\frac{\delta \MW^2}{\MW^2}\Bigr)
{}+ \frac{1}{2}\Bigl(m_{f,i}\DZfL_{ij}+\DZfRd_{ij}m_{f,j}\Bigr)
\Bigr] \earr\right.\notag\\[1ex]
& \phi^{+} \bar{u}_{i} d_{j} &&
\left\{ \barr{l}
\CR = -\frac{1}{\sqrt{2}s}\frac{1}{\MW}
\Bigl[V_{ij}m_{d,j}\Bigl(1 + \DZe -\frac{\delta s}{s} +
\frac{\delta m_{d,j}}{m_{d,j}} - \frac{1}{2}\frac{\delta \MW^2}{\MW^2}\Bigr) +
\delta V_{ij}m_{d,j}\\[1ex]
\hspace{4cm}
{}+\frac{1}{2}\sum_{k}\left(\DZuLd_{ik}V_{kj}m_{d,j}+V_{ik}m_{d,k}
\DZdR_{kj}\right)\Bigr] \\[1ex]
\CL = \frac{1}{\sqrt{2}s}\frac{1}{\MW}
\Bigl[m_{u,i}V_{ij}\Bigl(1 + \DZe -\frac{\delta s}{s} +
\frac{\delta m_{u,i}}{m_{u,i}} - \frac{1}{2}\frac{\delta \MW^2}{\MW^2}\Bigr) +
m_{u,i}\delta V_{ij} \\[1ex]
\hspace{4cm}
{}+\frac{1}{2}\sum_{k}\left(\DZuRd_{ik}m_{u,k}V_{kj}
+m_{u,i}V_{ik}\DZdL_{kj}\right)\Bigr]
\earr\right.\notag\\[1ex]
& \phi^{-} \bar{d}_{j} u_{i} &&
\left\{ \barr{l}
\CR = \frac{1}{\sqrt{2}s}\frac{1}{\MW}
\Bigl[V_{ji}^\dagger m_{u,i}\Bigl(1 + \DZe -\frac{\delta s}{s} +
\frac{\delta m_{u,i}}{m_{u,i}} - \frac{1}{2}\frac{\delta \MW^2}{\MW^2}\Bigr) +
\delta V_{ji}^\dagger m_{u,i}\\[1ex]
\hspace{4cm}
{}+\frac{1}{2}\sum_{k}\left(\DZdLd_{jk}V_{ki}^\dagger m_{u,i}
+V_{jk}^\dagger m_{u,k}\DZuR_{ki}\right)\Bigr]\\[1ex]
\CL = -\frac{1}{\sqrt{2}s}\frac{1}{\MW}
\Bigl[m_{d,j}V_{ji}^\dagger \Bigl(1 + \DZe -\frac{\delta s}{s} +
\frac{\delta m_{d,j}}{m_{d,j}} - \frac{1}{2}\frac{\delta \MW^2}{\MW^2}\Bigr) +
m_{d,j}\delta V_{ji}^\dagger \\[1ex]
\hspace{4cm}
{}+\frac{1}{2}\sum_{k}\left(\DZdRd_{jk}m_{d,k}V_{ki}^\dagger
+m_{d,j}V_{jk}^\dagger \DZuL_{ki}\right)\Bigr]
\earr\right. \notag\\[1ex]
& \phi^{+} \bar{\nu}_{i} l_{j} && 
\textstyle
\CR = -\frac{1}{\sqrt{2}s}\frac{m_{l,i}}{\MW}
\delta_{ij}\Bigl[ 1 + \DZe -\frac{\delta s}{s} +
\frac{\delta m_{l,i}}{m_{l,i}} - \frac{1}{2}\frac{\delta \MW^2}{\MW^2}
+\frac{1}{2}\left(\DZnLd_{ii}+\DZlR_{ii}\right)\Bigr],
\qquad
\CL = 0
\notag\\[1ex]
& \phi^{-} \bar{l}_{j} \nu_{i} && 
\textstyle
\CR = 0,
\qquad
\CL =  -\frac{1}{\sqrt{2}s}\frac{m_{l,i}}{\MW}
\delta_{ij}\Bigl[1 + \DZe -\frac{\delta s}{s} +
\frac{\delta m_{l,i}}{m_{l,i}} - \frac{1}{2}\frac{\delta \MW^2}{\MW^2}
+\frac{1}{2}\left(\DZlRd_{ii}+\DZnL_{ii}\right)\Bigr] 
\end{alignat}

\item
$\FV\FUbar\FU$ vertex:
\begin{align}
\label{FR:VUUa}
&\vcenter{\hbox{\includegraphics[page=19]{FRSMconvdiagrams.pdf}}}
= \ri eC k_{1\mu}
\\
\label{FR:VUUb}
&\barr[b]{ll}
V \bar{\FU}_{1} \FU_{2}   &  C  \\
\midrule
\FA \bar{u}^{\pm} u^{\pm}, \FW^{\mp} \bar{u}^{\mp} u^{\FA},\FW^{\pm}
\bar{u}^{\FA} u^{\mp}
&  \pm 1
\earr
\qquad
\barr[b]{ll}
V \bar{\FU}_{1} \FU_{2}   &  C  \\
\midrule
\FZ \bar{u}^{\pm} u^{\pm},\, \FW^{\mp} \bar{u}^{\mp} u^{\FZ},\, \FW^{\pm} \bar{u}^{\FZ} u^{\mp}
&  \mp \frac{c}{s}
\earr
\end{align}

\item
$\FS \FUbar\FU$ vertex:
\begin{align}
\label{FR:SUUa}
&\vcenter{\hbox{\includegraphics[page=20]{FRSMconvdiagrams.pdf}}}
= \ri eC
\\
\label{FR:SUUb}
&\barr[b]{ll}
S \bar{\FU}_{1} \FU_{2}   &  C  \\
\midrule
\FH \bar{u}^{\FZ} u^{\FZ} &  -\frac{1}{2s c^{2}}\MW\xi_Z
 \\[1ex]
\FH \bar{u}^{\pm} u^{\pm} &  -\frac{1}{2s}\MW\xi_W
\earr
\qquad
\barr[b]{ll}
S \bar{\FU}_{1} \FU_{2}   &  C  \\
\midrule
\chi \bar{u}^{\pm} u^{\pm} &  \mp \ri \frac{1}{2s}\MW\xi_W
\\[1ex]
\phi^{\pm} \bar{u}^{\FZ} u^{\mp} &   \frac{1}{2s c}\MW\xi_Z
\earr
\qquad
\barr[b]{ll}
S \bar{\FU}_{1} \FU_{2}   &  C  \\
\midrule
\phi^{\pm} \bar{u}^{\pm} u^{\FZ} & 
\frac{s^{2}-c^{2}}{2s c}\MW\xi_W
 \\[1ex]
\phi^{\pm} \bar{u}^{\pm} u^{\FA} &   \MW\xi_W
\earr
\end{align}

\end{myaitemize}

\section*{Feynman rules for the EWSM in the background-field formalism}
\label{app:FR_BFM}

We provide the Feynman rules in the BFM 't~Hooft gauge, as given in
\refeqs{eq:LGFph} and \refeqf{eq:gfbfm}, for an arbitrary quantum gauge
parameter $\xi=\xi_W=\xi_B$. In this gauge all propagators are
diagonal. We neglect quark mixing as in \citere{Denner:1994xt}, but
include the one-loop contributions of the counterterms from
the renormalization of physical parameters and fields in the vertices that
involve only background fields. While the Feynman rules for these
vertices are equivalent to those of the conventional formalism, the
counterterms have a much simpler structure. We do not include
counterterms for vertices involving quantum fields, as these are not
required for one-loop calculations. Furthermore, we assume 
$\DZfL=(\DZfL)^\dagger$ and $\DZfR=(\DZfR)^\dagger$.
Note that these conditions are not needed in the complex renormalization
as carried out in the CMS in \refse{se:CMS}, because
we rescale spinors and corresponding adjoint spinors by the same 
field renormalization constants, so that complex conjugated
field renormalization constants never appear in counterterms
[\cf\refeq{eq:CMSffieldren}].

The tadpole counterterms $\de t$ are given both in the FJTS
\cite{Fleischer:1980ub,Krause:2016oke,Denner:2016etu} and in the
(PRTS) scheme of \citere{Denner:1991kt}, as described in \refse{se:tadpoles}.
The respective tadpole terms are called $\dtFJ$ or $\dtPR$; whenever
$\de t$ is written without subscript FJTS or PRTS, this term
contributes in both schemes.
We note that the following Feynman rules do not only involve explicit
$\dtFJ$ terms, but also implicit $\dtFJ$ terms contained in the 
renormalization constant $\delta Z_{\FHhat}$
defined in \refeq{eq:delZB}. Whenever the $\delta Z_{\FHhat}$ 
term does not entirely originate from external Higgs or would-be
Goldstone fields, the explicit $\dtFJ$ term does not exactly
correspond to its counterterm in the conventional Feynman rules
listed above.

We do not list Feynman rules for the propagators of the background
fields and assume that the vertices are not affected by the gauge
fixing for the background fields, which holds for linear gauges.  The
gauge fixing for the background fields can be chosen independently
from the gauge-fixing of the quantum fields.  In a background 't~Hooft
gauge, the background-field propagators take the same form as the
propagators in the conventional gauge, given in
\refeqs{FR:Vprop}--\refeqf{FR:Fprop}, with $\xi_V$
replaced by the background gauge parameter $\xihat_V$. Note, however,
that it is sometimes preferable to use a more convenient gauge for the
background fields like the unitary gauge.

The propagators for the quantum field have the same form as those in
the conventional formalism given in \refeqs{FR:Vprop}--\refeqf{FR:Fprop} with
$\xi_V$ replaced by the quantum gauge parameter $\xi$.

\myparagraph{Vertices between background fields and symmetric counterterms}

\begin{myaitemize}
\item tadpole
\beq
\vcenter{\hbox{\includegraphics[page=1]{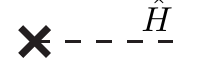}}}
= \ri\delta t 
\eeq

\item $\FVhat\FVhat$ counterterm
\begin{align}
&\vcenter{\hbox{\includegraphics[page=2]{FRSMBFMdiagrams.pdf}}}
= \ri \Bigl[(-g_{\mu\nu}k^2+ k_\mu k_\nu)C_{1} + g_{\mu\nu} C_{2} \Bigr]
\\
&\begin{array}[b]{ccccc} 
\FVhat_1\FVhat_2 & \FWhat^+\FWhat^- & \FZhat\FZhat & \FAhat\FZhat
& \FAhat\FAhat\\ \midrule
C_1 & \delta Z_{\FWhat} & \delta Z_{\FZhat\FZhat} & \frac{1}{2}
 \delta Z_{\FAhat\FZhat} & \delta Z_{\FAhat\FAhat} \\ [1ex]
C_2 & 
\de Z_{\FHhat}\MW^2 & \de Z_{\FHhat}\MZ^2 
& 0 & 0  \\[1ex]
\end{array}
\raisetag{4.2ex} 
\end{align}

\item $\FVhat\FShat$ counterterm
\begin{align}
&\vcenter{\hbox{\includegraphics[page=3]{FRSMBFMdiagrams.pdf}}}
= \ri k_\mu C  \delta Z_{\FHhat}
\\
&\barr[b]{cccc} 
\FVhat\FShat & \FWhat^\pm\phihat^\mp & \FZhat\chihat\\ \midrule
C & \pm\MW & \ri \MZ
\earr
\end{align}

\item $\FShat\FShat$ counterterm
\begin{align}
&\vcenter{\hbox{\includegraphics[page=4]{FRSMBFMdiagrams.pdf}}}
= \ri\Bigl[\delta Z_{\FHhat} k^{2} - C \Bigr]
\\
&\barr[b]{ccc} 
\FShat_1 \FShat_2 & \FHhat\FHhat & \chihat \chihat, \phihat
 \phihat\\ \midrule
C &  \delta Z_{\FHhat}\MH^2 + \delta \MH^2  -\frac{3e}{2s\MW}\dtFJ &
 - \frac{e}{2 s} \frac{\delta t}{\MW} 
\earr
\end{align}

\item $\FF\FFbar$ counterterm
\begin{align}
&\vcenter{\hbox{\includegraphics[page=5]{FRSMBFMdiagrams.pdf}}}
= \ri\Bigl[\DZfL\slashed{k}\omega_{-} + \DZfR\slashed{k}\omega_{+}
- C_{\rS} \Bigr]
\\
\label{fr:CFFct}
&C_{\rS} =\frac{1}{2} \left(\delta Z^{f,\rL} +
\delta Z^{f,\rR}\right) m_{f}  + \delta
m_{f}-m_{f}\frac{e}{2s\MH^2\MW}\dtFJ 
\end{align}

\item
$\FVhat\FVhat\FVhat\FVhat$ vertex:
\begin{align}
&\vcenter{\hbox{\includegraphics[page=6]{FRSMBFMdiagrams.pdf}}}
= \ri e^{2}C \Bigl[2g_{\mu\nu}g_{\rho\sigma} -
g_{\mu \sigma }g_{\nu\rho} - g_{\mu\rho}g_{\nu \sigma }\Bigr]
(1 + \delta Z_{\FWhat})
\\
\label{fr:Cvvvv}
&\barr[b]{ccccc} 
\FVhat_1\FVhat_2\FVhat_3\FVhat_4 & \FWhat^+\FWhat^+\FWhat^-\FWhat^- &
\FWhat^+\FWhat^-\FZhat\FZhat & \FWhat^+\FWhat^-\FAhat\FZhat &
\FWhat^+\FWhat^-\FAhat\FAhat\\ \midrule
C & \frac{1}{s^2} & -\frac{c^2}{s^2} & \frac{c}{s} & -1
\earr
\end{align}

\item
$\FVhat\FVhat\FVhat$ vertex:
\begin{align}
&\vcenter{\hbox{\includegraphics[page=7]{FRSMBFMdiagrams.pdf}}}
= \ri eC \Bigl[g_{\mu \nu }(k_{1}-k_{2})_{\rho}
+g_{\nu\rho}(k_{2}-k_{3})_{\mu}
+g_{\rho\mu}(k_{3}-k_{1})_{\nu}\Bigr]
(1 + \delta Z_{\FWhat})
\\
\label{fr:Cvvv} 
&\barr[b]{ccc}
\FVhat_1\FVhat_2\FVhat_3 & \FAhat\FWhat^+\FWhat^- & \FZhat\FWhat^+\FWhat^-\\
\midrule
C & 1 & - \frac{c}{s}
\earr
\end{align}

\item
$\FShat\FShat\FShat\FShat$ vertex:
\begin{align}
&\vcenter{\hbox{\includegraphics[page=8]{FRSMBFMdiagrams.pdf}}}
\textstyle
= \ri e^{2}C\left[1 + \frac{\delta \MH^2}{\MH^2} + \frac{e}{2 s\MW \MH^2}
                (\dtPR-2\dtFJ) + \delta Z_{\FHhat}
          \right]
\\
\label{fr:Cssss}
&\barr[b]{cccc}
\FShat_1\FShat_2\FShat_3\FShat_4 & \FHhat\FHhat\FHhat\FHhat,\,
\chihat\chihat\chihat\chihat &
\FHhat\FHhat\chihat\chihat,\, \FHhat\FHhat\phihat^+\phihat^-,\,
\chihat\chihat\phihat^+\phihat^- &
\phihat^+\phihat^-\phihat^+\phihat^- \\ \midrule
C & -\frac{3}{4 s^2} \frac{\MH^2}{\MW^2} &
-\frac{1}{4 s^2} \frac{\MH^2}{\MW^2} &
-\frac{1}{2 s^2} \frac{\MH^2}{\MW^2}
\earr
\end{align}

\item
$\FShat\FShat\FShat$ vertex:
\begin{align}
&\vcenter{\hbox{\includegraphics[page=9]{FRSMBFMdiagrams.pdf}}}
\textstyle
= \ri eC\left[1 + \frac{\delta \MH^2}{\MH^2} + \frac{e}{2 s \MW \MH^2}
(\dtPR-2\dtFJ) + \delta Z_{\FHhat}
      \right]
\\
&\barr[b]{ccc} 
\FShat_1\FShat_2\FShat_3 & \FHhat\FHhat\FHhat &
\FHhat\chihat\chihat,\, \FHhat\phihat^+\phihat^- \\\midrule 
C & -\frac{3}{2 s} \frac{\MH^2}{\MW} &
-\frac{1}{2 s} \frac{\MH^2}{\MW}
\earr
\end{align}

\item
$\FVhat\FVhat\FShat\FShat$ vertex:
\begin{align}\label{FR:VhVhShSha}
&\vcenter{\hbox{\includegraphics[page=10]{FRSMBFMdiagrams.pdf}}}
= \ri  e^{2}g_{\mu\nu}C(1 + \delta Z_{\FHhat})
\\
&\barr{cccccc} \strut
\FVhat_1\FVhat_2\FShat_1\FShat_2 & 
\FZhat\FZhat\FHhat\FHhat,\, \FZhat\FZhat\chihat\chihat &
\FWhat^+\FWhat^-\FHhat\FHhat,\, \FWhat^+\FWhat^-\phihat^+\phihat^-,\,
\FWhat^+\FWhat^-\chihat\chihat &
\FAhat\FAhat\phihat^+\phihat^- & \FZhat\FAhat\phihat^+\phihat^- &
\FZhat\FZhat\phihat^+\phihat^-\\ \midrule
C  & \frac{1}{2 c^2 s^2} & \frac{1}{2 s^2} & 2 &
-\frac{c^2 - s^2}{c s} & \frac{(c^2 - s^2)^2}{2 c^2 s^2}
\earr 
\notag\\[1ex]
\label{FR:VhVhShShb}
&\barr{cccccc}
\FVhat_1\FVhat_2\FShat_1\FShat_2 &
\FWhat^{\pm}\FAhat\phihat^{\mp}\FHhat &
\FWhat^{\pm}\FAhat\phihat^{\mp}\chihat &
\FWhat^{\pm}\FZhat\phihat^{\mp}\FHhat &
\FWhat^{\pm}\FZhat\phihat^{\mp}\chihat \\ \midrule
C & -\frac{1}{2 s} & \mp\frac{\ri}{2 s} &
-\frac{1}{2 c} & \mp \frac{\ri}{2c}
\earr 
\end{align}

\item
$\FVhat\FShat\FShat$ vertex:
\begin{align}
\label{FR:VhShSha}
&\vcenter{\hbox{\includegraphics[page=11]{FRSMBFMdiagrams.pdf}}}
= \ri eC(k_{1}-k_{2})_{\mu}(1 + \delta Z_{\FHhat})
\\
\label{FR:VhShShb}
&\barr{cccccc}
\FVhat\FShat_1\FShat_2 & \FZhat\chihat\FHhat & \FAhat\phihat^+\phihat^- &
\FZhat\phihat^+\phihat^- & \FWhat^{\pm}\phihat^{\mp}\FHhat &
\FWhat^{\pm}\phihat^{\mp}\chihat\\ \midrule
C & -\frac{\ri}{2 c s} & -1 & \frac{c^2 - s^2}{2 c s} &
\mp \frac{1}{2 s} & - \frac{\ri}{2 s}
\earr 
\end{align}

\item
$\FShat\FVhat\FVhat$ vertex:
\begin{align}
\label{FR:ShVhVha}
&\vcenter{\hbox{\includegraphics[page=12]{FRSMBFMdiagrams.pdf}}}
= \ri eg_{\mu\nu}C(1 + \delta Z_{\FHhat})
\\
\label{FR:ShVhVhb}
&\barr{ccccc}
\FShat\FVhat_1\FVhat_2 & \FHhat\FZhat\FZhat & \FHhat\FWhat^+\FWhat^- &
\phihat^{\pm}\FWhat^{\mp}\FAhat & \phihat^{\pm}\FWhat^{\mp}\FZhat  \\ \midrule
C & \frac{1}{c^2s} \MW & \frac{1}{s} \MW & -\MW &
-\frac{s}{c} \MW
\earr 
\end{align}

\item
$\FVhat\FFbar\FF$ vertex:
\begin{align}
&\vcenter{\hbox{\includegraphics[page=13]{FRSMBFMdiagrams.pdf}}}
= \ri e\gamma_{\mu}\Bigl[\CL\omega_{-} \left(1 + \delta Z^{F_1,\rL}\right) 
+ \CR \omega_{+}\Bigr(1 + \textstyle \frac{1}{2}\left(\de Z^{F_1,\rR} 
+ \de Z^{F_2,\rR}\right)\Bigr) \Bigr]
\\
\label{fr:CVff}
&\barr{cccc}
\FVhat\FFbar_{1} \FF_{2} & \FAhat\bar{f} f & \FZhat\bar{f} f &
\FWhat^+\bar{f}_u f_d ,\, \FWhat^-\bar{f}_d f_u\\ \midrule
C_{\rL} & - Q_f & \frac{I_{\rw, f}^3 - s^2 Q_f}{c s} &
\frac{1}{\sqrt{2} s}\\ 
C_{\rR} & - Q_f & -\frac{s}{c} Q_f & 0 
\earr  
\end{align}

\item
$\FShat\FFbar\FF$ vertex:
\begin{align}
&\vcenter{\hbox{\includegraphics[page=14]{FRSMBFMdiagrams.pdf}}}
\barr{l}
 = \ri e\biggl[ \CL\omega_{-}
  \left(1 + \frac{\delta m_{F_1}}{m_{F_1}}
  + \frac{1}{2} \delta Z^{F_2,\rL}
  + \frac{1}{2} \delta Z^{F_1,\rR}
  - \frac{e}{2s\MW\MH^2} \dtFJ
\right) \\
  \mbox{} \qquad + \CR\omega_{+}
  \left(1 + \frac{\delta m_{F_2}}{m_{F_2}}
  + \frac{1}{2} \delta Z^{F_1,\rL} +
  \frac{1}{2} \delta Z^{F_2,\rR}
  - \frac{e}{2s\MW\MH^2} \dtFJ
\right)
\biggr]
\earr
\\
\label{fr:Csff}
&\barr{ccccc} 
\FShat\FFbar_{1} \FF_{2} & \FHhat\bar{f} f & \chihat\bar{f} f &
\phihat^+\bar{f}_u f_d & \phihat^-\bar{f}_d f_u \\ \midrule
C_{\rL} & -\frac{1}{2 s} \frac{m_f}{\MW} &
-\ri \frac{1}{2 s} 2 I_{\rw,f}^3 \frac{m_{f}}{\MW} &
+\frac{1}{\sqrt{2} s} \frac{m_{f_u}}{\MW} &
-\frac{1}{\sqrt{2} s} \frac{m_{f_d}}{\MW} \\
C_{\rR} & -\frac{1}{2 s} \frac{m_f}{\MW} &
+\ri \frac{1}{2 s} 2 I_{\rw,f}^3 \frac{m_f}{\MW} &
-\frac{1}{\sqrt{2} s} \frac{m_{f_d}}{\MW} &
+\frac{1}{\sqrt{2} s} \frac{m_{f_u}}{\MW} 
\earr
\end{align}

\end{myaitemize}

In contrast to the conventional formalism, no counterterms are
present for the $\FZhat\FAhat\FHhat\FHhat$, $\FZhat\FAhat\chihat\chihat$,
$\FAhat\chihat\FHhat$, and $\FHhat\FZhat\FAhat$ couplings.

\myparagraph{Vertices involving quantum fields}

We provide the Feynman rules for vertices containing quantum fields in
lowest order, \ie we do not list the corresponding counterterms.  All
lowest-order vertices involving fermions have the usual form and are
not separately listed.  Since the gauge-fixing term is quadratic in
the quantum fields, 
only vertices containing exactly two quantum fields 
or containing ghost fields differ from the
conventional ones.  In the following, we list only those couplings for
which the generic form or actual insertion differ 
from the results in the conventional formalism.  
The vertices of types $\FVhat\FVhat
\FS\FS$, $\FShat\FShat \FV \FV$, $\FVhat \FS\FS$, 
$\FShat \FV\FV$, $\FV\FUbar\FU$, and $\FS\FUbar\FU$  
have the usual Feynman rules (with $\xi_\FZ,\xi_\FW\to \xi$).
Note that some of the insertions appearing in
the conventional couplings have no counterparts here; we list only the
non-vanishing insertions. 
%

\begin{myaitemize}
\item
$\FVhat\FVhat\FV\FV$ vertex: \quad
Here two types of Feynman rules exist.
\begin{align}
&\vcenter{\hbox{\includegraphics[page=15]{FRSMBFMdiagrams.pdf}}}
= \ri e^2 C \left[
2 g_{\mu\nu} g_{\rho\si} - g_{\mu\si} g_{\nu\rho} \left(1 - \frac{1}{\xiQ}\right)
- g_{\mu\rho} g_{\nu\si} \left(1 - \frac{1}{\xiQ}\right) \right]
\\
\label{fr:Cvvvqvq1}
&\barr[b]{ccccc} 
\FVhat_1\FVhat_2 \FV_3\FV_4 & 
\FWhat^\pm\FWhat^\pm \FW^\mp \FW^\mp &
\FZhat\FZhat \FW^+\FW^-,\, \FWhat^+\FWhat^-\FZ\FZ & 
\FAhat\FZhat \FW^+\FW^-,\, \FWhat^+\FWhat^-\FA\FZ & 
\FAhat\FAhat \FW^+\FW^-,\, \FWhat^+\FWhat^-\FA\FA \\ \midrule
C & \frac{1}{s^2} & -\frac{c^2}{s^2} & \frac{c}{s} & -1
\earr
\end{align}

\begin{align}
&\vcenter{\hbox{\includegraphics[page=15]{FRSMBFMdiagrams.pdf}}}
= \ri e^2 C \left[
2 g_{\mu\rho} g_{\nu\si} - g_{\mu\nu} g_{\rho\si} -
g_{\mu\si} g_{\nu\rho} \left(1 + \frac{1}{\xiQ}\right) \right]
\\
\label{fr:Cvvvqvq2}
&\barr[b]{ccccc} 
\FVhat_1\FVhat_2 \FV_3\FV_4 & \FWhat^+\FWhat^-\FW^+\FW^- &
\FWhat^{\pm}\FZhat \FW^{\mp}\FZ & 
\FWhat^{\pm}\FAhat \FW^{\mp}\FZ,\, \FWhat^{\pm}\FZhat \FW^{\mp}\FA & 
\FWhat^{\pm}\FAhat \FW^{\mp}\FA\\  \midrule
C & \frac{1}{s^2} & -\frac{c^2}{s^2} & \frac{c}{s} & -1
\earr
\end{align}

\item
$\FVhat\FV\FV$ vertex:
\begin{align}
&\vcenter{\hbox{\includegraphics[page=16]{FRSMBFMdiagrams.pdf}}}
=- \ri e C \left[
g_{\nu\rho} (k_3 - k_2)_{\mu} +
g_{\mu\nu} \left(k_2 - k_1 + \frac{k_3}{\xiQ}\right)_{\rho} +
g_{\rho\mu} \left(k_1 - k_3 - \frac{k_2}{\xiQ}\right)_{\nu} \right]
\raisetag{4ex}\\
\label{fr:Cvvqvq} 
&\barr[b]{ccc}
\FVhat_1 \FV_2 \FV_3 & \FAhat \FW^+\FW^-,\, \FWhat^+ \FW^- \FA,\, \FWhat^- \FA \FW^+ &
\FZhat \FW^+\FW^-, \,\FWhat^+ \FW^- \FZ,\, \FWhat^- \FZ \FW^+\\ \midrule
C & 1 & - \frac{c}{s}
\earr
\end{align}

\item
$\FShat\FShat\FS\FS$ vertex:
\begin{align}
&\vcenter{\hbox{\includegraphics[page=17]{FRSMBFMdiagrams.pdf}}}
= \ri e^{2}C
\\
\label{fr:Csssqsq}
&\barr[b]{ccccc}
\FShat_1 \FShat_2 \FS_3 \FS_4 & \FHhat\FHhat \FH\FH & \FHhat\FHhat\chi\chi &
\FHhat\chihat \FH\chi & \phihat^+\phihat^-\FH\FH,\, \FHhat\FHhat\phi^+\phi^- \\
&  \chihat\chihat\chi\chi & \chihat\chihat \FH\FH &
& \phihat^+\phihat^-\chi\chi,\, \chihat\chihat\phi^+\phi^- \\ \midrule
C & -\frac{3}{4 s^2} \frac{\MH^2}{\MW^2} &
- \frac{1}{4 s^2} \frac{\MH^2}{\MW^2} - \frac{\xiQ}{2 c^2 s^2} &
- \frac{1}{4 s^2} \frac{\MH^2}{\MW^2} + \frac{\xiQ}{4 c^2 s^2} &
- \frac{1}{4 s^2} \frac{\MH^2}{\MW^2} - \frac{\xiQ}{2 s^2}
\earr
\notag\\[1ex]
&\barr{ccccc} 
\FShat_1 \FShat_2 \FS_3 \FS_4 & \phihat^{\pm}\FHhat\phi^{\mp}\FH
& \phihat^+\phihat^-\phi^+\phi^- &
\phihat^{\pm}\phihat^{\pm}\phi^{\mp}\phi^{\mp} &
\phihat^{\pm}\FHhat\phi^{\mp}\chi\\
& \phihat^{\pm}\chihat\phi^{\mp}\chi
& & & \phihat^{\mp}\chihat\phi^{\pm}\FH  \\\midrule
C & - \frac{1}{4 s^2} \frac{\MH^2}{\MW^2} + \frac{\xiQ}{4 s^2} &
- \frac{1}{2 s^2} \frac{\MH^2}{\MW^2} - \frac{\xiQ}{4 c^2 s^2} &
- \frac{1}{2 s^2} \frac{\MH^2}{\MW^2} + \frac{\xiQ}{2 c^2 s^2} 
& \mp \frac{\ri \xiQ}{4 c^2} 
\earr 
\end{align}

\item
$\FShat\FS\FS$ vertex:
\begin{align}
&\vcenter{\hbox{\includegraphics[page=18]{FRSMBFMdiagrams.pdf}}}
= \ri eC
\\
&\barr{ccccccc} 
\FShat_1 \FS_2 \FS_3 & \FHhat \FH\FH & \FHhat\chi\chi &
\chihat \FH \chi & \FHhat \phi^+\phi^- 
& \phihat^{\pm}\phi^{\mp}\FH
& \phihat^{\pm}\phi^{\mp}\chi\\ \midrule
C & -\frac{3}{2 s} \frac{\MH^2}{\MW} &
-\frac{1}{2 s} \frac{\MH^2}{\MW} - \xiQ \frac{\MW}{c^2 s} &
-\frac{1}{2 s} \frac{\MH^2}{\MW} + \xiQ \frac{\MW}{2 c^2 s} &
-\frac{1}{2 s} \frac{\MH^2}{\MW} - \xiQ \frac{\MW}{s} 
& -\frac{1}{2 s} \frac{\MH^2}{\MW} + \xiQ \frac{\MW}{2 s}
& \mp i \xiQ \MW\frac{s}{2 c^2}
\earr
\end{align}

\item
$\FVhat\FV\FShat\FS$ vertex: 
\begin{align}\label{FR:VhVhShSh}
&\vcenter{\hbox{\includegraphics[page=19]{FRSMBFMdiagrams.pdf}}}
= \ri  e^{2}g_{\mu\nu}C
\\
&\barr{cccccccc} 
\FVhat_1 \FV_2\FShat_1 \FS_2  & \FZhat \FZ\FHhat \FH
& \FWhat^{\pm}\FW^{\mp}\FHhat\FH & \FWhat^{\pm}\FW^{\mp}\phihat^{\mp}\phi^{\pm}
& \FAhat \FA \phihat^{\pm}\phi^{\mp} & \FZhat \FA \phihat^{\pm}\phi^{\mp}
& \FZhat \FZ \phihat^{\pm}\phi^{\mp} 
& \FWhat^{\pm}\FA\FHhat\phi^{\mp}
\\
& \FZhat \FZ\chihat\chi & 
\FWhat^{\pm}\FW^{\mp}\chihat\chi & & &
\FAhat \FZ\phihat^{\pm}\phi^{\mp} &
& \FAhat \FW^{\pm}\phihat^{\mp}\FH 
\\  \midrule
C & \frac{1}{2 c^2 s^2} & \frac{1}{2 s^2} & \frac{1}{s^2} & 2
& -\frac{c^2 - s^2}{c s} & \frac{(c^2 - s^2)^2}{2 c^2 s^2} 
& -\frac{1}{s} 
\\ 
\earr
\notag\\[1ex]
&\barr{ccccccc} 
\FVhat_1 \FV_2\FShat_1 \FS_2 &
\FWhat^{\pm}\FA\chihat\phi^{\mp} &
\FWhat^{\pm}\FZ\phihat^{\mp}\FH &
\FWhat^{\pm}\FZ\phihat^{\mp}\chi &
\FWhat^{\pm}\FZ\FHhat\phi^{\mp} &
\FWhat^{\pm}\FZ\chihat\phi^{\mp} 
& \FWhat^\pm \FW^\mp\chihat\FH
\\
& 
\FAhat \FW^{\pm}\phihat^{\mp}\chi &
\FZhat \FW^{\pm}\FHhat \phi^{\mp} &
\FZhat \FW^{\pm}\chihat \phi^{\mp} &
\FZhat \FW^{\pm}\phihat^{\mp}\FH &
\FZhat \FW^{\pm}\phihat^{\mp}\chi
& \FWhat^\mp \FW^\pm\FHhat\chi
\\  \midrule
C & 
\mp\frac{\ri}{s} & -\frac{1}{2 c s^2} & \mp\frac{\ri}{2 c s^2} &
\frac{c^2 - s^2}{2 c s^2} & \pm \ri \frac{c^2 - s^2}{2 c s^2}
& \pm\frac{\ri}{2 s^2}
\\ 
\earr
\end{align}

\item
$\FVhat\FVhat\FS\FS$ vertex and $\FV\FV\FShat\FShat$: 
Feynman rules coincide with \refeqs{FR:VhVhShSha} and
\refeqf{FR:VhVhShShb} (apart from counterterms).

\item
$\FV\FShat\FS$ vertex:
\begin{align}
&\vcenter{\hbox{\includegraphics[page=20]{FRSMBFMdiagrams.pdf}}}
= \ri 2 eC k_{1\mu}
\\
&\barr{cccccccc} 
\FV\FShat_1 \FS_2 & \FZ\chihat \FH & \FZ\FHhat\chi & \FA\phihat^\pm\phi^\mp &
\FZ\phihat^\pm\phi^\mp & \FW^{\pm}\phihat^{\mp}\FH, \FW^{\mp}\FHhat\phi^\pm &
\FW^{\pm}\phihat^{\mp}\chi & \FW^{\pm}\chihat\phi^{\mp} \\  \midrule
C & -\frac{\ri}{2 c s}  & \frac{\ri}{2 c s} & \mp1 & \pm\frac{c^2 - s^2}{2 c s} &
\mp \frac{1}{2 s} & -\frac{\ri}{2 s} & \frac{\ri}{2 s} \\ 
\earr
\end{align}

\item
$\FVhat\FS\FS$ vertex: 
Feynman rules coincide with \refeqs{FR:VhShSha} and
\refeqf{FR:VhShShb} (apart from counterterms).

\item
$\FS\FVhat\FV$ vertex:
\begin{align}
&\vcenter{\hbox{\includegraphics[page=21]{FRSMBFMdiagrams.pdf}}}
= \ri eg_{\mu\nu}C
\\
\FRvskip
&\barr{cccccccc} 
\FS \FVhat_1 \FV_2 &\FH\FZhat \FZ & \FH\FWhat^{\pm}\FW^{\mp} & \chi\FWhat^\pm \FW^\mp &
\phi^{\pm}\FWhat^{\mp}\FA & \phi^{\pm}\FWhat^{\mp}\FZ & \phi^{\pm}\FZhat \FW^{\mp}
\\  \midrule
C & \frac{1}{c^2 s} \MW & \frac{1}{s} \MW & \mp\frac{\ri}{s} \MW &
- 2 \MW & \frac{c^2 - s^2}{c s} \MW & - \frac{1}{c s} \MW
\\ 
\earr
\end{align}

\item
$\FShat\FV\FV$ vertex: 
Feynman rules coincide with \refeqs{FR:ShVhVha} and
\refeqf{FR:ShVhVhb} (apart from counterterms).

\item
$\FVhat \FUbar\FU$ vertex:
\begin{align}
&\vcenter{\hbox{\includegraphics[page=22]{FRSMBFMdiagrams.pdf}}}
=  \ri e(k_{1}-k_2)_{\mu}C
\\
\label{fr:Cvgg}
&\barr{ccc} 
\FVhat\FUbar_1\FU_2 & \FAhat\bar{u}^{\pm} u^{\pm},\,
\FWhat^{\pm} \bar{u}^{\FA} u^{\mp},\, \FWhat^{\mp} \bar{u}^{\mp} u^{\FA} &
\FZhat\bar{u}^{\pm} u^{\pm},\, \FWhat^{\pm} \bar{u}^{\FZ} u^{\mp},\,
\FWhat^{\mp} \bar{u}^{\mp} u^{\FZ}  \\  \midrule
C & \pm{1} & \mp\frac{c}{s} \\ 
\earr
\end{align}

\item
$\FV \FUbar\FU$ vertex:
Feynman rules coincide with \refeqs{FR:VUUa} and \refeqf{FR:VUUb}.

\item
$\FVhat\FVhat \FUbar\FU$ vertex:
\begin{align}
&\vcenter{\hbox{\includegraphics[page=23]{FRSMBFMdiagrams.pdf}}}
= \ri e^2 g_{\mu\nu}C
\\
&\barr{ccccc} 
\FVhat_1\FVhat_2\FUbar_1\FU_2 & 
\FWhat^\pm\FWhat^\pm\bar{u}^{\pm}u^{\mp} &
\FWhat^+\FWhat^-\bar{u}^{\FA}u^{\FA}& \FWhat^+\FWhat^-\bar{u}^{\FA}u^{\FZ},\, 
\FAhat\FZhat\bar{u}^{\pm} u^{\pm} & \FWhat^+\FWhat^-\bar{u}^{\FZ}u^{\FZ}
\\
&&\FAhat\FAhat\bar{u}^{\pm} u^{\pm} & \FWhat^+\FWhat^-\bar{u}^{\FZ}u^{\FA} \hfill
& \FZhat\FZhat\bar{u}^{\pm} u^{\pm} \\ \midrule 
C & -\frac{2}{s^2} & 2 & -2\frac{c}{s} & 2\frac{c^2}{s^2} \\ 
\earr 
\notag\\[1ex]
&\barr{ccccc} 
\FVhat_1\FVhat_2\FUbar_1\FU_2 & 
\FWhat^+\FWhat^-\bar{u}^{\pm}u^{\pm} &
\FAhat\FWhat^\pm\bar{u}^{\pm} u^{\FA} & \FZhat\FWhat^\pm\bar{u}^{\pm} u^{\FA} ,\, 
\FAhat\FWhat^\pm\bar{u}^{\pm} u^{\FZ} & \FZhat\FWhat^\pm\bar{u}^{\pm} u^{\FZ} 
\\
&&\FAhat\FWhat^\pm\bar{u}^{\FA} u^{\mp} & \FZhat\FWhat^\pm\bar{u}^{\FA} u^{\mp} ,\, 
\FAhat\FWhat^\pm\bar{u}^{\FZ} u^{\mp} & \FZhat\FWhat^\pm\bar{u}^{\FZ} u^{\mp} 
\\ \midrule
C &  \frac{1}{s^2} & -1 & \frac{c}{s} & -\frac{c^2}{s^2} \\ 
\earr  \label{fr:Cvvgg}
\end{align}

\item
$\FVhat\FV \FUbar\FU$ vertex:
\begin{align}
&\vcenter{\hbox{\includegraphics[page=24]{FRSMBFMdiagrams.pdf}}}
= \ri e^2 g_{\mu\nu}C
\\
&\barr{ccccc} 
\FVhat_1 \FV_2\FUbar_1\FU_2 &
\FWhat^\pm \FW^\pm\bar{u}^{\pm}u^{\mp} &
\FWhat^\pm \FW^\mp\bar{u}^{\FA}u^{\FA}& \FWhat^\pm \FW^\mp\bar{u}^{\FA}u^{\FZ},\,
\FAhat \FZ\bar{u}^{\pm} u^{\pm} & \FWhat^\pm \FW^\mp\bar{u}^{\FZ}u^{\FZ}
\\
&&\FAhat \FA\bar{u}^{\pm} u^{\pm} & \FWhat^\pm\FW^\mp\bar{u}^{\FZ}u^{\FA},\,
\FZhat \FA\bar{u}^{\pm} u^{\pm} & \FZhat \FZ\bar{u}^{\pm} u^{\pm} \\ \midrule
C  & -\frac{1}{s^2} & 1 & -\frac{c}{s} & \frac{c^2}{s^2} \\ 
\earr
\notag\\[1ex]
&\barr{ccccc} 
\FVhat_1 \FV_2\FUbar_1\FU_2 &
\FWhat^\pm \FW^\mp\bar{u}^{\pm}u^{\pm} &
\FAhat \FW^\pm\bar{u}^{\pm} u^{\FA} & \FZhat \FW^\pm\bar{u}^{\pm} u^{\FA} ,\,
\FAhat \FW^\pm\bar{u}^{\pm} u^{\FZ} & \FZhat \FW^\pm\bar{u}^{\pm} u^{\FZ}
\\
&&\FWhat^\pm \FA\bar{u}^{\FA} u^{\mp} & \FWhat^\pm \FZ\bar{u}^{\FA} u^{\mp} ,\,
\FWhat^\pm \FA\bar{u}^{\FZ} u^{\mp} & \FWhat^\pm \FZ\bar{u}^{\FZ} u^{\mp}
\\ \midrule
C  &  \frac{1}{s^2} & -1 & \frac{c}{s} & -\frac{c^2}{s^2} \\ 
\earr
\end{align}

\item
$\FShat \FUbar\FU$ vertex:
\begin{align}
&\vcenter{\hbox{\includegraphics[page=25]{FRSMBFMdiagrams.pdf}}}
= \ri e C\xiQ
\\
\FRvskip
\label{fr:Csgg}
&\barr{ccccccc} 
\FShat\FUbar_1\FU_2 & \FHhat\bar{u}^{\FZ}u^{\FZ} & \FHhat\bar{u}^{\pm} u^{\pm}
& \phihat^{\pm} \bar{u}^{\pm} u^{\FA},\, \phihat^{\pm} \bar{u}^{\FA} u^{\mp} &
  \phihat^{\pm} \bar{u}^{\pm} u^{\FZ},\, \phihat^{\pm} \bar{u}^{\FZ} u^{\mp}
 \\  \midrule
C & -\frac{1}{c^{2}s}\MW & -\frac{1}{s}\MW &
\MW & \frac{s}{c}\MW \\ 
\earr
\end{align}

\item
$\FS \FUbar\FU$ vertex:
Feynman rules coincide with
\refeqs{FR:SUUa} and \refeqf{FR:SUUb} 
after setting \mbox{$\xi_{W,Z}\to\xi$}.

\item
$\FShat\FShat \FUbar\FU$ vertex:
\begin{align}
&\vcenter{\hbox{\includegraphics[page=26]{FRSMBFMdiagrams.pdf}}}
= \ri e^2 C \xiQ
\\
&\barr{ccccccc} 
\FShat_1 \FShat_2\FUbar_1\FU_2
& \FHhat\FHhat\bar{u}^{\FZ}u^{\FZ} & \FHhat\FHhat\bar{u}^{\pm} u^{\pm}
,\, \phihat^+\phihat^- \bar{u}^{\pm} u^{\pm}
& \phihat^+\phihat^-\bar{u}^{\FA}u^{\FA}
& \phihat^+\phihat^-\bar{u}^{\FA}u^{\FZ}
& \phihat^+\phihat^-\bar{u}^{\FZ}u^{\FZ}  \\
& \chihat\chihat \bar{u}^{\FZ}u^{\FZ} & \chihat\chihat \bar{u}^{\pm} u^{\pm}
\hfill  & & \phihat^+\phihat^-\bar{u}^{\FZ}u^{\FA}  &
\\ \midrule 
C & -\frac{1}{2c^{2}s^2} & -\frac{1}{2s^2} &
-2 & \frac{c^2-s^2}{cs} & - \frac{(c^2-s^2)^2}{2c^2s^2}  \\ 
\earr
\notag\\[1ex]
\label{fr:Cssgg}
&\barr{cccccc} 
\FShat_1\FShat_2\FUbar_1\FU_2
& \FHhat \phihat^\pm \bar{u}^{\pm}u^{\FA}
& \chihat \phihat^\pm \bar{u}^{\pm} u^{\FA}
& \FHhat \phihat^\pm \bar{u}^{\pm}u^{\FZ}
& \chihat \phihat^\pm \bar{u}^{\pm} u^{\FZ} \\
& \phihat^\pm \FHhat \bar{u}^{\FA}u^{\mp}
& \phihat^\pm \chihat \bar{u}^{\FA} u^{\mp}
& \phihat^\pm \FHhat \bar{u}^{\FZ}u^{\mp}
& \phihat^\pm \chi \bar{u}^{\FZ} u^{\mp}  \\  \midrule 
C & \frac{1}{2s} & \mp\frac{\ri}{2s}
& \frac{1}{2c} & \mp \frac{\ri}{2c}  \\ 
\earr
\end{align}

\item
$\FShat\FS \FUbar\FU$ vertex:
\begin{align}
&\vcenter{\hbox{\includegraphics[page=27]{FRSMBFMdiagrams.pdf}}}
= \ri e^2 C \xiQ
\\
&\barr{cccccccc} 
\FShat_1 \FS_2\FUbar_1\FU_2
& \FHhat \FH \bar{u}^{\FZ}u^{\FZ} & 
\FHhat \FH \bar{u}^{\pm} u^{\pm}
& \phihat^{\pm}\phi^\mp \bar{u}^{\pm} u^{\pm}
& \phihat^{\pm}\phi^\mp\bar{u}^{\FA}u^{\FA}
& \phihat^{\pm}\phi^\mp\bar{u}^{\FA}u^{\FZ}
& \phihat^{\pm}\phi^\mp\bar{u}^{\FZ}u^{\FZ} 
& \FHhat \phi^\pm \bar{u}^{\pm}u^{\FA}
\\
& \chihat\chi \bar{u}^{\FZ}u^{\FZ} & 
\chihat\chi \bar{u}^{\pm} u^{\pm}
& & & \phihat^{\pm}\phi^\mp\bar{u}^{\FZ}u^{\FA} & 
& \phihat^\pm \FH \bar{u}^{\FA}u^{\mp}
\\ \midrule 
C & -\frac{1}{4c^{2}s^2} & -\frac{1}{4s^2} & -\frac{1}{2s^2} & -1
& \frac{c^2-s^2}{2cs} & -\frac{(c^2-s^2)^2}{4c^2s^2} 
& \frac{1}{2s}
\earr
\notag\\[1ex]
&\barr{ccccccc} 
\FShat_1 \FS_2\FUbar_1\FU_2
& \chihat \phi^\pm \bar{u}^{\pm} u^{\FA}
& \FHhat \phi^\pm \bar{u}^{\pm}u^{\FZ}
& \FHhat \phi^\pm \bar{u}^{\FZ}u^{\mp}
& \chihat \phi^\pm \bar{u}^{\pm} u^{\FZ}
& \chihat \phi^\pm \bar{u}^{\FZ} u^{\mp} 
& \FHhat\chi\bar{u}^{\pm} u^{\pm} 
\\
& \phihat^\pm \chi \bar{u}^{\FA} u^{\mp}
& \phihat^\pm \FH \bar{u}^{\FZ}u^{\mp}
& \phihat^\pm \FH \bar{u}^{\pm}u^{\FZ}
& \phihat^\pm \chi \bar{u}^{\FZ} u^{\mp}
& \phihat^\pm \chi \bar{u}^{\pm} u^{\FZ} 
& \chihat \FH\bar{u}^{\mp} u^{\mp} 
\\  \midrule 
C & 
\mp\frac{\ri}{2s}
& -\frac{c^2-s^2}{4cs^{2}} & \frac{1}{4cs^2}
& \pm \ri\frac{c^2-s^2}{4cs^2} & \mp\frac{\ri}{4cs^2}  
& \mp\frac{\ri}{4s^2} 
\earr
\end{align}

\end{myaitemize}

\end{fleqn}


\section{Green functions and their generating functionals}
\label{se:conventions}

\newcommand{\phid}{\phi^\dagger}
\newcommand{\J}{J_{\phi}}
\newcommand{\Jd}{J_{\phid}}
In order to define our conventions for Green functions we use generic
charged bosonic fields $\phi$ with adjoints $\phid$ and generic
fermionic fields $\psi$ with adjoints $\psibar$.  General Green
functions, defined via vacuum-expectation values of time-ordered
products of field operators in the canonical formalism, are denoted as
\beq
\label{appGF}
G^{\phi\phid\cdots}(x,y,\ldots) =
\langle T\phi(x)\phid(y)\cdots\rangle  =
\frac{\delta}{\ri\delta\J(x)}
\frac{\delta}{\ri\delta\Jd(y)} \cdots
T [\J,\Jd] \Big|_{\J,\Jd,\ldots \equiv 0} =
\vcenter{\hbox{
\includegraphics[page=26]{diagrams.pdf}
}},
\eeq
where $T[\J,\Jd]$ is the {\it generating functional of Green functions}
represented by the functional integral
\beq
T[\J,\Jd] = N\,\int{\cal D}\phi\int{\cal D}\phi^\dagger\,
\exp\left\{
\ri\int\rd^4x\,\left[\cL(x) + \J(x)\phi(x) + \Jd(x)\phi^\dagger(x)\right]
\right\}, \qquad
T[0,0]=1,
\eeq
and $\J$ and $\Jd$ are the sources corresponding to $\phi$ and
$\phid$, respectively.  Because the field operator $\phi$ creates
antiparticles and annihilates particles, the particles and thus the
fields in $\langle T\phi(x)\phid(y)\cdots\rangle$ have to be
considered as outgoing.  The order of the field
indices equals the order of fields in the vacuum expectation value and
the order of corresponding derivatives. 
In the diagrams, the arrows indicate the flow of the particles, which
is opposite to the flow of antiparticles.

The transformation to momentum space is defined as
\beq
\label{appFT}
G^{\phi\phid\cdots}(x,y,\ldots) =
\int\frac{\rd^4p}{(2\pi)^4}\int\frac{\rd^4q}{(2\pi)^4}\cdots\, 
\exp{\left\{\ri(px + qy + \ldots) \right\}}\, 
(2\pi)^4\deuf(p+q+\ldots) \, G^{\phi\phid\cdots}(p,q,\ldots),
\eeq
where the momenta $p,q$ are incoming and the momentum-conservation
$\de$-function is extracted from the {\it momentum-space Green
  functions}.

The {\it generating functional for connected Green functions}
$G_{\rc}^{\phi\phid\cdots}(x,y,\ldots)$ is defined as
\beq
T_{\rm c}[\J,\Jd] = \ln T[\J,\Jd].
\eeq

{\it Truncated Green functions}, which appear in the calculation of
$S$-matrix elements, are denoted as
\beq
G_\trunc^{\phid\phi\cdots}(x,y,\ldots) 
= G^{\underline{\phid}\,\underline{\phi}\cdots}(x,y,\ldots)
= 
\langle T\underline{\phi}(x)\,\underline{\phid}(y)\cdots\rangle
=
\vcenter{\hbox{
\includegraphics[page=27]{diagrams.pdf}
}}.
\eeq
Underlining is also used to indicate {truncation} of single lines,
\beq\label{eq:Aconv_trunc}
G^{\cdots\phi\cdots}(\ldots,x,\ldots) = 
\int\rd^4y \sum_{\phi'} \,G^{\phi\phi'}(x,y)\,
G^{\cdots\underline{\phi'}\cdots}(\ldots,y,\ldots),
\eeq
where the r.h.s.\ involves a sum over all fields $\phi'$ that
can mix with $\phid$.
Note that the field indices in truncated Green functions denote incoming
particles or fields, \ie $\phid$ is one of the fields $\phi'$ 
(but $\phi$ in general not).

The {\it generating functional of vertex functions}
$\Gamma[\phi,\phid]$, also known as {\it effective action}, is defined by
\begin{align}
\ri\Gamma[\phi,\phid] ={}& 
 T_{\rm c}[\J,\Jd] -\ri \int\! \rd^4x\,  [\J(x)\phi(x) + \Jd(x)\phid(x)], 
\nn\\
\phi(x) ={}& \frac{\delta T_{\rm c}[\J,\Jd]}{\ri\delta\J(x)}, \qquad
\phid(x) = \frac{\delta T_{\rm c}[\J,\Jd]}{\ri\delta\Jd(x)}, 
\end{align}
and generates the vertex functions via
\beq
\Ga^{\phi\phid\cdots}(x,y,\ldots) =
\frac{\delta}{\delta\phi(x)}
\frac{\delta}{\delta\phid(y)}\cdots
\delta\Gamma[\phi,\phid] \Big|_{\phi,\phid,\ldots \equiv 0} .
\eeq
In the vertex functions, as in truncated Green functions,
the field arguments denote incoming quanta.  The Fourier
transformation of truncated functions or vertex functions is as in
\refeq{appFT}, \ie all momenta are incoming.  

Specifically, the 2-point functions are given by
\begin{align}
G^{\phi\phid}(x,y) ={}& \langle T\phi(x)\phid(y)\rangle 
=
\frac{\delta^2 T[\J,\Jd]}{\ri\delta\J(x)\ri\delta\Jd(y)} 
\Big|_{\J,\Jd,\ldots \equiv 0} , \nl
\Ga^{\phid\phi}(x,y) ={}&
\frac{\delta^2\Gamma[\phi,\phid]}{\delta\phid(x)\delta\phi(y)} 
\Big|_{\phi,\phid,\ldots \equiv 0}
\end{align}
and obey the relations
\beq
\int\rd^4 z\, \sum_{\phi'} \ri\Ga^{\phid\phi'}(x,z) G^{\phi'\phid{}^{\prime\prime}}(z,y) =
-\deuf(x-y)\delta_{\phid{}'\phid{}^{\prime\prime}},\qquad
\sum_{\phi'} \ri\Ga^{\phid\phi'}(k,-k) G^{\phi'\phid{}^{\prime\prime}}(k,-k) = -\delta_{\phid{}'\phid{}^{\prime\prime}},
\eeq
where the sum over the intermediate fields $\phi'$ appears in case of
mixing.

For Grassmann fields extra minus signs appear. The order of labels
corresponds to the order of left derivatives, and interchange of
labels of the Green functions leads to sign changes.  The Green
functions for fermion fields $\psi$ and $\psibar$ are defined as
\begin{align}\label{eq:GF_ferm}
G^{\psi\cdots}(x,\ldots) = 
\langle T\psi(x)\cdots\rangle 
= 
+\frac{\delta}{\ri\delta J_\psi(x)} \cdots
T[J_\psi,J_{\psibar}] \Big|_{J_\psi,\ldots \equiv 0},
\notag\\
G^{\psibar\cdots}(x,\ldots) = 
\langle T \psibar(x)\cdots\rangle 
= - \frac{\delta}{\ri\delta J_{\psibar}(x)} \cdots
T[J_\psi,J_{\psibar}] \Big|_{J_{\psibar},\ldots \equiv 0},
\end{align}
\ie each derivative with respect to $J_{\psibar}$ gets a minus sign,
and the source terms to be added to the Lagrangian have the form
$J_\psi \psi + \psibar J_{\psibar}$.

Truncation of Grassman fields is defined according to
\begin{align}\label{eq:Aconv_trunc2}
G^{\cdots\psi\cdots}(\ldots,x,\ldots) ={}& 
\int\rd^4y \sum_{\psibar'} \,G^{\psi\psibar'}(x,y)\,
G^{\cdots\underline{\psibar'}\cdots}(\ldots,y,\ldots), \notag\\
G^{\cdots\psibar\cdots}(\ldots,x,\ldots) ={}& 
\int\rd^4y \sum_{\psi'} \,
G^{\cdots\underline{\psi'}\cdots}(\ldots,y,\ldots)\,
G^{\psi'\psibar}(y,x).
\end{align}

The functionals are related via
\begin{align}
\ri\Gamma[\psi,\psibar] ={}& 
 T_{\rm c}[ J_\psi,J_{\psibar}] -\ri \int\! \rd^4x\,  
[J_\psi(x)\psi(x) + \psibar(x)J_{\psibar}(x)],
\nn\\
\psi(x) ={}& \frac{\delta T_{\rm c}[J_\psi,J_{\psibar}]}{\ri\delta J_\psi(x)}, \qquad
\psibar(x) = -\frac{\delta T_{\rm c}[J_\psi,J_{\psibar}]}{\ri\delta J_{\psibar}(x)},
\end{align}
and the vertex functions are defined by 
\begin{align}
\Ga^{\psi\cdots}(x,\ldots) =
-\frac{\delta}{\delta\psi(x)} \cdots
\Gamma[\psi,\psibar] \Big|_{\psi\ldots \equiv 0},
\qquad
\Ga^{\psibar\cdots}(x,\ldots) =
+\frac{\delta}{\delta\psibar(x)} \cdots
\Gamma[\psi,\psibar] \Big|_{\psibar\ldots \equiv 0},
\end{align}
\ie each derivative with respect to $\psi$ gets a minus sign. In this
way the tree-level vertex functions correspond to the usual Feynman rules.

The fermionic 2-point functions are given by
\begin{align}
G^{\psi\psibar}(x,y) ={}& \langle T\psi(x)\psibar(y)\rangle 
=
-\frac{\delta^2T[J_\psi,J_{\psibar}]}{\ri\delta J_\psi(x)\ri\delta J_{\psibar}(y)} 
\Big|_{J_\psi,J_{\psibar},\ldots \equiv 0} , \nl
\Ga^{\psibar\psi}(x,y) ={}&
{-}\frac{\delta^2\Gamma[\psi,\psibar]}{\delta\psibar(x)\delta\psi(y)} 
\Big|_{\psi,\psibar,\ldots \equiv 0}
\end{align}
and obey the relations
\begin{alignat}{3}
\int\rd^4 z\, \sum_{\psi'} \ri\Ga^{\psibar\psi'}(x,z) G^{\psi'\psibar''}(z,y) ={}&
-\deuf(x-y)\delta_{\psibar\psibar''}, \qquad &
 \sum_{\psi'} \ri\Ga^{\psibar\psi'}(k,-k) G^{\psi'\psibar''}(k,-k) ={}& -\delta_{\psibar\psibar''},
\notag\\
\int\rd^4 z\, \sum_{\psibar'}  G^{\psi\psibar'}(x,z)
\ri\Ga^{\psibar'\psi''}(z,y) ={}& -\deuf(x-y)\delta_{\psi\psi''}, \qquad &
 \sum_{\psibar'} G^{\psi\psibar'}(k,-k) \ri\Ga^{\psibar'\psi''}(k,-k) ={}& -\delta_{\psi\psi''},
\label{eq:ferm2ptrelations}
\end{alignat}
again with a sum over $\psi'$ or $\psibar'$ in case of mixing.

With these definitions, the lowest-order 2-point functions for
fermions read
\beqar
\Ga^{\psibar\psi}_{0,\al\be}(-p,p) =
(\slashed{p}-m)_{\al\be}=-\Ga^{\psi\psibar}_{0,\be\al}(p,-p),\nl
G^{\psi\psibar}_{0,\al\be}(-p,p) =
\frac{\ri(\slashed{p}+m)_{\al\be}}{p^2-m^2-\ri\veps} = -G^{\psibar\psi}_{0,\be\al}(p,-p). 
\eeqar

The conventions for Faddeev--Popov ghosts are equivalent to those for
the fermions.


\section{Ward identity for the on-shell \texorpdfstring{\boldmath $\FA\Ffbar\Ff$}{Aff} vertex }
\label{se:charge_WI}

In this appendix we give a derivation of the Ward identity
\refeqf{eq:WIchargeSM} for the on-shell $\FA\Ffbar\Ff$ vertex in the
SM that can be used to fix the charge renormalization constant from
self-energies at the one-loop level. While this Ward identity is often
used in the literature, it has to the best of our knowledge not been
derived from the symmetries of the EWSM, but only verified via direct
calculation of the one-loop diagrams.  We recall the situation in the
BFM where the QED-like relation~\refeqf{eq:dZeBFM} automatically fixes
the charge renormalization constant in terms of the background photon
self-energy to all orders, so that this complicated derivation is not
required.

We start from the {\em Lee identities}
\cite{Lee:1973hb,Lee:1973rb,KlubergStern:1974xv} for the EWSM as given
in \citere{Bohm:2001yx} (see also \citere{Aoki:1980ix}).%
\footnote{Alternatively, we could also derive \refeq{eq:WIchargeSM}
  starting from the BRS~invariance of the Green function $\langle T
  \ubar^A(x)\psi_f(x)\psibar_{\bar f}(z)\rangle$, which produces a
  Slavnov--Taylor identity for $\langle T \partial^\mu_x
  A_\mu(x)\psi_f(x)\psibar_{\bar f}(z)\rangle$.  The derivation
  proceeds similarly to the one described in the following, involves a
  smaller number of unphysical quantities, but is more cumbersome in
  the truncation of external lines.}  The Lee identities for the
vertex functional are a consequence of the invariance of the theory
under BRS transformations \refeqf{eq:BRSphf}.  They can be derived
from the Slavnov--Taylor identities for the functional of connected
Green functions via Legendre transformation to the effective action
$\Gamma$.

For linear {\it gauge-fixing functionals} like in \refeq{eq:gf},
or in the more generic form ($a=A,Z,\pm$,
$V^\FA=\FA$, $V^\FZ=\FZ$, $V^\pm=\FW^\pm$, $\phi^\FZ=\chi$)
\begin{equation} \label{44lingauge}
C^a = \partial_\mu V^{a,\mu}(x) 
+ \sum_{b=Z,\pm} \Phi^{ab} \phi^{b}(x) ,
\end{equation}
the Lee identities read
\begin{align} \label{eq:Leeid1}
0 ={}& \int\rd^4x \,\left[
\sum_{a=A,Z,\pm}\left(
\frac{\de\tilde\Ga}{\de V^a_\mu} \frac{\de\tilde\Ga}{\de \KV^{a,\mu}}
+ \frac{\de\tilde\Ga}{\de u^a} \frac{\de\tilde\Ga}{\de \Ku^a}
\right) + \sum_{a=Z,\pm}
\frac{\de\tilde\Ga}{\de \phi^a} \frac{\de\tilde\Ga}{\de \Kphi^a}
 +\frac{\de\tilde\Ga}{\de \FH} \frac{\de\tilde\Ga}{\de \KH}
+ \sum_{i}\, \Biggl(
\frac{\de\tilde\Ga}{\de \Kpsibar^i} \frac{\de\tilde\Ga}{\de \psibar_i}
+ \frac{\de\tilde\Ga}{\de \psi_i} \frac{\de\tilde\Ga}{\de \Kpsi^i}
\Biggr)
\right]\,, \\
0 ={}& 
\partial^{\mu} \frac{\de\tilde\Ga}{\de\KV^{a,\mu}}
+ \sum_{b=Z,\pm}\Phi^{ab} \frac{\de\tilde\Ga}{\de \Kphi^b}
+ \frac{\de\tilde\Ga}{\de \ubar^a},
\label{eq:Leeid2}
\end{align}
where 
\beq \label{eq:defgatilde}
\tilde\Ga = \Ga + \int\rd^4x\, \left(
\frac{1}{2\xi_\FA}C^\FA C^\FA +
\frac{1}{2\xi_\FZ}C^\FZ C^\FZ+
\frac{1}{\xi_\FW} C^+ C^-\right)
\eeq
is the vertex functional with the gauge-fixing term~\refeqf{eq:Lfix}
subtracted, and
$i$ runs over all 
fermion fields denoted generically by $\psi_i$.
The terms involving the sources $K$ of the BRS-transformed fields are
given by
\beq
\label{eq:BRS sources}
\cL_{\mbox{\scriptsize BRS-sources}} =
\sum_{a=A,Z,\pm} \left( \KV^{a,\mu} sV^a_\mu + \Ku^a su^a\right)
 + \sum_{a=Z,\pm} \Kphi^{a} s\phi^a 
+ \KH s\FH 
+ \sum_i \Kpsi^i s\psi_i + \sum_i (s\psibar_i) \Kpsibar^i,
\eeq
and the sources for the charged gauge-boson, Faddeev--Popov ghost, and
would-be Goldstone fields are related to the sources in the canonical
$\SU(2)_\rw$ basis via $K^\pm=(K^1\pm \ri K^2)/\sqrt{2}$.
Note that the sources $\KV^{a,\mu}$, $\Kphi^{a}$, $\KH$
are Grassmann-odd functions, while 
$\Ku^a$, $\Kpsi^i$,  $\Kpsibar^i$
are Grassmann even.  If functional derivatives w.r.t.\ the $K$'s are
taken, we label the corresponding Green functions and vertex functions
as follows,
\beq
G^{K\cdots}(x,\ldots) = \frac{\de}{\ri\de K(x)} \cdots T
\big|_{K,\ldots\equiv0}, \qquad
\Ga^{K\cdots}(x,\ldots) = \frac{\de}{\de K(x)} \cdots \Ga
\big|_{K,\ldots\equiv0}.
\eeq
The sign choice for $\Ga^{K\cdots}(x,\ldots)$ ensures that the
tree-level contributions of those functions can be read from the
Lagrangian in the same way as for ordinary vertex functions, where the
tree-level parts coincide with the corresponding Feynman rules (using
the conventions of \refapp{se:conventions}).  Denoting tree-level
parts with subscript zero, we give the tree-level parts of some
$\Ga^{K\cdots}(x,\ldots)$ in momentum space for later convenience,
\begin{align}
\label{eq:Gamma0Ku}
& \tilde\Gamma^{\KV^A u^A}_{0,\mu}(-k,k) = -\ri k_\mu, \qquad
\tilde\Gamma^{\KV^Z u^A}_{0,\mu}(-k,k) = 0, \qquad
\tilde\Gamma^{\Kchi u^A}_0(-k,k) = \tilde\Gamma^{\KH u^A}_0(-k,k) = 0,
\nl
& \tilde\Gamma^{\psibar_l\Kpsibar^k u^A}_0(\pbar,p,k) =
\tilde\Gamma^{\Kpsi^l \psi_k u^A}_0(\pbar,p,k) = -\ri Q_l e \de_{lk}, \qquad
\end{align}
which simply follow from inserting the respective BRS variations
\refeq{eq:BRSphf} into \refeq{eq:BRS sources}.

We start by taking functional derivatives of the Lee identity
\refeqf{eq:Leeid1} with respect to $\psibar_k$, $\psi_k$, and $u^A$,
where $\psibar_k$ and $\psi_k$ correspond to the same fermion species.
After transformation to momentum space, we obtain
\begin{align}\label{eq:AffWImom}
0={}& \sum_{a=A,Z}\tilde\Gamma^{V^a\psibar_k\psi_k}_\mu(k,\pbar,p) \,
     \tilde\Gamma^{\KV^a u^A,\mu}(-k,k) 
+  \sum_{S=\chi,\FH} \tilde\Gamma^{S\psibar_k\psi_k}(k,\pbar,p) \,
     \tilde\Gamma^{K_S u^A}(-k,k)
\notag\\&{}
-  \sum_i \tilde\Gamma^{\psibar_k\Kpsibar^i u^A}(\pbar,p,k)\,
    \tilde\Gamma^{\psibar_i\psi_k}(-p,p)
+ \sum_i \tilde\Gamma^{\psibar_k\psi_i}(\pbar,-\pbar) \,
     \tilde\Gamma^{\Kpsi^i \psi_k u^A}(\pbar,p,k) ,
\end{align}
where $\Kchi=K_\phi^\FZ$ and all quantities
$\tilde\Gamma^{\ldots}(\{p_i\})$ represent bare vertex functions
derived from the functional \refeqf{eq:defgatilde}, transformed to
momentum space with momentum-conservation $\de$-functions split off,
and with incoming momenta and labels denoting incoming fields.  Since
$\tilde\Gamma^{\ldots}=\Gamma^{\ldots}$ for vertex functions involving
fermion fields or sources $K$, we omit the tilde in the following
whenever possible.  Note that we do not get terms involving the
tadpole term $\Gamma^\FH(0)$ by definition, since we include tadpole
counterterms, which render $\Gamma^\FH(0)$ zero, in our definition of
unrenormalized vertex functions, as made explicit in
\refeq{eq:definition_se} for self-energies.  
The tadpole counterterms can be generated via a shift of the Higgs
field, which does not affect the Lee identities.
In \citeres{Aoki:1980ix,Bohm:2001yx} the identity \refeqf{eq:AffWImom}
is used for the renormalized vertex functions in order to proof charge
universality in the EWSM, \ie the fact that the (renormalized)
on-shell coupling of the photon to charged particles is independent of
the particle species. Here we use this identity for bare vertex
functions in order to derive an identity between the charge
renormalization constant and self-energies.
 
We insert the covariant decompositions 
\begin{equation}\label{eq:KDKu}
\Gamma^{\KV^a u^b}_\mu(-k,k) = -k_\mu \Gamma^{\KV^a u^b}(k^2), \qquad 
\Gamma^{K_S u^b}(-k,k) = \Gamma^{K_S u^b}(k^2), \qquad
S=\chi,\FH,
\end{equation}
take the derivative of \refeq{eq:AffWImom} with respect to $k^\mu$ for $p^\mu$
fixed and $\pbar^\mu=-p^\mu - k^\mu$, put $k_\mu=0$, sandwich the result 
between spinors $\ubar_k(p)\ldots u_k(p)$, and obtain:
\begin{align}\label{eq:AffWI}
0={}& {-\sum_{a=A,Z}} \ubar_k(p)\,\Gamma^{V^a\psibar_k\psi_k}_\mu(0,-p,p)\, u_k(p) \,
     \Gamma^{\KV^a u^A}(0) 
\nl&{}
+  \sum_{S=\chi,\FH} \ubar_k(p)\left[\frac{\partial}{\partial k^\mu} 
\Gamma^{S\psibar_k\psi_k}(k,-p-k,p)\right]_{k=0} u_k(p) \, \Gamma^{K_S u^A}(0)
\notag\\&{}
-  \sum_i \ubar_k(p)\left[\frac{\partial}{\partial k^\mu}\Gamma^{\psibar_k\Kpsibar^i u^A}(-p-k,p,k)\right]_{k=0}
    \Gamma^{\psibar_i\psi_k}(-p,p) \, u_k(p)
\notag\\&{}
+ \sum_i \ubar_k(p)\,\Gamma^{\psibar_k\psi_i}(-p,p)\, 
     \left[\frac{\partial}{\partial k^\mu}\Gamma^{\Kpsi^i \psi_k
         u^A}(-p-k,p,k)\right]_{k=0} u_k(p) 
\notag\\&{}
+ \sum_i \ubar_k(p)\left[\frac{\partial}{\partial p^\mu}\Gamma^{\psibar_k\psi_i}(-p,p)\right] 
     \Gamma^{\Kpsi^i \psi_k u^A}(-p,p,0)\,u_k(p). 
\end{align}
We have used the fact that the derivative of the functions
$\Gamma^{K_S u^A}(k^2)$ with respect to $k_\mu$ vanishes for
$k_\mu=0$.  In the following we evaluate \refeq{eq:AffWI} in one-loop
approximation line by line:
\begin{itemize}
\item
Defining the one-loop correction $\Lambda^{A\psibar_k\psi_k}$
to the $A\psibar_k\psi_k$ vertex as in \refeq{eq:RCE1},
the one-loop approximation of the first line of \refeq{eq:AffWI} reads
\begin{align}
\label{eq:affWIline1prelim}
& \ubar_k(p)\left[ 
eQ_k \ga_\mu \, \Gamma^{\KV^A u^A}(0) 
- e\Lambda^{A\psibar_k\psi_k}_\mu(0,-p,p)  \, \Gamma^{\KV^A u^A}_0(0) 
- e\ga_\mu(\varv_k-a_k\gamma_5)  \, \Gamma^{\KV^Z u^A}(0) 
\right] u_k(p)
\\
& {}= e \ubar_k(p)\left[ 
- \Lambda^{A\psibar_k\psi_k}_\mu(0,-p,p)  \, \Gamma^{\KV^A u^A}_0(0) 
+ Q_k \ga_\mu \, 
  \left(\Gamma^{\KV^A u^A}(0)+\frac{\sw}{\cw}\Gamma^{\KV^Z u^A}(0)\right) 
- 2a_k \ga_\mu \omega_-  \, \Gamma^{\KV^Z u^A}(0) 
\right] u_k(p),
\nn
\end{align}
where we have inserted the tree-level $V^a\psibar_k\psi_k$ couplings
and employed the vector and axial-vector couplings $\varv_k$ and $a_k$
defined in \refeq{vfaf}.  This can be further simplified by using
$\Gamma^{\KV^A u^A}_0(0)=\ri$ from \refeq{eq:Gamma0Ku} and by
exploiting the all-order relation
\beq
\label{eq:GammaKBuA}
\cw\Gamma^{\KV^A u^A}(k^2)+\sw\Gamma^{\KV^Z u^A}(k^2)=\ri\cw.
\eeq
This identity follows from the fact that $B_\mu$ is an abelian gauge field:
Using $sB_{\mu} = \partial_{\mu} (\cw u^{\FA}+ \sw u^{\FZ})$,
which can be read from \refeq{eq:BRSphf}, we can write
\beq
G^{K_V^{B}\ubar^a,\mu}(x,y) = \langle T\,sB^\mu(x)\,\ubar^a(y)\rangle
= \partial^{\mu}_x \left( \cw G^{u^A\ubar^a}(x,y)
+ \sw G^{u^Z\ubar^a}(x,y) \right).
\eeq
Truncating the outgoing antighost field $\ubar^a$
and going over to momentum space,
\beq
\Ga^{K_V^{B} u^b}_\mu(-k,k) =
-\sum_a G^{K_V^{B}\ubar^a}_\mu(-k,k) \,\ri\Ga^{\ubar^a u^b}(-k,k),
\eeq
with the help of \refeq{eq:ferm2ptrelations} directly produces
\refeq{eq:GammaKBuA}.  Inserting \refeq{eq:GammaKBuA} into
\refeq{eq:affWIline1prelim}, leads to
\beq
e \ubar_k(p)\left[ 
-\ri \Lambda^{A\psibar_k\psi_k}_\mu(0,-p,p)
+\ri Q_k \ga_\mu \, 
- 2a_k \ga_\mu \omega_-  \, \Gamma^{\KV^Z u^A}(0) 
\right] u_k(p).
\label{eq:affWIline1}
\eeq
The last term with $\Gamma^{\KV^Z u^A}(0)$ will be combined with
other terms below.
\item The second line of \refeq{eq:AffWI} does not contribute in
  one-loop approximation. According to \refeq{eq:Gamma0Ku},
  $\Gamma^{K_S u^A}(0)$ is of one-loop order, so that
  $\Gamma^{S\psibar_k\psi_k}$ is only needed at tree level.  Since the
  tree-level part $\Gamma^{S\psibar_k\psi_k}_0$ is constant, the whole
  contribution is at least of two-loop order.
\item
Lines three and four of \refeq{eq:AffWI} do not contribute at one loop
either, because the fermionic 2-point functions project the
external wave function $u_k(p)$ and $\ubar_k(p)$ to zero at tree level,
\beq
\Gamma^{\psibar_i\psi_k}_0(-p,p) u_k(p)=0, \qquad
\ubar_k(p)\,\Gamma^{\psibar_k\psi_i}_0(-p,p)=0.
\label{eq:Gamma0ffu}
\eeq
Since the tree-level parts $\Gamma^{\psibar_k\Kpsibar^i u^A}_0$ and
$\Gamma^{\Kpsi^i \psi_k u^A}_0$ are constant according to
\refeq{eq:Gamma0Ku}, the whole contributions are again zero at one loop.
\item
The fifth and last line of \refeq{eq:AffWI} produces a non-trivial
contribution. In order to derive it, we go back to \refeq{eq:AffWImom},
set $k=0$, $p^2=m_k^2$, and multiply it with $u_k(p)$ from the right.
Following the same reasoning as in the previous items and splitting
the fermionic 2-point functions into tree-level and one-loop
parts,
\beq
\Gamma^{\psibar_i\psi_j}(-p,p) 
= \Gamma^{\psibar_i\psi_j}_0(-p,p) + \Sigma^{\psibar_i\psi_j}(-p,p)
= (\slashed{p}-m_i)\de_{ij} + \Sigma^{\psibar_i\psi_j}(-p,p),
\eeq
at one loop this leads to 
\begin{align}
\label{eq:GammaKpsipsiuA1}
0={}& \Gamma^{\chi\psibar_k\psi_k}_0  \,u_k(p)\, \Gamma^{\Kchi u^A}(0)
+ \Gamma^{\FH\psibar_k\psi_k}_0 \,u_k(p)\, \Gamma^{\KH u^A}(0) 
\notag\\&{}
-  \sum_i \Gamma^{\psibar_k\Kpsibar^i u^A}(-p,p,0)\,
    \Gamma^{\psibar_i\psi_k}_0(-p,p) \,u_k(p)
+ \sum_i \Gamma^{\psibar_k\psi_i}_0(-p,p) \,
     \Gamma^{\Kpsi^i \psi_k u^A}(-p,p,0) \,u_k(p)
\notag\\&{}
-  \sum_i \Gamma^{\psibar_k\Kpsibar^i u^A}_0\,
    \Sigma^{\psibar_i\psi_k}(-p,p) \,u_k(p)
+ \sum_i \Sigma^{\psibar_k\psi_i}(-p,p) \,
     \Gamma^{\Kpsi^i \psi_k u^A}_0 \,u_k(p).
\end{align}
In this relation, the last two terms trivially cancel each other,
and the third term is zero according to \refeq{eq:Gamma0ffu}.
The second is zero as well, since it is the only term left after
multiplying \refeq{eq:GammaKpsipsiuA1} with $\ubar_k(p)$ from the
left and using $\ubar_k(p)\gamma_5 u_k(p)=0$.
Since $\ubar_k(p)\Gamma^{\FH\psibar_k\psi_k}_0 u_k(p)\ne0$, we get
\beq
\Gamma^{\KH u^A}(0) = 0.
\label{eq:GammaKHu}
\eeq
With these simplifications and
inserting the explicit form of 
\beq
\Gamma^{\chi\psibar_k\psi_k}_0 =
2\ri e a_k \frac{m_k}{\MZ} \gamma_5,
\eeq
\refeq{eq:GammaKpsipsiuA1} reads
\beq
\label{eq:GammaKpsipsiuA2}
0= 2\ri e a_k \frac{m_k}{\MZ} 
\,\gamma_5\,u_k(p) \, \Gamma^{\Kchi u^A}(0) 
+ (\slashed{p}-m_k) \,
     \Gamma^{\Kpsi^k \psi_k u^A}(-p,p,0) \,u_k(p).
\eeq
The function $\Gamma^{\Kpsi^k \psi_k u^A}(\pbar,p,k)$
receives one-loop contributions from the graphs 
shown in \reffi{fi:chargegraphs}.
\bfi
\centerline{
\includegraphics[page=18]{diagrams.pdf}\qquad 
\includegraphics[page=19]{diagrams.pdf}}
\caption{One-loop Feynman diagrams contributing to $\Gamma^{\Kpsi^i \psi_k u^A}$.}
\label{fi:chargegraphs}
\efi
Inspecting the BRS variations of fermions given in \refeq{eq:BRSphf},
we see that the graphs in \reffi{fi:chargegraphs} all involve
a chirality projection factor $\omega_-$ on the left, so that
Lorentz invariance admits us to write
\beq
\Gamma^{\Kpsi^k \psi_k u^A}(-p,p,0) \,u_k(p)= 
\left[ -\ri Q_k e
+ \omega_- \ri e \Lambda^{\Kpsi^k \psi_k u^A} \right] u_k(p)
\eeq
where $\Lambda^{\Kpsi^k \psi_k u^A}$ is a scalar constant,
since $\Gamma^{\Kpsi^k \psi_k u^A}(-p,p,0)$ can only depend on
$\slashed{p}$ and is projected on the spinor. 
Inserting this into \refeq{eq:GammaKpsipsiuA2}, we can calculate
$\Lambda^{\Kpsi^k \psi_k u^A}$ to
\beq
\Lambda^{\Kpsi^k \psi_k u^A}
= - \frac{2a_k}{\MZ} \, \Gamma^{\Kchi u^A}(0).
\eeq
With this result we can evaluate the last line of \refeq{eq:AffWI} to
\begin{align}
&  \ubar_k(p)\,\gamma_\mu \Gamma^{\Kpsi^k \psi_k u^A}(-p,p,0)\,u_k(p)
+ \sum_i \ubar_k(p)\left[\frac{\partial}{\partial p^\mu}\Sigma^{\psibar_k\psi_i}(-p,p)\right] 
     \Gamma^{\Kpsi^i \psi_k u^A}_0 \,u_k(p)
\nl
&{}=  -\ri e\,\ubar_k(p)
\left[ Q_k \,\gamma_\mu
+ \frac{2a_k}{\MZ} \, \Gamma^{\Kchi u^A}(0) \,\gamma_\mu\omega_- 
+ Q_k \left(
\frac{\partial}{\partial p^\mu}\Sigma^{\psibar_k\psi_k}(-p,p)\right)\right] u_k(p).
\label{eq:affWIline6}
\end{align}
\end{itemize}
In summary, the r.h.s.\ of \refeq{eq:AffWI} is given by the
sum of \refeqs{eq:affWIline1} and \refeqf{eq:affWIline6}, so that
the full identity reads
\beq
0 =\ri e \,\ubar_k(p)\left[ 
-\Lambda^{A\psibar_k\psi_k}_\mu(0,-p,p)
+ \frac{2a_k}{\MZ} \left(\ri\MZ \, \Gamma^{\KV^Z u^A}(0) 
- \Gamma^{\Kchi u^A}(0) 
\right)\gamma_\mu\omega_- 
- Q_k \left(\frac{\partial}{\partial p^\mu}\Sigma^{\psibar_k\psi_k}(-p,p)\right)
\right] u_k(p).
\label{eq:AffWIprelim}
\eeq
In a final non-trivial step, we express the combination of $\Gamma^{K
  u^A}$ vertex functions in terms of gauge-boson 2-point functions. To
this end, we take functional derivatives of \refeq{eq:Leeid1} w.r.t.\ 
a neutral gauge-boson field $V^b_\nu$ and a neutral ghost field $u^c$
to obtain
\beq
0 = 
\sum_{a=A,Z}
\tilde\Ga^{V^a V^b}_{\mu\nu}(k,-k) \, 
\Ga^{\KV^a u^c,\mu}(-k,k)
+\tilde\Ga^{\chi V^b}_\nu(k,-k) \, \Ga^{\Kchi u^c}(-k,k).
\eeq
Note that we have to keep the tilde on the 2-point
vertex functions without $K$~fields.
Specializing these relations to $b=Z$ and $c=A$ and introducing the
covariant decompositions \refeqf{eq:Lorentz_SE}, \refeqf{eq:GWphi},
and \refeqf{eq:KDKu}, leads to the one-loop relation
\beq
0 = \Ga^{AZ}_{\rL}(k^2) \, \Ga^{\KV^A u^A}_0(k^2)
+ \tilde\Ga^{ZZ}_{\rL,0}(k^2) \, \Ga^{\KV^Z u^A}(k^2)
+ \tilde\Ga^{\chi Z}_0(k^2) \, \Ga^{\Kchi u^A}(k^2),
\eeq
which can be further simplified with the tree-level results
\beq
\Ga^{\KV^A u^A}_0(k^2) = \ri, \qquad
\tilde\Ga^{ZZ}_{\rL,0}(k^2) = \MZ^2, \qquad
\tilde\Ga^{\chi Z}_0(k^2) = \ri\MZ.
\eeq
With the identification $\Ga^{AZ}_{\rL}(k^2)=-\Sigma^{AZ}_{\rL}(k^2)$,
we finally get
\beq
\Sigma^{AZ}_{\rL}(k^2) 
= \MZ\left(-\ri\MZ \Ga^{\KV^Z u^A}(k^2)
+  \Ga^{\Kchi u^A}(k^2)\right).
\eeq
For $k^2=0$ this relation, together with the identity
$\Sigma^{AZ}_{\rL}(0)=\Sigma^{AZ}_{\rT}(0)$ resulting from the
analyticity of $\Gamma^{AZ}(k,-k)$ at $k=0$, 
can be used to bring \refeq{eq:AffWIprelim}
into its final form
\beq
0 =\ubar_k(p)\left[ 
\Lambda^{A\psibar_k\psi_k}_\mu(0,-p,p)
+ \frac{2a_k}{\MZ^2} \Sigma^{AZ}_{\rT}(0)
\gamma_\mu\omega_- 
+ Q_k \left(\frac{\partial}{\partial p^\mu}\Sigma^{\psibar_k\psi_k}(-p,p)\right)
\right] u_k(p),
\label{eq:AffWIres}
\eeq
which is the Ward identity quoted in \refeq{eq:WIchargeSM}.

Upon inserting the Lorentz decompositions for the fermion self-energy
defined as in \refeq{eq:Lorentz_SE} and for the vertex function
\refeq{eq:RCE2} and using the Gordon identities and the definition
\refeqf{eq:CTF} of the fermion field renormalization constants 
$\delta Z^{\psi,\si}_{kk}$ in the OS scheme, this gives rise to the relations
\cite{Denner:1991kt,Bohm:2001yx}:
\begin{align}
0 ={}& \La_{\rV}(0)+\La_{\rS}(0)+\frac{a_k}{\MZ^2}\Si^{\FA\FZ}_{\rT}(0) 
-\frac{1}{2}Q_k\Bigl(\delta Z^{\psi,\rL}_{kk}+\delta Z^{\psi,\rR}_{kk}\Bigr),\notag\\
0 ={}& \La_{\rA}(0)+\frac{a_k}{\MZ^2}\Si^{\FA\FZ}_{\rT}(0) 
-\frac{1}{2}Q_k\left(\delta Z^{\psi,\rL}_{kk}-\delta Z^{\psi,\rR}_{kk}\right),
\label{eq:AffWIres2}
\end{align}
which link the vertex form factors at zero momentum transfer to
quantities defined in terms of self-energies.

For the derivation of the identities~\refeqf{eq:AffWIres} and
\refeqf{eq:AffWIres2} we just have used the Lee identities of the
$\SU(2)_\rw\times\U(1)_Y$ gauge symmetry within a general 't~Hooft
gauge and Lorentz invariance.  The arguments do not change if
additional fermions or scalars are added to the SM.  As long as the
gauge sector is not modified, the identity \refeqf{eq:AffWIres} and
thus the relation \refeqf{eq:DZE} between the renormalization of the
electromagnetic coupling and the gauge-boson self-energies in
\refse{se453eforc} are valid.

\section*{Acknowledgements}
\label{se:ackn}
\addcontentsline{toc}{section}{\nameref{se:ackn}}

AD acknowledges financial support by the German Federal Ministry for
Education and Research (BMBF) under contracts no.~05H15WWCA1 and
05H18WWCA1 and the German Research Foundation (DFG) under reference
numbers DE~623/5-1 and DE~623/6-1.  SD gratefully acknowledges support
by the BMBF under contracts no.~05H15VFCA1 and 05H18VFCA1, by the DFG
under reference numbers DI~784/3-2 and DI~784/4-1, and by the DFG
Research Training Group GRK 2044.  In the course of writing this
review, we have profited from many stimulating discussions and
fruitful collaborations with Th.~Hahn, V.~Hirschi, W.~Hollik, A.~Huss, 
A.~Kabelschacht, J.N.~Lang,
A.~M\"uck, G.~Passarino, S.~Pozzorini, Ch.~Schwinn, H.~Spiesberger, 
S.~Uccirati, and G.~Weiglein.
We thank Th.~Hahn, G.~Heinrich, V.~Hirschi, 
Ph.~Maierh\"ofer, R.~Pittau, and
S.~Pozzorini for feedback on a draft version of parts of \refse{se:virt}.


\addcontentsline{toc}{section}{References}


\end{document}